\documentclass[11pt,twoside,openany]{book}
\newcommand{\KBCPaperWidth}{8.25in}
\newcommand{\KBCPaperHeight}{11in}
\newcommand{\KBCInnerMargin}{0.64in}
\newcommand{\KBCOuterMargin}{0.64in}
\newcommand{\KBCTopMargin}{0.50in}
\newcommand{\KBCBottomMargin}{0.65in}
\usepackage[
  paperwidth=\KBCPaperWidth,
  paperheight=\KBCPaperHeight,
  inner=\KBCInnerMargin,
  outer=\KBCOuterMargin,
  top=\KBCTopMargin,
  bottom=\KBCBottomMargin,
  includehead,
  includefoot,
  heightrounded
]{geometry}

\usepackage{setspace}
\usepackage{csquotes}
\usepackage[backend=biber,style=apa,natbib=true]{biblatex}
\usepackage{amsmath,amssymb,mathtools}
\numberwithin{equation}{chapter}

\usepackage{booktabs,longtable,array,tabularx,ltablex}
\usepackage{lscape}
\usepackage{caption}
\usepackage{float}
\usepackage{enumitem}
\usepackage{tocloft}
\usepackage{xurl}
\usepackage[draft,bookmarks=false]{hyperref}
\usepackage{xcolor}
\usepackage{makeidx}
\makeindex
\usepackage{graphicx}
\usepackage{pdfpages}
\usepackage{titlesec}
\usepackage{fancyhdr}
\usepackage{etoolbox}
\usepackage{ragged2e}
\usepackage{newunicodechar}
\keepXColumns
\providecommand{\tightlist}{\setlength{\itemsep}{0pt}\setlength{\parskip}{0pt}}

\newunicodechar{≈}{\ensuremath{\approx}}
\newunicodechar{≤}{\ensuremath{\leq}}
\newunicodechar{≥}{\ensuremath{\geq}}
\newunicodechar{⊂}{\ensuremath{\subset}}
\newunicodechar{∈}{\ensuremath{\in}}
\newunicodechar{Σ}{\ensuremath{\Sigma}}
\newunicodechar{Δ}{\ensuremath{\Delta}}
\newunicodechar{Γ}{\ensuremath{\Gamma}}
\newunicodechar{φ}{\ensuremath{\phi}}
\newunicodechar{τ}{\ensuremath{\tau}}
\newunicodechar{ω}{\ensuremath{\omega}}
\newunicodechar{χ}{\ensuremath{\chi}}
\newunicodechar{ρ}{\ensuremath{\rho}}
\newunicodechar{·}{\ensuremath{\cdot}}
\newunicodechar{−}{\ensuremath{-}}
\newunicodechar{—}{, }
\newunicodechar{–}{--}
\newunicodechar{σ}{\ensuremath{\sigma}}
\newunicodechar{δ}{\ensuremath{\delta}}
\newunicodechar{γ}{\ensuremath{\gamma}}
\newunicodechar{λ}{\ensuremath{\lambda}}
\newunicodechar{η}{\ensuremath{\eta}}
\newunicodechar{μ}{\ensuremath{\mu}}
\newunicodechar{π}{\ensuremath{\pi}}
\newunicodechar{₀}{\ensuremath{_0}}
\newunicodechar{₁}{\ensuremath{_1}}
\newunicodechar{ℓ}{\ensuremath{\ell}}
\newunicodechar{≡}{\ensuremath{\equiv}}
\newunicodechar{⟹}{\ensuremath{\Longrightarrow}}
\newunicodechar{⟺}{\ensuremath{\Longleftrightarrow}}

\AtBeginEnvironment{longtable}{\footnotesize\setlength{\tabcolsep}{3pt}}

\usepackage{fancyvrb}
\DefineVerbatimEnvironment{Highlighting}{Verbatim}{commandchars=\\\{\}}

\newunicodechar{Θ}{\ensuremath{\Theta}}
\newunicodechar{ψ}{\ensuremath{\psi}}
\newunicodechar{∞}{\ensuremath{\infty}}
\newunicodechar{∫}{\ensuremath{\int}}
\newunicodechar{∀}{\ensuremath{\forall}}
\newunicodechar{∉}{\ensuremath{\notin}}
\newunicodechar{≠}{\ensuremath{\neq}}
\newunicodechar{→}{\ensuremath{\to}}
\newunicodechar{∑}{\ensuremath{\sum}}
\newunicodechar{≪}{\ensuremath{\ll}}
\newunicodechar{∃}{\ensuremath{\exists}}

\newunicodechar{∧}{\ensuremath{\wedge}}

\newunicodechar{∅}{\ensuremath{\emptyset}}

\newunicodechar{≻}{\ensuremath{\succ}}

\newunicodechar{ϕ}{\ensuremath{\phi}}

\newunicodechar{∂}{\ensuremath{\partial}}
\newunicodechar{̂}{\ensuremath{\hat{}}}
\newunicodechar{̃}{\ensuremath{\tilde{}}}
\newunicodechar{κ}{\ensuremath{\kappa}}
\newunicodechar{ξ}{\ensuremath{\xi}}
\newunicodechar{β}{\ensuremath{\beta}}
\newunicodechar{α}{\ensuremath{\alpha}}
\newunicodechar{ε}{\ensuremath{\varepsilon}}
\newunicodechar{′}{\ensuremath{\prime}}

\usepackage{ragged2e}
\usepackage{seqsplit}
\newcolumntype{L}[1]{>{\RaggedRight\hspace{0pt}\arraybackslash}p{#1}}
\newcolumntype{C}[1]{>{\Centering\hspace{0pt}\arraybackslash}p{#1}}
\newcolumntype{R}[1]{>{\RaggedLeft\arraybackslash}p{#1}}

\newcommand{\chapterhook}[1]{%
\begin{center}
{\large\itshape #1}
\end{center}
\vspace{0.6em}
}

\providecommand{\KPot}{\ensuremath{K^{\mathrm{Pot}}}}

\providecommand{\KYield}{\ensuremath{K^{\mathrm{Yield}}}}

\title{A Knowledge Theory of Capital:\\
\large The Value of Natural and Artificial Intelligence}
\author{Jeff Gardiner, DBA, MBA\\\normalsize Morgan Stanley}
\date{}

\begin{document}
\frontmatter
\clearpage
\thispagestyle{empty}
\thispdfpagelabel{copyright}
\phantomsection
\label{copyright}
\vspace*{0.36\textheight}
\begin{center}
{\small Copyright \textcopyright{} 2026 Jeffrey Gardiner. All rights reserved.}

\vspace{1.5em}
{\small A Knowledge Theory of Capital: The Value of Natural and Artificial Intelligence, Volume 1.}

\vspace{1.25em}
{\small ISBN: 979-8181342408}

\vspace{1.25em}
{\small DOI: \url{https://doi.org/10.48550/arXiv.2606.18288}}

\vspace{1.25em}
{\small Institutional association: Morgan Stanley.}

\vspace{1.25em}
{\footnotesize This work is written in the author's individual capacity. The views expressed are the author's own and do not represent Morgan Stanley or any affiliated entity.}

\vspace{3em}
{\footnotesize\itshape Cover motif: a recombination field rendered as an orbital system, representing productive knowledge as structured, mobile, relational, and governed.}
\end{center}
\clearpage

\pagenumbering{roman}
\setcounter{page}{1}
{\let\cleardoublepage\clearpage\maketitle}
\clearpage
\chapter*{Abstract}
\label{front:abstract}

Modern firms can command large market valuations from productive capacities that financial statements often record only weakly or not at all: software, datasets, model weights, user feedback, engineering routines, brands, institutional trust, platform access, tacit expertise, and public knowledge infrastructure. These capacities also explain several practical puzzles that now confront economists, investors, managers, and policy makers. Why can asset-light firms\index{asset-light firms} possess durable productive power? Why can an AI system improve faster when it controls the feedback generated by users? Why can an API closure, platform dependency, patent thicket, software lock-in, cybersecurity failure, or open-source maintainer shock reduce future productivity even when no factory, inventory, or physical asset has been destroyed? Why can a knowledge base lose reliability as it grows, when generation increasingly draws on unvalidated machine output? Why do productivity and accounting measures often miss the stock on which modern production increasingly depends?

This book begins with Adam Smith because his theory supplied the classical operating model of capitalist wealth: labour, stock, specialization, exchange, market extent, wages, profit, rent, and national wealth. Smith's model remains powerful. It explains why accumulated stock enables specialization, why specialization raises productive power, and why wider markets support deeper division of labour. But Smith wrote before knowledge became a dominant, separately governable, scalable, and measurable source of firm value. Long-run global income data also place him at the conceptual threshold of the modern growth regime: before the Smithian-industrial threshold, global GDP per capita was nearly flat for centuries; after 1820, it enters sustained compounding growth. Smith did not cause the break. He articulated mechanisms already emerging in Britain and gave later economists, policymakers, merchants, and reformers a language for specialization, capital accumulation, market extent, productivity, and exchange. His framework was built around capital that could be materially held, separated from workers, counted as stock, deployed in production, and linked to vendible output. Those assumptions strain when productive capacity resides in persons, organizations, software, data, platforms, models, professional standards, open-source communities, and public epistemic infrastructure.

The central claim is not that earlier economics ignored knowledge. Human capital theory, information economics, endogenous growth theory, resource-based theory, platform economics, commons theory, and intangible-capital research each explain important parts of the problem. The claim is that knowledge now requires an integrated capital theory. Productive knowledge can be embodied in people, disembodied into artefacts, institutionalized in organizations, sustained in commons, or maintained as public infrastructure. It can move among these forms. Each movement changes who can access it, who can exclude others from it, who can recombine it, who captures its yield, and whether future productive capacity expands or contracts.

This book develops this account as a conditional theory of knowledge-bearing capitalism (KBC). Its core mechanism is knowledge circulation: knowledge is generated, converted into governable stock, deployed, improved through feedback, enclosed or shared through institutions, and then used as an input to further knowledge generation. This theory separates generation from conversion. Generation explains how new productive knowledge arises through recombination, experimentation, discovery, invention, interpretation, and feedback. Conversion explains what happens after knowledge exists: how it becomes codified, institutionalized, accessed, restricted, priced, appropriated, or lost. The key institutional moment is first conversion, when newly generated knowledge first becomes a private asset, firm capability, public good, professional standard, platform dependency, or commons.

The main dynamic risk is that private accumulation can increase present productive power while narrowing the conditions for future generation. Intellectual property, trade secrecy, platform gating, API restriction, model-weight enclosure, feedback capture, and proprietary standards may be justified when they create incentives, quality control, and investment. Here, \emph{enclosure} means legal, contractual, technical, architectural, or capability-based restriction of access to knowledge-bearing stock; Chapter~\ref{ch:knowledge-bearing-stock} gives the formal definition. Enclosure can restrict two conditions on which future knowledge production depends: the recombination field, meaning the accessible set of knowledge inputs that other actors can combine into new productive knowledge, and the learning loop, meaning the feedback from use that improves later models, routines, and capabilities. In such cases, enclosure can increase the incumbent's private return while reducing the wider system's ability to generate alternative technologies, business models, research paths, and institutional capabilities. This book calls that pattern the Smithian inversion: the accumulation of knowledge-bearing capital that strengthens present production partly by weakening the future capacity to generate new productive knowledge.

The empirical status of this theory is limited and explicit. The formal results are conditional, not yet empirically estimated. This book states its primitives, separates established claims from extensions and conjectures, and identifies where measurement, calibration, and falsification are still required. Its claims would be weakened if knowledge enclosure did not measurably reduce recombination opportunities, if feedback exclusion did not create capability divergence, if open-source and public knowledge infrastructures were not material inputs to private productivity, or if market-to-book gaps and productivity shocks were adequately explained without knowledge-bearing stock, governance position, or dark-capital exposure. Capability divergence means that an actor controlling a feedback loop may improve faster while excluded actors improve more slowly or not at all. If two AI firms begin with similar models, but one captures user corrections, prompts, ratings, failures, and edge cases while the other lacks comparable feedback, the first may develop better training data, error detection, product tuning, and deployment routines. If controlled feedback loops do not produce such learning advantages, the feedback-capture claim is overstated.

The practical contribution is a framework for analysing the parts of modern wealth that conventional categories handle poorly. It provides a vocabulary for knowledge-bearing stock, the operative knowledge unit, conditional separability, first conversion, cognitive enclosure, feedback capture, knowledge impairment, and dark capital. It also develops a measurement approach that separates current use value, recombination option value, feedback-learning value, control value, and strategic option value, less expected knowledge loss. It treats measurement itself as a value-of-information problem, measuring where the expected decision improvement exceeds the cost. The purpose is not to replace existing valuation, accounting, growth, or competition models. It is to make visible the productive stock, governance conditions, and loss mechanisms those models often leave implicit.

The final argument is that the wealth of nations in a knowledge-bearing economy cannot be understood only by counting output, labour, physical capital, recognized intangible assets, or market exchange. It also depends on how societies generate, preserve, convert, govern, measure, and sometimes damage the knowledge-bearing stocks that make future production possible. The central economic question is therefore not only how capital accumulates. It is how productive knowledge becomes capital-like, who governs that conversion, who captures its yield, who bears its loss, and whether present enclosure widens or narrows the future wealth of nations.

\clearpage
\chapter*{How to Read This Book}
\label{front:how-to-read}
\addcontentsline{toc}{chapter}{How to Read This Book}

This book is a theoretical architecture and falsification programme for knowledge-bearing capitalism. It is not a completed empirical model, a finished accounting standard, or a calibrated welfare system. Its purpose is to define the economic objects, mechanisms, governance forms, measurement problems, and testable claims that arise when productive knowledge becomes stock-like, mobile across persons and artefacts, recombinable, governable, and imperfectly visible in existing accounts.

The governing spine of the argument is simple: knowledge-bearing capitalism creates wealth through the generation, first conversion, governance, deployment, feedback, and renewed future generation of productive knowledge-bearing stock; measurement makes that circulation visible, testable, and governable. Every major concept in this volume should serve that circulation. A concept earns its place when it explains how productive knowledge is generated, first converted into governable stock, placed under a governance form, deployed, improved or restricted through feedback, returned to future-generation conditions, measured, or lost. Material that mainly supports notation, proof, calibration, status discipline, or extended auditability belongs in the Technical Companion rather than in the main argument.

For that reason, the concepts in this volume are presented in three tiers. The spine concepts remain prominent: knowledge-bearing stock, the five forms of knowledge-bearing stock, generation and conversion, first conversion, governance forms, enclosure, feedback, dark capital, and measurement. Support concepts appear only when they solve a local explanatory problem: the Operative Knowledge Unit, K-CMM, knowledge potential, knowledge impedance, knowledge yield, boundary-case public operative units, and strategic payoff functions. Audit machinery is mainly preserved in the Technical Companion: theorem labels, formal-status registers, notation grammar, proof dependencies, and calibration mechanics. This tiering is not a reduction of this theory. It is a reader-burden rule: Volume 1 keeps the economic argument visible, while Volume 2 preserves the apparatus needed to audit it.

\noindent\textbf{Terminology discipline.} This volume uses \emph{governance} for the arrangements that determine access, exclusion, maintenance, transfer, recombination, and use of knowledge-bearing stock. Older or idiomatic uses of \emph{regime} are retained only where the legal, institutional, or formal context makes the term more precise. The dominant vocabulary is governance, governance form, governance transition, governance fit, and residence--governance pair.

The notation rule follows the same discipline. Acronyms are introduced once and then used only where they reduce rather than increase reader burden. The five knowledge forms, embodied knowledge (\(K^E\)), disembodied knowledge (\(K^D\)), institutionalized knowledge (\(K^I\)), commons knowledge (\(K^C\)), and public epistemic infrastructure (\(K^P\)), remain visible because they are part of the spine. The Operative Knowledge Unit appears mainly as a measurement bridge, not as the centre of this theory. Public epistemic infrastructure remains a core form, but public operative units are boundary cases rather than a new parallel architecture. Optional or future public-epistemic generation notation belongs in the Technical Companion.

Chapters 6 through 8 use a causal test rather than a taxonomy of enclosure types. They repeatedly ask five questions: what is enclosed, who is excluded, what recombination or learning path is lost, what the incumbent gains, and what evidence would falsify or weaken the claim. This keeps the argument from becoming a catalogue of enclosure types. The point is to show a mechanism: access restriction changes future generation by altering recombination fields, learning loops, capability accumulation, and measurement visibility.
The recurring caveats are centralized here so later chapters do not need to restate them at length. First, KBC is conditional theory: its formal claims hold under specified assumptions and require empirical calibration before numerical policy use. Second, not all enclosure is harmful: enclosure may fund creation, disclosure, quality control, security, coordination, or maintenance, and it becomes dynamically costly only when those benefits are outweighed by recombination loss, learning-loop exclusion, capability decay, or dark-capital exposure. Third, not all knowledge-bearing stock is capital: stock becomes capital-like only when it can yield productive services under the relevant access, permission, interoperability, capability, governance, and maintenance conditions. Fourth, the measurement apparatus is not a finished accounting standard: it is a decision and uncertainty-reduction framework that identifies what should be measured, why it matters, and what would count as evidence.

The formal claims should therefore be read conditionally. Where the book states a theorem, proposition, model, or diagnostic tool, it also marks the status of the object: established, extended, weakly specified, ordinal, empirically plausible, open, or conjectural. The standard of evaluation is not whether every coefficient is already estimated. It is whether the framework states its primitives, identifies its mechanisms, exposes its assumptions, and specifies what evidence would weaken, falsify, or force revision of its claims.

\noindent\textbf{Moral scope of the capital language.} KBC uses capital language analytically, not anthropologically. Persons are not capital, even when their embodied skill, judgement, and knowledge produce economically valuable services; to analyze what people know as capital is not to reduce who they are to it. Truth is not merely an asset, even when reliable knowledge improves productive capacity, and some knowledge is worth preserving because it is true, just, or identity-forming rather than because it raises measured yield. Productive value does not exhaust the value of knowledge, and this book does not claim that it does. Enclosure is not inherently theft, and openness is not inherently justice: the alternative to extractive control is rightly governed access, not the absence of boundaries, since some restriction protects persons, safety, security, or the integrity of what is known. The moral question is therefore not whether knowledge is owned, but whether a governance form honours legitimate labour, stewardship, truth, personhood, maintenance, access, accountability, and future generation, or whether it converts shared, embodied, public, or entrusted knowledge into extractive control that damages others' capacity to learn, build, heal, govern, or know what is true. The framework is diagnostic rather than redemptive: it can make such damage visible, but naming a disorder is not the same as curing it, and no scheme of knowledge governance should be mistaken for a remedy to the deeper causes of injustice.

\noindent\textbf{Three layers of claim, and what the tests can confirm.} It helps to separate three things this book asserts, because they carry different burdens of proof. The first is an \emph{architecture} claim: that knowledge-bearing capitalism can be described through a definite vocabulary and mechanism map, generation, first conversion, governance, deployment, feedback, enclosure, dark capital, and measurement. The second is a set of \emph{mechanism} claims: that particular processes, access restriction, feedback capture, commons depletion, institutional residue, operative-unit substitution, and governance transition, generate specific, testable predictions. The third is the \emph{grand frame}: the interpretive thesis that capitalism increasingly derives wealth from knowledge-bearing capital, and that this requires a revised theory of capital and of the future wealth of nations. These layers are not confirmed together. The empirical tests in this book bear on the mechanism claims: whether governance transitions contract recombination fields, whether feedback control concentrates capability, whether commons depletion lowers downstream productivity, whether bounded operative units can be compared across embodied and disembodied forms. A successful test corroborates a mechanism. It does not, by itself, prove that capitalism has entered a new stage, nor does it establish a completed theory of national wealth. The grand frame is an organizing interpretation that becomes more or less plausible as its mechanisms survive empirical confrontation, rival explanation, calibration, and negative cases. The reader should not treat the corroboration of a local mechanism as confirmation of the whole frame, and this book does not claim that it is.

\noindent\textbf{What the mathematics does and does not do.} The formal notation in this book carries more than one status, and the statuses should not be confused. Some expressions are \emph{object definitions}: they name and relate economic objects, such as the five knowledge forms, the operative unit, or the realization identity \(\KYield = \KPot \times \rho\). Some are \emph{diagnostic decompositions}: they organize what to measure without asserting a calibrated law. Some are \emph{conditional results}: they follow from stated assumptions and would change if those assumptions changed. Only a smaller subset are candidate \emph{empirical models}, and fewer still are \emph{calibrated estimates} tied to data. Most of the apparatus in Volume 1 is definitional and diagnostic; the conditional results, and their status, are recorded in the Technical Companion. The mathematics earns its place by making objects, mechanisms, and assumptions explicit and auditable, not by implying that every relation has been estimated. Where a relation has not been calibrated, the book says so; where notation does work that prose could do equally well, it is a summary, not a proof.

This volume is designed to be read in sequence. Chapters 1 and 2 establish the Smithian baseline and define knowledge-bearing stock. Chapters 3 and 4 separate knowledge generation from knowledge conversion. Chapter 5 explains the knowledge governance forms that govern knowledge after generation. Chapters 6 through 8 develop the main enclosure dynamics: cognitive enclosure, feedback capture, and strategic enclosure. Chapter 9 turns those mechanisms into a measurement and accounting problem. Chapters 10 and 11 translate this theory into governance design, empirical testing, calibration, and falsification. Chapter 12 restates the book's central claim about knowledge-bearing capital and the future wealth of nations.

Readers who want the economic argument can read the chapters straight through and treat formal passages as mechanism summaries rather than proof obligations. Readers who want the formal machinery should consult the Technical Companion, which contains the object register, extended equations, notation, proof archive, K-CMM details, calibration material, and open conjectures.

This division is deliberate. Volume 1 carries the argument as a book. Volume 2 makes the argument auditable without forcing every reader through the full proof apparatus.

\clearpage
\chapter*{Map of the Argument}
\label{front:argument-map}
\addcontentsline{toc}{chapter}{Map of the Argument}

\begin{center}
\Large\textbf{The knowledge-circulation spine}

\vspace{1.2em}

\normalsize
\[
\text{Generation}
\rightarrow
\text{First Conversion}
\rightarrow
\text{Governance}
\rightarrow
\text{Deployment}
\rightarrow
\text{Feedback}
\rightarrow
\text{Future Generation}
\]
\end{center}

\vspace{1em}

This is the simplest map of the book. Productive knowledge is generated, converted into governable stock, placed under a governance form, deployed in use, altered by feedback, and then returned to the conditions of future generation. The argument asks how wealth changes when that circulation is increasingly mediated by people, artefacts, organizations, commons, platforms, and public epistemic infrastructure.

\vspace{1em}

\begin{center}
\begin{tabularx}{0.92\textwidth}{@{}>{\raggedright\arraybackslash}p{0.34\textwidth}>{\raggedright\arraybackslash}X@{}}
\textbf{Question} & \textbf{Why it matters} \\
Where does knowledge reside? & Identifies whether productive capacity is embodied in people, encoded in artefacts, institutionalized in organizations, sustained in commons, or maintained as public infrastructure. \\
\addlinespace
How does it move? & Tracks conversion among forms, including codification, institutionalization, internalization, commons maintenance, and public standardization. \\
\addlinespace
Who governs it? & Locates the governance form that controls access, exclusion, maintenance, interoperability, permission, and use. \\
\addlinespace
Who captures its yield? & Separates productive service from ownership, control, rent capture, bargaining position, and distributional outcome. \\
\addlinespace
What future knowledge does it enable or suppress? & Tests whether present governance widens or narrows recombination fields, learning loops, experimentation, discovery, and capability accumulation. \\
\addlinespace
What do existing measures miss? & Turns invisible stock, lost options, capability decay, governance-position risk, and suppressed future generation into the measurement problem called dark capital. \\
\end{tabularx}
\end{center}

\vspace{1em}

The rest of the book elaborates this map. Chapters 1 and 2 define the object; Chapters 3 and 4 explain generation and conversion; Chapter 5 explains governance forms; Chapters 6 through 8 explain enclosure and feedback dynamics; Chapter 9 explains measurement failure; Chapters 10 and 11 translate this theory into governance and testing; Chapter 12 returns the map to the future wealth of nations.

\clearpage
\chapter*{Acknowledgements}
\label{front:acknowledgements}
\addcontentsline{toc}{chapter}{Acknowledgements}

This book has benefited from conversations across economics, technology, cybersecurity, accounting, organizational theory, and governance. I am especially grateful to colleagues and interlocutors who pressed the central question behind this book: whether knowledge is merely another input in existing economic theory, or whether its movement across persons, artefacts, firms, platforms, commons, and institutions requires a revised theory of capital.

The work also owes an intellectual debt to the economists, historians, theorists, and management scholars examined throughout this book. Smith, Machlup, Bell, Porat, Arrow, Romer, Weitzman, Hayek, Schumpeter, Becker, Mincer, Coase, Williamson, Ostrom, Heller and Eisenberg, Penrose, Teece, Grant, Barney, Wernerfelt, Peteraf\index{Peteraf, Margaret}, Lev, Haskel and Westlake, and others appear here not as authorities to be displaced, but as predecessors whose domain-specific insights make the present synthesis possible.

Any remaining errors, overstatements, omissions, or misreadings are my own. This book is intentionally framed as a theory-building work rather than a completed empirical settlement. Its claims should therefore be read as conditional, testable, and open to revision.

\clearpage
\tableofcontents
\clearpage
\phantomsection
\addcontentsline{toc}{chapter}{List of Tables}
{\makeatletter\let\addvspace\@gobble\makeatother\listoftables}
\clearpage
\phantomsection
\addcontentsline{toc}{chapter}{List of Figures}
{\makeatletter\let\addvspace\@gobble\makeatother\listoffigures}
\clearpage
\makeatletter\@mainmattertrue\makeatother
\pagenumbering{arabic}

\part{The Smithian Baseline and the New Object}
\chapter[The Smithian Baseline]{The Smithian Baseline: Labour, Stock, Specialization, and National Wealth}
\label{ch:smithian-baseline}

\chapterhook{When the Old Theory Meets the New Stock}

Smith enters here\index{Smith, Adam|textbf} as a measuring instrument, not as an object of antiquarian interest. His theory explained how labour, stock, specialization, exchange, and market extent could increase the annual produce\index{annual produce} of a nation\index{market extent}\index{national wealth}. The modern puzzle is that many of the productive stocks now driving firm value do not present themselves as looms, ships, mills, inventories, or improvements to land. They appear as software, datasets, model architectures, user feedback\index{software!opening anomaly}\index{datasets!opening anomaly}\index{model architectures}\index{user feedback!opening anomaly}, engineering routines\index{engineering routines}, platform access\index{platform access}, institutional trust\index{institutional trust}, and tacit expertise\index{tacit expertise}.

This shift is not an escape from production. It is a change in the residence, movement, and governance of productive capacity\index{productive capacity}. A firm with modest tangible assets may be valued as if it controls a large productive stock because its code, data, brand, governance position, and engineering routines are expected to generate future cash flows. An AI firm may improve faster than rivals not merely because it owns a model, but because deployment yields feedback signals that improve later models. A platform may concentrate power not only through price, scale, or network effects\index{network effects}, but by controlling the interfaces, standards, APIs, and data flows through which other firms reach users. Cybersecurity enters this theory as capital-preservation risk management: it preserves knowledge-capital value by protecting the exclusivity, integrity, availability, and productive use of knowledge-bearing stock. When that defence fails, the resulting breach can impair value far beyond remediation cost\index{breach remediation cost} because source code\index{source code}, model weights, customer data\index{model weights!cybersecurity failure}\index{customer data}\index{cybersecurity failure!model weights}, detection logic\index{detection logic}, or operating knowledge\index{operating knowledge} may be copied, poisoned, or made strategically legible to adversaries. An API closure, CUDA dependency, patent thicket, or open-source maintainer shock\index{API closure!opening anomaly}\index{CUDA dependency!opening anomaly}\index{patent thickets!opening anomaly}\index{open-source maintainer shock!opening anomaly} may impair productive capacity even when no factory burns, no employee is fired, and no recognized balance-sheet asset is written down.

Table~\ref{tab:early-market-cases} makes the terrain visible in contemporary markets.

\begin{center}
\captionof{table}{Market-facing cases and the knowledge-capital mechanisms they expose}
\label{tab:early-market-cases}
\footnotesize
\setlength{\tabcolsep}{3.5pt}
\renewcommand{\arraystretch}{0.92}
\begin{tabular}{@{}L{0.24\textwidth}L{0.72\textwidth}@{}}
\toprule
Case & Mechanism exposed \\
\midrule
Nvidia/CUDA\index{NVIDIA}\index{CUDA}\index{CUDA dependency}\index{platform dependency!CUDA} & Platform dependency and control over the recombination field: the practical field of tools, interfaces, libraries, documentation, optimization routines, and complementary capabilities that determines who can combine existing knowledge into new productive work. CUDA is not merely a chip feature; it is a developer environment that shapes which actors can productively recombine AI and high-performance-computing knowledge on Nvidia hardware \parencite{NvidiaCUDA2026}. \\
Microsoft/GitHub Copilot\index{GitHub Copilot}\index{GitHub}\index{Microsoft}\index{AI training!GitHub Copilot} & Software, public code, developer interaction, and institutional capability. Copilot shows how distributed developer knowledge can be converted into a model-mediated development tool, and how the development environment itself becomes a site of feedback capture and capability accumulation \parencite{GitHubCopilot2026}. \\
Reddit and X/Twitter API restrictions\index{Reddit API restrictions}\index{X/Twitter API restrictions}\index{API restriction!Reddit and X/Twitter} & Enclosure of ecosystem access. API terms\index{API terms}, rate limits\index{rate limits}, pricing, model-training restrictions\index{model-training restrictions}, and revocation rights\index{revocation rights} can narrow third-party access to platform data\index{platform data} and weaken the surrounding developer, research, moderation, and application ecosystem \parencite{RedditAPI2023,RedditTerms2023}. \\
Anthropic and OpenAI model-distillation disputes\index{Anthropic}\index{OpenAI}\index{model distillation}\index{model-output extraction} & Dark capital, model-output extraction, and feedback capture. Anthropic reported more than 16 million Claude\index{Claude} exchanges through approximately 24,000 fraudulent accounts\index{fraudulent API accounts} in alleged industrial-scale distillation\index{industrial-scale distillation} campaigns; OpenAI has made related allegations about unauthorized distillation through API access \parencite{AnthropicDistillation2026,ReutersOpenAI2026}. \\
Ingress NGINX\index{Kubernetes Ingress NGINX}\index{Ingress NGINX}\index{open-source maintainer shock!Ingress NGINX}\index{commons capability loss} & Commons capability loss. The Kubernetes project announced the retirement of Ingress NGINX after years of insufficient or barely sufficient maintainership, illustrating how a shared-access knowledge project, or commons, can remain legally open while losing productive capacity when the scarce embodied and institutionalized maintenance stock required to maintain it disappears \parencite{KubernetesIngress2025}. \\
AbbVie/Humira patent estate\index{Humira patent estate}\index{AbbVie}\index{patent thickets!Humira} & Biotech patent thickets and recombination suppression\index{recombination suppression}. Legal and policy analyses of Humira report a patent estate in the range of roughly 132 to 136 granted patents\index{patents}, a useful case for examining when protection becomes a dense access barrier around follow-on entry and recombination \parencite{Knox2022Humira,Zhou2023Humira}. \\
\bottomrule
\end{tabular}
\end{center}

These cases have a common structure. The economically relevant stock is real, but it is only partly visible in accounting. Its value depends on who can access it, who can interpret it, who can combine it with other stock, who can exclude rivals from it, and who captures the learning generated by its use. Standard valuation can observe some of the cash-flow consequence after the fact. It has more difficulty locating the mechanism before the fact: whether market value reflects genuine knowledge-bearing capacity, temporary monopoly\index{monopoly}, speculative premium, hidden dependency, or unpriced fragility. Standard productivity measurement\index{productivity measurement} faces the parallel problem at the economy level: output may rise or stagnate while the underlying knowledge conditions that make future output possible are widening, narrowing, decaying, or being transferred from commons to platforms.

This is the reason for beginning with Smith. Smith gave economics a powerful account of how labour, stock, specialization, exchange, and market extent generate national wealth. That account remains indispensable. But it was built around a world in which productive capital was more visibly separable, rival, material, vendible, and countable than the productive stock now driving large parts of the economy. The purpose of this chapter is therefore diagnostic. Before introducing new models, it reconstructs the assumptions that let Smith's machinery work, then asks which assumptions strain under knowledge-bearing capitalism (KBC), where the central productive stock is knowledge-bearing, non-rival, governable, recombinable, capability-dependent, and only partially measurable.

Modern capitalism increasingly produces wealth through assets that do not behave like the capital stock assumed by classical political economy. Software can be copied without being consumed. A dataset may gain value when aggregated across millions of observations and lose value when detached from the interpretive capability required to deploy it. Expertise may reside in persons, in organizational routines\index{organizational routines}, in trained models\index{trained models}, in platform architecture\index{platform architecture}s, or in institutional relationships\index{institutional relationships}, and it may migrate among these forms, gaining or losing productive power\index{productive power} depending on the institutional conditions governing each transition. Firms may be valued far above the book value of their tangible assets, and the productive capacities that explain that gap often remain poorly represented in accounting, in national statistics, and in the inherited categories of economic theory.

This gap between economic reality and economic measurement is not new. \textcite{BondCummins2000}\index{Bond and Cummins} identified it at the turn of the century as the central puzzle of ``new economy'' valuation: do market values above tangible book values reflect genuine intangible investment, or speculative overvaluation?\index{genuine intangible investment versus speculative overvaluation} \textcite{Lev2001}\index{Lev, Baruch} documented the same gap from the accounting side: intangible factors increasingly dominate wealth creation\index{wealth creation} in advanced economies, yet remain systematically underreported in financial statements. \textcite{Machlup1962}\index{Machlup, Fritz} had established, empirically, decades earlier that knowledge production (education, research, communications, information services) constituted a large and growing share of national income. \textcite{HaskelWestlake2018}\index{Haskel and Westlake} more recently gave that observation its modern architectural form: intangible assets\index{intangible assets} are scalable, synergistic, subject to spillovers\index{spillovers}, and characterized by sunk costs in ways that distinguish them from tangible capital and alter the competitive, distributional, and policy dynamics of advanced economies.

The literature has not missed this terrain. It has entered it through different doors. Bond and Cummins make the valuation puzzle visible. Lev makes the accounting problem visible. Machlup and Porat make knowledge and information activity measurable. Haskel and Westlake make the distinctive economic properties of intangible investment explicit. The question developed here is whether these domain-specific insights can be joined into a capital-system account of knowledge generation, conversion, enclosure, impairment, and measurement.

The contribution here is architectural: to connect established, scope-limited literatures (from information economics\index{information economics}, intangible capital measurement\index{intangible-capital accounting}, knowledge management, institutional economics\index{institutional economics}, Austrian knowledge theory\index{Austrian knowledge theory}, the resource-based theory of the firm\index{resource-based view}, and endogenous growth theory\index{endogenous growth theory}) into a mechanism-centred theory of knowledge circulation. The components exist in the literature. What is new is the systematic architecture: a generation model distinct from a conversion model, a governance-intervention framework built into pathway notation, a first-conversion zone that locates the institutional appropriability\index{appropriability} problem, and a unified treatment of enclosure's consequences for both existing stock and future generation.

Two predictions mark this architecture as more than a relabelling of its predecessors, and the book returns to them as its empirical signature. The first is a \emph{dual enclosure effect}\index{dual enclosure effect}: a governance transition can raise the enclosing actor's present yield while lowering the future generative capacity of excluded actors, by contracting their recombination field or excluding them from the learning loop. The second is \emph{institutional residue}\index{institutional residue}: productive knowledge that persists in routines, standards, and governance systems beyond the embodied knowledge of the individuals present. Neither follows from intangibles accounting, endogenous growth, or the resource-based view on its own, and both are developed and made testable in Chapter~\ref{chapter-11-testing-knowledge-bearing-capitalism}.

KBC's predecessors form a convergence rather than a single lineage. Endogenous growth, information economics, institutional economics, RBV/KBV, dynamic capabilities\index{dynamic capabilities}, IT economics, intangible-capital research, complementary-assets work, and Hubbard's measurement discipline each explain part of the problem. KBC integrates them by treating productive knowledge as stock that moves across forms, changes yield under governance, becomes dark to accounting, and either widens or narrows future recombination and learning.

The claim is therefore not priority, but integration: KBC asks what follows when these predecessor insights are treated as properties of one capital system rather than as separate literatures. Chapter~\ref{chapter-11-testing-knowledge-bearing-capitalism} returns to this claim in the predecessor audit and tests which parts of the synthesis are established, extended, synthesized, novel, or speculative.

Endogenous growth theory already made knowledge central to growth. \textcite{Arrow1962}\index{Arrow, Kenneth} showed learning-by-doing, and \textcite{Romer1986,Romer1990}\index{Romer, Paul} showed how non-rival ideas\index{non-rivalry} generate increasing returns\index{increasing returns}. KBC does not contest that result. Its contribution is different: it disaggregates knowledge into embodied, disembodied, institutionalized, commons, and public epistemic forms; explains how knowledge moves among those forms; shows how governance forms condition productive yield; identifies enclosure-driven impairment and foregone generation; and explains why accounting and productivity statistics often miss both the stock and its loss.

Information economics asks what follows when information is costly, asymmetric, incomplete, or strategically revealed \parencite{Akerlof1970,Stiglitz1985}\index{Akerlof, George}\index{Stiglitz, Joseph}. KBC asks a different but related question: what follows when productive capacity is increasingly constituted by accumulated knowledge-bearing stock that can reside in persons, artefacts, firms, platforms, commons, and public infrastructure; move across those forms; depreciate; be enclosed; and generate future productive capacity.\index{knowledge-bearing stock}

Put more sharply: information economics changed the information assumption; human-capital theory changed the labour-quality assumption; intangible-capital theory changed the investment boundary; KBC changes the capital assumption. It asks what follows if productive knowledge is not merely information, skill, or recognized intangible asset, but stock-like, mobile across forms, recombinable, capability-dependent, governable, and imperfectly measured.

\textcite{Smith1776} remains the necessary point of departure because \emph{The Wealth of Nations} supplied capitalism's classical architecture\index{classical political economy}\footnote{All references to Smith in this book refer to the 1776 first edition entry in the reference list unless otherwise indicated. The analysis uses book-level textual locations only where page-level precision is not required.}: labour, stock, specialization, exchange, market extent, wages, profit, rent, and national wealth. Smith's model is not a historical curiosity. It is the inherited operating system (the set of default assumptions about what wealth is, how it is produced, what capital does, and what labour contributes) that subsequent economics extended, qualified, and disputed without fully replacing. Knowledge-bearing capitalism does not refute Smith's mechanism. It reveals the assumptions built into that mechanism, assumptions that were invisible to Smith because they were not contingencies in his world but background conditions. Making those assumptions visible is the first task of this book, because revising a theory requires knowing precisely what is being revised.

The long-run income record makes Smith's placement sharper. Global average GDP per capita does not merely grow more slowly before Smith and the Industrial Revolution. It is nearly flat for centuries. In the long-run series summarized in Table~\ref{tab:gdp-growth-regime-comparison}, global GDP per capita rises from about \$1,100 in year 1 to \$1,400 in 1700, a total increase of only 27 percent over 1,699 years. It then rises from \$1,400 in 1700 to \$1,500 in 1820, only 7 percent over 120 years. After 1820, the series changes character: it rises from \$1,500 to \$21,405 by 2024, a total increase of 1,327 percent over 204 years. The 1700--1820 interval already shows acceleration relative to the pre-1700 baseline: its approximate annual growth rate is about four times higher than the earlier long-run rate. The inflection therefore should not be read as appearing suddenly in 1820. Rather, 1820 is the first clearly measured point at which an earlier generational transformation becomes visible in the long-run data \parencite{OWIDGDP2026}.

\enlargethispage{2\baselineskip}
This interpretation matters for KBC. The GDP series cannot isolate \emph{The Wealth of Nations} as a discrete causal event, and no aggregate GDP series could. But the convergence of timing, mechanism, and institutional evidence is theoretically significant. Smith wrote at the threshold of the modern growth regime, describing and legitimating the mechanisms of division of labour, capital accumulation, market extent, productivity, and exchange that were already nascent in the industrial districts of Britain. His theoretical articulation gave those mechanisms vocabulary, coherence, and policy consequence. The growth that becomes clearly visible by 1820 was not caused by a book in any simple sense. Smith's framework helped make the emerging regime intelligible and legitimate: it gave economists, policymakers, merchants, and reformers a language for specialization, capital accumulation, market extent, productivity, and exchange. That is precisely the kind of productive service that knowledge-bearing capital can yield: it changes how actors interpret production, coordinate expectations, justify institutions, and reorganize future productive capacity.

\noindent\begin{minipage}{\textwidth}
\captionof{table}{Global GDP per capita growth before and after the Smithian-industrial threshold}
\label{tab:gdp-growth-regime-comparison}
\footnotesize
\setlength{\tabcolsep}{3.5pt}
\renewcommand{\arraystretch}{0.92}
\begin{tabular}{@{}L{0.18\textwidth}L{0.24\textwidth}L{0.30\textwidth}L{0.20\textwidth}@{}}
\toprule
Period & GDP per capita & Total growth & Approximate annual growth \\
\midrule
1--1700 & \$1,100 $\to$ \$1,400 & +27\% over 1,699 years & 0.014\% per year \\
1700--1820 & \$1,400 $\to$ \$1,500 & +7\% over 120 years & 0.058\% per year \\
1820--2024 & \$1,500 $\to$ \$21,405 & +1,327\% over 204 years & 1.31\% per year \\
1850--2024 & \$1,800 $\to$ \$21,405 & +1,089\% over 174 years & 1.43\% per year \\
1950--2024 & \$4,600 $\to$ \$21,405 & +365\% over 74 years & 2.10\% per year \\
\bottomrule
\end{tabular}
\par\vspace{0.25em}
\emph{Note.} Growth rates are compound annual approximations calculated from the long-run global GDP per capita series summarized in the table. The table is not evidence for KBC; it identifies why Smith is the correct baseline for asking how capital theory changes at growth thresholds.
\end{minipage}

\begin{figure}[!htbp]
\caption[Global Average GDP per Capita Over the Long Run]{Global Average GDP per Capita Over the Long Run}
\label{fig:gdp-per-capita-long-run}
\centering
\vspace{-0.25\baselineskip}
\includegraphics[width=0.94\textwidth,trim=0 230bp 0 0,clip]{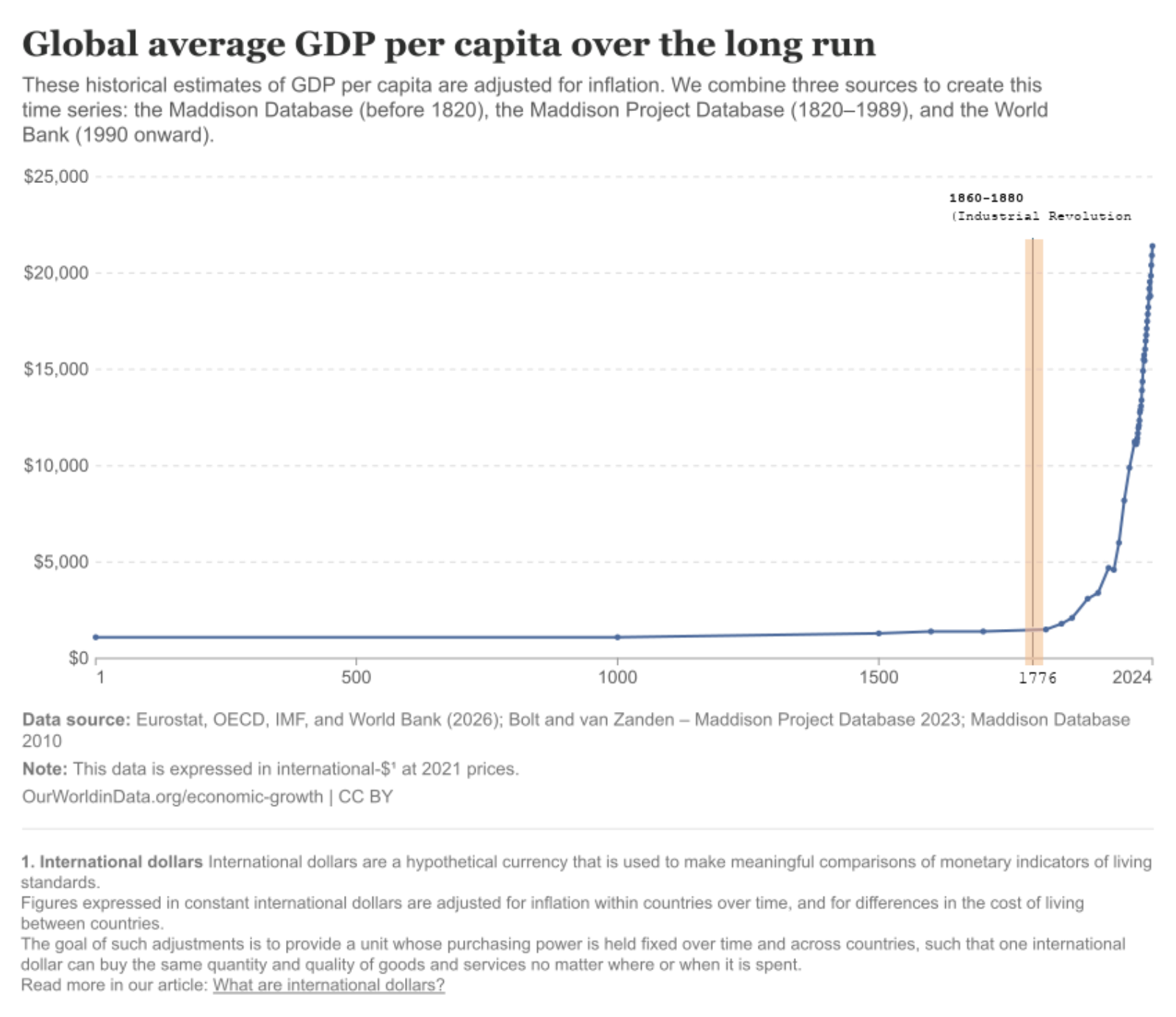}
\parbox{0.94\textwidth}{\footnotesize\emph{Note.} Adam Smith's \emph{The Wealth of Nations} was printed in 1776. The Industrial Revolution is dated here to 1760--1780, so Smith's book appeared during the early industrial threshold. Its intellectual and institutional influence was not immediate; it began to shape society after roughly one generation.}
\par\vspace{-0.45\baselineskip}
\end{figure}

This historical break makes Smith more than background context. It makes him the right baseline for the book. Smith gave a theory of the mechanisms that later became visible as a growth regime, and his framework helped make intelligible the institutional and commercial imagination through which those mechanisms were understood. KBC asks the successor question: if Smith wrote at the threshold of a capitalism organized around specialization, material stock, and industrial accumulation, what conceptual threshold are we now crossing as value moves increasingly into knowledge-bearing stock, artificial intelligence, data, models, commons, platforms, and public epistemic infrastructure?

The book's fixed-point question follows from this diagnosis: the key question is not whether knowledge is enclosed or shared by nature, but under what institutional conditions knowledge becomes a private asset, a firm capability, a public good, a professional standard, a platform dependency, or a commons.\index{enclosed knowledge versus shared knowledge}\index{private asset versus public good}\index{firm capability versus commons} Knowledge does not arrive in the economy already labelled with its governance form. The same discovered mechanism becomes a patented compound under pharmaceutical IP law\index{pharmaceutical patenting}, a shared clinical protocol under professional stewardship norms, or a public infrastructure under open-science funding, depending on the institutional conditions at the moment of first conversion. Understanding those conditions and tracing their consequences for both wealth creation and its distribution is the work of this theory.

The caveats established at the front of the book govern the rest of the argument. The claim is not that knowledge automatically becomes capital, that enclosure is generally bad, or that the measurement framework is already a finished accounting system. The claim is narrower and more testable: when productive knowledge becomes stock-like and moves across persons, artefacts, organizations, commons, and public infrastructure, its economic effects depend on the conditions under which it is generated, converted, governed, deployed, fed back into future knowledge, measured, or lost.

This theory, therefore, separates two problems.\index{knowledge generation}\index{knowledge conversion} The Knowledge Generation Model (KGM) explains how new knowledge-bearing stock arises, through recombination, experimentation, discovery, invention, interpretation, and feedback from prior deployment. The Knowledge Conversion Matrix (KCM) explains what happens once knowledge exists: how it becomes separable, codified, institutionalized, enclosed, diffused, priced, or appropriated under different governance forms. The two models meet in the first-conversion zone, where newly generated knowledge-bearing stock first acquires a governable form and becomes subject to claims of ownership, access, or public use.\index{first conversion} Chapter 2 introduces this vocabulary. Chapters 3 and 4 develop the two models in full.

This chapter closes on its structural claim: knowledge-bearing capitalism can increase present productive power by enclosing, scaling, and improving knowledge-bearing stock, while simultaneously suppressing the wider recombination fields from which future productive knowledge would otherwise emerge. That is a Smithian problem. It operates through accumulation, not against it. The task of this chapter is to establish the baseline from which the anomaly becomes visible.

\section{Smith's Operating Model of Wealth}\label{smiths-operating-model-of-wealth}

\textcite{Smith1776} defined wealth as ``the annual produce of the land and labour of the society.'' Not the stock of gold in a nation's vaults, not the balance of trade in the mercantilist sense, not the exclusive productivity of agriculture in the physiocratic sense, but productive output per year. This was a polemical move against two inherited frameworks simultaneously. Against mercantilism, Smith argued that precious metals are valuable only because they facilitate exchange, and exchange facilitates production, and production is the actual source of wealth. Against the physiocrats, he argued that manufacturing labour is equally productive when it adds value to the subject on which it is bestowed. The wealth of a nation is not what it hoards; it is what it produces.

The definition is deceptively simple. Its analytical weight comes from what it implies about where to look to understand wealth creation. If wealth is annual produce, then increasing national wealth means increasing productive output per year. The question becomes: what increases productive output? Smith's answer was specific, mechanistic, and organized around a causal chain that begins with the division of labour.

\begin{quote}
``The greatest improvement in the productive powers of labour, and the greater part of the skill, dexterity, and judgment with which it is anywhere directed, or applied, seem to have been the effects of the division of labour\index{division of labour}.''
\end{quote}

This opening claim of Book I is not merely an observation. It is a theoretical commitment: the primary source of productivity is specialization, not the natural fertility of land, not the cunning of merchants, not the accumulation of money in itself, but the decomposition of production into specialized operations performed by workers who do nothing else.

The pin factory illustrates the claim with unusual precision. A single worker unacquainted with the trade ``could scarce, perhaps, with his utmost industry, make one pin in a day, and certainly could not make twenty.'' Organize ten workers across eighteen distinct operations (drawing out the wire, straightening it, cutting it, pointing it, grinding the top for receiving the head, making the head, whitening, papering) and they produce approximately forty-eight thousand pins per day. The productive multiplier is not merely additive. Specialization raises productive power by several orders of magnitude.

Smith identifies three distinct sources of this productive increase. First, repetition of a single operation increases the worker's dexterity in performing it. Second, eliminating time lost to switching between operations increases the proportion of working time spent on actual production. Third (and most consequential for this theory that follows) workers confined to a few operations are more likely to discover improved methods or machines for their specific task. The division of labour generates technical progress endogenously, not as an exogenous shock but as a consequence of the intensive attention that specialization directs at narrow problems.

Smith's division of labour is, for skilled productive labour\index{productive labour}, the division of embodied productive knowledge\index{division of knowledge}. This is not a loose analogy. The pin-maker's increased dexterity is not merely faster motion; it is accumulated domain-specific knowledge made operational through repeated practice. Smith saw the productive service as labour productivity. KBC identifies the capital source of that service as embodied knowledge capital. The worker who draws the wire, the worker who cuts it, and the worker who heads the pin are not merely assigned to different operations; each becomes the bearer of a different operative knowledge unit within the pin-making domain. Smith arrived at the correct productivity insight by a route that concealed the asset doing the productive work. This is a recoverable limitation in the classical vocabulary: what appeared as specialized labour was also domain-specific knowledge-bearing capital in deployment.

This third mechanism connects the division of labour to capital accumulation. The machines and improvements that specialization generates are themselves forms of stock, fixed capital, in Smith's later terminology. Division of labour, therefore, not only increases output by raising productive power; it generates the capital stock that further increases productive power in subsequent periods. The model is self-reinforcing: specialization raises productivity, productivity raises output, output enables saving, saving enables the accumulation of stock, stock maintains labour and enables further specialization, and further specialization raises productivity again. National wealth expands through the compounding of these effects across time and across an increasingly extended market. Smith's mechanism is a growth engine, not a static equilibrium.

\section{Stock, Labour, and the Separability Assumption}\label{stock-labour-and-the-separability-assumption}

Before a worker can be employed in specialized production, someone must maintain them during the production period. The wool merchant cannot wait months for the cloth to sell before feeding the weavers. The accumulated provision that maintains labour during production is therefore the essential enabling condition for any production beyond simple subsistence. As Smith states directly: ``The accumulation of stock is previously necessary\ldots{} for carrying on this great improvement in the productive powers of labour.''

Smith divides stock into three categories. The first is stock reserved for immediate consumption (food, clothing, furniture) which yields no revenue because it is consumed without being put to productive use. The second is fixed capital: machines, tools, buildings, improvements to land, and ``the acquired and useful abilities of all the inhabitants or members of the society.'' The third is circulating capital: money, provisions in transit, raw materials, and finished goods in the process of sale. Fixed capital yields revenue ``without changing masters or circulating further''; a machine increases the weaver's productive power without the machine itself being sold. Circulating capital yields revenue precisely by changing hands, cloth is sold, the revenue is used to buy more wool, the wool becomes more cloth.

What matters most for the theory developed in these volumes is a feature of Smith's account that he does not thematize as an assumption because it was not visible to him as a contingency. Capital in Smith's model is overwhelmingly separable from the workers who employ it. A loom, a ship, a mill, a building, these exist independently of any particular worker. They can be owned by an employer who has no skill in operating them. They can be transferred through sale. They outlast any individual labourer's working life and can be maintained by replacing workers without replacing the capital itself. The machine does not leave when the machinist does.

Separability is what makes stock easily ownable, transferable, and accumulable by someone other than the worker. A loom, ship, or mill can generate revenue for its owner independent of any particular person's continued presence. If productive capacity were inseparable from the worker (if it left when the worker left, could not be owned by the employer, and could not be transferred through sale or bequest) it would not function as employer-held capital in Smith's framework. It might be productive, but its revenue-generating potential could not be appropriated by anyone other than its inseparable possessor. Separability is therefore a condition of easy appropriability, and easy appropriability is one condition of capital accumulation.

Smith does recognize one partial exception. Among his categories of fixed capital, he includes ``the acquired and useful abilities of all the inhabitants or members of the society,'' noting that their acquisition ``always costs a real expense, which is a kind of capital fixed and realized\ldots{} in his person.'' This is as close as Smith comes to anticipating human capital theory. But he does not develop the point as an anomaly. The skill of a trained craftsman functions, in his model, as productive stock analogous to a machine: the investment in training is a real cost, the resulting skill generates higher wages, and those higher wages are the return on that fixed capital. The deeper question (whether the knowledge embedded in skilled workers can be separated from them, owned by their employers, transferred to successors, or accumulated independently as a firm-level asset) Smith largely leaves unexamined. In his economy, it largely could not be. The craftsman's knowledge left with him; what remained was the physical capital he had operated.

Smith's theory of productive and unproductive labour reinforces this structure. Productive labour ``adds to the value of the subject upon which it is bestowed'' and ``fixes and realizes itself in some particular subject or vendible commodity, which lasts for some time at least after that labour is past.'' Unproductive labour (the services of domestic servants, lawyers in most of their activities, actors, musicians)``generally perishes in the very instant of its performance.'' The criterion is not whether the labour is useful, difficult, or socially valued. The criterion is whether it leaves behind a durable, vendible residue\index{vendible residue} that exists independently of the labourer after the labour is complete.

This is the vendible-residue criterion, and it has a precise implication for this theory: Smith's model of productive wealth creation requires that labour produce something separable from itself. The pin is separable from the pin-maker. The cloth is separable from the weaver. The improvement to the land is separable from the labourer who made it. This is why productive labour is the engine of accumulation in Smith's framework, its output can be traded, taxed, reinvested, and accumulated as stock. The residue is what makes accumulation possible.

The physician's cure, the lawyer's advice, the musician's performance, these perish in the moment of delivery and cannot be accumulated, stored, traded, or reinvested as capital. Their value is real; their social importance is beyond dispute. But in Smith's framework they are consumption rather than production in the economic sense, because they do not generate a durable, separable residue that can be put back into the productive circuit.

This distinction will not survive intact into knowledge-bearing capitalism. The lawyer's contract template, once written, can be deployed ten thousand times. The physician's diagnostic protocol, once codified, can be instantiated across a hospital network\index{hospital networks}. The musician's recorded performance, once captured, can be distributed at near-zero marginal cost to millions of listeners. The question is not whether knowledge work can leave a durable residue, it manifestly can. The question is whether that residue is recognized, measurable, appropriable, and accumulable in the way that Smith's vendible commodity is. The short answer, as Lev and others have documented, is that it generally is not: the accounting frameworks\index{accounting recognition} designed to capture vendible residues were not designed for this kind of durable output\index{accounting recognition versus economic reality}\index{economic reality!versus accounting recognition}, and the gap between economic reality and accounting recognition is structural, not merely a matter of technical lag.

\section{Value in Use and Value in Exchange: The Half Smith Set Aside}\label{value-in-use-and-value-in-exchange}

Smith opens his account of price by naming a distinction and then, quietly, declining to pursue one half of it. The word \emph{value}\index{value!use versus exchange}, he observes, carries two meanings: the utility of an object, its \emph{value in use}\index{value in use}, and the power of purchasing other goods that its possession conveys, its \emph{value in exchange}\index{value in exchange}. He dramatizes the gap with the paradox that has carried his name ever since. Nothing is more useful than water, yet it will purchase scarcely anything; a diamond has hardly any use, yet a great quantity of other goods may be had in exchange for it \parencite{Smith1776}. Having drawn the distinction, Smith makes a decision that organizes everything after it. He treats value in exchange as the tractable object of political economy and sets value in use aside.

For Smith's problem the decision was not merely defensible but correct. His subject was rival, vendible commodities: goods used up in consumption, owned by one party to the exclusion of others, and brought to market to be sold. For such goods exchange-value is the measure that does the analytical work, and the apparatus assembled around it (labour as the real measure of price, natural price resolving into wages, profit, and rent, and market price gravitating toward natural price as effectual demand rises and falls) is sufficient. Value in use sets an outer bound on what anyone would pay and otherwise stands in the background while exchange-value is measured, exchanged, and divided. The vendible-residue criterion examined in the previous section is the same choice seen from the production side: productive labour counts because it leaves an exchangeable residue, and use that leaves no such residue is set outside the productive circuit.

This book begins where that decision begins to cost us. Knowledge-bearing capitalism does not, in its most consequential cases, follow the structure on which the decision rested. Productive knowledge often yields its wealth along a different path: it amplifies the use-value of other activities, diffuses and recombines, and appears as exchange-value only later and elsewhere, in the price of goods it never itself became. A search algorithm may be free at the point of use while it reorganizes labour, discovery, advertising, and the matching of buyers to sellers across an entire economy. A theorem may carry no market price and become a durable input to cryptography, engineering, and computation. An open-source library may be free to copy and productive across thousands of firms at once. A public standard may be freely accessible and yet coordinate whole markets. In each case the wealth is evident and the price is not, because the wealth appears first as use-value and only indirectly, if at all, as exchange-value.

This is the diamond-water paradox returning, and returning as the central problem rather than a curiosity. Water, in Smith's example, has immense use-value and almost no exchange-value. The most productive knowledge has the same structure, but on the plane of production rather than of personal consumption, and so it cannot be left in the background. Smith could set value in use aside because exchange-value was sufficient for analysing rival, vendible commodities; knowledge-bearing capitalism cannot set it aside, because its most productive assets may carry high use-value, little or no direct exchange-value, and large indirect effects on the exchange-value of other goods. That is the buried problem that surfaces whenever wealth becomes non-rival\index{non-rivalry}.

An immediate objection must be met. Knowledge is not consumed; how then can it constitute wealth at all? The answer is to distinguish consumption from productive service. A machine is capital because it yields a flow of productive services, and it does so while wearing out. Knowledge yields productive services without being used up: it is capital-like in that it conditions the productivity of other factors, and unlike a machine in that using it does not consume it. Its contribution is the increment it causes in other processes, the same stock of knowledge raising the output, speed, reach, precision, coordination, and trust of the labour and capital with which it is combined. Knowledge of consumer behaviour is not consumed as a final good; it raises the yield of the labour of selling, the accuracy of targeting, the fit of product to demand. It creates value by changing the productivity and the direction of other labour. Knowledge, in short, is use-value that acts on production before it appears, if it appears, as exchange-value\index{knowledge as amplifier}.

If that is how knowledge produces wealth, a further question follows immediately: how does any of it become exchange-value at all, given that knowledge is non-rival and naturally resists a price? Here the classical argument meets the governance argument that occupies much of this book. Where a good is non-rival, exchange-value is not given by nature; it is manufactured by controlling access. Non-rival use-value is converted into a scarcity-like condition, and only then into rent, by institutions that decide who may use the knowledge and on what terms. This is the inversion at the centre of knowledge-bearing capitalism: knowledge does not become exchangeable because it is naturally scarce, but because institutions make access to it scarce\index{governed access!conversion of use-value to exchange-value}. Intellectual property, paywalls, application programming interfaces, platforms, proprietary standards, trade secrets, and controlled model access are not incidental legal detail; they are the machinery that converts non-rival use-value into private exchange-value. Where Smith's merchants and manufacturers controlled capital and access to markets, the gatekeepers of knowledge capital control access to productive use-value and its conversion into exchange-value.

It would be a mistake, however, to read governance as a synonym for enclosure, or to hear in this the claim that governance is merely extractive. Governance is the conversion mechanism, and enclosure is only one of its forms. Other governance preserves shared use-value rather than privatizing it: the contribution rules, stewardship, and durability by which a commons sustains itself \parencite{Ostrom1990}, the open licences and foundations that keep a codebase common, the public standards that hold a market together. Other governance conditions use-value rather than converting it, through the maintenance of quality, safety, interoperability, and trust on which reliable knowledge depends. And the mechanism runs in both directions: just as governance can convert non-rival use-value into private exchange-value by closing access, it can release exchange-value-locked knowledge back into shared use-value by opening it. The variable is not whether knowledge is governed, but in which direction and in which form, and whether the same act that makes knowledge privately capturable also reduces the diffusion and recombination that would have produced more use-value later, a trade whose worth is conditional: enclosure can be the precondition of production where appropriability is what funds it, and the suppressor of production where it fences off a commons that already exists.

The same move sharpens Smith's deepest structural principle, examined in the section that follows. The division of labour, he argues, is limited by the extent of the market: a wider market can absorb more specialized output and so justifies finer specialization. Knowledge-bearing capitalism supplies the analogue, and it is structural rather than metaphorical. The recombination of labour and knowledge is limited by the extent of accessible knowledge\index{extent of accessible knowledge}. Where a larger market permits a finer division of labour, a larger field of accessible knowledge permits a wider recombination of inputs, methods, datasets, code, examples, standards, and feedback; and because access is the thing institutions can narrow, enclosure contracts the field of recombination directly. The empirical chapters return to this contraction and its consequences as the theory's testable signature.

The use-value that knowledge carries, and that exchange-value records only partially, is not beyond measurement merely because it is unpriced. Where a good is given away, its use-value is revealed by what users would pay to keep it, or accept to give it up; large-scale willingness-to-pay studies of free digital goods have found values far in excess of anything recorded in price or in the national accounts \parencite{Brynjolfsson2019}\index{willingness to pay}\index{consumer surplus}. The measurement problem this book returns to is therefore not that knowledge wealth is immeasurable, but that the inherited instruments measure the exchange-value side and leave the use-value side, where much of the wealth resides, in the dark\index{dark capital!unconverted use-value}.

The departure from Smith can therefore be stated without condescension to him, and that is the point. Smith's political economy makes exchange-value analytically central because rival commodities become wealth through vendible output, labour-commanded exchange, and price. Knowledge-bearing capitalism reopens the use-value problem that Smith deliberately bracketed, because productive knowledge may have little or no direct exchange-value while greatly increasing the productive power, reach, specialization, and recombination capacity of labour and capital. Its value appears indirectly, through amplification, and institutionally, through governed access. Compressed to a single sentence: Smith set aside value in use because value in exchange was sufficient for rival commodities; knowledge-bearing capitalism forces value in use back into political economy because its central productive assets may be free to copy, difficult to price, and yet decisive for the wealth of nations.

\section{Yield and Generativity: What Use-Value Produces}\label{yield-and-generativity}

If use-value rather than exchange-value is where much of the wealth of knowledge resides, the question Smith answered for exchange returns in a new form. Exchange-value has a native quantity, price, that emerges at a definite event, the transaction, in a definite arena, the market. What are the corresponding quantity, event, and arena for use-value? The answer reorganizes much of what follows.

Begin with the event. The primitive of exchange is the transaction: a rival good changes hands, and the act conserves, the same thing now held by someone else. The primitive of productive use is different. To use knowledge productively is to recombine it: one or more knowledge units, embodied or disembodied, combine under some capability and governance condition to produce a service, an outcome, or a new knowledge unit. A citation is one piece of disembodied knowledge re-encoded into another. An interface call is one knowledge-bearing system composed with another. A diagnosis is embodied expertise combined with codified medical knowledge and a particular case. Strip the surfaces and each is the same event: a recombination\index{recombination!as the primitive of use}. Productive use is recombination under capability and governance conditions, and unlike the transaction it does not conserve, it generates, because its output becomes an input available for further recombination \parencite{Weitzman1998,Arthur2009}.

That difference forces the two roles of price apart. Price is powerful because a clearing market fuses them into a single number: it measures the value of the unit and it signals to everyone how to act. The recombination field has no clearing moment, and so the two roles separate. The value produced at a recombination is its \emph{yield}\index{yield!productive service at recombination}: the productive service the combination renders. The visible trace the recombination leaves, citation, dependency, adoption, reuse, is its \emph{generativity}\index{generativity!visible recombination signal}. Actors coordinate on generativity because generativity is what they can see; yield is mostly unseen, produced in use, often delayed, and frequently unattributable among the several inputs that made it. Where price is value and signal at once, knowledge use gives value as yield and signal as generativity, and the two need not agree.

The gap between them is where knowledge wealth is mismeasured, and it has two signs. When yield exceeds visible generativity, real productive service is produced but under-signalled, unpriced, unattributed, or never observed: this is dark capital\index{dark capital!yield exceeds generativity}, the wealth the ledger cannot see. When visible generativity exceeds yield, the trace outruns the service: a paper widely cited but unreplicable, a method fashionable but weak, a library heavily depended upon yet fragile, a benchmark that coordinates a field without improving it, a claim that travels precisely because it alarms rather than because it is true. Call this signal-inflated stock\index{signal-inflated knowledge stock}. The knowledge economy therefore does not only hide valuable things; it also amplifies visible but low-yield ones, and a theory that noticed only the first would be half a theory.

This is why validation is an economic institution and not a bureaucratic friction. Peer review, replication, audit, certification, clinical trial, and provenance check all perform one function: they work to make generativity track yield, to keep the visible signal an honest estimate of the unseen value\index{validation!coupling signal to yield}. A governance gate is justified to the extent that it improves that tracking; it becomes extractive when it estimates yield from origin, status, or ownership instead, which is the failure later chapters name and test. The central problem of a knowledge economy is therefore not simply more access or more enclosure. It is the design of institutions under which the signal a knowledge unit emits comes to track the service it actually renders, without suppressing the high-yield recombinations that have not yet announced themselves.

Two sentences carry the inversion. Exchange-value is monetized at the point of transfer; use-value is generated at the point of recombination. Transfer conserves, recombination generates, and that is the deepest reason the half of value Smith set aside cannot stay set aside: in an economy whose primitive event creates rather than moves, the bracketed half is the half that compounds.

\section{Division of Labour and the Extent of the Market}\label{division-of-labour-and-the-extent-of-the-market}

Smith's most elegant theoretical proposition is that the division of labour is limited by the extent of the market. A maker of nails in a remote village cannot specialize entirely in nail-making if local demand for nails supports only a fraction of a working day. A porter in a large city can specialize entirely in portage because the demand for such services across a dense population is sufficient to keep one person fully employed at that single task. As markets expand (through improvements in transport, the removal of trade barriers, the growth of population, and the extension of commercial networks) specialization deepens. Deeper specialization raises productive power. Higher productive power generates more output, which enables further trade, which extends markets further. The growth of commercial society is a positive feedback between market extent and productive specialization.

This mechanism assigns a particular role to exchange. Exchange is not merely the distribution of what is already produced; it is the enabling condition of production itself. Without the capacity to exchange pins, no pin factory is economically viable. Without the capacity to exchange cloth across wide markets, the weaving trade cannot specialize to the degree that raises its productivity. Smith's critique of mercantilist trade restrictions is partly a moral argument about freedom and partly a productivity argument: every restriction on exchange limits the extent of the market, which limits the division of labour, which limits productive power, which limits the annual produce of the nation.

The market also serves as the coordination mechanism that enables the division of labour to function across a society of strangers. ``It is not from the benevolence of the butcher, the brewer, or the baker, that we expect our dinner, but from their regard to their own interest.'' Coordination among strangers is achieved not through benevolence or hierarchy but through exchange: price signals inform each producer of what others need without requiring any producer to know the others' circumstances or care about their welfare\index{welfare}. The price mechanism in Smith's model carries informational weight, though Smith does not theorize it primarily as an information system. That formulation belongs to Hayek a century and a half later. For Smith, the more fundamental point is that market exchange is what enables the division of labour to persist: without exchange, each producer would need to produce everything they consume, and specialization would be impossible.

Capital accumulation\index{capital accumulation} enters this picture as the mechanism that expands the capacity of the market to support specialization. An employer with accumulated stock can maintain more workers, invest in more fixed capital, and enable more intensive division of labour within a firm. As the stock of a nation grows through saving (through the abstention from current consumption) the nation can support a larger workforce employed in more deeply specialized production. The rate of capital accumulation therefore governs the rate at which productivity can increase.

In Smith's model, the extent of the market is in the first instance a function of geography and population: transport cost, trade barriers, and the density of demand. This makes market extent a relatively straightforward variable, large markets support more specialization than small ones, and improvements to transport or the removal of trade barriers are productivity policies because they expand the effective market. For knowledge goods, this relationship requires revision. The formal theory introduces the concept of effective field extent to capture the full set of conditions that determine how widely knowledge-bearing stock can be discovered, accessed, permitted, interoperably combined, and deployed by capable actors.\footnote{The formal notation introduced later represents this as \(S_a\): effective field extent for actor \(a\). It depends not only on market size \(M\) and marginal distribution cost \(MC_d\), but also on field access \(F_{a,t}\), complementary capability \(C_a\), interoperability\index{interoperability} \(I_a\), platform/API access \(P_a\), and legal access conditions\index{legal access conditions} \(L_a\). The point in Chapter 1 is conceptual rather than algebraic: for Smith's rival goods, market size dominates; for non-rival knowledge goods\index{non-rivalry versus excludability}\index{rival goods versus non-rival knowledge goods}, governance and capability conditions may dominate.} For non-rival knowledge goods, a stock with near-zero marginal distribution cost may still have a narrow effective field if access is legally restricted, platform-gated, or technically non-interoperable. This is the mechanism by which cognitive enclosure, meaning legal or technical control over who may access, use, or recombine knowledge-bearing stock, can restrict specialization even in the absence of conventional market-size constraints.

Smith's framework for revenue confirms this structure. All income in his model ultimately resolves into wages, profit, and rent, the returns to labour, to accumulated capital, and to land respectively. These are not merely accounting categories; they reflect the three productive inputs that Smith recognizes as jointly generating the annual produce. Wages compensate labour for productive effort. Profit compensates the capitalist for accumulated stock put to productive risk. Rent compensates the landlord for access to natural productive resources. Knowledge does not appear as a separate category of productive input with a corresponding revenue stream. In Smith's economy, knowledge is either embedded in labour and compensated through wages, or embodied in machines and compensated through capital's profit share. The possibility that knowledge itself might constitute a primary productive input with its own appropriable revenue stream is not a category available in his framework.

This Smithian mechanism supplies the constructive market intuition that this book later tests: privately directed stock, when deployed under competitive conditions, can increase public wealth by widening specialization, deepening exchange, and raising annual produce. Knowledge-bearing capitalism does not reject that intuition. It asks where its domain conditions change. A later chapter formalizes this as the Smithian inversion: privately rational enclosure can strengthen present production while weakening the conditions for future knowledge generation\index{private return versus social return}\index{private incentive versus social optimum}\index{Smithian inversion!private versus social return}, so that private best responses produce an equilibrium that does not maximize social wealth. Its precise statement, and how it differs from monopoly deadweight loss\index{monopoly!deadweight loss}, the anticommons, and Schumpeterian rents\index{rents!Schumpeterian}, is given in Chapter~8.

\section{The Knowledge-Economy Pressure Points}\label{the-knowledge-economy-pressure-points}

Smith's model is coherent, mechanistically rich, and historically appropriate to the economy he was describing. The question for this book is not whether the model was right for 1776 (it largely was) but whether its inherited assumptions hold when knowledge, software, data, institutional capability, and intangible assets become primary sources of productive wealth. Seven assumptions in the Smithian model face pressure under knowledge-bearing capitalism. They are considered here in order of increasing theoretical depth, from the more tractable to the more structurally significant.

\subsection{Assumption 1: Capital is separable from the worker}\label{assumption-1-capital-is-separable-from-the-worker}

Smith's fixed capital (machines, buildings, tools) exists independently of any particular worker. The productive stock of the pin factory persists when any given pin-maker leaves. The capital is appropriable by the employer, transferable through sale, and accumulable as an independent asset because it does not leave with its operators.

The pressure point: a substantial and growing portion of modern productive capacity is embodied in persons, teams, and relational networks that cannot be straightforwardly separated from the people and relationships in which they live. The senior surgeon's operative judgment, the software architect's intuition about system design, the research team's shared understanding of an unsolved problem, the organizational culture\index{organizational culture} that allows a firm to coordinate complex projects rapidly, these are productive assets in any reasonable economic sense. They generate higher output, command premium prices, and constitute genuine sources of competitive advantage. But they reside, at least in part, in people and relationships from which they cannot be extracted unchanged.

This book refers to this as the conditional separability problem. Knowledge-bearing productive capacity becomes separable not by nature but through institutional achievement: employment contracts, non-compete agreements\index{non-compete agreements}, codification programmes, documentation standards, training systems, and knowledge management architectures. The question is not whether embodied knowledge can ever be separated from the person or team who exercises it (it often can, with effort and partial loss) but whether separability is the natural default or an institutional accomplishment. In Smith's model, the separability of capital from labour is a background condition that requires no theoretical attention. In knowledge-bearing capitalism, it is an achievement that may be costly, incomplete, and, in some cases, impossible without sacrificing productive capacity itself.

\subsection{Assumption 2: Productive labour leaves a durable vendible residue}\label{assumption-2-productive-labour-leaves-a-durable-vendible-residue}

Smith's vendible-residue criterion distinguishes labour that generates accumulable wealth from labour that is economically consumed in the moment of performance. The productive residue is what can be traded, accumulated, and reinvested.

The pressure point: much of the most economically valuable work in modern economies produces residues that are durable but not straightforwardly vendible in Smith's sense, or vendible but not recognized as such in conventional accounting. The software engineer's codebase\index{codebase}, the data scientist's trained model, the consultant's engagement methodology, the clinician's diagnostic protocol, these are durable residues of skilled labour that can generate revenue across many deployments. But they often fail either the accounting recognition criteria\index{recognition criteria} for intangible assets or the market pricing mechanisms that would reveal their productive value. The residue exists and is economically significant; the apparatus for its recognition and measurement is inadequate.

The deeper issue is that the residue of knowledge work may not be a discrete, separable object at all. The experienced clinician's judgment improves with each patient encounter, but the improvement is distributed across embodied knowledge rather than deposited in any separable artefact. The software team's collective ability to build reliable systems improves with each project, but this institutional learning does not appear on a balance sheet. The residue is real (it generates economic value\index{economic value} that bond markets and equity markets may partially price through spreads, multiples, acquisition premia, or valuation discounts) but it lacks the separability and legibility that Smith's vendible commodity possesses. Smith's criterion requires that productive labour leave behind something that can be picked up and traded independently. Knowledge work often leaves behind something that can be amplified but not straightforwardly detached.

\subsection{Assumption 3: Division of labour decomposes productive activity and differentiates embodied knowledge capital}\label{assumption-3-division-of-labour-decomposes-productive-activity-and-differentiates-embodied-knowledge-capital}

Smith's division of labour does not merely divide physical tasks. For skilled productive labour, it is also the division of knowledge-bearing capital. Each specialized operation creates, deepens, and localizes a domain-specific stock of skill, dexterity, judgement, and procedural knowledge. Smith identified the resulting productive-service yield and attributed it to labour productivity. KBC supplies the missing asset category: the yield arises from embodied knowledge capital deployed through labour. The correction is precise rather than merely historical: Smith saw the productivity effect, but his framework lacked a category for the capital source when that source was inseparable from the person.

The pressure point: many high-value knowledge tasks resist this kind of decomposition without losing what makes them valuable. The literary editor's judgment about whether a paragraph coheres cannot be decomposed into a checklist of steps. The epidemiologist's interpretation of ambiguous outbreak data cannot be fully proceduralized without losing the tacit pattern recognition that years of experience have built. What can be decomposed in knowledge work is often automatable; what resists decomposition is precisely the interpretive capacity that constitutes the productive core.

This does not mean that knowledge work cannot be divided. It can and is: legal research is separated from legal argument; software testing is separated from software development; radiological image reading is separated from clinical diagnosis. The same pattern appears in market-facing settings such as radiology AI triage\index{radiology AI example!division of knowledge}, drug-discovery pipelines, software architecture review, fraud analytics, and cybersecurity detection engineering, where specialized outputs must be recombined before they become economically valuable. But the division of knowledge work operates differently from Smith's pin factory decomposition. Rather than decomposing a single operation into simpler operations, it distributes interpretation among specialists who must recombine their outputs to produce the valuable result. The productive unit is not the isolated operation but the recombination of interpretations, which requires communication, shared vocabulary, trust, and integrative capacity that Smith's model does not theorize. The economic question is not how to decompose tasks into simpler steps, but how to maintain the capacity for recombination that makes specialized interpretation valuable.

\subsection{Assumption 4: Market extent governs the degree of specialization}\label{assumption-4-market-extent-governs-the-degree-of-specialization}

In Smith's model, market breadth determines how specialized any given trade can become. Wide markets support deep specialization; narrow markets force producers to be generalists. The relationship between market size and specialization is governed by the cost of serving each additional customer.

The pressure point is often misstated as if non-rivalry alone produces digital concentration. That is too compressed. Several channels operate at once, and they should be kept analytically separate.

Smith's market-extent proposition strains because these channels alter different parts of the specialization mechanism. Fixed-cost scalability and non-rivalry explain why a knowledge-bearing output can travel widely once produced. Network effects and switching costs\index{switching costs} explain concentration of demand. Data feedback explains product improvement through use. Governance and access control explain why the effective field may be much narrower than the potential market.

\begin{center}
\captionof{table}{Concentration channels in knowledge and digital markets}
\label{tab:concentration-channels}
\begin{tabular}{@{}L{0.30\textwidth}L{0.66\textwidth}@{}}
\toprule
Channel & What it explains \\
\midrule
Fixed-cost scalability & Why software or model output can serve many users at low marginal cost after high first-copy, development, or training costs. \\
Non-rivalry & Why use by one actor does not consume the knowledge stock or prevent simultaneous use by others. \\
Network effects & Why user value rises with more users, developers, complements, or shared standards. \\
Data feedback & Why deployment can improve the product by generating errors, labels, use patterns, edge cases, and performance signals. \\
Switching costs & Why users may remain locked in even when alternatives exist. \\
Governance/access control & Why incumbents can restrict access to the recombination field through IP, API control, licensing, platform architecture, or interoperability limits\index{interoperability limits}. \\
\bottomrule
\end{tabular}
\end{center}

KBC does not treat these channels as a single mechanism. Its distinctive emphasis falls on the last two: feedback capture and governed recombination-field access.\index{recombination field} Feedback capture determines who learns from deployment, as in model-improvement and model-distillation disputes. Governed recombination-field access determines who can use existing knowledge-bearing stock as input to future generation, as in CUDA dependency, Reddit/X-style API restrictions, patent thickets, or open-source dependency shocks\index{CUDA dependency!recombination field}\index{Reddit API restrictions!recombination field}\index{X/Twitter API restrictions!recombination field}\index{patent thickets!recombination field}\index{open-source dependency shocks}. The full implications are developed through the concept of effective field extent introduced in Section 1.3, the Knowledge Conversion Matrix in Chapter 4, and the enclosure dynamics of Chapters 6 and 7.

\subsection{Assumption 5: Stock is primarily material or money-mediated}\label{assumption-5-stock-is-primarily-material-or-money-mediated}

Smith's three categories of stock (consumption goods, fixed capital, and circulating capital) are all material or money-denominated. His inclusion of ``acquired and useful abilities'' in fixed capital is an early gesture toward human capital, but the category is undertheorized, and the assets are not separately measurable in his framework.

The pressure point: in advanced economies, the most economically significant stock is often non-material. Organizational routines that enable rapid complex coordination, brand equity\index{brand equity} that generates consumer trust and pricing power, proprietary datasets\index{proprietary datasets} that enable prediction and personalization, trained models that distil interpretive expertise across millions of cases, institutional relationships that reduce transaction costs\index{transaction costs}, these are productive stock in any meaningful economic sense. They are accumulated through investment, generate revenue, depreciate, and can be lost. But they do not appear in national accounts or firm balance sheets in any systematic way, because the accounting frameworks that govern measurement were designed for the material and money-mediated stock that Smith's model presupposes.

The consequence is not merely a measurement gap. It is a gap in integrated capital theory and measurement. Existing literatures explain important parts of non-material productive stock, but they do not provide a unified account of its forms, conversion, governance, depreciation, impairment, and valuation. Smith's categories are insufficient, not because they were wrong for his time, but because they were designed for a different kind of economy.

\subsection{Assumption 6: Revenue resolves into wages, profit, and rent}\label{assumption-6-revenue-resolves-into-wages-profit-and-rent}

Smith's three revenue categories are meant to be exhaustive. All income in a market economy ultimately derives from wages, profit, or rent; all productive inputs are compensated through one or more of these revenue streams. The model leaves no clean mechanism for a revenue stream corresponding to governed access to knowledge itself as a primary productive input.

The pressure point: knowledge-bearing capitalism generates revenue forms that do not sit cleanly in Smith's tripartite resolution. Intellectual property rent\index{rents!intellectual property} (income derived from governed access to knowledge-bearing stock, where the return may exceed the contemporaneous productive contribution required to maintain or deploy that stock) is not profit in Smith's sense, which is the return to risk-bearing and active management of deployed capital. Platform rent\index{rents!platform} (income extracted from intermediating access to a network of users, sellers, and data without producing the underlying goods or services) is similarly anomalous. Both forms are closer to rent than to profit, but they attach not to land but to legally constructed exclusivity\index{excludability} over non-rival goods and network architectures. The issue is not that economics lacks the word rent; it is that the source of rent is governed access to non-rival knowledge-bearing stock. Smith's tripartite framework lacks a clean mechanism for non-rival, legally governed, platform-mediated knowledge rents because it was built for an economy in which the primary productive inputs were labour, material capital, and land.

\subsection{Assumption 7: Wealth expands through specialization, exchange, and accumulation}\label{assumption-7-wealth-expands-through-specialization-exchange-and-accumulation}

Smith's growth mechanism is fundamentally optimistic. Market exchange coordinates specialization. Specialization raises productive power. Accumulated stock enables further specialization. The extent of the market expands through trade. Each step reinforces the others, and the result is the progressive wealth of commercial society. There is no systemic mechanism in Smith's model by which wealth creation could undermine the conditions for future wealth creation, no internal tension between the private accumulation of capital and the public conditions for productive growth.

The pressure point: knowledge-bearing capitalism introduces a mechanism of generative suppression that Smith's model cannot accommodate. When knowledge-bearing stock is enclosed (legally restricted, technically locked, or institutionally sequestered) the stock generates private rent while simultaneously narrowing the recombination field from which future knowledge would otherwise emerge. A patent protects the patent holder's revenue while restricting others' use of the patented knowledge, including in recombinations that would produce new knowledge. A proprietary dataset confers predictive capability on its holder while denying the training inputs that would enable competitors or public research\index{public research to private IP}ers to develop comparable or complementary capabilities. An AI system trained on widely shared knowledge and deployed with enclosed access accumulates improvements from each user interaction while distributing none of that improvement back to the community whose knowledge constitutes its inputs.

Smith's division of labour and market coordination\index{market coordination} are mechanisms that increase productive power without depleting the conditions for further increase. The pin factory's operation does not reduce the future supply of wire or workers' ability to specialize. This does not apply to all enclosure; it applies when exclusion suppresses downstream recombination or feedback more than it increases creation incentives. Enclosure of non-rival knowledge stock is then of a different kind: it creates legal and technical exclusion\index{technical exclusion} around a non-rival good, generating private revenue while reducing the total productive potential of the recombination field. At sufficient scale and duration, the very process of accumulating knowledge capital suppresses the conditions from which future knowledge capital would otherwise arise.

This is a Smithian problem that arises not from exploitation within the production process; rather, it arises when property-like exclusion\index{property rights} is applied to non-rival knowledge goods. The result is that the mechanism of accumulation (Smith's engine of national wealth) can, under knowledge-bearing capitalism, simultaneously increase present productive power and reduce the generative conditions for future productive power. That is the central anomaly that the remaining chapters of this book are built to explain. Chapter 8 returns to the strategic version of this problem: the firms best positioned to prevent generative suppression may also have the strongest private incentives to practise it.

\section{Coda: The Departures Required}\label{coda-the-departures-required}

These seven pressure points do not uniformly invalidate Smith's model. Some of his mechanisms survive largely intact. The general proposition that market extent governs the degree of specialization remains broadly true, even if non-rival goods and network effects add dynamics that Smith's framework does not contain, and even if the operative variable is better understood as effective field extent than as market size in the conventional sense. The proposition that capital accumulation enables further productive specialization remains broadly true, even if the nature of the capital being accumulated has changed. The proposition that exchange coordinates specialization across a society of strangers through price signals remains broadly true, even if prices are increasingly imperfect signals for knowledge goods whose value is contextual, cumulative, and non-rival\index{price signal versus knowledge value}\index{price signals!limits for knowledge goods}.

Smith supplies the production-side baseline. Bastiat\index{Bastiat, Frederic} supplies the exchange-side warning: value appears through services rendered and received, while error arises when visible effects are counted and displaced alternatives ignored. KBC uses Smith to reconstruct specialization and stock, and Bastiat to discipline service, obstacle, and the unseen without treating him as a knowledge-economy theorist.

What does not survive intact is the specific set of assumptions about the nature of capital and productive labour that Smith's mechanisms presuppose. Capital is not reliably separable from the people whose judgement and experience make it productive in knowledge-bearing capitalism. Productive labour does not reliably leave behind a vendible material residue. The division of labour in knowledge work is partly a division and recombination of interpretive capacity\index{division of labour versus division of knowledge}\index{division of knowledge!versus division of labour}, not merely a decomposition of physical tasks. Stock is not primarily material or money-denominated. Revenue does not cleanly resolve into wages, profit, and rent when knowledge rents and platform rents are major income categories. The alignment between private incentive and social optimum may weaken when non-rival knowledge goods can be controlled through access restriction. And accumulation of knowledge capital at scale introduces the possibility that private wealth creation suppresses the public conditions for future wealth creation, a structural dynamic with no analogue in Smith's model.

Beneath these particular departures lies a single one. Smith built political economy on value in exchange and set value in use aside, a choice that rival, vendible commodities made not only workable but right. Knowledge-bearing capitalism forces the bracketed half back into account, because its wealth often lives in use-value that governance must convert before exchange-value, and the instruments built around exchange-value, can record it.

These departures require a revised theory, not a repudiation of Smith. The Smithian machine of specialization, exchange, accumulation, and national wealth describes a real and important set of mechanisms. What knowledge-bearing capitalism requires is an extension of that machine to account for a different kind of stock, a different kind of productive residue, a different kind of accumulation, and a different relationship between private appropriation\index{appropriation} and the public conditions of future generation.

That extension begins with the question of what kind of thing knowledge-bearing stock is, what makes it different from the machines, buildings, and provisions that constitute Smith's capital, what forms it takes, how it moves among those forms, and what institutional conditions determine whether it becomes a private asset, a firm capability, a platform dependency, or a commons. Chapter 2 builds this vocabulary. The earlier chapters in the theoretical literature suggest a tripartite taxonomy of embodied, disembodied, and institutionalized knowledge capital, and that tripartite distinction remains a useful first approximation. The formal theory developed here requires five forms rather than three, because the governance form and depreciation dynamics of knowledge commons knowledge capital and public epistemic capital differ structurally from those of institutionalized knowledge held within firms. That distinction, and the reasons it matters, is the subject of Chapter 2.

\chapter[Knowledge-Bearing Stock]{Knowledge-Bearing Stock: Forms, Separability, and the First-Conversion Problem}
\label{ch:knowledge-bearing-stock}

\chapterhook{The Bridge Between Human and Artificial Knowledge Capital}

In the terms of Chapter 1, this chapter is about the substrate of the conversion. Before governance can turn use-value into exchange-value, the use-value must reside somewhere, in a person, an artefact, a routine, a commons, or a piece of public infrastructure, and it is the form of that residence that decides how easily the knowledge can be separated, owned, excluded, and priced. The forms catalogued here are the forms in which knowledge holds its use-value before any gate decides what becomes of it.

It also fixes what later chapters call yield. A knowledge-bearing stock is capital not because it exists but because it renders productive service when used, and that service, the yield of the stock at the point of recombination, is the use-value quantity the rest of this book tracks. The forms set out here are the forms in which yield is stored before it is realized.

When a radiologist's diagnostic judgement\index{radiology AI example!diagnostic judgement} is partly embedded in an AI model, what has moved? The radiologist has not sold a machine in the ordinary sense. The hospital has not merely purchased software in the ordinary sense. Yet some portion of a formerly embodied diagnostic capability has been encoded, trained, tested, deployed, and made available through an artefact that can operate apart from the original expert.

The same problem appears elsewhere. When an engineer's expertise becomes a software tool, when a firm's informal operating know-how becomes a repeatable workflow, when open-source code\index{open-source code to managed cloud service} becomes a managed cloud service\index{managed cloud service}\index{open-source software!managed cloud service}\index{cloud service}, or when user behaviour\index{user behaviour} becomes model-training data\index{model-training data}, economic capacity changes form. A firm's email system\index{email system!as knowledge-bearing stock}\index{coordination costs}\index{institutional memory} gives a mundane example: it converts transient human communication into durable, searchable, governable knowledge-bearing stock that lowers coordination costs, preserves institutional memory, supports accountability, enables recombination, and sustains organizational capability. Email is valuable less as messaging software than as knowledge-bearing infrastructure\index{email system!as knowledge-bearing infrastructure}: it preserves requests, decisions, approvals, records, relationships, warnings, and commitments so they can be searched, transmitted, audited, governed, and reused. In KBC terms, email partly externalizes embodied judgement into disembodied records\index{email system!externalization of embodied knowledge}, contributes to institutionalized knowledge when retained inside routines and governance processes\index{email system!institutionalized knowledge}, and supports recombination when prior negotiations, client issues, technical fixes, incident responses, or policy interpretations become inputs to later work\index{email system!recombination value}. It also creates dark-capital exposure\index{email system!dark-capital exposure}\index{dark capital!email systems}: the same system that preserves organizational memory can concentrate sensitive data, privileged communications, credentials, trade secrets, and operational dependencies. A breach, deletion, retention failure, or search failure can destroy value even when no physical asset is harmed. Email is therefore not valuable because it merely stores messages. It is valuable because it maintains the productive relation among people, records, decisions, trust, workflow, and future action. Productive knowledge moves from persons into artefacts, routines, platforms, models, databases, organizational structures, or legal rights. It may become easier to scale, easier to own, easier to exclude, easier to licence, easier to measure, or easier to mismeasure.

These movements are now central to firm value. A company may depend on software it did not write, data it does not fully understand, routines that survive only because a small group of employees maintains them, or an AI model whose value depends on continuous user feedback\index{AI systems!user feedback}\index{user feedback!AI model}. A firm may lose productive capacity when key personnel leave, when documentation decays, when an API closes\index{API closure!productive capacity loss}, when an open-source maintainer exits\index{open-source maintainer shock!Chapter 2 example}, when a model is no longer updated, or when a dataset loses provenance. Yet much of this productive capacity may not appear clearly on the balance sheet, even when investors, managers, and competitors know that it matters.

\begin{center}
\fbox{\begin{minipage}{0.92\textwidth}
\small
\textbf{Running case: API closure/access restriction.}\index{API closure!running case}\index{access restriction!API running case} A platform API is a useful first case because the knowledge-bearing stock is identifiable: a structured interface to platform-held data, user activity, documentation, developer routines, and operational feedback. Before closure, downstream actors may treat the API as part of their accessible recombination field. After closure or repricing, the same stock remains technically present but the governance condition changes: access, permission, feedback, and recombination become conditional on the platform's rules.
\end{minipage}}
\end{center}

The puzzle is therefore not simply that modern economies use more information. It is that productive knowledge now appears in people, artefacts, institutions, commons, and public systems whose economic properties differ sharply. A worker's tacit judgement, a software library\index{software library}, a trained model, a patent, a professional standard, an organizational routine, a public dataset, and an open-source project may all support production. But they do not behave alike. They differ in where the knowledge resides, how it persists, whether it can be separated from the person, team, community, or setting in which it first became productive, who can access it, who can exclude others from it, how it depreciates, and whether it can be recombined into future value.

Data therefore requires a boundary condition. Data is one important form of knowledge-bearing stock, but it is not identical with knowledge-bearing stock as such. Data may be non-rival, tradeable, depreciating, and generated as a by-product of transactions. Ideas, software, routines, professional judgement, institutional procedures, model weights, embodied expertise\index{model weights!as knowledge-bearing stock}\index{embodied expertise}, and institutional capability obey related but not identical dynamics. A dataset may carry signals; an idea may specify instructions; software may operationalize a procedure; model weights may encode learned parameters; embodied expertise may reside in trained judgement; and institutional capability may persist in roles, routines, standards, and coordination practices. KBC therefore treats data economics as a central predecessor and special case, not as this whole theory \parencite{FarboodiVeldkamp2021, JonesTonetti2020, LambrechtTucker2017}\index{Farboodi and Veldkamp}\index{Jones and Tonetti}\index{Lambrecht and Tucker}.

Consider the radiology example more closely. A radiologist does not merely possess information about images. She possesses a trained diagnostic capability: pattern recognition, anatomical knowledge, probabilistic judgement, familiarity with false positives, awareness of clinical context, and the ability to decide whether an observed feature is medically significant. That capability is embodied in a person. It is not simply a file, a rule, or a fact.

Now suppose a hospital deploys an AI model that can perform part of that diagnostic task. The model does not become a radiologist in the full professional sense. It does not assume legal responsibility, understand the patient's total medical history, participate in clinical judgement as a human professional, or carry the same institutional obligations. But it may perform a bounded operation that was previously performed by embodied expert judgement: identify, classify, rank, or flag features in medical images with sufficient reliability for a defined use.

That is the conceptual bridge. The relevant economic question is not whether the model ``knows'' in the same way the radiologist knows. The question is whether a bounded productive operation once carried by embodied expertise has been externalized into a disembodied artefact\index{embodied versus disembodied knowledge}\index{embodied knowledge@embodied knowledge (\ensuremath{K^E})!versus disembodied knowledge}\index{disembodied knowledge@disembodied knowledge (\ensuremath{K^D})!versus embodied knowledge} that can be deployed, scaled, governed, improved, licensed, audited, or substituted into a workflow.

This movement creates the unit problem. What exactly has moved from worker to model? Not the whole person. Not the whole profession. Not all medical judgement. What has moved is a bounded operative capability: a specific diagnostic operation under specified conditions, with specified inputs, quality thresholds, risks, and governance rules.

This is why Chapter 2 introduces the Operative Knowledge Unit, or OKU.\index{Operative Knowledge Unit (OKU)|textbf}\index{OKU|see{Operative Knowledge Unit (OKU)}} An OKU is the smallest deployable operation-bearing unit of productive knowledge in a defined domain. It is not a universal unit of all knowledge. It is a practical unit for asking whether a productive operation is performed through a person, a team, a routine, a software system, a model, a commons, or an institution.

Before introducing the formal vocabulary, Table~\ref{tab:ch2:roadmap} previews the concepts that Chapter 2 will use. The purpose is not to add terminology for its own sake, but to give the reader a map of the economic problems each term solves.

\begingroup
\small
\setlength{\tabcolsep}{3pt}
\renewcommand{\arraystretch}{1.14}
\sloppy
\par\addvspace{0.8\baselineskip}\noindent
\begin{longtable}{@{}L{0.20\textwidth}L{0.26\textwidth}L{0.28\textwidth}L{0.20\textwidth}@{}}
\caption{Roadmap of Chapter 2 Concepts}\label{tab:ch2:roadmap}\\
\toprule\noalign{}
\begin{minipage}[b]{\linewidth}\raggedright Concept\end{minipage} &
\begin{minipage}[b]{\linewidth}\raggedright Plain-language meaning\end{minipage} &
\begin{minipage}[b]{\linewidth}\raggedright Practical question\end{minipage} &
\begin{minipage}[b]{\linewidth}\raggedright Example\end{minipage} \\
\midrule\noalign{}
\endfirsthead
\toprule\noalign{}
\begin{minipage}[b]{\linewidth}\raggedright Concept\end{minipage} &
\begin{minipage}[b]{\linewidth}\raggedright Plain-language meaning\end{minipage} &
\begin{minipage}[b]{\linewidth}\raggedright Practical question\end{minipage} &
\begin{minipage}[b]{\linewidth}\raggedright Example\end{minipage} \\
\midrule\noalign{}
\endhead
\bottomrule\noalign{}
\endlastfoot
\textbf{Knowledge-bearing stock} & A durable residence or operative form of productive knowledge. & Where does productive capacity reside? & Software, model, dataset, routine. \\
\textbf{Knowledge-bearing capital} & Knowledge-bearing stock that yields productive services\index{productive services} over time. & Does this stock actually produce economic value? & A deployed AI model that improves diagnostic throughput. \\
\textbf{Operative Knowledge Unit, OKU} & A task-denominated unit of productive knowledge. & What exactly is being substituted, transferred, or automated? & AI reading a scan for a specified abnormality. \\
\textbf{Separability} & Whether knowledge can operate apart from its original person, team, artefact, organization, or setting. & Can the firm retain or redeploy it after the worker leaves? & Documented process, trained model, reusable codebase. \\
\textbf{Five knowledge forms} & The main locations where productive knowledge resides. & Is the knowledge embodied, encoded, organizational, shared, or infrastructural? & Expert skill, code, routine, open-source project, public standard. \\
\textbf{Knowledge potential, impedance, and yield} & Possible value, barriers to use, and realized productive service. & Why does valuable-looking knowledge sometimes produce little value? & Dataset with high potential but poor provenance. \\
\textbf{First conversion} & The first institutional assignment of newly generated knowledge. & Who gets the initial claim? & Patent, publication, trade secret, public dataset. \\
\textbf{Governance form} & The governance form around the stock. & Who can access, modify, exclude, monetize, or maintain it? & Commons, platform, firm capability, private IP. \\
\textbf{Knowledge-stock life cycle} & How knowledge stock emerges, matures, decays, or becomes impaired. & What causes productive knowledge to lose value? & Documentation decay, model drift, maintainer exit. \\
\textbf{KGM / KCM link} & Generation explains how new stock arises; conversion explains what happens after it exists\index{generation versus conversion}\index{knowledge generation!versus knowledge conversion}\index{knowledge conversion!versus knowledge generation}. & Are we analysing creation, movement, control, or appropriation? & New model design enters employment contract, platform, or IP governance. \\
\end{longtable}
\endgroup

The table also signals a boundary. Chapter 2 does not yet present the full generation or conversion models. It supplies the vocabulary those models require. Chapter 3 asks how new knowledge-bearing stock is generated. Chapter 4 asks how existing knowledge-bearing stock is transformed, accessed, governed, enclosed, appropriated, or shared. The concepts introduced here are therefore not decorative definitions. They are the objects the later models operate on.

The radiology case shows why ordinary categories are insufficient. ``Labour'' names the human performance of the task, but not the knowledge-bearing operation embedded in that performance. ``Software'' names the artefact, but not the productive diagnostic capability it carries. ``Human capital'' captures investment in the radiologist, but not the movement from embodied judgement into a deployable model. ``Intangible asset'' may capture some future benefit, but not the specific operation being transferred, substituted, governed, or recombined. The OKU is introduced to identify that operation.

Inherited economic vocabulary captures parts of this problem, but not the whole. Human capital theory\index{human capital theory}\index{human capital} explains investments in persons. Intangible-asset theory\index{intangible assets} explains some nonphysical claims to future benefit. Intellectual-property theory explains legal exclusion\index{legal exclusion}. Knowledge-management theory\index{knowledge management} explains some managerial practices. Resource-based and knowledge-based theories of the firm\index{resource-based view}\index{knowledge-based view} explain firm-level capability. Austrian knowledge theory\index{Austrian knowledge theory} explains dispersed knowledge and coordination. Endogenous growth theory\index{endogenous growth theory} explains the importance of non-rival ideas. Each is useful. None by itself gives a unified vocabulary for tracking productive knowledge as it moves among persons, artefacts, firms, platforms, commons, and public infrastructure.

Chapter 1 established seven pressure points where Smith's assumptions strain under knowledge-bearing capitalism\footnote{In this book, \emph{A Knowledge Theory of Capital: The Value of Natural and Artificial Intelligence} names the book and its analytical programme; knowledge-bearing capitalism names the theoretical concept being analysed, the form of capitalism in which wealth increasingly depends on the generation, conversion, governance, and recombination of knowledge-bearing stock.}. Understanding those pressure points requires vocabulary that Smith's framework does not supply and that existing economic theory supplies in domain-specific ways. Chapter 2 supplies that vocabulary. Its central object is \textbf{knowledge-bearing stock}: accumulated artefacts, routines, systems, rights, records, capabilities, or infrastructures that carry productive knowledge without being identical to knowledge itself. Chapter 2 distinguishes knowledge-bearing stock from knowledge-bearing capital\index{knowledge-bearing stock!distinguished from knowledge-bearing capital}, introduces the unit problem through the Operative Knowledge Unit, separates embodied, disembodied, institutionalized, commons, and public epistemic forms, and defines the conditions under which knowledge becomes separable, governable, productive, depreciable, and measurable.

The purpose is not to multiply terminology. It is to make visible what otherwise remains blurred: where productive knowledge resides, when it becomes capital-like, how it changes form, who controls it, how it yields services, and what is lost when it is converted, enclosed, depleted, or mismeasured.

Established literatures already explain parts of the problem: human capital explains skills held and exercised by people; intangible-capital research explains recognition and measurement; IP explains legal exclusion; platform and data economics explain coordination, lock-in\index{lock-in}, and data-enabled learning; RBV/KBV explain firm capability; endogenous growth explains non-rival ideas; Austrian and information economics explain dispersed knowledge and asymmetric exchange. KBC uses these as boundary conditions, then tracks the movement of productive knowledge across persons, artefacts, firms, platforms, commons, and public infrastructure.

\section{Knowledge-Bearing Stock}\label{knowledge-bearing-stock}
\index{knowledge-bearing stock|textbf}\index{knowledge capital|textbf}\index{knowledge-bearing capital|see{knowledge capital}}

\noindent Firms, investors, and policymakers often know that software, datasets, routines, models, standards, and employee expertise matter, but inherited categories do not say cleanly what kind of economic object these are. Calling them ``knowledge'' is too vague; calling them ``capital'' too quickly hides where the knowledge lives, who governs it, and what service flow it can produce.\index{knowledge versus capital} The first task is therefore to distinguish knowledge itself from the stocks that bear, store, operationalize, or restrict it\index{knowledge versus knowledge-bearing stock}\index{knowledge-bearing stock!distinguished from knowledge}.

The practical anomaly is that productive capacity now moves across people, artefacts, firms, platforms, commons, and public infrastructure. A senior engineer's judgement may become a deployment tool; a radiologist's pattern recognition may become a model-mediated triage process; open-source code may become a managed cloud service; user behaviour may become model-training stock. In each case, something economically real has moved, but inherited vocabulary often names only one side of the movement.

Existing terms are useful but partial: human capital is person-anchored, intangible assets are accounting-bound, IP is legal-control focused, RBV/KBV are firm-level, and endogenous growth is aggregate. Knowledge-bearing stock gives these frameworks a common object. The term does not replace those inherited frameworks; it gives them a common object. Knowledge-bearing stock means accumulated artefacts, records, systems, routines, rights, capabilities, or infrastructures that store, encode, transmit, operationalize, restrict, or amplify knowledge without being identical to knowledge itself.

\noindent This distinction prevents conceptual inflation. Not every stock of knowledge-bearing material is capital in the economic sense. A dataset, model, patent, routine, or manual may exist without producing useful services. For this theory to remain disciplined, it must distinguish where productive knowledge resides from the smaller subset of knowledge-bearing stocks that actually yield productive services.

The distinction between \emph{stock} and \emph{capital} follows Smith deliberately. \textbf{Knowledge-bearing stock} is the residence or operational form of productive knowledge: an artefact, record, routine, model, dataset, right, capability, process, commons, or infrastructure. \textbf{Knowledge-bearing capital} is such stock when it yields, or is credibly positioned to yield, productive services under governance and capability conditions. The capital-services discipline\index{capital services} is the limiting rule: stock is not capital merely because it contains knowledge\index{stock versus capital|textbf}\index{knowledge-bearing stock!versus knowledge-bearing capital}\index{knowledge-bearing capital!versus knowledge-bearing stock}; it becomes capital when productive services can actually be drawn from it.

This boundary map explains why a new object is needed before the formal models begin. Without it, the analysis would keep switching among person-level, firm-level, legal, accounting, platform, and aggregate vocabularies without a stable object to track.

\textbf{Truth-dependence qualification.} Not every information-bearing stock qualifies as knowledge-bearing capital. Durable knowledge-bearing capital must be truth-tracking, reliable, validated, and action-guiding in context. False or unreliable models, datasets, protocols, or AI outputs may have exchange or speculative value, but they create epistemic risk rather than durable productive capacity. The Technical Companion, Appendix B states the formal condition; The Technical Companion, Appendix D incorporates reliability into K-CMM. Chapter~3 models how \(\tau_i\) can fall endogenously when generation recycles unvalidated machine output, the recursive truth-decay mechanism.

\section{The Operative Unit of Knowledge Capital}\label{the-operative-unit-of-knowledge-capital}

Return to the radiology case\index{radiology AI example!OKU bridge}. The economic object that moves from worker to model is not radiology as a profession, the whole person, or all clinical judgement. It is a bounded diagnostic operation: for example, detecting, classifying, ranking, or flagging a specified abnormality in a specified image type, under specified performance thresholds, clinical workflow conditions, and liability rules\index{liability rules}.

This is a task-equivalence problem before it is a notation problem. A model does not need to become a radiologist to substitute for part of a radiologist's work. It must perform the same defined productive operation with sufficient fidelity in the relevant domain, over the relevant time horizon, in the relevant context, and under a judgement standard that users, regulators, clinicians, or institutions accept.

The \textbf{Operative Knowledge Unit (OKU)} names that bounded operation-bearing unit. An OKU is a \textbf{decision-relative, domain-bounded, equivalence-specified} unit of productive knowledge. It does not create a universal cardinal unit\index{cardinal aggregation problem} of knowledge. It specifies the task, domain, actor, performance threshold, and equivalence condition under which embodied, disembodied, institutionalized, commons, or public epistemic knowledge can be compared for a particular decision. In operational terms, an OKU is the smallest deployable operation-bearing unit of productive knowledge capable of engaging a domain problem and generating domain-specific yield under stated conditions. It lets this theory ask what is being performed, substituted, transferred, automated, governed, or valued.

\begin{figure}[!htbp]
\caption[OKU comparison: bounded task, not whole-person equivalence]{OKU comparison: bounded task, not whole-person equivalence}
\label{fig:ch2:oku-turing-comparison}
\centering
\includegraphics[width=0.92\textwidth]{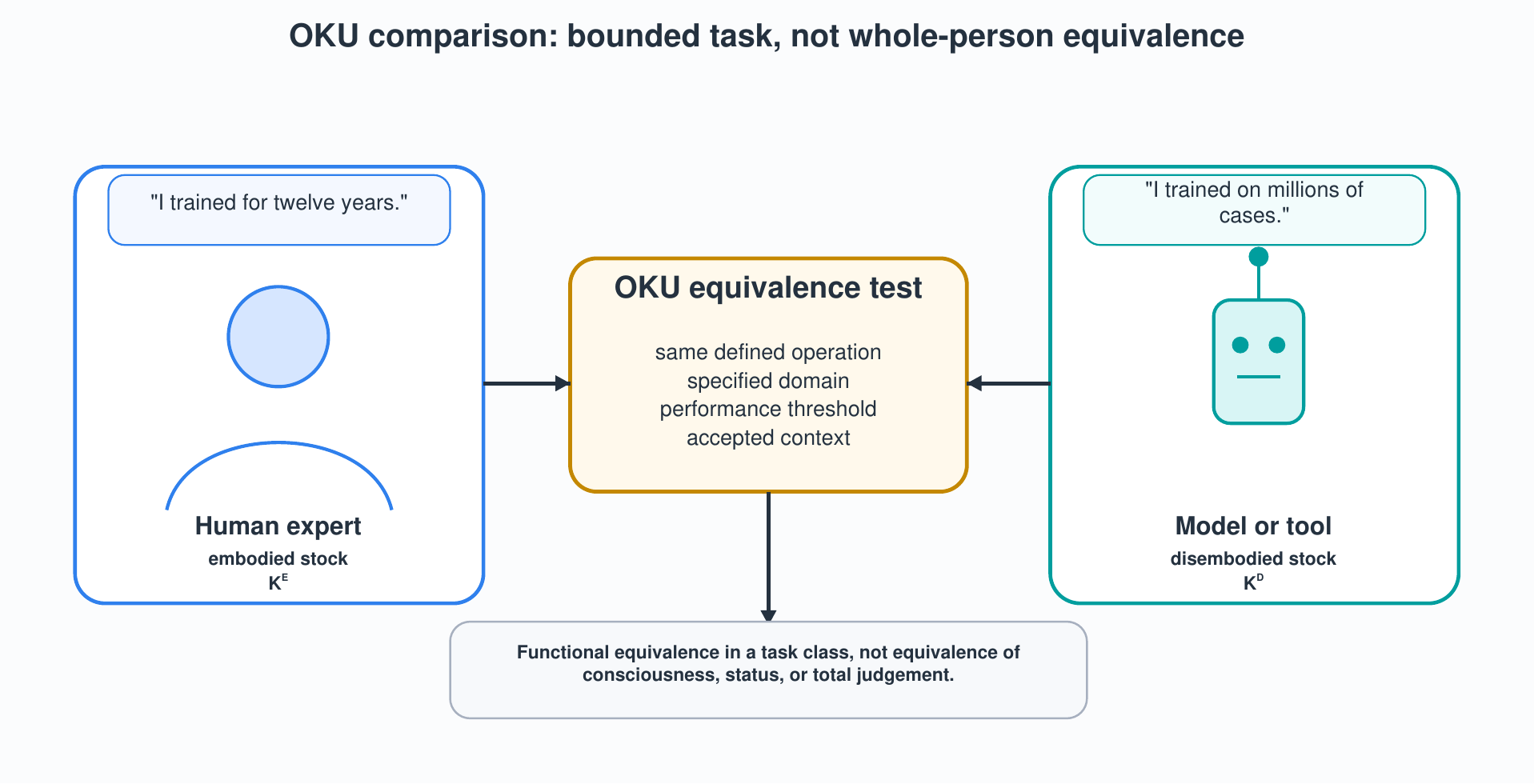}
\par\smallskip\noindent\footnotesize\emph{Note.} The OKU comparison is deliberately narrower than a full Turing-style equivalence claim. It asks whether a bounded operation can be performed under a specified task class, domain, performance threshold, and accepted context.
\end{figure}

The elementary unit of knowledge-bearing capital is therefore not the smallest item of information. A fact, formula, code snippet, procedure, dataset entry, or isolated skill may be knowledge-bearing stock, but it is not yet an operative unit unless it can be deployed as an independently productive operation.\index{fragment versus operative unit}\index{operative unit!versus fragment} The OKU gives Smith's ordinary economic fact a bounded denominator: the pin-maker, surgeon, mechanic, programmer, radiologist, model, routine, or institutional protocol is economically relevant because it can perform a productive operation in a domain.

The formal concept is:

\begin{equation}
\text{OKU}^{x}_{z, d, t}
\label{eq:ch2:oku-x-z-d-t}
\end{equation}

where \(x\) denotes the K-form \(E, D, I, C\). Public epistemic infrastructure \(K^P\) is treated separately because it is usually system-level infrastructure rather than an independently deployable task unit; specific public epistemic objects may nevertheless become boundary-case \(OKU^P\)-like units when they perform bounded operations across actors. Here \(z\) denotes the specific person, artefact, system, structure, or governance arrangement, \(d\) denotes the domain, and \(t\) denotes time. The notation is registered in the Technical Companion, Appendix B, §B.14.

\subsection[Embodied operative units and the embodied unit denominator]{\texorpdfstring{Embodied operative units: \(OKU^E\) and the \(1K^E\) denominator}{Embodied operative units and the embodied unit denominator}}\label{embodied-operative-units-okue-and-the-1ke-denominator}

For embodied knowledge capital, the operative unit is one person-equivalent:

\begin{equation}
\text{OKU}^{E}_{a, d, t}
\label{eq:ch2:oku-e-a-d-t}
\end{equation}

where \(a\) is the embodied actor, \(d\) is the domain, and \(t\) is time. One surgeon, one programmer, one clinician, or one analyst may each constitute an embodied operative knowledge unit within a defined domain. This does not imply that all persons are equally productive or that one surgeon is commensurable with one programmer. It means only that, within the defined domain, the person's skill and judgement can independently perform the specified operation.

The base embodied unit (the unit denominator) may also be written:

\begin{equation}
1K^E_{d, t}
\label{eq:ch2:1k-e-d-t}
\end{equation}

meaning one person-equivalent of embodied knowledge capital in domain \(d\) at time \(t\). The distinction between the unit denominator and a specific operative unit matters: \(1K^E_{d, t}\) is the standard; \(OKU^E_{a, d, t}\) is a particular person deploying that standard within a specific context.

The denominator does not eliminate heterogeneity. A senior surgeon and a junior surgeon may each count as one embodied operative unit, but they will not generate the same yield. That is a quality and productivity problem, not a unit-definition problem. Quality, reliability, contextual fit, and yield are handled separately through the K-yield coefficients \(\omega\), \(\chi\), and \(\rho\), and through the K-CMM measurement architecture.

This is analogous to dimensional analysis. A metre does not make steel and air equivalent in economic or physical significance; it defines a length dimension. Likewise, \(1K^E_{d, t}\) defines a domain-specific embodied denominator. It does not collapse all domains into one commensurable substance.

\subsection[Disembodied operative units]{\texorpdfstring{Disembodied operative units: \(OKU^D\)}{Disembodied operative units}}\label{disembodied-operative-units-okud}

For disembodied knowledge capital, the operative unit is the smallest functionally unified system capable of independently engaging the domain problem:

\begin{equation}
\text{OKU}^{D}_{m, d, t}
\label{eq:ch2:oku-d-m-d-t}
\end{equation}

where \(m\) is the model, module, software system, dataset-interface, template, protocol artefact, or other bounded disembodied system. Functional unity determines the boundary. A component that cannot independently generate domain yield is a fragment of the larger \(OKU^D\). A component that can independently engage a domain problem and generate yield may itself be counted as a separate operative unit.

This boundary rule prevents artificial subdivision. A software platform is not many operative units merely because it contains many files or modules. The test is independent domain performance, not architectural decomposition.

\subsection{Turing-style equivalence as a cross-form bridge}\label{turing-style-equivalence-as-a-cross-form-bridge}

Turing-style equivalence supplies a bridge between embodied and disembodied operative units. A bounded disembodied operative unit achieves functional equivalence with an embodied operative unit when qualified judges cannot reliably distinguish their domain performance under specified conditions:

\begin{equation}
\text{OKU}^{D}_{m, d, t}
\equiv_{F, c, \tau, J}
\text{OKU}^{E}_{a, d, t}
\label{eq:ch2:oku-d-m-d-t-2}
\end{equation}

This means equivalence holds for function \(F \subset d\), through channel \(c\), over evaluation interval \(\tau\), under judge population \(J\). The equivalence is conditional, not absolute. It applies to a defined task class under defined constraints, not to the whole person, occupation, or domain.

The Turing equivalence test does not define the unit. It tests whether a \(K^D\) unit has achieved functional equivalence with a \(K^E\) unit. The unit comes first; equivalence testing is performed against it.

This matters for AI substitution. A radiology AI system may achieve functional equivalence with an \(OKU^E\) for preliminary image classification without achieving it for clinical judgement, patient-facing communication, liability-bearing diagnosis, or ambiguous escalation. The task-class limitation is essential: substitution does not occur at the level of an occupation unless all material task classes in that occupation independently cross the equivalence threshold.

The Turing-style bridge therefore provides a principled way to express some \(K^D\) capacity in embodied person-equivalent terms without claiming universal cardinal aggregation. It supports claims such as:

\[
OKU^D_{m, d, t}
\equiv_{F, c, \tau, J}
OKU^E_{a, d, t}
\]

It does not support a claim such as:

\[
K^{D}_{\mathrm{model}} = nK^E
\]

Such a claim is not valid unless the domain, task class, evaluation\index{valuation} channel, interval, judge population, and calibration procedure are specified.

\subsection{Limits and caveats of the OKU bridge}\label{limits-and-caveats-of-the-oku-bridge}

The OKU bounds the unit problem; it does not eliminate measurement difficulty. It identifies the operation being counted and compared, but it does not determine the unit's value, quality, durability, truth-reliability, or cross-domain exchange rate. A cardiac-surgery OKU and a mechanic-diagnostics OKU are both economically real, but they are not immediately commensurable.

The OKU also does not create a universal unit of all knowledge. It does not make commons knowledge capital or public epistemic infrastructure easy to measure, and public operative units\index{public operative units} remain boundary cases. It does not remove judgement from domain definition, since someone must still specify the domain, operation, context, fidelity threshold, and judgement standard that make the unit meaningful. Nor does it yet provide empirical coefficients for quality, capability, reliability, substitution, or yield. Its contribution is narrower but important: it gives this theory a disciplined unit grammar for later measurement.

The OKU is not the knowledge equivalent of the dollar. It does not create a universal cardinal unit of all knowledge. It is a unit grammar: it identifies the bounded operative knowledge object whose productive services may later be compared, substituted, valued, or aggregated under specified conditions. Chapter~9 returns to the separate problem of monetary commensuration and explains why valuation, aggregation, calibration, and accounting recognition must be distinguished.

The OKU identifies the operative unit whose productive services may later be compared or valued.\index{unit definition versus valuation} It does not imply that knowledge-bearing stocks add arithmetically. Because non-rival stocks often create value through complements, recombination, and learning loops, later valuation must often occur at the portfolio level. Chapter~9 introduces the Knowledge Portfolio Value Function for this purpose.

The OKU bridge asks whether a unit can perform a specified operation. It does not, by itself, establish whether the operation is economically wanted. A radiology AI may perform a diagnostically equivalent operation, but yield depends on whether that operation meets a live clinical, institutional, legal, or workflow need. Functional equivalence must still become productive service.

The KBC framework therefore distinguishes two problems and treats only the first as bounded by OKU notation:

\begingroup
\small
\setlength{\tabcolsep}{3pt}
\renewcommand{\arraystretch}{1.12}
\sloppy
\par\addvspace{0.8\baselineskip}\noindent
\begin{longtable}{@{}L{0.27\textwidth}
L{0.33\textwidth}
L{0.34\textwidth}@{}}
\caption{Operative Unit and Cardinal Aggregation Problems}\label{tab:ch2:operative-unit-problems}\\
\toprule\noalign{}
\begin{minipage}[b]{\linewidth}\raggedright
Problem
\end{minipage} & \begin{minipage}[b]{\linewidth}\raggedright
Question
\end{minipage} & \begin{minipage}[b]{\linewidth}\raggedright
Status in this framework
\end{minipage} \\
\midrule\noalign{}
\endfirsthead
\toprule\noalign{}
\begin{minipage}[b]{\linewidth}\raggedright
Problem
\end{minipage} & \begin{minipage}[b]{\linewidth}\raggedright
Question
\end{minipage} & \begin{minipage}[b]{\linewidth}\raggedright
Status in this framework
\end{minipage} \\
\midrule\noalign{}
\endhead
\bottomrule\noalign{}
\endlastfoot
\textbf{Operative-unit problem} & What is the smallest independently operative bearer of productive knowledge under specified domain, time, fidelity, context, and judgement conditions? & Bounded for \(K^E\), \(K^D\), and \(K^I\); provisional for \(K^C\); boundary-case treatment for \(K^P\) through \(OKU^P\)-like public objects \\
\textbf{Cardinal aggregation problem} & How can heterogeneous knowledge-bearing stocks be summed across domains, forms, actors, and governance forms? & Not solved; treated as calibration, weighting, and valuation work \\
\end{longtable}
\endgroup

The operative unit sits at the third level of the following hierarchy:

\begingroup
\small
\setlength{\tabcolsep}{3pt}
\renewcommand{\arraystretch}{1.12}
\sloppy
\par\addvspace{0.8\baselineskip}\noindent
\begin{longtable}{@{}L{0.27\textwidth}
L{0.33\textwidth}
L{0.34\textwidth}@{}}
\caption{Levels of Knowledge-Capital Unit Analysis}\label{tab:ch2:unit-analysis-levels}\\
\toprule\noalign{}
\begin{minipage}[b]{\linewidth}\raggedright
Level
\end{minipage} & \begin{minipage}[b]{\linewidth}\raggedright
Question
\end{minipage} & \begin{minipage}[b]{\linewidth}\raggedright
KBC function
\end{minipage} \\
\midrule\noalign{}
\endfirsthead
\toprule\noalign{}
\begin{minipage}[b]{\linewidth}\raggedright
Level
\end{minipage} & \begin{minipage}[b]{\linewidth}\raggedright
Question
\end{minipage} & \begin{minipage}[b]{\linewidth}\raggedright
KBC function
\end{minipage} \\
\midrule\noalign{}
\endhead
\bottomrule\noalign{}
\endlastfoot
Fragment & What item of knowledge exists? & Fact, rule, code snippet, skill element, method \\
Carrier & Where does knowledge reside? & Person, model, software system, routine, institution, commons \\
\textbf{Operative Knowledge Unit} & \textbf{Can the unit independently engage a domain problem and generate yield?} & \textbf{Defines the elementary unit} \\
Yield & What productive output does the unit generate? & Requires quality, capability, truth-reliability, and realization coefficient\index{realization coefficient}s \\
Equivalence & Can another K-form substitute for it? & Requires functional equivalence testing \\
Valuation & What is the unit worth? & Requires market, institutional, accounting, or social valuation \\
\end{longtable}
\endgroup

This hierarchy prevents a category error. The operative unit is not supposed to solve valuation or perfect measurement. It identifies the bounded object being valued. Critics who object that the OKU framework does not determine how much a surgeon's knowledge is worth are asking valuation to do the work of unit definition.

Capital theory\index{capital theory} has long faced this aggregation problem, and the Cambridge Capital Controversy\index{Cambridge capital aggregation problem} showed that it is deeper than a missing convention. Reswitching and capital-reversing established that there is in general no way to define an aggregate quantity of heterogeneous physical capital\index{physical capital} independent of the distribution of income and the prices and rate of return that the aggregate is then supposed to explain \parencite{Samuelson1966, CohenHarcourt2003}\index{Samuelson, Paul}\index{Cohen and Harcourt}. The difficulty is conceptual, not merely institutional: even physical capital has no distribution-free cardinal measure\index{distribution-free capital measure}. This cuts two ways for the present theory, and both must be stated. As a \emph{shield}, it means knowledge-bearing capital is no worse off than physical capital: the absence of a clean cardinal unit is the normal condition of capital theory, not a special embarrassment of knowledge, so the lack of a standardized knowledge-service unit is no reason to deny knowledge-bearing stock capital-good status\index{capital goods}. As a \emph{threat}, it means the measurement apparatus this book develops, the K-CMM valuation of Chapter~9 and the Technical Companion, inherits the same incoherence: any aggregate it reports is valuation-relative and price-relative, not a distribution-free magnitude\index{price-relative valuation}. The book accordingly does not claim a cardinal total of knowledge capital. Its formal results rest on within-domain orderings\index{within-domain ordering}, ratios, and marginal comparisons (Chapter~11), and K-CMM is presented throughout as conditional, governance-indexed valuation\index{governance-indexed valuation} under a stated price environment, not as the recovery of an objective scalar. The Cambridge result is therefore not a reason to abandon knowledge-capital valuation but a discipline on what such valuation can claim, the same discipline physical-capital theory has worked under since the 1960s.

Domain specificity is the same kind of problem. Monetary systems handle different currencies through denominations, exchange rates, purchasing-power comparisons, settlement institutions, and market conversion. Knowledge-capital units require the corresponding mechanisms: domain-specific prices, wages, training costs, certification barriers, risk-reduction estimates, performance standards, scarcity measures\index{scarcity}, substitution tests, and expected productive-service yields. Cross-domain incommensurability is therefore not a defect of this theory. It is the ordinary condition that gives rise to domain-specific pricing and knowledge-capital exchange-rate analogues.

\subsection[Institutionalized Operative Knowledge Units]{\texorpdfstring{Institutionalized Operative Knowledge Units: \(OKU^I\)}{Institutionalized Operative Knowledge Units}}\label{institutionalized-operative-units-okui}

For institutionalized knowledge capital, the operative unit is structurally rather than personally independent:

\begin{equation}
\text{OKU}^{I}_{s, d, t}
\label{eq:ch2:oku-i-s-d-t}
\end{equation}

where \(s\) is a routine, protocol, role-system, governance arrangement, standard, or organizational subsystem. An institutionalized operative unit is the smallest organizational arrangement capable of producing repeatable domain value independently of any particular person occupying it. Unlike \(K^E\), whose independence is person-dependent, \(K^I\) independence is structural: the unit persists across personnel changes and governs behaviour through roles, routines, standards, or organizational memory.

\(K^I\) is not merely aggregated \(OKU^E\). A hospital's surgical protocol, for example, encodes collective learning accumulated across many practitioners, patients, administrators, regulators, insurers, and adverse events. No individual fully carries it. It persists through structure, governs behaviour across personnel changes, and preserves productive knowledge beyond the tenure of any particular person. For this reason:

\begin{equation}
\text{OKU}^{I}_{s, d, t}
\neq
\sum_a \text{OKU}^{E}_{a, d, t}
\label{eq:ch2:oku-i-s-d-t-2}
\end{equation}

The institutional unit is not the sum of the embodied units currently present in the organization. It is a structurally persistent knowledge arrangement that shapes what those embodied units can do. This inequality is not only conceptual: it grounds the institutional-residue prediction tested in Chapter~11, the one prediction in this book that cannot be formed without the operative unit as a unit.

This distinction is foundational for understanding the commons-depletion mechanism. When incumbents extract \(OKU^E\) (key persons) from a commons or institution, the \(OKU^I\) (the governance and maintenance structures those persons sustained) does not automatically follow. It may persist, degrade, or collapse depending on how deeply it was embedded in organizational form rather than in individuals. What is extracted and what remains are analytically distinct events.

\subsection[Commons Operative Knowledge Units]{\texorpdfstring{Commons Operative Knowledge Units: \(OKU^C\)}{Commons Operative Knowledge Units}}\label{commons-operative-units-okuc}

For commons knowledge capital, the operative unit is provisionally defined as:

\begin{equation}
\text{OKU}^{C}_{g, d, t}
\label{eq:ch2:oku-c-g-d-t}
\end{equation}

where \(g\) is a self-sustaining contributor-governance structure capable of maintaining shared domain knowledge. This formulation is stronger than merely saying ``community'' or ``commons stock.'' The unit is not the shared artefact alone. It is the smallest maintenance-capable governance structure that can preserve, update, and discipline the shared stock.

A Wikipedia\index{Wikipedia} article is not, in general, an \(OKU^C\) unless it has a self-sustaining maintenance community and governance pattern. A well-maintained open-source subsystem with stable maintainers, contribution rules, review processes, error-correction mechanisms, and independent domain utility may qualify.

The open problem is calibration: contributor depth, review capacity, release cadence\index{release cadence}, and continuity under member turnover. This calibration question is the \(K^C\) analogue of the quality-within-domain problem for \(K^E\), and it belongs to the measurement programme of Chapter 11 rather than to the unit definition itself.

\subsection[Public epistemic infrastructure and boundary-case operative units]{Public epistemic infrastructure and boundary-case operative units}
\label{public-operative-units-and-public-epistemic-infrastructure}

Public epistemic infrastructure \(K^P\) is a core knowledge form, but it is usually not itself an Operative Knowledge Unit. It is the background infrastructure that makes operative knowledge definable, comparable, trustworthy, auditable, and transferable. A specific public output may function as a boundary-case operative unit when it performs a bounded epistemic or governance operation across actors: a NIST standard\index{NIST standards}\index{standards bodies!NIST}\index{standards maintenance}, an ISO control requirement, a legal test\index{legal classifications}\index{courts!legal tests} from a Supreme Court judgment, a statistical classification\index{statistical classifications}\index{public datasets}, or a measurement protocol.

The distinction should remain bounded. NIST, ISO, supreme courts, statistical agencies, metrology bodies, public research agencies, sport and market-governance bodies, and standards communities\index{standards governance}\index{standards bodies} are normally \(K^P\)-generating institutions or processes, not operative units. Their accumulated outputs form public epistemic stock. Only a specific output becomes an operative-unit-like object, and only when it performs a bounded classification, measurement, validation, coordination, trust, or governance function. In such boundary cases, the notation may be written as:

\begin{equation}
\text{OKU}^{P}_{p, d, t}
\label{eq:ch2:oku-p-p-d-t}
\end{equation}

where \(p\) is a public epistemic output, such as a NIST standard, ISO standard\index{ISO standards}\index{standards bodies!ISO}, FIA rule, NHL rule, legal test, statistical classification, or measurement protocol, that performs a bounded operative function in domain \(d\) at time \(t\). The rule for this theory is therefore simple: \(K^P\) is a core form; public operative units are boundary cases; future formalization of public-epistemic generation belongs in the Technical Companion.

\subsection{Formal status of the OKU framework}\label{formal-status-of-the-oku-framework}

\noindent\fbox{%
\begin{minipage}{0.94\linewidth}
\small
\textbf{Formal-status note.} The OKU framework is a bounded operationalization, not a universal meter for all knowledge. Its purpose is to identify operative units within specified domains, times, contexts, fidelity standards, and judgement standards. The later formal results therefore rely on orderings, ratios, and within-domain comparisons rather than on a single cardinal measure of total knowledge capital. In plain terms, the argument does not require the claim that a firm has exactly 5,000 units of knowledge capital. It requires the weaker and more defensible claim that, under stated conditions, governance changes can reduce accessible recombination fields, widen capability gaps, lower trajectory diversity\index{trajectory diversity}, or make strategic enclosure privately rational while socially costly under Chapter 8's welfare specification. The full notation and proof structure are developed in the later mechanism chapters and the Technical Companion, Appendix B. The operative-unit decomposition also earns its keep empirically rather than only expositionally: the inequality \(\text{OKU}^{I}\neq\sum_a\text{OKU}^{E}\) yields the institutional-residue prediction of Chapter~11, which cannot be formed without classifying operative units by residence.
\end{minipage}%
}

\subsection{What the OKU framework bounds and what it does not}\label{what-the-oku-framework-resolves-and-what-it-does-not}

The framework does not resolve the unit problem completely. It bounds the problem at the operative-unit level for \(K^E\), \(K^D\), and \(K^I\) by specifying the operation performed, the domain, the time, the fidelity standard, the context, and the relevant judgement standard. It provides a principled definition, three convergent rationales (Turing equivalence structure, human capital theory's individual basis, and the framework's own deployability condition), and a decision procedure (independent domain performance) for distinguishing units from fragments across all three forms.

What remains unresolved is quality calibration within domain, aggregation across domains, commons sustainability thresholds, public-operative boundary cases, cross-form conversion beyond defined equivalence tests, and economic valuation. These are measurement problems. They are not solved by the OKU, and they should not be hidden by it.

The framework therefore moves from the broad claim that the unit problem is unsolved to the narrower claim that the OKU bounds the KBC unit problem at the operative-unit level for \(K^E\), \(K^D\), and \(K^I\). It does so by denominating knowledge in terms of operations performed within specified domains, times, fidelity standards, contexts, and judgement standards. Calibration, aggregation, commons unitization, public-operative boundary cases, and valuation remain open measurement questions.

That is the methodological position used throughout this book: OKU is a bounded operationalization, not a claim of perfect measurability.

\emph{Detailed operative-unit notation is registered in the Technical Companion, Appendix B. The main point in this chapter is conceptual: operative units provide a bounded measurement bridge, not a universal meter for all knowledge.}

\subsection{Revealed Operative-Unit Ratios}\label{revealed-operative-unit-ratios}

The clearest sign that the operative unit is more than a notational convenience is that organizations already budget by it, without ever computing it cardinally. Firms staff knowledge work to ratios: an information-technology function of roughly one professional per some tens of supported users or systems, and a security function sized as a fraction of that, on the order of one specialist per five to ten IT staff, or equivalently five to ten percent of the IT headcount \parencite{ISC2WorkforceStudy2024}. The exact figures vary by sector, regulatory exposure, and automation level and require local calibration: a high-touch or regulated environment may run closer to one IT professional per ten supported users, a heavily automated one far leaner. The figures here are illustrative industry rules of thumb, not constants.

What matters for the theory is the \emph{form} of these conventions. A coverage ratio\index{revealed operative-unit ratios}\index{coverage ratios} is a statement of how many embodied operative units \(\text{OKU}^{E}\) a domain needs to deliver a defined productive service at a given load. The organization does not measure \(\text{OKU}^{E}\) on a cardinal scale; it converges, through experience, benchmarking, and incident history, on a ratio that encodes the operative-unit intensity of the task. The ratio is the firm's implicit operative-unit count. This is precisely the case in which an institutional convention substitutes for a missing cardinal unit, the same move by which physical capital was made countable through accounting and market conventions rather than through any intrinsic measure.

The layered staffing structure adds a second observation. The nested ratio, supported users served by IT operative units, those IT units in turn maintained and protected by security operative units, is a span-of-maintenance hierarchy: some operative units exist to preserve the productive capacity of others. This is the operative-unit reading of cybersecurity as capital-preservation risk management (Chapter~1) and of capability-bounded codification (Proposition~B): disembodied stock yields productive services only up to the embodied and institutionalized capability available to maintain it, and the maintenance ratio\index{maintenance ratios} is the operational expression of that bound.

Two qualifications keep the claim disciplined. A coverage ratio may be a genuine technical coefficient, in the input--output sense of operative-knowledge units per unit of supported work, or merely a focal-point convention sustained by imitation, budget norms, or regulation; distinguishing the two is an empirical question, not a definitional one, and it is the same ``discovered or conventional?'' problem that attaches to capital aggregation generally. And the ratio is not a constant: because a disembodied unit can substitute for an embodied one at the task-class level, the framework predicts that coverage ratios should drift as \(\text{OKU}^{D}\) absorbs routine operations, a prediction developed as a test in Chapter~11.

\section{The Five-Form Taxonomy (Knowledge Forms)}\index{five-form taxonomy|textbf}\label{the-five-form-taxonomy}

The knowledge-bearing stock of an economy at any given time can be classified into five analytically distinct forms. When these forms are employed or positioned to produce future value, they function as knowledge-bearing capital. These five labels are a main-text classification, not an additive partition. As the residence--governance ontology later in this section makes explicit, two of them, \(K^{C}\) and \(K^{P}\), are governance-dominant and are typically composed of embodied, disembodied, or institutionalized material held under a particular governance arrangement; the same open-source library is therefore both \(K^{D}\) (a disembodied artefact) and \(K^{C}\) (commons-governed). The five forms consequently overlap and must never be summed as though they were disjoint. The object that does admit a clean additive identity is the residence--governance cell\index{residence--governance cell} \(K^{r, g}_{a, t}\), in which every stock has exactly one residence \(r\in\{E, D, I\}\) and one governance arrangement \(g\):

\begin{equation}
K_t=\sum_a\sum_{r, \,g}K^{r, g}_{a, t}, \qquad r\in\{E, D, I\}
\label{eq:ch2:k-t}
\end{equation}

Here \(r\) is the residential form, embodied (E), disembodied (D), or institutionalized (I), and \(g\) is the governance arrangement (private, firm, platform, professional, commons, or public-epistemic). Equation~\ref{eq:ch2:k-t} adds up all the classified knowledge-stock cells. Because each stock is placed into only one residence--governance category, the same knowledge is not counted twice. That makes the total well defined, even though the theory is not yet claiming perfect measurement of every stock. This notation identifies where different knowledge stocks are located across actors, forms, and time. It is a classification map\index{knowledge-stock state-space address}, not yet a formula for measuring total knowledge or estimating how knowledge produces output. It does not assume that very different kinds of knowledge can already be added together on a single common numerical scale. The five-form classification of this section is the main-text overlay on that partition: embodied, disembodied, and institutionalized capital are the three residences, while commons (\(K^{C}\)) and public-epistemic (\(K^{P}\)) capital are governance-dominant groupings that re-sort the same cells by how they are governed, which is why they overlap the residences and cannot be added to them. In coordinate terms, \(K^{r, g}_{a, t}\) is a coordinate-addressed stock cell: a way of locating knowledge-bearing stock by actor, residence, governance, and time. These coordinates are classificatory addresses, not yet cardinal magnitudes in a continuous or Euclidean field. This residence--governance cell is the primary additive object; the five forms are the non-additive classification layered on top of it. The recombination field \(F_{a, t}\) is therefore not the same object as total \(K_t\). It is the subset of the knowledge-stock state space reachable by actor \(a\) under access, interoperability, governance, and capability constraints. The notation follows the grammar used throughout this book: the main symbol identifies the object class; a right superscript identifies form, type, or mechanism; the subscript identifies actor, stock, domain, or time; and roman-text superscripts identify measured properties or valuation states. The Technical Companion, Appendix A provides the full notation register. These five forms differ in their persistence logic, their separability properties, their depreciation dynamics, and their susceptibility to different governance forms.

\begin{figure}[htbp]
\caption[Five Knowledge Forms: Residence, Separability, and Governance Dependence]{Five Knowledge Forms: Residence, Separability, and Governance Dependence}
\label{fig:ch2:five-forms-separability}
\centering
\includegraphics[width=0.98\textwidth]{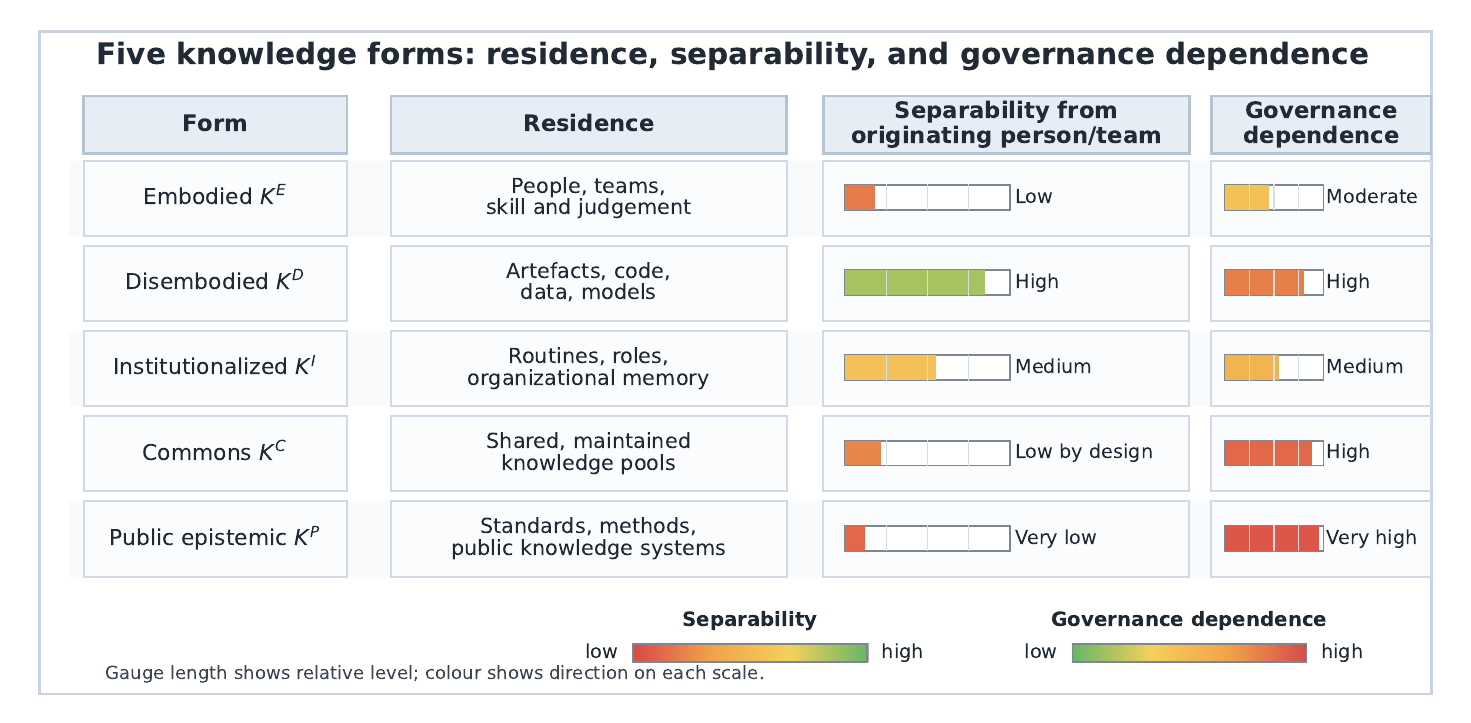}
\par\smallskip\noindent\footnotesize\emph{Note.} The figure summarizes the five forms introduced in this section through the residence--governance ontology. The separability profile is not intrinsic to the object alone; it depends on law, technical stack, documentation, maintenance capability, and governance support.
\end{figure}

The taxonomy is economic, not merely classificatory. Each form answers a different practical question about productive capacity:

\begin{center}
\small
\begin{tabular}{@{}p{0.24\textwidth}p{0.11\textwidth}p{0.53\textwidth}@{}}
\toprule
Form & Symbol & Economic question \\
\midrule
Embodied & \(K^{E}\) & What leaves when people leave? \\
Disembodied & \(K^{D}\) & What can be copied, owned, licensed, scaled, or enclosed? \\
Institutionalized & \(K^{I}\) & What persists through turnover? \\
Commons & \(K^{C}\) & What is shared but maintenance-dependent? \\
Public epistemic & \(K^{P}\) & What makes knowledge production possible across the whole economy? \\
\bottomrule
\end{tabular}
\end{center}

The forms therefore identify different economic vulnerabilities. Embodied knowledge creates departure risk; disembodied knowledge creates ownership, scaling, licensing, and enclosure questions; institutionalized knowledge creates organizational-continuity questions; commons knowledge creates maintenance and contributor-depth questions; public epistemic knowledge creates system-wide infrastructure and underinvestment questions.

A preliminary version of the taxonomy employed a tripartite distinction: embodied, disembodied, and institutionalized. That tripartite distinction remains a useful first approximation and is not discarded here. The formal theory requires five forms because the governance form and the depreciation dynamics of knowledge held in commons (\(K^{C}\)) and public epistemic capital (\(K^{P}\)) differ structurally from those of knowledge institutionalized within private firms (\(K^{I}\)). This is not a taxonomic nicety; it is the distinction on which Proposition B's commons-depletion corollary, Proposition C's recombination-field analysis, and the knowledge-governance-forms framework all depend. A theory that collapses \(K^{C}\) and \(K^{P}\) into \(K^{I}\) cannot explain how commons erode, why public epistemic capital is systematically underinvested, or why the recombination field available to the broader economy differs from the knowledge accessible within any particular firm.

\textbf{\(K^{E}\), Embodied knowledge capital.}\index{embodied knowledge@embodied knowledge (\ensuremath{K^E})|textbf} The economic question is: what leaves when people leave? Productive knowledge embodied in particular people or tightly bonded teams. It includes tacit knowledge, expertise, judgment, and the accumulated interpretive capacity that individuals develop through experience. Its persistence is fragile: it depreciates when key people leave, retire, or become unavailable, and it cannot be transferred intact to successors. Its separability is low: it cannot be detached from the people whose judgement and experience make it productive without loss of productive power.

\textbf{\(K^{D}\), Disembodied knowledge capital.}\index{disembodied knowledge@disembodied knowledge (\ensuremath{K^D})|textbf} The economic question is: what can be copied, owned, licensed, scaled, or enclosed? Productive knowledge externalized into separable artefacts: software, codified procedures, datasets, trained models, patents, documented protocols\index{documented protocols}, design specifications\index{design specifications}. Its persistence depends on the continued relevance and maintenance of the artefact; software that is not maintained depreciates rapidly; a well-documented procedure can retain value for decades. Its separability is high: once externalized, it can in principle be owned, transferred, sold, licensed, and deployed across multiple contexts independently of any particular person.

\textbf{\(K^{I}\), Institutionalized knowledge capital.}\index{institutionalized knowledge@institutionalized knowledge (\ensuremath{K^I})|textbf} The economic question is: what persists through turnover? Productive knowledge embedded in roles, routines, governance systems, standards, and trust networks within organizations: organizational culture, standard operating procedures, accreditation\index{accreditation} systems, peer review\index{peer review} norms within institutions, professional ethics frameworks, and the shared understanding that enables complex coordination within firms. Its persistence is medium to high: it survives personnel turnover because it is embedded in structures and roles rather than persons. Its separability is medium to low: it is accessible through the institution and can be partially transferred through acquisition, but it cannot be sold as a discrete asset.

\textbf{\(K^{C}\), Knowledge commons.}\index{commons knowledge@commons knowledge (\ensuremath{K^C})|textbf}\index{knowledge commons|see{commons knowledge (\ensuremath{K^C})}} The economic question is: what is shared but maintenance-dependent? A knowledge commons is a pool of productive knowledge that many actors can use without exhausting it, but which still requires ongoing human, institutional, and technical maintenance to remain useful. In economic terms, \(K^{C}\) is productive knowledge held in institutional arrangements that maintain shared access while governing contribution and use: open-source software, open-access scientific literature\index{open-access scientific literature}, Creative Commons content\index{Creative Commons content}, Wikipedia, shared clinical-practice guidelines\index{clinical-practice guidelines} maintained by professional communities, and similar structures. Its productive value is intrinsically non-rival: use by one actor does not diminish availability to others. Its persistence depends on the collective maintenance burden: continued voluntary contribution, governance infrastructure, and the embodied and institutionalized capability stock required to maintain, curate, and develop the commons. When that maintenance capacity is withdrawn through defunding, the departure of key contributors, or platform displacement\index{platform displacement}, \(K^{C}\) depreciates even though its formal legal status may be unchanged. Its separability is low by design: the commons is constituted by collective access, and individual appropriation of the commons as a private asset is typically what the governance structure is designed to prevent.

\textbf{\(K^{P}\), Public epistemic capital.}\index{public epistemic infrastructure@public epistemic infrastructure (\ensuremath{K^P})|textbf}\index{public epistemic capital|see{public epistemic infrastructure (\ensuremath{K^P})}} The economic question is: what makes knowledge production possible across the whole economy? Productive knowledge held in state-funded, collectively provisioned, or open-standards-based knowledge infrastructure that makes knowledge creation, validation, transmission, and recombination possible: universities\index{universities} and public research laboratories\index{public research laboratories}\index{public epistemic infrastructure!research laboratories}, peer-review systems\index{peer-review systems}\index{public epistemic infrastructure!peer review}, national statistical systems\index{national statistical systems}\index{Statistics Canada}, open measurement standards\index{measurement standards}, shared scientific instrumentation\index{scientific instrumentation}, public data archives\index{public data archives}, and open technical standards\index{open technical standards}. \(K^{P}\) is the systemic enabling condition for knowledge generation more broadly: it constitutes the infrastructure on which \(K^{E}\), \(K^{D}\), \(K^{I}\), and \(K^{C}\) all depend to varying degrees. Its persistence is governed by political and funding decisions rather than by market signals; it is therefore subject to depreciation\index{depreciation} through budget cuts, privatization, or institutional erosion in ways that do not generate the market-price signals that would alert private actors to the loss of productive capacity. Its separability is very low: it is collectively maintained and individually inaccessible as a private asset, though portions of it can be enclosed through privatization.

\noindent\textit{Distinguishing \(K^{C}\) from \(K^{P}\).}\index{commons versus public epistemic infrastructure|textbf}\index{commons knowledge@commons knowledge (\ensuremath{K^C})!versus public epistemic infrastructure}\index{public epistemic infrastructure@public epistemic infrastructure (\ensuremath{K^P})!versus commons knowledge} Commons knowledge stock is shared, usable, and maintained by a contribution community. Public epistemic infrastructure is deeper: standards, scientific methods, measurement systems, protocols, and public knowledge institutions that make reliable knowledge production possible in the first place. The difference is economic rather than merely institutional: \(K^{C}\) is a maintained pool that actors can draw upon, while \(K^{P}\) is part of the enabling infrastructure that makes trustworthy knowledge creation, validation, and recombination possible across the wider economy.

{\scriptsize
\begingroup
\scriptsize
\setlength{\tabcolsep}{3pt}
\renewcommand{\arraystretch}{1.12}
\sloppy
\par\addvspace{0.8\baselineskip}\noindent
\begin{longtable}{@{}L{0.15\textwidth}
C{0.06\textwidth}
L{0.26\textwidth}
L{0.16\textwidth}
L{0.11\textwidth}
L{0.15\textwidth}@{}}
\caption{Five-Form Taxonomy of Knowledge-Bearing Stock}\label{tab:ch2:five-form-taxonomy}\\
\toprule\noalign{}
\begin{minipage}[b]{\linewidth}\raggedright
Form
\end{minipage} & \begin{minipage}[b]{\linewidth}\raggedright
Symbol
\end{minipage} & \begin{minipage}[b]{\linewidth}\raggedright
Definition
\end{minipage} & \begin{minipage}[b]{\linewidth}\raggedright
Persistence logic
\end{minipage} & \begin{minipage}[b]{\linewidth}\raggedright
Separability
\end{minipage} & \begin{minipage}[b]{\linewidth}\raggedright
Depreciation driver
\end{minipage} \\
\midrule\noalign{}
\endfirsthead
\toprule\noalign{}
\begin{minipage}[b]{\linewidth}\raggedright
Form
\end{minipage} & \begin{minipage}[b]{\linewidth}\raggedright
Symbol
\end{minipage} & \begin{minipage}[b]{\linewidth}\raggedright
Definition
\end{minipage} & \begin{minipage}[b]{\linewidth}\raggedright
Persistence logic
\end{minipage} & \begin{minipage}[b]{\linewidth}\raggedright
Separability
\end{minipage} & \begin{minipage}[b]{\linewidth}\raggedright
Depreciation driver
\end{minipage} \\
\midrule\noalign{}
\endhead
\bottomrule\noalign{}
\endlastfoot
Embodied & \(K^{E}\) & Productive knowledge in persons or bonded teams & Fragile; person-specific & Low & Personnel departure, retirement \\
Disembodied & \(K^{D}\) & Productive knowledge in separable artefacts & Artefact maintenance & High & Obsolescence, maintenance failure\index{maintenance failure} \\
Institutionalized & \(K^{I}\) & Productive knowledge in organizational roles, routines, governance & Survives personnel turnover; structure-dependent & Medium-low & Organizational disruption, acquisition, dissolution \\
Commons & \(K^{C}\) & Productive knowledge in shared-access collectively maintained arrangements & Collective maintenance burden & Low by design & Contributor withdrawal, governance failure, platform displacement \\
Public epistemic capital & \(K^{P}\) & Productive knowledge in public research, standards, statistical, educational, and validation infrastructure\index{validation infrastructure} & Political and funding decisions & Very low & Budget cuts, privatization, institutional erosion \\

\end{longtable}
\endgroup
}

\subsection{Residence and Governance Beneath the Five Forms}\label{sec:ch2:residence-governance-pairing}
\index{residence--governance pair|textbf}

The five-form taxonomy is the practical main-text classification used throughout this book, but the residence--governance cell \(K^{r, g}_{a, t}\) is the primary formal object beneath it, and the only one that aggregates without double-counting.\index{five-form taxonomy versus residence--governance ontology} The five-form taxonomy should not be mistaken for a claim that all five forms are ontologically identical, nor for an additive partition. Its deeper structure is two-dimensional. Some forms are primarily residential: they identify where productive knowledge is housed. Others are governance-dominant: they identify the conditions under which knowledge is accessed, maintained, trusted, recombined, and made usable across actors. Institutionalized knowledge is the bridge case because the institution is both a place where knowledge resides and a governance structure that makes that knowledge repeatable.\index{institutionalized knowledge@institutionalized knowledge (\ensuremath{K^I})!bridge case}

A firm is not merely a place where production happens. It is a governance structure that makes divided knowledge repeatable, coordinated, durable, and productive.\index{firm!as knowledge-coordination structure}\index{theory of the firm!KBC proposition} Institutionalized knowledge explains why firms can be more than collections of workers or contracts. The firm stores productive knowledge in routines, roles, standards, escalation paths, quality controls, incentives, records, and learned coordination. It converts the division of labour into repeatable productive capacity.

\begin{quote}
\textbf{KBC Theory-of-Firm Proposition.\index{KBC Theory-of-Firm Proposition|textbf}} Firms exist partly because productive knowledge is not fully reducible to individual skill or separable artefacts. Some productive knowledge becomes valuable only when embedded in routines, roles, authority structures, incentives, workflows, trust relations, and feedback systems. The firm is therefore a residence--governance system for converting specialized knowledge into repeatable productive services.
\end{quote}

Institutionalized knowledge connects \(K^I\) directly to the theory of the firm. The firm is not only a production unit or a nexus of contracts, but a knowledge-coordination structure that converts divided labour and specialized embodied knowledge into durable productive capability. In Smithian terms, institutionalized knowledge is what allows the division of labour to become repeatable output rather than merely fragmented skill.\index{Smith, Adam!division of labour}\index{division of labour!institutionalized knowledge}

\emph{Originality status: synthesized/extended.} The main predecessors are Smith's division of labour, Coase's firm as an alternative to market contracting, Williamson's governance and transaction-cost theory, Penrose's firm as a bundle of productive resources, Nelson and Winter's routines, and Grant's knowledge-based theory of the firm. KBC's added move is to place this inside the \(K^E\to K^I\) and residence--governance architecture.\index{Coase, Ronald}\index{Williamson, Oliver}\index{Penrose, Edith}\index{Nelson and Winter}\index{Grant, Robert}\index{knowledge-based theory of the firm}

\begin{quote}
\textbf{Residence--Governance Pairing Principle\index{Residence--Governance Pairing Principle}.} A knowledge-bearing stock is economically specified only when both its residential form and governing arrangement are identified. Residence locates productive knowledge; governance determines the conditions under which it can be accessed, maintained, separated, transferred, excluded, trusted, recombined, and converted into productive services. Governance here is descriptive rather than celebratory: it does not imply competent or normatively desirable governance. Governance arrangements may be effective, weak, opaque, fragmented, captured, underfunded, or failing.
\end{quote}

\begin{quote}
\textbf{Governance-Produced Separability Corollary\index{Governance-Produced Separability Corollary}.} Separability is not intrinsic to knowledge-bearing stock. It is produced, strengthened, weakened, or destroyed by legal, contractual, technical, organizational, platform, commons, and public-epistemic governance.
\end{quote}

A compact way to read the corollary is as a separability diagnostic rather than a calibrated theorem:
\[
\operatorname{Sep}(K^{r, g}_{a, t})=f(A, T, M, \mathcal{E}, I, C),
\]
where separability depends on access (A), transferability (T), maintenance (M), enforceability (\(\mathcal{E}\)), interoperability (I), and capability (C). The expression does not claim an already estimated function. It records the conditions that make separated knowledge usable apart from its originating carrier.

Table~\ref{tab:ch2:five-form-deeper-ontology} states the resulting classification. The point is not to abandon the five forms, but to clarify what each form primarily identifies.

{\scriptsize
\begingroup
\scriptsize
\setlength{\tabcolsep}{4pt}
\renewcommand{\arraystretch}{1.15}
\sloppy
\par\addvspace{0.8\baselineskip}\noindent
\begin{longtable}{@{}C{0.12\textwidth}L{0.78\textwidth}@{}}
\caption{Five Forms Under the Residence--Governance Ontology}\label{tab:ch2:five-form-deeper-ontology}\\
\toprule\noalign{}
\begin{minipage}[b]{\linewidth}\raggedright
Form
\end{minipage} & \begin{minipage}[b]{\linewidth}\raggedright
Status under the deeper ontology
\end{minipage} \\
\midrule\noalign{}
\endfirsthead
\toprule\noalign{}
\begin{minipage}[b]{\linewidth}\raggedright
Form
\end{minipage} & \begin{minipage}[b]{\linewidth}\raggedright
Status under the deeper ontology
\end{minipage} \\
\midrule\noalign{}
\endhead
\bottomrule\noalign{}
\endlastfoot
\(K^{E}\) & Residence-dominant: embodied in people, teams, skill, judgement, and experience. \\
\(K^{D}\) & Residence-dominant: encoded in artefacts, data, software, models, and documents. \\
\(K^{I}\) & Hybrid: the institution is both residence and governance structure; it houses knowledge in routines, roles, systems, and memory while supplying authority, incentives, maintenance, and continuity. \\
\(K^{C}\) & Governance-dominant: shared-access, collectively maintained knowledge stock, often composed of embodied, disembodied, and institutionalized elements. \\
\(K^{P}\) & Governance-dominant: public epistemic infrastructure for measurement, validation, standards, law, protocols, statistics, and trust, often composed of embodied, disembodied, institutionalized, and commons elements. \\
\end{longtable}
\endgroup
}

The earlier tripartite taxonomy \((K^{E}, K^{D}, K^{I})\) effectively collapsed commons knowledge stock \((K^{C})\) and public epistemic infrastructure \((K^{P})\) into institutionalized knowledge \((K^{I})\). That approximation is useful for firm-level analysis, but it fails once the theory asks how knowledge is maintained, governed, recombined, or underprovided beyond the boundaries of a single organization. The five-form taxonomy remains useful because it keeps the reader's attention on recognizable knowledge stocks; the residence--governance ontology explains why those stocks differ in separability, depreciation, value capture\index{value capture}, and capital formation.

For clarity of thought and ease of learning, the five forms remain the main vocabulary of this book. The deeper technical notation is introduced in the Technical Companion as \(K^{r, g}_{a, t}\), where \(r\) identifies residential form and \(g\) identifies the governing arrangement. Under that notation, \(K^{C}\) and \(K^{P}\) should be read as shorthand for governance-dominant knowledge stocks: \(K^{C} \approx K^{r, g_C}_{a, t}\) and \(K^{P} \approx K^{r, g_P}_{a, t}\). They may contain embodied, disembodied, and institutionalized components, but their defining feature is commons or public-epistemic governance.

\subsection{Demand by Knowledge Form}\label{demand-by-knowledge-form}

The five-form taxonomy also clarifies what economic actors are actually demanding. They do not demand ``knowledge'' as a vague abstraction. They demand productive service flows from knowledge-bearing stock in a form that can be hired, bought, licensed, accessed, governed, maintained, or converted. Table~\ref{tab:ch2:demand-by-knowledge-form} makes this demand visible across the five forms.

{\scriptsize
\begingroup
\scriptsize
\setlength{\tabcolsep}{4pt}
\renewcommand{\arraystretch}{1.15}
\sloppy
\par\addvspace{0.8\baselineskip}\noindent
\begin{longtable}{@{}L{0.23\textwidth}C{0.16\textwidth}L{0.51\textwidth}@{}}
\caption{Demand by Knowledge Form}\label{tab:ch2:demand-by-knowledge-form}\\
\toprule\noalign{}
\begin{minipage}[b]{\linewidth}\raggedright
What is demanded
\end{minipage} & \begin{minipage}[b]{\linewidth}\raggedright
KBC form
\end{minipage} & \begin{minipage}[b]{\linewidth}\raggedright
Productive service demanded
\end{minipage} \\
\midrule\noalign{}
\endfirsthead
\toprule\noalign{}
\begin{minipage}[b]{\linewidth}\raggedright
What is demanded
\end{minipage} & \begin{minipage}[b]{\linewidth}\raggedright
KBC form
\end{minipage} & \begin{minipage}[b]{\linewidth}\raggedright
Productive service demanded
\end{minipage} \\
\midrule\noalign{}
\endhead
\bottomrule\noalign{}
\endlastfoot
Skilled worker & \(K^{E}\) & Judgement, diagnosis, execution, domain skill \\
Software\index{software} & \(K^{D}\) & Automation, computation, reproducibility \\
Dataset\index{datasets} & \(K^{D}\) & Prediction, classification, inference, training input \\
Organizational routine\index{routines} & \(K^{I}\) & Repeatable coordination and institutional memory \\
Open-source library\index{open-source software}\index{software!open-source} & \(K^{C}\) & Reusable code, shared maintenance, recombination input \\
Scientific publication\index{public epistemic infrastructure@public epistemic infrastructure (\ensuremath{K^P})!scientific publication}\index{standards} & \(K^{P}/K^{C}\) & Validation, discovery input, epistemic reference \\
AI answer & \(K^{D}\)-mediated service & Synthesis, classification, drafting, decision support \\
\end{longtable}
\endgroup
}

The table is not a decorative classification. It links the stock taxonomy to demand: every entry asks what productive service the actor expects to receive, and through which form of knowledge-bearing stock that service is delivered.

\textbf{Notation.} Throughout this book, \(K^{E}\), \(K^{D}\), \(K^{I}\), \(K^{C}\), and \(K^{P}\) denote stocks of each form. Subscripts denote actor (a), time (t), or knowledge type (j) where needed: \(K^{D}_{a, t}\) is the disembodied knowledge-bearing stock held by actor a at time t. Total social knowledge-bearing stock at time t is \(K_t=\sum_a\sum_{r, g}K^{r, g}_{a, t}\), summed across actors and over the disjoint residence--governance cells; the five-form labels are a non-additive classification of those cells and are never summed directly. This notation is schematic: it marks the analytical location of stocks across actors, residence, and governance, rather than asserting that all such stocks are already measurable in a common unit.

\begin{figure}[!htbp]
\caption[Where did the knowledge go? Residence, governance, and productive service]{Where did the knowledge go? Residence, governance, and productive service}
\label{fig:ch2:knowledge-migration-map}
\centering
\includegraphics[width=0.95\textwidth]{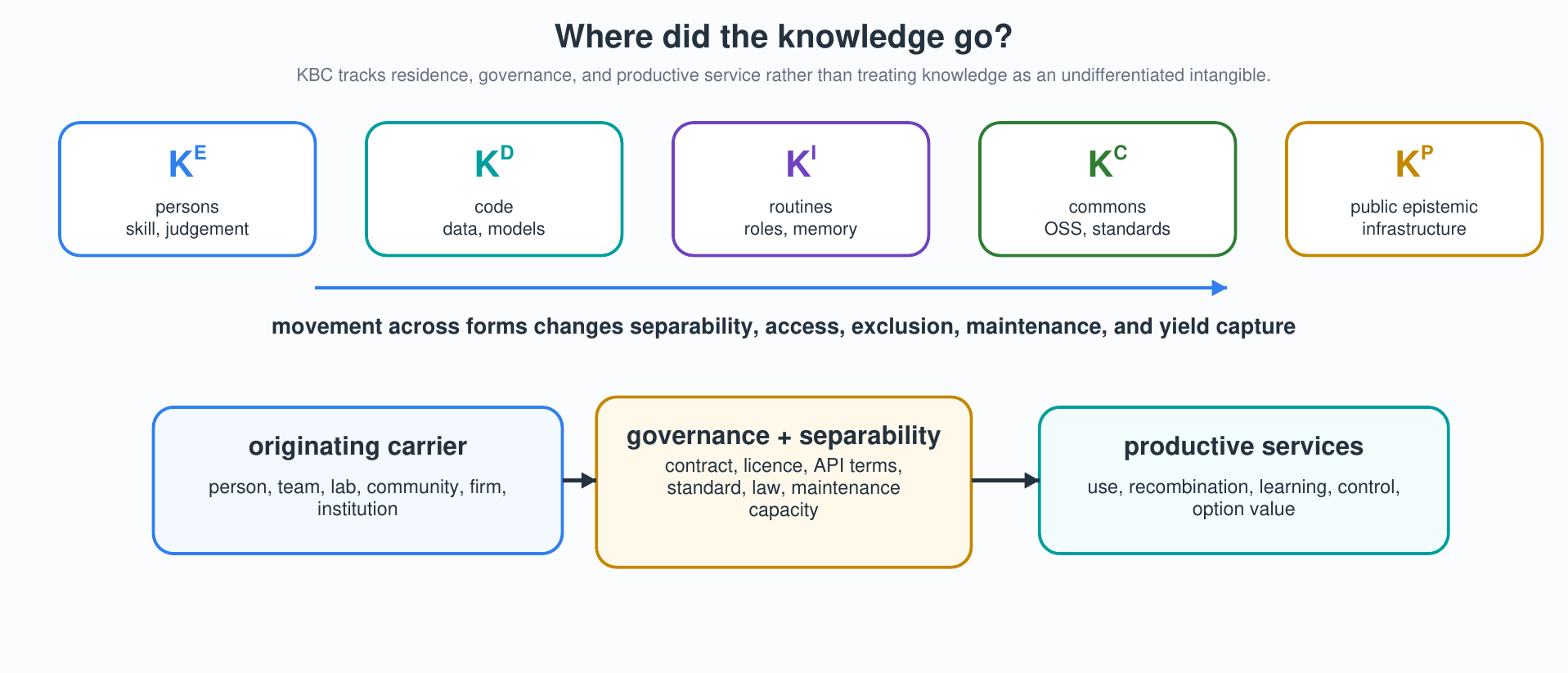}
\par\smallskip\noindent\footnotesize\emph{Note.} The same productive knowledge can move across residence forms and governance arrangements. The economic question is not only where the knowledge is carried, but which conditions make it separable, usable, maintainable, and capable of yielding productive services.
\end{figure}

\begin{quote}
\textbf{Enclosure.}\label{def:enclosure} Enclosure is the legal, contractual, technical, architectural, or capability-based restriction\index{enclosure|textbf} of access to knowledge-bearing stock, reducing the ability of other actors to use, modify, recombine, learn from, or build upon it. Enclosure may be formal (patent, copyright, trade secret) or operational (API restriction, platform gating, non-compete, trade secrecy, deliberate obfuscation\index{API restriction}\index{platform gating}\index{trade secrecy}). This theory distinguishes enclosure from legitimate appropriability: the question is not whether returns are captured but whether the terms of capture reduce productive capacity available to others.\index{enclosure versus legitimate appropriability}
\end{quote}

\textbf{The tripartite approximation\index{tripartite taxonomy versus five-form taxonomy}\index{five-form taxonomy!versus tripartite taxonomy}.} The earlier tripartite taxonomy \((K^{E}, K^{D}, K^{I})\) effectively collapsed commons knowledge stock \((K^{C})\) and public epistemic infrastructure \((K^{P})\) into institutionalized knowledge \((K^{I})\). This approximation is adequate for analysing knowledge conversion within a single firm or dyadic relationship, because those analyses rarely depend on distinguishing between what an institution controls privately and what a broader community governs collectively. It fails as soon as the analysis involves commons, public research, epistemic infrastructure, or multi-party governance, which includes virtually all analyses of recombination, spillovers, generative suppression, and the institutional conditions for knowledge generation. The five-form framework is not a more elaborate version of the same theory; it is the version in which those phenomena can be stated precisely.

\textbf{Why \(K^{C}\) and \(K^{P}\) are structurally distinct.} \(K^{C}\) and \(K^{P}\) are not residuals or subtypes of \(K^{I}\). They are governance-dominant forms whose economic logic cannot be reduced to internal institutional knowledge. First, governance: \(K^{I}\) is governed by the institution that holds it (the firm, the professional body); \(K^{C}\) is governed collectively by its contributor and user community; \(K^{P}\) is governed by political, public-funding\index{public funding conditions}, professional, scientific, legal, statistical, and standards-setting processes. Second, depreciation: \(K^{I}\) depreciates when the organizational structures that carry it are disrupted; \(K^{C}\) depreciates when collective maintenance capacity is withdrawn; \(K^{P}\) depreciates when public funding, institutional credibility, validation systems, or epistemic infrastructure is reduced. These are structurally different depreciation dynamics with different policy implications and different responses to enclosure. Distinguishing them is not taxonomic pedantry; it is the precondition for a theory of knowledge governance rather than merely a theory of knowledge management.

\section{The Conditional Separability Axiom}\label{sec:ch2:conditional-separability-axiom}\index{Conditional Separability Axiom|textbf}

The most important conceptual move in the taxonomy is to treat separability as relational rather than intrinsic. Separability is not an inherent property of knowledge. It is separability from a person, team, community, organization, platform, or institutional setting under particular legal, technical, contractual, documentary, and capability conditions. Physical capital often appears naturally separable because the supporting conditions are already familiar: property law, title, contracts, maintenance markets, technical standards, and accounting conventions. A loom can be owned by someone who does not operate it because the institutional world around the loom makes that separation stable. Knowledge does not usually begin in that condition. It begins in persons, teams, communities, documents, routines, datasets, models, institutions, platforms, and contexts. Whether it can become capital depends on the conditions under which it can be separated from those original bearers or settings without destroying its productive force.

The \textbf{Conditional Separability Axiom} states: knowledge-bearing stock becomes knowledge-bearing capital only when it can be separated sufficiently from a specified originating person, team, community, organization, platform, context, or institutional setting to be preserved, accessed, governed, deployed, transferred, reused, enclosed, or recombined under supporting legal, technical, contractual, documentary, institutional, and capability conditions. Separability is therefore not what makes capital capital. It is one condition that can make knowledge-bearing stock easier to retain, own, transfer, scale, govern, or accumulate as capital-like stock. The relation is conditional among four elements: the knowledge-bearing stock, the bearer or setting from which it is being separated, the actor seeking to use or control it, and the governance arrangement that makes the separation usable.

\noindent\textbf{What separability does and does not decide.} This axiom should not be read to mean that only separable knowledge is productive, or that inseparable embodied knowledge is not capital in any sense. Inseparable embodied and tacit knowledge, the surgeon's operative judgement, a team's shared understanding of an unsolved problem, is knowledge-bearing stock, and it is capital-like in the productive sense whenever it yields productive services, even when it cannot be cleanly alienated from the person or relationship that carries it. What separability governs is not productive reality but \emph{appropriability and transferability}: the routes by which knowledge-bearing stock can be owned as a discrete asset, priced, transferred, enclosed, collateralized, or institutionalized. The axiom's ``becomes capital'' should therefore be read in that asset sense, stock that can be held and moved as separable, appropriable capital, not as a denial that inseparable stock produces value. High-separability stock, such as much \(K^D\) under strong governance, can be held and traded as a discrete asset; low-separability stock, such as much \(K^E\), \(K^C\), and \(K^P\), can generate enormous productive value that is simply harder for any party other than its bearer to appropriate. Separability thus determines the governance and monetization routes available to a stock, not whether it produces value.

Separability failure is visible in ordinary business cases. Software\index{software!separability failure} may be legally owned but economically fragile if no maintainers understand it. A dataset\index{datasets!provenance} may exist as an asset file but have low productive value if its provenance, definitions, consent history, or collection conditions are missing. An AI model may be technically complete but non-deployable without inference infrastructure, monitoring, workflow integration, security controls, and domain oversight. A patent may grant exclusion while producing little value without manufacturing, regulatory, distribution, marketing, or complementary technical assets. An acquired firm may appear to transfer routines, customer knowledge, or engineering capability, yet those routines may disappear when key personnel leave. An open-source project may remain legally open while its productive capacity decays because maintainers burn out. In each case, knowledge-bearing stock exists, but the supporting conditions for useful separability are missing or failing.

Several implications follow directly from the axiom. First, the same knowledge-bearing stock may be separable under one institutional governance arrangement and inseparable under another. A drug mechanism\index{drug mechanism} discovered in a publicly funded university laboratory may or may not give rise to a separable patent depending on the funding agreement, the technology transfer office's practice, the researcher's employment contract, and the funder's IP policy. The same mechanism discovered under corporate R\&D funding enters a different first-conversion governance arrangement: employer IP ownership, confidentiality obligations\index{confidentiality obligations}, and trade-secret protections may govern it before a patent is filed. Identical knowledge; different separability conditions; different capital trajectories.

Second, the degree of separability achievable at a given cost determines how knowledge is governed and monetized. High-separability knowledge, such as \(K^{D}\) under strong IP governance, can be owned, licensed, traded, and accumulated as a discrete asset. Low-separability knowledge, such as much \(K^{E}\), \(K^{C}\), and \(K^{P}\), may generate enormous value through wages, competitive advantage, public productivity, or shared generative capacity, but that value cannot easily be appropriated by parties other than the persons, communities, or institutions that sustain it. The institutional effort required to increase separability (codification, documentation, legal assignment, standards, infrastructure, and complementary capability development) is itself a productive investment.

Third, enclosure often functions by sustaining separability after it has been achieved. A patent, licence, platform API, access contract, trade-secret governance, or employment IP assignment does not merely allocate an existing asset. It helps maintain the conditions under which knowledge remains separable from its originating context and appropriable by a particular actor. Without those conditions, the same knowledge may diffuse, decay, be reabsorbed into a commons, or require the return of embodied capability that the separating actor no longer possesses.

The non-rival character of knowledge creates the fundamental tension that Arrow identified\index{Arrow, Kenneth}: information goods tend to diffuse beyond the producer's control, which creates underinvestment incentives that exclusivity rights are designed to address, but granting exclusivity creates static and dynamic efficiency losses. The Conditional Separability Axiom adds a prior question: under what conditions does the knowledge become separable enough to be capital at all, and who has the power to establish those conditions at the first-conversion zone?

\section{Knowledge Potential, Impedance, and Yield}\label{knowledge-potential-impedance-and-yield}\index{knowledge potential|textbf}\index{knowledge impedance|textbf}\index{knowledge yield|textbf}

\noindent Apparent knowledge value and usable knowledge value often diverge. Many firms possess knowledge-bearing stock that appears valuable but produces little usable value. A dataset may be large but poorly documented; a patent may be legally strong but commercially irrelevant; a model may be technically impressive but hard to deploy. The distinction between potential, impedance, and yield explains why the same stock can be valuable to one actor and nearly useless to another.

The five-form taxonomy classifies what knowledge-bearing stock is. The Knowledge Potential, Impedance, and Yield triad characterizes what it can do, and what obstructs it from doing that. Put in decision language, knowledge potential asks what value might be present. Knowledge impedance asks what blocks that value from being realized. Knowledge yield asks what productive service actually flows from the stock under present conditions. The three concepts are formally defined here; they will be deployed analytically throughout Chapters 3 and 4.

This order matters for economists, investors, firms, accountants, and policymakers because it separates three decisions that are often confused. First, is there latent value worth investigating? Second, what access, capability, governance, provenance, or deployment barrier prevents that value from being used? Third, after those barriers are considered, what service flow is actually being produced now? The triad is therefore not merely descriptive vocabulary. It is a decision sequence for moving from apparent value to blocked value to realized productive service.

The same distinction also turns the triad into a management and valuation tool. A dataset may have high potential, high impedance, and low current yield: the data may contain valuable patterns, but poor provenance, missing metadata, privacy restrictions, or weak modelling capability may prevent the firm from using it productively. A well-maintained software library may have only moderate potential but low impedance and high yield because it is documented, interoperable, trusted, and already integrated into production. A patent may have high control value but low productive yield if the firm lacks the complementary manufacturing, regulatory, distribution, or technical assets required to turn legal exclusion into productive service. In each case, the analyst should not ask only whether the stock is valuable. The analyst should ask whether its value is latent, blocked, or actually flowing.

At the highest level, KBC values knowledge-bearing stock by asking a capital-services question: what productive services does this stock yield, compared with substitute or complementary stocks, under a given governance and capability conditions? The generic valuation identity is:

\begin{equation}
V(K_i)=\mathbb{E}\left[\sum_{t=0}^{H}\frac{\text{productive services yielded by }K_i}{(1+r)^t}\right].
\label{eq:ch2:generic-capital-services-valuation}
\end{equation}

This is a schematic capital-services identity\index{capital-services identity}, not the full KBC valuation model. The full version is actor-specific, governance-specific, capability-conditioned, loss-adjusted, and not yet calibrated on a common empirical scale.

The same capital-services logic supplies the general demand form. Actors demand knowledge-bearing capital because they expect it to yield productive-service flow\index{productive-service flow}s. The object of demand is not knowledge in the abstract, but usable productive knowledge in a form the actor can access, govern, hire, buy, rent, licence, build, use, or convert. A deliberately simplified general demand function is:

\begin{equation}
D_{a, t}(K_i^x)=f\left(\mathbb{E}[PS_{a, i}], P_i, C_{a, i}, g_i, \mathrm{Sub}_i, U_i\right), \quad x\in\{E, D, I, C, P\}.
\label{eq:ch2:simplified-general-knowledge-demand}
\end{equation}

Here \(D_{a, t}(K_i^x)\) is actor \(a\)'s demand at time \(t\) for knowledge-bearing stock \(K_i\) in form \(x\); \(\mathbb{E}[PS_{a, i}]\) is the expected productive-service flow from the stock to the actor; \(P_i\) is the relevant price, wage, licence fee, subscription cost, build cost, acquisition cost, or access cost; \(C_{a, i}\) is the complementary capability required to use the stock; \(g_i\) is the governance form governing stock \(i\); \(\mathrm{Sub}_i\) is substitutability by other knowledge forms, labour, machine capital, AI systems, or external services; and \(U_i\) is uncertainty about validity, durability, relevance, provenance, reliability, or legal status. The equation is intentionally minimal and schematic. Its purpose is to distinguish the variables that govern demand for knowledge-bearing stock, not to claim immediate econometric calibration. The expanded demand model belongs in advanced valuation, strategy, or governance analysis; this book begins with the simple form so the demand side of this theory remains empirically usable.

\emph{Notation note:} The five-form taxonomy reserves \(K^{I}\) for institutionalized knowledge capital and \(K^{P}\) for public epistemic capital. To avoid collision, the productivity triad uses \(K^{\mathrm{pot}}\) for knowledge potential, \(K^{\mathrm{imp}}\) for knowledge impedance, and \(K^{\mathrm{yield}}\) for knowledge yield. Here \(K\) remains the object class; \(E\), \(D\), \(I\), \(C\), and \(P\) are right superscripts identifying forms; and \(\mathit{pot}\), \(\mathit{imp}\), and \(\mathit{yield}\) are label superscripts identifying measured properties or valuation states rather than additional forms.

\textbf{Knowledge Potential (\(K^{\mathrm{pot}}\)).} For a given knowledge-bearing stock \(K^{form}_{a, t}\) held by actor a at time t, the knowledge potential represents the productive output that the stock is capable of generating under the most favourable institutional conditions: full access, full complementary capability, no impedances. Formally, \(K^{\mathrm{pot}}\) is defined by the range of productive tasks the stock enables, the breadth of recombinations it can enter, and the rate at which it can generate new knowledge through the KGM mechanisms. A large and varied \(K^{D}\) stock with broad recombination potential has high \(K^{\mathrm{pot}}\). A \(K^{E}\) stock held by a single specialist has high \(K^{\mathrm{pot}}\) within that specialist's domain and low \(K^{\mathrm{pot}}\) outside it. \(K^{P}\) (public epistemic capital) typically has very high \(K^{\mathrm{pot}}\) because it enables the productive deployment of all other forms.

\textbf{Knowledge Impedance (\(K^{\mathrm{imp}}\)).} Knowledge impedance is what resists the conversion or productive use of existing stock, the friction between potential and realized output. Impedances may be legal (patent thickets, non-compete agreements, export controls), technical (non-interoperable formats, access-restricted platforms, proprietary standards), cognitive (lack of the complementary \(K^{E}\) required to interpret or deploy \(K^{D}\)), organizational (institutional barriers to cross-functional recombination), or economic (high transaction costs\index{transaction costs} for licensing, high cost of the capability required to use the stock). Impedance is always relative to a specific conversion event: a particular actor attempting a particular use of a particular stock. What is high impedance for one actor (a small firm without IP counsel attempting to license a complex patent portfolio) may be low impedance for another (a large firm with the same capability).

\textbf{Knowledge Yield (\(K^{\mathrm{yield}}\)).} Knowledge yield is the actual productive output generated per unit of knowledge-bearing stock in a given period, given the prevailing impedances, the complementary capability of the deploying actor, and demand alignment\index{demand alignment}, denoted by the want-match\index{want-match coefficient} term \(m^{\mathrm{want}}_{a, j, t}\): the degree to which the stock matches a live productive want. Formally, for actor a deploying knowledge of type j at time t:

\begin{equation}
K^{\mathrm{yield}}_{a, j, t}=\omega_j \cdot \chi_{a, j, t} \cdot \rho_{a, j, t} \cdot m^{\mathrm{want}}_{a, j, t}
\label{eq:ch2:knowledge-yield}
\end{equation}

where \(\omega_{j}\) is the weight of knowledge type j in productive output (how much this knowledge contributes to value if fully deployed), \(\chi_{a, j, t}\) is the access fraction\index{access fraction} (the share of the available stock of type j that actor a can reach at time t, with 0 = fully blocked and 1 = unrestricted access), \(\rho_{a, j, t}\) is the realization coefficient: the degree to which actor a can actually turn accessible knowledge potential in stock j into productive yield at time t, based not merely on whether the knowledge has been converted into a form but on whether the actor has the access, permission, capability, interoperability, governance conditions, and maintenance support needed to realize its productive services, and \(m^{\mathrm{want}}_{a, j, t}\in[0, 1]\) is the demand-alignment or want-match coefficient: the degree to which knowledge type j, available to actor a at time t, matches a live productive want, problem, task, or decision context. This expression is schematic as well: it separates weight, access, realization, and demand alignment so the management problem can be seen, without asserting that the four terms are already observable without further measurement design. The multiplicative form is the limiting case of perfect complementarity\index{complementarity} (a Leontief composition), in which weight, access, realization, and demand alignment cannot substitute for one another. In other words, knowledge value requires the joint presence of importance, access, capability, and demand. These are complements, not substitutes: the equation assumes that all required conditions must be present together. A weakness in one condition cannot be fully offset by strength in another. It is used here because it makes the management problem legible, not because the four inputs are known to be strict complements; where they are partially substitutable, an actor can offset a weak term by routing toward second-best inputs, so realized yield falls less than the product implies. The main book uses a strict version of the model for clarity. Volume~2 checks whether the theory's conclusions still point in the same direction when the strict ``all factors must work together'' complementarity assumption is weakened.

The distinction between realization coefficient and demand alignment is important. \(\rho_{a, j, t}\) measures realization capability: can the actor interpret, integrate, maintain, and deploy the stock? \(m^{\mathrm{want}}_{a, j, t}\) measures demand alignment: does the stock address a productive problem the actor actually has? A technically capable actor may have full access to a stock and still generate little current yield if the stock does not match any live productive want. Conversely, a modest stock can produce high yield when it matches an urgent task, decision, workflow, or regulatory need.

\emph{In plain terms:} knowledge yield is the fraction of a stock's potential value that an actor can actually realize in context. High stock value (ω), unrestricted access (χ = 1), strong complementary capability (ρ close to 1), and strong want-match (\(m^{\mathrm{want}}\) close to 1) produce yield close to potential. Enclosure drives access toward zero: as \(\chi_{a, j, t}\to 0\), \(K^{\mathrm{yield}}\to 0\) in the perfectly-complementary limit, regardless of how valuable the stock is, how capable the actor is, or how well the stock would otherwise fit the actor's problem; where the actor can substitute toward second-best inputs, the fall is bounded rather than total. Capability loss drives ρ toward zero: an actor who loses the \(K^{E}\) required to interpret and deploy a stock loses yield even if access remains open. Demand misalignment drives \(m^{\mathrm{want}}\) toward zero: an actor may be able to use the stock, but current yield remains low if the stock solves no live productive problem. Any one of these conditions can reduce realized yield sharply.

The relationship between potential, impedance, and yield can be stated simply: \(K^{\mathrm{pot}}\) is the ceiling; \(K^{\mathrm{imp}}\) reduces χ and ρ; weak demand alignment reduces \(m^{\mathrm{want}}\); \(K^{\mathrm{yield}}\) is the realized output below that ceiling. An actor with high \(K^{\mathrm{pot}}\) (large stock), low \(K^{\mathrm{imp}}\) (few access barriers), high ρ (strong complementary capability), and high \(m^{\mathrm{want}}\) (strong problem-fit) achieves \(K^{\mathrm{yield}}\) close to \(K^{\mathrm{pot}}\). An actor with high \(K^{\mathrm{pot}}\) but high \(K^{\mathrm{imp}}\), low ρ, or low \(m^{\mathrm{want}}\) achieves a \(K^{\mathrm{yield}}\) far below the ceiling. The resulting productive gap is a private loss to the excluded, under-capable, or demand-misaligned actor and a potential social loss when foregone recombination exceeds incentive, quality, security, or investment benefits. The productive gap hurts the actor who cannot access, use, or align the knowledge. It becomes a wider economic loss when the blocked uses would have generated more value for society through recombination, learning, competition, or future innovation than society gains from keeping the knowledge restricted.

\textbf{Connecting the triad to the propositions.} The \(K^{\mathrm{pot}}\)/\(K^{\mathrm{imp}}\)/\(K^{\mathrm{yield}}\) framework connects directly to the core propositions of this theory. Proposition A (codification as reciprocal bargaining transformation) operates through changes in \(K^{\mathrm{imp}}\), \(K^{\mathrm{yield}}\), and the appropriability conditions governing the resulting stock. In plain terms, codification reshapes bargaining power by changing the usability, productive yield, and control conditions of knowledge. It may reduce dependence on the original knowledge-holder, but it can also create new forms of leverage for that holder if they retain scarce capability, authorship, maintenance authority, or marketable expertise. Codification changes the form, residence, and governance of embodied knowledge. The bargaining result depends on whether codification is substitutionary, amplificatory, reputational, collaborative, institutionalizing, or coercive. In substitutionary and coercive modes, firm-controlled \(K^{D}\) or \(K^{I}\) may reduce the worker's access to, control over, or yield from the codified operative unit. In amplificatory, reputational, collaborative, and institutionalizing modes, codification may raise worker \(K^{\mathrm{yield}}\), firm \(K^{\mathrm{yield}}\), or joint surplus through better tools, attribution, higher-skill roles, durable routines, or shared governance. The point is not that codification mechanically transfers bargaining power\index{bargaining power} from worker to firm. It changes the stock's residence, separability, access conditions, and realization conditions, so the distributional outcome depends on the governing mode and on who retains the complementary \(K^{E}\) and \(K^{I}\) required to maintain, interpret, and deploy the codified system.

Workers often cooperate with codification for economically rational reasons. Documentation, mentoring, process-building, and systemization can produce reputation, promotion, reduced operating burden, professional pride, and new higher-order roles. A senior engineer who documents an architecture may lose some monopoly over tacit know-how\index{monopoly}, but may also become the architect, reviewer, teacher, or standard-setter for the system. A clinician who helps codify a diagnostic workflow may reduce repetitive work while increasing status, governance authority, or supervisory scope. The bargaining question is therefore not whether workers resist codification. It is whether codification substitutes for the worker's operative unit, amplifies it, gives it attribution, embeds it in a collaborative routine, institutionalizes it into a higher-value role, or extracts it coercively under asymmetric control. The key question is what codification does to the worker's productive capability: does it replace the worker, make the worker more productive, give the worker credit, turn individual know-how into a team routine, move the worker into a more valuable role, or capture the worker's knowledge under unequal control?

Proposition B (capability-bounded codification) operates through \(\rho\): disembodied \(K^{D}\) generates productive value only up to the limit of the \(\rho\) available to deploy it; without the complementary \(K^{E}\) and \(K^{I}\), even high-\(K^{\mathrm{pot}}\) disembodied stock generates low \(K^{\mathrm{yield}}\). Propositions C and D (generative suppression and feedback-enclosure) operate through the aggregate recombination field: enclosure reduces \(\chi\) for non-incumbents, lowering their \(K^{\mathrm{yield}}\); it simultaneously raises the incumbent's \(\rho\) through the learning loop (Proposition D), widening the yield gap. The \(K^{\mathrm{pot}}\)/\(K^{\mathrm{imp}}\)/\(K^{\mathrm{yield}}\) triad is therefore not a separate framework grafted onto the propositions; it is the productivity vocabulary in which the propositions are stated precisely.

\noindent\textbf{Bastiat\index{Bastiat, Frederic}ian service and visibility.} A low current \(m^{\mathrm{want}}_{a, j, t}\) in the yield expression above implies low current yield, not necessarily zero capital value\index{current yield versus option value}\index{option value!versus current yield}. Knowledge-bearing stock may still carry recombination option value\index{recombination option value|textbf}, learning-loop option value\index{learning-loop option value|textbf}, control value, strategic option value\index{strategic option value|textbf}, or future-use value\index{future-use value} if future demand alignment is plausible. Bastiat's service theory therefore disciplines current yield without eliminating option value. Bastiat forces the analysis to ask whether a knowledge stock is actually rendering a useful service now, but he does not deny that the stock may also have future strategic value. The service theory keeps the current-value claim honest: a knowledge stock has present yield only when it renders a useful service. But this does not rule out option value, because the same stock may still create future opportunities for recombination, licensing, learning, defence, or innovation. Smith supplies this book's production-side baseline: labour, stock, specialization, market extent, and accumulation. Bastiat supplies a complementary exchange-side discipline. Value is not located in the object alone but in the service rendered and received. Economic analysis must therefore ask not only whether a knowledge-bearing stock exists, but whether it reduces effort, resolves a want, improves capability, or expands future productive possibilities. Bastiat's further distinction between the seen and the unseen explains why many knowledge effects are structurally difficult to measure: the visible item is often the transaction, expenditure, licence fee, or booked cost; the unseen item is the capability formed\index{visible value versus unseen loss}\index{seen versus unseen}\index{Bastiat, Frederic!seen versus unseen}, recombination not attempted, learning loop not entered, or common benefit not realized.

Bastiat supplies the counterfactual discipline KBC requires: economic analysis must count not only the visible asset, transaction, rent, or protected industry, but also displaced production, foregone recombination, suppressed learning, and unrealized common benefit. His obstacle theory further distinguishes productive knowledge services, which reduce effort per unit of satisfaction, from institutionally created access friction, which multiplies compensable obstacles without expanding productive capacity.

\section{What Is the Value of Information?}\index{value of information|textbf}\label{sec:ch2:value-of-information}

The subtitle of this book asks a direct question that should be answered before the technical machinery accumulates: what is the value of natural and artificial intelligence? In this theory, the answer begins one level deeper: intelligence is economically valuable when it performs, improves, preserves, governs, or scales productive knowledge operations. Information by itself is not yet capital. Stored data, expressed claims, model output, documents, or signals become economically valuable only when they can be interpreted, trusted, accessed, maintained, deployed, recombined, or used to prevent loss by a capable actor under a specific governance arrangement.

The value of information is therefore not its mere possession, storage, or expression. Information has economic value when it can yield productive services, improve decisions, enable future recombination or learning, create a defensible governance position, or prevent productive loss under specified access, capability, maintenance, demand, and governance conditions. This is why the framework speaks primarily of \emph{knowledge-bearing stock}: the economically relevant object is not isolated information, but information, skill, data, routines, models, standards, or institutions organised so that productive services can be drawn from them.

A compact KBC valuation expression is:

\[
V^{K}_{i}
=
CUV_{i}
+
ROV_{i}
+
LLOV_{i}
+
COV_{i}
+
SOV_{i}
-
EKL_{i}.
\]

\index{expected knowledge loss (EKL)}\index{expected knowledge loss (EKL)!value decomposition}\index{EKL|see{expected knowledge loss (EKL)}}
Here \(CUV\) is current use value\index{current use value|textbf}: the productive service the stock yields now. \(ROV\) is recombination option value: the value of being able to combine the stock with other stocks in future contexts. \(LLOV\) is learning-loop option value: the value of feedback from use that improves later models, routines, decisions, or capabilities. \(COV\) is control or governance option value\index{control value}\index{governance option value}: the value created by being able to govern access, exclusion, maintenance, quality, attribution, permission\index{permission rules}, or transfer. \(SOV\) is strategic option value: the value of preserving future strategic possibilities, including entry, defence, adaptation, bargaining, or platform positioning. \(EKL\) is expected knowledge loss: the expected value destroyed through degradation, false stock, obsolescence, capability loss, access loss, cyber compromise, governance failure, or suppressed future generation.

This formula is not an accounting rule and it is not a claim that every term can already be measured precisely. It is a plain-language decomposition of value. It says that knowledge-bearing stock is valuable because it can produce current services, preserve future options, accelerate learning, establish governance positions, support strategic flexibility, and avoid loss. It also says that apparent knowledge value must be reduced by expected loss. A dataset with poor provenance, a model no one can safely deploy, a patent without complementary capability, a brittle software dependency, or a commons without maintainers may have high apparent potential but low realised value after impedance and loss are considered.

Natural intelligence and artificial intelligence are valuable when they perform or enable productive knowledge operations. Natural intelligence is valuable through embodied judgement, interpretation, learning, creativity, coordination, responsibility, and context-sensitive action. Artificial intelligence is valuable when some of those operations are externalised into models, software, data systems, or automated workflows that can be scaled, governed, recombined, improved, or deployed. The economic question is therefore not whether intelligence is human or artificial, but how productive knowledge operations are embodied, externalised, governed, and converted into durable capability. The radiology example earlier in this chapter illustrates the point. The model does not acquire the whole radiologist. It may acquire or approximate a bounded operative capability: a task-specific pattern of diagnostic discrimination that can be tested, deployed, monitored, and improved under defined conditions.

The economic problem is therefore not to maximise the private value of information in isolation. That would be too narrow. A regime that maximises private information value may do so through monopoly, surveillance, coercive extraction, excessive secrecy, or enclosure that suppresses future recombination. The better question is how knowledge-bearing stock should be governed so that it produces current services, future recombination, learning, renewal, and public or private productive capacity without creating greater losses through exclusion, fragility, degradation, false stock, or suppressed future generation.

In KBC, the value of knowledge is the expected value of the productive services, future options, learning effects, governance positions, and avoided losses generated by knowledge-bearing stock under specified actor, access, capability, maintenance, demand, and governance conditions. Put more simply, the value of knowledge is its capacity, under specific governance and capability conditions, to yield present productive services, enable future recombination and learning, and prevent productive loss. That is the general answer to the book's subtitle. Chapter~9 develops the measurement apparatus needed to decide when such value is material, when it is merely apparent, and when reducing uncertainty about it is itself worth the cost.

\section{Knowledge-Stock Life-Cycle Profiles}\label{knowledge-stock-life-cycle-profiles}
\index{knowledge depreciation}\index{knowledge impairment}

The five-form taxonomy identifies where productive knowledge resides. The potential-impedance-yield triad identifies how much of that stock can be productively realized by a given actor. A third lens is needed for time: the life-cycle profile of a knowledge-bearing stock. A knowledge-stock life-cycle profile\index{knowledge-stock life-cycle profile} is the trajectory through which a knowledge-bearing stock is formed, embodied, codified, institutionalized, recombined, enclosed, depreciated, revalued, impaired, or renewed over time.

The life-cycle lens prevents two common errors. First, it prevents this theory from treating knowledge-bearing stock as if it only accumulates. Some stock decays, some becomes obsolete, some loses context, some becomes more valuable when new tools make it usable, and some loses value when exclusivity or interpretive capability disappears. Second, it prevents measurement from confusing current use value with future option value. A dataset, protocol, model, or routine may have low present yield but high future value if the surrounding capability, tooling, or institutional governance changes.

\noindent\textit{Example: passive appreciation.} A firm may hold an old dataset of medical images. In 2015, the dataset is useful but limited: available models, graphics processors, annotation methods, and diagnostic protocols\index{diagnostic protocols} restrict what can be extracted from it. By 2026, better AI models\index{AI models}, better GPUs, improved annotation methods, and stronger diagnostic protocols may make the same dataset more productive, even though the dataset itself has not changed. Its value rises because new complementary stocks have entered the actor's accessible recombination field. This is not ordinary depreciation in reverse. A truck usually loses productive value as it ages; knowledge-bearing stock can appreciate when surrounding knowledge, tools, standards, or capabilities make it newly usable. The deeper point is that the value of knowledge-bearing stock is relational, not intrinsic: it depends not only on what the stock is, but on what other stocks it can be combined with, whether the actor can access them, and whether the actor has the capability to use them.

The depreciation profile differs by knowledge form.\index{embodied knowledge@embodied knowledge (\ensuremath{K^E})}\index{disembodied knowledge@disembodied knowledge (\ensuremath{K^D})}\index{institutionalized knowledge@institutionalized knowledge (\ensuremath{K^I})}\index{commons knowledge@commons knowledge (\ensuremath{K^C})}\index{public epistemic infrastructure@public epistemic infrastructure (\ensuremath{K^P})} Embodied knowledge \(K^{E}\) depreciates through non-use, retirement, turnover, or obsolescence. Disembodied knowledge \(K^{D}\) depreciates through technical decay, context loss, cyber compromise, or incompatibility. Institutionalized knowledge \(K^{I}\) depreciates through reorganization, leadership change, or process breakdown. Commons knowledge \(K^{C}\) depreciates through maintainer depletion\index{maintainer depletion}, even where formal access remains open. Public epistemic infrastructure \(K^{P}\) depreciates through underfunding, politicization, fraud, or trust erosion. The point is not simply that knowledge assets lose value over time. It is that the form of residence determines the likely impairment\index{impairment} pathway, which is why Chapter 9 treats hidden knowledge-capital loss as a dark-capital problem rather than as ordinary depreciation alone.

\begingroup
\small
\setlength{\tabcolsep}{4pt}
\renewcommand{\arraystretch}{1.12}
\begin{longtable}{@{}L{0.19\textwidth}L{0.30\textwidth}L{0.43\textwidth}@{}}
\caption{Knowledge-Stock Life-Cycle Profile}\label{tab:ch2:knowledge-stock-life-cycle-profile}\\
\toprule
Stage & Physical-capital analogue & Knowledge-capital difference \\
\midrule
\endfirsthead
\toprule
Stage & Physical-capital analogue & Knowledge-capital difference \\
\midrule
\endhead
\bottomrule
\endlastfoot
Formation & Investment & Learning, discovery, invention, interpretation, or codification creates stock. \\
Deployment & Use & Use may improve the stock through feedback rather than merely consume it. \\
Recombination & Redeployment & Value may rise when the stock enters new contexts or new pairings. \\
Enclosure & Ownership or control & Control may fund investment and quality, but may also narrow future fields. \\
Depreciation & Wear & Depreciation may take the form of decay, obsolescence, context loss, or regeneration failure. \\
Revaluation & Appraisal or resale & Dormant knowledge may appreciate when new tools, questions, or governance arrangements make it usable. \\
Impairment & Damage or loss & Copying, poisoning\index{poisoning}, exfiltration, capability departure, or access loss can impair the value of knowledge-bearing stock even when the original file or artefact remains. \\

\end{longtable}
\endgroup

These five measurement properties prevent KBC measurement from collapsing into valuation alone. A knowledge-bearing stock may need to be measured as a stock, a service flow, a risk exposure\index{risk exposure}, a trajectory, or a counterfactual non-formation event.

\begingroup
\small
\setlength{\tabcolsep}{4pt}
\renewcommand{\arraystretch}{1.12}
\begin{longtable}{@{}L{0.22\textwidth}L{0.28\textwidth}L{0.42\textwidth}@{}}
\caption{Measurement Properties of Knowledge-Bearing Stock}\label{tab:ch2:measurement-properties}\\
\toprule
Measurement property & Question answered & Example \\
\midrule
\endfirsthead
\toprule
Measurement property & Question answered & Example \\
\midrule
\endhead
\bottomrule
\endlastfoot
Stock level\index{measurement!stock level} & How much knowledge-bearing stock exists? & Size, coverage, completeness, or capability base of a dataset, model, routine, or expert team. \\
Service yield\index{measurement!service yield} & What productive service does it generate? & Time saved, error reduction, throughput, decision quality, resilience, revenue support. \\
Loss exposure\index{measurement!loss exposure} & What value is at risk? & Platform dependency, exfiltration, capability loss, maintainer failure, licence withdrawal. \\
Trajectory\index{measurement!trajectory} & Is the stock improving, decaying, or changing governance? & Model drift, documentation decay, maintainer decline, learning-loop improvement. \\
Foregone formation\index{measurement!foregone formation} & What knowledge did not form because conditions were suppressed? & Lost recombinations, unattempted experiments, blocked learning, degraded commons maintenance. \\
\end{longtable}
\endgroup

The first three properties are already familiar in mature cybersecurity and resilience programmes. Asset and data inventories approximate stock-level measurement\index{stock-level measurement}; asset criticality and risk ratings approximate loss-exposure measurement\index{loss-exposure measurement}; and business-impact analysis, uptime expectations, RTO, and RPO approximate service-yield requirements\index{service-yield measurement}. KBC extends this familiar measurement logic to knowledge-bearing stock and adds two properties that are usually weaker or absent: trajectory\index{trajectory measurement} and foregone formation\index{foregone-formation measurement}. The Technical Companion, Appendix D, operationalizes these properties through the KBC-AIE measurement sequence, which determines when further measurement is worth its cost.

A schematic life-cycle accounting identity\index{knowledge-stock life-cycle accounting identity} can therefore be written as:

\begin{equation}
K^{x}_{i, t+1}
=
K^{x}_{i, t}
+G^{x}_{i, t}
+\mathrm{Conv}^{x}_{i, t}
+\mathrm{Rec}^{x}_{i, t}
-\mathrm{Dep}^{x}_{i, t}
-\mathrm{Imp}^{x}_{i, t}
+\mathrm{Rev}^{x}_{i, t}.
\label{eq:ch2:knowledge-stock-lifecycle-placeholder}
\end{equation}

Equation~\ref{eq:ch2:knowledge-stock-lifecycle-placeholder} is a placeholder identity. Its purpose is qualitative: to distinguish formation, conversion, recombination, depreciation, impairment, renewal, and revaluation without implying that these flows are already calibrated as an aggregate stock equation.

\section{The First-Conversion Zone}\label{the-first-conversion-zone}
\index{first conversion|textbf}

The same discovery can become a patent, trade secret, preprint, public dataset, professional standard, or commons contribution\index{IP policy!first conversion}\index{public datasets!first conversion}\index{professional standards!first conversion} depending on the first institutional conditions it encounters. A drug mechanism discovered under public funding by a university researcher enters the first-conversion zone with one set of conditions: the relevant technology transfer provisions, the researcher's employment contract, the funder's IP policy\index{technology transfer}\index{IP policy!technology transfer}\index{employment contracts!first conversion}\index{funding policy!IP}, and the researcher's own decisions about disclosure. The same mechanism discovered under corporate R\&D funding enters the zone under different conditions: employer IP ownership, confidentiality obligations, trade secret protections, and patent strategy decisions. The same mechanism discovered independently by an individual researcher outside institutional employment enters under yet another set of conditions: personal ownership subject to voluntary disclosure or non-disclosure, with no employer claim and no institutional governance requirement. Identical knowledge; different governance trajectories through the same first-conversion zone.

The First-Conversion Zone is the institutional moment or interval in which newly generated knowledge-bearing stock first acquires a governable form and becomes subject to claims of ownership, access, enclosure, sharing, or public use. It is not a single transaction or a defined point in a linear sequence but a zone of first-conversion events, the decisions, processes, and institutional conditions by which newly generated stock is externalized, codified, institutionalized, or enclosed for the first time.

The relevant object is not every movement of every idea. The relevant object is a materially significant governance transition: a movement that changes expected cash flows, productive capacity, access, control, risk exposure, bargaining power, recombination opportunity, competition, or public value. This materiality filter\index{first-conversion materiality filter}\index{materiality} prevents this theory from becoming a catalogue of ordinary idea movement. KBC is concerned with conversions large enough to alter valuation, management, governance, or public-economic consequences.\index{ordinary idea movement versus material governance transition}

The concept addresses a gap that standard theories of knowledge creation and knowledge use do not clearly fill: the question of when and how the governance of new knowledge is determined. Generation does not settle ownership. The fact that a researcher has discovered something tells us that the discovery has occurred; it does not tell us whether the resulting knowledge becomes a private asset, a public good, a firm capability, or a professional standard. That determination happens in the first-conversion zone, and it depends entirely on the institutional conditions surrounding the generation event, not on any property of the knowledge itself. This is the Conditional Separability Axiom applied to the moment of first conversion.

The practical consequence of the First-Conversion Zone concept is that it makes visible an intervention point that conventional discussions of knowledge economics tend to obscure. In most accounts, intellectual property law governs knowledge that has already been converted, patents are filed after invention, copyright attaches to fixed expression, trade secrets are protected after they exist. The First-Conversion Zone concept identifies the interval before those legal instruments attach, in which the institutional context determines what governance instrument will be available and who will hold it. Understanding that zone, who is in it, under what institutional conditions, with what resources and obligations, is the prerequisite for understanding why identical generation processes produce different distributional outcomes.

The five-form taxonomy applies to the first-conversion zone with full force. Newly generated knowledge does not enter the conversion cycle in a form-neutral state; it enters as some initial combination of the five forms, and the initial form partly determines the institutional conditions of first conversion. Knowledge generated through embodied \(K^{E}\) (a researcher's judgment) enters the zone as \(K^{E}\) first; whether it subsequently becomes \(K^{D}\) (through codification), \(K^{I}\) (through organizational embedding), \(K^{C}\) (through open publication), or \(K^{P}\) (through public dissemination) depends on the institutional conditions in that zone. The trajectory is not predetermined by the act of generation.

\section{Knowledge Governance Forms}\label{knowledge-regimes}
\index{governance form|textbf}\index{governance transition}

The same knowledge-bearing stock may pass through several different governance forms across the course of its conversion cycle. The knowledge governance form governing a given stock at any point is determined not by any intrinsic property of the knowledge but by the institutional conditions in place at that stage of conversion. This is the governance-level application of the Conditional Separability Axiom: just as separability is relational rather than intrinsic, so governance is contextual rather than fixed.

Governance forms are therefore a diagnostic tool, not merely a taxonomy. Given a knowledge-bearing stock, the practical question is: which governance form currently governs it, and what would change if it moved to another governance form? The answer determines who can access it, use it, maintain it, monetize it, recombine it, exclude others from it, or convert it into another form.

The following transition questions operationalize the point. They ask not merely what label applies to a stock, but what economic consequence follows when the governing arrangement changes.

{\small
\begingroup
\setlength{\tabcolsep}{5pt}
\renewcommand{\arraystretch}{1.15}
\par\addvspace{0.8\baselineskip}\noindent
\begin{longtable}{@{}L{0.38\textwidth}L{0.54\textwidth}@{}}
\caption{Governance Transitions as Diagnostic Questions}\label{tab:ch2:governance-transition-questions}\\
\toprule\noalign{}
Transition & Economic question \\
\midrule\noalign{}
\endfirsthead
\toprule\noalign{}
Transition & Economic question \\
\midrule\noalign{}
\endhead
\bottomrule\noalign{}
\endlastfoot
Worker expertise \(\rightarrow\) firm routine/software & Did embodied knowledge become firm-retained capital? \\
Public research \(\rightarrow\) private patent & Did public epistemic capital become proprietary rent? \\
Open-source commons \(\rightarrow\) platform service & Did commons value become platform rent? \\
User behaviour \(\rightarrow\) AI model improvement & Did distributed user activity become enclosed model capability? \\
Firm routine \(\rightarrow\) public standard & Did private capability become public epistemic infrastructure? \\
\end{longtable}
\endgroup
}

Six governance forms are analytically significant. They are not exhaustive categories; they are ideal types that help diagnose the governance conditions operating on any given piece of knowledge-bearing stock at any given moment.

{\scriptsize
\begingroup
\scriptsize
\setlength{\tabcolsep}{3pt}
\renewcommand{\arraystretch}{1.12}
\sloppy
\par\addvspace{0.8\baselineskip}\noindent
\begin{longtable}{@{}L{0.14\textwidth}
L{0.18\textwidth}
L{0.27\textwidth}
L{0.17\textwidth}
L{0.18\textwidth}@{}}
\caption{Knowledge Governance Forms: Governance Mechanisms and Economic Character}\label{tab:ch2:knowledge-regimes}\\
\toprule\noalign{}
\begin{minipage}[b]{\linewidth}\raggedright
Governance form
\end{minipage} & \begin{minipage}[b]{\linewidth}\raggedright
Governance mechanism
\end{minipage} & \begin{minipage}[b]{\linewidth}\raggedright
Economic character
\end{minipage} & \begin{minipage}[b]{\linewidth}\raggedright
Primary form(s)
\end{minipage} & \begin{minipage}[b]{\linewidth}\raggedright
Vulnerability
\end{minipage} \\
\midrule\noalign{}
\endfirsthead
\toprule\noalign{}
\begin{minipage}[b]{\linewidth}\raggedright
Governance form
\end{minipage} & \begin{minipage}[b]{\linewidth}\raggedright
Governance mechanism
\end{minipage} & \begin{minipage}[b]{\linewidth}\raggedright
Economic character
\end{minipage} & \begin{minipage}[b]{\linewidth}\raggedright
Primary form(s)
\end{minipage} & \begin{minipage}[b]{\linewidth}\raggedright
Vulnerability
\end{minipage} \\
\midrule\noalign{}
\endhead
\bottomrule\noalign{}
\endlastfoot
Private IP & Patents, copyright, trade secret & Knowledge rent; institutionally created excludability\index{excludability!institutionally created} & \(K^{D}\) & Enforcement cost; patent expiry \\
Firm capability & Employment contract, non-compete, organizational knowledge & Institutionalized knowledge capital; not marketable as discrete asset & \(K^{I}\), \(K^{E}\) & Personnel departure; organizational disruption \\
Platform dependency & ToS, API access, data architecture & Platform captures user-generated signals and controls the improved capability & \(K^{D}\), \(K^{I}\) & Platform strategic decisions; regulatory change \\
Professional stewardship & Accreditation, peer review, liability, ethical standards\index{ethical standards} & Institutionalized norms governing interpretation and attribution & \(K^{I}\), \(K^{E}\) & Credentialing capture; professional fragmentation \\
Commons & Open licence, foundation governance, contribution norms & Non-rival surplus; collective maintenance burden & \(K^{C}\) & Contributor withdrawal; governance failure; commons-depletion mechanism \\
Public infrastructure & State funding, open standards, shared measurement & Epistemic infrastructure; precondition for generation & \(K^{P}\) & Budget cuts; privatization; institutional erosion \\
\end{longtable}
\endgroup
}

The diagnostic question (which governance form governs this stock at this stage of the conversion cycle?) is a tool, not a classification. Any given piece of knowledge-bearing stock may shift from one governance form to another as it moves through the cycle. A drug mechanism discovered in a publicly funded laboratory (public infrastructure, \(K^{P}\)) becomes a patented compound (private IP, \(K^{D}\)), is manufactured under licence by a firm (firm capability, \(K^{I}\)), and enters clinical guidelines (professional stewardship, \(K^{I}\)). The governance form governing the stock at any moment is determined by the institutional conditions at that stage, not by any property of the knowledge itself.

The five-form taxonomy and the governance-form typology connect as follows: any given governance form primarily governs one or two forms of knowledge-bearing stock, as the table above indicates. Commons governance governs \(K^{C}\); public-infrastructure governance governs \(K^{P}\). The connection matters for the commons-depletion corollary: an incumbent can enclose \(K^{C}\) not by acquiring control of the knowledge itself (which may be formally open-licensed) but by hiring away the \(K^{E}\) and \(K^{I}\) (the embodied expertise and the organizational capacity) that the commons depends on for maintenance, governance, and development. The governance form governing \(K^{C}\) does not change; the capability stock required to operate that governance form is depleted. The formal legal status of the commons is preserved while its productive capacity is extinguished.

This governance-cycling property has distributional implications that static accounts of knowledge economics tend to miss. The revenue generated at each stage (the licensing fees from private IP, the profits from firm capability, the managed access fees from platform dependency) accrues to different parties depending on which institutional actors control the relevant governance mechanism at that stage. Understanding the distributional consequences of knowledge-bearing capitalism therefore requires tracking economically material knowledge-bearing stock when it crosses a governance boundary and changes who can access, use, govern, monetize, maintain, or exclude others from it. The practical question is not where all knowledge went, but whether a transition changed productivity, access, rents, risk, competition, public value, bargaining power, or future generation capacity.

\noindent\textit{Why this matters for the next chapters.} Chapter 2 defines the objects. It does not yet explain how new knowledge-bearing stock is generated or how existing stock moves through governance pathways. Those are separate questions: the Knowledge Generation Model explains creation; the Knowledge Conversion Matrix explains transformation, access, and distribution after stock exists.

\section*{What this lets us see}
\addcontentsline{toc}{section}{What this lets us see}

The purpose of this chapter has not been to multiply terminology. It has been to make visible five questions that conventional categories often blur: where productive knowledge resides, whether it can be separated from the person, artefact, organization, community, or setting in which it resides, which governance form governs it, what productive service it yields, and what is lost when it is converted, enclosed, depleted, or mismeasured.

This matters because those questions change how familiar economic problems are diagnosed:

\begin{itemize}
\item \textbf{Valuation:} the relevant asset may not be the recorded artefact but the unrecorded knowledge-bearing stock, complementary capability, access position, or learning option that allows the artefact to yield future service.
\item \textbf{Labour bargaining:} codification does not mechanically transfer value from worker to firm; it changes the form, residence, and governance of embodied knowledge, with different bargaining outcomes depending on whether codification substitutes for, amplifies, institutionalizes, or reputationally rewards the worker's contribution.
\item \textbf{AI substitution:} the economically relevant question is not whether a model imitates a task, but whether embodied judgement has been transformed into disembodied or institutionalized capability under conditions that allow substitution, augmentation, monitoring, or new dependence.
\item \textbf{Accounting invisibility:} economically material knowledge-bearing stock may produce cash flows, risk reduction, option value, or fragility without appearing as a recognized balance-sheet asset.
\item \textbf{Platform power:} platforms do not merely intermediate exchange; they can govern access to data, users, feedback loops, APIs, standards, and recombination fields, thereby converting distributed activity into proprietary capability.
\item \textbf{Commons depletion:} a formally open stock can lose productive capacity when the maintainers, institutions, norms, or complementary capabilities required to keep it usable are exhausted or enclosed elsewhere.
\item \textbf{Cyber and operational impairment:} knowledge-bearing capital can be impaired by data corruption, loss of provenance, inaccessible systems, compromised models, broken workflows, or degraded trust, even when physical assets remain intact.
\item \textbf{Public infrastructure:} public epistemic capital, including standards, measurement systems, scientific methods, protocols, and knowledge institutions, is not background scenery. It is productive infrastructure whose erosion changes the economy's capacity to generate reliable knowledge.
\end{itemize}

Interlude I now turns from definition to interpretation. If Chapter 2 has named the objects, the interlude asks why those objects are difficult to see inside inherited economic language. That interpretive bridge prepares the next two mechanism chapters. Chapter 3 asks how new knowledge-bearing stock is generated. Chapter 4 asks what happens after such stock exists: how it is converted, accessed, governed, enclosed, or appropriated.

\chapter*{Interlude I: Knowledge Stock Dynamics}
\addcontentsline{toc}{chapter}{Interlude I: Knowledge Stock Dynamics}
\label{interlude:knowledge-stock-dynamics}
\emph{Decay, Obsolescence, Regeneration, and Appreciation under Enclosure}

\section*{Why Stock Dynamics Are Necessary}\label{why-stock-dynamics-are-necessary}

The preceding chapters establish what knowledge-bearing stock is, how it is generated through the interaction of search, interpretation, and articulation, and how it is converted across the five forms of the taxonomy, embodied (\(K^{E}\)), disembodied (\(K^{D}\)), institutionalized (\(K^{I}\)), commons knowledge capital (\(K^{C}\)), and public epistemic capital (\(K^{P}\)). What the framework has not yet confronted is time. Without a theory of how knowledge-bearing stock changes in value over the course of its existence, the model implies a kind of false permanence: stock, once created and converted, simply accumulates. Productive capacity compounds without limit. The balance sheet, even if redesigned\index{balance-sheet redesign} to recognize intangible assets, merely grows.

This implication is obviously false. A codebase that is not maintained becomes a liability. A clinical protocol that is not updated against new evidence loses clinical validity. An organizational routine whose practitioners have retired carries diminishing productive power. A trained model whose training data no longer reflects the current distribution of inputs begins to produce unreliable outputs. A commons whose governance community has been hired away by an incumbent firm retains its licence and its repository while losing the interpretive, review, and coordination capacity that made it productive.

This theory of knowledge stock dynamics developed in this section addresses this gap. It introduces three analytic levels (value-change processes, causal pathways, and governance conditions) and uses them to explain how knowledge-bearing stock loses, preserves, transfers, and gains productive value over time. It then shows that these dynamics are not symmetric across actors. Under enclosed knowledge governance forms, the value trajectories of incumbent and excluded actors diverge and compound in ways that constitute the concentration mechanism of knowledge-bearing capitalism.

\section*{The Life-Cycle Lens}\label{the-life-cycle-lens}

Chapter 2 defines the knowledge-stock life-cycle profile: formation, deployment, recombination, enclosure, depreciation, revaluation, impairment, and renewal. This interlude supplies the dynamic interpretation of that profile. The same stock can appreciate in one relation and depreciate in another; it can remain physically present while its productive yield collapses; it can become more valuable when new tools make old knowledge newly recombinable; and it can be impaired when adversaries, platforms, or departing specialists change the actor's access, capability, or exclusivity position. The life-cycle lens therefore treats knowledge-capital dynamics as actor-to-stock relations over time, not as automatic stock accumulation.

\begin{figure}[htbp]
\caption[Knowledge-Stock Life-Cycle as Actor-to-Stock Relation]{Knowledge-Stock Life-Cycle as Actor-to-Stock Relation}
\label{fig:interlude:lifecycle-actor-relation}
\centering
\includegraphics[width=\textwidth]{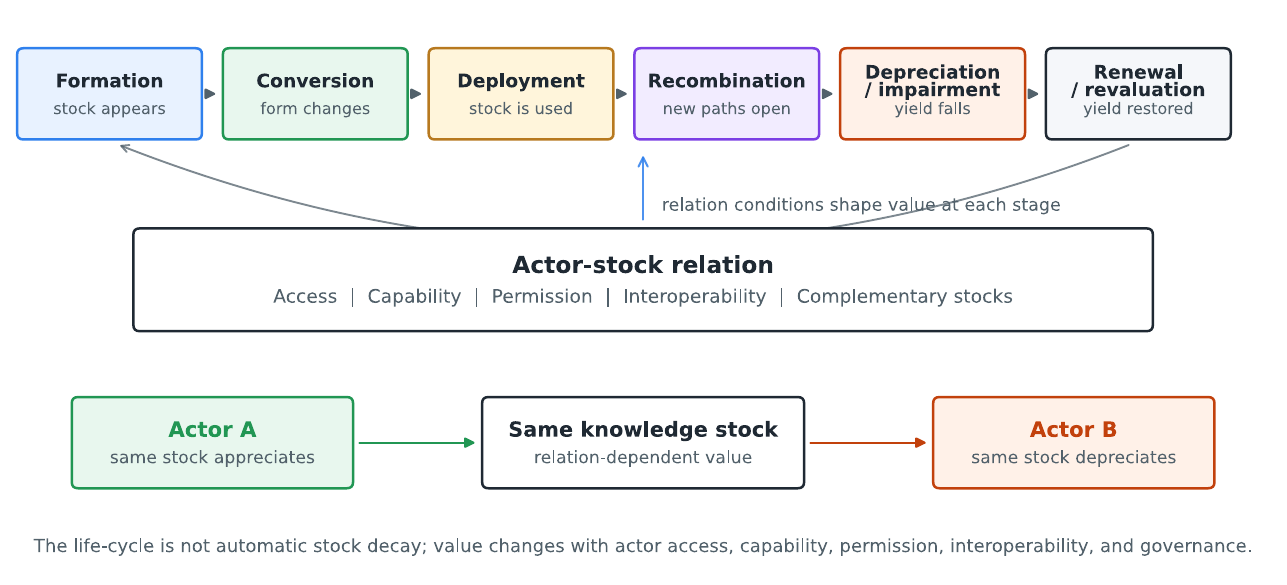}
\footnotesize\emph{Note.} The life-cycle is not an automatic depreciation schedule. The same stock can appreciate for one actor and depreciate for another because value depends on access, capability, permission, interoperability, and complementary stocks.
\end{figure}

\section*{The Analytic Architecture: Three Levels}\label{the-analytic-architecture-three-levels}

Inherited accounts of depreciation, from both economics and accounting, apply a single mechanism (the passage of time combined with use) to a defined asset class and express the result as a rate. This works tolerably well for physical capital, whose productive decline through metal fatigue, material degradation, and technological obsolescence is roughly measurable in physical terms. It fails for knowledge-bearing stock for two related reasons.

First, because knowledge is non-rival, it does not wear out through use. A codebase is not degraded by being executed. A dataset is not depleted by being queried. A protocol is not consumed by being followed. Whatever productive loss occurs must arise from something other than use-consumption. Second, the mechanisms through which knowledge-bearing stock loses productive value are heterogeneous. Code rot, the emergence of a superior competing model, the withdrawal of API access, and the departure of the governance community are four distinct phenomena requiring distinct analytical treatment. Compressing them into a single depreciation rate produces false precision.

The solution is a three-level analytic architecture. Level 1 identifies value-change processes: what happens to the productive value of a stock. Level 2 identifies causal pathways: what initiates or accelerates those processes. Level 3 identifies governance conditions: the institutional arrangements that enable, prevent, or shape the pathways. Real cases will often require all three levels to describe adequately, and a sequence of pathways may be required to account for what appears as a single event.

\subsection*{Level 1: Value-Change Processes}\label{level-1-value-change-processes}

The six value-change processes below describe what happens to productive value. They are not mutually exclusive in application (a real case may involve several simultaneously) but they are conceptually distinct in mechanism.

\begingroup
\small
\setlength{\tabcolsep}{3pt}
\renewcommand{\arraystretch}{1.12}
\sloppy
\par\addvspace{0.8\baselineskip}\noindent
\begin{longtable}{@{}L{0.20\textwidth}
L{0.24\textwidth}
L{0.31\textwidth}
L{0.19\textwidth}@{}}
\caption{Value-Change Processes in Knowledge Stock Dynamics}\label{tab:int1:value-change-processes}\\
\toprule\noalign{}
\begin{minipage}[b]{\linewidth}\raggedright
Process
\end{minipage} & \begin{minipage}[b]{\linewidth}\raggedright
Definition
\end{minipage} & \begin{minipage}[b]{\linewidth}\raggedright
Physical-capital analogue
\end{minipage} & \begin{minipage}[b]{\linewidth}\raggedright
Accounting recognition
\end{minipage} \\
\midrule\noalign{}
\endfirsthead
\toprule\noalign{}
\begin{minipage}[b]{\linewidth}\raggedright
Process
\end{minipage} & \begin{minipage}[b]{\linewidth}\raggedright
Definition
\end{minipage} & \begin{minipage}[b]{\linewidth}\raggedright
Physical-capital analogue
\end{minipage} & \begin{minipage}[b]{\linewidth}\raggedright
Accounting recognition
\end{minipage} \\
\midrule\noalign{}
\endhead
\bottomrule\noalign{}
\endlastfoot
\textbf{Decay} & Loss of productive value through failure to maintain the capability conditions required to interpret, update, deploy, or govern the stock; the artefact persists while the productive relation erodes & Partial: physical wear, but mechanism differs; knowledge decay is capability-relative, not material & Rarely recognized; impairment applies only to identifiable assets with external indicators of value loss \\
\textbf{Obsolescence} & Loss of relative productive value through frontier advancement in the same domain, without modification of the stock itself; the stock is intact but no longer at the productive frontier & Yes: technological obsolescence of machinery and equipment & Partial: depreciation schedules approximate expected obsolescence; relative obsolescence (frontier-based) is invisible to accounting \\
\textbf{Access impairment} & Decline in productive value for an excluded actor class resulting from change in the legal, contractual, technical, price, or institutional conditions governing their actor-to-stock relation; the stock remains intact for the controlling party & No direct analogue: a machine cannot be impaired for one party while fully productive for another & Not recognized; accounting tracks assets, not actor-to-stock relations; access impairment leaves no accounting trace \\
\textbf{Regeneration} & Recovery or enhancement of productive value through maintenance, updating, governance repair, or reactivation of capability conditions that had eroded & Maintenance and repair (partial) & Expensed rather than capitalized in most accounting treatments; the restored value is not recognized on the balance sheet \\
\textbf{Passive revaluation} & Appreciation of existing stock through emergence of new complementary knowledge, infrastructure, techniques, or institutional conditions, without modification of the stock itself & No direct analogue: physical capital does not spontaneously appreciate through others' investments & Not recognized; accounting does not mark assets upward for exogenous complementary investment by third parties \\
\textbf{Active recombination} & Appreciation through intentional combination of existing stocks from different domains or forms, producing new productive capability that exceeds the sum of the inputs & Innovation and R\&D investment (partial) & Only if a new separately identifiable IP asset is created; the combinatorial gain accruing to prior stocks is typically unrecognized \\
\end{longtable}
\endgroup

Two structural features of this table deserve immediate emphasis. First, passive revaluation and active recombination reveal an asymmetry that distinguishes knowledge-bearing stock from physical capital: existing knowledge can appreciate without being newly created or replaced. Physical capital generally depreciates monotonically. A machine does not spontaneously regain productive capacity because the economic environment changes. Knowledge-bearing stock can, and frequently does, gain productive value when new complements emerge, when computation becomes cheaper, or when regulatory or market conditions shift.

Second, access impairment is not equivalent to deterioration of the stock. A dataset that becomes paywalled has not decayed. A standard that has become proprietary has not become obsolete. What has changed is the actor-to-stock relation, the bundle of legal, technical, and institutional conditions that determine whether an actor can read, use, modify, copy, redistribute, audit, or build upon the stock. This distinction between stock deterioration and relation change is analytically critical and is developed further in Section 4.

\subsection*{Level 2: Causal Pathways}\label{level-2-causal-pathways}

Causal pathways are what initiate or accelerate value-change processes. They are not themselves value-change processes. Capability displacement (an engineer leaving a firm, a maintainer community being hired away) is not itself decay. It is the initiating event that creates conditions under which decay accelerates. Recognizing this distinction allows the framework to describe sequential dynamics: a single initiating pathway can set in motion a chain of value-change processes whose combined effect is much larger than any single Level 1 entry. Access withdrawal and capability displacement are later formalized in the Knowledge Conversion Matrix (Chapter 4) as {[}A{]} (access) and {[}T{]} (transformation) mechanisms respectively: what the interlude treats as causal pathways in the dynamics of stock value, the KCM treats as governance mechanisms in the conversion of existing stock.

\begingroup
\small
\setlength{\tabcolsep}{3pt}
\renewcommand{\arraystretch}{1.12}
\sloppy
\par\addvspace{0.8\baselineskip}\noindent
\begin{longtable}{@{}L{0.20\textwidth}
L{0.24\textwidth}
L{0.31\textwidth}
L{0.19\textwidth}@{}}
\caption{Causal Pathways in Knowledge Stock Dynamics}\label{tab:int1:causal-pathways}\\
\toprule\noalign{}
\begin{minipage}[b]{\linewidth}\raggedright
Pathway
\end{minipage} & \begin{minipage}[b]{\linewidth}\raggedright
Definition
\end{minipage} & \begin{minipage}[b]{\linewidth}\raggedright
Initiates or accelerates
\end{minipage} & \begin{minipage}[b]{\linewidth}\raggedright
Illustrative instance
\end{minipage} \\
\midrule\noalign{}
\endfirsthead
\toprule\noalign{}
\begin{minipage}[b]{\linewidth}\raggedright
Pathway
\end{minipage} & \begin{minipage}[b]{\linewidth}\raggedright
Definition
\end{minipage} & \begin{minipage}[b]{\linewidth}\raggedright
Initiates or accelerates
\end{minipage} & \begin{minipage}[b]{\linewidth}\raggedright
Illustrative instance
\end{minipage} \\
\midrule\noalign{}
\endhead
\bottomrule\noalign{}
\endlastfoot
\textbf{Capability displacement} & Movement of embodied (\(K^{E}\)) or institutionalized (\(K^{I}\)) capability from one actor or governance arrangement to another, severing the maintenance relation to remaining disembodied (\(K^{D}\)), commons (\(K^{C}\)), or public (\(K^{P}\)) stock & Decay; acceleration of obsolescence; incumbent regeneration & Commons maintainers hired by an incumbent firm; founding team departure following acquisition \\
\textbf{Access withdrawal} & Change in legal, contractual, technical, price, or interface conditions governing an actor's relation to stock, such that the actor can no longer use, modify, contribute to, or build upon the stock on prior terms & Access impairment; delayed decay through severed distributed maintenance & API closure; dataset paywalling; open standard proprietization; model weight terms-of-service restriction \\
\textbf{Feedback capture} & Diversion of deployment-generated signals: errors, corrections, usage patterns, edge cases, ground-truth labels: from distributed or open collection to exclusive incumbency & Incumbent regeneration and active recombination; excluded-actor obsolescence relative to frontier & Terms-of-service capture of user interaction data; proprietary fine-tuning on deployment feedback; closed evaluation benchmarks \\
\textbf{Commons underinvestment} & Failure to fund, govern, or maintain the institutional infrastructure required to keep commons knowledge (\(K^{C}\)) or public epistemic capital (\(K^{P}\)) productive, without any single actor's deliberate enclosure & Decay; obsolescence acceleration; erosion of regeneration capacity & Chronic underfunding of open-source maintainer communities; withdrawal of public research infrastructure support; governance community attrition \\
\textbf{Frontier movement} & Advance in the productive frontier of a knowledge domain, which produces passive revaluation for actors who can access the new frontier and relative obsolescence for actors who cannot & Passive revaluation (for actors with access); obsolescence acceleration; access impairment (where frontier is enclosed) & New model generation displacing prior generation; new clinical evidence superseding prior protocol; new computational paradigm obsolescing prior infrastructure \\
\textbf{Retrospective enclosure} & Extension or reassertion of proprietary rights over stock whose productive value has appreciated through subsequent public or third-party investment in complements & Access impairment; capture of passive revaluation gains by incumbent rather than distributing them & Patent term extension; copyright extension on culturally and technically productive works; re-proprietization of open standards after adoption \\
\end{longtable}
\endgroup

Three sequential dynamics deserve particular attention because they recur throughout this book's case analyses.

\emph{Access withdrawal triggering capability displacement and then decay.} When a platform closes an API or a dataset is paywalled, the actor class that previously built on or maintained that stock loses both its access and its practical standing to contribute to maintenance. The community of users who were also validators, annotators, correctors, or integrators is severed from the stock. The stock persists nominally, but its maintenance pathway has been severed, and decay begins.

\emph{When the productive capability needed to use or improve a knowledge asset becomes concentrated in one incumbent, that incumbent gets stronger while challengers' related capabilities decay.} When maintainers of a commons migrate to a single incumbent employer, the incumbent gains embodied (\(K^{E}\)) and institutionalized (\(K^{I}\)) capability. The commons (\(K^{C}\)) loses it. Incumbent stock may improve while commons stock depreciates. These are not independent events; they are two expressions of one capability movement, visible only because the five-form taxonomy keeps track of both sides.

\emph{Frontier movement producing opposite effects for actors with and without access.} An advance at the frontier of a knowledge domain creates passive revaluation opportunities for actors who can access and combine with the new knowledge. For actors without access (whether because the frontier is enclosed, priced beyond reach, or technically incompatible with their infrastructure) the same frontier movement accelerates relative obsolescence of their existing stock. The event is identical; its distributional consequences are determined by the access-governance arrangement.

\subsection*{Level 3: Governance Conditions}\label{level-3-governance-conditions}

Governance conditions are the institutional arrangements that enable, prevent, or shape the causal pathways described at Level 2. They determine who can move what capability where, and on what terms. In first-conversion terms, they are the arrangements through which generated knowledge is paired with residence and governance, \(G \rightarrow K^{r,g}\), before later deployment, feedback, or depreciation dynamics unfold.

\begingroup
\small
\setlength{\tabcolsep}{3pt}
\renewcommand{\arraystretch}{1.12}
\sloppy
\par\addvspace{0.8\baselineskip}\noindent
\begin{longtable}{@{}L{0.27\textwidth}
L{0.33\textwidth}
L{0.34\textwidth}@{}}
\caption{Governance Conditions for Knowledge Stock Dynamics}\label{tab:int1:governance-conditions}\\
\toprule\noalign{}
\begin{minipage}[b]{\linewidth}\raggedright
Condition
\end{minipage} & \begin{minipage}[b]{\linewidth}\raggedright
Definition
\end{minipage} & \begin{minipage}[b]{\linewidth}\raggedright
Effect on causal pathways
\end{minipage} \\
\midrule\noalign{}
\endfirsthead
\toprule\noalign{}
\begin{minipage}[b]{\linewidth}\raggedright
Condition
\end{minipage} & \begin{minipage}[b]{\linewidth}\raggedright
Definition
\end{minipage} & \begin{minipage}[b]{\linewidth}\raggedright
Effect on causal pathways
\end{minipage} \\
\midrule\noalign{}
\endhead
\bottomrule\noalign{}
\endlastfoot
\textbf{IP regime} & Legal rules governing ownership, duration, licensing, and enforcement of rights over knowledge artefacts & Determines whether access withdrawal and retrospective enclosure are available as strategies; shapes the scope of feedback capture; governs whether passive revaluation gains are retained or distributable \\
\textbf{Labour market governance} & Non-compete clauses, non-disclosure agreements, and mobility constraints on workers & Enables or limits capability displacement; determines whether embodied knowledge (\(K^{E}\)) can migrate with its bearer and under what conditions \\
\textbf{Open-licence architecture} & Open-source, Creative Commons, and analogous licences governing rights to use, modify, redistribute, and build upon disembodied stock & Constrains access withdrawal and retrospective enclosure for licensed works; sustains distributed maintenance pathways; determines conditions under which commons stock (\(K^{C}\)) can remain productively accessible \\
\textbf{Data portability and interoperability mandates} & Regulatory requirements governing actors' rights to access, export, and use data generated through deployment & Shapes feedback capture scope; determines whether deployment signals are exclusively retained by the deploying actor or are partially accessible to others; governs the depth of feedback enclosure \\
\textbf{Commons governance institutions} & Formal and informal institutions governing contribution, maintenance, review, stewardship, and evolution of shared knowledge stocks (\(K^{C}\)) and public epistemic capital (\(K^{P}\)) & Determines resilience of commons to capability displacement and underinvestment; shapes regeneration capacity; governs whether distributed maintenance can sustain productive value against incumbent competition for maintainer time \\
\textbf{Competition and antitrust framework} & Rules governing market concentration, merger review, and abuse of dominant position & Determines whether enclosure strategies face competitive discipline; shapes whether access impairment of excluded actors is subject to remedy; governs whether feedback capture creates contestable or non-contestable advantages \\
\end{longtable}
\endgroup

This three-level architecture means that a complete analysis of any knowledge stock event will typically identify a governance condition that enabled a causal pathway that produced one or more value-change processes. A non-compete clause weakens the labour market pathway through which capability displacement might otherwise occur. An open-source licence creates a governance condition under which feedback and contribution remain distributed rather than enclosed. A data portability mandate governs whether feedback capture is available only to the deploying firm or accessible to others.

\section*{The Depreciating Object: Actor-to-Stock Relations}\label{the-depreciating-object-actor-to-stock-relations}

The most important conceptual move in the knowledge stock dynamics model is to specify correctly what is depreciating. The tempting answer is: the stock itself. A model, dataset, codebase, or protocol ages, becomes stale, loses relevance, falls behind the frontier. But this answer is incomplete in a distinctive way.

Because knowledge is non-rival, the encoded content of a knowledge artefact does not wear out through use or passage of time alone. A mathematical theorem may remain as true today as when it was first written. The archived dataset contains the same records it always did. The documented protocol is still readable. The asset did not disappear. The system that makes the asset valuable deteriorated. What deteriorates is not necessarily the artefact itself, but the capability system around it: the maintainers, interpreters, routines, infrastructure, permissions, trust, feedback access, and recombination opportunities that determine whether the stock produces value or merely sits in storage.\index{artefact persistence versus productive relation}\index{ordinary depreciation versus productive-relation impairment}\index{capability system}\index{context required to use knowledge-bearing stock}

This insight is already present in Chapter 2's Proposition B (Capability-Bounded Codification), which states that disembodied knowledge capital (\(K^{D}\)) can create value only up to the limit of the embodied (\(K^{E}\)) and institutionalized (\(K^{I}\)) capability available to maintain, interpret, update, govern, and redeploy it. The stock dynamics model converts that static bound into a dynamic principle: capability does not merely set a floor on productive value at a point in time; it determines the rate and direction of value change over time.

The implication is that knowledge stock analysis must track not only the artefact (which accounting and IP law already track, however imperfectly) but the actor-to-stock relation: the bundle of legal, technical, institutional, and capability conditions that determine whether a given actor can productively use the stock. This relation can change while the artefact remains intact. Access withdrawal changes the relation without touching the artefact. Capability displacement changes the relation by removing the human infrastructure required to maintain it. These are the processes that produce what this framework calls invisible depreciation: productive value loss that does not register in accounting valuations, IP records, or formal access indicators.

\section*{Access Withdrawal and Access Impairment}\label{access-withdrawal-and-access-impairment}

The distinction between access withdrawal (a causal pathway) and access impairment (a value-change process) resolves a problem that has recurred throughout the preceding analysis. When a dataset is paywalled, a standard becomes proprietary, or an API is closed, the stock has not decayed in any physical sense. But its productive value for excluded actors has declined. The earlier formulation (that ``permission or usability changes'') was too weak to carry the theoretical weight the concept requires.

Access withdrawal is a causal pathway in which the legal, contractual, technical, price, interface, or institutional conditions governing an actor's relation to knowledge-bearing stock change so that the actor can no longer read, use, modify, audit, copy, redistribute, combine, or build upon the stock on prior terms. Access impairment is the resulting value-change process: the decline in productive value for the excluded actor or actor class.

This distinction is important because access withdrawal need not immediately deteriorate the stock itself. A paywalled dataset may initially remain intact and even continue to be updated by the controlling party. But access withdrawal frequently triggers later capability displacement and decay by severing the community of contributors, validators, annotators, maintainers, and integrators from the maintenance pathway. The distributed community that was cleaning, correcting, and extending the dataset no longer has standing to do so. The dataset stales. What began as a redistribution of access becomes, in time, a degradation of the stock.

Access impairment is therefore the depreciation-side expression of enclosure. It connects the stock dynamics model directly to the enclosure analysis developed in Chapter 6 and explains why enclosure is harmful not only as a present restriction on use but as a dynamic threat to the productive capacity of the stock itself.

\section*{Appreciation: Passive Revaluation and Active Recombination}\label{appreciation-passive-revaluation-and-active-recombination}

The inclusion of passive revaluation and active recombination as value-change processes is not a supplement to the depreciation concept. It is a structural requirement. Without them, the model is asymmetric in a way that misrepresents the distinctive dynamics of knowledge-bearing capital.

Physical capital depreciates monotonically. A machine does not spontaneously regain productive value because the economic environment changes. Knowledge-bearing stock does. This is a consequence of its cumulative and combinatorial character: existing knowledge can become more productive when new complementary knowledge, infrastructure, techniques, or institutional conditions emerge. The appreciation process is not confined to recently created knowledge. Dormant mathematical results become newly productive when computation becomes cheap enough to operationalize them. Archived datasets gain research value when new methods can extract previously inaccessible patterns. Legacy expertise in infrastructure systems becomes valuable when those systems prove unexpectedly durable.

The governance implications of passive revaluation merit particular attention. When old proprietary stock passively revalues because of subsequent complementary investments made by others (by public research programmes, by the open-source community, by infrastructure providers) the intellectual property regime may grant the stock's holder appreciation that depends entirely on others' contributions. The original grant of rights did not include, and in many cases could not have anticipated, this appreciation. This creates what might be called a retrospective enclosure problem: IP duration combined with passive revaluation allows enclosure of old stock to capture value generated by subsequent public and private investment in complements. This is a specific policy implication that follows from the appreciation concept and that has no clean parallel in physical capital theory, where assets do not spontaneously appreciate through others' investments.

\section*{Compounding Asymmetry under Enclosure}\label{compounding-asymmetry-under-enclosure}

The stock dynamics model's most important theoretical result is not any of the individual value-change processes or causal pathways taken in isolation. It is the observation that enclosed governance arrangements produce a specific pattern of simultaneous, mutually reinforcing divergence in value trajectories across actor classes. This is the concentration mechanism of knowledge-bearing capitalism.

Four effects operate simultaneously, not as alternatives. Under enclosed governance arrangements:

\begin{itemize}
\tightlist
\item
  \textbf{Generative suppression} (Proposition C): excluded actors' recombination field is narrowed, reducing their rate of new knowledge generation.
\item
  \textbf{Suppressed appreciation} (Proposition C2): excluded actors' existing stocks fail to reach their passive-revaluation potential because the frontier complements required to activate that potential are inaccessible.
\item
  \textbf{Feedback acceleration} (Proposition D, Component 1): incumbents use deployment-captured feedback to improve their stocks faster than external actors can observe or replicate.
\item
  \textbf{Exclusion-induced depreciation} (Proposition D, Component 2): excluded actors' stocks age relative to the frontier because they lack access to the refresh pipelines that would slow relative obsolescence.
\end{itemize}

These effects compound: the numerator of incumbent advantage rises through active recombination and feedback-driven regeneration; the denominator falls through excluded-actor depreciation and suppressed appreciation. The compounding is not arithmetic but structural, each period of enclosure sets the conditions for a larger gap in the next period.

\section*{Proposition C: Generative Suppression\index{Proposition C!stock-dynamics restatement}}\label{proposition-c-generative-suppression-interlude}

Generative suppression is the first external effect of enclosure. When a governance form restricts access to knowledge-bearing stock that would otherwise enter excluded actors' recombination fields, those actors do not merely lose use of an existing stock. They lose possible inputs into future generation. In KBC terms, enclosure reduces \(D_u(F_{a,t})\), narrowing the useful diversity of the accessible recombination field and thereby reducing the rate at which new knowledge-bearing stock can be generated. Chapter 6 develops this mechanism formally.

\section*{Proposition C2: Suppressed Appreciation\index{Proposition C2!suppressed appreciation}\index{suppressed appreciation!Proposition C2}}\label{proposition-c2-suppressed-appreciation}

Proposition C2 makes a related but distinct claim about the value of existing stock.

\textbf{Proposition C2 (Suppressed Appreciation).} Let actor \(a\) hold knowledge-bearing stock \(S_a\) at time \(t\). Let \(F_{a,t}\) denote the set of knowledge stocks accessible to \(a\) for combination at time \(t\). Let \(V(S_a,F_{a,t})\) denote the productive value of \(S_a\) given access to field \(F_{a,t}\). If enclosure reduces \(F_{a,t}\) to \(F'_{a,t}\subset F_{a,t}\), then \(V(S_a,F'_{a,t})\leq V(S_a,F_{a,t})\), with strict inequality when the excluded stocks \(F_{a,t}\setminus F'_{a,t}\) are complementary to \(S_a\) in production.

The distinction between Proposition C and Proposition C2 matters both theoretically and for policy. Proposition C concerns future generation: enclosure reduces what will be created. Proposition C2 concerns the present value of existing stock: enclosure reduces what already exists in value relative to what it could have been worth. An excluded actor may possess datasets, methods, protocols, or institutional knowledge that are genuinely valuable but that cannot realize their full productive potential because the frontier models, APIs, or complementary techniques required to operationalize them are inaccessible. The enclosure harm is therefore not confined to future exclusion from a larger knowledge base. It is also a present depreciation of existing capability.

\section*{Proposition D: Feedback Acceleration and Exclusion-Induced Depreciation\index{Proposition D!feedback acceleration and exclusion-induced depreciation}\index{feedback acceleration}\index{exclusion-induced depreciation}}\label{proposition-d-feedback-acceleration-and-exclusion-induced-depreciation}

The enclosure-driven concentration mechanism is composed of two analytically separable dynamics that have previously been conflated. The first component concerns incumbent acceleration: enclosed feedback lets the incumbent improve faster. The second concerns challenger depreciation: excluded actors' stocks age relative to the frontier because they lack access to refresh pipelines. These are distinct mechanisms with different policy implications.

\textbf{Proposition D, Component 1 (Feedback Acceleration).} Let incumbent \(i\) deploy knowledge-bearing stock \(K^D_i\) and capture deployment feedback \(\varphi_i\) exclusively. Let the stock update function be \(K^D_{i,t+1}=K^D_{i,t}+g(\varphi_i,t)\), where \(g(\cdot)\) is increasing in feedback volume and quality. If a competing actor \(j\) has \(K^D_{j,t}\approx K^D_{i,t}\) at time 0 but \(\varphi_j=0\) (no deployment feedback), then \(\lVert K^D_{i,t}-K^D_{j,t}\rVert\) is non-decreasing in \(t\), and strictly increasing when \(g(\varphi_i,t)>0\).

\textbf{Proposition D, Component 2 (Exclusion-Induced Depreciation).} Let the productive value of stock \(K^D_j\) be defined relative to the frontier \(F_t\) of the knowledge domain. If \(F_t\) advances through frontier movement and \(j\) lacks access to frontier stocks for combination or refresh, then the relative productive value \(V(K^D_j)/V(F_t)\) declines monotonically in \(t\), even where \(K^D_j\) itself is unchanged.

The research question that follows from this split is: under what conditions does incumbent advantage arise primarily from Feedback Acceleration rather than Exclusion-Induced Depreciation, and vice versa? This matters for policy because the remedies differ. If the problem is Feedback Acceleration, policy instruments include feedback portability mandates, data-sharing obligations, interoperability requirements, and competitive remedies targeting incumbents' use of deployment data. If the problem is Exclusion-Induced Depreciation, policy instruments include public infrastructure investment, commons regeneration funding, open-standards mandates, and access to frontier refresh pathways for excluded actors.

\subsection*{Conditional Hypotheses on Component 1/Component 2 Dominance}\label{conditional-hypotheses-on-component-1component-2-dominance}

Two preliminary conditional hypotheses organize the research question. They should be understood as candidates for empirical investigation, not established findings.

\emph{Hypothesis: Feedback-Acceleration Dominance.} Where a sector exhibits rapid frontier advancement, high feedback density, high model or process update frequency, and high tacit capability intensity, incumbent advantage will arise primarily through feedback acceleration (Component 1). In these conditions, the gap between captured and uncaptured deployment experience is large and growing quickly, and tacit interpretation of that experience compounds the advantage. Machine learning applications in perception and prediction, drug discovery, and cyber threat intelligence are candidate sectors.

\emph{Hypothesis: Exclusion-Depreciation Dominance.} Where a sector depends heavily on shared infrastructure, open standards, public protocols, or distributed maintainer communities, and where the frontier advances more slowly, incumbent advantage will arise more through exclusion-induced depreciation (Component 2). In these conditions, capability displacement from commons to incumbent, and the resulting erosion of distributed maintenance capacity, is the dominant mechanism. Open-source infrastructure, clinical practice protocols, scientific instrumentation, and public-sector digital standards are candidate sectors.

A complicating factor is that many important sectors exhibit both conditions simultaneously. Machine learning, for example, combines rapid frontier movement (favouring Component 1) with massive dependence on open-source libraries, pre-trained models, public benchmarks, and academic knowledge production (which creates Component 2 vulnerability). The Component 1/Component 2 balance in such sectors may shift over time as enclosure matures: early stages may be Component 2-dominated, as incumbent advantage comes from capturing and concentrating the distributed commons; later stages may be Component 1-dominated, as captured feedback creates an improving advantage independent of the original commons appropriation. This dynamic trajectory, if observable, would itself constitute evidence for the compounding asymmetry proposition.

\section*{Master Proposition: Knowledge Stock Dynamics}\label{master-proposition-knowledge-stock-dynamics}

\textbf{Master Proposition (Knowledge Stock Dynamics).} Let \(K_{a,t}\) denote the productive value of knowledge-bearing stock held by actor a at time t, across the five forms \(K^{E}\), \(K^{D}\), \(K^{I}\), \(K^{C}\), \(K^{P}\). Let Γ denote the governance form (IP rules, labour market governance, open-licence architecture, data portability mandates, commons governance institutions, and competition framework). Then:

\begin{enumerate}
\def\labelenumi{(\roman{enumi})}
\item
  \textbf{Non-monotonicity.} Under any governance form Γ, \(K_{a,t}\) is not monotonically decreasing in t. Knowledge-bearing stock can appreciate without being newly created, through passive revaluation when new complementary stocks emerge in the accessible field \(F_{a,t}\).
\item
  \textbf{Capability-relative dynamics.} The rate and direction of change in \(K_{a,t}\) depend not only on the artefact properties of the stock but on the actor-to-stock relation: the bundle of legal, technical, institutional, and embodied-capability conditions that determine whether actor a can maintain, interpret, update, deploy, and combine the stock. Changes in this relation (through capability displacement, access withdrawal, or governance change) alter \(K_{a,t}\) without necessarily altering the stock artefact.
\item
  \textbf{Governance-conditioned asymmetry.} Let \(i\) denote actors who control the enclosed knowledge stock: incumbents who set access terms, capture deployment feedback, and hold governance position over the stock. Let \(j\) denote actors excluded from those terms: actors who cannot use, modify, maintain, or combine with the enclosed stock on prior or equivalent terms. Under enclosed governance forms (\(\Gamma_{enc}\)), the expected value-change trajectories of \(i\) and \(j\) diverge:
\end{enumerate}

\begin{quote}
E{[}\(\Delta K_{i,t}\) \textbar{} \(\Gamma_{enc}\){]} \textgreater{} E{[}\(\Delta K_{j,t}\) \textbar{} \(\Gamma_{enc}\){]}
\end{quote}

\emph{In plain terms:} under enclosed governance forms, incumbents systematically gain more from any given change in the knowledge stock than excluded actors do. This is not a market outcome driven by superior products or greater effort, it is a structural consequence of asymmetric access, asymmetric feedback capture, and asymmetric capability accumulation. The gap does not require bad faith; it follows from the mechanics of enclosure even when incumbents act within the law and excluded actors work as hard as they can.

The divergence is driven simultaneously by feedback acceleration (Proposition D, Component 1), exclusion-induced depreciation (Proposition D, Component 2), suppressed appreciation (Proposition C2), and generative suppression (Proposition C). Under open governance forms (\(\Gamma_{open}\)), this divergence is attenuated because access impairment is constrained, capability displacement yields less exclusive advantage, and deployment feedback is not exclusively capturable.

\begin{enumerate}
\def\labelenumi{(\roman{enumi})}
\setcounter{enumi}{3}
\item
  \textbf{Compounding.} The governance-conditioned asymmetry compounds across periods. Because incumbent regeneration in period t increases \(K_{i,t+1}\), and because \(K_{i,t+1}\) is an input to feedback generation and recombination in t+1, each period of enclosure sets the initial conditions for a larger asymmetry in the next period. The compounding is structural rather than parametric: it follows from the architecture of the generation, conversion, and dynamics model, not from assumptions about actor strategy.
\item
  \textbf{Invisible depreciation.} A subset of the productive value loss in \(K_{a,t}\) (specifically the loss attributable to access impairment and to capability displacement without asset transfer) is not recognized by accounting valuations, intellectual property records, or formal access indicators. This invisible depreciation is not only a standard pricing failure; it is a governance-conditioned measurement and governance failure. Markets and institutions cannot price or remedy losses they cannot observe.
\end{enumerate}

\emph{Formal derivation note.} Parts (iii) and (iv) follow from Propositions C, C2, and D (Components 1 and 2) under the condition that the enclosed governance arrangement satisfies the enclosure conditions in Chapters 6 and 8. Part (v) follows from the accounting recognition analysis developed in Chapter 9. Part (i) is established by construction in the Level 1 taxonomy (Section 2 above): passive revaluation is a value-change process, and its existence is sufficient to establish non-monotonicity. Part (ii) is established by the actor-to-stock relation analysis in Section 3.

\section*{Originality Boundary: Positioning against Predecessor Literature}\label{originality-boundary-positioning-against-predecessor-literature}

The concepts developed in this section draw on a rich body of predecessor scholarship. Intellectual honesty requires that the framework position itself carefully against this literature rather than overclaiming novelty.

The decay and obsolescence concepts build on evolutionary economics and dynamic-capabilities theory, which recognize that productive capability degrades if not maintained and reconfigured \parencite{NelsonWinter1982,TeecePisanoShuen1997,Teece2007}. The appreciation concept has partial precedents in endogenous growth theory and in the knowledge spillover literature, which model the cumulative character of knowledge production \parencite{Romer1990}. The commons dynamics build on commons-governance and open-source-governance literatures \parencite{Ostrom1990,VonHippel2005,BaldwinVonHippel2011}. The invisible depreciation problem is related to the intangible-asset accounting literature and to the broader critique of GAAP's treatment of knowledge assets \parencite{Lev2001,HaskelWestlake2018}.

Across this literature, the common limitation is the firm boundary. Dynamic capabilities, combinative capabilities, the knowledge-based theory of the firm, evolutionary economics, and institutional path dependence are all primarily theories of how firms or institutions manage their own knowledge assets over time. What this framework adds is the cross-actor, cross-form dimension: capability movement in one actor's embodied (\(K^{E}\)) or institutionalized (\(K^{I}\)) stock can depreciate the productive value of another actor's disembodied (\(K^{D}\)), institutionalized, or commons-based (\(K^{C}\)) stock, even where the displacing actor does not legally acquire, own, or directly control the depreciating stock. The commons maintainer hired by an incumbent has not transferred any IP. No assets appear on any balance sheet as having changed hands. Yet the productive value of the commons has declined. This invisible, cross-actor depreciation is the framework's primary contribution.

\section*{Testable Implications}\label{testable-implications}

The propositions developed in this section are not merely conceptual. They make specific empirical predictions that distinguish the framework from simpler accounts of inequality, market concentration, or human capital mobility. Two candidate hypotheses are developed here as the first targets of empirical investigation. Both are stated with explicit acknowledgment of identification challenges.

\subsection*{The Maintainer Concentration Hypothesis}\label{the-maintainer-concentration-hypothesis}

If the commons depreciation mechanism is real, open-source projects should show systematic health deterioration following sharp increases in maintainer concentration, particularly where that concentration results from the migration of core maintainers to a single commercial employer.

The identification challenge is serious. Simple correlations between maintainer concentration and health decline are insufficient because concentration may be endogenous to prior decline: niche or ageing projects may naturally become concentrated as contributors exit for reasons unrelated to any enclosure mechanism. The hypothesis is therefore stated in event-study form: projects experiencing exogenous shocks to maintainer concentration (through documented commercial acquisition of maintainer time, documented hiring events, or organizational restructuring) should show differential health deterioration compared to projects with similar prior trajectories that did not experience equivalent shocks. Candidate health metrics include: commit frequency, issue-resolution latency, dependency update lag, security-vulnerability response time, and contributor count.

The falsification condition is correspondingly specific: the commons depreciation mechanism is materially weakened if projects experiencing exogenous maintainer concentration shocks show no differential deterioration in health metrics relative to comparable unexposed projects, after controlling for matching variables. If highly concentrated projects show no subsequent health decline under exogenous shock conditions, the capability-displacement-to-decay pathway is weaker than this theory predicts.

\subsection*{The Invisible Depreciation Hypothesis}\label{the-invisible-depreciation-hypothesis}

The framework's core originality claim (that productive value loss occurs invisibly when the nominal stock persists while the capability system erodes) requires its own empirical implication. The hypothesis is that accounting valuations, IP records, and formal access indicators systematically underestimate productive value loss in cases involving capability displacement without asset transfer.

The acquisition case is particularly tractable because post-acquisition performance can be compared against pre-acquisition valuations. If the invisible depreciation mechanism is real, acquirers who retain founding teams should systematically outperform acquirers who do not, controlling for deal price, target characteristics, and acquirer integration capability. The gap between retained-team and departed-team outcomes would provide an estimate of the value attributable to the capability system that accounting and IP valuation could not capture.

\section*{A Note on Measurement}\label{a-note-on-measurement}

This section is primarily theoretical. It identifies causal mechanisms through which knowledge-bearing stock loses, preserves, transfers, and gains productive value. Measurement is a secondary and more difficult task: estimating the magnitude, rate, visibility, and distributional consequences of these mechanisms across governance arrangements, sectors, and actor classes.

The candidate indicators proposed in the preceding section (commons health, maintainer concentration, refresh rate, regeneration capacity, capability extraction, and recombination-field breadth) are not yet fully operationalized measures. They are research directions, and the framework should not be read as claiming that these constructs are currently measurable with the precision the propositions require. What the framework does claim is that the constructs refer to real phenomena, that those phenomena have determinate effects on productive value, and that the development of adequate measurement instruments is a tractable research programme.

The aggregation problem is a related and harder challenge. The Master Proposition implies something like a net stock dynamic: gross generation plus gross regeneration plus appreciation minus decay minus obsolescence minus access impairment. But knowledge stocks are heterogeneous, non-fungible, context-dependent, and governance-conditioned. A codebase, a clinical protocol, a trained model, a maintainer community, and a trust-governance arrangement are not summed in any obvious common unit. The formula is coherent as a conceptual scaffold for a defined knowledge domain, not as an aggregate accounting equation. The unit of productive knowledge value remains unresolved, and the framework is not yet an aggregate macroeconomic model. That resolution remains a task for subsequent theoretical and empirical development.

\section*{Conclusion: The Dynamic Core of the Theory}\label{conclusion-the-dynamic-core-of-the-theory}

Knowledge-bearing capitalism does not merely accumulate knowledge-bearing stock. It must continuously regenerate the capability conditions that keep that stock productive. Where those conditions decay, move, become enclosed, or fail to refresh, productive value can disappear while the nominal stock remains fully visible to accounting, IP law, and formal access indicators. Where new complements arise, old knowledge can appreciate. This non-monotonic, capability-relative behaviour is what distinguishes knowledge stock from physical capital and what requires the framework developed in this section.

The framework's central claim can now be stated with precision. Knowledge-bearing stock does not merely accumulate or depreciate. Its productive value moves through a governance-conditioned system of decay, obsolescence, access impairment, regeneration, passive revaluation, and active recombination, driven by causal pathways that include capability displacement, access withdrawal, feedback capture, and commons underinvestment, under governance conditions that determine who can move what capability, on what terms, with what institutional standing. Under enclosed governance arrangements, these mechanisms combine to produce compounding asymmetry: incumbent stocks rise through captured feedback and active recombination; excluded-actor stocks fail to appreciate, depreciate relative to the frontier, and lose the regeneration capacity that would have kept them productive. The productive gap widens not because incumbents possess superior knowledge in any simple sense, but because the governance conditions under which knowledge-bearing stock changes value have been arranged to systematically favour one class of actors over another.

This is the dynamic core of this theory. It connects the static taxonomy of stock forms in Chapter 2 to the generation model in Chapter 3, the conversion analysis in Chapter 4, and the governance analysis that follows in Chapter 5. Without it, the framework describes what knowledge-bearing capitalism is. With it, the framework begins to describe what knowledge-bearing capitalism does over time, and to whom.

\part{Mechanisms}
\chapter{The Knowledge Generation Model}
\index{knowledge generation|textbf}\index{Knowledge Generation Model (KGM)|textbf}\index{KGM|see{Knowledge Generation Model (KGM)}}
\label{ch:knowledge-generation-model}

\chapterhook{Generation Before Conversion}

Generation is where use-value is made. Chapter 1 argued that knowledge produces wealth first as use-value and only later, through governance, as exchange-value; this chapter takes the first half of that claim seriously and asks how the use-value comes to exist at all, by what recombination, learning, and discovery, before any institution decides whether to enclose it, preserve it, or let it diffuse.

\section{Scope and Boundary}\label{scope-and-boundary}

Economists often encounter AI systems\index{AI systems!generation model}, software platforms, R\&D pipelines, and open-source ecosystems after they have already become assets, products, firms, or markets. Chapter 3 moves one step earlier. It asks how new productive knowledge-bearing stock arises before it is converted into property, firm capability, platform dependency, commons, or accounting value.

The Knowledge Generation Model (KGM) accounts for the production of new productive knowledge-bearing stock. It is structurally separate from the Knowledge Conversion Matrix (KCM), which models the movement and governance of existing stock.

\begingroup
\small
\setlength{\tabcolsep}{3pt}
\renewcommand{\arraystretch}{1.12}
\sloppy
\par\addvspace{0.8\baselineskip}\noindent
\begin{longtable}{@{}L{0.27\textwidth}
L{0.33\textwidth}
L{0.34\textwidth}@{}}
\caption{KGM and KCM Boundary Conditions}\label{tab:ch3:kgm-kcm-boundary-conditions}\\
\toprule\noalign{}
\begin{minipage}[b]{\linewidth}\raggedright
Model
\end{minipage} & \begin{minipage}[b]{\linewidth}\raggedright
Core question
\end{minipage} & \begin{minipage}[b]{\linewidth}\raggedright
Mechanisms
\end{minipage} \\
\midrule\noalign{}
\endfirsthead
\toprule\noalign{}
\begin{minipage}[b]{\linewidth}\raggedright
Model
\end{minipage} & \begin{minipage}[b]{\linewidth}\raggedright
Core question
\end{minipage} & \begin{minipage}[b]{\linewidth}\raggedright
Mechanisms
\end{minipage} \\
\midrule\noalign{}
\endhead
\bottomrule\noalign{}
\endlastfoot
Knowledge Conversion Matrix (KCM) & How does existing productive knowledge move, become separable, become governed, and generate appropriability effects? & Transformation, access, distributional \\
Knowledge Generation Model (KGM) & How does new productive knowledge arise? & Recombination (\(G^{R}\)), Experimentation (\(G^{X}\)), Discovery (\(G^{D}\)), Invention (\(G^{N}\)), Judgment/Interpretation (\(G^{J}\)), Learning Feedback Loop (\(G^{L}\)); candidate: Dialectical Generation (\(G^{DIA}\)) \\
Interface & When does newly generated stock enter the conversion cycle? & Generation output → conversion input; appropriability of new stock is the first conversion question \\
\end{longtable}
\endgroup

\textbf{Why the boundary matters.} A model that absorbs generation into conversion becomes harder to test unless generation and conversion effects are separately identified. Keeping the models separate makes each testable on its own terms.

\section{Growth Boundary: Beneath the Productivity Residual}\label{growth-boundary-beneath-the-productivity-residual}

KBC does not reject growth theory's use of aggregate productivity terms. It asks what institutional and knowledge-stock mechanisms sit beneath those terms. Solow's growth framework made the residual visible\index{Solow residual}\index{Solow, Robert} as the portion of growth not explained by measured capital and labour, and Solow later described that residual as a measure of ignorance about the growth process \parencite{Solow1988}. Endogenous growth theory then brought ideas inside the model\index{endogenous growth theory}. KBC adds a further mechanism layer: ideas and knowledge-bearing stock do not merely exist as an aggregate scalar. They are generated, converted, recombined, enclosed, deployed, and fed back through actor-specific fields.

A schematic growth-boundary expression is:

\begin{equation}
A_{t+1}
=
A_t+\,\Phi\left(G^R_t,G^L_t,K^E_t,K^D_t,K^I_t,K^C_t,K^P_t,\pi_t\right).
\label{eq:ch3:growth-boundary-placeholder}
\end{equation}

\noindent\emph{Where:} $A_t$ is the aggregate productivity state at time $t$; $A_{t+1}$ is the next-period productivity state; $\Phi(\cdot)$ is the KGM mechanism-composition function; $G^R_t$ is recombination generation; $G^L_t$ is learning-feedback generation; $K^E_t$, $K^D_t$, $K^I_t$, $K^C_t$, and $K^P_t$ are, respectively, embodied, disembodied, institutionalized, commons, and public epistemic knowledge-bearing stocks at time $t$; and $\pi_t$ is the governance form conditioning access, deployment, enclosure, and feedback at time $t$.

Equation~\ref{eq:ch3:growth-boundary-placeholder} is a schematic placeholder, not a replacement for growth accounting. It is not yet estimable. It identifies mechanisms that Chapter~11 later proposes to proxy. It is a placeholder for a mechanism layer beneath the productivity term. It says that measured productivity growth may depend on recombination generation, learning feedback, embodied capability, disembodied stock, institutionalized capability, commons knowledge stock, public epistemic infrastructure, and the governance form that determines access to those stocks.

\section{Core Question}\label{core-question}

Under what conditions does the combination, deployment, modification, or interpretation of existing knowledge-bearing stock (or direct encounter with the natural world) produce a productive capability that was not present in any of the inputs?

\section{The Canonical KGM Mechanism Set}\label{the-canonical-kgm-mechanism-set}

The KGM operates with six canonical generation mechanisms\index{generation mechanisms|textbf} and one named candidate mechanism. The mechanism set for the present manuscript is recombination, experimentation, discovery, invention, interpretive judgement, and learning from feedback\index{recombination!generation mechanism|textbf}\index{experimentation!generation mechanism|textbf}\index{discovery!generation mechanism|textbf}\index{invention!generation mechanism|textbf}\index{interpretation!generation mechanism|textbf}\index{feedback learning|textbf}. Dialectical generation remains a named candidate mechanism\index{dialectical generation|textbf}, not yet part of the canonical composition function.

\begingroup
\scriptsize
\setlength{\tabcolsep}{3pt}
\renewcommand{\arraystretch}{1.12}
\sloppy
\par\addvspace{0.8\baselineskip}\noindent
\begin{longtable}{@{}L{0.14\textwidth}
L{0.18\textwidth}
L{0.27\textwidth}
L{0.17\textwidth}
L{0.18\textwidth}@{}}
\caption{Canonical Knowledge Generation Mechanism Set}\label{tab:ch3:canonical-kgm-mechanisms}\\
\toprule\noalign{}
\begin{minipage}[b]{\linewidth}\raggedright
Symbol
\end{minipage} & \begin{minipage}[b]{\linewidth}\raggedright
Chapter 3 label
\end{minipage} & \begin{minipage}[b]{\linewidth}\raggedright
Order
\end{minipage} & \begin{minipage}[b]{\linewidth}\raggedright
Direct KGM-to-KCM bridge?
\end{minipage} & \begin{minipage}[b]{\linewidth}\raggedright
Notes
\end{minipage} \\
\midrule\noalign{}
\endfirsthead
\toprule\noalign{}
\begin{minipage}[b]{\linewidth}\raggedright
Symbol
\end{minipage} & \begin{minipage}[b]{\linewidth}\raggedright
Chapter 3 label
\end{minipage} & \begin{minipage}[b]{\linewidth}\raggedright
Order
\end{minipage} & \begin{minipage}[b]{\linewidth}\raggedright
Direct KGM-to-KCM bridge?
\end{minipage} & \begin{minipage}[b]{\linewidth}\raggedright
Notes
\end{minipage} \\
\midrule\noalign{}
\endhead
\bottomrule\noalign{}
\endlastfoot
\(G^{R}\) & Recombination & First-order & No & Governed by recombination field \(D_{u}(F_{a,t})\); formal specification: \(G^{R}_{a,t+1}\) = \(\lambda_{a}\) $\cdot$ \(R_{a,t}^{\eta}\) $\cdot$ \(D_{u}(F_{a,t})^{\mu}\) \\
\(G^{X}\) & Experimentation & First-order & No & Produces meta-knowledge about the conditions under which knowledge creates value. Examples: A/B testing, model ablation, clinical trials, semiconductor process testing \\
\(G^{D}\) & Discovery & First-order & No & Less cleanly specified by \(D_{u}(F_{a,t})\) than recombination, but often conditioned by instruments, datasets, laboratories, research communities, and public epistemic infrastructure. Examples: CRISPR, protein-structure discovery, astronomical detection, open scientific datasets \\
\(G^{N}\) & Invention & First-order & No & Less cleanly specified by \(D_{u}(F_{a,t})\) than recombination; conditioned by access to prior techniques, materials, problem definitions, capability, and IP governance at the KCM interface. Examples: transformer\index{AI training!transformer ecosystem} architecture, mRNA platform, lithium-ion battery, lithography \\
\(G^{J}\) & Judgment/Interpretation & First-order / bridge & Yes: \(K^{E}\) → judgment/output → \(K^{D}\) or \(K^{I}\) (KCM Cell 1) & Smithian bridge mechanism; embodies both initial application and iterative refinement. Examples: radiology AI, legal review, investment committee judgement, cybersecurity triage \\
\(G^{L}\) & Learning/feedback & Second-order in sequence; first-order in capability change once feedback exists & No & Requires prior deployment before feedback can exist; once feedback exists, it directly changes \(\widetilde{C}_a\). Examples: RLHF, post-deployment model improvement, production telemetry, user correction loops \\
\(G^{DIA}\) & Dialectical generation & Candidate: not in canonical composition function & No & Conditional on formal resolution of the \(\Theta^-\) counter-frame notation and the \(\mathcal{T}_{DIA}(T,O,M,H)\) inverted-U\index{inverted-U condition} specification. Examples: peer review, red-team challenge, investment committee dissent, Popper\index{Popper, Karl}ian refutation, Socratic elenchus \\
\end{longtable}
\endgroup

\textbf{Composition function.} The total generation output for actor \(a\) at time \(t\) is governed by the weakest-commitment aggregate:

\begin{quote}
\begin{equation}
\begin{aligned}
G^{\mathrm{tot}}_{a,t+1}
&=\Phi(\cdot) \\
&=\chi_{\pi}\,\widetilde{C}_{a,t}^{\beta}\,\widehat{D}_{u}(F_{a,t})^{\nu}
\Bigl[
&\quad w_R G^R_{a,t+1}+w_X G^X_{a,t+1}+w_D G^D_{a,t+1} \\
&\quad +w_N G^N_{a,t+1}+w_J G^J_{a,t+1}+w_L G^L_{a,t+1}
\Bigr]^{\rho}.
\end{aligned}
\label{eq:ch3:g-a-t-1}
\end{equation}
\end{quote}

\textbf{Where:} \(G^{\mathrm{tot}}_{a,t+1}\) is total new productive knowledge generated by actor \(a\) for period \(t+1\); \(\Phi(\cdot)\) is the KGM composition function; \(\chi_{\pi}\) is the governance-specific access and permission multiplier; \(\widetilde{C}_{a,t}\) is dynamic capability; \(\widehat{D}_{u}(F_{a,t})\) is normalized useful diversity of the accessible field; the \(w_k\) are non-negative mechanism weights that sum to one; \(\beta\) and \(\nu\) are capability and field-access elasticities; and \(0<\rho\le 1\) imposes weak concavity or sub-additivity.

\(G^{DIA}\) is discussed below as a named candidate mechanism but remains outside the canonical \(\Phi(\cdot)\) composition until its formal specification is resolved.

\section{Information-Technology Market Structure as the Applied Setting}\label{information-technology-market-structure-as-the-applied-setting}

Information-technology industries are the clearest applied setting for the KGM because they combine high fixed costs, low marginal distribution costs, switching costs, standards, systems effects, network effects, and strong intellectual-property questions. \textcite{VarianFarrellShapiro2004} treat these features as central to the economics of information technology. KBC does not infer the full theory from IT alone, but IT makes visible why knowledge-bearing stock cannot be treated as ordinary physical stock. A software platform, model, standard, or dataset may be costly to create and maintain, yet cheap to distribute once produced. Its value depends not only on copies sold, but on who can access it, recombine it, learn from its deployment, and control the standards or interfaces around it.

This market-structure grounding clarifies the near-zero marginal-cost point. The claim is not that knowledge goods are costless. First-copy creation, maintenance, security, quality control, documentation, governance, and complementary capability may be expensive. The anomaly is that marginal distribution cost can fall far below the first-copy and capability costs that make the stock valuable. That gap makes scale, standards, switching costs, and learning-loop capture decisive for knowledge generation.

The transformer ecosystem illustrates the point. Early AI progress depended on a wide recombination field of papers, code, libraries, benchmarks, datasets, and distributed engineering labour, including the transformer architecture, open-source machine-learning frameworks, model hubs, and shared benchmarks \parencite{Vaswani2017,Paszke2019,Abadi2016,Wolf2020,Wang2018GLUE}. Later enclosure did not merely change who captured value; it changed which actors could participate in the next generation cycle.

\section{Why Recombination is the Most Directly Specified Mechanism}\label{why-recombination-is-the-governing-mechanism}

The six mechanisms are not symmetric in their present formal treatment. The framework gives \(G^{R}\) the only closed formal specification (the \(G^{R}_{a,t+1}\) = \(\lambda_{a}\) $\cdot$ \(R_{a,t}^{\eta}\) $\cdot$ \(D_{u}(F_{a,t})^{\mu}\) formula) while the others receive qualitative characterization. That asymmetry requires justification, because this theory would be weakened by the objection that recombination was simply chosen as the formal focus because it made the enclosure model tractable.

That asymmetry is contained procedurally, not merely conceded. The closed form gives \(G^R\) analytical, not empirical, priority: under the identification protocol of Chapter~11 (\S\ref{111-identification-protocol}), every suppression, feedback, and welfare claim is identified through treated-versus-control access shocks, not through the closed form. Recombination is therefore the most \emph{specified} mechanism without being the privileged \emph{evidence}. The formula sharpens calibration once a result is established by quasi-experiment, and the objection that recombination was chosen because it made the enclosure model tractable is answered by routing identification away from the tractable object.

The justification is not that recombination is more important in some absolute sense than discovery or invention. A world without Newtonian mechanics, the germ theory of disease, or the discovery of DNA would have impoverished knowledge generation far more than a world with lower recombination rates. The justification is more specific: recombination is the mechanism most directly specified in this chapter because it can be modelled through field magnitude, useful diversity, access, capability, and governance conditions. It is therefore the mechanism through which the institutional form of capitalism can be connected most directly to observable changes in knowledge production.

Consider the contrast. Discovery and invention are less cleanly specified by \(D_{u}(F_{a,t})\) than recombination, but they are still often conditioned by access to instruments, datasets, laboratories, materials, prior techniques, research communities, and epistemic infrastructure. A latent truth about the world is not itself enclosed by a patent, but access to the experimental infrastructure, datasets, chemical substrates, journals, laboratories, and research networks through which discovery becomes likely can be restricted. Invention is goal-directed and closely connected to IP law at the KCM interface, but it often depends on prior techniques, materials, standards, problem definitions, and complementary capability that can be opened, licensed, or enclosed. Experimentation (\(G^{X}\)) depends on access to substrates for experiment, but the productive step (systematic variation of conditions) is actor-internal once the substrate is accessible. Judgment and interpretation (\(G^{J}\)) are primarily conditioned by embodied capability, although their quality can still depend on access to cases, records, standards, and comparison classes.

Recombination is different in a specific way: \(D_{u}(F_{a,t})\) is an argument of \(G^{R}\) directly. The generation rate is a function of the diversity and breadth of the accessible field. This is not merely an upstream condition but a parameter in the formula itself. When enclosure reduces \(D_{u}(F_{a,t})\), it reduces \(G^{R}\) in the same period through the formal specification. No other mechanism in this chapter is currently specified with field diversity as a direct argument. For the others, enclosure operates through chains of upstream effects that are empirically real but analytically less tractable.

There is also a non-rivalry argument. Recombination draws on non-rival\index{non-rivalry} knowledge goods: \(K^{D}\), \(K^{C}\), and \(K^{P}\) can be simultaneously present in many actors' fields. When a non-rival \(K^{D}\) element is enclosed, all actors who might have recombined it lose access simultaneously, while the enclosing actor retains it. The aggregate \(G^{R}\) loss is distributed across many excluded actors, none of whom caused it. This is the structure that makes the T2 aggregate loss result (\(\Delta G^{R}_{agg}\) \textgreater{} 0) conditionally demonstrable: non-rivalry means the incumbent gains nothing from the excluded actors' lost use, so there is no automatic compensating gain to offset the distributed loss. For mechanisms that draw on rival inputs, enclosure more often produces a transfer from excluded actors to the incumbent. The lost-use property of recombination-field enclosure makes \(G^{R}\) the mechanism through which the social cost of enclosure can be stated most cleanly, holding creation incentives, disclosure, quality control, security, and investment benefits constant.

The claim follows the caveat stated in the front matter: enclosure is dynamically costly only when lost recombination and learning-loop benefits exceed creation, disclosure, quality-control, security, or investment gains.

The \(G^{L}\) (feedback loop) mechanism rivals \(G^{R}\) in its governance sensitivity. Chapter 7 shows that exclusive deployment access can concentrate \(G^{L}\) in incumbents while denying it to excluded actors. But \(G^{L}\) requires prior deployment, and its governance sensitivity is about deployment access rather than recombination-field diversity. It is the mechanism through which enclosure affects learning from use; \(G^{R}\) is the mechanism through which enclosure affects the generation possibilities of excluded actors. Both matter, but they are distinct causal pathways.

This is the reason this theory treats recombination as the most directly specified mechanism: not because recombination is the source of all important knowledge, but because it is the mechanism for which the relationship between governance conditions and generation rates is most direct, most formal, and most amenable to the suppression-ratio analysis stated formally in T2. The other mechanisms are real, and this theory claims them all; but \(G^{R}\) is the one where the causal pathway from enclosure to lost knowledge production runs shortest and can be shown conditionally in the formal model.

\subsection[AI turns knowledge reach into economic return]{AI Turns Knowledge Reach into Economic Return: The Transformer Ecosystem as Recombination Field}\label{a-worked-demonstration-the-transformer-ecosystem-as-recombination-field}

The claim that \(G^{R}\) is the most directly specified KGM mechanism because it is formally conditioned by governance, access, and field diversity can be traced in the development of large language models between 2017 and 2023. The case is not chosen to illustrate a pathology; it is chosen because it shows the mechanism working first in the open direction (wide \(D_{u}\) enabling distributed \(G^{R}\) across many actors) and then beginning to narrow, which allows both effects to be observed in sequence.

\textbf{The knowledge-bearing stocks.} In 2017, the foundational elements for transformer-based language modelling were almost entirely in the accessible commons of the recombination field. Vaswani et al.'s ``Attention Is All You Need'' contributed the attention mechanism and openly specified the transformer architecture \parencite{Vaswani2017}. Residual connections, positional encodings, and layer normalization drew on a series of publicly available predecessor papers. PyTorch and TensorFlow\index{TensorFlow} were released as open-source \(K^{D}\) under permissive licences \parencite{Paszke2019,Abadi2016}. Large pre-training corpora (Common Crawl, Wikipedia, BookCorpus), model libraries, and shared benchmarks were accessible to researchers with computing infrastructure \parencite{Wolf2020,Wang2018GLUE,Rajpurkar2016SQuAD}. The GATE conditions for a wide population of actor types were substantially satisfied: \(A_{a,i,t}\) (accessibility) was high across universities, research labs, and technically capable independent groups; \(I_{a,i,t}\) (institutional permission) was granted by arXiv preprint norms and open-source licences; \(P_{a,i,t}\) (interoperability) was supported by standardized Python APIs and common hardware; \(C_{a,i,t}\) (complementary capability) was achievable for research groups with graduate-level ML expertise.

\textbf{The pre-change recombination field.} The field available to a competent ML research group in 2019 was heterogeneous in the most productive sense: it contained elements from linguistics (tokenization theory), information theory (perplexity metrics), statistical learning, cognitive science (attention as a model of human processing), systems software (distributed training), and empirical machine translation. No actor held monopoly control over the foundational published methods, although compute, data, talent, and distribution were already unevenly concentrated. \(D_{u}(F_{a,t})\) was high because the field spanned distant domains that had never been systematically combined at scale. The optimal combination distance \(d^*_{a}\) for this population was close to many of the available elements: researchers in ML, NLP, computational linguistics, and cognitive science were all positioned near \(d^*\) for different subsets of the available stock.

\textbf{What \(G^{R}\) produced.} From this open field, multiple distinct generation events occurred in rapid succession: BERT\index{BERT}, GPT-2, T5, RoBERTa\index{RoBERTa}, ALBERT, DistilBERT, and dozens of variants built by university and independent groups \parencite{Devlin2018BERT,Radford2019GPT2,Raffel2020T5,Liu2019RoBERTa}. These were not incremental improvements on a single trajectory; they were actor-indexed recombination events. BERT's masked-language-modelling approach and GPT-2's left-to-right autoregressive approach drew on the same base stocks but combined them in distinct ways, generating distinct productive outputs. Each trajectory was made possible by the diversity of the accessible field: because published results were non-rival, each trajectory generated stocks that entered the accessible field and could be recombined into the next round.

The formal expression is direct: high \(D_{u}(F_{a,t})\) across this period, combined with broad actor-population access (high \(R_{a,t}\) across many actor types), produced \(G^{R}\) output well above what any single incumbent's generation function could have produced alone. The T8 result is visible in the period's structure: many trajectories, distributed across actor types, each advancing a distinct approach to the same foundational problem. This is the high-trajectory-count condition (\(N_{traj}\)(\(\pi\)₀) large) that enclosure subsequently begins to reduce.

\textbf{Why this illustrates direct specification of \(G^{R}\).} The transformer ecosystem illustrates the theoretical claim of the preceding section: \(G^{R}\) is not more important than discovery or invention in some absolute sense, but it is the mechanism most directly conditioned by governance conditions, specifically \(D_{u}\). What made the 2017--2020 period productive was not the quality of any single discovery but the breadth and diversity of the accessible field that allowed many actors to recombine the same foundational elements into distinct outputs. A counterfactual world in which the attention mechanism had been patented, the pre-training corpora enclosed under data licences, and the initial models released only through proprietary APIs would have had the same underlying discoveries but substantially lower \(G^{R}\) output, because \(D_{u}(F_{a,t})\) would have been narrowed at the elements with the highest productive weight.

\textbf{The enclosure gradient.} Beginning approximately 2022, the governance conditions of the field began to shift. GPT-4's technical report explicitly withheld details of model size, architecture, and training data\index{AI training!training data} \parencite{OpenAI2023GPT4}. Gemini's technical report introduced a frontier multimodal model family without open frontier-weight release, and Anthropic's Claude family was documented through system or model cards and API/product access rather than open frontier weights \parencite{GeminiTeam2023,Anthropic2024Claude3}. The weights of the most capable frontier systems were not released for external use, research, or extension. The GATE conditions for the previously productive actor population began to fail at the \(A_{a,i,t}\) and \(I_{a,i,t}\) dimensions: accessibility declined (proprietary API only, no weight access), and institutional permission was restricted to the enclosing actors' own researchers for the most productive stocks. Field diversity \(D_{u}(F_{a,t})\) for the actor population outside the frontier incumbents began to fall, not because the elements disappeared from the world, but because they exited the accessible recombination field.

\textbf{What standard theory sees.} A market in which incumbents with large compute budgets and training-data scale have competitive advantages. The appropriate framework is entry barriers (capital requirements), economies of scale (training compute), and possibly network effects (user base). Policy interest centres on market concentration and whether dominant incumbents are engaging in anticompetitive conduct against rivals.

\textbf{What KBC foregrounds.} The standard account identifies the distributional problem (who wins), while standard market-concentration analysis underweights the generative problem, what is no longer being generated. The trajectory count \(N_{traj}\) is not captured by ordinary concentration metrics. Academic research groups, independent labs, and smaller commercial actors who had contributed distinct trajectories in 2018--2021, each representing an actor-indexed recombination event that this theory identifies as difficult to replace, are progressively unable to work at the frontier because the highest-productive-weight field elements have migrated into enclosed governance arrangements. The suppression ratio \((\sigma_a^R)^{\eta}\cdot(D_{u}\;\mathrm{ratio})^{\mu}\) is falling for this population: \(R_{a,t}\) falls as frontier models exit the accessible field, and \(D_{u}(F_{a,t})\) falls as the elements with the highest productive weights for cross-disciplinary recombination enter enclosure. None of this appears directly in market concentration statistics. It is precisely the generative cost, the counterfactual trajectories that the open-field period would have produced, that KBC foregrounds.

\emph{Smithian departure:} Smith's benign accumulation logic assumes that productive accumulation expands the conditions for further productive accumulation. The transformer ecosystem illustrates both directions. The open-field period confirms the logic: distributed investment in foundational research widened the recombination field and increased the productive power of all actors who could access it. The enclosure gradient shows the inversion: private accumulation of frontier model capability narrows the recombination field for the actor population that produced the foundational investment. Present productive power grows; future generative diversity contracts.

\section{Natural and Artificial Intelligence in the KGM}\label{natural-and-artificial-intelligence-in-the-kgm}

This chapter can also be read as part of an economics of natural and artificial intelligence. \textbf{Natural intelligence} means embodied human judgement, skill, interpretation, tacit knowledge, professional expertise, organizational learning, and community-maintained knowledge. \textbf{Artificial intelligence} means disembodied or platform-mediated knowledge-bearing capital encoded in models, software, training systems, inference engines, stochastic systems, and feedback loops.

The KGM explains how both forms generate new knowledge-bearing stock. Natural intelligence generates primarily through judgment (\(G^{J}\)), interpretation, experimentation (\(G^{X}\)), discovery (\(G^{D}\)), invention (\(G^{N}\)), and recombination (\(G^{R}\)) across embodied (\(K^{E}\)), institutionalized (\(K^{I}\)), and commons (\(K^{C}\)) forms. Artificial intelligence generates by recombining encoded stock (\(G^{R}\) applied to \(K^{D}\) and \(K^{P}\)), producing probabilistic outputs, transforming feedback into revised models (\(G^{L}\) at scale), and scaling inference across many contexts simultaneously.

The theoretical question is not whether artificial intelligence replaces natural intelligence, but how natural intelligence is converted into artificial, institutionalized, disembodied, commons-based, or platform-mediated intelligence, and who controls the resulting stock at first conversion. That question is the mandate of the KCM. The KGM's task is to explain the generation events on either side of that conversion.

\begin{center}
\fbox{\begin{minipage}{0.92\linewidth}
\textbf{Boxed proposition: AI Multiplies Capability More Than It Replaces It}

Artificial intelligence does not create equal value for every actor who can access it. Its yield depends on embodied judgement, institutional capability, domain knowledge, and the capacity to evaluate and deploy its outputs. In KBC terms, AI is not only a substitution technology. Under capability-conditioned use, it becomes a multiplier of embodied and institutionalized knowledge capital.
\end{minipage}}
\end{center}

\textbf{The epistemic filter requirement for AI-generated stock.} AI systems generate candidate knowledge-bearing stock, not automatically knowledge-bearing capital. Their outputs are epistemic mixtures: true claims, false claims, unverifiable claims, misleading compressions, useful patterns, and synthetic noise. This mixture does not disqualify AI-generated output from eventually becoming knowledge-bearing capital, it identifies the condition under which it does: validation, provenance, reliability testing, and contextual adequacy assessment must precede economic capitalization, meaning conversion into economically productive knowledge-bearing stock rather than mere accounting recognition. Unvalidated AI output is information-bearing stock. It enters the KGM/KCM cycle as potential generation input, but its productive value as knowledge-bearing capital depends on its epistemic reliability coefficient \(\tau_{i}\) (Axiom B.2, Technical Companion, Appendix B). A hallucinated result that enters a firm's research protocol as accepted fact may appear productive in the short run; it will generate \(EKL_{false}\), epistemic knowledge loss, in proportion to the probability that downstream actions are based on it when it fails. The KGM's generation mechanisms (\(G^{R}\), \(G^{X}\), \(G^{D}\), \(G^{N}\), \(G^{J}\), \(G^{L}\)) apply to AI-generated stock exactly as they apply to human-generated stock, with \(\tau_{i}\) as a modifier on the productive value of the output at the point of first conversion.

\section{Six Canonical Generation Mechanisms}\label{six-canonical-generation-mechanisms}

\subsection[Recombination generation]{\texorpdfstring{\(G^{R}\): Recombination}{Recombination generation}}\label{gr-recombination}

Existing knowledge-bearing stock from distinct domains is combined to produce new stock whose productive capability neither domain possessed alone. Formally:

\begin{quote}
\(G^{R}_{a,t+1}\) = \(\lambda_{a}\) $\cdot$ \(R_{a,t}^{\eta}\) $\cdot$ \(D_{u}(F_{a,t})^{\mu}\)
\end{quote}

where \(\lambda_{a}\) is actor \(a\)'s recombination productivity, \(R_{a,t}\) is the magnitude or scope of distinct knowledge stocks available for combination, \(D_{u}(F_{a,t})\) is the useful diversity of the actor's recombination field, and \(\eta\), \(\mu\) \textgreater{} 0 are elasticities. The useful-diversity term is defined locally as:

\begin{quote}
\begin{equation}
D_{u}(F_{a,t}) = -\sum_{j \in F_{a,t}} p_{a,j,t} \log p_{a,j,t}
\label{eq:ch3:d-u-f-a-t}
\end{equation}
\end{quote}

The productive-weight shares are:

\begin{quote}
\begin{equation}
p_{a,j,t} = \frac{\omega_j \chi_{a,j,t}}{\sum_{k \in F_{a,t}} \omega_k \chi_{a,k,t}}
\label{eq:ch3:p-a-j-t}
\end{equation}
\end{quote}

\textbf{Where:} \(p_{a,j,t}\) is the productive-weight share of knowledge stock \(j\) in actor \(a\)'s accessible field at time \(t\); \(\omega_j\) is the productive weight or relevance of stock \(j\); and \(\chi_{a,j,t}\) is actor \(a\)'s usable access or effective usability of stock \(j\) at time \(t\). In plain terms, a field is more useful when the actor has usable access to productive stocks that are not all from the same narrow source or domain. \textbf{Functional-form note.} The Shannon form is a modelling choice, not a derived necessity, and because \(D_{u}\) is one of the three keystone parameters of this theory the choice does real work that should be made explicit. Equation~\ref{eq:ch3:d-u-f-a-t} is maximized when accessible productive weight is spread evenly across stocks and falls as the field concentrates on a few stocks; it therefore rewards breadth and balance, so that a field dominated by one high-value stock scores low even when that stock is valuable. Whether generation rewards balance (many complementary inputs) or concentration (a few high-value inputs) is an empirical question, so \(D_{u}\) is best read as one member of the Hill-number family of effective-variety measures, \[ D_{u}^{(q)}(F_{a,t}) = \Bigl(\sum_{j\in F_{a,t}} p_{a,j,t}^{\,q}\Bigr)^{1/(1-q)},\qquad q\ge 0, \] the effective number of productive stocks of order \(q\). Orders \(q<1\) reward the mere presence of high-value stocks and are less punishing of dominance, while \(q>1\) weight the dominant stocks more heavily. The qualitative results that run through \(D_{u}\) (enclosure that removes accessible high-weight stocks lowers useful diversity, and field contraction lowers \(G^{R}\)) hold for every \(q\ge 0\); only the curvature and magnitude of the effect change. The order \(q\) is therefore an object for calibration (Chapter~11), and the bare entropy used here is the logarithm of the order-\(q\to1\) Hill number, the analytically convenient baseline of this family, not a claim that uniformity is uniquely generative.

The field-diversity term means \(G^{R}\) is not purely additive: it depends on conditions (\(F_{a,t}\)) that also affect other mechanisms.

\textbf{Key property.} Value scales with the breadth and heterogeneity of the knowledge base available to combine, not only with depth in any single domain. A chip-design team, a compiler specialist, and an ML-workload engineer can each contribute deep domain expertise. But the most valuable recombination often appears when an actor or organization can hold the interfaces among chip architecture, compiler behaviour, and workload demand together, because the productive combination is visible only across the boundary among those fields.

\textbf{The recombination field.} Recombination requires that knowledge be discoverable, interoperable, trusted, and institutionally permissible to combine. The enabling conditions (open standards, translation capacity, legal access, cognitive diversity, shared vocabulary, modularity, incentive alignment) constitute the recombination field \(F_{a,t}\): the set of knowledge stocks that actor a can practically and legally access and combine at time t. Formally, a knowledge stock \(K_i\) is in actor a's recombination field if and only if all four admission conditions are satisfied:

\begin{quote}
\(K_i\) $\in$ \(F_{a,t}\) $\Longleftrightarrow$ \(A_{a,i,t}\) $\cdot$ \(I_{a,i,t}\) $\cdot$ \(C_{a,i,t}\) $\cdot$ \(P_{a,i,t}\) \textgreater{} 0
\end{quote}

where \(A_{a,i,t}\) is accessibility (can actor a reach \(K_i\)?), \(I_{a,i,t}\) is institutional permission (is combination legally and contractually permitted?), \(C_{a,i,t}\) is complementary capability (does actor a have the \(K^{E}\) and \(K^{I}\) required to work with \(K_i\)?), and \(P_{a,i,t}\) is practical interoperability (is \(K_i\) in a form that actor a can combine with other stocks in the field?). If any one condition is zero, the stock is excluded from the field regardless of its potential productive value. This binary admission rule is the strict perfect-complement limit applied to a single stock; at the level of the whole field, partial substitutability means that excluding one stock more often raises the cost of recombination than removes it absolutely, because the actor can route toward a second-best input. The field-level suppression results built on this rule are therefore bounded by the degree of substitutability rather than catastrophic, a bound the function-class audit in Volume~2 (Appendix~F) makes precise. Enclosure operates primarily through A and I; capability decay operates through C\index{capability decay}; standards failure and architectural lock-in operate through P\index{architectural lock-in}\index{standards failure}. This is the structural environment of the KGM's most directly specified mechanism and is developed further in the section below.

\textbf{Limiting factors.} Combinatorial space is vast; only a small fraction of possible combinations produce positive value. What determines which combinations are attempted: standards and interoperability, trust and attribution norms, legal access to both knowledge stocks, cognitive capacity to hold both domains, and the incentive to combine rather than defend existing stock against combination.

\textbf{Natural and artificial intelligence.} Recombination is the mechanism by which AI systems produce outputs that no single training source could produce alone. The productive novelty in large-model outputs can be interpreted economically as recombination of encoded patterns across domains, recombination at scale, at speed, and across a field wider than any individual natural intelligence could hold. The boundary question (Is AI training recombination or conversion?) is formally deferred to KCM Cell 4, but the criterion used here is: recombination has occurred when the resulting stock enables a productive task that neither input stock could enable alone.

\textbf{Source lineage.} \textcite{Romer1990,Romer1994}\index{Romer, Paul} on non-rival ideas and combinatorial growth; \textcite{Weitzman1998}\index{Weitzman, Martin} on recombination as the engine of economic growth; \textcite{Arthur2009}\index{Arthur, W. Brian} on technology as combination of existing techniques; \textcite{Schumpeter1934}\index{Schumpeter, Joseph} on new combination as the definition of innovation. \textcite{CohenLevinthal1990} on absorptive capacity: the ability to recognize, assimilate, and apply external knowledge depends on prior related knowledge stock. This motivates the field-breadth reading of \(D_{u}(F_{a,t})\) rather than supplying its functional form, since absorptive capacity concerns prior related knowledge enabling uptake; \(D_{u}\) proxies the breadth and balance of the accessible related-knowledge base, not an entropy of weight shares as such.

\subsection[Experimental generation]{\texorpdfstring{\(G^{X}\): Experimentation}{Experimental generation}}\label{gx-experimentation}

Existing knowledge is deliberately deployed in a novel context to test whether it produces valuable outcomes. The experiment generates new knowledge about the conditions under which existing knowledge creates value, the boundaries of its applicability, and the domains where it fails.

\textbf{Key property.} Experimentation produces meta-knowledge, knowledge about knowledge. A firm that has run many structured experiments does not only know what it learned from each experiment; it knows what kinds of experiments tend to yield results in what kinds of settings. This institutional learning capacity is itself a form of knowledge-bearing stock (\(K^{I}\), high durability, difficult to transfer).

\textbf{Examples.} Practical cases include A/B testing in digital products, model ablation in machine learning, clinical trials in medicine, and semiconductor process testing in advanced manufacturing.

\textbf{Limiting factors.} Experimentation requires willingness to accept failure, institutional permission to attempt novel things, measurement capacity to detect results, and the ability to preserve and transmit what is learned. Firms and institutions that penalize failure suppress experimentation even when they nominally encourage it.

\textbf{Source lineage.} \textcite{Arrow1962}\index{Arrow, Kenneth} on learning by doing; \textcite{NelsonWinter1982}\index{Nelson and Winter} on organizational routines and the evolutionary theory of the firm; the economics of R\&D and uncertainty.

\subsection[Discovery generation]{\texorpdfstring{\(G^{D}\): Discovery}{Discovery generation}}\label{gd-discovery}

New knowledge is encountered rather than constructed, latent properties of the world become accessible that were not previously known. The relevant property or relation existed before it was known; what changes is human access to it.

\textbf{Key property.} Discovery cannot be planned at the level of the specific find. It can be made more likely by investing in the structural conditions that increase encounter probability: diverse inquiry across many domains, open publication of intermediate results, cross-disciplinary conversation, freedom to investigate questions whose productive applications are not yet known. Discovery often depends on epistemic infrastructure, even when firms optimize search within that infrastructure.

\textbf{Examples.} Practical cases include CRISPR-related biological discovery, protein-structure discovery, astronomical detection, and open scientific datasets that make unexpected patterns visible.

\textbf{Knowledge-bearing capitalism implication.} This is why the vocabulary of epistemic infrastructure, knowledge commons (\(K^{C}\)), and public epistemic capital (\(K^{P}\)) is not peripheral to this theory. Discovery often occurs where knowledge is openly shared, where researchers can build on each other's work with low transaction costs, and where the recombination field is wide. Enclosure of knowledge stocks can restrict the recombination field and reduce the probability of discovery, although proprietary investment may also fund costly instruments, laboratories, datasets, and research programmes. The enclosure analysis in Chapter 6 is therefore also a conditional theory of discovery suppression.

\textbf{Source lineage.} \textcite{Merton1973} on the sociology of science; \textcite{Bush1945} on the relation between basic research and applied knowledge; \textcite{Ostrom1990} and the broader open-science literature on shared-access knowledge institutions.

\subsection[Inventive generation]{\texorpdfstring{\(G^{N}\): Invention}{Inventive generation}}\label{gn-invention}

New knowledge-bearing stock is deliberately constructed through the application of existing knowledge to solve a defined problem. Invention differs from recombination in emphasis because it is goal-directed: the target capability is specified before the generation process begins. It is less cleanly specified by \(D_{u}(F_{a,t})\) than recombination, but it is still conditioned by access to instruments, datasets, laboratories, materials, prior techniques, research communities, and epistemic infrastructure.

\textbf{Key property.} Invention requires both existing knowledge (to draw on) and a defined problem structure (to direct the search). Neither alone is sufficient. A researcher with deep knowledge but no defined problem may discover without inventing. A firm with a defined market need but insufficient knowledge may commission invention without being able to execute it.

\textbf{Examples.} Practical cases include the transformer architecture, mRNA platform technologies, lithium-ion battery designs, and lithography systems.

\textbf{Appropriability at the interface.} Invention is the generation mechanism most directly connected to IP law. Patents and trade secrets govern inventions; copyright governs expressive and software artefacts that may carry disembodied knowledge. The KCM interface question for invention is therefore well-established: who has a claim over the resulting stock, and under what conditions does that claim hold?

\textbf{Source lineage.} Schumpeter on entrepreneurial combination; the economics of patent systems; Mokyr on the knowledge economy of the Industrial Revolution.

\subsection[Interpretive generation]{\texorpdfstring{\(G^{J}\): Judgment/Interpretation}{Interpretive generation}}\label{gj-judgmentinterpretation}

Existing knowledge-bearing stock (primarily embodied capability \(K^{E}\)) is applied to data, events, context, or anomaly to produce new productive understanding. \(G^{J}\) encompasses both the initial application of embodied capability to novel context and the refinement of future judgment through repeated acts of interpretation. Both are aspects of one mechanism: each act of judgment generates productive output that may subsequently be codified, and each act simultaneously improves the capacity for future judgment.

\textbf{Direct KGM-to-KCM bridge?.} \(G^{J}\) is the Smithian bridge mechanism: \(K^{E}\) applied to context generates new productive understanding (the judgment, diagnosis, design, classification, or insight) that may enter the conversion cycle as new knowledge-bearing stock. Formally: \(K^{E}\) → judgment/output → \(K^{D}\) or \(K^{I}\) (KCM Cell 1). This is the junction at which KGM output becomes KCM input. The KGM produces the judgment; the KCM governs what happens to it next.

\textbf{Key property: initial application.} The senior clinician who interprets a patient's test results; the analyst who reads market patterns; the engineer who diagnoses an anomalous system behaviour, all are using embodied knowledge to generate new knowledge (the diagnosis, the market insight, the fault identification) that may subsequently be codified. Interpretation outputs become new knowledge-bearing stock only when they are externalized. An insight held in a clinician's mind depreciates when the clinician retires. An insight codified into a diagnostic protocol enters the conversion cycle and may be scaled, enclosed, or shared.

\textbf{Examples.} Practical cases include radiology AI review, legal document review, investment committee judgement, and cybersecurity triage.

\textbf{Key property: iterative refinement.} Repeated application of embodied capability to analogous problems improves future judgment quality. The experienced diagnostician's second hundred cases are read more productively than the first, not only because the clinician has more information but because the interpretive faculty itself has been refined through iterated application. This is the refinement dimension of \(G^{J}\): the mechanism is self-improving, and the improvement is embodied in \(K^{E}\), not yet in \(K^{D}\).

\textbf{Tacit knowledge and the interpretation problem.} Interpretation capacity is often tacit and context-specific. It cannot be fully codified without loss of productive power, the expert's judgment is not just the conclusion but the process of reaching it under uncertainty. This is why the Capability-Bounded Codification Principle (Proposition B of Chapter 2) applies to interpretation: codifying the output of interpretation generates knowledge-bearing stock, but the stock's productive value depends on the ongoing availability of the interpretation capability that can maintain, apply, and update it.

\textbf{The Smithian bridge: full statement.} Smith's pin factory decomposes physical production into specialized tasks: dexterity increases, switching time falls, and machinery follows. The productive unit is a manual operation. In knowledge-bearing capitalism, the productive unit is often not a manual task but an act of interpretation. This is not merely the observation that services are productive. It is sharper: interpretation is the knowledge-economy analogue to skilled productive labour, except that its output is a judgment, diagnosis, model, design, insight, or classification that may subsequently become knowledge-bearing stock, rather than a physical object that enters exchange directly.

Smith gives the pressure point rather than the full authority: once productive skill can be decomposed, encoded, transferred, and redeployed, the division of labour becomes a division and conversion of embodied productive knowledge. Knowledge-bearing capitalism decomposes and recombines interpretive capacity: clinicians, analysts, engineers, researchers, and designers apply embodied knowledge to ambiguous inputs and generate judgments that can later be codified, scaled, enclosed, or shared. The analogy to the pin factory is disciplined but not exact: just as the factory's output was pins that could be traded, the interpretation economy's output is judgment that can be converted into knowledge-bearing stock that can be owned, deployed, and enclosed by the firm that employs the interpreter. The economic problem is not that Smith ignored services, but that he could not foresee that the durable residue of skilled service (the judgment once codified into a protocol, the insight once encoded in a model) would become the primary accumulation vehicle of a later capitalism.

\textbf{Natural intelligence connection.} \(G^{J}\) is the central generation mechanism of natural intelligence. The professional expertise, tacit skill, organizational judgement, and community-maintained interpretation that constitute natural intelligence operate primarily through this mechanism. When AI systems encode the output of repeated human judgment (when patterns of expert interpretation become training data) they are converting the residue of \(G^{J}\) from \(K^{E}\) into \(K^{D}\). This conversion, and who controls it at first conversion, is the central appropriability question of knowledge-bearing capitalism.

\textbf{Source lineage.} \textcite{Smith1776} on productive labour and the durable residue of skilled work; \textcite{Hayek1945}\index{Hayek, Friedrich} on the use of dispersed knowledge; \textcite{Polanyi1966}\index{Polanyi, Michael} on tacit knowledge; \textcite{Penrose1959}\index{Penrose, Edith} on the distinction between objective and experiential knowledge.

\textbf{Interpretive judgement and AI substitution.} In measurement-bridge terms, the embodied operative unit is the person or team whose accumulated judgement is being exercised; interpretive judgement is the generative act. The person or team holds the accumulated interpretive capacity, and the judgement mechanism deploys it against a domain problem and generates new knowledge-bearing stock.

A model and a person may be economically comparable only within a defined task class, fidelity threshold, cost condition, and time horizon. The relevant question is therefore not whether \(K^{D}\) equals \(K^{E}\) in general, but whether a disembodied system can reproduce the observable output of embodied judgment for a specified function, channel, evaluation interval, and judge standard. When that bounded condition holds, \(K^{D}\) has captured an observable bridge function formerly performed by \(K^{E}\), but only for the defined task class. It does not imply full equivalence of the person, role, profession, or domain. AI substitution occurs at the task-class level: a system may achieve \(G^{J}\)-output equivalence for report-drafting or preliminary image classification without achieving it for clinical judgment, patient-facing communication, or ambiguous case escalation.

Chapter 11 returns to this as an empirical test of AI feedback-capture and task-capability substitution.

\begin{quote}
\textbf{Technical note.} The formal Turing-style OKU equivalence relation is stated in Chapter 2 and registered in the Technical Companion, Appendix B. It is not reproduced here because the main Chapter 3 point is conceptual: interpretive outputs can be transferred from embodied judgement into disembodied artefacts only under bounded task, context, fidelity, and judgement conditions.
\end{quote}

\section[Learning feedback loop]{\texorpdfstring{\(G^{L}\): The Learning Feedback Loop}{Learning feedback loop}}\label{gl-the-learning-feedback-loop}
\index{learning loop|textbf}

\(G^{L}\) is not a primary origin mechanism. It is the feedback loop through which conversion and deployment activity modifies future knowledge-bearing stock and productive capability. The structural reason for this classification: \(G^{R}\), \(G^{X}\), \(G^{D}\), \(G^{N}\), and \(G^{J}\) can produce new knowledge-bearing stock independently of prior conversion activity. \(G^{L}\) requires a prior deployment event, it operates on the outputs of the conversion cycle, not in parallel with it.

\textbf{Dual-order classification.} Learning feedback is second-order because it requires prior deployment. Once feedback exists, however, it becomes a first-order driver of capability change. No deployment event, no feedback, no \(G^{L}\) input. Once feedback exists, \(G^{L}\) modifies the capability stock \(\widetilde{C}_a\) directly and at the same structural level as \(G^{R}\). The dynamic capability equation is:

\begin{equation}
\widetilde{C}_{a,t+1}=(1-\delta_C)\cdot\widetilde{C}_{a,t}+\gamma\cdot G^{R}_{a,t}+\ell\cdot G^{L}_{a,t}
\label{eq:ch3:capability-state}
\end{equation}

\textbf{Where:} \(\widetilde{C}_{a,t+1}\) is actor \(a\)'s dynamic capability stock in the next period; \(\widetilde{C}_{a,t}\) is actor \(a\)'s current dynamic capability stock; \(\delta_C\) is the passive depreciation rate of capability; \(\gamma\) is the realization coefficient from recombination generation into capability growth; \(\ell\) is the realization coefficient from learning feedback into capability growth; \(G^{R}_{a,t}\) is recombination generation for actor \(a\) at time \(t\); and \(G^{L}_{a,t}\) is learning-feedback generation for actor \(a\) at time \(t\).

This equation supplies the capability term in the residence--governance--capability triad. Governance makes a stock economically actionable; capability determines whether a particular actor can actually convert that actionable stock into productive services. A knowledge stock may be residentially available and legally accessible while yielding little capital service if \(\widetilde{C}_{a,t}\), or the GATE-C capability component used in K-CMM, is weak.

\(G^{L}\) appears alongside \(G^{R}\) as a direct capability modifier.

The same logic applies at the level of domain-specific embodied knowledge capital. Learning loops deepen \(K^{E}\), not only models, artefacts, or organizational routines. A simple skill-formation branch is:

\begin{equation}
K^{E}_{d,t+1}
=
K^{E}_{d,t}
+
\lambda_d \cdot \mathrm{Practice}_{d,t}(F_{a,t})
-
\delta^{E}_{d}K^{E}_{d,t}.
\label{eq:ch3:practice-generated-ke}
\end{equation}

\textbf{Where:} \(K^{E}_{d,t+1}\) is embodied knowledge capital in domain \(d\) in the next period; \(K^{E}_{d,t}\) is current embodied knowledge capital in domain \(d\); \(\lambda_d\) is the domain-specific learning coefficient from practice; \(\mathrm{Practice}_{d,t}(F_{a,t})\) is practice in domain \(d\) at time \(t\), conditioned by actor \(a\)'s accessible field \(F_{a,t}\); and \(\delta^{E}_{d}\) is the depreciation rate of embodied knowledge capital in domain \(d\).

Repeated domain practice increases embodied knowledge capital, but only when actors have access to the field of problems, tools, feedback, and interpretive conditions required for learning. The term \(\mathrm{Practice}_{d,t}(F_{a,t})\) therefore matters: practice is not mere repetition in isolation. It is field-dependent engagement with domain problems. When enclosure narrows \(F_{a,t}\), it can reduce not only current recombination but also the future formation of \(K^{E}_{d}\). This formulation applies most directly to domains where practice requires access to live problems, tools, data, feedback, or interpretive conditions. Some embodied skills improve through repetition even without broad recombination-field access. This is the skill-formation branch of the learning-loop mechanism.

In prose, the loop is straightforward: generation produces new stock; conversion governs and deploys it; use generates feedback; feedback modifies capability; revised capability then re-enters future generation and conversion.

The loop is closed: conversion activity generates deployment experience; experience generates feedback; feedback modifies capability stock; revised stock re-enters the conversion cycle or feeds back into generation mechanisms.

\textbf{Mechanism.} Deployment of knowledge-bearing stock generates experience. Experience generates feedback signals. Feedback signals are absorbed and processed by agents, institutions, or systems capable of distinguishing informative signal from noise. Processed feedback modifies existing stock or triggers generation of revised stock.

\textbf{Examples.} Practical cases include RLHF, post-deployment model improvement, production telemetry, and user correction loops.

\textbf{Compounding implication.} Incumbents with larger deployed knowledge bases generate more feedback inputs, which accelerates stock revision, which widens productive advantage over time. This is a compounding dynamic endogenous to knowledge-bearing capitalism: knowledge deployment generates knowledge improvement, which enables further deployment. Enclosure of knowledge stock restricts this loop for non-incumbents while intensifying it for incumbents. This is the formal basis of Proposition D1 (Feedback Acceleration) developed in Interlude I.

\textbf{Appropriability at the loop.} Feedback-generated improvements raise the same appropriability questions as original training. Who owns the improvement generated by RLHF applied to a deployed AI model? Who owns the revised clinical protocol generated by post-deployment outcomes data? Who owns the updated algorithm trained on platform interaction data? The learning loop is not institutionally neutral; it inherits the appropriability structure of the conversion events that generate its inputs.

\textbf{Natural and artificial intelligence.} \(G^{L}\) is the mechanism by which natural intelligence is continuously converted into artificial intelligence improvement inside enclosed platforms and model systems. Users supply natural intelligence signals (corrections, continuations, preferences, refusals, work patterns, domain-specific responses) that the deploying system converts into revised \(K^{D}\) through model fine-tuning, reward modelling, or preference learning. The value of that conversion, and whether it accrues to the users who generated the feedback or to the incumbent that controls the deployed system, is the central learning-loop appropriability question. It is developed fully in Chapter 7.

\textbf{Source lineage.} \textcite{Arrow1962} on learning by doing; \textcite{Penrose1959} on experiential knowledge; \textcite{ChristianoEtAl2017}\index{Christiano et al.} on RLHF as the contemporary AI instance of feedback-driven model revision.

\section{The Recombination Field}\label{the-recombination-field}
\index{recombination field|textbf}

Section~\ref{gr-recombination} defined the recombination field as the set of knowledge stocks that actor \(a\) can practically and legally access and combine at time \(t\). This section operationalizes that definition. The recombination field is not itself a generation mechanism; it is the institutional, technical, and social environment that makes \(G^{R}\) possible. In plain terms, \(F_{a,t}\) asks which knowledge-bearing stocks are actually reachable, permitted, usable, trustworthy, and feedback-generating for a specific actor at a specific time.

\begin{longtable}{@{}L{0.13\textwidth}L{0.20\textwidth}L{0.16\textwidth}L{0.18\textwidth}L{0.18\textwidth}@{}}
\caption{Operational Components of the Recombination Field}\label{tab:ch3-recombination-field-operational-components}\\
\toprule
Component & Practical question & Example & Observable proxy & Failure mode \\ 
\midrule
\endfirsthead
\toprule
Component & Practical question & Example & Observable proxy & Failure mode \\ 
\midrule
\endhead
Access & Can the actor reach the stock? & API, journal, repository & availability, licence, paywall & exclusion \\ 
Permission & Is use legally allowed? & licence, IP right & terms of use & litigation risk \\ 
Interoperability & Can it combine with other stock? & standard, file format & API compatibility & fragmentation \\ 
Capability & Can the actor use it? & skilled team, compute & headcount, compute budget & unusable access \\ 
Trust/reliability & Is it safe to use? & validated dataset & provenance, audit trail & poisoned or low-quality stock \\ 
Feedback & Can use improve future stock? & RLHF, telemetry & feedback-capture rights & blocked learning loop \\ 
\bottomrule
\end{longtable}

\textbf{Formal representation.} The field diversity term \(D_{u}(F_{a,t})\) in the \(G^{R}\) specification captures recombination field breadth. \(F_{a,t}\) is the actor-specific, time-indexed specification of the effective field extent\index{effective field extent} \(S_a\) introduced in Chapter 1: what Chapter 1 names as the knowledge-economy analogue to Smith's market extent, Chapter 3 formalizes as the recombination field available to actor a at time t. Because \(D_{u}(F_{a,t})\) also enters the \(\Phi(\cdot)\) composition as a shared enabling condition, enclosure that narrows the recombination field for one actor class affects not only \(G^{R}\) but potentially \(G^{D}\) and \(G^{J}\) as well. This is the interaction-effect entry point: mechanisms that appear formally separable in a list are in practice conditionally linked through shared enabling conditions. This coupling is also where the multiplicative composition does the most work: a shared \(D_u\) gate transmits a single enclosure shock across \(G^{R}\), \(G^{D}\), and \(G^{J}\) only if the mechanisms enter \(\Phi(\cdot)\) as complements. The robustness of the cross-mechanism claim is therefore exactly the gated-complementarity case of the function-class audit (Volume~2, Appendix~F): under an additive or strongly substitutable \(\Phi\) the shock stays local to \(G^{R}\), whereas under gated complementarity it propagates as described here. The claim should accordingly be read as holding in the gated-complementarity case, with the audit recording where it fails.

\textbf{KCM connection.} Cognitive enclosure narrows the recombination field by restricting access to knowledge-bearing stock that would otherwise be available for combination. Epistemic infrastructure (public research, open standards, shared measurement systems, commons knowledge (\(K^{C}\)), and public epistemic capital (\(K^{P}\))) widens the field. Proposition C (Generative Suppression, Chapter 6) operates formally through reduction of \(D_{u}(F_{a,t})\): enclosure narrows the field diversity term, which reduces \(G^{R}\) output and, under the weakest-commitment \(\Phi(\cdot)\) specification in §3.13, lowers the gated aggregate generation rate for the affected actor class. If later evidence rejects that aggregate specification, the claim falls back to the mechanism-wise result: enclosure suppresses the relevant \(G^k\) channels whose inputs depend on field breadth, access, capability, or feedback. The KCM and the recombination field are therefore in a dynamic relationship: the broader the recombination field, the richer the inputs to the KGM; the more pervasive enclosure, the narrower the generation base. This is also an instance of the Smithian inversion: the enclosing actor's individually rational strategy (restricting access to appropriate returns) may exceed the social optimum once Chapter~8's welfare specification is applied, by suppressing one or more generation mechanisms of the knowledge economy as a whole.

\section{Recursive Truth-Decay: Endogenous Reliability in the Generation Cycle}\index{recursive truth-decay equation family|textbf}\index{truth decay}\label{recursive-truth-decay}
\index{recursive truth-decay|textbf}

\index{Truth-Dependence Axiom|textbf}
Axiom~B.2 introduced the epistemic reliability coefficient \(\tau_i\in[0,1]\) as a modifier on the productive value of knowledge-bearing stock, the degree to which the stock is truth-tracking, validated, provenance-supported, and defeater-resistant. So far \(\tau_i\) has entered statically: a stock arrives with a reliability, and low reliability is charged as expected knowledge loss \(EKL_{false}\) (Chapter~9). But in a generation cycle increasingly fed by its own outputs, reliability is not exogenous. The \(\tau\) of newly generated stock depends on the \(\tau\) of the stock it was generated from. When the learning loop \(G^{L}\) returns deployment output into the recombination field \(F_{a,t}\) that feeds \(G^{R}\), and a rising share of that field is itself prior, unvalidated machine output, reliability degrades endogenously. This section models that degradation. It is the mechanism by which the artificial-intelligence half of this book's subject becomes a source of capital risk in its own right, not merely a faster generator.

\textbf{The recursion.} Partition the generation input pool at round \(t\) into a grounded share \(s_t\), stock anchored to human judgement, measurement, or real-world outcome with reliability \(\tau_H\), and a synthetic share \(1-s_t\), prior machine output whose current reliability is \(\tau_t\). Let \(\delta_\tau\in[0,1]\) be the per-generation transmission loss (compression, hallucination leakage, error propagation) incurred when output is produced from inputs without correction, and let \(v_t\in[0,1]\) be the validation intensity, the share of output re-anchored to ground truth through testing, outcome feedback, provenance verification, or expert review. The reliability of the synthetic pool evolves as
\[
\tau_{t+1} = v_t\,\tau_H + (1-v_t)(1-\delta_\tau)\bigl[\,s_t\,\tau_H + (1-s_t)\,\tau_t\,\bigr].
\]
The first term is the re-grounded fraction; the bracket is the reliability of the input pool, a mixture of grounded and synthetic stock; the factor \((1-\delta_\tau)\) is transmission loss on the unvalidated remainder. This is a linear recursion \(\tau_{t+1}=A\,\tau_t+B\) with \(A=(1-v)(1-\delta_\tau)(1-s)\) and \(B=\tau_H\bigl[v+(1-v)(1-\delta_\tau)s\bigr]\).

\textbf{The collapse result.} With constant coefficients the recursion has the stable fixed point
\[
\tau^{*}=\frac{\tau_H\bigl[v+(1-v)(1-\delta_\tau)s\bigr]}{1-(1-v)(1-\delta_\tau)(1-s)},
\]
since \(0\le A<1\) whenever any validation, grounding, or transmission loss is present. Three boundary cases fix intuition. Under full validation, \(v=1\), \(\tau^{*}=\tau_H\): reliability is held at ground truth no matter how much synthetic stock circulates. Under grounded inputs without validation, \(s=1,\ v=0\), \(\tau^{*}=(1-\delta_\tau)\tau_H\): one generation of transmission loss, then anchored. But under unvalidated self-consumption, \(s\to0\) and \(v\to0\), \(B\to0\) and \(\tau_{t+1}=(1-\delta_\tau)\tau_t\), so
\[
\tau_t=(1-\delta_\tau)^{t}\,\tau_0\;\longrightarrow\;0.
\]
When generation feeds on its own unvalidated output, reliability decays geometrically toward zero: the stock degrades from knowledge-bearing capital to information-bearing noise. This is the formal counterpart of what the machine-learning literature reports as model collapse \parencite{Shumailov2024} and model-autophagy disorder \parencite{Alemohammad2023}, and it instances the data-processing inequality \parencite{CoverThomas2006}: processing cannot manufacture truth-content that grounding did not supply.

\textbf{The grounded-anchor condition.} The recursion makes the sustainability condition explicit. Reliability stays above collapse only if \(B>0\), that is, only if \(v>0\) or \(s>0\): some positive flow of grounded anchor, external validation or grounded input, must persist. As machine output comes to dominate the accessible corpus (\(s\) falling) and validation remains costly (\(v\) falling), \(\tau^{*}\) falls, and the knowledge base drifts from the capital-like state toward the information-bearing state. Grounding is therefore not a one-time input but a maintenance flow, the epistemic analogue of the Maintenance-Gap Condition of Chapter~5: a non-rival knowledge base remains capital only while enough grounded validation is reinvested to offset transmission loss. The deeper reason the anchor cannot lie inside the cycle is that validation is ultimately a judgement about correspondence to the world, and \emph{there is no algorithm for truth}. An algorithm can check internal consistency, reproduce a benchmark, or flag a defeater it was trained to recognize, but the final adjudication of whether a claim tracks reality rests on embodied judgement, measurement against outcomes, or institutional validation, the \(K^{E}\), \(K^{I}\), and \(K^{P}\) that sit outside the model. This is the precise sense in which embodied knowledge capital is irreplaceable in the artificial-intelligence economy: it is the non-substitutable judge that supplies the grounded anchor, the validation intensity \(v\) and the ground-truth reliability \(\tau_H\), on which the whole recursion depends. It is also why the Turing-style equivalence of Chapter~2 is bounded: a disembodied unit can reach task-class equivalence on operations whose ground truth is cheap to check, but truth-adjudication itself is the residue that resists full codification. Remove that judge from the loop and \(\tau\) has nothing to converge toward but its own prior output, and decays.

\textbf{What it couples to.} The mechanism is exactly \(G^{L}\) feeding contaminated stock back into \(G^{R}\): deployment output re-enters the recombination field as generation input, and the recursion governs the reliability of what comes out. Three couplings follow. First, \(\tau\)-decay is an \emph{endogenous depreciation} of knowledge-bearing capital, distinct from the obsolescence and maintenance-failure pathways of Chapter~2: a stock can be perfectly available and well maintained yet lose productive value because its truth-content has eroded. Second, falling \(\tau\) raises \(EKL_{false}\), so truth-decay generates dark risk (Chapter~9), measurable as the expected loss from acting on degraded stock. Third, and most consequentially, it sharpens the capability divergence of Chapter~7. An incumbent that controls a live, grounded feedback loop sustains high \(s\) and high \(v\) and holds its \(\tau\) near \(\tau_H\); actors excluded from that loop fall back on scraped or synthetic corpora with low \(s\) and costly \(v\) and suffer \(\tau\)-decay. The reliability gap, not only the volume gap, widens the capability gap \(\Delta_t\). Recursive truth-decay is thus the epistemic-quality channel of the double-acting asymmetry (\S\ref{the-double-acting-asymmetry}) made dynamic: feedback capture is also reliability capture.

\textbf{Provisional vocabulary.} \emph{Epistemic depreciation} names the loss of productive value through declining \(\tau\) rather than through wear, obsolescence, or access loss; \emph{recursive truth-decay} names its specific driver, the endogenous fall in \(\tau\) when generation recycles unvalidated output. Existing vocabulary is insufficient because ``depreciation'' in capital theory tracks physical or technological wear, not erosion of truth-content, and ``model collapse'' names the engineering phenomenon without locating it as a capital-depreciation pathway with a sustainability condition. Misuse risk: the mechanism does \emph{not} say synthetic data is always degrading. Curated or validated synthetic data raises \(v\) and can preserve or improve \(\tau\); the claim concerns \emph{unvalidated} recycling as \(s\) and \(v\) fall. Recommendation: retain; calibration of \(\delta_\tau\), \(s\), and \(v\) is open.

\textbf{Originality.} The engineering phenomenon is established \parencite{Shumailov2024,Alemohammad2023} and the information-theoretic backbone is classical \parencite{CoverThomas2006}. KBC's contribution is to recast reliability as an endogenous, depreciating property of knowledge-bearing \emph{capital} circulating in the generation cycle, to state the grounded-anchor sustainability condition, and to couple truth-decay to dark risk and to the capability-divergence engine of the Smithian inversion. The underlying decay result is Established; the capital-theoretic integration is Synthesized and potentially Novel as architecture. Chapter~11 develops the test; the falsifier is a domain in which reliability is invariant to synthetic-input share once quality filtering is held fixed.

\textbf{Observability caveat.} The mechanism is falsifiable in design, but its variables are fragile to measure: \(\tau\) and the synthetic-input share \(1-s\) become harder to observe as provenance collapses and synthetic and organic content blur. Chapter~11 therefore carries the empirical claim as \emph{Testable but observability-fragile} and names candidate provenance proxies, watermarking, content-credential standards, and dataset audits, whose coverage is itself declining.

\section{Source Lineage}\label{source-lineage}

\begingroup
\small
\setlength{\tabcolsep}{3pt}
\renewcommand{\arraystretch}{1.12}
\sloppy
\par\addvspace{0.8\baselineskip}\noindent
\begin{longtable}{@{}L{0.43\textwidth}
L{0.51\textwidth}@{}}
\caption{Source Lineage of KGM Mechanisms}\label{tab:ch3:source-lineage-kgm}\\
\toprule\noalign{}
\begin{minipage}[b]{\linewidth}\raggedright
Component
\end{minipage} & \begin{minipage}[b]{\linewidth}\raggedright
Primary sources
\end{minipage} \\
\midrule\noalign{}
\endfirsthead
\toprule\noalign{}
\begin{minipage}[b]{\linewidth}\raggedright
Component
\end{minipage} & \begin{minipage}[b]{\linewidth}\raggedright
Primary sources
\end{minipage} \\
\midrule\noalign{}
\endhead
\bottomrule\noalign{}
\endlastfoot
\textbf{\(G^{R}\): Recombination generation} & \textcite{Romer1990,Romer1994}; \textcite{Weitzman1998}; \textcite{Arthur2009}; \textcite{Schumpeter1934} \\
\textbf{\(G^{X}\): Experimental generation} & \textcite{Arrow1962}; \textcite{NelsonWinter1982} \\
\textbf{\(G^{D}\): Discovery generation} & \textcite{Merton1973}; \textcite{Bush1945}; \textcite{Ostrom1990}; broader open-science literature \\
\textbf{\(G^{N}\): Inventive generation / novel construction} & \textcite{Schumpeter1934}; \textcite{Mokyr2002}; \textcite{Arrow1962}; broader patent-economics literature \\
\textbf{\(G^{J}\): Interpretive generation / judgement} & \textcite{Smith1776}; \textcite{Hayek1945}; \textcite{Polanyi1966}; \textcite{Penrose1959} \\
\textbf{\(G^{L}\): Learning-loop generation} & \textcite{Arrow1962}; \textcite{Penrose1959}; \textcite{ChristianoEtAl2017} \\
\textbf{\(G^{DIA}\): Dialectical generation (candidate)} & \textcite{Kuhn1962}\index{Kuhn, Thomas}; \textcite{Schumpeter1942}; \textcite{Lakatos1978} on scientific research programmes \\
\textbf{Recombination field} & \textcite{Romer1990}; \textcite{Hayek1945}; \textcite{Bell1973}; \textcite{Ostrom1990} \\
\textbf{NI/AI bridge} & Original synthesis: this book's integration of natural and artificial intelligence within the KGM/KCM architecture \\
\end{longtable}
\endgroup

\section[Candidate Mechanism (Non-Canonical): Dialectical Generation]{Candidate Mechanism (Non-Canonical): Dialectical Generation}\label{candidate-mechanism-gdia-dialectical-generation}

\textbf{Status decision.} This book makes a definite choice: the canonical generation set is the six mechanisms developed above (\(G^{R}, G^{X}, G^{D}, G^{N}, G^{J}, G^{L}\)), and dialectical generation is \emph{excluded} from the canonical composition function \(\Phi(\cdot)\), retained here only as a candidate for future formalization, which is why it is placed among the open questions rather than in the mechanism set.

\emph{Dialectical generation is preserved as a named candidate mechanism in Chapter 3. It is not part of the canonical KGM composition function in this book. The mechanism remains because contradiction-driven generation identifies a distinct source of productive knowledge; formal inclusion in \(\Phi(\cdot)\) depends on further specification.}

\textbf{The fused-mode reading.} Dialectical generation is the paradigm of the Fused Generation Proposition (\S\ref{fused-generation-proposition}): the case in which a single generative act co-realizes stock-production and admissibility-revision, structured opposition changing which assumptions remain admissible at the same time as it yields the revised hypothesis. This sharpens the candidate's status rather than upgrading it. What is settled is the \emph{existence and location} of the mechanism, a real, fused-mode generative act, not a vague or under-specified seventh channel. What remains open is only its \emph{separate formal specification}: the magnitude and curvature of the relation, the counter-frame notation \(\Theta^-\), the tension function \(\mathcal{T}_{DIA}(T,O,M,H)\), and the inverted-\(U\). That is why \(G^{DIA}\) stays a named candidate outside \(\Phi(\cdot)\) and is not promoted into the composition function.

Some knowledge is generated not by simple recombination, experimentation, discovery, invention, or interpretation alone, but by structured opposition. A claim, model, or proposed course of action is challenged by an antithesis that exposes missing evidence, defective assumptions, invalid inference, or internal contradiction. Analysis then identifies what must be preserved, rejected, or reframed. The synthesis becomes a new hypothesis rather than a final conclusion. This mechanism is treated here as candidate rather than canonical because its formal representation remains unresolved, but it is preserved because it captures a distinct source of knowledge generation: productive contradiction.

The mechanism does not treat antithesis as logical negation. In the notation used here, \(\Theta_{a,t}\) denotes an operative thesis or frame, while \(\Theta^-_{a,t}\) denotes an antithetical counter-frame: contradiction, anomaly, failed prediction, critique, resistant reality, or opposing claim. The minus sign marks dialectical opposition, not formal negation.

Hegel\index{Hegel} supplies one lineage through determinate negation \parencite{Hegel1977}, but the mechanism does not depend on a crude thesis-antithesis-synthesis template. Popper's conjectures and refutations \parencite{Popper1963}, Argyris and Sch\"{o}n's double-loop learning \parencite{ArgyrisSchon1978}, Kuhn's anomaly-driven paradigm crisis \parencite{Kuhn1962}, Peirce's abduction from surprising facts \parencite{Peirce1931}, and Socratic elenchus \parencite{Vlastos1983} all identify forms of knowledge generation in which a claim is improved because it is challenged. KBC treats this as contradiction-driven hypothesis generation.

\textbf{Mechanism distinction.} Recombination combines compatible or complementary knowledge-bearing stock. Dialectical generation changes which assumptions, elements, claims, or hypotheses remain admissible after structured opposition exposes a defect in the original position. \(G^{R}\) operates on complementarity; \(G^{DIA}\) operates on productive resistance.

\textbf{Boundary test.} Dialectical generation applies only when the output resolves, reframes, or productively preserves a contradiction between partially incompatible frames. If the output merely combines distant but compatible complements, the mechanism is recombination rather than dialectical generation.

\textbf{Applied case.} A red-team exercise that changes an AI safety protocol is dialectical generation when the challenge exposes a false assumption in the original safety claim, analysis identifies what must be preserved or rejected, and the revised protocol becomes a new hypothesis for safer deployment. The value is not merely that two knowledge stocks have been combined. The value arises because structured opposition changes which assumptions remain admissible.

\begin{center}
\fbox{%
\begin{minipage}{0.92\linewidth}
\textbf{Candidate formalization, not yet canonical.} The following expression is retained as a formal sketch only. It is not part of the canonical \(\Phi(\cdot)\) composition function.
\begin{equation}
G^{DIA}_{a,t+1}=\lambda^{DIA}_a\cdot\mathcal{T}_{DIA}(T,O,M,H)
\label{eq:ch3:g-dia-a-t-1}
\end{equation}
where \(\mathcal{T}_{DIA}(T,O,M,H)\) captures a proposed tension function across theoretical, organizational, methodological, and historical dimensions. The inverted-U specification remains unresolved.
\end{minipage}}
\end{center}

\textbf{Why \(G^{DIA}\) is not \(G^{R}\).} Recombination combines stocks whose combination is productive precisely because the domains are compatible or complementary. Dialectical generation operates on partially incompatible frames. The productive output arises from the resistance between them: the challenged assumption, the exposed contradiction, the rejected inference, or the reframed hypothesis. These cannot be captured as recombination of compatible complements without losing the mechanism's distinctive character.

\textbf{Status and inclusion condition.} The mechanism is included in Chapter 3 because it identifies a real form of knowledge generation. It is a named candidate mechanism; formal specification remains unresolved. \(G^{DIA}\) is therefore preserved as a named candidate mechanism, but its formal equation remains non-canonical and outside the \(\Phi(\cdot)\) composition function until the tension function \(\mathcal{T}_{DIA}(T,O,M,H)\) and the inverted-U specification are formally disciplined.
\subsection{Inclusion Proposal: What Canonical Status Would Require}\label{gdia-inclusion-proposal}

Candidate status does not mean the mechanism is unspecifiable. This subsection states precisely what would discipline \(\mathcal{T}_{DIA}\) and the inverted-U enough for \(G^{DIA}\) to enter the canonical composition function \(\Phi(\cdot)\), and the conditions under which that inclusion would be earned. It is an inclusion proposal, not a claim of canonical status: until its identification conditions are met, \(G^{DIA}\) remains outside \(\Phi(\cdot)\).

\textbf{1. The tension function as a normalized index.} Let the four dimensions of \(\mathcal{T}_{DIA}\) be measured frame-incompatibilities: theoretical \(T\) (conflicting predictions or models), organizational \(O\) (conflicting goals or incentives), methodological \(M\) (incompatible evidence standards), and historical \(H\) (accumulated anomaly relative to prior expectation). Aggregate them into a single tension index
\[
\zeta = \mathrm{norm}\!\left(w_T\,T + w_O\,O + w_M\,M + w_H\,H\right)\in[0,1],
\]
with non-negative weights and a normalizing link \(\mathrm{norm}(\cdot)\). This is no less specified than useful diversity \(D_u\) or the capability gate: an index with named, separately proxied components.

\textbf{2. The inverted-U, derived rather than asserted.} Dialectical yield rises with contradiction but falls when contradiction outstrips the adjudicative capacity available to resolve it. Let \(A_{\mathrm{adj}}\) be that capacity, the shared evidence standards, arbitration, peer-review, and red-team institutions that convert opposition into tested knowledge, and let \(\mathrm{Res}(\zeta,A_{\mathrm{adj}})\) be a resolvability factor decreasing in \(\zeta/A_{\mathrm{adj}}\) (for instance \(e^{-\zeta/A_{\mathrm{adj}}}\)). Then
\[
G^{DIA}_{a,t+1}=\lambda^{DIA}_a\,\cdot\,\zeta\,\cdot\,\mathrm{Res}(\zeta,A_{\mathrm{adj}}).
\]
This product is single-peaked by construction: zero at \(\zeta=0\) (no contradiction to resolve), tending to zero as \(\zeta\) grows large relative to \(A_{\mathrm{adj}}\) (incommensurable frames, opposition without adjudication), with an interior maximum at \(\zeta^{*}\) that shifts upward as \(A_{\mathrm{adj}}\) rises. The inverted-U is therefore a result, not a stipulation, and its peak carries economic content: greater adjudicative capacity lets a field metabolize sharper contradiction. The capacity \(A_{\mathrm{adj}}\) is supplied by institutionalized and public-epistemic capital (\(K^{I}\), \(K^{P}\)) and by embodied judgement (\(K^{E}\)), the same non-substitutable judge that anchors validation in \S\ref{recursive-truth-decay}: there is no algorithm that resolves a contradiction without a standard against which to resolve it.

\textbf{3. A non-overlapping place in \(\Phi(\cdot)\).} The boundary test against \(G^{R}\) becomes a formal partition of the combination space by frame-compatibility. Low-\(\zeta\) pairs (compatible complements, high \(D_u\)) feed recombination; high-\(\zeta\) pairs (incompatible frames) feed dialectical generation; the highest \(\zeta\) (incommensurable) feeds neither. \(G^{DIA}\) then enters \(\Phi(\cdot)\) as its own channel, capability- and diversity-gated like the others, with no risk of re-counting recombination, because the two mechanisms operate on disjoint regions of \(\zeta\).

\textbf{4. Identification.} Object validity precedes calibration (Chapter~11). The proposal is testable only with proxies: \(\zeta\) from expert-prediction divergence, replication or anomaly rates, and conflicting-method prevalence; \(A_{\mathrm{adj}}\) from the presence and strength of adjudicative institutions; and \(G^{DIA}\) output from structured-opposition processes, red-team exercises, adversarial collaboration, peer review, litigation, investment-committee dissent, measured as new admissible hypotheses or assumption revisions rather than as recombinations.

\textbf{5. The falsifier, and the price of admission.} \(G^{DIA}\) earns canonical status only if dialectical output is non-monotone in \(\zeta\) (a monotone relation defeats the inverted-U), if the high-\(\zeta\) case generates knowledge beyond what \(G^{R}\) predicts (otherwise it collapses into recombination), and if the peak shifts with \(A_{\mathrm{adj}}\). There is also a structural cost: the inverted-U is a non-monotone form, the congestion member of the function-class audit battery (Volume~2, Appendix~F), so admitting \(G^{DIA}\) would commit the \(\Phi(\cdot)\) family to that member for this channel, and the enclosure and inversion results would have to be re-audited for survival under it. Until these conditions are met the mechanism is specified but not admitted: a disciplined candidate, not a canonical channel.

\section{The Fused Generation Proposition}\index{fused generation proposition|textbf}\index{Aufhebung@\emph{Aufhebung}!fused generation}\label{fused-generation-proposition}

The six canonical mechanisms are stated above as if each generative act produced stock within a fixed admissibility space, the standing set of assumptions, frames, and evidence standards under which generation runs. For most acts that idealization is harmless. For some it is not, and the architecture should name the exception rather than leave it implicit.

\textbf{Proposition (Fused Generation).} In some generative acts the existing generators co-realize two distinct operations, stock-production and admissibility-revision, in a single act with no clean sequential decomposition. The act produces new knowledge-bearing stock and, in the same movement, revises which assumptions remain admissible for subsequent generation. There is no prior step in which the admissibility space is revised and a generator then runs inside it; the revision and the production are one act.

This is the generation-side analogue of the Simultaneous Conversion Proposition (\S\ref{simultaneous-conversion-proposition}). There, transformation and distributional appropriation, two distinct conversion operations, fuse in one operation, so that governance designed for their sequential separation misclassifies the event. Here, stock-production and admissibility-revision, two distinct generative operations, fuse in one act, so that any account assuming a fixed admissibility space misclassifies it. The parallel is exact in form: in both cases two distinct operations are co-realized in a single act, and in both cases the analytical work is done by naming the fusion, not by adding a mechanism.

\textbf{Paradigm cases.} Recombination across incommensurable frames (\(G^{R}\) where the combined stocks carry conflicting assumptions, so combining them forces a revision of what is admissible); interpretation that resolves a contradiction (\(G^{J}\) adjudicating between partially incompatible frames); and a discovery that overturns a standing axiom (\(G^{D}\) whose content is the deletion of a previously admissible assumption). In each, the new stock and the revised admissibility space are the same product under two descriptions.

\textbf{What it is and is not.} The proposition adds no mechanism. The canonical set remains the six, and \(\Phi(\cdot)\) is untouched: its function class, its arguments, and the theorem system that runs on it are all unchanged. Fused generation is a \emph{mode} in which the existing generators can operate, not a seventh generator and not an operator outside \(\Phi(\cdot)\). It is the generation-side counterpart to a conversion-side proposition, and like that proposition it earns its place by removing a misclassification, not by enlarging the ontology.\footnote{Originality status: synthesized, using the book's own Simultaneous Conversion Proposition as the template. The claim is architectural, that admissibility-revision and stock-production can co-realize in one generative act, not a new generative primitive.}

\section{Open Questions}\label{open-questions}

\textbf{1. \(\Phi(\cdot)\) calibration.} The composition function is now weakly specified as a capability-gated and useful-diversity-gated aggregate. The remaining open task is empirical calibration of \(\chi_{\pi}\), \(w_k\), \(\beta\), \(\nu\), and \(\rho\), together with tests of whether the aggregate form outperforms mechanism-wise modelling.

\textbf{2. Recombination and AI training.} The boundary between \(G^{R}\) (generative recombination producing new capability) and KCM Cell 4 (transformation of existing encoded stock into a compressed representation) remains contested for AI model training. Working criterion: recombination has occurred when the resulting stock enables a productive task that neither input stock could enable alone. Operationally, test whether the trained system enables a task capability unavailable from any input stock alone. This criterion is not yet formally operationalized.

\textbf{3. \(G^{DIA}\) formalization.} The tension function \(\mathcal{T}_{DIA}(T,O,M,H)\) and the inverted-U specification require formal development before \(G^{DIA}\) can join the canonical composition function. Candidate empirical instances (paradigm shifts in science, Schumpeterian creative destruction at the knowledge level, productive methodological debates) can be documented while the formal specification remains open.

\textbf{4. Recombination field as measurable variable.} What empirical indicators best capture recombination field breadth \(D_{u}(F_{a,t})\)? Candidates: patent citation diversity across technology classes; academic co-authorship across disciplines; open-standard adoption rates; cross-sector licensing activity; AI training corpus heterogeneity. Development of this measure is a priority for Chapter 11 (Empirical Calibration, Falsification, and Measurement).

\textbf{5. Arena-specific parameterization.} A remaining empirical question is whether knowledge generation differs systematically across scientific, technological, organizational, platform, and professional arenas. The practical issue is whether the same KGM variables can be proxied across arenas, or whether each arena requires separate empirical parameterization.

\section{The Interface with the Knowledge Conversion Matrix}\label{the-interface-with-the-knowledge-conversion-matrix}

Newly generated knowledge-bearing stock enters the conversion cycle at the point where it receives an economically consequential pairing of residence and governance. In shorthand, first conversion is \(G \rightarrow K^{r,g}\): generated knowledge becomes stock located in some residential form and governed under some arrangement that determines access, maintenance, separation, transfer, exclusion, trust, and use.

\begin{figure}[H]
\caption[Knowledge Circulation: KGM, First Conversion, KCM, and Feedback]{Knowledge Circulation: KGM, First Conversion, KCM, and Feedback}
\label{fig:ch3:kgm-kcm-interface-loop}
\centering
\includegraphics[width=\textwidth]{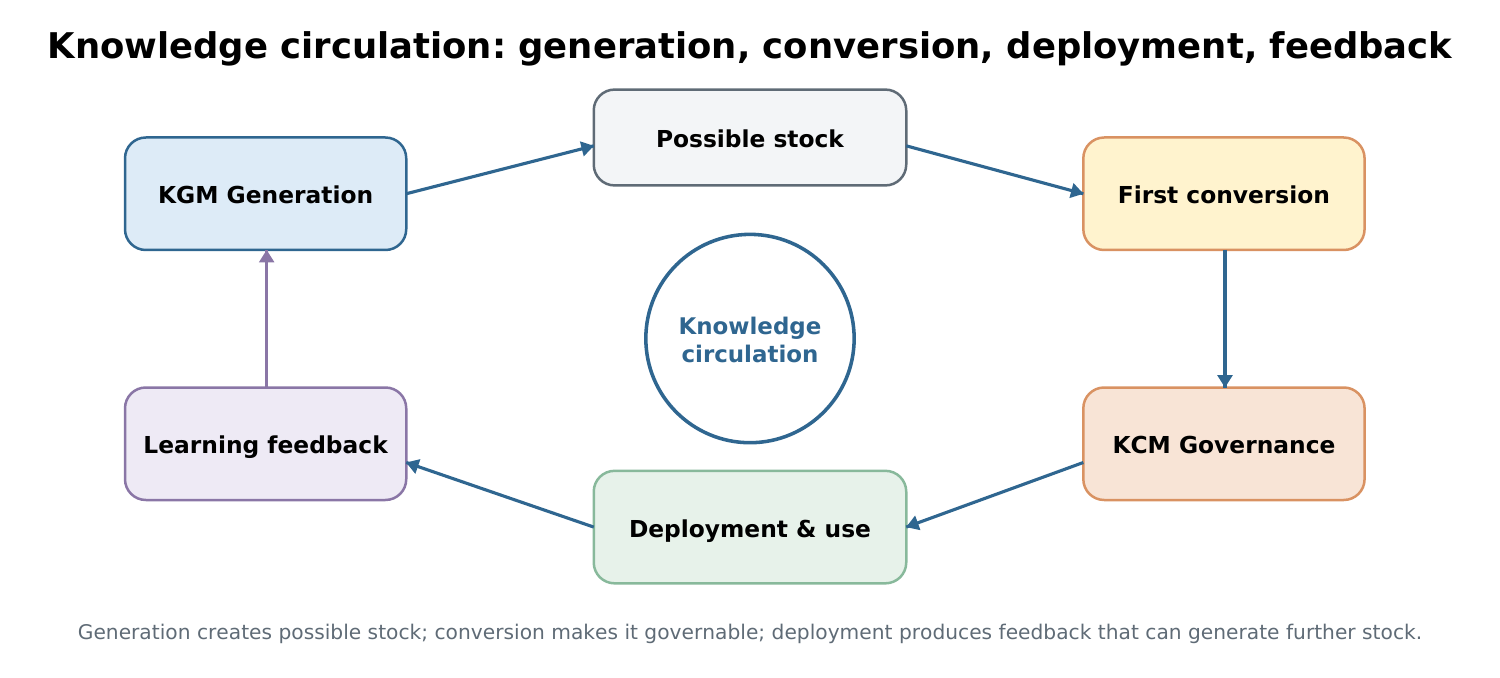}
\par\smallskip\noindent\footnotesize\emph{Note.} Generation creates possible stock; first conversion pairs that stock with residence and governance. Conversion then determines access, exclusion, yield capture, and future feedback. Deployment feeds back into later generation through learning loops.
\end{figure}

\textbf{The first conversion question} for any newly generated stock is therefore twofold: where does it reside, and who has an appropriability claim over it?

\begin{figure}[!htbp]
\caption[The first-conversion zone]{The first-conversion zone}
\label{fig:ch3:first-conversion-zone}
\centering
\includegraphics[width=0.95\textwidth]{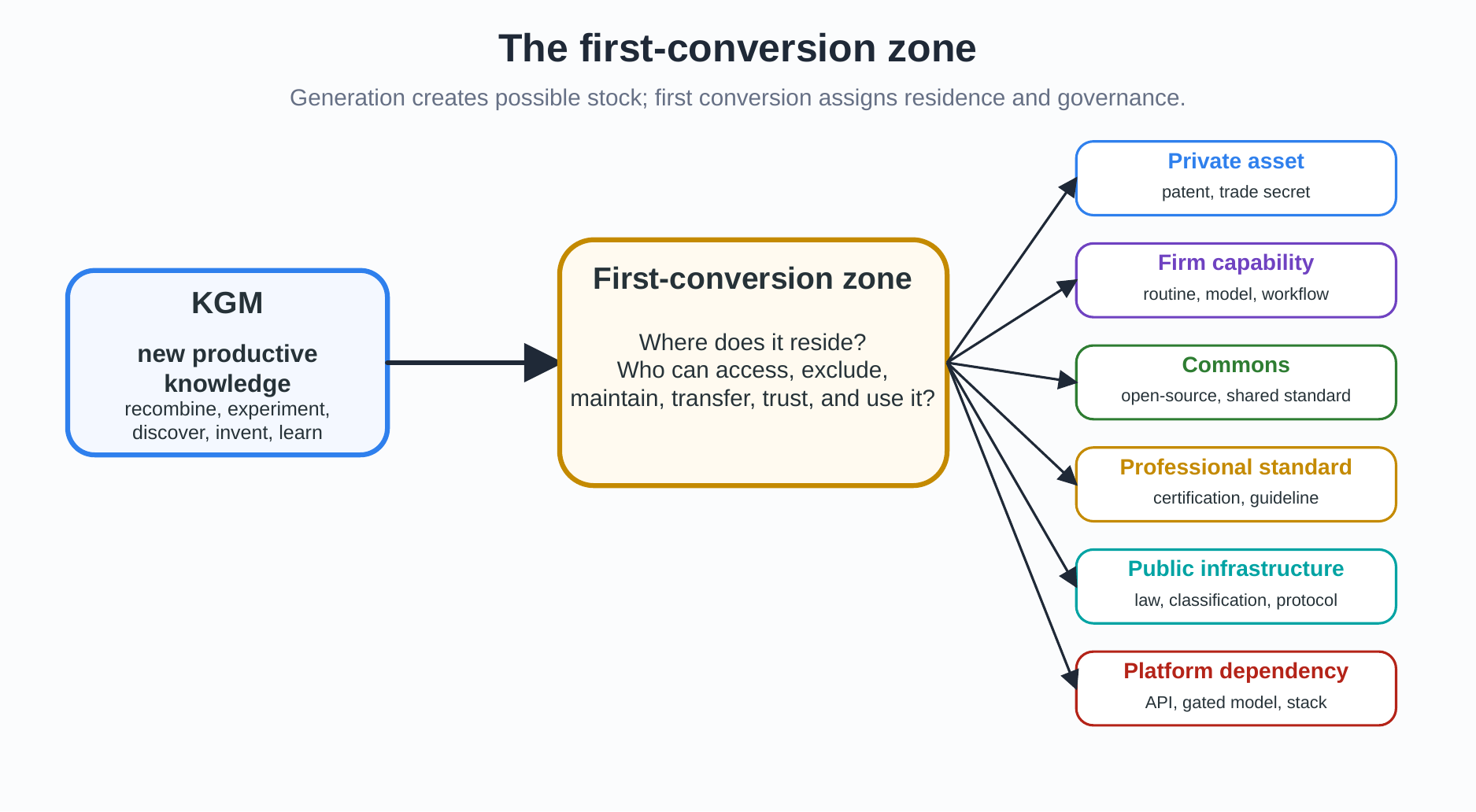}
\par\smallskip\noindent\footnotesize\emph{Note.} Generation creates possible knowledge-bearing stock, but first conversion assigns residence and governance. This is where newly generated stock becomes a private asset, firm capability, commons contribution, professional standard, public infrastructure, platform dependency, or another governed form.
\end{figure}

These questions are never automatically answered by the act of generation. They depend on who performed the generation work (individual, team, firm, public institution, open community); what governance form applies (employment contract, research funding terms, open-source licence, public commissioning, independent creation); and what residence the new stock takes (embodied, disembodied, institutionalized, or a governance-dominant commons/public-epistemic configuration).

The Conditional Separability Axiom applies to newly generated stock with the same force as to existing stock: generation does not automatically produce separability or appropriability. Separability is an institutional achievement, not a natural property of knowledge. A researcher who discovers a mechanism in a publicly funded university laboratory does not automatically create a private IP claim. The appropriability of the generated stock depends on the institutional conditions surrounding the generation event, and those conditions are what the KCM maps.

\textbf{Generation of public epistemic infrastructure.} Public epistemic infrastructure \(K^P\) is not generated by ``the public'' in the abstract. It is generated when scientific, professional, judicial, statistical, standards-based, or commons-based processes convert dispersed knowledge into stable public epistemic objects: standards, classifications, protocols, benchmarks, doctrines, evidentiary rules, measurement systems, and authoritative interpretations. This is one reason dialectical hypothesis generation matters. Courts, standards bodies\index{standards maintenance}, peer-review systems, public classification processes, and other public-epistemic institutions often generate public knowledge through structured opposition, review, objection, revision, and authoritative settlement. In Volume 1 this remains a bounded generation claim about \(K^P\), not a new formal mechanism; the optional notation for public-epistemic generation is reserved for the Technical Companion.

\textbf{The natural/artificial intelligence interface question.} When natural intelligence generates codified output (when a clinician's judgment becomes a protocol, when an engineer's diagnosis becomes a specification) who has the first appropriability claim? When artificial intelligence generates revised stock (when RLHF produces an improved model) who owns the improvement? These are not new questions added by the NI/AI vocabulary; they are the KGM--KCM interface questions restated in terms that connect directly to the contemporary knowledge economy.

\textbf{First-conversion zone.} The interface is not a single moment but a zone of first-conversion events, the decisions and processes by which newly generated stock is paired with residence and governance: externalized, codified, institutionalized, enclosed, shared, or made public. The KGM produces the stock; the KCM governs what happens to it next. Whether generated stock becomes knowledge-bearing \emph{capital}, the productive subset of stock that is deployed, maintained, and yielding value, is the KCM's question, not the KGM's: generation creates possible stock; first conversion pairs it with residence and governance; capability and deployment determine whether it enters productive use.

\chapter{The Knowledge Conversion Matrix}
\index{knowledge conversion|textbf}\index{Knowledge Conversion Matrix (KCM)|textbf}\index{KCM|see{Knowledge Conversion Matrix (KCM)}}
\label{ch:knowledge-conversion-matrix}

\chapterhook{How Knowledge Changes Form, Control, and Appropriability}

This chapter is the conversion itself. Where Chapter 1 named governance as the engine that turns non-rival use-value into appropriable exchange-value, the conversion matrix sets out the moves that engine can make, the routes by which the same productive knowledge is hired, bought, rented, trained, absorbed, or drawn from a commons, each route a different way of converting use-value into a different distribution of control, price, and future recombination.

AI training is the clearest contemporary case of knowledge conversion, but Chapter 4's argument is broader: expertise, routines, datasets, commons, platform behaviour, and institutional capability all become economically significant when they change form, control, and appropriability.

This chapter explains why the same productive knowledge can have different economic consequences depending on how it is converted. A diagnostic service can be hired as expert labour, bought as software, rented through an API, absorbed through acquisition, drawn from a commons, trained into a model, or embedded in an organizational routine. Each route changes ownership, bargaining power, dependency, valuation, maintenance burden, and future recombination possibilities.

The conversion problem is closest to transaction-cost economics\index{transaction-cost economics}, appropriability theory\index{appropriability theory}, absorptive-capacity theory\index{absorptive-capacity theory}, open innovation\index{open innovation}, and the knowledge-based theory of the firm\index{knowledge-based view} \parencite{Coase1937,Williamson1985,TeecePisanoShuen1997,CohenLevinthal1990,VonHippel2005,BaldwinVonHippel2011,Grant1996}\index{Teece, David}\index{Cohen and Levinthal}\index{von Hippel, Eric}\index{Baldwin and von Hippel}\index{Grant, Robert}. These literatures explain contracting, learning, capability, openness, and firm knowledge; the KCM adds the conversion moment where knowledge changes form, control, separability, and appropriability. The KCM extends these traditions by locating the institutional moment at which knowledge first becomes governable stock.

\section{Demand Routes}\label{sec:ch4:demand-routes}

The same productive service can be demanded through different conversion routes: hiring an expert, buying software, licensing an API, acquiring a firm, using a commons, training a model, or building an internal routine. These are not merely procurement choices. They determine whether the demanded knowledge service is embodied, disembodied, institutionalized, commons-based, platform-mediated, or converted into another form.

This means that demand and conversion are analytically inseparable. An actor demanding diagnosis, prediction, coordination, design, or decision support must also choose a route through which that service will be obtained. Hiring an expert primarily accesses embodied knowledge capital (\(K^{E}\)); buying software primarily accesses disembodied knowledge capital (\(K^{D}\)); acquiring a firm buys a bundle of embodied, disembodied, and institutionalized capability (\(K^{E}+K^{D}+K^{I}\)); using a commons draws on collectively maintained knowledge capital (\(K^{C}\)); and licensing an API or platform service often rents access to provider-controlled \(K^{D}\), \(K^{I}\), and feedback infrastructure without transferring the underlying stock. The conversion route therefore shapes not only cost and access, but also ownership, maintenance burden, dependency, appropriability, and future recombination options.

\begin{center}
\fbox{\begin{minipage}{0.92\textwidth}
\small
\textbf{Running case: API closure/access restriction.}\index{API closure!KCM running case} In the KCM, an API relationship is not merely a service contract. It is a conversion route through which one actor rents access to provider-governed \(K^D\), \(K^I\), data flows, documentation, and feedback infrastructure without receiving the underlying stock. An API closure is therefore a governance transition: the conversion pathway changes from accessible platform-mediated use to restricted, priced, or revoked access. The empirical question is whether downstream recombination, capability maintenance, or product formation changes after that transition.
\end{minipage}}
\end{center}

\section{Fixed-Point Sentence}\label{fixed-point-sentence}

\begin{quote}
The key question is not whether knowledge is enclosed or shared by nature, but under what institutional conditions knowledge becomes a private asset, a firm capability\index{firm capability}, a public good, a professional standard, a platform dependency, or a commons.
\end{quote}

\section{The Conditional Separability Axiom}\label{sec:ch4:conditional-separability-axiom}

In this chapter, conditional separability means separability from a person, firm, commons, platform, or institution under specific legal, technical, and capability conditions.

\begin{quote}
Productive knowledge becomes capital-like not merely when it is codified, but when it can be separated from its originating person, team, community, or institution under a supporting governance arrangement of law, infrastructure, standards, documentation, enforcement, and complementary capability. Cognitive enclosure (restriction of access, use, recombination, or learning from knowledge-bearing stock) often functions by sustaining this separability against spillover, imitation, decay, or reabsorption.
\end{quote}

Separability is an institutional achievement, not a natural property of knowledge. The same embodied knowledge (\(K^{E}\)) can remain non-separable in one governance context and become separable \(K^{D}\) or \(K^{I}\) in another through the same act (codification or contract\index{codification!governance conditions}) executed under different institutional conditions. This is why conversion analysis requires governance analysis: the conversion pathway and its appropriability outcome depend on the governance conditions surrounding the conversion event.\footnote{Originality status: synthesized. The predecessor materials include SECI, transaction-cost economics, appropriability theory, and KBV; Chapter 4's contribution is to make separability, governance form, and conversion pathway part of one KCM architecture, not to claim priority for conditionality or codification.}

\section{First-Conversion Zone}\label{first-conversion-zone}
\index{first conversion}\index{Knowledge Conversion Matrix (KCM)!first-conversion zone}

\textbf{Definition.} The institutional moment or interval in which newly generated knowledge-bearing stock first receives an economically consequential residence--governance pairing\index{residence--governance pairing}, \(G \rightarrow K^{r,g}\), and becomes subject to claims of ownership, access, enclosure, sharing, or public use.

\textbf{Why it matters.} Generation does not settle residence, governance, or ownership. A discovery, interpretation, invention, or recombination must pass through institutional conditions that determine where the resulting stock resides and whether it becomes a private asset, firm capability, public good, professional standard, platform dependency, or commons. The first-conversion zone is where the Conditional Separability Axiom first applies to new stock: the stock does not become capital-like by being generated, but by being paired with governance conditions that can make it separable, maintainable, transferable, trusted, and usable.

\textbf{Examples.} A researcher discovers a drug mechanism in a publicly funded laboratory. The discovery is new knowledge-bearing stock. Whether it enters the conversion cycle as a patent (\(K^{D}\), enclosed), a preprint (\(K^{P}\), public), a trade secret (\(K^{D}\), firm-enclosed), or public data (\(K^{C}\), commons) depends on the governance form at the moment of first conversion, the funding agreement, employment contract, institutional IP policy, and publication norms that apply. Two chemically identical discoveries can have materially different appropriability trajectories depending on the first-conversion conditions surrounding them.

\textbf{Relation to the KCM.} The matrix governs what happens once stock enters a conversion pathway. The first-conversion zone is the zone immediately upstream, where the residence--governance pairing is initially chosen or imposed. Governance interventions in the first-conversion zone (open-access mandates for publicly funded research, employee invention assignment rules, platform terms governing user-generated content) shape the entire subsequent conversion trajectory. The KGM generates stock; the KCM governs whether and how that stock becomes knowledge-bearing capital, the productive subset of stock that is deployed, maintained, and yielding value under a given governance form.

\section{Scope}\label{scope}

The Knowledge Conversion Matrix models the movement and governance of \textbf{existing} knowledge-bearing stock across the five forms of the taxonomy: embodied (\(K^{E}\)), disembodied (\(K^{D}\)), institutionalized (\(K^{I}\)), commons knowledge capital (\(K^{C}\)), and public epistemic capital (\(K^{P}\)). It asks: once productive knowledge exists in one of these forms, how does it move into another, become separable, become governed, become enclosed or diffused, and generate appropriability or valuation effects?

A KCM cell should not be read only as movement from one knowledge form to another. It also records a governance transition. The same disembodied stock may have different economic consequences under private IP governance, employment contract governance, platform terms, commons governance, or public epistemic governance. The KCM therefore tracks both residence transitions and governance transitions: what residential movement occurs, and what governance movement occurs?

\textbf{Intellectual lineage.} The KCM descends from the SECI model of \textcite{NonakaTakeuchi1995}, the socialization, externalization, combination, and internalization cycle through which tacit and explicit knowledge convert among forms within organizations. The KCM departs from SECI in three respects. First, it extends the conversion framework beyond organizational boundaries to include commons, platforms, and public epistemic capital as a conversion endpoint. Second, it treats access and distributional mechanisms as analytically distinct from transformation mechanisms, which SECI does not. Third, it foregrounds governance and appropriability at each conversion stage, rather than treating conversion as a primarily managerial or cognitive process.

\textbf{KCM/KGM boundary.} Echoing Chapter 3, newly generated knowledge-bearing stock enters conversion when it receives a residence--governance pairing and becomes subject to conversion mechanisms. The first conversion question is therefore where the generated stock resides and who has an appropriability claim over it. Generative mechanisms (recombination (\(G^{R}\)), experimentation (\(G^{X}\)), discovery (\(G^{D}\)), invention (\(G^{N}\)), judgment/interpretation (\(G^{J}\)), and learning feedback (\(G^{L}\))) are excluded from this chapter and treated in Chapter 3. They appear here only at the interface where newly generated stock enters the conversion cycle, and at Cell 4b where the learning feedback loop (\(G^{L}\)) is the direct recipient of conversion output.

\textbf{Residence--governance diagnostic.} In this notation, AI training can be read as \(K^{E/D,g_{\mathrm{author/user/commons/public}}} \rightarrow K^{D,g_{\mathrm{platform/private}}}\): embodied and disembodied contributor knowledge, held under authorial, user, commons, or public governance, is converted into privately or platform-governed model stock. Open-source platform capture can be read as \(K^{D,g_C} \rightarrow K^{D/I,g_{\mathrm{platform}}}\): the code may remain disembodied and formally open, but practical use and maintenance shift toward platform-governed service and institutional capability. A cybersecurity breach can be read as \(K^{D,g_{\mathrm{private}}} \rightarrow K^{D,g_{\mathrm{unauthorized\ shared\ control}}}\): the stock may remain in the victim's possession, but the governance position changes because unauthorized actors now share control, use, or recombination possibilities.

\begin{figure}[H]
\caption[Knowledge Conversion Matrix Anchor Cells and Pathways]{Knowledge Conversion Matrix Anchor Cells and Pathways}
\label{fig:ch4:kcm-cells-schematic}
\centering
\includegraphics[width=\textwidth]{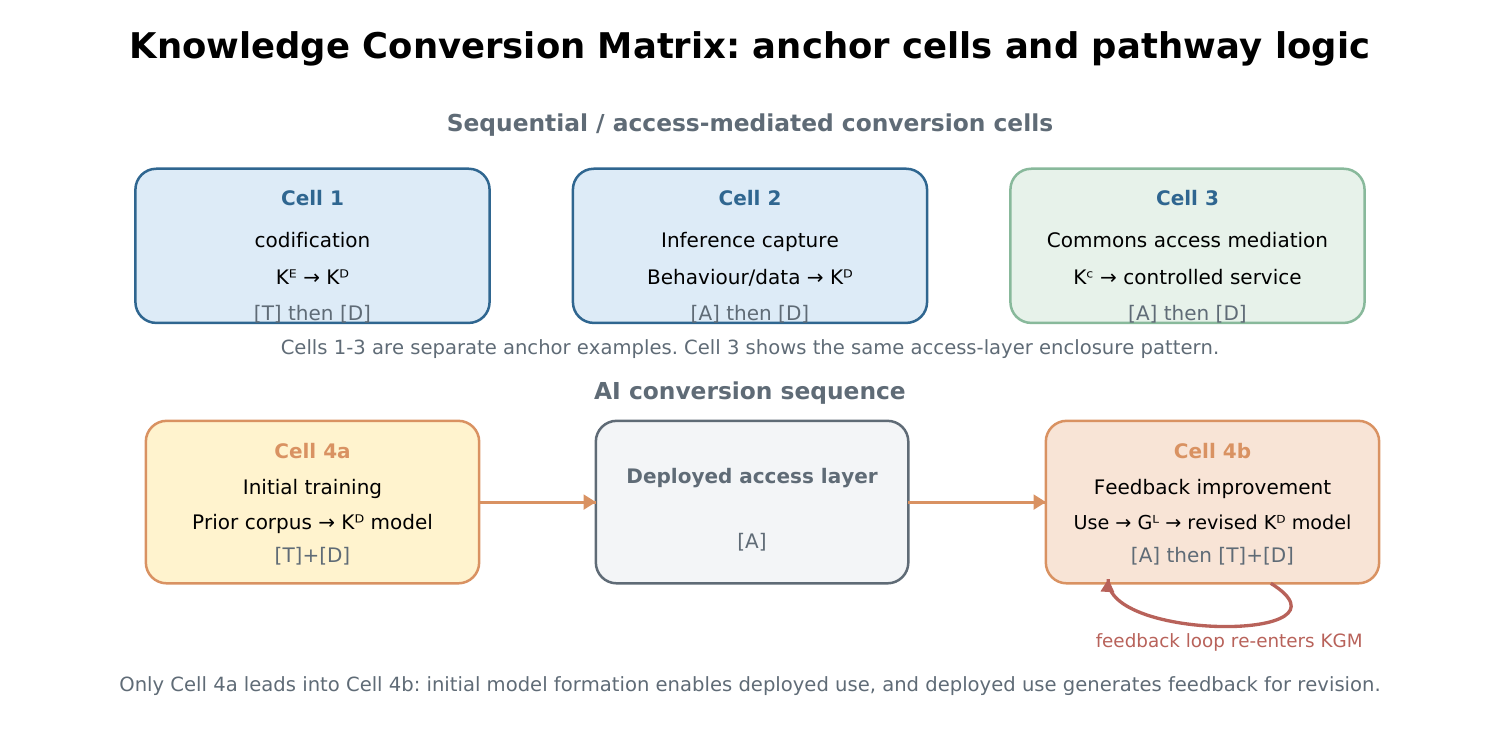}
\footnotesize\emph{Note.} The schematic compresses the anchor cells developed in the chapter. Cells 1--3 are separate anchor examples of sequential or access-mediated conversion. Cell 3 does not directly feed Cell 4b. Cell 4a leads to deployed access, and deployed access enables Cell 4b's feedback loop. Sequential pathways expose possible governance intervention points; simultaneous pathways collapse transformation and appropriation into the same operation.
\end{figure}

\section{Mechanism Taxonomy}\label{mechanism-taxonomy}

\begingroup
\small
\setlength{\tabcolsep}{3pt}
\renewcommand{\arraystretch}{1.12}
\sloppy
\par\addvspace{0.8\baselineskip}\noindent
\begin{longtable}{@{}L{0.20\textwidth}
L{0.24\textwidth}
L{0.31\textwidth}
L{0.19\textwidth}@{}}
\caption{Knowledge Conversion Mechanism Taxonomy}\label{tab:ch4:conversion-mechanism-taxonomy}\\
\toprule\noalign{}
\begin{minipage}[b]{\linewidth}\raggedright
Type
\end{minipage} & \begin{minipage}[b]{\linewidth}\raggedright
Symbol
\end{minipage} & \begin{minipage}[b]{\linewidth}\raggedright
Function
\end{minipage} & \begin{minipage}[b]{\linewidth}\raggedright
Mechanisms included
\end{minipage} \\
\midrule\noalign{}
\endfirsthead
\toprule\noalign{}
\begin{minipage}[b]{\linewidth}\raggedright
Type
\end{minipage} & \begin{minipage}[b]{\linewidth}\raggedright
Symbol
\end{minipage} & \begin{minipage}[b]{\linewidth}\raggedright
Function
\end{minipage} & \begin{minipage}[b]{\linewidth}\raggedright
Mechanisms included
\end{minipage} \\
\midrule\noalign{}
\endhead
\bottomrule\noalign{}
\endlastfoot
Transformation\index{Knowledge Conversion Matrix (KCM)!transformation mechanism} & \texttt{{[}T{]}} & Changes the form or residence of existing knowledge across \(K\)-forms & Externalization, codification, institutionalization, internalization \\
Access\index{Knowledge Conversion Matrix (KCM)!access mechanism} & \texttt{{[}A{]}} & Governs who can use or reach existing knowledge in its current \(K\)-form & Enclosure, diffusion, licensing, commons governance, platform access \\
Distributional\index{Knowledge Conversion Matrix (KCM)!distributional mechanism} & \texttt{{[}D{]}} & Determines who captures value from existing knowledge or its conversion & Extraction, rent capture, profit-sharing, attribution, bargaining \\
Generative & --- & Produces new knowledge-bearing stock & \textbf{Excluded. See Chapter 3 (Knowledge Generation Model).} \\
\end{longtable}
\endgroup

\section{Pathway Notation}\label{pathway-notation}

Sequential:\index{Knowledge Conversion Matrix (KCM)!sequential conversion} \texttt{{[}T{]}\ →\ {[}A{]}\ →\ {[}D{]}} Mechanisms occur in stages. Governance can intervene between steps.

Simultaneous:\index{Knowledge Conversion Matrix (KCM)!simultaneous conversion} \texttt{{[}T{]}\ +\ {[}D{]}} Transformation and appropriation occur in the same operation. No governance intervention point between transformation and appropriation. Reserve for cases where this absence is the analytical point.

\textbf{Governance intervention points.} The pathway notation makes governance opportunities explicit. In Cell 1, the sequential gap between transformation and distributional appropriation is where employment law, IP assignment negotiation, collective bargaining, and non-compete limits can operate. In Cell 2, the gap at the access layer is where data portability\index{data portability} requirements, terms-of-service constraints, privacy regulation, and competition law can alter the appropriability structure before inference value is captured. In Cell 3, the gap at the enclosure layer is where open-access mandates, contribution requirements, foundation governance, and licence design\index{licence design} can constrain rent capture. Cell 4a has no equivalent gap: the transformation and distributional appropriation occur in the same operation, which is why copyright law, designed around more sequential categories of copying, access, distribution, infringement, and remedy, can struggle with AI training as an economic conversion event. Cell 4b reintroduces a sequential structure but at the access layer, where \texttt{{[}A{]}} governs the conversion before feedback is captured.

\section{How to Read a KCM Cell: The Six-Layer Architecture}\label{six-layer-architecture}

\begingroup
\small
\setlength{\tabcolsep}{3pt}
\renewcommand{\arraystretch}{1.12}
\sloppy
\par\addvspace{0.8\baselineskip}\noindent
\begin{longtable}{@{}L{0.43\textwidth}
L{0.51\textwidth}@{}}
\caption{Six-Layer Architecture of the Knowledge Conversion Matrix}\label{tab:ch4:six-layer-kcm-architecture}\\
\toprule\noalign{}
\begin{minipage}[b]{\linewidth}\raggedright
Layer
\end{minipage} & \begin{minipage}[b]{\linewidth}\raggedright
Content
\end{minipage} \\
\midrule\noalign{}
\endfirsthead
\toprule\noalign{}
\begin{minipage}[b]{\linewidth}\raggedright
Layer
\end{minipage} & \begin{minipage}[b]{\linewidth}\raggedright
Content
\end{minipage} \\
\midrule\noalign{}
\endhead
\bottomrule\noalign{}
\endlastfoot
1. Knowledge movement & What \(K\)-form moves into what other \(K\)-form? \\
2. Conditional separability & Separable from whom or what, and under what institutional, technical, legal, documentary, contractual, or capability conditions? \\
3. Pathway notation & Sequential \texttt{{[}T{]}\ →\ {[}A{]}\ →\ {[}D{]}} or simultaneous \texttt{{[}T{]}\ +\ {[}D{]}} \\
4. Arena & Worker--firm / Platform ecosystem / Commons--incumbent / AI training / Nation/IP governance / Profession--institution \\
5. Focal anomaly & The one or two consequences existing frameworks least adequately explain \\
6. Testable proposition & What observable pattern would support or weaken the claim? \\
\end{longtable}
\endgroup

\section{Core Propositions A--E}\label{core-propositions-ad}

The propositions that follow are generated by the Six-Layer Architecture. Each proposition identifies a recurring conversion pattern by asking what knowledge moves, what makes it conditionally separable, which pathway governs the conversion, which arena is involved, what anomaly inherited frameworks under-specify, and what observable pattern would support or weaken the claim. Propositions C and D state the conditions under which enclosure suppresses generation; Proposition E states the conditions under which it enables generation. The three are not in competition: they are the negative and positive terms of a single net-effect criterion\index{C/D/E net-effect condition}\index{net-effect criterion}, resolved formally in Chapter~8.

\textbf{Proposition A, Codification as Reciprocal Bargaining Transformation\index{Proposition A!codification as reciprocal bargaining transformation}\index{codification!Proposition A}.} Codification changes the bargaining relation and yield allocation between a person or team whose knowledge is being codified and an organization. It may reduce worker leverage when embodied expertise becomes separable firm-controlled stock, but it may also raise the value of workers who retain scarce complementary expertise, integration capability, client trust, tacit judgement, reputation portability, or external market access.

Codification shifts bargaining power not because knowledge has merely been written down, but because an Operative Knowledge Unit has moved, been reproduced, or been placed under a new governance form. A firm that documents an employee's practice may still depend on that employee. A firm that converts the practice into a repeatable protocol, independently deployable model, or platform-mediated system has altered the bargaining structure. Conversely, a worker who uses codification to publish expertise, build portable reputation, sell tools, maintain client access, or occupy a higher\mbox{-}order integration role may preserve or increase their bargaining position.

\textbf{Boundary against the one-way transfer reading.} Proposition A does
not claim that codification always reduces worker leverage or always
transfers value to the firm. Codification can be substitutionary,
amplificatory, reputational, collaborative, institutionalizing, or
coercive. The bargaining result depends on outside options, residual
interpretive capability, compensation, reputation portability,
firm-specific context, and whether the codified unit creates new higher\mbox{-}order roles in maintenance, training, interpretation, governance, or
system stewardship.

\begingroup
\small
\setlength{\tabcolsep}{3pt}
\renewcommand{\arraystretch}{1.12}
\sloppy
\par\addvspace{0.8\baselineskip}\noindent
\begin{longtable}{@{}L{0.27\textwidth}
L{0.33\textwidth}
L{0.34\textwidth}@{}}
\caption{Codification Modes and OKU-Level Effects}\label{tab:ch4:codification-modes-oku-effects}\\
\toprule\noalign{}
\begin{minipage}[b]{\linewidth}\raggedright
Codification mode
\end{minipage} & \begin{minipage}[b]{\linewidth}\raggedright
OKU-level effect
\end{minipage} & \begin{minipage}[b]{\linewidth}\raggedright
Bargaining implication
\end{minipage} \\
\midrule\noalign{}
\endfirsthead
\toprule\noalign{}
\begin{minipage}[b]{\linewidth}\raggedright
Codification mode
\end{minipage} & \begin{minipage}[b]{\linewidth}\raggedright
OKU-level effect
\end{minipage} & \begin{minipage}[b]{\linewidth}\raggedright
Bargaining implication
\end{minipage} \\
\midrule\noalign{}
\endhead
\bottomrule\noalign{}
\endlastfoot
Documentary & Converts elements of \(K^{E}\) into \(K^{D}\) fragments, but does not create an independently operative \(OKU^{D}\). & Limited substitution effect. The worker remains the operative performer. \\
Amplificatory & Adds \(K^{D}\) support to an existing \(OKU^{E}\), increasing its productivity without replacing it. & May strengthen the worker, team, or firm depending on control over the augmentation layer. \\
Substitutionary & Converts task-specific \(K^{E}\) into a functionally equivalent \(OKU^{D}\). & Worker leverage falls where the firm no longer requires the person as the unique independently operative performer. \\
Institutionalizing & Converts embodied practice into \(OKU^{I}\): routines, protocols, role-systems, or governance arrangements. & Firm capability persists across personnel changes; worker-specific bargaining power may decline unless the worker captures new roles, reputation, compensation, or external market access. \\
Platformizing & Converts \(K^{E}\), \(K^{D}\), or \(K^{I}\) into platform-controlled operative capacity behind an access layer. & Platform captures governance position by controlling access, feedback, rules, and improvement conditions. Creates or deepens the platform-dependency governance (Chapter 5). \\
\end{longtable}
\endgroup

Documentary codification produces fragments. Amplificatory codification enhances an existing operative unit. Substitutionary, institutionalizing, and platformizing codification can produce new operative units. The distributional question is therefore not whether codification occurs, but which conversion mode governs it and who controls the resulting operative unit.

\textbf{The substitutionary condition} can be stated formally using the OKU framework:

\begin{quote}
\[
\mathrm{OKU}^{D}_{m,d,t} \equiv_{F,c,\tau,J} \mathrm{OKU}^{E}_{a,d,t} \quad \Longrightarrow \quad \text{substitutionary codification condition is met for } F
\]
\end{quote}

Here, \(F \subset d\). The task-class limitation is essential. An OKU is a task-defined unit of productive knowledge operation, not a universal unit of knowledge. Substitution does not occur at the level of the entire occupation unless all material task classes cross the equivalence threshold independently. When a bounded \(K^{D}\) system achieves functional equivalence with an embodied operative unit for a defined task class \emph{F}, the employer no longer requires that embodied actor as the unique independently operative performer for that task class. This does not imply full occupational replacement.

\textbf{Platformizing codification} is distinct from substitutionary codification. Substitutionary codification asks whether \(K^{D}\) \emph{can perform} a task formerly requiring \(K^{E}\). Platformizing codification asks \emph{who controls the conditions} under which the task can be performed, the access layer, improvement loop, data flow, permission structure, and governance arrangements. The platform may not merely replace labour; it may control the conditions under which both embodied and disembodied operative units can generate yield.

The practical implication is that codification can create joint surplus, bargaining transfer, or both, depending on whether the converted unit remains attached to the worker, becomes firm-controlled stock, or creates new complementary roles.\footnote{A compact diagnostic version is: worker cooperation is favoured when \((B^{codification})_{\mathrm{worker}}-(L^{leverage})_{\mathrm{worker}}>0\); firm codification is favoured when \((V^{codification})_{\mathrm{firm}}=\Delta (K^{D})_{\mathrm{firm}}+\Delta (K^{I})_{\mathrm{firm}}-C_{codification}-C_{trust,loss}-C_{maintenance}-C_{fidelity,decay}>0\). These conditions are diagnostic rather than a new formal proof in the main chapter; the notation belongs with the Technical Companion-level KCM and valuation apparatus.}

\textbf{Proposition B, Capability-Bounded Codification\index{Proposition B!capability-bounded codification}\index{capability-bounded codification}.} Codified knowledge is not self-activating. Disembodied knowledge capital (\(K^{D}\)) can create productive value only up to the limit of the embodied (\(K^{E}\)) and institutionalized (\(K^{I}\)) capability available to maintain, interpret, update, govern, and redeploy it. Rapid depreciation of \(K^{D}\) therefore increases the relative value of the \(K^{E}\) and \(K^{I}\) required for its regeneration. This is the static version of the dynamic principle developed in Interlude I. The asset did not disappear. The system that makes the asset valuable deteriorated. In this book, that system is called the \emph{capability system}\index{capability system}\index{context required to use knowledge-bearing stock}: the people, routines, infrastructure, permissions, trust, feedback, and complementary knowledge required to make a knowledge artefact useful. Formally, it is mainly carried by embodied and institutionalized knowledge stock (\(K^{E}\) and \(K^{I}\)). It bounds current \(K^{D}\) productivity and helps determine the direction of \(K^{D}\) value change over time.

\textbf{Proposition C, Generative Suppression\index{Proposition C!generative suppression}\index{generative suppression!Proposition C}.} \emph{(Stated here; formally developed with proof mechanisms in Chapter 6.)} Cognitive enclosure reduces not only the diffusion of existing knowledge-bearing stock but also the generation of new knowledge by narrowing recombination fields\index{cognitive enclosure} (reducing \(D_{u}(F_{a,t})\)), restricting experimentation on enclosed stock, weakening the learning loops that depend on widely deployed knowledge, and concentrating feedback signals within incumbent systems rather than distributing them across a broader ecosystem.

Proposition C operates through two structurally distinct effects:

\begingroup
\small
\setlength{\tabcolsep}{3pt}
\renewcommand{\arraystretch}{1.12}
\sloppy
\par\addvspace{0.8\baselineskip}\noindent
\begin{longtable}{@{}L{0.20\textwidth}
L{0.24\textwidth}
L{0.31\textwidth}
L{0.19\textwidth}@{}}
\caption{Proposition C: Value-Capture and Generation-Suppression Effects}\label{tab:ch4:core-propositions-a-d}\\
\toprule\noalign{}
\begin{minipage}[b]{\linewidth}\raggedright
Effect
\end{minipage} & \begin{minipage}[b]{\linewidth}\raggedright
Model location
\end{minipage} & \begin{minipage}[b]{\linewidth}\raggedright
Time horizon
\end{minipage} & \begin{minipage}[b]{\linewidth}\raggedright
Economic meaning
\end{minipage} \\
\midrule\noalign{}
\endfirsthead
\toprule\noalign{}
\begin{minipage}[b]{\linewidth}\raggedright
Effect
\end{minipage} & \begin{minipage}[b]{\linewidth}\raggedright
Model location
\end{minipage} & \begin{minipage}[b]{\linewidth}\raggedright
Time horizon
\end{minipage} & \begin{minipage}[b]{\linewidth}\raggedright
Economic meaning
\end{minipage} \\
\midrule\noalign{}
\endhead
\bottomrule\noalign{}
\endlastfoot
Value-capture effect & KCM & Short to medium term & Enclosure captures rents or appropriability from existing stock \\
Generation-suppression effect & KCM → KGM & Medium to long term & Enclosure narrows future recombination, experimentation, learning, or discovery possibilities \\
\end{longtable}
\endgroup

The value-capture effect is the conventional enclosure critique: the enclosing actor appropriates rents that would otherwise flow to contributors, users, or the commons. The generation-suppression effect is the stronger claim: enclosure does not only redistribute existing value but reduces the future stock of knowledge-bearing capital available to the economy as a whole. An enclosure event can produce value-capture without generation-suppression (a patent on a minor variation of existing stock). It can also produce both simultaneously (enclosure of foundational research tools, training corpora, or commons \(K^{C}\) that would otherwise seed independent recombination). Distinguishing the two effects matters for both theory and policy.

The mechanism: enclosure restricts access to existing stock → the recombination field narrows because fewer knowledge-bearing stock forms can be legally or practically combined → experimentation on enclosed stock is limited to the enclosing firm → learning loops that require deployment at scale are inaccessible to non-incumbents → future knowledge generation is concentrated in the same institutions that control present knowledge. The generation-suppression effect operates formally through reduction of \(D_{u}(F_{a,t})\): enclosure reduces field diversity, which depresses \(G^{R}\) output and, under the weakly specified \(\Phi(\cdot)\) aggregator, lowers the gated aggregate generation rate for excluded actors (Chapter 3). If the aggregate form fails empirical calibration, the KCM claim remains mechanism-wise: the affected generation channels must be tested separately rather than inferred through a single total generation rate.

\emph{Testable implications.} Sectors with stronger enclosure governance states should show lower cross-domain recombination, lower rates of independent experimentation, and weaker spillover-driven innovation than comparable sectors with wider access-governance arrangements, controlling for R\&D investment. Open-standard ecosystems should produce more diverse recombination pathways than closed proprietary ecosystems of comparable scale. Platform lock-in should reduce learning diversity because feedback is absorbed by the platform rather than distributed across the contributing ecosystem.\footnote{Originality status: synthesized / extended. The access-restriction and anticommons problems are established\index{anticommons blockage}; KBC's extension is to state them as generative-suppression mechanisms operating through recombination-field narrowing, experimentation restriction\index{experimentation restriction}, and learning-loop restriction.}

\textbf{Commons-Depletion Corollary.} Where non-rival commons (\(K^{C}\)) depend on embodied (\(K^{E}\)) or institutionalized (\(K^{I}\)) capability stock for maintenance, governance, and regeneration, incumbents can enclose the commons without owning the underlying knowledge-bearing stock by extracting or hiring away the capability stock required for commons self-governance. This depletes \(K^{C}\) knowledge impedance (\(K^{\mathrm{imp}}\), the capability system that keeps the commons productive) while increasing the value of the incumbent's managed access layer. Cognitive enclosure is therefore not only legal; it can be operational, infrastructural, and human-capability-based.

\subsection{Proposition D: Feedback-Enclosure}\index{Proposition D!feedback-enclosure}\index{feedback-enclosure!Proposition D}\label{sec:ch4:proposition-d-feedback-enclosure}\index{feedback capture}
\index{feedback enclosure}

When access to knowledge-bearing stock is enclosed, the enclosing actor captures not only rents from existing stock but also the feedback generated by deployment, thereby increasing its rate of knowledge revision relative to excluded competitors.

\textbf{Mechanism:} \(\text{Enclosure}\to\text{concentrated deployment}\to\text{concentrated feedback}\to G^L\to\text{improved stock}\to\text{stronger enclosure position}\)

Each cycle of improvement reinforces the incumbent's appropriability position, making subsequent enclosure easier to maintain and the underlying stock more valuable to extract rents from. The incumbent does not merely own today's knowledge stock; it captures a privileged improvement trajectory for that stock.

\textbf{Distinction from Proposition C.} Proposition C (Generative Suppression) describes what enclosure does to the generation capacity of \emph{third parties}: their recombination field narrows, their experimentation is restricted, and their learning loops are severed. Proposition D describes what enclosure does to the generation capacity of \emph{the enclosing actor itself}: it primarily captures the improvement feedback stream, accelerating its own stock revision. Both are effects of the same \texttt{{[}A{]}} access mechanism. Together they explain why self-reinforcing enclosure can be individually rational while being socially suppressive: the incumbent's accelerated improvement is purchased at the cost of suppressed generation system-wide.

\textbf{Joint prediction.} Strong enclosure produces fewer but faster-improving knowledge trajectories. It concentrates improvement while suppressing diversity. This is the central systemic cost of knowledge-bearing enclosure: not merely a redistribution of existing knowledge-bearing value, but a reorganization of the geography of future knowledge generation. This is also the clearest instance of the Smithian inversion in the framework: each enclosing actor pursues an individually rational strategy (capturing the improvement trajectory of its stock) that diverges from the system-wide generation optimum, under Chapter 8's welfare specification by suppressing the generation and diversity of knowledge available to the system as a whole. Formally, Volume 2 gives the audit trail for this governance-conditioned asymmetry: E{[}\(\Delta K_{i,t}\) \textbar{} \(\Gamma_{enc}\){]} \textgreater{} E{[}\(\Delta K_{j,t}\) \textbar{} \(\Gamma_{enc}\){]}, where Proposition D is the primary mechanism driving the incumbent term and Proposition C drives the excluded-actor term.

\textbf{Why existing vocabulary is insufficient.} Network effects captures user-side compounding value but not the institutional capture of improvement feedback from restricted access. Learning by doing captures improvement through use but does not describe how access governance concentrates that improvement to one actor. The Feedback-Enclosure Proposition connects these through the KCM's conversion architecture: enclosure is an access mechanism \texttt{{[}A{]}}, and its consequence is that the learning loop (\(G^{L}\)) running from deployment back to the KGM is primarily captured within the enclosing institution, unless portability, interoperability, licensing, audit, collective governance, or compensation mechanisms preserve an external claim.\footnote{Originality status: the feedback-learning advantage is established/extended (\textcite{HagiuWright2023}; \textcite{FarboodiVeldkamp2021}); feedback-enclosure as a conversion pathway is synthesized/extended; the integration of feedback capture with excluded-field suppression and trajectory narrowing is potentially novel as an architecture. See the originality audit, \S\ref{originality-audit-and-predecessor-boundaries}.}

\subsection{Proposition E: Appropriability-Enabled Generation}\index{Proposition E!appropriability-enabled generation}\index{appropriability-enabled generation}\label{sec:ch4:proposition-e-appropriability-enabled-generation}

Under specified conditions, enclosure \emph{increases} knowledge generation and maintenance rather than suppressing it. Restricting access to knowledge-bearing stock can raise the expected private return to costly generation (inventive \(G^{N}\) and experimental \(G^{X}\)), coordinate the complementary assets\index{complementary assets} required to deploy it, enable the quality control and standardization that make stock trustworthy enough to function as capital (the truth-dependence condition of Chapter~2), protect disclosure-sensitive experimentation whose value would collapse under premature diffusion, and fund the embodied (\(K^{E}\)) and institutionalized (\(K^{I}\)) capability that non-rival stock requires for maintenance against the depreciation pathways of Interlude~I. Proposition E is the appropriability counterpart to Propositions C and D: the same \texttt{{[}A{]}} access mechanism that narrows third-party fields (C) and concentrates feedback (D) can also be the condition under which some stock is generated, validated, or maintained at all. Proposition E does not weaken C or D; it supplies the offsetting term that C and D, taken alone, omit.

\begingroup
\small
\setlength{\tabcolsep}{3pt}
\renewcommand{\arraystretch}{1.12}
\sloppy
\par\addvspace{0.8\baselineskip}\noindent
\begin{longtable}{@{}L{0.24\textwidth}L{0.44\textwidth}L{0.26\textwidth}@{}}
\caption{Generative Channels of Enclosure (Proposition E)}\label{tab:ch4:proposition-e-channels}\\
\toprule\noalign{}
\begin{minipage}[b]{\linewidth}\raggedright Channel \end{minipage} & \begin{minipage}[b]{\linewidth}\raggedright Mechanism and lineage \end{minipage} & \begin{minipage}[b]{\linewidth}\raggedright When it dominates \end{minipage} \\
\midrule\noalign{}
\endfirsthead
\toprule\noalign{}
\begin{minipage}[b]{\linewidth}\raggedright Channel \end{minipage} & \begin{minipage}[b]{\linewidth}\raggedright Mechanism and lineage \end{minipage} & \begin{minipage}[b]{\linewidth}\raggedright When it dominates \end{minipage} \\
\midrule\noalign{}
\endhead
\bottomrule\noalign{}
\endlastfoot
Incentive / expected return & Appropriability raises the private return to costly invention (\(G^{N}\)) and experimentation (\(G^{X}\)); the \(B_{\mathrm{incentive}}(T)\) term of Chapter~8 \parencite{Arrow1962,Schumpeter1942}, and the standard Nordhaus optimal-patent analysis. & High fixed generation cost; the stock would not otherwise be produced. \\
Complementary-asset coordination & Exclusion aligns the \(K^{E}\), \(K^{I}\), and physical complements required to convert stock into yield \parencite{TeecePisanoShuen1997,Williamson1985}. & Deployment requires co-specialized assets that fragmented access would strand. \\
Quality control and standardization & Stewardship and curation sustain the truth-dependence condition of Chapter~2: validated, maintained stock is knowledge-bearing capital, whereas unmaintained stock is merely information-bearing. & Reliability is decisive and unvalidated diffusion would propagate epistemic risk. \\
Disclosure protection & Temporary exclusion protects experimentation and discovery whose value collapses once revealed (the Arrow information paradox; the trade-secret rationale). & The stock is disclosure-sensitive and imitation is cheap after disclosure. \\
Maintenance funding & Appropriable rent funds the rival \(K^{E}\)/\(K^{I}\) maintenance that non-rival stock still requires (Proposition B; Interlude~I), forestalling the commons-depletion decay pathway. & Maintenance capability is scarce and unfunded openness would let the stock decay. \\
\end{longtable}
\endgroup

\textbf{The net-effect welfare condition.} The welfare question is therefore not whether enclosure suppresses or generates, but the sign of its net marginal effect. Let \(\mathcal{G}^{+}(T)\) be the marginal generation-and-maintenance gain from extending the enclosure term \(T\) (the channels above, of which \(B_{\mathrm{incentive}}(T)\) is the leading component) and \(\mathcal{L}^{-}(T)\) the marginal recombination, feedback, and field-narrowing loss (Propositions C and D, carried by Chapter~8's cost terms):
\[
\mathcal{G}^{+}(T)=B_{\mathrm{incentive}}(T)+B_{\mathrm{coord}}(T)+B_{\mathrm{qual}}(T)+B_{\mathrm{disc}}(T)+B_{\mathrm{maint}}(T),
\]
\[
\mathcal{L}^{-}(T)=C_{T2}(T)+C_{T6}(T)+C_{T7}(T)+C_{T8}(T).
\]

\begin{figure}[!htbp]
\caption[C/D/E Net-Effect Condition Over the Enclosure Term]{C/D/E Net-Effect Condition Over the Enclosure Term}
\label{fig:ch4:cde-net-effect-enclosure}
\centering
\vspace{-0.25\baselineskip}
\includegraphics[width=0.92\textwidth]{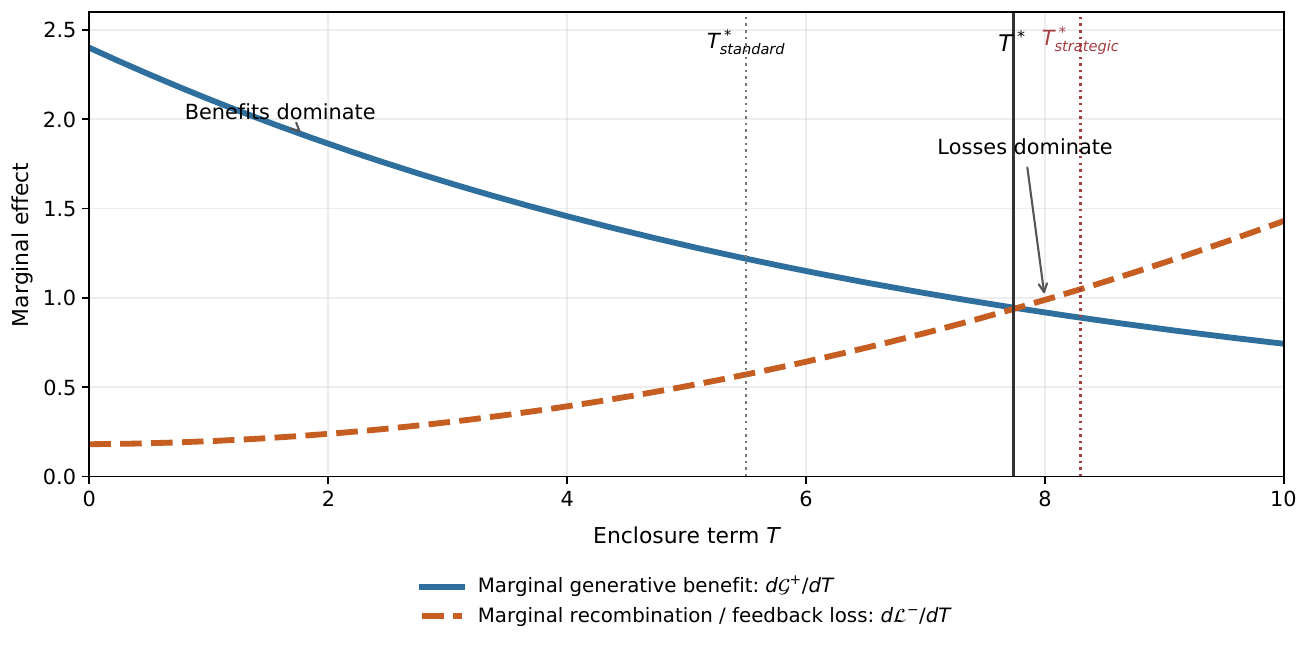}
\footnotesize\emph{Note.} The diagram visualizes Proposition E's correction to a one-sided enclosure story. The socially optimal enclosure term balances marginal generation-and-maintenance benefits against marginal recombination, feedback, and field-narrowing losses. Strategic over-enclosure arises only under the M5.T1 conditions, when private choice pushes the term beyond the social balance point.
\par\vspace{-0.45\baselineskip}
\end{figure}

Enclosure is generative-positive at the margin when \(\mathrm{d}\mathcal{G}^{+}/\mathrm{d}T>\mathrm{d}\mathcal{L}^{-}/\mathrm{d}T\) and suppressive when the inequality reverses; the socially optimal enclosure term \(T^{*}\) is the interior point where \(\mathrm{d}\mathcal{G}^{+}/\mathrm{d}T=\mathrm{d}\mathcal{L}^{-}/\mathrm{d}T\). Chapter~8's over-enclosure result (\(T^{*}_{\mathrm{strategic}}>T^{*}\)) is the claim that private optimization pushes \(T\) past that balance once the recombination multiplier \(M_{\mathrm{rec}}\) is positive, not the claim that \(\mathcal{L}^{-}\) always dominates \(\mathcal{G}^{+}\). Where \(\mathcal{G}^{+}\) dominates over the relevant range, high fixed generation cost, strong disclosure sensitivity, thin recombination demand, scarce maintenance capability, Smith's benign accumulation dynamic holds and enclosure raises net generation. Where \(\mathcal{L}^{-}\) dominates, foundational stock, high recombination intensity, commons-dependent maintenance, learning-loop concentration, the Smithian inversion of Chapter~12 holds. Stating the conclusion as the sign of an inequality between two formalized vectors is what keeps the welfare result earned rather than assumed: the costs of enclosure are not built into the premises, they are weighed against an equally explicit benefit side.

\emph{Testable implications.} Sectors with high fixed generation cost, strong disclosure sensitivity, and low cross-domain recombination intensity should show higher generation and maintenance under stronger appropriability (Proposition E dominates). Sectors built on foundational, recombination-intensive stock should show lower generation under stronger appropriability (Propositions C and D dominate). \emph{Weakened by:} evidence that generation and maintenance fail to rise with appropriability even where fixed costs and disclosure sensitivity are high, for example open-science or open-source ecosystems matching proprietary output in such sectors \parencite{BoldrinLevine2008}, which would push the net effect toward suppression across the board and collapse the case for the positive channels.\footnote{Originality status: established/extended. The incentive channel is Arrow--Nordhaus and Schumpeterian; complementary-asset coordination is Teece and Williamson; disclosure protection is the Arrow information paradox and trade-secret economics; the quality-control and maintenance channels draw on the professional-stewardship and commons-maintenance literatures (Chapter~5; Interlude~I). KBC's contribution is to disaggregate the benefit side symmetrically with the cost side and place both inside one net-effect condition with \(M_{\mathrm{rec}}\) and \(D_{u}(F_{a,t})\), so that the welfare conclusion of the enclosure chapters is derived rather than presupposed.}

\section{Simultaneous Conversion Proposition}\index{simultaneous conversion proposition}\label{simultaneous-conversion-proposition}

Where transformation and distributional appropriation occur in the same operation, governance forms designed for sequential copying, access, use, or distribution are likely to misclassify the event unless they are adapted to simultaneous conversion.

Sequential conversion creates defined governance intervention points\index{sequential conversion|textbf} between mechanism stages. Simultaneous conversion collapses those stages\index{simultaneous conversion|textbf} into a single operation, making the standard remedies of licensing, consent, and per-use compensation harder to map cleanly onto the conversion event at individual-work scale, without collective, ex ante, or statutory mechanisms.

\emph{Applied to AI training (Cell 4a).} Copyright law is designed around more sequential categories of creation, copying, access, distribution, infringement, and remedy. AI model training collapses transformation (statistical extraction from the corpus) and distributional appropriation (transfer of the resulting stock to the AI firm) into one technical process. The governance intervention point that copyright is designed to occupy does not exist in the same clean sequential form. This is a claim about where an economic governance-intervention point exists in the conversion architecture, not a prediction about how any particular jurisdiction will rule; the underlying legal questions remain actively contested.

\emph{Applied generally.} Conversion processes that simultaneously transform knowledge and transfer appropriability are likely to strain governance frameworks built for sequential copying, access, use, or distribution. This is not simply a lag; it is a structural mismatch between governance architecture and conversion architecture.\index{governance architecture versus conversion architecture}\index{regulatory lag versus structural mismatch}\footnote{Originality status: synthesized / potentially novel in the simultaneous-conversion architecture, especially as applied to AI model training. The claim is architectural and economic rather than a settled legal doctrine.}

A generation-side analogue holds in the KGM: just as transformation and distributional appropriation can fuse in one conversion operation, stock-production and admissibility-revision can co-realize in one generative act (the Fused Generation Proposition, \S\ref{fused-generation-proposition}).

\section{Inclusion Rule}\index{inclusion rule|textbf}\label{inclusion-rule}

A cell belongs in the matrix only if the combination of knowledge movement, conversion mechanism, agent arena, and governance form produces a \textbf{distinctive anomaly, valuation problem, governance problem, bargaining-power shift, or appropriability question that inherited vocabulary does not adequately isolate.} The matrix is selective, not exhaustive.

\section{Cell 1: Employee Expertise → Firm-Owned Stock}\index{embodied-to-disembodied conversion|textbf}\index{codification|textbf}\label{cell-1-employee-expertise-firm-owned-stock}

\textbf{\(K\)-formal expression:} \textgreater{} \(K^{E}_{worker}\) → \((K^{D})_{\mathrm{firm}}\) / \((K^{I})_{\mathrm{firm}}\) through {[}T: codification{]} → {[}D: IP assignment / employment contract{]}

\textbf{Smithian bridge.} Cell 1 is the modern conversion of Smithian skill. What Smith saw as specialized labour productivity appears here as embodied knowledge capital \(K^{E}\) being transformed into disembodied stock \(K^{D}\) or institutionalized capability \(K^{I}\). The conversion question is not merely whether the firm documents worker expertise\index{worker expertise to firm routine/software}, but who captures the resulting productive-service flow after embodiment becomes separable stock.

\textbf{Example.} A senior engineer's debugging practice may become code comments, scripts, incident runbooks, model-training examples, or automated remediation tools. The economic question is not whether the worker helped the firm, but how much of the future service flow remains attached to the worker and how much becomes separable firm stock.\footnote{Cell 1 originality status: extended / boundary condition to Becker and KBV. The contribution is not that worker skill, training, or firm knowledge matter, but that codification can move embodied capability across a separability and governance boundary, changing bargaining and yield allocation.}

\textbf{1. Knowledge movement.} Embodied expertise (\(K^{E}\): person-bound skill, tacit judgement, expert practice, contextual understanding) converts into disembodied firm-owned stock (\(K^{D}\): code, documentation, protocols, datasets, procedural artefacts, AI training material) and/or institutionalized firm capability (\(K^{I}\): embedded in organizational routines, process designs, and operational norms controlled by the firm). The worker's knowledge yield (\(K^{\mathrm{yield}}\)) on their \(K^{E}\) may fall if codification enables substitution; it may temporarily rise if the worker retains scarce complementary \(K^{E}\) that has not yet been codified.

\textbf{2. Separability conditions.} Person-bound knowledge (\(K^{E}\)) becomes separable from its originator through employment contract, IP assignment, work-for-hire doctrine, documentation mandates, and, increasingly, AI training on employee-generated output. Separability depends on legal enforcement and the existence of codification infrastructure. Knowledge that resists formalization (contextual judgement, adaptive expertise, relational understanding of a specific institutional environment) remains non-separable regardless of contractual intent. The appropriability boundary of the firm advances with each successful codification event and retreats where the codified \(K^{D}\) depreciates beyond the firm's capacity to maintain it without the originating worker.

\textbf{3. Pathway notation.} \texttt{{[}T:\ externalization/codification{]}\ →\ {[}D:\ extraction{]}} Sequential. A governance intervention point exists between transformation and distributional appropriation: the worker could in principle negotiate retention of rights, a royalty, or a participation claim over the resulting stock. In practice this intervention point is rarely occupied because employment terms are set before the specific knowledge is created and because the codification event is rarely named as such in the employment relation, it appears as documentation, training, or process improvement.

\textbf{4. Arena.} Worker--firm. Agents: employees, contractors, consultants, gig workers; firms and their legal and human-resource functions; professional bodies and unions where present.

\textbf{5. Focal anomaly.} The general/specific training distinction of \textcite{Becker1964} provides the most relevant human-capital predecessor for Cell 1. Becker holds that general training (skills productive across many employers) yields returns to workers in competitive markets, because firms cannot recapture training investment when trained workers leave for higher wages. Specific training (skills whose productivity advantage is concentrated in the training firm) produces split returns: both the worker and the firm hold a stake in the relation, because neither can fully realize the value without the other's continued participation. This framework holds most clearly when productive skill remains embodied in the worker and permissioned through the worker's continued participation; the worker's exclusivity over embodied skill is what enforces the split-return bargain. KBC's extension begins when AI-assisted codification converts observable patterns of judgment, correction, reformulation, and expertise into separable firm-controlled \(K^{D}\) or \(K^{I}\), changing the bargaining relation between the worker, the firm, and the codified residue of work. What was previously embodied capability (the patterns of expert inference that made a worker valuable) can become, through training-data capture, a firm-specific model that no longer requires the originating worker's participation. The worker's exclusivity over that productive pattern does not transfer with it; it dissolves at the first-conversion moment. Becker's framework is not wrong within the conditions it presupposes. KBC identifies the institutional boundary at which those conditions no longer hold.

Labour economics more broadly treats codification as a productivity mechanism and does not systematically describe the simultaneous change in the appropriability structure. The worker who documents their process does not merely improve firm efficiency, they convert their own \(K^{E}\) into firm-owned \(K^{D}\), altering the governance of the yield from that embodied stock. Proposition A (Codification as Reciprocal Bargaining Transformation)\index{codification!reciprocal bargaining transformation} formalizes this: codification is not only a knowledge-management event; it is a transformation of the bargaining relation, with distributional consequences that depend on codification mode. Cell 1 does not show that codification always transfers productive advantage from worker to firm. It shows that codification changes the form, residence, and governance of embodied knowledge. Depending on the mode, substitutionary, amplificatory, reputational, collaborative, institutionalizing, or coercive, the resulting \(K^{D}\) or \(K^{I}\) may substitute for the worker, amplify the worker, create reputational value, raise worker-firm joint surplus, institutionalize durable role-based capability, or coercively transfer capability without adequate compensation. The distributional question is not whether codification occurs but which mode governs the conversion and who captures the resulting knowledge yield (productive service flow from knowledge-bearing stock).

\textbf{6. Testable proposition.} \emph{H1 (substitutionary):} Workers in high-IP-assignment sectors with predominantly substitutionary codification governance arrangements should show lower wage growth than comparably skilled workers in low-IP-capture sectors, controlling for industry productivity. Firms that systematically codify worker expertise into directly substitutable \(K^{D}\) should show lower wage bills relative to revenue in subsequent periods, and faster substitution of codified roles with cheaper hires or automation. Workers whose tacit \(K^{E}\) successfully resists codification should command persistent wage premiums over peers whose output is more readily formalized.

\emph{H1-upside (amplificatory):} Where codification creates governance, training, architecture, or system-stewardship roles for the originating worker, codification is associated with wage premia, promotion probability, or retention incentives rather than wage compression. Workers in roles that require ongoing interpretation, maintenance, and contextual governance of codified systems should show stronger wage trajectories than workers whose roles were fully substituted.

\emph{H2 (joint-surplus):} Where codification is collaborative or amplificatory, firm productivity rises without corresponding evidence of worker displacement, wage compression, or role degradation. Firms with explicit knowledge-sharing and co-authorship frameworks in their codification programmes should show higher joint productivity gains than firms with extractive codification structures.

\emph{Weakened by:} evidence that codification raises wages proportionally through productivity gains in substitutionary governance arrangements; evidence that workers in high-documentation sectors show no systematic appropriability change; evidence that AI training on employee output does not reduce employment or wages in affected roles; evidence that amplificatory codification fails to generate the higher\mbox{-}order roles Proposition A predicts.

\section{Cell 2: Institutional Practice → Platform Data Asset}\label{cell-2-institutional-practice-platform-data-asset}

\textbf{Example.} A marketplace platform may observe seller pricing, consumer search patterns, refund rates, delivery failures, and conversion behaviour. Those observations can become platform inference stock, even when no individual seller experiences the conversion as a transfer of capital.\footnote{Cell 2 originality status: synthesized / extended. The platform, data, privacy, and information-economics literatures already analyse behavioural data capture and platform advantage; KBC's contribution is to treat the event as conversion of distributed institutional practice into platform-controlled inference stock.}

\textbf{\(K\)-formal expression:} \textgreater{} \(K^{I}_{distributed}\) / \(K^{C}_{behavioural}\) → \((K^{D})_{\mathrm{platform}}\) through {[}T: instrumentation/capture{]} → {[}A: ToS enclosure{]} → {[}D: inference value appropriation{]}

\textbf{1. Knowledge movement.} Distributed institutionalized knowledge (\(K^{I}\): firm operational practices, organizational routines, transaction patterns) and behavioural commons knowledge (\(K^{C}\): user behavioural sequences, market-condition signals, preference distributions, distributed knowledge about commercial decisions and market conditions) converts into platform-owned disembodied stock (\(K^{D}\): structured data assets and inference capacity). The platform's knowledge yield (\(K^{\mathrm{yield}}\)) on the resulting \(K^{D}\) is high because inference, predictive and targeting capacity, is productive at scale; the contributors' knowledge yield on their surrendered \(K^{I}\) and \(K^{C}\) does not directly increase unless compensation, attribution, licensing, access, or data-rights mechanisms intervene.

\textbf{2. Conditional separability.} Institutionalized knowledge (\(K^{I}\)) is embedded in firm practices and user behaviours and is not individually separable from its participants. Platform instrumentation converts distributed patterns into structured \(K^{D}\) via API architecture, sensor integration, and terms-of-service acceptance. The critical separability event is not the capture of individual records but the generation of inference (the platform's predictive and targeting capacity) which is separable from any individual contributor's identity and not addressed by data-protection frameworks designed for personal information. The appropriability boundary is drawn by platform terms of service\index{platform terms of service} rather than by contribution to value creation.

\textbf{3. Pathway notation.} \texttt{{[}T:\ operational/behavioural\ capture{]}\ →\ {[}A:\ platform\ enclosure\ under\ ToS{]}\ →\ {[}D:\ appropriation\ of\ inference\ value{]}} Sequential. A governance intervention point exists at \texttt{{[}A{]}}: mandatory data portability, licensing requirements, or participant data rights could alter the appropriability structure before inference value is fully captured. This is the point where regulatory intervention is architecturally feasible, which is why it is also the point most contested in data-economy policy debates.

\textbf{4. Arena.} Platform ecosystem. Agents: platform (controlling data architecture and inference capacity); firms providing operational activity data; users; API customers; advertisers; regulators (privacy, competition, data portability).

\textbf{5. Focal anomaly.} Property law more readily addresses ownership, control, or protection of personal data than the economic ownership of inference derived from institutional and behavioural patterns. The individual-contribution/aggregate-value problem is structural: no single firm's or user's contribution generates a viable property claim over \((K^{D})_{\mathrm{platform}}\), but the aggregate is the source of the platform's inference capability. Existing frameworks operate on the individual-contribution model and do not cleanly generate aggregate-contribution claims. The appropriability boundary is therefore drawn not by contribution but by the terms-of-service architecture the platform controls unilaterally.

\textbf{6. Testable proposition.} Platform firm market-to-book ratios should correlate with behavioural and operational data quality and exclusivity more strongly than with physical capital, in sectors where platform inference is a primary product, controlling for network effects, scale, infrastructure, brand, switching costs, ad-market position, and conventional monopoly power. Platforms that lose access to behavioural data (through privacy regulation, app tracking restrictions, or cookie deprecation) should show proportional declines in advertising revenue tracking the signal lost, after the same controls. Jurisdictions with mandatory data portability should show lower platform concentration relative to jurisdictions without, again controlling for network effects, switching costs, and scale economies. \emph{Weakened by:} evidence that platform value derives primarily from network effects and infrastructure rather than data, such that substituting different user populations produces equivalent inference capability; evidence that the service-for-data exchange is economically fair when full surplus is counted.

\section{Cell 3: Commons Code → Enclosed Commercial Asset}\index{open-source software!commons code}\label{cell-3-commons-code-enclosed-commercial-asset}

\textbf{Example.} An open-source project may remain legally open while its practical use shifts toward a managed cloud service controlled by an incumbent. The commons is not owned, but the access layer, maintenance layer, and user feedback channel may become privately mediated.\footnote{Cell 3 originality status: established-to-extended, with KBC-specific commons-depletion framing. Open-source capture, cloud-managed-service mediation, maintainer scarcity, and defensive licensing are established discourses; KBC's extension is to specify how access-layer control and capability extraction\index{capability extraction} can deplete the commons capability stock without changing the formal licence.}

\textbf{\(K\)-formal expression:} \textgreater{} Primary residence--governance transition: \(K^{D,g_C} \rightarrow K^{D/I,g_{\mathrm{platform}}}\) through {[}A: access-layer enclosure{]} → {[}D: knowledge rent capture{]}. Secondary capability transition: \((K^{E})_{\mathrm{maintainers}} \rightarrow (K^{I})_{\mathrm{firm}}\) through {[}T: capability extraction{]}.

\textbf{1. Knowledge movement.} Disembodied knowledge-bearing stock under commons governance (\(K^{D,g_C}\), conventionally summarized as \(K^{C}\): open licence; freely usable, modifiable, distributable) converts without change in legal ownership into platform-governed service and institutional capability (\(K^{D/I,g_{\mathrm{platform}}}\)) generating knowledge rent for an incumbent. The primary conversion is not of the code itself but of the actor-to-stock relation (the concept developed in Interlude I §3): from open access to access-layer-mediated access controlled by the incumbent. A secondary conversion moves embodied maintainer capability (\(K^{E}\)) into institutionalized firm capability (\(K^{I}\)): the incumbent hires core maintainers, absorbing the knowledge required for commons self-governance into firm-owned stock. The \(K^{C}\) knowledge impedance (\(K^{\mathrm{imp}}\): the distributed capability system that keeps the commons productive) is thereby depleted.

\textbf{2. Conditional separability.} Commons code (\(K^{C}\)) is legally non-separable from any single party; it belongs to no one and may be copied by anyone. Enclosure does not transfer legal ownership. Separability shifts operationally: the incumbent controls the complementary access layer (cloud hosting, managed services, enterprise support, proprietary extensions, distribution networks) through which the commons is practically accessed. De facto enclosure occurs without de jure ownership. The secondary capability-extraction mechanism operates through the hiring of key maintainers, transferring the \(K^{E}\) needed for commons governance into the firm, simultaneously giving the incumbent the capability to maintain the commons stock and depriving the community of the capacity for self-governance. This is Proposition B (Capability-Bounded Codification) weaponized: the incumbent extracts the capability stock the commons requires for its own regeneration, accelerating \(K^{C}\) depreciation as described in Interlude I.

\textbf{3. Pathway notation.} Primary: \texttt{{[}A:\ de\ facto\ enclosure\ via\ access-layer\ control{]}\ →\ {[}D:\ knowledge\ rent\ capture{]}} Sequential. Governance can intervene at \texttt{{[}A{]}} by requiring open access to managed services, mandating proportional contribution, or prohibiting anti-competitive extension strategies. Secondary: \texttt{{[}T:\ extraction\ of\ $K^E$\ maintainer\ capability\ →\ $K^I$\ institutionalization\ into\ firm{]}}\index{institutionalization}\index{commons depletion} This secondary pathway depletes the \(K^{C}\) knowledge impedance, accelerating the depreciation of the commons stock and strengthening the primary enclosure.

\textbf{4. Arena.} Commons--incumbent. Agents: open-source contributors (volunteers, foundation employees, corporate contributors); foundations and governance bodies; commercial incumbents (cloud providers, enterprise software firms); users (firms and developers dependent on open-source stacks); maintainers (often employed simultaneously by foundations and commercial firms).

\textbf{5. Focal anomaly.} Welfare economics explains the deadweight loss from monopoly but not why enclosure of a legally free and forkable commons persists when the underlying \(K^{C}\) can be copied at zero cost. Under simple free-entry assumptions, forkability should discipline rents; KBC adds why that discipline can fail even when the code is legally open. Access-layer control, enterprise support, integration cost, switching costs, reputation, network effects, scale, and maintainer absorption can weaken the fork threat without requiring IP ownership. The additional KBC mechanism is self-reinforcing: extraction of \(K^{E}\) can accelerate \(K^{C}\) depreciation, reducing the credibility of the fork threat that would otherwise discipline the incumbent.

\textbf{6. Testable proposition.} Cloud providers hosting major open-source databases should generate revenues comparable to the original vendors without proportional code contribution. Open-source projects should adopt commercial-restriction licences measurably after cloud incumbents begin offering managed services, not before. Commons communities that lose key maintainers (\(K^{E}\)) to commercial incumbents should show declining maintenance health (commit frequency, issue resolution time, security vulnerability rates) relative to communities that retain maintainer independence. \emph{Weakened by:} evidence that commercial enclosure increases adoption and commons contribution rather than depleting it; evidence that strong copyleft enforcement (GPL-governed projects) successfully resists cloud enclosure without licence changes; evidence that fork events reliably restore commons independence within economically significant timeframes.

\section{Cells 4a and 4b: Framing}\label{cells-4a-and-4b-framing}

The 4a/4b split separates two structurally different conversion events that share an arena but differ in temporal structure, source of contribution, and appropriability problem:

\begingroup
\scriptsize
\setlength{\tabcolsep}{3pt}
\renewcommand{\arraystretch}{1.12}
\sloppy
\par\addvspace{0.8\baselineskip}\noindent
\begin{longtable}{@{}L{0.14\textwidth}
L{0.18\textwidth}
L{0.27\textwidth}
L{0.17\textwidth}
L{0.18\textwidth}@{}}
\caption{Cells 4a and 4b: AI Training and Feedback Conversion}\label{tab:ch4:cells-4a-4b}\\
\toprule\noalign{}
\begin{minipage}[b]{\linewidth}\raggedright
Cell
\end{minipage} & \begin{minipage}[b]{\linewidth}\raggedright
\(K^x\) movement
\end{minipage} & \begin{minipage}[b]{\linewidth}\raggedright
Temporal structure
\end{minipage} & \begin{minipage}[b]{\linewidth}\raggedright
Source of contribution
\end{minipage} & \begin{minipage}[b]{\linewidth}\raggedright
Appropriability problem
\end{minipage} \\
\midrule\noalign{}
\endfirsthead
\toprule\noalign{}
\begin{minipage}[b]{\linewidth}\raggedright
Cell
\end{minipage} & \begin{minipage}[b]{\linewidth}\raggedright
\(K^x\) movement
\end{minipage} & \begin{minipage}[b]{\linewidth}\raggedright
Temporal structure
\end{minipage} & \begin{minipage}[b]{\linewidth}\raggedright
Source of contribution
\end{minipage} & \begin{minipage}[b]{\linewidth}\raggedright
Appropriability problem
\end{minipage} \\
\midrule\noalign{}
\endhead
\bottomrule\noalign{}
\endlastfoot
4a: Initial training & \(K^{D}\) + \(K^{I}\) + \(K^{C}\) + \(K^{P}\) → \(K^{D}_{\mathrm{model}}\) & Historical event & Pre-existing corpus created before the model & Who had a claim over knowledge used before the model existed? \\
4b: Feedback improvement & Natural intelligence feedback → revised \(K^{D}_{\mathrm{model}}\) via \(G^{L}\) & Continuous loop & Endogenous user interaction generated through ongoing use of the enclosed system & Who has a claim over improvement generated through ongoing use? \\
\end{longtable}
\endgroup

Cell 4b is structurally more enclosed than 4a: the contributor cannot separate use from contribution. The act of using the enclosed system is the act of generating the improvement signal. In 4a, contributors could in principle have withheld their stock before training occurred; in 4b, the extraction is inseparable from access.

\textbf{Natural and artificial intelligence frame.} Both cells are instances of a single deep conversion: natural intelligence (\(K^{E}\) people's expertise, \(K^{I}\) organizational knowledge, \(K^{C}\) community-maintained knowledge, \(K^{P}\) public knowledge) being converted into artificial intelligence (\(K^{D}_{\mathrm{model}}\): disembodied or platform-mediated knowledge-bearing capital encoded in model weights). Cell 4a performs this conversion from the historical corpus of human knowledge production. Cell 4b performs it continuously, from ongoing human interaction and correction. The distributional question in both cells is whether the value of this natural-to-artificial conversion accrues to the people and communities who generated the converted knowledge, or to the incumbent that controls the deployed system at first conversion.

\section{Cell 4a: Content/Cultural Knowledge → AI Model Weights (Initial Training)}\index{model weights!AI training}\index{AI training!model weights}\label{cell-4a-contentcultural-knowledge-ai-model-weights-initial-training}

\textbf{Example.} A corpus of writing, code, images, documentation, public research, and open material can be statistically encoded into model weights. The issue is not ordinary copying alone; it is whether transformation and appropriability transfer occur in the same operation.\footnote{Cell 4a originality status: synthesized / potentially novel in simultaneous-conversion architecture. The legal, copyright, data, and AI-training debates are established; KBC's contribution is to frame initial training as a conversion event in which transformation and appropriability transfer can collapse into one technical operation.}

\textbf{\(K\)-formal expression:} \textgreater{} Residence--governance transition: \(K^{E/D,g_{\mathrm{author/user/commons/public}}} \rightarrow K^{D,g_{\mathrm{platform/private}}}_{\mathrm{model}}\) through {[}T: statistical encoding{]} + {[}D: appropriability transfer to AI firm{]}. The older five-form shorthand is \(K^{D}+K^{I}+K^{C}+K^{P} \rightarrow K^{D}_{\mathrm{model}}\), but the residence--governance notation makes the governance movement explicit.

\textbf{1. Knowledge movement.} Multi-form knowledge-bearing stock, embodied authorial practice (\(K^{E}\): creative judgement, stylistic decision-making, disciplinary expertise); disembodied text, code, and images (\(K^{D}\): individually protected by copyright); institutionalized professional convention (\(K^{I}\): genre standards\index{standards!AI training}, editorial practices, disciplinary norms encoded in the corpus); commons knowledge (\(K^{C}\): open-licensed work, publicly available datasets\index{datasets!AI training}); and public epistemic capital (\(K^{P}\): scientific literature, government data, publicly funded research outputs), is consolidated simultaneously into a new class of platform/private-governed disembodied knowledge-bearing stock: trained model weights\index{model weights|textbf} (\(K^{D,g_{\mathrm{platform/private}}}_{\mathrm{model}}\)). The residence movement is toward disembodied model stock; the governance movement is from authorial, user, commons, or public governance contexts into platform/private governance. The conversion is simultaneous: {[}T{]} and {[}D{]} occur in the same operation with no governance intervention point between them. The knowledge yield (\(K^{\mathrm{yield}}\)) on the resulting \(K^{D}_{\mathrm{model}}\) flows primarily to the AI firm unless licensing, attribution, compensation, access, or collective-governance mechanisms preserve an external claim; the knowledge yield on the contributing stocks does not directly increase for contributors unless such mechanisms intervene.

\textbf{2. Conditional separability.} Each source \(K\)-form carries a different separability profile. Embodied authorial practice (\(K^{E}\)) is non-separable from individual creators. Disembodied text and code (\(K^{D}\)) are separable under copyright, but licensing for AI training is legally contested. Institutionalized professional convention (\(K^{I}\)) is distributed across communities with no clear rights-holder. Commons knowledge (\(K^{C}\)) and public epistemic capital (\(K^{P}\)) are non-exclusively accessible but their governance for AI training purposes is unresolved. The resulting model weights (\(K^{D}_{\mathrm{model}}\)) are separable from all contributors simultaneously. This separability is sustained by two institutional conditions: legal uncertainty about how training maps onto copyright categories under applicable law and technical opacity (individual contributions cannot be attributed or traced within the weight space, so no compensation mechanism can be applied retrospectively).

\textbf{3. Pathway notation.} \texttt{{[}T:\ statistical\ encoding\ of\ multi-form\ \(K^x\)\ corpus{]}\ +\ {[}D:\ appropriability\ transfer\ to\ AI\ firm{]}} Simultaneous. The transformation of source material and the appropriation of the resulting weights occur through the same operation. There is no clean sequential governance intervention point between them. \emph{Note on generative component.} AI training may also constitute recombination (\(G^{R}\)), producing capabilities not present in any source. This generative dimension is deferred to Chapter 3's interface discussion. The present cell concerns the conversion and appropriability structure of the process.

\textbf{4. Arena.} AI training. Agents: AI firms (training, deploying, commercializing); contributors (writers, artists, programmers, journalists, researchers, often without knowledge of inclusion); publishers and media organizations; open-source developers; API customers; courts and regulators determining legal classification of training.

\textbf{5. Focal anomaly.} The individual-contribution/aggregate-value gap. Copyright operates at the level of individual works (\(K^{D}\)). No single contributor's work is sufficient to generate a viable property claim, each work is a statistically negligible fraction of the training corpus. But the aggregate of \(K^x\) contributions is the source of \(K^{D}_{\mathrm{model}}\)'s productive capability. Existing IP frameworks do not provide a clean general mechanism for aggregate-contribution claims. Chapter 4 treats this gap as structural for economic analysis, while leaving legal resolution to courts, legislatures, and regulators. The simultaneous pathway removes the governance intervention point that sequential conversion would provide, making the standard legal remedies of licensing, consent, and per-use compensation harder to map onto the conversion event at individual-work scale, without collective, ex ante, or statutory mechanisms. The most significant knowledge conversion in contemporary capitalism proceeds under a governance framework designed for a materially different conversion type.

\textbf{6. Testable proposition.} AI firm valuations should correlate with training corpus quality and heterogeneity (across \(K^{D}\), \(K^{I}\), \(K^{C}\), \(K^{P}\) sources) more strongly than with equivalent computational investment alone, reflecting the underpriced knowledge-bearing contribution of training data. Models trained on curated, licensed, high-quality corpora should show performance advantages that cannot be explained by corpus size, suggesting knowledge quality (not only quantity) drives value. Jurisdictions that establish contributor licensing requirements for AI training should produce measurably different model development strategies: smaller models trained on licensed data, or models trained outside those jurisdictions. \emph{Weakened by:} evidence that model capabilities derive primarily from architecture and training methodology rather than corpus specificity, that substituting different corpora of comparable size produces equivalent productive capability; evidence that open models trained on public-domain corpora achieve performance comparable to models trained on proprietary content.

\section{Cell 4b: User Feedback → Improved AI Model Weights}\index{user feedback!KCM Cell 4b}\index{feedback capture!user feedback}\label{cell-4b-user-feedback-improved-ai-model-weights}

\textbf{Example.} In Cell 4b, use and contribution converge. Prompts, corrections, ratings, edits, rejections, completions, and expert feedback may improve the system that mediates access. The contributor often cannot use the system without producing some improvement signal.\footnote{Cell 4b originality status: extended from data-economics; potentially novel only when paired with excluded-field suppression and trajectory narrowing. Data-enabled learning and feedback advantage are established/extended; KBC's narrower claim is the integrated architecture linking feedback capture, KCM access governance, Proposition C's excluded-field suppression, and T8's narrowing of alternative trajectories.}

\textbf{\(K\)-formal expression:} \textgreater{} Natural intelligence feedback (\(K^{E}\) signals from use) → revised \(K^{D}_{\mathrm{model}}\) through {[}A: platform-mediated deployment{]} → {[}T: feedback capture{]} + {[}D: appropriability transfer{]} → \(G^{L}\) → revised \(K^{D}_{\mathrm{model}}\)

\textbf{1. Knowledge movement.} User interaction data (corrections, preference ratings, prompt-response pairs, RLHF signals, engagement patterns) generated through use of an enclosed AI system, converts into revised disembodied model weights\index{model weights!feedback revision} (revised \(K^{D}_{\mathrm{model}}\)). Unlike Cell 4a, where the source corpus pre-existed the platform and contributors could in principle withhold it, the feedback in 4b is inseparable from ordinary use: participation in the enclosed system \emph{is} contribution to its improvement.

In natural/artificial intelligence terms: users supply natural intelligence signals (their \(K^{E}\) expertise, judgment, correction, and interpretive capacity expressed through interaction) and those signals are absorbed by the platform's \(G^{L}\) feedback loop to improve artificial intelligence (\(K^{D}_{\mathrm{model}}\)). Cell 4b is therefore the paradigmatic natural-to-artificial intelligence conversion, and it is continuous, endogenous, and inseparable from access. Users do not merely consume \(K^{D}_{\mathrm{model}}\) outputs; through prompts, corrections, continuations, refusals, preferences, and task-specific use patterns, they supply \(K^{E}\) signals that the system converts into revised \(K^{D}_{\mathrm{model}}\) through \(G^{L}\). The contribution is frequently inseparable from use, which is why Cell 4b is more than data extraction: it is learning-loop capitalization of natural intelligence.

\textbf{2. Conditional separability.} Individual user interactions (\(K^{E}\) signals) are technically separable by platform logging from the person who generated them, the platform captures them automatically as a condition of use. The aggregate improvement signal generated by millions of interactions is inseparable from any individual contributor: no single interaction is attributable and none generates a viable claim. Separability of improved \(K^{D}_{\mathrm{model}}\) from contributing users is sustained by the same mechanisms as 4a, legal uncertainty about how sequential copyright categories apply to training and technical opacity, but applied to improvement rather than initial training. The terms of service that grant the platform rights to use interactions for model improvement are typically accepted before any specific interaction occurs, without disclosure of the improvement value that will be extracted.

\textbf{3. Pathway notation.} \texttt{[A]} \(\to\) \texttt{[T]+[D]} \(\to G^L \to K^{D}_{\mathrm{model}}\) The access mechanism \texttt{{[}A{]}} is the precondition: enclosure of the deployed model creates the conditions under which \(K^{E}\) feedback signals are generated exclusively for the incumbent. The feedback capture and appropriability transfer are simultaneous within the \(G^{L}\) cycle: the interaction and its incorporation into the improvement pipeline occur in the same governed relationship.

\emph{Note on KCM/KGM junction.} Cell 4b is the primary KCM/KGM junction: the conversion output (deployed \(K^{D}_{\mathrm{model}}\)) generates the KGM's \(G^{L}\) learning input, which produces revised stock that re-enters the KCM. The Feedback-Enclosure Proposition (Proposition D) applies directly: enclosure of \(K^{D}_{\mathrm{model}}\) at the access layer (\texttt{{[}A{]}}) primarily captures the \(G^{L}\) feedback loop for the incumbent, concentrating improvement while excluding competitors.

\textbf{4. Arena.} AI training (feedback phase). Agents: AI firms (capturing, processing, and incorporating feedback); users (generating \(K^{E}\) feedback through interaction, often without awareness of its improvement role); RLHF contractors and preference labelers; API customers whose deployments generate additional feedback signal; regulators beginning to address ongoing data capture.

\textbf{5. Focal anomaly.} The endogenous compounding problem. Unlike 4a, where the source knowledge was external and prior to the platform, in 4b the improvement signal is generated \emph{inside} the enclosed system by users who have no alternative path to the service. The individual-contribution/aggregate-value gap is structurally identical to 4a, but the extraction mechanism is more enclosed: users cannot participate in the system without contributing \(K^{E}\) signals to its improvement. Participation and contribution are identical; yet participation generates no claim.

The Feedback-Enclosure Proposition (Proposition D) operates with maximum force here: enclosure concentrates deployment, deployment concentrates \(K^{E}\) feedback, concentrated feedback accelerates \(G^{L}\) improvement of \(K^{D}_{\mathrm{model}}\), improved model increases user adoption, increased adoption deepens enclosure. The system can become increasingly self-reinforcing because ordinary use supplies many of the improvement signals that sustain the cycle.

\textbf{6. Testable proposition.} AI firms with larger deployed user bases should show faster \(K^{D}_{\mathrm{model}}\) improvement rates than firms with equivalent initial training investment but smaller deployment, controlling for compute, model size, architecture, research staff, post-training data, release cadence, and benchmark selection; the performance gap should widen over time rather than converge. Platforms subject to mandatory interoperability or data-sharing requirements should show slower improvement rates than closed equivalents, reflecting the redistribution of \(K^{E}\) feedback signal, after the same controls. Users who generate high-value corrections or refinements (power users, domain experts, RLHF labelers) should show no differential compensation relative to low-value users despite generating disproportionate improvement signal. \emph{Weakened by:} evidence that model improvement rates are primarily determined by researcher-directed fine-tuning rather than user feedback; evidence that deployment scale above a threshold does not further predict improvement rate.

\section{Summary Tables}\label{sec:ch4:summary-table}

\begingroup
\scriptsize
\setlength{\tabcolsep}{2pt}
\renewcommand{\arraystretch}{1.12}
\sloppy
\par\addvspace{0.8\baselineskip}\noindent
\begin{longtable}{@{}L{0.10\textwidth}L{0.18\textwidth}L{0.12\textwidth}L{0.19\textwidth}L{0.19\textwidth}L{0.16\textwidth}@{}}
\caption{KCM Cells 1--3}\label{tab:ch4:kcm-summary-a}\\
\toprule
Cell & Practical conversion & Pathway & Focal anomaly & Main economic consequence & What would weaken the claim \\
\midrule
\endfirsthead
\toprule
Cell & Practical conversion & Pathway & Focal anomaly & Main economic consequence & What would weaken the claim \\
\midrule
\endhead
\bottomrule
\endlastfoot
Cell 1 & Employee expertise becomes documentation, tools, routines, models, or firm-owned operating stock. & \texttt{[T]} \(\to\) \texttt{[D]} & Codification changes bargaining and yield allocation, not merely productivity. & Worker leverage may fall or rise depending on codification mode, complementary expertise, compensation, reputation, and external market access. & Codification consistently raises worker compensation, creates portable reputation, or leaves no measurable change in bargaining position. \\
Cell 2 & Platform participants' behaviour becomes platform inference stock. & \texttt{[T]} \(\to\) \texttt{[A]} \(\to\) \texttt{[D]} & Behavioural contribution becomes aggregate inference without ordinary sale or explicit capital transfer. & Platform control of access and terms can draw the appropriability boundary independently of contribution. & Data portability, interoperability, bargaining, or regulation preserves contributor claims or prevents inference capture from becoming durable advantage. \\
Cell 3 & Commons knowledge remains legally open while practical access shifts to an incumbent-controlled service or capability. & \texttt{[A]} \(\to\) \texttt{[D]} & Enclosure can occur without ownership by controlling the access, maintenance, service, or feedback layer. & Incumbents may capture yield while weakening the commons capability stock needed for regeneration. & Forking, foundation governance, open-access mandates, or distributed maintainer capacity prevents effective private mediation. \\
\end{longtable}
\endgroup

\begingroup
\scriptsize
\setlength{\tabcolsep}{2pt}
\renewcommand{\arraystretch}{1.12}
\sloppy
\par\addvspace{0.8\baselineskip}\noindent
\begin{longtable}{@{}L{0.10\textwidth}L{0.18\textwidth}L{0.12\textwidth}L{0.19\textwidth}L{0.19\textwidth}L{0.16\textwidth}@{}}
\caption{KCM Cells 4a--4b}\label{tab:ch4:kcm-summary-b}\\
\toprule
Cell & Practical conversion & Pathway & Focal anomaly & Main economic consequence & What would weaken the claim \\
\midrule
\endfirsthead
\toprule
Cell & Practical conversion & Pathway & Focal anomaly & Main economic consequence & What would weaken the claim \\
\midrule
\endhead
\bottomrule
\endlastfoot
Cell 4a & Prior human, cultural, technical, and public knowledge is encoded into model weights. & \texttt{[T]+[D]} & Transformation and appropriability transfer occur in the same technical operation. & Sequential governance categories can struggle to assign licensing, attribution, compensation, and control to the conversion event. & Model capability is explained mainly by architecture and compute rather than source knowledge, or ex ante licensing/governance preserves external claims. \\
Cell 4b & User interaction, correction, preference, and expert feedback improve deployed AI systems. & \texttt{[A]} \(\to\) \texttt{[T]+[D]} \(\to G^L\) & Use and contribution converge; participation generates improvement signal. & Enclosed deployment primarily captures the learning loop for the incumbent and may widen capability divergence over time. & Improvement rates are unrelated to deployment feedback\index{deployment feedback}, or interoperability, portability, audit, licensing, or compensation mechanisms preserve external claims. \\
\end{longtable}
\endgroup

\section{What This Lets Us See}\label{sec:ch4:what-this-lets-us-see}

The KCM shows that the economic significance of knowledge-bearing stock depends not only on what knowledge exists, but on how it is converted. The same knowledge can become worker skill, software, platform dependency, commons infrastructure, model weights, or institutional capability. Each conversion route changes who can access the stock, who must maintain it, who can recombine it, who captures its yield, and what future generation becomes possible. Chapter 5 now turns from conversion pathways to the governance forms that stabilize, distort, or fail to sustain those outcomes.

\chapter{The Governance of Knowledge-Bearing Stock}
\index{governance form}
\label{ch:governance-of-knowledge-bearing-stock}
\label{ch:knowledge-regimes}

\chapterhook{Who Gets to Decide What Knowledge Becomes}

If governance is the conversion mechanism, this chapter is its anatomy. Chapter 1 distinguished governance that encloses use-value into private exchange-value, governance that preserves it as a commons, and governance that conditions it through standards and trust; here those forms are set out as the regimes through which a society decides not merely who owns knowledge but what its knowledge is permitted to become.

The governance-form analysis draws on commons governance\index{commons theory}, intellectual-property economics\index{intellectual-property economics}, information economics, platform economics\index{platform economics}, and open-collaborative innovation\index{open-collaborative innovation} \parencite{Ostrom1990,Ostrom1999,HellerEisenberg1998,BessenMeurer2008,ShapiroVarian1999,RochetTirole2003,VonHippel2005}\index{Ostrom, Elinor}\index{Heller and Eisenberg}\index{Shapiro and Varian}\index{Rochet and Tirole}. Its purpose is not to choose one governance form as inherently superior, but to ask which governance form fits which knowledge stock at which stage of conversion. In this chapter, the six governance forms are private IP governance, firm-capability governance, platform-dependency governance, professional-stewardship governance, commons governance, and public epistemic governance. Governance remains a descriptive term: it may be competent, weak, fragmented, opaque, captured, underfunded, or failing.

This chapter asks who gets to govern knowledge after it is generated, and what that decision does to future productive capacity. The machinery matters only because governance allocation\index{governance allocation} changes access, exclusion, recombination, yield capture, maintenance, and future learning.

The governance test is not whether knowledge should be open or enclosed. The test is whether the assigned governance form fits the economic function of the knowledge stock. Table~\ref{tab:ch5:governance-fit-test} gives the practical diagnostic used throughout the chapter.

\begingroup
\small
\setlength{\tabcolsep}{4pt}
\renewcommand{\arraystretch}{1.15}
\par\addvspace{0.8\baselineskip}\noindent
\captionof{table}{Governance-Fit Test\index{governance-fit test|textbf}}\label{tab:ch5:governance-fit-test}
\begin{tabularx}{\textwidth}{@{}L{0.28\textwidth}X@{}}
\toprule
Governance-fit factor & Practical question \\
\midrule
Incentive need & Does exclusivity materially increase creation, validation, or investment? \\
Recombination importance & Does future generation depend on broad access? \\
Maintenance burden & Who keeps the stock updated, secure, documented, and usable? \\
Quality/security requirement & Which governance form best preserves reliability, privacy, accountability, and safety? \\
Renewal/depreciation risk & Will the stock decay if the governance form does not support renewal? \\
\bottomrule
\end{tabularx}
\par\addvspace{0.8\baselineskip}
\endgroup

The fixed-point question has been the book's organizing principle since Chapter 1: the key question is not whether knowledge is enclosed or shared by nature, but under what institutional conditions knowledge becomes a private asset, firm capability, public good, professional standard, platform dependency\index{platform dependency!governance form}, or commons. Chapters 2 through 4 built the analytical map required to address that question directly: the five-form \(K^x\) taxonomy (\(K^{E}\), \(K^{D}\), \(K^{I}\), \(K^{C}\), \(K^{P}\)), the Conditional Separability Axiom, the first-conversion zone, the Knowledge Generation Model, and the Knowledge Conversion Matrix with its pathway notation, six-layer cell architecture, and five anchor cells. This chapter uses that map to explain how institutional conditions determine what knowledge becomes.

The answer cannot be that open governance forms are always superior or that enclosed governance forms are always pathological. Enclosure can supply investment incentives, quality control, accountability, and incumbent learning speed. Openness can widen recombination fields, lower entry barriers, and preserve trajectory diversity. The governance question is therefore comparative and conditional: which governance form produces the strongest net knowledge-capital effect for this stock, at this stage, under these capability and governance conditions? Proposition~E (\S\ref{sec:ch4:proposition-e-appropriability-enabled-generation}) gives this comparison its formal form: the generative channels of enclosure (incentive, coordination, quality control, disclosure protection, and maintenance, the vector \(\mathcal{G}^{+}\)) are weighed against the recombination, feedback, and field-narrowing losses of Propositions~C and~D (the vector \(\mathcal{L}^{-}\)). Governance fit is the judgement of which side of \(\mathrm{d}\mathcal{G}^{+}/\mathrm{d}T \gtrless \mathrm{d}\mathcal{L}^{-}/\mathrm{d}T\) a given stock falls on, not a standing preference for openness or enclosure. This is a diagnostic comparison, not yet an empirically calibrated test.

The answer is not in the nature of the knowledge itself. The Conditional Separability Axiom established that separability is an institutional achievement, not an intrinsic property. The same drug mechanism can enter a private IP governance form or a public epistemic governance form depending on who funded its discovery, under what contractual terms, and through what disclosure pathway. The same software can be firm capability (\(K^{I}\)), a commons (\(K^{C}\)), or a platform dependency depending on its licence, its governance structure, and whether the platform that deploys it controls the access layer. The same interpretive judgment (\(K^{E}\) output) can generate records, codified protocols, data rights, trained artefacts, or work product assigned to an organization, while the originating worker retains embodied expertise unless contract, law, or institutional rules specify otherwise. Governance allocation is determined not by the character of the knowledge but by the institutional conditions that govern its generation, its first conversion, and its subsequent transitions through the conversion cycle.

Markets price expected returns under a governance form and influence incentives, bargaining power, and investment, but legal, contractual, organizational, and institutional rules formally allocate which governance arrangement governs the knowledge stock. The question of whether a given piece of knowledge is governed as private IP or as a commons is not settled by price signals alone. It is shaped by contract, funding source, employment status, publication timing, and the relative distribution of institutional authority, contractual leverage, and enforcement capacity among the parties present at the first-conversion zone. Governance allocation is a political economy problem: it is resolved through institutional design, contractual negotiation, legislative choice, and the strategic decisions of individuals, firms, platforms, and states.

This chapter proceeds as follows. Section 5.1 states the governance allocation problem in its full theoretical form. Section 5.2 analyses the six governance forms as institutional configurations with enabling conditions, \(K\)-form effects, and displacement vulnerabilities. Section 5.3 traces governance-form cycling\index{governance-form cycling} through four historical examples. Section 5.4 identifies the three primary determinants of first-conversion governance allocation. Section 5.5 analyses the structural asymmetry between open and enclosed governance forms. Section 5.6 connects governance form to recombination field breadth \(D_{u}(F_{a,t})\), the formal link between governance-form analysis and the Knowledge Generation Model. Chapter~\ref{cognitive-enclosure-and-generative-suppression} then examines the generative effects of enclosure, while Chapter~\ref{strategic-enclosure-and-the-smith-nash-problem} examines when individually rational governance choices become strategic over-enclosure.

\section{The Governance Allocation Problem}\label{the-governance-allocation-problem}\label{the-regime-allocation-problem}

The governance allocation problem arises from two properties of knowledge-bearing stock that previous chapters established and that in combination create a governance question market price signals cannot fully resolve.

New Institutional Economics shows that institutions and transaction costs shape whether exchange, contracting, and production are efficient \parencite{North1990,Williamson1985}\index{Williamson, Oliver}. KBC's governance-form conditioning is a related but distinct mechanism: for non-rival, form-shifting knowledge-bearing stock, governance does not merely determine whether exchange occurs or transaction costs fall. It conditions the productive yield of the stock itself by shaping who can access, interpret, recombine, update, and learn from it.

The first property is non-rivalry. One person's use of an idea does not diminish its availability to others. This means that any restriction on access (any governance form that excludes some parties from using knowledge-bearing stock) depends on institutional, legal, contractual, or technical conditions rather than on consumption of the underlying knowledge. The excludability that private IP, platform dependency, and firm capability governance forms maintain is institutionally created, not a direct consequence of the resource being depleted by use. A qualification is necessary. The scarcity of access to non-rival \(K^{D}\), \(K^{C}\), and \(K^{P}\) is institutionally produced; the complementary capabilities required to generate, validate, maintain, and deploy that stock (residing in \(K^{E}\) and \(K^{I}\)) remain materially and organizationally scarce. Enclosure analysis applies to access restriction; scarcity analysis of the capability system required to use, maintain, and improve the stock requires separate treatment.\index{capability system}

The second property is institutional dependence. The governance arrangement that governs a piece of knowledge-bearing stock is not determined by any natural property of the knowledge; it is determined by the institutional conditions in place at the moment of first conversion and at each subsequent governance transition. These conditions are the product of deliberate institutional choices (legislative, contractual, organizational) that could in principle have been made differently. The same \(K^{D}\) could have been governed under a different governance form had the institutional conditions been different.

Together, these two properties mean that the governance allocation decision is a genuine institutional choice with distributional consequences that exceed what market price signals can fully resolve on their own. Markets price and influence governance choices, but institutions formally allocate them. The price signal reflects expected returns under the current or anticipated governance form, not the social value of the knowledge under all alternative governance forms. A drug compound priced at monopoly rent under the private IP governance form generates a different price signal from the same compound priced at marginal cost under a compulsory licence\index{compulsory-access remedy}. Price therefore cannot by itself adjudicate between these governance forms because the price is partly a consequence of the governance form, not a neutral guide to which governance form should apply.

The Schumpeterian defence of the private IP governance form addresses part of this problem: without the prospect of supranormal returns from temporary monopoly, private investment in knowledge generation would fall below the socially optimal level. This is a genuine argument with genuine force for the specific case of goal-directed invention (\(G^{N}\)) where private investment is the primary generation mechanism and where the knowledge generated would not otherwise be produced. The KBC objection is not to appropriability as such. It is to governance mismatch\index{governance mismatch}: enclosure becomes inefficient when the protected stock is central to third-party recombination, when the enclosure term exceeds the incentive need, when the stock was generated from commons or public inputs without reciprocal maintenance, or when platform dependency converts temporary incentive protection into durable access control.

The governance allocation problem is therefore not a single problem but a family of problems that arise at different stages of the knowledge conversion cycle, under different generation mechanisms, and among different institutional parties. What is common to all instances of the problem is that the institutional conditions at the first-conversion zone determine the appropriability trajectory of the resulting stock for everything that follows, and those conditions are institutionally contingent rather than naturally fixed.

This yields Chapter 5's central diagnostic concept: \textbf{governance-fit failure\index{governance-fit failure}}. A governance-fit failure occurs when the property-rights\index{property rights}, access, incentive, coordination, quality-control, and renewal governance arrangement assigned to a knowledge-bearing stock does not match that stock's generative requirements. An overly open governance form may fail because it cannot fund maintenance, validation, or security. An overly enclosed governance form may fail because it suppresses recombination, blocks learning, or converts temporary incentive protection into durable access control. A fragmented IP governance form may fail because too many exclusion rights prevent assembly of the knowledge inputs required for downstream generation. The governance task is therefore not to rank openness above enclosure, or enclosure above openness, but to test fit.

The firm is one possible solution to this allocation problem, not the whole answer. Resource-based and knowledge-based theories correctly treat the firm as a site where heterogeneous resources and specialist knowledge are integrated. KBC accepts that firm-level insight but generalizes the allocation question: which knowledge-bearing stocks should be integrated inside firms, which should be protected by IP, which should remain in commons, which should be held as public epistemic capital, and which should be governed by professional stewardship or platform dependency? In that sense, firm integration is a subset of governance allocation.

\index{productive enclosure benefits}\index{dynamic enclosure harms}\index{enclosure!benefits and harms}\index{enclosure!comparative governance-fit test}
\begingroup
\small
\setlength{\tabcolsep}{4pt}
\renewcommand{\arraystretch}{1.12}
\begin{longtable}{@{}L{0.30\textwidth}L{0.31\textwidth}L{0.31\textwidth}@{}}
\caption{Enclosure Benefits and Dynamic Enclosure Harms}\label{tab:ch5:enclosure-benefits-harms}\\
\toprule
Governance question & Productive enclosure benefit & Dynamic enclosure harm \\
\midrule
\endfirsthead
\toprule
Governance question & Productive enclosure benefit & Dynamic enclosure harm \\
\midrule
\endhead
\bottomrule
\endlastfoot
Creation incentive & Temporary exclusion may fund discovery, disclosure, experimentation, and risky investment. & Excessive scope or duration can suppress follow-on discovery and reduce independent experimentation. \\
Quality and reliability & Controlled access may support validation, curation, documentation, security review, and standards compliance. & Control can become a gatekeeping device that blocks alternative validation, replication, or repair. \\
Maintenance and renewal & Proprietary revenue can finance updates, support, monitoring, and responsible stewardship. & Closure can sever distributed maintenance, contributor learning, and outside error discovery. \\
Security and misuse prevention & Restriction may reduce adversarial use, privacy leakage, unsafe deployment, or integrity loss. & Security rationales can be overextended into durable exclusion from legitimate recombination or audit. \\
Coordination and interoperability & Central governance can stabilize interfaces, versioning, compatibility, and user expectations. & Platform control can convert interoperability into dependency, lock-in, or unilateral access withdrawal. \\
Appropriability and investment capture & Enclosure can make investment recoverable when free-riding would otherwise prevent production. & Appropriability can become rent extraction when private gain exceeds the lost value of recombination, learning, and capability formation. \\
\end{longtable}
\endgroup

Table~\ref{tab:ch5:enclosure-benefits-harms} makes the governance-fit standard operational. The rule is comparative: exclusion can increase total productive capacity by funding creation, quality, security, coordination, or maintenance, but it can also suppress future generation by narrowing recombination fields, learning loops, challenger capability, commons maintenance, or dark-capital visibility.

\clearpage

\section{The Six Governance Forms: Enabling Conditions, Knowledge Effects, and Vulnerabilities}\label{the-six-governance-forms-enabling-conditions-knowledge-effects-and-vulnerabilities}\label{the-six-regimes-enabling-conditions-knowledge-effects-and-vulnerabilities}
\vspace{-1.3\baselineskip}

The six governance forms are institutional configurations, not taxonomic labels. Each governance form requires specific enabling conditions to be established and maintained; each has a distinctive effect on the form of the governed stock, the yield available to different actor classes, and the vulnerabilities through which the stock is displaced or degraded. The table below summarizes the configurations; the analysis that follows treats each in depth.

\begingroup
\scriptsize
\setlength{\tabcolsep}{3pt}
\renewcommand{\arraystretch}{1.12}
\sloppy
\par\addvspace{0.8\baselineskip}\noindent
\addtocounter{table}{-1}
\begin{longtable}{@{}L{0.14\textwidth}
L{0.18\textwidth}
L{0.27\textwidth}
L{0.17\textwidth}
L{0.18\textwidth}@{}}
\caption{Six Governance Forms of Knowledge-Bearing Stock: Effects and Vulnerabilities}\label{tab:ch5:six-knowledge-governance-forms}\\
\toprule\noalign{}
\begin{minipage}[b]{\linewidth}\raggedright
Governance form
\end{minipage} & \begin{minipage}[b]{\linewidth}\raggedright
Primary \(K\)-form governed
\end{minipage} & \begin{minipage}[b]{\linewidth}\raggedright
\(K\)-form of enclosing claim
\end{minipage} & \begin{minipage}[b]{\linewidth}\raggedright
Effect on \(K^{\mathrm{yield}}\) for non-incumbents
\end{minipage} & \begin{minipage}[b]{\linewidth}\raggedright
Primary displacement pathway
\end{minipage} \\
\midrule\noalign{}
\endfirsthead
\toprule\noalign{}
\begin{minipage}[b]{\linewidth}\raggedright
Governance form
\end{minipage} & \begin{minipage}[b]{\linewidth}\raggedright
Primary \(K\)-form governed
\end{minipage} & \begin{minipage}[b]{\linewidth}\raggedright
\(K\)-form of enclosing claim
\end{minipage} & \begin{minipage}[b]{\linewidth}\raggedright
Effect on \(K^{\mathrm{yield}}\) for non-incumbents
\end{minipage} & \begin{minipage}[b]{\linewidth}\raggedright
Primary displacement pathway
\end{minipage} \\
\midrule\noalign{}
\endhead
\bottomrule\noalign{}
\endlastfoot
Private IP governance & \(K^{D}\) (patents, copyright); \(K^{E}\) (trade secret adjacent) & \(K^{D}\) (exclusive rights) & Can reduce to zero within protected scope & Expiry; invalidation; compulsory licence; open-source substitution \\
Firm-capability governance & \(K^{E}\) (embodied expertise); \(K^{I}\) (organizational routines) & \(K^{I}\) (organizational embedding) & Inaccessible; practical bar not legal & Key-person departure; reverse engineering; acquisition failure \\
Platform-dependency governance & \(K^{I}\) (behavioural patterns); \(K^{C}\) (community knowledge) & \(K^{D}\) (inference and API control) & Can reduce to zero outside platform terms & Regulatory interoperability; data portability \\
Professional-stewardship governance & \(K^{E}\) (credentialed expertise) & \(K^{E}\) (accreditation gate) & Restricted in production; widened post-publication & Credentialism capture; AI disintermediation \\
Commons governance & \(K^{C}\) (open-licensed stock); \(K^{P}\) (public knowledge) & None (open access) & High when governance, quality, interoperability, and maintenance are strong & Free-rider accumulation; capability extraction; platform layer addition \\
Public epistemic governance & \(K^{P}\) (foundational knowledge); \(K^{C}\) (shared systems) & None (non-appropriable) & Foundational for broad recombination and validation; many governance forms depend on it to varying degrees & Defunding; Bayh-Dole enclosure; standards capture\index{standards governance!capture}\index{standards capture} \\
\end{longtable}
\noindent\emph{Note.} Private IP compresses distinct instruments; scope, duration, disclosure, exceptions, and independent creation differ by instrument.
\endgroup

\subsection{Private IP Governance}\index{IP policy!private IP governance}\index{private IP governance}\label{private-ip}

The private IP governance form requires three enabling conditions: a legal framework (patent, copyright, or trade secret) that creates exclusive rights over a defined scope of \(K^{D}\) use; enforcement capacity sufficient to sustain those rights against infringement; and a first-mover advantage at the first-conversion zone sufficient to establish the rights before the \(K^{D}\) has diffused beyond the reach of legal enclosure.\footnote{This discussion compresses patent, copyright, trade-secret, database-right, licensing, and related exclusion mechanisms. These instruments differ in disclosure requirements, duration, scope, exceptions, exhaustion, independent creation, and enforceability. Claims about private IP should therefore be read as governance form-level claims unless a specific legal instrument is named.} Without enforcement capacity, legal rights are nominal rather than effective. Without first-mover advantage at first conversion, the rights may be contested by prior art claims or by competitors who independently generated similar stock. Examples include mRNA vaccine patents, CRISPR licensing disputes, smartphone patent thickets, and standards-essential patents.

The private IP governance form's effect on the recombination field is proportional to the breadth and duration of its exclusivity. A narrow patent on a specific molecular compound restricts a small domain of potential combinations. A broad platform patent on a fundamental interaction paradigm, or a copyright over a foundational training corpus, may restrict entire classes of combination. The duration matters for the same reason: IP rights that persist for twenty years (patent) or the author's life plus seventy years (copyright) remove the enclosed \(K^{D}\) from recombination for periods that may exceed the productive life of the knowledge itself.

The private IP governance form is displaced by four mechanisms: expiry of the legal term, which moves the \(K^{D}\) into the public domain (\(K^{P}\)); legal invalidation through prior art challenges; compulsory licensing imposed by competition authorities or public health emergencies; and open-source displacement, when a commons-governed (\(K^{C}\)) functional equivalent reduces the private IP's market value to the point where exclusivity is no longer economically productive.

\subsection{Firm-Capability Governance}\label{firm-capability}

The firm capability governance form requires that productive \(K^{E}\) and \(K^{I}\) knowledge be embedded in the firm's organizational structure in ways that make it difficult to replicate from the outside. The embedding is achieved through employment contracts with IP assignment clauses, which vest disembodied outputs (\(K^{D}\)) in the firm; through organizational routines that distribute \(K^{I}\) across roles rather than persons, so that departure of individuals does not carry the entire capability; through secrecy norms and non-compete agreements that prevent \(K^{E}\) from flowing to competitors; and through complementary capability investments that make the firm's specific configuration of \(K^{E}\), \(K^{D}\), and \(K^{I}\) more productive than any component would be in isolation. Examples include AI lab teams, semiconductor design groups, quant-finance desks, and failed acquisition integration, where the acquired capability does not survive organizational transfer.

The firm capability governance form's effect on the recombination field is different from private IP: it narrows the field by practical inaccessibility rather than legal prohibition. The \(K^{I}\) that constitutes firm capability is not publicly disclosed (as patents require through the disclosure bargain) and is not legally excluded in a way that creates a clear scope of restriction. It is simply unavailable for recombination because no actor outside the firm has access to it, and the firm has no incentive to share it. This makes the field-narrowing effect of firm capability harder to measure and harder to govern than the field-narrowing effect of private IP, but not less real.

The firm capability governance form is displaced by key-person departure, the Cell 1 mechanism in reverse, where the \(K^{E}\) that the firm depends on exits through personnel turnover, returning to the embodied form and reappearing in a competitor or an independent venture. Acquisition can transfer firm capability but often fails to do so because the \(K^{I}\) organizational embedding, the routines, trust networks, and governance systems, does not transfer with the personnel.

\subsection{Platform-Dependency Governance}\label{platform-dependency}

The platform dependency governance form requires an API architecture that controls access to aggregated \(K^{D}\) and \(K^{I}\) capability that participants cannot readily replicate outside the platform; terms of service that capture the value of participant contributions without sharing it; and network effects that make the platform more valuable to each participant as more participants join, thereby raising switching costs. The governance form does not require ownership of the underlying \(K^{C}\) or \(K^{P}\) knowledge that participants generate or that the platform aggregates. It requires control of the access layer \texttt{{[}A{]}} that makes the aggregated knowledge reachable. Examples include Reddit API pricing, X/Twitter API restrictions, Apple App Store rules, and managed cloud services built around open-source tools.

The platform dependency governance form's effect on the recombination field is to control access to collectively generated knowledge that participants have produced but cannot use independently of the platform's infrastructure. The \((K^{D})_{\mathrm{platform}}\) asset (aggregated interaction data, inference capability derived from that data, the network of relationships among participants) is non-rival in the sense that one actor's use does not by itself consume the underlying knowledge stock. Its broader availability is nevertheless constrained by privacy, consent, security, abuse-prevention, latency, quality-control, and competitive-investment considerations. The platform dependency governance form converts this stock into a managed resource whose availability is conditioned on the platform operator's terms. The field is narrowed not by ordinary consumption of the knowledge but by the architecture and governance of its access layer.

The platform dependency governance form's principal displacement pathway is regulatory intervention imposing interoperability or data portability requirements. Its effectiveness depends on whether interoperability requirements\index{interoperability requirements} can be designed to include the inference capability embedded in the \(K^{D}_{\mathrm{model}}\) constructed from aggregated behavioural signals, rather than merely the raw \(K^{I}\) data layer that constitutes a less productive subset of the platform's knowledge stock.

\subsection{Professional-Stewardship Governance}\label{professional-stewardship}

The professional stewardship governance\index{professional stewardship governance} form requires accreditation systems that certify who has the \(K^{E}\) required to interpret and apply the governed stock appropriately; liability frameworks that create incentives for quality and disclosure; peer review systems that validate the content of the \(K^{D}\) or \(K^{C}\) stock; and ethical norms that govern attribution, conflicts of interest, and the duty to correct errors. The governance form does not restrict access to the knowledge in the same way as private IP or platform dependency; it restricts the authority to interpret and apply the knowledge to those who have demonstrated the \(K^{E}\) required to do so responsibly. Examples include medical licensing, accounting standards, engineering accreditation, peer-review norms, and professional disciplinary systems.

This governance form has a distinctive effect on the recombination field: it narrows who can authoritatively contribute to the knowledge stock while widening access to the stock once contributed. Scientific knowledge governed by peer review is restricted in its production (only methodologically competent researchers can contribute) but freely accessible once published (\(K^{P}\) or \(K^{C}\)). Medical knowledge governed by clinical guidelines is restricted in its application (only licensed practitioners can prescribe) but publicly available in its content. The governance form trades production restriction for quality assurance, which may widen the effective field by increasing the reliability of the \(K^{D}\) and \(K^{P}\) available for recombination.

The professional stewardship governance form is vulnerable to three displacement pathways: credentialism capture, where accreditation requirements become barriers to entry that protect incumbents rather than quality standards that protect the public; regulatory arbitrage; and AI-enabled disintermediation, where systems trained on professional \(K^{D}\) and \(K^{E}\) generate outputs that substitute for professional interpretation without engaging the accountability mechanisms of the governance form. A more adversarial reading must be stated plainly. Professional accreditation systems combine a genuine quality-assurance function with a restriction function whose effect is indistinguishable, in distributional terms, from a cartel. The governance form's legitimacy is a function of its governance, not its label.

\subsection{Commons Governance}\index{commons governance}\index{commons maintenance!governance form}\label{commons}\index{commons knowledge@commons knowledge (\ensuremath{K^C})!governance}\index{commons governance}

The commons governance requires an open licence that preserves shared access to \(K^{C}\) while governing use; governance institutions that manage contribution norms, conflict resolution, and quality standards; a sufficient community of \(K^{E}\) contributors to sustain the collective maintenance burden; and active protection from enclosure mechanisms, particularly the capability extraction pathway of Proposition B's commons-depletion corollary. The commons is not a residual or a default; it is an achieved institutional configuration that requires ongoing investment in \(K^{E}\) and \(K^{I}\) governance capacity to maintain. Examples include Linux, Log4j, OpenSSL/Heartbleed, Wikipedia governance, and OpenStreetMap.

The commons governance's effect on the recombination field is directly positive: a well-governed \(K^{C}\) of accessible, interoperable, trusted knowledge-bearing stock widens the field in proportion to its breadth and quality. The Linux kernel, the OpenStreetMap database, Wikipedia, and the public domain of expired IP (\(K^{P}\)) all represent forms of commons-governed stock that contribute to the recombination field. A poorly governed or depleted \(K^{C}\) narrows the effective field even though it remains formally open, because the quality and accessibility of the stock have deteriorated below the threshold at which productive recombination is reliably possible.

The commons governance faces three principal displacement pathways: free-rider accumulation\index{free-rider accumulation}; capability extraction by incumbents\index{commons depletion!capability extraction} (\(K^{E}\) maintainers hired away, depleting the commons' maintenance capability and thereby increasing the commons' knowledge impedance per Proposition B's corollary); and platform dependency layer addition\index{platform dependency!layer addition}, where a platform operator deploys \(K^{C}\) stock as the technical basis for a managed service, drawing contributor community and governance \(K^{I}\) into the platform's orbit.

\subsection{Public Epistemic Governance}\index{public epistemic governance}\index{public epistemic infrastructure!governance form}\label{public-infrastructure}\index{public epistemic infrastructure@public epistemic infrastructure (\ensuremath{K^P})!governance}\index{public epistemic infrastructure generation|textbf}

The public epistemic governance form requires funding and governance sufficient to prevent exclusive private appropriation from determining access; open publication or disclosure requirements that ensure the resulting knowledge enters \(K^{C}\) or \(K^{P}\); governance institutions that maintain quality, accessibility, and cumulativeness; and political consensus sufficient to sustain investment across electoral cycles. Examples include GPS, TCP/IP, HTTP, weather data, public genomic databases\index{datasets!public datasets}, and metrology standards\index{standards!metrology}.

Public epistemic infrastructure is not merely one governance-form option among others. It supplies standards, measurements, research bases, datasets, and institutions that make other governance forms productive. Private actors can build parts of this layer when complements are profitable, but they usually cannot capture the full social return from maintaining it. The public epistemic governance form exists because the recombination field requires shared foundations that are difficult to sustain through private appropriability alone.

Public infrastructure is also a source of \(K^P\), not only a preservation governance form.Standards bodies, courts, statistical agencies, metrology systems, public research institutions, and standards communities convert dispersed technical, legal, scientific, professional, and statistical knowledge into public epistemic objects that other actors can rely on. In Volume 1 this claim remains bounded: \(K^P\) is a core form, while specific public standards, doctrines, tests, classifications, or measurement rules are treated only as boundary-case operative objects when they perform a defined cross-actor function.

The public epistemic governance form is displaced by four mechanisms: defunding; enclosure of outputs (the Bayh-Dole dynamic \parencite{BayhDole1980}, publicly funded \(K^{P}\) converted to private IP); standards capture; and gradual substitution, when platform dependency slowly replaces \(K^{P}\) as the de facto access layer for foundational knowledge.

\subsection{Distributed-Yield Maintenance Gap}\label{distributed-yield-maintenance-gap}

Commons governance and public epistemic governance can be read as equilibrium-like governance arrangements, but not necessarily as optimal equilibria. They are settled patterns in which private actors, public institutions, and collective contributors allocate access, reliance, free-riding, contribution, maintenance, and under-maintenance. Sometimes the equilibrium is productive and stable. Sometimes it is underfunded, fragile, captured, or exploitative. The point is not that \(K^C\) and \(K^P\) sit outside private interest. Private actors often depend on shared or public knowledge stocks whose value exceeds what any one actor can fully appropriate.

The central asymmetry is therefore not simply underproduction. It is under-maintenance of already existing knowledge-bearing stock whose service yield is widely harvested while the maintenance burden is only partially internalized. For \(x\in\{C,P\}\), the stock law can be written schematically as
\[
\Delta K^x_t = G^x_t + M^x_t - \delta^x_t K^x_t.
\]
Here \(G^x_t\) is new commons or public-epistemic generation, \(M^x_t\) is maintenance, validation, updating, governance, stewardship, and security work, and \(\delta^x_tK^x_t\) is depreciation through under-maintenance, institutional erosion, contributor exhaustion, obsolete standards, degraded trust, or loss of interpretive capability.

The stock-maintenance equilibrium condition is
\[
G^x_t+M^x_t=\delta^x_tK^x_t,
\qquad x\in\{C,P\}.
\]
Sustainability requires at least the weak inequality
\[
G^x_t+M^x_t\geq \delta^x_tK^x_t.
\]
This is deliberately not stated as ``yield equals depreciation.'' Depreciation is stock erosion, while yield is a service flow. Yield matters only when some portion of it is converted back into replenishment, validation, governance, and maintenance.

Let \(Y^x_t\) denote the yield generated by commons or public epistemic stock and let \(s^x_t\) denote the share of that yield effectively recaptured for maintenance and renewal. A financing condition for sustainability is therefore
\[
s^x_tY^x_t\geq \delta^x_tK^x_t.
\]
\textbf{Maintenance-Gap Condition.} A non-rival knowledge stock is sustainable only while reinvested yield covers depreciation, \(s^x_t Y^x_t \geq \delta^x_t K^x_t\). Because non-rival use lets many actors harvest the same stock without any of them bearing its upkeep, this condition can fail silently while the stock still appears abundant. The gap between broad harvest and thin reinvestment is the structural source of commons and public-epistemic fragility; it underlies the dark-risk analysis of Chapter~9 and the commons-preservation instruments of Chapter~10.

Knowledge-bearing capitalism can nevertheless permit a distributed-yield maintenance gap\index{distributed-yield maintenance gap}:
\[
Y^x_t\gg M^x_t,
\qquad x\in\{C,P\},
\]
where the yield from commons or public epistemic stock is broadly harvested while maintenance reinvestment remains weak. This is why \(K^C\) and \(K^P\) can appear abundant while becoming fragile. Non-rival access allows many actors to harvest the same stock, but non-rivality does not eliminate depreciation. The stock remains productive only if enough of the distributed yield is converted back into maintenance, validation, governance, and renewal.

\section{Governance-Form Cycling}\label{governance-form-cycling}
\index{governance transition|textbf}

The governance form allocated to a piece of knowledge-bearing stock at first conversion is not permanent. A stock can and commonly does pass through multiple governance forms across the course of its conversion history. Each transition changes the governance conditions, the appropriability structure, the yield available to different actor classes, and the capability burden required to keep the stock productive. Understanding governance-form cycling is therefore as important as understanding the governance forms themselves.

Governance form cycling is not an exception to this theory. It is one of the normal dynamics of knowledge-bearing capitalism. A stock may begin as public epistemic capital, enter commons experimentation, become enclosed through firm-specific implementation, return as a public standard, and later be revalued by new tools or depreciated by lost capability. The simplified pathway
\[
K^P \rightarrow K^C \rightarrow K^D_{\mathrm{firm}} \rightarrow K^I_{\mathrm{platform}} \rightarrow K^C/K^P
\]
records governance form movement only. It does not by itself settle whether the movement is efficient, legitimate, or welfare-improving. That judgement depends on governance fit\index{governance fit}, materiality, and the effect on future generation.

\textbf{The pharmaceutical sequence} is the most thoroughly documented instance of governance-form cycling. Basic research funded by public institutions identifies the mechanism of a disease process (\(K^{P}\), public infrastructure). A compound that modulates that mechanism is discovered and patented, often by a university or biotechnology firm drawing on the publicly funded basic research (\(K^{D}\), private IP). The manufacturing process is developed and optimized by the patent-holder (\(K^{I}\), firm capability). Clinical trial data forms the basis for regulatory approval and becomes the evidentiary foundation for clinical practice guidelines (\(K^{D}\) governed by professional stewardship). After patent expiry, generic manufacturers produce the compound at marginal cost, and the clinical guidelines enter the medical \(K^{C}\) or \(K^{P}\) (commons and public infrastructure). This does not deny the private investment required for clinical trials, regulatory approval, scale-up, manufacturing, distribution, and pharmacovigilance. The full cycle may span thirty to fifty years and involve transitions through all six governance forms.

\textbf{The software sequence} illustrates the specific vulnerability of \(K^{C}\) to platform dependency displacement. A firm develops proprietary software under employment contracts (\(K^{I}\), firm capability). Facing competitive pressure or strategic interest in ecosystem development, it releases the software under an open-source licence (\(K^{C}\), commons). A platform operator deploys the open-source software as the technical foundation of a managed cloud service, adding proprietary APIs and complementary features (\((K^{D})_{\mathrm{platform}}\), platform dependency above \(K^{C}\) base). The nominal \(K^{C}\) persists but is effectively displaced by the platform dependency layer above it, as contributor \(K^{E}\) is absorbed into the platform ecosystem.

\textbf{The AI training sequence} is one of the most consequential current instances of governance-form cycling because it involves some of the largest commercially relevant \(K^x\) stocks and can produce especially durable enclosure. Open academic research, publicly funded datasets, and community-generated discourse (governed under \(K^{P}\) and \(K^{C}\) governance forms) enter the first-conversion zone as training inputs. The simultaneous conversion of Cell 4a produces model weights (\(K^{D}_{\mathrm{model}}\)) owned by the training organization (private IP) and deployed through a controlled API (platform dependency). Model value also depends on architecture, compute, curation, RLHF, evaluation, safety systems, infrastructure, and engineering routines. The feedback loop of Cell 4b deepens the enclosure by concentrating \(G^{L}\) improvement within the incumbent's system. The resulting governance form (private IP combined with platform dependency, compounding through \(G^{L}\) feedback) can produce especially durable enclosure, because it combines legal exclusivity with architectural access control and endogenous improvement.

\textbf{The standards sequence} shows how the professional stewardship governance form is vulnerable to private IP capture when a participating firm claims standards-essential patent rights over its technical contributions to a standard. The FRAND (fair, reasonable, and non-discriminatory) licensing obligation is intended to preserve the \(K^{P}\) or \(K^{C}\) character of the standard while compensating the patent-holder. In practice, disputes about what constitutes a fair and reasonable royalty have, in several cases, effectively converted a professional stewardship governance form into a private IP governance form with a nominal FRAND constraint.

\section{Three Primary Determinants of First-Conversion Governance Allocation}\label{three-primary-determinants-of-first-conversion-governance-allocation}

The first-conversion zone is where the governance form trajectory of newly generated \(K^x\) stock is initially determined. Chapter 4 showed how it operates in each of the five anchor cells. This section identifies the three primary institutional determinants of first-conversion governance allocation.

\subsection{Funding source at generation}\label{funding-source-at-generation}

The funding source at generation is one of the most powerful determinants of first-conversion governance form. Publicly funded research often enters the first-conversion zone with a public-use expectation, although statutory terms, grant terms, and contracts may permit private appropriation. Privately funded research often enters with an expectation of private control, although the allocation of rights depends on employment contracts, R\&D funding terms, disclosure obligations, and applicable law. Mixed funding creates contested appropriability structures that require explicit contractual determination of who holds which rights over what portions of the resulting \(K^x\) stock.

The funding source determination is made before the generation event: the institutional conditions of the funding (the grant terms, the employment contract, the research agreement) set the appropriability structure before the knowledge exists. This is why the first-conversion zone is a zone of first-conversion events rather than a single moment: the legal and contractual conditions that govern governance allocation are typically established upstream of the generation act itself.

\subsection{Employment and contractual status of the generator}\label{employment-and-contractual-status-of-the-generator}

An employee generating knowledge under a standard employment contract may assign IP rights to the employer through an IP assignment clause. Subject to jurisdiction, contract terms, invention-assignment law, moral rights, academic policies, and scope-of-employment rules, this can vest the resulting \(K^{D}\) or \(K^{I}\) in the firm capability or private IP governance form before the stock exists, as a condition of employment rather than as a negotiation over a specific piece of knowledge. An independent researcher, contractor, or open-source contributor working outside a standard employment relationship may retain personal ownership of the resulting stock unless explicit contractual assignment has occurred.

The employment contract's pre-assignment of IP rights is the primary mechanism by which Cell 1's pathway is established: {[}T{]} → {[}D{]} with the {[}D{]} element pre-structured by the contract rather than determined ex post. This is also why the distributional dynamics of Proposition A operate at the point of employment negotiation rather than at the point of codification: the worker who accepts a standard IP assignment clause has already assigned the distributional rights to the future \(K^{D}\) before it is generated. Whether the resulting codification is substitutionary, amplificatory, reputational, collaborative, institutionalizing, or coercive depends on how subsequent role design, compensation structures, and governance arrangements are specified, conditions that are often underspecified in the initial employment contract.

Professional status complicates the employment-status determination. A physician employed by a hospital may generate records, codified protocols, data rights, trained artefacts, or work product that the hospital controls under employment, data-governance, or institutional rules, while the physician retains embodied diagnostic expertise unless contract, law, or professional regulation specifies otherwise. Clinical guidelines that the physician contributes to as a member of a medical society are governed under a professional stewardship governance form that may be partially or wholly independent of the employment relationship. The same person may generate knowledge under two different governance form conditions depending on whether they are acting in their employment or their professional capacity.

\subsection{Timing and conditions of disclosure}\label{timing-and-conditions-of-disclosure}

The timing and conditions of the first public disclosure of newly generated \(K^x\) stock determine whether it can subsequently be enclosed under private IP or must remain in a more open governance form. Public disclosure before an IP claim is filed can function as prior art that limits later enclosure, depending on jurisdiction, timing, subject matter, and claim scope. Knowledge published in an open-access journal enters \(K^{P}\) or \(K^{C}\) immediately.

The disclosure timing decision is a genuine institutional choice with lasting consequences for governance allocation. Open-access mandates imposed by funding agencies (requiring that publicly funded research be published in open-access form within a specified period) are an attempt to shift the disclosure timing decision from the private interest of the researcher or institution to the public interest in open access. Their effectiveness depends on their coverage, their timing, and their enforcement.

\subsection{The knowledge-bearing firm and the theory of firm boundaries}\index{firm boundaries}\label{the-knowledge-bearing-firm-and-the-theory-of-firm-boundaries}

The three determinants above (funding source, employment status, and disclosure timing) jointly identify why the firm is the critical institutional location for first-conversion governance allocation in knowledge-bearing capitalism. They also bear on a question that the three determinants do not explicitly address: why do knowledge-intensive firms employ researchers rather than contracting them?

The transaction-cost\index{transaction-cost economics} tradition gives one answer \parencite{Coase1937,Williamson1985}\index{Coase, Ronald}: market contracting for knowledge-specific assets generates hold-up risk and measurement problems that hierarchical governance resolves more efficiently. \textcite{Wernerfelt2016}\index{Wernerfelt, Birger}'s adaptation and specialization theory gives a second: the firm is better than markets at governing adaptation under relationship-specific investment and at capturing the joint surplus that distinct productive specializations create when combined. Both rationales are correct within their scope.

KBC adds a third rationale that is specific to knowledge-bearing capitalism and that operates at the first-conversion zone rather than in the transaction or adaptation phase. Employment contracts can establish IP assignment that places newly generated \(K^{D}\) and \(K^{I}\) inside the firm at the moment of first conversion, before any market transaction over the resulting knowledge stock could occur and before the worker has leverage over its appropriability. Subject to jurisdiction, contract terms, invention-assignment law, moral rights, academic policies, and scope-of-employment rules, an independent contractor may retain personal ownership of the \(K^{D}\) they generate unless explicit assignment has been negotiated over a specific piece of knowledge, while an employee may already have assigned relevant rights as a condition of employment. The firm employs the researcher not only because market contracting of specific assets is inefficient or because specialization economies require hierarchical governance, but because employment is the institutional mechanism by which newly generated knowledge-bearing stock is assigned to the firm-capability or private-IP governance form at first conversion, before its productive value can be separately priced.

This third rationale (the \emph{appropriability rationale for employment}) operates alongside and often upstream of transaction-cost and specialization rationales: it helps determine who holds the right to assign newly generated knowledge to a governance form, and therefore who influences the conditions under which the other two rationales apply. It also identifies an institutional asymmetry that transaction-cost and specialization theories can understate: the firm's position at the first-conversion zone is not merely an efficiency advantage; it is a governance advantage, the ability to influence the initial governance form assignment of knowledge whose productive value is not yet known and whose governance trajectory it may be difficult to reverse.

\section{The Structural Asymmetry Between Open and Enclosed Governance Forms}\label{the-structural-asymmetry-between-open-and-enclosed-governance-forms}

One of this theory's most important observations is the structural asymmetry between the ease of moving from open to enclosed governance forms and the difficulty of moving in the reverse direction. Knowledge-bearing capitalism can drift toward enclosure in commercially valuable fields where private enclosure returns are concentrated, open-governance-form maintenance is underfunded, and counter-institutional investment is weak. The relevant evidence would include licence changes, API closures, standards capture, patent thickets, proprietary forks, managed-service capture of open-source projects, and reduced public or commons maintenance\index{commons maintenance!underfunding}\index{public epistemic infrastructure!underinvestment}. The intellectual lineage of this observation runs through the distinction of \textcite{DasguptaDavid1994}\index{David, Paul}\index{Dasgupta and David} between the open ``republic of science\index{republic of science}\index{science policy}'' (governed by Mertonian norms\index{Mertonian norms} of communalism, universalism, and priority disclosure) and the proprietary science system governed by commercial incentives and confidentiality. Dasgupta and David identified the institutional conditions that sustain each equilibrium and the pressures that drive movement between them. The present analysis extends their framework to include platform dependency and simultaneous conversion as additional enclosure mechanisms they did not consider.\footnote{A further Bastiatian implication is that challengers harmed by enclosure may not abolish the enclosure logic once they gain scale. They may instead reproduce it under their own control. In KBC terms, an entrant's payoff function may shift from access-seeking to access-controlling as scale, user lock-in, data advantage, and feedback capture increase. This observation is not treated here as a separate theorem; it is a political-economy tendency that qualifies the optimism of challenger-led openness.}

The political-economy mechanism is self-reinforcing. Once enclosure produces private knowledge rent, some of that rent can be reinvested in the legal, regulatory, standards, contractual, or platform-governance conditions that preserve the enclosure. The institutional distribution is therefore not merely an external constraint on knowledge governance forms; it is partly shaped by the rents\index{rents} those governance forms generate. This is the Bastiatian extension to enclosure drift: beneficiaries of compensable obstacles may acquire both the incentive and the capacity to maintain the obstacles from which they benefit. The Bastiatian vocabulary here is used analytically, to name a counterfactual-accounting discipline (the seen versus the unseen), and does not import a general normative verdict on enclosure; Proposition~E and the net-effect condition keep that verdict conditional.

Moving from an open governance form to an enclosed governance form is often easier when licences, grant terms, public-domain status, and community governance do not prevent private layering. A \(K^{C}\) commons can be forked by a firm that develops the fork proprietary, adding an access layer \texttt{{[}A{]}} that prevents others from doing what the firm has done. A \(K^{P}\) stock can be enclosed through Bayh-Dole-style IP assignment when a publicly funded institution files patents over its discoveries. Each transition requires institutional action, but the institutional resources required (filing a patent, establishing an API with controlled access, asserting a copyright claim) are privately investible and privately incentivized.

Moving from an enclosed governance form to an open governance form is much harder. It requires either legal expiry of the exclusivity, legal invalidation through expensive prior art litigation, compulsory licensing through regulatory intervention, or deliberate abandonment by the rights-holder. None of these mechanisms is automatic.

Bastiat's obstacle theory clarifies why this asymmetry matters for value theory, not only for institutional design. Wealth rises when the satisfaction obtained from an activity increases relative to the effort required to obtain it. A knowledge governance form is therefore productive when it lowers effort per unit of productive satisfaction, improves capability at equal effort, or preserves necessary quality, safety, disclosure, or investment incentives. It becomes rent-extractive when it creates or preserves compensable access friction without increasing productive satisfaction.

This yields a Service-Friction Diagnostic for distinguishing productive service from access control and obstacle monetization.

\begingroup
\small
\setlength{\tabcolsep}{4pt}
\renewcommand{\arraystretch}{1.15}
\par\addvspace{0.8\baselineskip}\noindent
\captionof{table}{Service, Access, and Friction Rents}\label{tab:ch5:service-access-friction-rents}
\begin{tabularx}{\textwidth}{@{}L{0.20\textwidth}L{0.44\textwidth}X@{}}
\toprule
Rent type & Clean test & KBC interpretation \\
\midrule
Service rent & Is payment tied to productive service, maintenance, implementation, expertise, or improvement? & Genuine knowledge yield. \\
Access rent & Is payment tied mainly to controlled permission to use a stock? & Ambiguous; may be justified or rent-extractive. \\
Friction rent & Is payment tied to obstacles, lock-in, switching costs, or strategically imposed difficulty? & Enclosure rent when not matched by productive service. \\
\bottomrule
\end{tabularx}
\par\addvspace{0.4\baselineskip}
\noindent\emph{Note.} Rent is used here broadly; strict economic rent applies where payment exceeds the cost of productive service.
\par\addvspace{0.8\baselineskip}
\endgroup

Bastiat supplies the classical obstacle intuition; Baumol\index{Baumol, William} supplies the modern allocation frame. The diagnostic asks whether entrepreneurial and institutional effort is reducing productive friction or monetizing it.

This asymmetry has a collective action structure. The private return to enclosure is concentrated in the enclosing actor. The cost of enclosure (the narrowed recombination field, the suppressed future generation, the reduced \(K^{\mathrm{yield}}\) available to all potential users of the enclosed knowledge) is distributed across all potential future beneficiaries. This is the same structure identified in the Interlude I governance-conditioned asymmetry result: under enclosed governance forms (\(\Gamma_{enc}\)), E{[}\(\Delta K_{i,t}\) \textbar{} \(\Gamma_{enc}\){]} \textgreater{} E{[}\(\Delta K_{j,t}\) \textbar{} \(\Gamma_{enc}\){]}, where incumbents (i) systematically accumulate higher expected value change than excluded actors (j), and the gap compounds across periods.

This conditional enclosure drift is not determined by technology alone. It is institutionally determined by the distribution of institutional authority, contractual leverage, and enforcement capacity among parties who have different incentives with respect to the governance allocation decision. Changing that direction requires changing the institutional conditions at the first-conversion zone: modifying the funding terms that allow public investment to be enclosed, strengthening open-access mandates, investing in knowledge-commons governance capacity, and designing regulatory instruments that address platform dependency and simultaneous conversion events in ways that existing frameworks cannot.

Open-source software, open-access research, open standards, and functional \(K^{C}\) commons all persist and generate substantial productive value. But they persist because deliberate institutional investment in governance and protection counteracts the conditional enclosure tendency, not because the open equilibrium is self-sustaining in the face of the enclosure incentives. The governance question is not whether open governance forms are possible but what institutional investment is required to sustain them.

\section{Governance-Form Effects on Recombination Field Breadth}\label{governance-form-effects-on-recombination-field-breadth}
\index{recombination field}

The governance-form analysis of §5.2 can now be connected formally to the Knowledge Generation Model of Chapter 3. The recombination field diversity term \(D_{u}(F_{a,t})\) governs \(G^{R}\) output and enters the weakly specified \(\Phi(\cdot)\) aggregator as a shared enabling condition. This is a proxy architecture, not a calibrated measure; \(\omega_j\), \(\chi_{a,j,t}\), and the \(\Phi(\cdot)\) parameters require sector-specific operationalization. It is formally defined as the productive-weight entropy of the accessible field:

\begin{equation}
D_u(F_{a,t})=-\sum_{j\in F_{a,t}}p_{a,j,t}\log p_{a,j,t}
\label{eq:ch5:d-u-f-a-t}
\end{equation}

where
\begin{equation}
p_{a,j,t}=\frac{\omega_j\chi_{a,j,t}}{\sum_{k\in F_{a,t}}\omega_k\chi_{a,k,t}}
\label{eq:ch5:p-a-j-t}
\end{equation}

The term \(p_{a,j,t}\) is the productive-weight share of stock \(j\) in actor \(a\)'s accessible field at time \(t\), \(\omega_j\) is the productive weight of type \(j\) (how much that knowledge contributes to value if fully deployed), and \(\chi_{a,j,t}\) is the access fraction (the share of type \(j\) that actor \(a\) can reach). \(D_{u}\) is maximized when all accessible stocks are equally productive and equally accessible, the most diverse possible field. It falls toward zero as the field collapses toward a single dominant stock, regardless of how large that stock is. Diversity is valuable only when the accessible stocks are productive, reliable, and complementary; entropy alone is not welfare. The entropy form is the order-\(q\to1\) baseline of the Hill-number family introduced at the \(G^{R}\) definition in Chapter~3; the order \(q\) controls how far concentration on a few high-value stocks is penalized and is left for calibration.

\begingroup
\small
\setlength{\tabcolsep}{3pt}
\renewcommand{\arraystretch}{1.12}
\sloppy
\par\addvspace{0.8\baselineskip}\noindent
\addtocounter{table}{-1}
\begin{longtable}{@{}L{0.20\textwidth}
L{0.24\textwidth}
L{0.31\textwidth}
L{0.19\textwidth}@{}}
\caption{Governance-Form Effects on Recombination Field Breadth}\label{tab:ch5:governance-form-effects-recombination-field}\\
\toprule\noalign{}
\begin{minipage}[b]{\linewidth}\raggedright
Governance form
\end{minipage} & \begin{minipage}[b]{\linewidth}\raggedright
Effect on \(D_{u}(F_{a,t})\)
\end{minipage} & \begin{minipage}[b]{\linewidth}\raggedright
Mechanism
\end{minipage} & \begin{minipage}[b]{\linewidth}\raggedright
Generation consequence
\end{minipage} \\
\midrule\noalign{}
\endfirsthead
\toprule\noalign{}
\begin{minipage}[b]{\linewidth}\raggedright
Governance form
\end{minipage} & \begin{minipage}[b]{\linewidth}\raggedright
Effect on \(D_{u}(F_{a,t})\)
\end{minipage} & \begin{minipage}[b]{\linewidth}\raggedright
Mechanism
\end{minipage} & \begin{minipage}[b]{\linewidth}\raggedright
Generation consequence
\end{minipage} \\
\midrule\noalign{}
\endhead
\bottomrule\noalign{}
\endlastfoot
Private IP governance & Reduces: proportional to breadth and duration of exclusivity & Excluded \(K^{D}\) stocks unavailable for legal combination & \(G^{R}\) output falls for excluded actors; generation concentrated in IP-holding actors \\
Firm-capability governance & Reduces: for actors outside the firm & \(K^{I}\) and \(K^{D}\) embedded in firm inaccessible to external recombination & Recombination field of non-firm actors is narrower by the range of firm-held \(K^{I}\) \\
Platform-dependency governance & Reduces for non-platform actors; may increase for platform operator & \(K^{I}\) and \(K^{C}\) aggregated into \((K^{D})_{\mathrm{platform}}\) accessible only on platform terms & Platform operator gains access to the widest recombination field; excluded actors have the narrowest \\
Professional-stewardship governance & Mixed: narrows who can produce; may widen what is accessible & \(K^{E}\) gate restricts contributors; open publication widens \(K^{D}\) and \(K^{P}\) accessible to all & Net effect depends on governance quality: strong stewardship with open publication widens the effective field; credentialism capture narrows it \\
Commons governance & Increases: proportional to breadth and quality of governance & \(K^{C}\) accessible for legal combination; wide field of interoperable stocks & \(G^{R}\) output highest under well-governed \(K^{C}\); poorly governed commons narrows effective field below nominal breadth \\
Public epistemic governance & Foundational: sets the baseline field breadth on which all other governance forms operate & \(K^{P}\) and \(K^{C}\) provide the interoperable, cumulative, accessible layer that makes field diversity possible & Without \(K^{P}\) investment, \(D_{u}(F_{a,t})\) contracts for all actors regardless of other governance choices \\
\end{longtable}
\endgroup

Three implications follow from this table.

\textbf{First, governance choice is generation policy.} Because \(D_{u}(F_{a,t})\) governs \(G^{R}\) output and enters the weakly specified \(\Phi(\cdot)\) composition function, the institutional decision about which governance form governs a piece of \(K^x\) stock is simultaneously a decision about the future rate of knowledge generation for the actors operating in that field. This is Proposition C's generation-suppression effect stated at the governance form level: enclosure is not merely a distributional choice about who captures the returns to existing stock; it is a generative choice about which mechanism-specific channels can produce future stock.

\textbf{Second, the six governance forms are not equivalent in their aggregate effects on \(D_{u}(F_{a,t})\).} The commons and public epistemic governance forms are both field-widening; private IP and platform dependency are field-narrowing for non-incumbents; firm capability is field-narrowing by inaccessibility; professional stewardship is mixed. Holding creation incentives and quality effects constant, a more enclosed governance form portfolio narrows accessible diversity for non-incumbents relative to a portfolio weighted toward well-governed commons and public epistemic infrastructure. This is an empirically testable aggregate-suppression hypothesis, not a calibrated law: total \(G^{R}\) output should vary with the overall governance form portfolio, but the magnitude and any translation into \(G^{\mathrm{tot}}\) depend on the estimated \(\Phi(\cdot)\) parameters and may need to be tested mechanism by mechanism.

\textbf{Third, the foundational position of public infrastructure is now formally visible.} \(K^{P}\) is a baseline \(K\)-form for broad recombination, validation, measurement, and interoperability across governance forms. When \(K^{P}\) investment falls (through defunding, Bayh-Dole enclosure, or platform dependency substitution), \(D_{u}(F_{a,t})\) contracts for all actors regardless of their individual governance choices. An actor with access to the widest private IP and platform dependency stocks cannot fully or efficiently compensate for a depleted \(K^{P}\) foundation, because the foundational layer provides shared vocabulary, measurement systems, and open standards that make diverse stocks interoperable and therefore combinable. This is the formal reason that public epistemic infrastructure is not merely one governance-form option among six: it supplies the shared conditions that allow \(D_{u}(F_{a,t})\) to have productive breadth.

\section{Public Epistemic Governance as Foundation, Not Mere Option}\label{epistemic-infrastructure-as-foundation-not-governance-form-option}

The public epistemic governance form occupies a special position in this theory. Treating it as one governance option among six understates its significance. It is a foundational layer for broad recombination, validation, measurement, and interoperability across governance forms.

Public epistemic infrastructure is not merely one governance-form option among others. It supplies standards, measurements, research bases, datasets, and institutions that make other governance forms productive. Private actors can build parts of this layer when complements are profitable, but they usually cannot capture the full social return from maintaining it. Arrow's appropriability problem is especially severe for foundational \(K^{P}\), because broadly useful knowledge often creates value beyond the actor that funds, maintains, or discloses it. Public epistemic infrastructure depreciates when public datasets become obsolete, standards bodies\index{standards maintenance} decay, trained researchers become scarce, replication systems weaken, public statistics degrade, or metrology is underfunded.

The positive argument: private IP and platform dependency generate returns only because they draw on a foundational \(K^{P}\) and \(K^{C}\) base that public epistemic governance and commons governance built. Drug patents build on basic research in biochemistry, pharmacology, and cell biology that was publicly funded. Software platforms build on open standards\index{standards!open standards} (TCP/IP, HTTP, SSL) developed through public and academic investment. AI systems build on mathematical foundations, linguistic corpora, and research literature that the scientific community produced and shared under \(K^{P}\) and \(K^{C}\) governance forms. The productive value of enclosed \(K^{D}\) is not self-generated; it is derived from the recombination of foundational knowledge that open governance forms made available.

When public investment in the foundational layer falls, \(D_{u}(F_{a,t})\) contracts for the economy as a whole. For a period, private IP and platform dependency can continue to generate returns by drawing down the foundational \(K^{P}\) base that prior public investment produced. But as the foundational layer depreciates without replenishment, the combinatorial space from which new private \(K^{D}\) could emerge contracts, and the returns to enclosure begin to fall. The process is slow and its effects are diffuse, which is precisely why the collective action problem of §5.5 tends to prevail: the short-term private returns to enclosure remain positive even as the long-term social costs of foundational layer depletion accumulate.

Epistemic infrastructure is therefore not merely a policy preference among alternatives; it is a shared productive foundation for knowledge-bearing capitalism. An economy that progressively encloses its \(K^{D}\) while underinvesting in the \(K^{P}\) and \(K^{C}\) that regenerate the recombination field is drawing down a productive asset that private accumulation often uses but does not fully rebuild. This is the Smithian inversion as a structural risk: where enclosure incentives are strong and open-governance-form maintenance is weak, present gains can narrow future generation.

\section{Coda: The Governance Implication and Handoff to Chapters 6 and 7}\label{coda-the-governance-implication-and-handoff-to-chapter-6}

The governance allocation analysis completes the theoretical answer to the fixed-point question. The conditions that determine whether knowledge becomes a private IP asset, firm capability, platform dependency, professional standard, commons-governed stock, or public epistemic infrastructure are institutional: funding source, contractual status at generation, disclosure timing, access architecture, and governance investment in maintaining open governance forms against conditional enclosure drift. These conditions are chosen, through legislative design, contractual negotiation, regulatory intervention, and strategic decision. They are not determined by technology alone, not economically inevitable, and not immune to deliberate revision.

This theory's distributional implication follows. The distribution of productive capability in a knowledge-bearing economy is determined not only by generation capacity, but also by institutional conditions governing conversion rights. Smith's model assumed that market exchange and capital accumulation would coordinate the distribution of productive capacity through the price mechanism. Knowledge-bearing capitalism requires an additional layer of analysis: the governance allocation layer, where institutional authority, contractual leverage, and enforcement capacity among the parties present at the first-conversion zone influence which governance form applies, and therefore which party captures the resulting productive value.

\subsection*{What this lets us see}
\addcontentsline{toc}{subsection}{What this lets us see}

Chapter 5 changes the question. For a knowledge-intensive firm, the relevant question is not only what knowledge stock it holds, but which governance form governs that stock. For platform dependency or IP valuation, the question is not only whether returns exist, but whether they arise from productive service, access control, or friction. For innovation policy, the question is not only whether knowledge should be open or enclosed, but which governance form best supports creation, maintenance, recombination, quality, and renewal.

Understanding that layer (and deliberately designing institutions that maintain wide recombination fields, functional commons, and robust epistemic infrastructure against conditional enclosure drift) is the governance problem that this theory has been built to identify. Chapters 6 and 7 address the most developed contemporary enclosure dynamic: the cognitive and feedback-capture mechanisms through which knowledge-bearing incumbents accumulate productive capability while narrowing the generative conditions available to everyone else.

\part{Dynamics of Enclosure}
\chapter[Cognitive Enclosure]{Cognitive Enclosure and Generative
Suppression}\label{cognitive-enclosure-and-generative-suppression}
\index{cognitive enclosure|textbf}

\chapterhook{The Future That Enclosure Prevents}

Here the dynamic trade-off of Chapter 1 becomes the chapter's whole subject. The act that converts use-value into private exchange-value by enclosing it can also foreclose the recombination and learning that would have generated more use-value later. Cognitive enclosure is conversion read forward in time: what is captured now set against what is prevented from ever being generated.

In the vocabulary developed later, this is suppressed generativity\index{generativity!suppressed}: enclosure does not merely lower the yield of what already exists, it removes recombination edges that would otherwise have formed, so the lost value never registers as a measured loss because it was never generated. That is distinct from knowledge left dormant for want of a complement; suppression is a governance act, and unlike dormancy it can be reversed by de-enclosure.

\emph{How Enclosed Governance Arrangements Narrow the Recombination Field and Suppress
Future Knowledge Generation}

The practical issue is whether an access restriction prevents others from building the next product, method, dataset, or model.

The enclosure analysis begins from the established economics of non-rival ideas, invention incentives, anticommons\index{anticommons theory}, patent failure, and open-source production \parencite{Arrow1962, Romer1990, Heller1998, HellerEisenberg1998, JaffeLerner2004, BoldrinLevine2008, Eghbal2020, Nagle2019}\index{Heller and Eisenberg}. This chapter's specific contribution is to connect enclosure to the actor-specific recombination field rather than to access costs alone.

Chapter 5 established that governance allocation is a political economy
problem with structural consequences: the six governance forms governing
knowledge-bearing stock differ in who controls conversion pathways,
whose productive potential is realised, and whether the recombination
field is widened or narrowed for actors outside the governing coalition.
It also established a structural asymmetry: knowledge-bearing capitalism
drifts toward enclosure in the absence of deliberate counter-institutional
investment, because the private returns to enclosure are concentrated
while the social costs are distributed.

This drift should not be read as a claim that enclosure is always socially wasteful. Some enclosure protects investment, supports disclosure bargains, funds costly validation, and gives firms a reason to build and maintain knowledge-bearing stock. The claim is narrower: once knowledge becomes a recombination input, the same enclosure that strengthens private incentives can also reduce the accessible field from which future knowledge is generated. This chapter measures that second effect without denying the first.

A related welfare guard applies throughout Chapter 6. Recombination generation, denoted \(G^R\), is an intermediate productive-capacity variable, not a welfare measure. More \(G^R\) is not automatically better. Its welfare value depends on the downstream use of the generated knowledge, its social value, its risk profile, the institutional context in which it is deployed, and whether the knowledge at issue falls into harmful-knowledge exclusions. The formal claim in this chapter is therefore narrower than a universal welfare claim: enclosure can suppress recombination capacity for excluded actors, and that suppression is economically material when the affected inputs are central, non-redundant, socially valuable, and not harmful.

The causal story is straightforward. What is enclosed is access to knowledge-bearing stock as a recombination input, experimental substrate, observational evidence, or interpretive resource. Those excluded may include rivals, complementors, researchers, maintainers, users, entrants, and commons participants. What is lost is not merely use of an existing object, but the recombinations, tests, interpretations, standards, fixes, and adjacent inventions that would have drawn on that object. The incumbent gains control over access, timing, price, interoperability, disclosure, and dependence. The claim would be weakened if enclosure events did not measurably reduce excluded recombination fields, useful diversity, or downstream combinations, or if the incentive, quality, privacy, security, or coordination benefits of enclosure systematically outweighed the lost generative paths.

Chapter 6 asks what those social costs are. The question has two parts:
what enclosure does to actors outside the enclosing relationship (the
external effect, developed here), and what it does to the enclosing
actor's own knowledge accumulation trajectory (the internal effect,
developed in Chapter 7). The two parts must be held together: it is
their joint operation, generation suppressed externally, accelerated
internally, that produces the structural outcome T8 names as the
signature of a knowledge economy under strong enclosure.

Standard static welfare analysis of IP addresses a version of the external-effect question
but often frames it as a distributional problem: who gets the value of the
enclosed knowledge, at what static welfare cost. This chapter does not claim that
innovation economics\index{innovation economics}, information economics, or industrial organization lack
dynamic concepts. Its narrower claim is that KBC isolates a recombination-field
contraction channel: enclosure is not merely a distributional event; it is also a
generative event when it removes inputs from the field through which future knowledge
is produced. Its consequences can extend forward through the knowledge conversion
cycle and compound across successive rounds of generation.

The anticommons tradition supplies one predecessor\index{anticommons blockage|textbf} for this claim, but its mechanism must be distinguished from KBC's enclosure mechanism. \textcite{Heller1998} identifies the anticommons problem, and \textcite{HellerEisenberg1998} apply it to biomedical research: excessive fragmentation of property rights can block productive use even when individual holders behave rationally. KBC's enclosure mechanism is structurally distinct: it concerns concentration rather than fragmentation. Heller's anticommons arises when too many holders each have veto rights over use; KBC's enclosure arises when too few holders concentrate access to formerly shared field elements, suppressing third-party recombination and generation capacity.

The anticommons case can still be restated in KBC terms. When too many actors hold exclusion rights over upstream knowledge inputs, downstream actors cannot assemble the permission bundle required to generate new combinations. KBC restates the blocked object as recombination generation. The loss is not only a failed transaction over existing IP; it is the absence of future combinations that would have existed under a better-fitted governance arrangement:
\[
G^R_{\mathrm{actual}} < G^R_{\mathrm{potential}}
\quad \text{when} \quad
X_{\mathrm{rights}} \uparrow,
\ C_{\mathrm{assembly}} \uparrow,
\ R_{\mathrm{field}} \downarrow.
\]
This expression is a diagnostic placeholder rather than a calibrated law. It records one predecessor mechanism to be measured: fragmented exclusion rights raise assembly cost and reduce the effective field from which recombination can proceed. The distinct KBC mechanism analysed below is concentrated enclosure of formerly shared field elements.

A second guard is equally important. Enclosure may be socially useful where it funds costly creation, preserves quality, assigns accountability, protects sensitive knowledge, or coordinates investment that open governance arrangements would underprovide. It becomes inefficient only when marginal enclosure benefit is outweighed by recombination loss, learning-loop concentration, capability loss, or trajectory suppression:
\[
MB_{\mathrm{enclosure}} \gtrless MC_{\mathrm{rec}}+MC_{\mathrm{learning}}+MC_{\mathrm{diversity}} .
\]
This inequality is also a diagnostic placeholder. Its purpose is to prevent Chapter 6 from treating enclosure as intrinsically bad. The relevant question is whether the enclosure fits the stock and the stage of its knowledge life-cycle.

\section{Cognitive Enclosure Defined}\label{cognitive-enclosure-defined}
\index{cognitive enclosure}

\paragraph{Causal role.} This section identifies the enclosed object: access to knowledge-bearing stock as an input to recombination, experiment, observation, or interpretation. The causal test is whether restricting that access changes who can build from the stock, not merely who can buy or consume it.

Enclosure, as used here, is the institutional process by which
knowledge-bearing stock that was previously accessible to a wide range
of actors for use as recombination input, as experimental
substrate, as observational evidence, or as interpretive resource, is
brought under conditions of restricted, mediated, or conditional access.
Those restrictions may support investment, quality control, security, or
accountability, but they also reduce the ability of actors outside the
governing coalition to use the stock as a recombination input. \emph{(First-use definition of enclosure appears in Chapter~2, Section~2.2:
the legal, contractual, technical, architectural, or capability-based
restriction of access to knowledge-bearing stock, reducing the ability
of other actors to use, modify, recombine, learn from, or build upon it.
That definition names the mechanisms; the present chapter analyses their
economic consequences.)}

The definition has several features that distinguish it from the
standard IP-law notion of enclosure.

First, it applies across all five \(K\)-forms, but not every legal or institutional
arrangement in these categories is enclosure in the KBC sense. \(K^E\) (embodied
knowledge) may be enclosed through employment IP assignment clauses, confidentiality
duties, or restrictive covenants where those arrangements materially prevent workers
from carrying trained judgement into legitimate competing, complementary, or research
uses. \(K^D\) (disembodied knowledge) can be enclosed through patent, copyright,
trade-secret, database, licence, or API\index{API access!cognitive enclosure} access restrictions when those restrictions
remove usable field elements from excluded actors. \(K^I\) (institutionalized knowledge)
may be enclosed when organizational routines, protocols, or standards are kept inside
a firm, consortium, or credentialled group rather than being made available to the
broader field under terms that permit recombination. \(K^C\) (commons knowledge capital)
can be impaired or enclosed indirectly through extraction of governance capacity,
maintainer labour, or interoperability control, depleting the commons from within
without requiring legal seizure of the knowledge base itself. \(K^P\) (public epistemic
capital) can be weakened when the institutional conditions for public knowledge
production are undermined: research funding is withdrawn, open-access mandates are
reversed, standards processes become inaccessible, or public dataset access is
restricted.

Second, enclosure is a process, not a state. A stock may move between
accessible and restricted conditions multiple times over its conversion
history, as Chapter 5's governance-cycling analysis showed. What Chapter 6
adds is the recognition that each enclosure event, and the duration of
each enclosed interval, can impose costs that are not simply reversed when
the enclosure ends or the stock transitions to a more open governance arrangement.

Third, cognitive enclosure specifically names the form of enclosure that
operates through the knowledge production and recombination cycle, not
merely through the pricing or distribution of existing stock. A text
copyright encloses a fixed cultural product; cognitive enclosure of a
training corpus encloses the substrate for generating the next
generation of language capability. The target of cognitive enclosure is
the generative potential of the knowledge base, not only its current
productive value.

A final boundary condition keeps the mechanism distinct from ordinary IP exclusion
and from Heller's anticommons. Ordinary IP exclusion may be justified as a disclosure,
investment, coordination, or quality-control bargain. Anticommons blockage arises
when too many right-holders create veto points that prevent downstream assembly.
KBC's recombination-field impairment is the narrower claim that a governance change
removes usable field elements from excluded actors and thereby reduces their future
generation capacity. These mechanisms can overlap, but they should not be treated as
identical.

\textbf{Definition 6.1} (Cognitive enclosure): An institutional process
\(\pi_0\) → \(\pi_1\) that removes a set \(E \subset F_{a, t}(\pi_0)\) of
knowledge-bearing stocks from the accessible recombination field of
actor class \(A_{ex}\), while the enclosing actor class \(A_{inc}\) retains \(E\) in
its field:

\begin{equation}
F_{a, t}(\pi_1)=F_{a, t}(\pi_0)\setminus E\setminus \mathrm{Cascade}(E, a, \tau)\qquad\text{for }a\in A_{ex}
\label{eq:ch6:f-a-t-pi-1}
\end{equation}
\begin{equation}
F_{a, t}(\pi_1)=F_{a, t}(\pi_0)\qquad\text{for }a\in A_{inc}
\label{eq:ch6:f-a-t-pi-1-2}
\end{equation}

where \(\mathrm{Cascade}(E, a, \tau)\) is the set of non-enclosed stocks that
fall below the field admission threshold at time \(t+\tau\) as a
secondary consequence of enclosing \(E\), through the interoperability
and capability cascades of the recombination field (Chapter 3).

The cascade term is the formal expression of Chapter 6's central
claim: enclosure removes more from the accessible field than the
enclosed stocks themselves. The legal, architectural, or institutional
removal of \(E\) initiates secondary contractions that extend the field
restriction beyond the explicit boundary of the enclosure act.

\section{From Governance Allocation to Enclosure
Dynamics}\label{from-regime-allocation-to-enclosure-dynamics}

\paragraph{Causal role.} This section converts Chapter 5's governance-form taxonomy into a mechanism. The issue is not which governance form a stock occupies in the abstract, but whether a governance transition excludes actors from a recombination path they previously could use.

Chapter 5 analysed governance allocation as a problem of first-conversion
zone governance: which institutional conditions at the point where
knowledge-bearing stock first becomes economically productive determine
which governance form it enters. That analysis established three findings
relevant to Chapter 6.

The first is that private IP, firm capability, and platform dependency
governance arrangements can narrow \(D_u(F_{a, t})\), especially for non-incumbents
when access, licensing, interoperability, or capability conditions are restrictive. The
Section~5.6 analysis showed this formally: each of the three enclosed governance arrangements
reduces the recombination field breadth available to actors outside the
governing coalition, while commons and public infrastructure governance arrangements
widen it. Professional stewardship governance arrangements are mixed, depending on
whether professional governance maintains open access or restricts it to
credentialled practitioners.

The second finding is that governance allocation drifts toward enclosure in
the absence of deliberate counter-institutional investment. The
structural asymmetry of Section~5.5 can be stated as:
\begin{equation}
E[\Delta K_{i, t} \mid \Gamma_{\mathrm{enc}}] > E[\Delta K_{j, t} \mid \Gamma_{\mathrm{enc}}],
\label{eq:ch6:governance-conditioned-asymmetry}
\end{equation}
where \(i\) denotes incumbents and \(j\) denotes excluded actors. This
asymmetry means that enclosure compounds advantage over time. Actors who
control the enclosure boundary capture more of the improvement in the
enclosed stock; actors outside the boundary lose access to the field
elements that would allow them to close the gap.

The third finding is that governance cycling is asymmetric. The sequences in
Section~5.3 show knowledge-bearing stock moving from more open to more enclosed
governance arrangements more readily than in the reverse direction: from public
infrastructure to private IP (pharmaceutical); from commons to platform
dependency (software); from \(K^C\) to \((K^{D})_{\mathrm{platform}}\) (AI training
data). These transitions are easier to complete than their reversals
because enclosure is institutionally entrenched, legal rights once
granted are difficult to revoke, while the re-opening of enclosed
stock often requires expiry, policy, litigation, standards pressure,
strategic opening, market entry, or technological substitution.

Chapter 6 picks up where Chapter 5's static governance allocation analysis
leaves off. The governance allocation problem asks which governance form a stock
enters. The enclosure dynamics problem asks what happens during the
interval of enclosure, to the stocks outside the enclosure boundary,
to the actors excluded from it, and to the generation capacity of the
knowledge economy as a whole.

\section[The formal channel: useful diversity of the recombination field]{\texorpdfstring{The Formal Channel: \(D_u(F_{a, t})\)}{The formal channel: useful diversity of the recombination field}}\label{the-formal-channel-d_uf_at}
\index{recombination field}

\paragraph{Causal role.} This section names the lost pathway: useful diversity in the recombination field. The empirical question is whether enclosure removes inputs that are central, non-redundant, and usable by excluded actors.

The formal channel through which enclosure suppresses knowledge
generation is the recombination field diversity term \(D_u(F_{a, t})\).
Chapter 3 established the generation function for recombination:

\begin{equation}
G^R_{a, t+1}=\lambda_a\cdot R_{a, t}^{\eta}\cdot D_u(F_{a, t})^{\mu}
\label{eq:ch6:g-r-a-t-1}
\end{equation}

where \(R_{a, t}=\sum_{j\in F_{a, t}} w_{a, j}\) is the field
magnitude (the sum of productive weights of accessible stocks),
\(D_u(F_{a, t})\) is the useful diversity of the accessible field,
\(\eta \in (0, 1)\) and \(\mu\) \textgreater{} 0 are the generation
elasticities, and \(\lambda_a\) is the actor-specific generation
productivity parameter.

Enclosure of \(E \subset F_{a, t}(\pi_0)\) reduces \(G^R\) through two
distinct channels operating simultaneously.

\textbf{Channel 1, Field magnitude reduction}: The productive weight
\(w_{a, j}=\omega_j\cdot\chi_{a, j}\cdot\rho_{a, j}\) of each enclosed stock is
removed from \(R_{a, t}\). If the enclosed set \(E\) contains stocks with
positive productive weight for excluded actors, \(R_{a, t}(\pi_1)<R_{a, t}(\pi_0)\). The reduction in field magnitude
directly reduces the base recombination rate.

\textbf{Channel 2, Useful diversity reduction}: \(D_u(F_{a, t})\)
measures the entropy of the productive-weight distribution across
accessible stocks. The enclosure of frontier stocks, those at or near
the optimal combination distance \(d^*_{a}\) from excluded actors'
existing knowledge, removes the highest-weight elements from the
field. Removing high-weight elements both reduces \(D_u\) mechanically
(fewer elements contribute to the entropy sum) and reduces it
compositionally (the remaining distribution is less diverse because the
most distinctive, highest-productivity elements have been removed).

The combined effect on the generation rate for excluded actors is the
suppression ratio from T2\index{T2 Governance Suppression!suppression ratio}:

\begin{equation}
\frac{G^R_{a, t}(\pi_1)}{G^R_{a, t}(\pi_0)}
= (\sigma_a^R)^{\eta}
\cdot
\left(\frac{D_u(\pi_1)}{D_u(\pi_0)}\right)^{\mu}
< 1
\label{eq:ch6:frac-g-r-a-t-pi-1-g-r-a-t-pi-0}
\end{equation}

where \(\sigma_a^R=R_{a, t}(\pi_1)/R_{a, t}(\pi_0)<1\) is the field magnitude ratio. The ratio falls below one
whenever enclosure reduces either field magnitude or useful diversity,
and it falls furthest when enclosure is concentrated at the elements
with the highest productive weight for excluded actors, precisely the
frontier stocks that the enclosure incentive is strongest to protect.

In plain terms, the suppression ratio measures how much recombination-generation
capacity remains after enclosure reduces the size of the accessible field and the
useful diversity of what remains. It is not a price measure. It asks whether excluded
actors still have enough relevant, varied, and usable inputs to generate new
knowledge-bearing stock.

\textbf{The cascade extension}: Definition 6.1 includes \(\mathrm{Cascade}(E, a, \tau)\) in the field contraction. The cascade operates through two
mechanisms established in the recombination field model. The
interoperability cascade: if \(K_j \in E\) was required for technical
interoperability with \(K_k\in F\setminus E\) (common standards, API dependencies,
format requirements), then \(I_{a, k, t+1}\) falls below the admission
threshold, removing \(K_k\) from the accessible field even though \(K_k\) was
not enclosed. The capability cascade: if \(\widetilde{C}_{a, t}\) falls
below the productive threshold for using \(K_k\) (because the generation
activity that maintained and developed that capability relied on access
to E), then \(K_k\) effectively exits the productive field through
capability degradation rather than access restriction. Both cascades
mean that the field contraction from enclosure of \(E\) exceeds the direct
field contraction by \(\lvert E\rvert\cdot\text{(cascade multiplier)}\).

\emph{Smithian departure:} Smith's critique of trade restriction was
framed primarily as a price and exchange argument: monopoly raises prices above
competitive levels, reducing exchange and harming consumers. The T2 suppression
ratio is not a price argument. No price need rise, and no consumer need pay more
for an enclosed knowledge stock that excludes them. What falls is the rate at
which new knowledge-bearing stock is generated. Smith's framework does not contain
this channel; later innovation and information economics address parts of it, but
KBC isolates the recombination-field mechanism by which a cost accrues to a
generation event that never occurs.

\section{Frontier Enclosure and the Understatement of
Suppression}\label{frontier-enclosure-and-the-understatement-of-suppression}

\paragraph{Causal role.} This section asks whether the loss extends beyond current access. Frontier enclosure matters when exclusion prevents actors from discovering adjacent possibilities, not only when it blocks an already known use.

The suppression ratio \(\left(\sigma_a^R\right)^{\eta}\cdot(D_u\;\mathrm{ratio})^{\mu}\) measures the fraction of an excluded actor's
generation capacity that remains after enclosure. These suppression results take their sharpest form under the multiplicative (Leontief-style) composition in which access, useful diversity, capability, and interoperability act as complements, so that a shortfall in one cannot be offset by abundance in another; where those inputs are substitutable, the realized suppression is bounded rather than near-total. Volume~2's function-class audit (Appendix~F) records which results retain their sign across the admissible composition forms. The ratio is derived from
average field conditions: it computes the weighted magnitude and
diversity of what remains accessible relative to what was accessible
before. That averaging conceals a systematic understatement.

Enclosure is not distributed uniformly across the knowledge base. The
incentive to enclose is positively correlated with productive value, 
actors often enclose stocks with high private value and high expected option
value for others. Productive value is concentrated at the frontier: recently
generated, technically advanced, broadly applicable knowledge-bearing
stock whose domain distance from many actors' existing bases is close to
\(d^*_{a}\). These frontier stocks appear in the recombination field
with the highest productive weights \(w_{a, j}=\omega_j\cdot\chi_{a, j}\cdot\rho_{a, j}\). They are simultaneously the elements most worth enclosing
and the elements whose loss most reduces \(D_u(F_{a, t})\).

The result is a systematic bias in what enclosure removes. The
accessible field after enclosure is not a random sample of the prior
field with some elements removed; it is the prior field with its
highest-productive-weight elements selectively extracted. For excluded
actors, the combinations that remain available in \(F_{a, t}\)(\(\pi_1\))
are precisely those furthest from the productive frontier, 
combinations with older, more distant, lower-productivity stocks. The
suppression ratio applied to average field weights therefore understates
the generative loss, because the average masks the concentrated removal
of the highest-option-value inputs.

\textbf{Corollary C.1 (Frontier Enclosure):} Let \(\bar{\omega}\) be the
mean productive weight of elements in the enclosed set \(E\), and let
\(\bar{\omega}_F\) be the mean productive weight of the full
pre-enclosure field \(F_{a, t}\)(\(\pi_0\)). When enclosure is
concentrated at the frontier, \(\bar{\omega}\) \textgreater{}
\(\bar{\omega}_F\), the enclosed elements are above-average in
productive weight. The realized suppression of \(G^R\) therefore exceeds
the prediction of the average-field suppression ratio by a factor
increasing in \(\bar{\omega}\)/\(\bar{\omega}_F\) and in the
diversity-elasticity \(\mu\). Formally:

\begin{equation}
\Delta G^R_{actual}\geq\Delta G^R_{suppression\ ratio}\qquad\text{when }\bar{\omega}>\bar{\omega}_F
\label{eq:ch6:delta-g-r-actual}
\end{equation}

The inequality grows tighter as enclosure concentrates more heavily at
the frontier and as \(\mu\) increases; the more diversity-sensitive
generation is, the more the loss of high-weight frontier diversity
compounds. This is a directional, diagnostic claim unless \(\bar{\omega}\),
\(\bar{\omega}_F\), and \(\mu\) are empirically estimated. The corollary is
not a refinement of the suppression model; it identifies a bias in the
direction of that model's error: average-field calculations can understate
the suppression when the incentive to enclose is highest precisely at the
elements whose removal causes the largest generative loss.

\textbf{Implication 1, T5 (Enclosure Efficiency Loss)\index{T5 Enclosure Efficiency Loss!frontier enclosure implication}.} T5
states the social optimum \(T^*\) from \(B_incentive\)(T) =
\(C_{enclosure}\)(T). \(C_{enclosure}\) includes \(C_{T2}\), the formal
integral of the per-period \(G^R\) suppression. When the suppression
ratio is applied at average field weights, \(C_{T2}\) is computed on the
assumption that enclosure removes a random element from the field.
Corollary C.1 states that this understates \(C_{T2}\) for enclosure
concentrated at high-weight stocks, which describes most economically
significant enclosure, because high-value knowledge is precisely what IP
protection is designed for. The social welfare optimum \(T^*\) will tend to be shorter under the theorem's assumptions than T5's already-conservative estimate implies: the over-enclosure region can extend further than average-field calculations show when high-weight frontier stocks are disproportionately enclosed.

\textbf{Implication 2, M5.T1 (Strategic Over-Enclosure).} The gap
\(T^*_{\mathrm{strategic}}\) − \(T^*\) is increasing in \(M_{rec}\), the
recombination multiplier that captures how central the enclosed stock is
to excluded actors' productive fields. When enclosure is concentrated at frontier stocks, \(M_{rec}\) will tend to be high because frontier stocks are the ones with the highest productive weights for other actors. Chapter 8's derivation of \(T^*_{\mathrm{strategic}}\) should therefore be understood as conservative in the same direction where the theorem's assumptions hold: the divergence between what incumbents choose and what is socially optimal may be larger than average-field estimates suggest.

\textbf{Implication 3, Chapter 9 (Dark Capital).} The
governance-position value term \(\delta\cdot E[v(R_i(\pi, t))]\), which
captures the productive surplus from a firm's recombination-field access, will tend to be most valuable for firms positioned at the frontier. The governance position that controls frontier stocks carries a \(\delta\) term reflecting the expected surplus from future high-weight combinations, not average combinations. The Accounting Shadow will therefore tend to be larger for frontier-positioned incumbents, and the fragility shadow \(\Omega_i\) will tend to be larger for firms whose recombination field is most dependent on frontier stocks held under enclosure. These terms are defined formally in Chapters 8 and 9; the present point is only that frontier dependence raises recombination loss.

The frontier effect is not a refinement of the suppression model; it is a recurring feature of the political economy of enclosure. Incumbents have reason to enclose the frontier because that is where private value is highest. This is also where generative suppression\index{generative suppression|textbf} can be most severe. The mechanisms that make frontier enclosure\index{frontier enclosure|textbf} privately rational, high expected future returns, are also the mechanisms that can make it socially costly.

\section{Proposition C: Generative
Suppression}\label{proposition-c-generative-suppression}

\paragraph{Causal role.} Proposition C states the chapter's mechanism in one line: enclosure suppresses future generation when it removes productive inputs from the accessible recombination field of excluded actors. The claim stands or falls on observable contraction in field access, useful diversity, or downstream combinations.

\textbf{Proposition C (Generative Suppression)}: The enclosure of
knowledge-bearing stock into governance arrangements that restrict third-party access
reduces the rate at which third parties generate new knowledge-bearing
stock by recombination with, experimentation on, or interpretation of
the enclosed material. The effect is not confined to the distribution of
the enclosed stock's immediate value. It extends forward through the
knowledge conversion cycle, contracting \(D_u(F_{a, t})\) and thereby
suppressing \(G^R\) for excluded actors across successive rounds of the
generation-conversion cycle. Enclosure is therefore a generative event,
not merely a distributional one, and its consequences compound over
successive rounds.\footnote{A further application of the same mechanism reaches adjudicative capacity. \(A_{\mathrm{adj}}\), the shared evidence standards, peer review, red-team capacity, and arbitration that convert opposition into tested knowledge, is supplied by embodied, institutional, and public-epistemic capital (\(K^{E}, K^{I}, K^{P}\)) and is therefore enclosable. Enclosing or capturing it suppresses admissibility-revising generation, the fused-mode case of \S\ref{fused-generation-proposition}, for the excluded, exactly as access enclosure suppresses recombination: this is Proposition~C applied to the adjudicative subset. \emph{Falsifier:} capture of adjudicative institutions yields no measurable decline in excluded actors' rate of assumption-correction or frame-revision.}

The generative effect differs from the distributional effect that
standard static IP analysis addresses. Standard static welfare analysis asks who gets
the value of the enclosed stock, a question about the distribution of
existing value. Proposition C asks what new knowledge would have been
generated from the enclosed base had it remained accessible, and is
therefore about a counterfactual flow of future knowledge production.
The two effects can move in opposite directions: expanding access may
simultaneously reduce distributional inefficiency and increase
generative efficiency, even if it reduces the private incentive to
invest in creation.

\noindent\textbf{Observable local suppression versus inferred aggregate suppression.} Two versions of this claim must be kept apart, because they are observable to very different degrees. The \emph{local} version is the testable one: around a specific enclosure, does the accessible recombination field of excluded actors contract, and do observable downstream combinations, entrant activity, follow-on work, or dependent capability decline relative to a credible counterfactual? That is a bounded, falsifiable claim, and Chapter~\ref{chapter-11-testing-knowledge-bearing-capitalism} specifies designs, event studies, difference-in-differences, and synthetic controls around governance transitions, that can test it. The \emph{aggregate} version, systemic generative suppression across an economy, is not a directly observed object. Foregone generation is, by construction, a counterfactual: knowledge that would have formed but did not leaves no positive trace. The aggregate claim is therefore an \emph{inference} built up from many local cases that survive testing, not a quantity read off the data. This book treats it accordingly: Proposition~C is asserted and tested at the local level, and the systemic pattern is offered as the cumulative interpretation those local results would or would not support. No single case, including the canonical natural experiments, establishes the system-wide claim on its own.

The generative effect compounds in a way the distributional effect does
not. A distributional inefficiency from enclosure produces a first-order static welfare
loss while the enclosure applies; when the enclosure expires, the first-order
static loss stops, although ordinary dynamic effects may persist. Generative
suppression in one period produces a second-order loss: the
knowledge that would have been generated from the enclosed base, and
that would itself have served as input to the next generation round, is
not created. Because knowledge is cumulative, each round draws from
and contributes to the stock, the suppression in one period narrows
\(D_u\) in all subsequent periods.

Five mechanisms operate the suppression.

\textbf{Recombination exclusion}: The legal or architectural barrier
that prevents third parties from using enclosed \(K^D\), \(K^I\), or
\(K^C\) as input to their own generation processes. A patent over a core
mechanism removes it from excluded actors' \(F_{a, t}\); an API that
withholds the data layer removes the platform's accumulated
\((K^{D})_{\mathrm{platform}}\). In each case, the enclosure reduces both \(R_{a, t}\)
and \(D_u(F_{a, t})\) for excluded actors, suppressing \(G^R\) through
both channels of the formal mechanism.

\textbf{Disclosure suppression}: Patent enclosure includes a disclosure
bargain, public notice of the enclosed space in exchange for the
exclusivity grant, that partially limits generative suppression by
allowing excluded actors to know the boundary of the enclosure and to
design around it. Trade secrecy and firm capability enclosure carry no
analogous requirement. Third parties often cannot know the full boundary of what they are missing
from their recombination field; the \(F_{a, t}\) contraction may be invisible
to them. Platform dependency enclosure can be more severe where it encloses
not only \((K^{D})_{\mathrm{platform}}\) but the observation of attempted
recombination, the signal that, in an open system, would be the
primary mechanism by which the productive frontier of user demand is
identified. Each user interaction with a platform generates knowledge
about where productive recombination is being sought; that knowledge
flows into the enclosed training and inference cycle, invisible to
actors outside. The most productive knowledge the platform holds is not
in its raw data or model weights but in its integrated understanding of
where the useful diversity of user demand lies.

\textbf{Experimentation restriction}: The \(G^X\) mechanism (Chapter 3)
depends on access to the knowledge-bearing stock being varied. Enclosure
of a drug compound, a training corpus, or a platform's recommendation
algorithm prevents excluded researchers from running systematic
experiments to identify the stock's as-yet-unknown properties. The
restriction can be more severe at the frontier, where the most productive
experimentation is exploration of properties that the generating actor
has not yet fully characterized, properties that distributed
experimental inquiry may identify faster than single-actor
exploration.

\textbf{Observation blockade}: Arrow's learning-by-doing mechanism
generates knowledge as a by-product of production; in open knowledge
systems, these improvement signals can be shared across actors and, when
reporting channels, standards, and maintenance capacity exist, combined to
produce collective improvement faster than any single actor's internal learning. Enclosure converts distributed learning into
concentrated learning: deployment exclusively through a proprietary
platform or trade-secret process generates feedback visible only to the
incumbent. Third parties cannot contribute to or draw from the shared
learning pool. The observation blockade\index{observation blockade} does not prevent knowledge
generation; it removes it from the accessible social learning process,
concentrating it within the enclosing actor, an effect whose internal
consequences are developed in Chapter 7.

\textbf{Skill-formation suppression}: Enclosure can also suppress the formation
of future domain-specific embodied knowledge capital. This is the point at which enclosure affects not only artefact access but the future
formation of embodied productive knowledge. If an actor loses access to
the field of problems, tools, examples, audit opportunities, feedback, imitation,
and recombination attempts through which practice becomes learning, then the
actor does not merely lose access to an existing stock. The actor loses part of
the practice field required to form future \(K^E_d\). This is the Smithian skill
channel inside Proposition C: specialized skill deepens through repeated practice
on a domain field, and field restriction can therefore narrow the future supply of
embodied domain capability.

\begin{equation}
\frac{\partial K^E_{d, t+1}}{\partial F_{a, t}} > 0.
\label{eq:skill-formation-field-access}
\end{equation}

Equation~\eqref{eq:skill-formation-field-access} is not a new theorem. Here \(d\)
denotes the domain, so \(K^E_{d, t+1}\) is next-period embodied capability in that
domain. The equation is a corollary of field restriction and learning-loop capture:
repeated domain practice increases embodied knowledge capital only when actors have
access to the problems, tools, feedback, and interpretive conditions required for
learning.

\emph{Smithian departure:} Smith's analysis of trade restriction focused
on monopoly as a distributional problem: prices above competitive levels
reduce exchange and harm consumers. Proposition C identifies a different
channel. Enclosure of knowledge-bearing stock can reduce not only who can
use the enclosed good but how much of the next generation of goods can be
produced at all. Smith's framework does not contain this channel; later
innovation and information economics address parts of it, but KBC isolates
the recombination-field mechanism. The welfare cost is not only the triangle
under a demand curve; it is the growth in the recombination field that does
not occur, the trajectories that are not pursued, and the knowledge that is
not generated.

\begin{center}
\fbox{\begin{minipage}{0.92\textwidth}
\small
\textbf{Running case: API closure/access restriction.}\index{API closure!cognitive enclosure running case} Chapter 6 uses the API case to show the recombination-field channel. The stock is not destroyed; the interface, data stream, documentation, and accumulated developer know-how may still exist. What changes is the excluded actor's usable field: prior integrations, experiments, monitoring tools, research pipelines, and complements lose access to a high-weight recombination input. The KBC prediction is field contraction, not merely a price increase.
\end{minipage}}
\end{center}
\section{A Worked Demonstration: Platform API Enclosure and Field
Contraction}\label{a-worked-demonstration-platform-api-enclosure-and-field-contraction}

\paragraph{Causal role.} The API case translates the mechanism into a market sequence: an interface is enclosed, complementors and users are excluded, alternative products and workflows disappear, and the platform gains control over dependency, timing, and value capture.

The mechanisms described in Sections~6.3--6.5 can be traced concretely through
the progressive restriction of the Twitter/X API between 2020 and 2023.
The case is a clean instance of a broader pattern also visible, with different
mechanisms, in platform, pharmaceutical, and AI enclosure events. Its
value here is that it makes visible what the suppression ratio, the
cascade, and the frontier effect actually mean for a population of real
actors.

\textbf{The knowledge-bearing stock and the pre-enclosure field.}
Depending on tier and endpoint, Twitter's API provided structured
programmatic access to portions of Twitter/X's public interaction stream
\parencite{TechCrunch2023TwitterAPI, Wired2023TwitterAPIPrices, MurtfeldtEtAl2024TwitterAPI}. This distinctive \(K^D\) stock included public discourse,
attention signals, geographic distributions, interaction patterns, and
network relationships. It had properties that made it valuable as a
recombination input rather than merely as a consumption good. It was a
frontier element in recombination fields across disciplines and industries:
academic researchers combined Twitter data with linguistic corpora, health
surveillance data, and electoral records to study misinformation, public
health, and political polarization. Journalists combined it with document
datasets and event timelines to investigate public figures and organizations.
Software developers combined it with location data, recommendation engines,
and notification systems to build tools for crisis monitoring, customer
service, and content discovery. The productive weight \(w_{a, j}\) of the
Twitter API element was high across a wide range of actors because its
combination distance \(d_{a, j}\) was close to \(d^*_{a}\) for many of
them, recent, technically compatible, broadly applicable.

Before restriction, the GATE conditions were largely satisfied for a
large actor population. Access was available at a free or low-cost tier.
Institutional permission was granted by the Terms of Service.
Interoperability was high, because the API was built to standard REST
architecture. Complementary capability was broadly held: the API was
accessible to engineers with modest technical preparation. The result
was a high-\(\chi\) element in the recombination field of a large and diverse
actor population.

\textbf{The governance change and its mechanisms.} Between 2020 and
2023, Twitter progressively tightened API access through rate-limit
reductions, deprecation of endpoints, and, following the 2022 ownership
change, a sharp repricing. Contemporary reporting described the move from
free or low-cost access to new free, basic, and enterprise tiers; the basic
tier was reported at \$100 per month, while enterprise packages reported to
academic users began at \$42,000 per month and rose to \$210,000 per month
for larger packages \parencite{TechCrunch2023TwitterAPI, Wired2023TwitterAPIPrices}. By mid-2023, the access and permission dimensions of the GATE coefficient had
collapsed for much of the prior academic and civil-society actor population:
\(A_{a, i, t}\) fell toward zero for institutions without commercial budgets,
and \(P_{a, i, t}\) became conditional on payment terms that excluded many
research and public-interest actors
\parencite{Nature2023TwitterAPIResearch, CJR2023TwitterResearch, MurtfeldtEtAl2024TwitterAPI}.

\textbf{The mechanisms that activate.} Proposition C's five mechanisms
operated simultaneously. Recombination exclusion was immediate:
researchers who had built workflows combining Twitter data with other
\(K^D\) stocks found those workflows non-functional. Disclosure
suppression intensified: the Terms of Service changes occurred faster
than excluded actors could redesign around them, preventing the
boundary-mapping that patent enclosure at least formally provides.
Experimentation restriction was severe for precisely the fields at the
frontier of social science methodology: natural-language processing,
network analysis, and computational social science depend on large,
real-time, diverse social-interaction corpora of the kind Twitter's API
uniquely provided. Observation blockade was structurally new: the
deployment-scale feedback from user interactions, what content
spreads, what generates replies, what patterns signal coordinated
behaviour, became exclusively available to the platform, removing the
distributed observation pool that had enabled distributed improvement of
social-media analysis methods.

\textbf{The cascade.} The field contraction extended beyond the Twitter
API element itself. Tools and research pipelines built on the Twitter
API became worthless or required complete reconstruction; the \(K^E\) of
engineers and researchers who had specialized in Twitter-data
methodology depreciated as their primary substrate became inaccessible;
\(K^I\) held by institutions whose research capacity depended on the
combined stack, social science departments, public health monitoring
programmes, fact-checking organizations, was impaired as the
organizational routines governing that capacity became inoperable. The
interoperability cascade removed further elements from \(F_{a, t}\):
datasets, models, and tools that had been built and validated on Twitter
data became difficult to benchmark, compare, or extend once the data
stream was inaccessible. The total field contraction exceeded the direct
removal of the API element.

\textbf{The frontier effect in this case.} The Twitter API was a
frontier element for social science, computational journalism, and
platform analysis, fields where productive weight was concentrated at
recent, high-volume, high-diversity social-interaction data. The
suppression ratio calculated on average field conditions would
understate the loss, because the element removed was not average: it was
among the highest-weight elements in the recombination fields of the
affected actor population. The academic and civil society actors who had
depended on it were operating at \(d_{a, j}\) close to \(d^*_{a}\), the
distance at which the surplus function \(\psi_a\)(d) is maximized. The welfare
loss from the enclosure may therefore be larger than an average-field
calculation would suggest where the field-weight assumptions hold.

\textbf{What standard static welfare analysis sees.} Standard static welfare analysis of the API
repricing would assess a price increase by a platform with market power,
identify users who valued access between \$0 and \$42,000 as the
deadweight loss population, and note that the commercial users who
remained were surplus-plus-transfer. The API closure would appear\index{API closure} as a
distributional event (money moved from users to platform) with a
calculable static efficiency loss.

\textbf{What KBC adds.} Standard static welfare analysis is not designed to capture
the recombination-field contraction channel. The academic
researchers who lost access did not merely lose a consumption good; they
lost a primary input to their knowledge-generation process. The
counterfactual knowledge, the research that would have been
conducted, the methods that would have been developed, the patterns that
would have been discovered, does not appear in any welfare ledger. It
is foregone knowledge capitalization: \(K^D\) that would have been
generated, \(K^E\) that would have been developed, \(K^I\) in research
institutions that would have been built. The suppression ratio
\((\sigma_a^R)^{\eta}\cdot(D_u\ \mathrm{ratio})^{\mu}\) for social
science researchers fell materially; the \(D_u(F_{a, t})\) of their
accessible field fell because a high-weight frontier element was
removed; and the cascade removed further elements. None of this appears
in a standard static welfare analysis. The over-enclosure question, whether the pricing was above \(T^*\), becomes
more explicit in KBC by separating \(C_{T2}\), \(C_{T7}\), \(C_{T6}\),
and \(C_{T8}\).

\emph{Smithian departure.} Smith's accumulation mechanism assumes that
privately directed stock increases productive capacity broadly. In this
case, the enclosure of a productive infrastructure that had been an
accessible recombination input converted a distributed generative
resource into a private capture mechanism. The productive capacity of
the incumbent did not fall; the productive capacity of a large
distributed research and development community did. Smith's framework does not contain this channel; later innovation and information
economics address parts of it, but KBC isolates the recombination-field mechanism.
The goods involved are not rival, the enclosure did not happen through price
competition, and the welfare cost accrues not as higher consumer prices but as
suppressed knowledge generation that never enters any market. The model of the following
chapter (Chapter 7) will show the internal counterpart: while excluded
actors' generation capacity fell, the incumbent's learning-loop capture
was accelerating.

\section{Suppressed Appreciation and Proposition
C2}\label{suppressed-appreciation-and-proposition-c2}

\paragraph{Causal role.} This section tracks the valuation consequence of the lost path. Suppressed appreciation is present when excluded actors would have increased the value of a stock, standard, tool, or field through recombination but cannot do so under the enclosure.

Proposition C identifies generative suppression: enclosure reduces the
rate at which excluded actors generate new knowledge-bearing stock.
Proposition C2 identifies suppressed appreciation\index{suppressed appreciation|textbf}: enclosure reduces the
productive value of non-enclosed knowledge-bearing stock already held by
excluded actors. Suppressed appreciation is a theoretically useful
valuation and risk concept, but its magnitude requires domain-level
estimation of complementarity, field dependence, and the relevant
governance conditions.

\textbf{Proposition C2 (Suppressed Appreciation)}: When
knowledge-bearing stock \(K_a\) held by actor a requires access to enclosed
stock \(K_i\in E\) as a complementary field element, the enclosure of \(K_i\)
reduces v(\(K_a\), a, \(\pi\)), the productive value that actor a can
realize from \(K_a\) under governance state \(\pi\). The suppression of \(K_a\)'s value
does not require that \(K_a\) itself be enclosed; it follows from \(K_i\)'s
removal from \(F_{a, t}\).

A stock's value depends not only on its standalone use but on the surplus
it can generate with accessible complements. The value function, following
the pairwise surplus model:

\begin{equation}
v(K_a, a, \pi)=v_0(K_a, a)+E[S(K_a, F_{a, t}(\pi))]
\label{eq:ch6:v-k-a-a-pi}
\end{equation}

where \(v_0\) is \(K_a\)'s standalone productive value and E{[}S{]} is the
expected recombination surplus from the accessible field. Enclosure of
\(K_i\) removes \(K_i\)'s contribution to E{[}S{]}: \(SA(K_a, a, E)=\sum_{K_j\in E}\omega_j\cdot\chi_{a, j}\cdot\psi_a(d_{a, j})\cdot\rho_{a, j}>0\)
when \(E\) contains stocks with positive complementarity to \(K_a\).

The cascade extension amplifies the direct loss. The interoperability
cascade removes \(K_k\in F\setminus E\) from \(F_{a, t}\) when \(K_k\) depends on \(K_i\)
for technical integration, so v(\(K_a\), a, \(\pi_1\)) may fall by more
than SA(\(K_a\), a, E) once the cascade runs. T7 (Suppressed Appreciation\index{T7 Suppressed Appreciation!cognitive enclosure},
the relevant formal proof file) gives the formal audit trail and states
the cascade amplification term.

Suppressed appreciation has a systematic pattern across the five \(K^x\)
forms. \(K^E\) stocks, embodied in people's expertise and judgment, 
are suppressed when the training environments, cases, or empirical bases
from which that expertise is calibrated are enclosed. A physician's
diagnostic \(K^E\) can lose productive value where clinical workflow becomes
materially dependent on proprietary model outputs and those models are
enclosed; under those conditions, the physician may be less able to
interpret, challenge, or refine model outputs without access to the model's
structure. The claim does not apply where comparable substitutes, transparent
audit tools, or independent clinical standards preserve the physician's effective
field. \(K^D\) held by small firms and independent
researchers is suppressed when platform-dependency governances enclose the
data infrastructure on which their \(K^D\) was developed or through
which it was deployed. \(K^I\) held by communities of practice is
suppressed when professional standards are enclosed within credentialling
bodies that restrict the knowledge base practitioners can draw on. This is
suppression only where the restriction exceeds quality, liability, and
safety justification.

The suppression is differentially severe across the KGM generation
mechanisms. This ranking is qualitative; it identifies proximity to the
suppression channel, not measured magnitude. Recombination (\(G^R\)) is most directly suppressed, because
recombination exclusion\index{recombination exclusion} reduces both inputs to the generation function
simultaneously. Experimentation (\(G^X\)) is suppressed where the
enclosed stock is the substrate for productive experiments. Discovery
(\(G^D\)) is partially suppressed where the phenomena being observed are
mediated by proprietary infrastructure; observations of algorithmically
curated environments require access to the curation architecture to be
correctly interpreted. Invention (\(G^N\)) is suppressed through
cumulative blocking: improvement of enclosed current-generation
knowledge is restricted to the enclosing incumbent. Judgement and
interpretation (\(G^J\)) are least directly suppressed, but they are
still suppressed when actors lose access to the cases, anomalies, model
behaviour, empirical environments, and contradictory evidence through
which judgement is calibrated.

Proposition C2 therefore completes the external suppression argument.
Enclosure does not merely prevent excluded actors from using the enclosed
stock. It can also reduce the productive value of the knowledge-bearing
stock they already hold by severing that stock from the complementary
field elements required for appreciation, interpretation, experimentation,
and recombination. Chapter 7 turns to the internal counterpart of the
same process: the enclosing actor does not merely restrict others'
learning. It concentrates the feedback loop through which its own stock
improves. Chapter~\ref{strategic-enclosure-and-the-smith-nash-problem}
then asks when these enclosure incentives become strategically rational
but socially costly, and Chapter~\ref{chapter-11-testing-knowledge-bearing-capitalism}
returns to Proposition C as a testable empirical claim.

\chapter[Feedback-Enclosure]{Feedback-Enclosure and Learning-Loop
Capture}\label{feedback-enclosure-and-learning-loop-capture}
\index{feedback enclosure|textbf}\index{learning loop}

\chapterhook{AI, Platforms, and the Economics of Cumulative Learning}

The learning loop is use-value regenerating itself. Chapter 1 held that knowledge wealth lives in flow rather than in a static stock; a feedback loop is that flow made cumulative, each use producing the data that improves the next. To enclose the loop is to convert a self-renewing source of use-value into a private asset, and this chapter follows what that does to the parties left outside it.

In feedback-intensive knowledge markets, deployment is not only a way to deliver an existing product. Deployment is also a way to learn. A model, platform, software tool, professional system, or AI service that reaches users generates correction signals, usage traces, failure reports, behavioural responses, ratings, preferences, interaction histories, and performance observations. If those signals are captured and incorporated into the next improvement cycle, use becomes a source of capital accumulation.

The economic mechanism is direct. Deployed systems generate feedback; feedback improves the system; controlled feedback lets the incumbent improve faster than actors who can see only the visible product. A rival may imitate the interface, copy a feature, or observe the market outcome, but it does not automatically receive the private stream of errors, corrections, prompts, searches, complaints, retries, ratings, and workflow adaptations that teaches the deployed system how to improve. The competitive advantage is therefore not only possession of a stock. It is cumulative access to the learning stream produced by that stock in use. This learning-stream advantage is the most distinctive mechanism in this book's account of knowledge-bearing capitalism: unlike the access restrictions of Chapter~6, it is a supply-side compounding effect that ordinary remedies such as interoperability and data portability do not reach, a point developed in Proposition~D and carried into the competition analysis of Chapter~10.

\begin{center}
\fbox{\begin{minipage}{0.92\linewidth}
\textbf{Boxed proposition: The Firms That Learn Fastest Win}

In feedback-intensive knowledge markets, firms that learn fastest can compound capability faster than firms that only imitate the visible product.
\end{minipage}}
\end{center}

This proposition is intentionally narrower than a general claim that the largest firm always wins. Scale matters only when deployment produces useful feedback, when the actor can capture that feedback, and when the feedback can be incorporated into the next version of the stock. Under those conditions, feedback capture becomes a capital-accumulation mechanism: current deployment improves future capability, and future capability increases the quality and volume of the next feedback stream.

This chapter builds on learning-by-doing, absorptive capacity, dynamic capabilities, organizational learning, and AI preference-learning literatures \parencite{Arrow1962,CohenLevinthal1990,NelsonWinter1982,Teece2007,NonakaTakeuchi1995,ChristianoEtAl2017}. Its extension is to treat exclusive access to learning signals as a capital-accumulation mechanism.

\emph{How Enclosed Deployment Concentrates Feedback, Grows
\(\widetilde{C}_{a}\), and Accelerates Incumbent Trajectories}

Chapter 6 established the external effect of cognitive enclosure:
enclosure reduces \(D_u(F_{a,t})\) for excluded actors, suppresses
\(G^R\) through both field magnitude and useful diversity channels, and
extends the field contraction through the interoperability and
capability cascades. The social cost, aggregated in T5 as
\[
C_{\mathrm{enclosure}}(T)=C_{T2}+C_{T7}+C_{T6}+C_{T8},
\]
can be substantially larger than standard IP welfare analysis identifies when enclosure suppresses not only current access, but also recombination, feedback learning, capability accumulation, and trajectory diversity\index{trajectory narrowing}.

Chapter 6 is, however, incomplete in a specific way. It specifies what
enclosure removes from excluded actors without specifying what it gives
to the enclosing incumbent. Enclosure is not only a subtraction; it is
simultaneously an institutional mechanism for concentrating productive
capacity. What the observation blockade removes from the distributed
social learning process, it adds to the incumbent's private learning
process. What the experimentation restriction forecloses for outsiders,
it reserves for insiders. The external suppression is the shadow of a
corresponding internal acceleration.

Chapter 7 develops the internal effect: Proposition D,
feedback-enclosure. When knowledge-bearing stock is deployed within an
enclosed governance arrangement, the feedback signals that deployment generates are
captured within the enclosure, flow into the \(G^L\) mechanism, raise
\(\widetilde{C}_{a}\) for the enclosing actor, and accelerate trajectory
improvement while excluded actors may fall behind relative to the incumbent when substitute feedback, capability maintenance, and recombination access are insufficient.
Together with the external suppression of Chapter 6, this produces the
joint result that T8 names: fewer but faster-improving knowledge
trajectories, the empirical signature of a knowledge economy
operating under strong enclosure.

The formal bridge is \(G^L\), the Knowledge Generation Model's learning-loop mechanism. \(G^L\) denotes knowledge generated when deployment produces feedback that is captured, interpreted, and incorporated into later capability. T6 formalizes the resulting learning-loop capture\index{T6 Learning-Loop Capture!formalization}\index{learning-loop capture!T6}\index{learning-loop capture|textbf} claim inside this chapter: feedback capture can produce deployment differentials, capability divergence, productive-range divergence, generation-function divergence, capability traps as limiting cases under strong assumptions, and recovery lag\index{recovery lag}s after access is reopened. The Technical Companion, Appendix C, §C.8.3, further formalizes this learning-speed mechanism as Derived Proposition~C.1, the Dynamic Recombination-Range Recursion. The associated recursion equation shows why present access to the recombination field, useful diversity, and learning feedback jointly determine future productive range. Thus exclusion does not merely reduce current access; it can reduce future capability.

This is not a simple welfare ranking. The feedback advantage may represent genuine productive improvement, better error correction, safer deployment, and faster iteration by the incumbent. The moderation rule is that enclosure has dual effects: it can increase private incentive and incumbent learning speed while reducing third-party recombination fields, trajectory diversity, and future generative capacity. Whether the net effect is desirable depends on stock centrality, redundancy, complementarity, capability distribution, duration, and feedback capture.

The causal story is equally direct. What is enclosed is the feedback stream generated through deployment: corrections, usage traces, error reports, behavioural responses, ratings, preferences, interaction histories, and performance signals. The excluded actors may observe the product but cannot observe, reuse, or incorporate the private learning signals that improve it. What is lost is the cumulative learning path that would have allowed rivals, users, commons contributors, or downstream complementors to improve their own models, routines, or capabilities. The incumbent gains faster learning, error correction, adaptation, model improvement, and capability-gap expansion. The claim would be weakened if comparable alternative feedback channels remained available, if feedback capture improved the incumbent without narrowing rival learning opportunities, if capability rather than feedback fully explained the advantage, or if no measurable trajectory divergence followed the enclosure event.

\paragraph{Observable indicators.} The feedback-capture claim is testable only if it is connected to observable proxies. Evidence supporting the mechanism would include high user-interaction volume, privileged telemetry access, or explicit feedback-use disclosure\index{feedback-use rights}s; faster release cadence, benchmark improvement, error reduction, or safety-patch learning relative to comparable rivals; widening frontier gaps between the incumbent and alternative systems; API closures, telemetry restrictions, or terms-of-service changes that redirect learning signals into a controlled channel; and declining third-party extensions, forks, plugins, or compatible tools after access is restricted. Evidence weakening the claim would include comparable learning by excluded actors through alternative telemetry, open datasets, substitute user bases, shared benchmarks, public research, interoperability, hiring, or capital investment. Evidence falsifying the strong version would be a case where feedback is enclosed, feedback is material to capability improvement, excluded actors lack substitute channels, and yet no relative capability divergence, recombination loss, or trajectory concentration follows. Governance interventions are observable as well: audit rights\index{audit rights}, portability obligations, contractual data-use limits, feedback-use opt-outs, or mandated disclosure of learning-loop use should reduce the divergence if feedback governance is the relevant mechanism.

\section{Feedback-Enclosure Defined}\label{feedback-enclosure-defined}
\index{feedback enclosure}

\paragraph{Causal role.} This section identifies the enclosed object as a learning stream rather than a finished asset. The key question is who can observe, reuse, and incorporate the feedback generated by deployment.

Feedback-enclosure occurs when the feedback generated by deployment is captured by the actor controlling the access layer and incorporated into the next improvement cycle. The object enclosed is not only the model, code, data, or platform. It is the learning stream produced by use.

Formally, feedback-enclosure is the institutional process by which the learning
signals generated through the deployment of knowledge-bearing stock are
captured within the enclosing governance arrangement rather than distributed across the
accessible knowledge base. The definition has three elements that must
be held separately.

The first element is deployment. Knowledge-bearing stock generates
learning signals only through deployment, through interaction with
real-world conditions, users, problems, and failure modes that no
pre-deployment evaluation can fully anticipate. Deployment is the
generative act that reveals what the knowledge stock can and cannot do,
at what rates and in what contexts. Without deployment at sufficient
scale to encounter the productive tails of the use distribution, the
feedback signal is too thin to drive systematic improvement.

The second element is capture. Not all deployment generates learning
signals that the deploying actor can capture and use. Open deployment, such as releasing a model under an open licence, generates
signals from a distributed user population, but the developer observes
only what users voluntarily report unless telemetry, reporting, benchmark, or governance systems are deliberately built; the private signals embedded in use
patterns, reformulations, and failure-recovery strategies are not
systematically available by default. Enclosed deployment, through a proprietary
API, a controlled platform interface, or an enterprise integration with
data sharing provisions, allows the enclosing actor to observe the
full pattern of use interactions, capture the feedback signals those
interactions generate, and convert them into improvements in the
deployed system.

The third element is the institutional loop. Capture alone does not
produce feedback-enclosure; the captured signals must flow back into the
system's improvement process. The feedback-enclosure mechanism is
complete only when the captured signals are incorporated into the
training and refinement process that governs the next deployment cycle.
The institutional loop connects deployment (signal generation) \(\to\)
enclosure (signal capture) \(\to\) improvement (signal incorporation) \(\to\) next
deployment (signal generation again). The loop is self-reinforcing: each
cycle starts from a higher capability baseline than the last, generates
a richer signal from a more capable deployment, and produces a larger
improvement than the previous cycle could have produced. Each iteration
deepens the advantage.

\textbf{Definition 7.1} (Feedback-enclosure): An institutional
arrangement in which (i) knowledge-bearing stock \(K_i\) is deployed
exclusively through an access layer controlled by the enclosing actor
class \(A_inc\); (ii) the feedback signals \(F_{dep}\) generated by that
deployment, performance observations, use patterns, interpretive
corrections, distributional data, are captured within the enclosure;
and (iii) captured signals flow into the \(G^L\) mechanism that improves
\(K_i\) or raises \(\widetilde{C}_{inc}\). Excluded actor class \(A_ex\)
loses access to the feedback generated by the enclosed deployment stream
and therefore cannot observe, contribute to, or benefit from the incorporation
of those signals. Excluded actors may still learn from substitute, adjacent,
public, or independently generated feedback channels.

\emph{Incumbent}, throughout this chapter, refers to an actor that
controls a knowledge-bearing stock, access layer, or feedback stream in
a position sufficient to shape the conversion conditions faced by other
actors. The term is positional, not evaluative: incumbency is defined by
structural position relative to the feedback loop, not by the normative
character of the decisions that established that position.

\section{From Access Control to Learning-Loop
Capture}\label{from-access-control-to-learning-loop-capture}

\paragraph{Causal role.} This section shows the conversion from access control to capability divergence. Access control matters because it channels deployment through the incumbent, concentrating feedback volume, feedback quality, and improvement rights.

The connection between access control and learning-loop capture is not
immediate; it runs through deployment scale. Access control determines
who deploys the enclosed stock; deployment scale determines how much
feedback is generated; feedback volume and quality determine how much
improvement each training cycle can achieve. The chain is:

\begin{equation}
\begin{aligned}
\text{access control} &\to \text{deployment exclusivity}
\to \text{feedback volume and quality} \\
&\to G^L \to \Delta\widetilde{C}_a
\to \text{trajectory acceleration}
\end{aligned}
\label{eq:ch7:access-control}
\end{equation}

Access control without deployment is inert: a firm that holds exclusive
rights to a stock but does not deploy it at scale generates no feedback
and captures no learning. This is the standard patent hold-out case, 
exclusivity without use, which generates the \(C_{T2}\) welfare cost of
Chapter 6 without generating the internal acceleration of Chapter 7.
Feedback-enclosure requires both exclusivity and deployment.

Deployment without feedback capture is Arrow's distributed
learning-by-doing: many producers each accumulate experience, and the
aggregate learning is reflected system-wide. Arrow's formulation is
expansive; Proposition D identifies what happens to Arrow's mechanism
under enclosure. The same accumulation of productive experience occurs,
but it is concentrated within the enclosing actor rather than
distributed across the production system. Learning-by-doing is
preserved; its social distribution is suppressed.

Feedback capture without the loop back to improvement is data hoarding
rather than learning-loop capture. A firm that collects interaction data
but does not use it systematically to improve the deployed system
accumulates a large dataset without generating the capability growth
that Proposition D describes. The mechanism requires closure of the
loop: captured signals must become inputs to the next training or
refinement cycle. Where that loop is closed, as it is in systems that
use reinforcement learning from human feedback, continuous fine-tuning
from deployment data, or systematic capability evaluation against
live-use failure modes, the feedback-enclosure mechanism is active.

The feedback signal generated by deployment has four analytically distinct types, each contributing differently to \(G^L\). Table~\ref{tab:ch7-feedback-types} separates these signals so the mechanism does not collapse into a general claim that ``feedback matters.''

\begin{table}[H]
\centering
\caption{Four feedback types in feedback-enclosure}
\label{tab:ch7-feedback-types}
\begin{tabularx}{\textwidth}{@{}L{0.24\textwidth}X@{}}
\toprule
\textbf{Feedback type} & \textbf{Contribution to \(G^L\)} \\
\midrule
Performance feedback & Reveals where and how the deployed system succeeds and fails against real-world inputs, calibrating capability against the actual use distribution rather than only against pre-specified evaluation sets. \\
Use-pattern feedback & Reveals what capabilities users actually demand, in what proportions and combinations, including uses the developers did not anticipate and that expose the system's productive envelope. \\
Interpretive feedback & Captures the revealed judgments users provide through acceptance, correction, or rejection of system outputs, including standards that cannot be fully specified in advance. \\
Distributional feedback & Reveals where outputs cluster across the possibility space: what the system can reliably produce, where its gaps are, and where the boundary conditions define the frontier of productive extension. \\
\bottomrule
\end{tabularx}
\end{table}

Each type generates a distinct \(G^L\) contribution; together they constitute the feedback signal \(F_{dep}\) in FP-10.

\subsection[Boundary: data-enabled learning and the data economy]{Boundary: Data-Enabled Learning and the Data Economy}\label{boundary-data-enabled-learning}

This boundary extends to the modern data-economics literature, which formalizes parts of the learning-loop mechanism directly and must therefore bound the originality of Proposition~D. \textcite{HagiuWright2023}\index{Hagiu and Wright} model dynamic competition between firms that improve their products by learning from customer data, distinguishing \emph{across-user} learning, in which a firm pools data across many customers, from \emph{within-user} learning, in which a firm improves its product for a given customer through that customer's repeated use. They identify three features that separate data-enabled learning from classical learning-by-doing: learning raises customers' willingness to pay rather than lowering marginal cost; improvement can be specific to an individual customer; and the product can improve while it is still being consumed. \textcite{FarboodiVeldkamp2021}\index{Farboodi and Veldkamp} formalize the firm-level data feedback\index{data feedback} loop in which more data raises product quality, which raises production and transactions, which generate more data, and show that this loop can produce a firm growth trap in which a data-poor firm stays poor. \textcite{JonesTonetti2020} establish that data nonrivalry generates increasing returns, that firms hoard data for fear of creative destruction, and that this hoarding produces an inefficient use of a nonrival good.

Proposition~D accepts each of these results. The incumbent-side compounding advantage of exclusive deployment feedback is, in capital-theoretic dress, the mechanism \textcite{HagiuWright2023} and \textcite{FarboodiVeldkamp2021} already describe; the firm growth trap is a recognizable predecessor of the capability trap developed in Section~\ref{the-formal-channel-gl-and-widetildec_a} (\(\widetilde{C}_{ent}\!\to\!\widetilde{C}_{min}\)); and the hoarding inefficiency of nonrival data is the static counterpart of the dynamic suppression this book develops. The within-/across-user distinction is adopted here directly: Cell~4b feedback, in which the act of using the enclosed system is the act of generating the improvement signal, is the within-user case in its most enclosable form, because the contributor cannot separate use from contribution.

These predecessors also discipline the welfare claim. Data-enabled learning may generate efficient outcomes under some competitive conditions; the existence of a private feedback loop is not, by itself, evidence of social loss. The KBC claim is narrower: feedback-enclosure becomes socially costly when the private feedback loop also excludes rival learning, narrows recombination fields, blocks superior entrants, suppresses trajectory diversity, or shifts data externalities onto users or third parties. In that case the relevant harm is not learning by the incumbent, but the governance condition that converts learning into excluded-field suppression and a reduction in future generative alternatives.

What none of these accounts models is the consequence of feedback capture for actors \emph{outside} the loop. \textcite{HagiuWright2023} characterize the advantage that accrues to the learning firm in a bounded competitive setting; \textcite{FarboodiVeldkamp2021} trace a firm's own life-cycle; \textcite{JonesTonetti2020} evaluate the static allocation of a nonrival input. Proposition~D's distinctive object is the conversion event itself: deployment feedback is a governance-mediated conversion of natural-intelligence interaction (\(K^{E}\) signals) into firm-controlled knowledge-bearing stock, captured at the access layer~\texttt{[A]}, whose enclosure narrows the recombination field \(D_u(F_{a,t})\) for an excluded actor class (Proposition~C) and lowers the trajectory count \(N_{traj}\) for the system as a whole (T8). The contribution claimed here is therefore not the feedback-learning advantage, which is established, but its integration with excluded-field suppression and trajectory narrowing into a single conversion architecture.

This integration also inherits a built-in falsifier. \textcite{LambrechtTucker2017}\index{Lambrecht and Tucker}, applying the resource-based view, argue that amassing data alone is not a durable competitive advantage, because data is rarely rare or inimitable, substitutes exist, and standalone value is low; durable advantage requires complementary organizational capability. This is not a refutation of Proposition~D but a confirmation of Section~\ref{the-formal-channel-gl-and-widetildec_a} and of Proposition~B (Capability-Bounded Codification): feedback-enclosure compounds an advantage only where the enclosing actor also holds the complementary \(K^{E}\) and \(K^{I}\) required to close the loop, captured here through \(\widetilde{C}_{a}\). The boundary condition yields a sharp falsifier. If the feedback advantage reliably dissipates wherever the complementary capability is commoditized or freely available, then the strong form of Proposition~D fails, and feedback-enclosure reduces to ordinary data hoarding without trajectory consequences.\footnote{Originality status: feedback-learning advantage established/extended (\textcite{HagiuWright2023}; \textcite{FarboodiVeldkamp2021}); feedback-enclosure as a conversion pathway synthesized/extended; feedback capture plus excluded-field suppression plus trajectory narrowing potentially novel as integrated architecture.}

\section[The formal channel: feedback learning and dynamic capability]{\texorpdfstring{The Formal Channel: \(G^L\) and \(\widetilde{C}_{a}\)}{The formal channel: feedback learning and dynamic capability}}\label{the-formal-channel-gl-and-widetildec_a}

\paragraph{Causal role.} This section names the lost pathway as feedback learning. The excluded actor loses not only access to the stock but participation in the learning loop that would have raised its future capability.

The formal mechanism connecting feedback-enclosure to trajectory
acceleration runs through two objects introduced in Chapter 3: the
\(G^L\) (Learning Feedback Loop) mechanism and the dynamic capability
\(\widetilde{C}_{a}\).

\textbf{\(\boldsymbol{G^L}\) learning-loop equation}: The \(G^L\) mechanism formalizes the
learning-by-doing process that Arrow identified.\footnote{FP-10 is the Technical Companion's proof-system label for the feedback-capture premise; it is not a separate Volume 1 axiom.}

\begin{equation}
G^L_{a,t}=\alpha_a\cdot Dp_{a,t}\cdot F^{dep}_{a,t}\cdot\phi_a
\label{eq:ch7:g-l-a-t}
\end{equation}

where \(\alpha_{a}\) is actor a's absorptive capacity, the fraction
of available feedback that the actor's existing knowledge base can
productively integrate; \(Dp_{a,t}\) is deployment scale at time t (the
volume of interactions generating feedback); \(F_{dep,a,t}\) is the
quality and volume of the feedback signal from deployment; and
\(\phi_{a}\) is the feedback capture rate, the fraction of generated
feedback that the actor's institutional arrangements allow it to observe
and incorporate. Empirical proxies include active users, interaction volume, labelled correction events, retained sessions, evaluation failures, telemetry coverage, audit logs, and the share of captured feedback incorporated into model, product, or process updates.

\begin{figure}[!htbp]
\caption[Feedback capture as a learning flywheel]{Feedback capture as a learning flywheel}
\label{fig:ch7:feedback-capture-flywheel}
\centering
\makebox[\textwidth][c]{\includegraphics[width=1.06\textwidth]{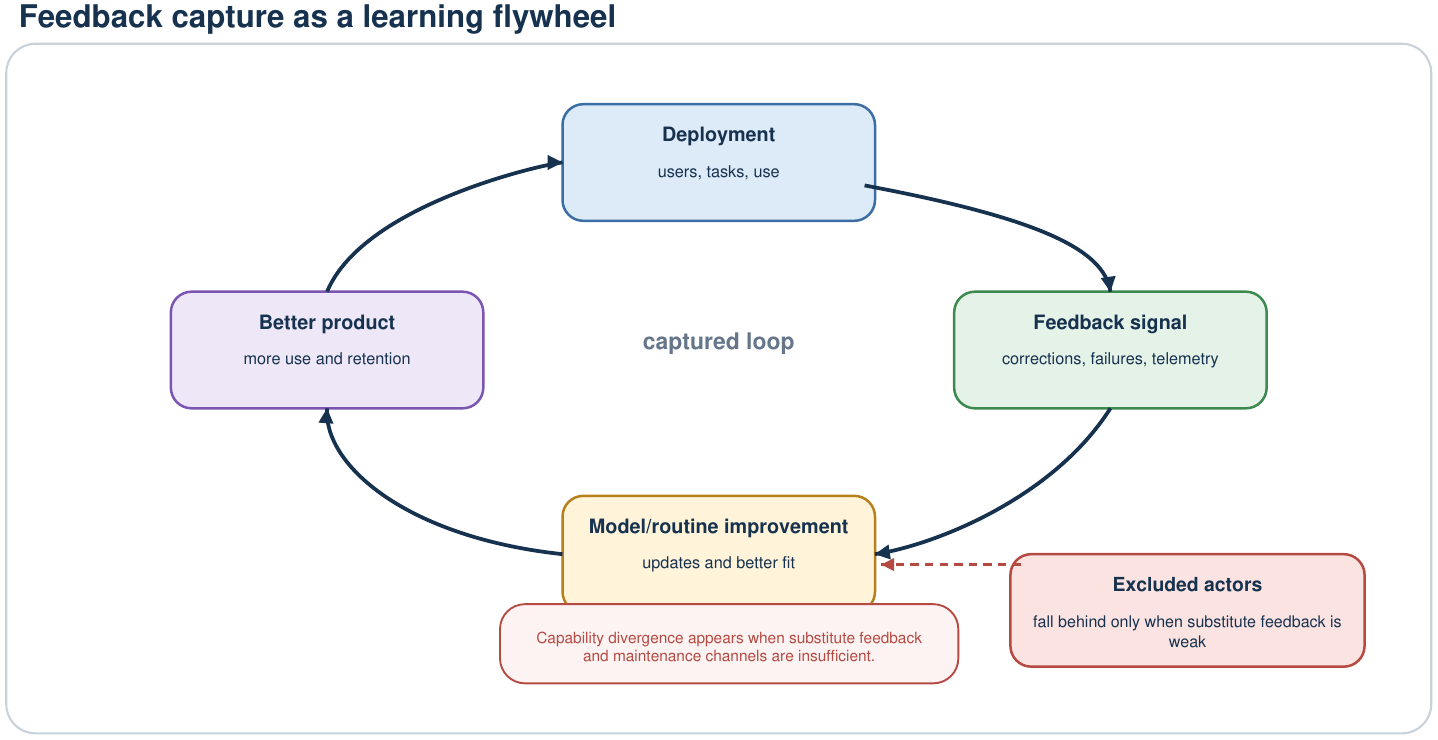}}
\par\smallskip\noindent\footnotesize\emph{Note.} Captured deployment feedback can improve the next model, routine, or process, which can attract more deployment and generate more feedback. Excluded actors fall behind only when substitute feedback and capability-maintenance channels are insufficient.
\end{figure}

Under feedback-enclosure, the deployment differential is direct for the enclosed stock: the
enclosing actor inc has \(Dp_{inc}(K_i)>0\) while the
excluded actor ent has \(Dp_{ent}(K_i)=0\) for that stock. Therefore, for the enclosed deployment stream:

\begin{equation}
G^L_{inc,t}=\alpha_{inc}\cdot Dp_{inc}\cdot F^{dep}_{inc}\cdot\phi_{inc}>0
\label{eq:ch7:g-l-inc-t}
\end{equation}
\begin{equation}
G^{L,K_i}_{ent,t}=0\qquad\text{for the enclosed deployment stream during the enclosure}
\label{eq:ch7:g-l-ent-t}
\end{equation}

The excluded actor does not experience learning-loop suppression in the
sense that all learning capacity is reduced to zero; it experiences
learning-loop denial\index{learning-loop denial} in the narrower sense that this feedback source is
structurally unavailable. Its total learning term may remain positive if
substitute, adjacent, public, or independently generated feedback channels exist.

\textbf{The dynamic capability state equation}: Chapter 3 introduced
\(\widetilde{C}_{a}\) as the actor's dynamic capability, the
knowledge endowment that conditions productive range, field magnitude,
and recombination output. The state equation established in Chapter 3
(and grounded formally in Model 4):

\begin{equation}
\widetilde{C}_{a,t+1}=(1-\delta_C)\widetilde{C}_{a,t}+\gamma G^R_{a,t}+\ell G^L_{a,t}
\label{eq:ch7:widetilde-c-a-t-1}
\end{equation}

where \(\delta_C\) ∈ (0,1) is the capability depreciation rate,
\(\gamma\) ∈ (0,1{]} is the contribution of recombination generation to
capability growth, and \(\ell\) ∈ (0,1{]} is the contribution of
deployment learning to capability growth (with \(\ell\) ≤ \(\gamma\),
since deployment learning deepens existing range more than it extends
reach to genuinely new domain distances).

Under feedback-enclosure, the two paths diverge. For the enclosing
incumbent:

\begin{equation}
\widetilde{C}_{inc,t+1}=(1-\delta_C)\widetilde{C}_{inc,t}+\gamma G^R_{inc,t}+\ell G^L_{inc,t}
\label{eq:ch7:widetilde-c-inc-t-1}
\end{equation}

Both generation mechanisms contribute positively;
\(\widetilde{C}_{inc}\) grows or at minimum is sustained above the
depreciation threshold. For the excluded actor:

\begin{equation}
\widetilde{C}_{ent,t+1}=(1-\delta_C)\widetilde{C}_{ent,t}+\gamma G^R_{ent,t}+\ell G^{L,alt}_{ent,t}
\label{eq:ch7:widetilde-c-ent-t-1}
\end{equation}

With \(G^{L,K_i}_{ent}=0\) for the enclosed deployment stream and \(G^R_{ent}\) suppressed through the field
contraction of Chapter 6, the remaining learning term \(G^{L,alt}_{ent,t}\) may still sustain the actor if alternative telemetry, open datasets, a substitute user base, hiring, capital investment, shared benchmarks, public research, or interoperability provide usable substitutes. Divergence becomes strong only when these substitute channels are insufficient to offset depreciation. If (\(\gamma\) · \(G^R_{ent,t}\)) + (\(\ell\) · \(G^{L,alt}_{ent,t}\)) \textless{}
\(\delta_C\) · \(\widetilde{C}_{ent,t}\), \(\widetilde{C}_{ent}\)
may decline each period. As \(\widetilde{C}_{ent}\) declines, the optimal combination distance \(d^*_{ent}\) contracts, narrowing the useful
field \(F^u_{ent}\) and further suppressing \(G^R_{ent}\). The downward process is therefore conditional, not automatic: capability decay narrows
the productive field only when substitute learning and capability maintenance fail to compensate for the enclosed-stream loss. This limiting case is the capability trap,
formalized in T6.5.

\textbf{The productive range divergence}: As \(\widetilde{C}_{inc}\)
grows, \(d^*_{inc,t}=\alpha\widetilde{C}_{inc,t}/\beta\) increases: the incumbent can productively reach knowledge
elements at greater distance from its existing base, widening its useful
field and increasing both \(R_{inc,t}\) and \(D_u(F_{inc,t})\). If
\(\widetilde{C}_{ent}\) contracts, \(d^*_{ent}\) decreases: the excluded
actor's productive reach narrows toward its own knowledge base, reducing
both \(R_{ent,t}\) and \(D_u(F_{ent,t})\). The relevant claim is relative productive-range divergence, not automatic absolute collapse. Feedback-enclosure amplifies the field contraction of Chapter 6 only when the loss of enclosed-stream feedback is not adequately replaced: Chapter 6 removes elements
from \(F_{a,t}\) through the access restriction; Chapter 7 can narrow the
useful range of elements within \(F_{a,t}\) through capability decay under the specified insufficiency condition.

\section{Proposition D:
Feedback-Enclosure}\label{sec:ch7:proposition-d-feedback-enclosure}

\paragraph{Causal role.} Proposition D states the internal acceleration mechanism: enclosed feedback increases incumbent capability while rivals lose the enclosed-stream learning path. The claim is false if comparable feedback is available outside the enclosure or if no capability divergence follows.

\textbf{Proposition D (Feedback-Enclosure)}: When knowledge-bearing
stock is deployed within an enclosed governance arrangement, the feedback signals
generated by that deployment are captured within the enclosure rather
than distributed across the accessible knowledge base. The capture is
self-reinforcing: as the incumbent's \(\widetilde{C}\) grows through the
\(G^L\) mechanism, the gap between the enclosed trajectory and
accessible alternatives widens, increasing the informational advantage
the incumbent holds and reducing the probability that alternatives can
close the gap through independent effort. Feedback-enclosure therefore
makes enclosure not merely self-sustaining but self-deepening: each
improvement cycle generates additional feedback that justifies continued
enclosure, while simultaneously making the return to open alternatives
more costly and less likely.

\textbf{The self-reinforcing structure}: The claim of self-reinforcement
requires elaboration, because self-reinforcing mechanisms are easy to
assert and difficult to bound. The mechanism operates through the
interaction of three components. First, \(G^L_{inc}\) \textgreater{} 0
raises \(\widetilde{C}_{inc}\), which widens \(d^*_{inc}\) and expands
the useful field. Second, the wider useful field increases \(R_{inc}\)
and \(D_u(F_{inc})\), raising \(G^R_{inc}\). Third, higher \(G^R_{inc}\)
contributes further to \(\widetilde{C}_{inc,t+1}\) through the
\(\gamma\) term. The loop is: \(G^L\to\widetilde{C}\) up, \(d^*\) up, field wider, \(G^R\) up, and \(\widetilde{C}\) further up. This is a
compounding loop, not merely an additive one.

The self-reinforcement has a natural ceiling: \(\widetilde{C}_{a,t}\) ≤
\(\widetilde{C}_{max}\) is a boundary condition of the capability system
(Model 4, M4.2). The incumbent's trajectory acceleration does not
compound indefinitely; it decelerates as \(\widetilde{C}_{inc}\)
approaches the capability ceiling. But in knowledge-bearing capitalism
as described, this ceiling is not typically binding in the medium run,
because the frontier of what constitutes advanced capability advances as
new knowledge-bearing stock is generated. An incumbent improving its
\(\widetilde{C}\) along an active knowledge frontier does not face a
fixed ceiling; it faces a ceiling that moves outward as the frontier
advances, and its feedback capture is precisely the mechanism that moves
its capability toward the frontier as fast as the frontier moves.

\textbf{The asymmetry with Chapter 6}: Proposition C (Chapter 6)
operates through a pull mechanism: enclosure pulls field elements out of
excluded actors' accessible range, reducing their generation capacity.
Proposition D operates through a push mechanism: feedback capture pushes
the incumbent's capability upward, raising its productive range and
generation output. The two mechanisms are related but analytically
distinct, and their interaction is not simply additive. As
\(\widetilde{C}_{inc}\) grows and \(d^*_{inc}\) expands, the incumbent's
productive field extends into territory that was previously beyond its
useful range, including territory where its feedback-generating
deployment provides information about user demand at the frontier. This
expansion is driven by the incumbent's own improvement, not by access
restriction, and it takes place independently of whether the enclosure
of specific field elements (the Proposition C mechanism) is operating.
The two mechanisms together narrow the gap between incumbent and
excluded actors' generation capacity more than either alone.

\textbf{Feedback capture vs.~network effects}: Feedback capture is
distinct from network effects and must not be conflated with them,
because the distinction determines the governance response. Network
effects are demand-side: more users make the technology more valuable to
each user, through the coordination externality that Shapiro\index{Shapiro and Varian} and Varian\index{Varian, Hal}
analysed and that underpins much of the platform economics literature.
Feedback capture is supply-side: more deployment generates more
feedback, which improves the system, which attracts more deployment.
Network effects can in principle be addressed through interoperability:
if users can reach each other across platforms, the coordination
externality is achieved without platform monopoly. Feedback capture
cannot be addressed through interoperability in the same way, because
the productive asset at stake is not access to the user network but the
knowledge generated by observing and learning from the user network. A
data portability mandate that requires the incumbent to share raw
interaction logs does not transfer the improvement knowledge the
incumbent has generated from analysing those logs: that knowledge is
encoded in model weights, architectural refinements, and training
procedure improvements that are not in the input data. The governance
challenge of feedback capture requires governance of the
deployment-feedback-training cycle, not merely of the initial training
inputs.

\section[Conditional learning-loop capture result]{T6: Conditional Learning-Loop Capture Result\index{conditional learning-loop capture result|textbf}}\label{t6-learning-loop-capture}

\paragraph{Causal role.} T6 formalizes why feedback capture compounds. The causal burden is to show that captured signals improve the next stock, that improved stock attracts further deployment, and that further deployment generates better signals.

This book-level claim of Proposition D is formalized in Theorem T6
(Conditional Learning-Loop Capture), whose full proof is in
the relevant formal proof file. This section states
the practical result of T6 and its role in this theory.

T6 converts the \(G^L \to \widetilde{C}_{a}\) channel from a
qualitative claim into a fully traceable formal result, using the
dynamic capability system of Model 4 in which \(\widetilde{C}_{a}(t)\)
is an endogenous state variable governed by the state equation of §7.3.
The result is six conditional claims for all \(\tau\geq 1\) under sustained
enclosure governance state \(\pi_1\). T6 is conditional on informative feedback, sufficient absorptive capacity, weak substitute feedback, and incorporation into future capability. These are divergence claims, not claims that excluded actors lose all learning. Strong divergence is weakened when excluded actors can access alternative telemetry, open datasets, a substitute user base, hiring, capital investment, shared benchmarks, public research, or interoperability:

\textbf{T6.1 (Deployment differential)}: \(G^L_{inc,t}\) \textgreater{}
0 and \(G^{L,K_i}_{ent,t}=0\) for the enclosed deployment stream during the enclosure period. This
follows directly from FP-10 and the \(Dp_{ent}(K_i)=0\) condition of
Definition 7.1. The claim does not imply that the excluded actor has no learning at all; it means the excluded actor receives no learning feedback from the enclosed deployment stream.

\textbf{T6.2\index{T6.2 Capability Divergence} (Capability divergence)}: The capability gap \(\Delta_t=\widetilde{C}_{inc,t}-\widetilde{C}_{ent,t}\) grows over the enclosure period when the incumbent's captured feedback and recombination gains exceed the excluded actor's substitute feedback, recombination, and capability-maintenance gains. The gap widens because the incumbent's
\(\widetilde{C}\) grows through both \(G^R\) and \(G^L\), while the
excluded actor's \(\widetilde{C}\) is sustained by \(G^R\), substitute learning channels, and capability-maintenance investments. The strong form requires these alternatives to be insufficient when Chapter 6's field contraction is severe.
This result is the formal counterpart to the governance-conditioned
asymmetry of Interlude I: \(\mathbb{E}[\Delta K_{i,t}\mid\Gamma_{enc}]>\mathbb{E}[\Delta K_{j,t}\mid\Gamma_{enc}]\) now has a
specific mechanism (the \(G^L\) differential) not merely a
governance-conditioned claim.

\textbf{T6.3 (Productive range divergence)}: \(d^*_{inc,t}\)
\textgreater{} \(d^*_{inc,0}\), while \(d^*_{ent,t}\) may fall below
\(d^*_{ent,0}\) for \(t\geq 1\) if substitute feedback and capability maintenance are insufficient. The incumbent's productive reach expands as
\(\widetilde{C}_{inc}\) grows; the excluded actor's relative productive reach contracts only when \(\widetilde{C}_{ent}\) decays. The two movements are not
symmetric: the incumbent's expansion is driven by a compounding loop,
while the excluded actor's contraction is driven by the interaction of
the loss of \(G^{L,K_i}_{ent}\) from the enclosed stream with the field contraction of Chapter 6.

\textbf{T6.4 (Generation function divergence)}: \(G^R_{inc,t}\) /
\(G^R_{ent,t}\) \textgreater{} \(G^R_{inc,0}\) / \(G^R_{ent,0}\) for \(t\geq 1\). The generation rate ratio diverges over time when the incumbent's captured-feedback advantage is not offset by substitute feedback, public inputs, or capability investment; the incumbent's
recombination output then grows relative to the excluded actor's. This is the
formal expression of the ``faster-improving trajectories'' part of the
T8 joint prediction: the incumbent improves faster not only because it
has exclusive access to the enclosed frontier but because its
\(\widetilde{C}\) grows while the excluded actor's \(\widetilde{C}\) grows more slowly, stagnates, or contracts under the stated insufficiency conditions.

\textbf{T6.5 (Capability trap)}: \(\widetilde{C}_{ent,t}\to\widetilde{C}_{min}\) as \(t\to\infty\) only if \(G^{L,K_i}_{ent,t}=0\), substitute learning \(G^{L,alt}_{ent,t}\) is insufficient, capability-maintenance investment fails, and
\(G^R_{ent,t}\to 0\). The capability trap is a limiting case, not the ordinary case: if the
field contraction of Chapter 6 is severe enough to drive \(G^R_{ent}\)
toward zero, and substitute learning channels cannot offset capability depreciation, the state equation
approaches \(\widetilde{C}_{ent,t+1}\approx(1-\delta_C)\widetilde{C}_{ent,t}\), which is a contracting sequence converging to
the capability floor \(\widetilde{C}_{min}\). Below
\(\widetilde{C}_{min}\), the excluded actor cannot productively use any
element of its remaining field; it has lost the capability to generate
even from accessible stocks. This outcome is not predicted where alternative telemetry, open datasets, a substitute user base, hiring, capital investment, shared benchmarks, public research, or interoperability sustain \(G^{L,alt}_{ent}\) and \(G^R_{ent}\).

\textbf{T6.6 (Recovery lag)}: When enclosure ends at \(t_R\) and governance
reverts to \(\pi_0\), the capability gap does not immediately close. The
gap at \(t_R + \tau\) is:

\begin{equation}
\Delta(t_R+\tau)=\Delta(t_R)(1-\delta_C)^\tau+
\sum_{s=0}^{\tau-1}
\left[(\gamma\Delta G^R_s+\ell\Delta G^L_s)(1-\delta_C)^{\tau-1-s}\right]
\label{eq:ch7:delta-t-r-tau}
\end{equation}

The gap closes at rate \(1-\delta_C\) per period after reversion,
but the initial gap \(\Delta(t_R)\) may be large, the recovery duration
\(\tau_R\) is increasing in the length of the enclosure period, and the
capability-trap limiting case (T6.5) means that longer enclosures can leave deeper
starting gaps when substitute feedback and capability maintenance are insufficient. This is the formal basis for the \(C_{T6b}\) cost
component in T5: the welfare cost of enclosure extends beyond the
enclosure term itself through the recovery lag.

\emph{Smithian departure:} Smith's capital accumulation is a story in
which any actor's successful investment expands the productive
conditions for others, more capital maintains more labour, more
labour produces more output, more output generates further saving. T6 is
the structural inversion of this story for knowledge-bearing capital
under enclosure: the incumbent's capability can grow through captured deployment feedback (T6.1--T6.4) while excluded actors may fall behind, and may decay when substitute feedback, capability maintenance, and recombination access are insufficient (T6.5). Both effects are
caused by the same mechanism, the deployment-feedback loop, which
concentrates rather than distributes learning. In Smith's model, one
actor's accumulation does not reduce another actor's productive capacity
through the accumulation mechanism itself. In T6, it does.

\emph{Smithian departure:} Smith's division of labour produces
productive diversity as a by-product of specialization, more tasks,
more skills, more routes to productive participation in the market
economy. T8 predicts the opposite under strong enclosure: enclosure
reduces \(N_{traj}\), the count of actor-indexed approaches to the
knowledge frontier, not because specialization fails but because the
feedback-capture advantage incumbents accumulate forecloses the
conditions under which diverse generation trajectories could be
sustained. Smith's division of labour was the engine of expanding
productive participation. T8's trajectory contraction is the engine of
contracting participation in knowledge generation, a narrowing of
productive diversity precisely at the frontier where new generation is
most consequential.

T6 is Chapter 7's counterpart to Chapter 6's Section~6.3: where Section~6.3 showed
how enclosure reduces \(D_u(F_{a,t})\) and suppresses \(G^R\) for
excluded actors, T6 shows how enclosure raises \(\widetilde{C}_{inc}\)
through \(G^L\) and accelerates \(G^R_{inc}\) for the enclosing actor.
The two results together establish the capability divergence that T8
formalizes as the ``fewer but faster'' prediction.

\section{AI Systems as the Paradigmatic
Case}\label{ai-systems-as-the-paradigmatic-case}

\paragraph{Causal role.} AI systems make the mechanism visible\index{AI systems!feedback enclosure} because deployment itself produces training, correction, preference, safety, and performance signals. The incumbent gains model improvement and capability acceleration; excluded actors lose comparable learning data and iteration speed.

Chapter 4's anchor Cell 4b, the fifth anchor cell of the Knowledge
Conversion Matrix, is the paradigmatic formal case of
feedback-enclosure: \([\text{Feedback}\to\text{revised }K^D\text{ weights}]\), following
from the simultaneous conversion of Cell 4a. Its pathway notation is
\([A]\to[T]+[D]\to G^L\) loop, in which an access-mediated event
(user query through the incumbent's API) triggers a simultaneous
transformation and distributional appropriation (the interaction updates
model weights under conditions controlled by the incumbent) that loops
back into the next deployment cycle.

Cell 4b captures the structural feature that makes AI system deployment
one of the most concentrated instances of feedback-enclosure in current
knowledge-bearing capitalism. In many architecturally integrated AI
systems, use and contribution are difficult to separate. Where terms of
service, telemetry, and model-update architecture permit feedback
incorporation, ordinary use can also generate improvement signals for
the provider. When a user queries such a deployed AI system, the user
uses the system, consuming its outputs, relying on its capability, and
pursuing the goal for which they accessed it. At the same time, the
query, reformulations, acceptance or rejection of outputs, and the
continuation or abandonment of the session may generate feedback signals
that flow into the system's training or revision pipeline. The KBC claim
is therefore conditional: use becomes contribution only where governance
and architecture allow deployment feedback to be captured and
incorporated into future capability.

The self-reinforcing structure of Cell 4b combines three mechanisms that
individually produce lock-in and collectively produce a form that
compounds through time. The Cell 4a simultaneous conversion creates
a deep initial enclosure, combining legal exclusivity over model weights
with architectural access control through the API. The Proposition D
feedback capture concentrates the deployment-generated learning signal
within the enclosure, continuously improving the enclosed model. And the
path dependence that Arthur's increasing returns analysis identifies, 
early improvement advantages compound into capability gaps across
subsequent improvement cycles, amplifies the divergence described in T6.2: the capability gap grows, and the growing gap makes
displacement increasingly difficult.

The AI training cycle makes the \(G^L\) term in the state equation
particularly large. For a deployed language model receiving billions of
queries per day, \(Dp_{inc}\) is enormous; the diversity of \(F_{dep}\)
is broad, covering the full use distribution; and \(\alpha_{inc}\), 
absorptive capacity, is high because the model's existing \(K^D\)
provides a rich prior for interpreting and incorporating feedback. The
\(G^L\) contribution to \(\widetilde{C}_{inc,t+1}\) in systems of this
scale is not a marginal increment; it is a substantial term in the state
equation, capable of driving \(\widetilde{C}_{inc}\) upward faster than
the depreciation term erodes it even if \(G^R\) were suppressed. This is
why the observation blockade mechanism of Chapter 6 is most
consequential in the platform-dependency governance: what is blocked from
excluded actors is not merely access to the model weights (the
\((K^{D})_{\mathrm{platform}}\)) but the feedback signal that makes each version of
the model better than the last.

\textbf{Mutual amplification versus one-way conversion.} LLM platforms sharpen this feedback-enclosure problem because expert use can convert \(K^E_{\mathrm{expert}}\) into \(K^D_{\mathrm{model}}\) and then into platform capability. The governance question is whether the learning loop returns capability to the expert actor or is enclosed by the platform. In the mutual-amplification case, expert feedback improves the model while the expert also receives better outputs, retained local learning, or improved institutional capability:

\[
K^E_{\mathrm{expert}}
\rightarrow
K^D_{\mathrm{model}}
\rightarrow
G^L_{\mathrm{actor}} + G^L_{\mathrm{platform}}.
\]

In the one-way conversion case, expert judgement, correction, contradiction, and refinement improve the platform's model or institutional stock while little capability returns to the expert actor:

\[
K^E_{\mathrm{expert}}
\rightarrow
K^D_{\mathrm{model}}
\rightarrow
K^I_{\mathrm{platform}},
\qquad
G^L_{\mathrm{platform}}>0,
\quad
G^{L,\mathrm{returned}}_{\mathrm{actor}}\approx 0.
\]

The difference is not whether user interaction creates feedback. The difference is who receives the capability gain from the feedback. Where the loop is reciprocal, the relation is mutually amplifying. Where the loop is enclosed, the relation becomes one-way conversion of embodied expertise into platform-controlled disembodied and institutionalized knowledge capital. This completes the feedback-enclosure point: the actor who monopolizes deployment feedback also monopolizes the antithetical counter-frame stream from which dialectical generation can occur, while the expert whose contradictions train the system may receive little frame-revising capability in return.

The governance implication follows directly from the distinction between
network effects and feedback capture. Policy focused on data portability
requiring incumbents to share training data or interaction logs with
competitors, addresses recombination exclusion (Proposition C) but
does not address feedback capture (Proposition D), because the
productive knowledge the incumbent generates from its deployment is not
stored in the raw data; it is generated through the interaction between
the data, the deployed model, and the continuous flow of live-deployment
feedback. Governance of feedback capture requires governance of the
deployment-feedback-training loop itself, not merely of its inputs.

\paragraph{Governance instruments.} The practical implication is limited and institutional rather than programmatic: feedback-enclosure is not solved by a general duty to share models, data, or interaction logs. It is governed by instruments that make the learning loop visible, contestable, or bounded. Feedback-use disclosure makes learning-loop capture observable\index{feedback-use disclosure}; audit rights let customers, contracting parties, or regulators verify whether deployment feedback is incorporated into later systems; enterprise data-use restrictions prevent customer activity from silently becoming training input for the vendor; feedback portability standards\index{feedback portability standards} let users move improvement-relevant signals where such movement is technically feasible and privacy-preserving; opt-in or opt-out rules determine whether use implies contribution to future model capability; benchmark disclosure makes capability divergence more observable; and competition-review logging preserves evidence of feedback-based exclusion. These instruments should also include safety, privacy, cybersecurity, anti-abuse, and trade-secret exceptions\index{trade secrecy}, because the point is not overbroad mandatory sharing. The point is to govern the conditions under which use becomes enclosed learning.

\paragraph{Non-AI illustration: cybersecurity telemetry.} The same mechanism operates outside generative AI. Endpoint-security vendors, cloud providers, identity platforms, and managed-detection services improve partly because deployed systems generate incident telemetry: attempted intrusions, false positives, anomalous logins, privilege-escalation patterns, malware behaviours, failed detections, containment outcomes, and analyst corrections. When that telemetry is visible only to the provider controlling the access layer, the provider can incorporate it into future signatures, detection rules, identity-risk scores, automated playbooks, and response routines. Customers and rival providers may receive a better product, but they may not receive the underlying learning stream that explains why the product improves.

This is feedback-enclosure rather than ordinary data ownership. The relevant asset is not merely the logs, alerts, or malware samples; it is the conversion of deployed incident feedback into improved detection capability. Divergence is not automatic: excluded actors may still learn from public vulnerability reports, shared indicators of compromise, open-source detection rules, government advisories, substitute telemetry, or their own incident base. But where one provider observes a much broader live threat surface than its rivals or customers, and where contractual or technical arrangements prevent others from auditing or reusing that learning stream, the Chapter 7 mechanism applies without any generative-AI example. The cybersecurity case also explains why the strategic-stealth problem in Section~\ref{strategic-stealth-and-learning-loop-denial} matters: hidden exploitation denies defenders the feedback needed to revise controls, while concentrated defensive telemetry can accelerate the actor that sees the widest incident field.

\subsection[A worked demonstration: ChatGPT deployment and learning-loop divergence]{\texorpdfstring{A Worked Demonstration: ChatGPT Deployment and the \(G^L\) Divergence}{A worked demonstration: ChatGPT deployment and learning-loop divergence}}\index{ChatGPT}\index{OpenAI!ChatGPT}\index{AI systems!ChatGPT}\label{a-worked-demonstration-chatgpt-deployment-and-the-gl-divergence}

Section 7.6 identifies AI systems as the paradigmatic case of
feedback-enclosure. The ChatGPT/open-weight\index{open-weight models} comparison is not a
controlled test. The actors differ in compute, capital, distribution,
safety infrastructure, talent, brand, organizational capability, and
user base. It is useful here only because it makes the feedback
architecture visible. The period following ChatGPT's deployment
(November 2022) and the simultaneous development of open-weight
alternative systems (LLaMA series, Mistral, and their derivatives)
therefore provides an illustrative natural contrast, not an empirical
identification strategy. The point is not that feedback capture alone
explains capability differences, or that one class of model is generally
superior. The point is that different governance positions around
deployment feedback create different opportunities to convert use into
future capability.

\textbf{The knowledge-bearing stocks.} At the point of ChatGPT's
release, the two populations drew from a common ancestry. GPT-3.5, the
base for the initial ChatGPT deployment, derived from the transformer
architecture whose foundational elements were established in the open
field described in Chapter 3's worked demonstration. LLaMA (February
2023, Meta AI) was a comparable architecture trained at comparable scale
with publicly documented methods. The Chinchilla scaling laws \parencite{Hoffmann2022} had been published; both populations knew how to trade off
parameters and training tokens for optimal performance; architecture
differences were incremental rather than fundamental. What differed was
deployment position and the feedback-enclosure arrangement that
deployment position enabled.

\textbf{The governance change.} ChatGPT accumulated approximately 100
million users in its first two months of operation (January 2023), a
deployment scale unprecedented for a language model. Under Definition
7.1, this scale was the primary determinant of \(G^L\): \(Dp_{inc,t}\)
the deployment scale entering the \(G^L\) equation, was enormous.
Each user interaction generated feedback signals: which queries users
continued, which outputs they corrected, which reformulations they
tried, which sessions they abandoned. These signals constituted a
real-time measure of the system's productive envelope, where it
succeeded, where it failed, and where the frontier of user demand lay.
Under OpenAI's architecture, those signals were captured through the
API's instrumentation and fed back into subsequent fine-tuning, RLHF
update cycles, and model alignment calibration. \(\phi_{inc}\), the
feedback capture rate, was high by construction: an API-mediated
deployment with contractual terms governing data use gave the incumbent
full observation of the feedback stream.

For the LLaMA population, the arrangement was structurally different.
LLaMA weights were released for research use; downstream fine-tuning was
possible; community contributors created instruction-tuned variants
(Alpaca, Vicuna, and others) from small synthetic datasets. But the
fine-tuning signals available to this population were thin relative to
the incumbent's live-deployment signal: synthetic datasets could not
replicate the heterogeneous real-world feedback from 100 million users
pursuing 100 million distinct tasks. \(Dp_{ent,t}\) for community
fine-tuners was orders of magnitude smaller than \(Dp_{inc,t}\) for
OpenAI. \(G^L_{ent,t}\) was not zero, it was positive through
community fine-tuning experiments, but it was structurally bounded
below the incumbent's \(G^L\).

\textbf{The KBC mechanism.} The formal channel is exactly as Section~\ref{the-formal-channel-gl-and-widetildec_a}
specifies. For the incumbent:

\begin{quote}
\(\widetilde{C}_{inc,t+1}=(1-\delta_C)\widetilde{C}_{inc,t}+\gamma G^R_{inc,t}+\ell G^L_{inc,t}\)
\end{quote}

Both \(\gamma\) and \(\ell\) terms contribute positively: \(G^R\)
through continued model development drawing on the growing research
literature, \(G^L\) through deployment-scale feedback incorporation. For
the open-weight population:

\begin{quote}
\(\widetilde{C}_{ent,t+1}=(1-\delta_C)\widetilde{C}_{ent,t}+\gamma G^R_{ent,t}+\ell G^L_{ent,t}\)
\end{quote}

With \(G^L_{ent,t}\ll G^L_{inc,t}\), the excluded population's
capability growth depended primarily on \(G^R\), on architectural and
training innovations it could develop and publish, while the
incumbent's capability growth drew on both. The state equation predicts
a widening capability gap \(\Delta_t=\widetilde{C}_{inc,t}-\widetilde{C}_{ent,t}\) over the enclosure period when the incumbent's feedback-learning gains exceed the rival population's recombination, substitute-feedback, and capability-maintenance gains, consistent with T6.2. Performance comparisons between closed deployed systems and open-weight alternatives are therefore treated here as suggestive only. They are consistent with the feedback-governance mechanism, but they do not isolate it from differences in compute, capital, distribution, talent, architecture, data, safety systems, and organizational capability.

\textbf{Value created, captured, and made invisible.} The value captured
by the incumbent is not monopoly rent on a static product. It is the
capability improvement generated by deployment-scale feedback, 
improvement encoded in model weights and alignment data that do not
appear on any balance sheet as a separately recognized asset. The
\(G^L\) contribution to \(\widetilde{C}_{inc,t}\) is dark capital:
economically real (it drives the capability gap), not recognized in any
accounting category (it is not capitalized R\&D, not an identifiable
intangible, not a recognized training-data asset), and not visible to
any investor who reads a standard financial report.

The feedback-learning value unavailable to the open-weight population is the \(G^L\)
contribution its capability state might have received under equivalent deployment-scale feedback. The recovery lag (T6.6) is
directly applicable: even if incumbents released all fine-tuning
datasets, the accumulated alignment calibration embedded in current
model weights (the processed output of the feedback signal) would
not transfer. The capability gap is not undone by data sharing; it is
undone only by closing the feedback-capture asymmetry.

\textbf{What standard theory sees.} Scale economies in compute;
first-mover advantage in user acquisition; proprietary training data.
Policy interest centres on data access, particularly training-data
provenance and licensing, and on whether the dominant actor is using
its market position to unfairly exclude competitors.

\textbf{What KBC sees that standard theory misses.} The
deployment-feedback loop is not a data asset. It is a \(G^L\) mechanism
that continuously improves \(\widetilde{C}_{inc}\) while
\(\widetilde{C}_{ent}\) falls behind, not because of any single
data-access restriction, but because of the structural arrangement that
gives \(Dp_{inc}\) its scale. A data-portability mandate, the
standard policy intervention for data concentration problems, 
addresses the initial training input (\(K^D\) in the pre-deployment
configuration) while leaving the feedback architecture intact. The
\(G^L\) differential persists because the mechanism that drives it is
not only the training data but the deployment arrangement that generates
the feedback. This is the governance lesson of the example. Data-access
reform may affect the initialization of capability, but feedback-loop
governance determines whether use itself becomes a continuing source of
incumbent capability. That distinction is not available to any
theoretical framework that does not distinguish \(G^R\) from \(G^L\) as
separate and formally distinct generation mechanisms. It is precisely
this distinction that Chapter 7's governance implication depends on.

\emph{Smithian departure:} Smith's model has no mechanism for a form of
capital investment whose primary productive consequence is the
generation of feedback signals that improve the investing actor's
capability while making the same deployment-scale signals unavailable to
other actors on equivalent terms. Deployment-scale feedback is not
analogous to any accumulation mechanism Smith analysed: it is not
saving, not the division of labour, and not the extension of the market.
It is a form of productive investment whose distributional logic is
bifurcation rather than expansion: present accumulation can increase the
incumbent's capability while making the same accumulation path harder
for the excluded population to reproduce.

\section{Strategic Stealth and Learning-Loop
Denial}\label{strategic-stealth-and-learning-loop-denial}

\paragraph{Causal role.} This section tracks the hidden version of the mechanism. The loss is difficult to see because outsiders observe the product but not the private feedback stream that explains why the incumbent's trajectory accelerates.

The mechanisms described in Sections~7.3--7.6 can operate through pure market
structure without any deliberate strategic action: an incumbent with a
large user base, a closed API, and an active training pipeline will
capture feedback and improve faster than excluded actors even if no one
in the firm explicitly intends to deny learning to competitors. But the
structural advantage creates an incentive for deliberate strategic
action that deepens the feedback-enclosure beyond what market structure
alone would produce.

Strategic stealth is the practice of deploying knowledge-bearing stock
in ways that minimize the information available to excluded actors about
the stock's properties, capabilities, and improvement trajectory. A firm
that discloses its model architecture provides information that helps
excluded actors replicate, extend, or circumvent the model's
capabilities; a firm that treats its architecture as a trade secret
forces excluded actors to infer the architecture's properties from
external behaviour, which is slower and less accurate. Strategic stealth
extends beyond architecture: the decision not to publish benchmark
results, not to release evaluation datasets, not to disclose fine-tuning
methods, and not to publish improvement trajectories all reduce the
information available to excluded actors about where the enclosed
system's productive frontier lies and how fast it is moving.

This claim requires a legitimacy boundary. Some feedback concealment is
legitimate when it protects privacy, cybersecurity, anti-abuse systems,
safety, trade secrets, legal compliance, or investment incentives. It
becomes analytically problematic when concealment prevents affected
actors from knowing that their use is improving a system they cannot
access, audit, contest, or benefit from. The KBC point is therefore not
that all concealed feedback systems are suspect. The point is narrower:
concealment becomes part of feedback-enclosure when it prevents affected
or excluded actors from observing, evaluating, or governing the learning
stream that their activity helps produce.

Learning-loop denial is the active counterpart: deliberate architectural
or institutional choices that prevent excluded actors from observing the
feedback signals generated by the enclosed deployment, even feedback
that excluded actors could in principle infer from observing external
behaviour. The most significant form is platform architecture that
prevents third-party observation of user interaction patterns: if
excluded actors could observe which queries users are making to the
incumbent's system, which outputs users accept or correct, and how the
system's behaviour changes over time in response to the correction
pattern, they could infer much of the feedback signal that the incumbent
is capturing. Platform architectures that prevent this observation, 
through opaque APIs, encrypted interaction logs, and contractual
prohibitions on interaction recording, actively deny the feedback
signal to excluded actors rather than merely retaining it.

The strategic dimension creates a connection between Proposition D and
the Model 5 analysis of strategic enclosure equilibria (Chapter 8,
the relevant formal proof file). In the
strategic enclosure framework, incumbents choose not just what to
enclose but how to deploy the enclosure to maximize the trajectory
divergence advantage described in T6.2. The Smithian-inversion concern is not that every privately rational feedback-capture strategy is socially
harmful. It is that, under the T5/T8 conditions, privately rational
capture may concentrate learning inside one trajectory while weakening
substitute learning paths, independent validation, and trajectory
diversity. The incumbent internalizes the benefit of accelerated
improvement (higher \(\widetilde{C}_{inc}\)) but may not internalize the
cost of excluded actors' relative capability loss or, in the limiting
case, capability decay (lower \(\widetilde{C}_{ent}\)) or the trajectory
diversity loss (lower \(N_{traj}\)).

The strategic stealth\index{strategic stealth} and learning-loop denial practices also have
implications for the invisibility of the depreciation mechanism.
Interlude I identified invisible depreciation, productive value loss
unobservable to accounting, IP records, and access indicators, as a
central feature of knowledge-bearing stock dynamics. Strategic stealth can make the depreciation not merely invisible to accounting systems but
actively concealed from excluded actors: they cannot observe whether their
\(\widetilde{C}\) is declining relative to the incumbent's, cannot
observe the rate of the incumbent's improvement, and cannot assess the
magnitude of the capability gap they are trying to close. The exclusion
from the feedback loop and the exclusion from observing the feedback
loop are compounding forms of epistemic disadvantage.

\section{Joint Result with Fewer but Faster
Trajectories}\label{joint-result-with-chapter-6-fewer-but-faster-trajectories}

\paragraph{Causal role.} This section combines the two enclosure effects: Chapter 6 removes third-party recombination paths, while Chapter 7 accelerates the incumbent's learning path. Under strong enclosure, and where excluded trajectories cannot replace the lost recombination and feedback channels, the causal outcome can be fewer viable trajectories, but faster improvement inside the enclosed trajectory.

With the external effect of Chapter 6 and the internal effect of Chapter
7 both established, the joint prediction of the framework can be stated
in conditional form. It is the result that T8 names: under strong
enclosure, and where excluded trajectories cannot replace the lost
recombination and feedback channels, knowledge-bearing capitalism can
generate fewer but faster-improving knowledge trajectories. Aggregate
\(G^R\) can fall despite incumbent acceleration when lost trajectories
and suppressed diversity exceed incumbent gains. 

\textbf{Trajectory diversity is a net term, not a free good.} A fall in \(N_{traj}\) is not by itself a social cost. Convergence on a dominant design can be efficient: it ends duplicated parallel effort and can unlock standardization and network benefits that fragmented variety forecloses. The path-dependence\index{path-dependence literature} literature cuts both ways, showing both that markets can lock in to inferior standards and foreclose better alternatives \parencite{David1985,Arthur1989} and that premature or excessive variety can waste resources before a dominant design emerges. The welfare-relevant quantity is therefore the \emph{net} trajectory effect: the option value of the foreclosed trajectories, the expected value of the alternative paths enclosure removes weighted by the probability that one of them would have outperformed the dominant path, minus the coordination, standardization, and duplication-avoidance savings from convergence. A lower \(N_{traj}\) is a social cost only when that option value exceeds those savings. This is the condition under which \(C_{T8}\) is positive; it is an empirical question, not a property of variety as such.

\textbf{The external mechanism (Chapter 6)} produces fewer trajectories.
Enclosure of \(E\subset F_{a,t}(\pi_0)\) reduces \(D_u(F_{a,t})\) for
excluded actors, suppresses \(G^R\) through the suppression ratio
\((\sigma_a^R)^{\eta}\cdot(D_u\ \mathrm{ratio})^{\mu}\), and extends
the contraction through the cascade. The trajectory count
\(N_{traj}(\pi_1)<N_{traj}(\pi_0)\) because
excluded actors lose actor-indexed recombination paths that the
incumbent's retention of the same abstract pair (\(K_i\), \(K_j\)) cannot
substitute, distinct trajectories arise from distinct knowledge
contexts \(K_a\), and enclosure eliminates the contexts.

\textbf{The internal mechanism (Chapter 7)} produces faster trajectories
for incumbents. Feedback capture raises \(\widetilde{C}_{inc}\) through
\(G^L\), widens \(d^*_{inc}\), expands \(R_{inc}\) and \(D_u(F_{inc})\),
and raises \(G^R_{inc}\). At the same time, excluded actors'
\(\widetilde{C}_{ent}\) may stagnate or decay when enclosed-stream \(G^L\) is unavailable, substitute feedback is insufficient, and \(G^R\) is
suppressed, narrowing \(d^*_{ent}\) and contracting both \(R_{ent}\)
and \(D_u(F_{ent})\). The generation rate ratio \(G^R_{inc}\) /
\(G^R_{ent}\) diverges over time (T6.4).

\textbf{The net result}: T8 Lemma T8.3 states formally that the
aggregate loss from excluded actors' trajectory contraction exceeds the
aggregate gain from incumbent acceleration, under the conditions of
T2.4, the distributed \(G^R\) loss exceeds the concentrated
Δ\(G^R_{inc}\). This is the critical constraint on the Schumpeterian
argument: Schumpeter is right that incumbents accelerate along their
surviving trajectories, and Arrow is right that the diversity of
trajectories falls. T8's contribution is to show that both are
simultaneously consistent and that the welfare-relevant quantity, 
aggregate knowledge generation capacity, falls net. The
Schumpeter-Arrow ambiguity is not a disagreement about facts; it is a
disagreement about which dimension of the same phenomenon matters for
welfare.

The joint prediction, fewer trajectories, faster within-trajectory
improvement, lower aggregate \(G^R\), is the empirical signature of a
knowledge economy under strong enclosure. Three observable patterns
follow. First, sectors with stronger enclosure and significant feedback
capture should show fewer distinct knowledge trajectories, lower
recombination diversity, more concentrated patterns of citation and
technical building, than comparably complex sectors without strong
enclosure. Second, the rate of improvement within dominant trajectories
should be faster in high-enclosure, high-feedback-capture sectors, as
incumbents' concentrated \(G^L\) accelerates their advancement. Third,
the rate of trajectory replacement, the frequency with which new
entrants displace incumbents by introducing fundamentally different
approaches, should be lower in these sectors, because new
trajectories cannot be established without access to the frontier that
enclosure blocks and feedback capture accelerates past.

The prediction is falsifiable. If strong enclosure does not produce
trajectory concentration, if recombination diversity remains high
despite significant IP coverage and closed deployment, the mechanism
described in Chapter 6 is not operating at the scale this theory claims.
If strong feedback capture does not produce faster within-trajectory
improvement relative to accessible alternatives, if incumbents
improve no faster than open-access competitors with similar starting
capabilities, Proposition D is not the primary driver of capability
divergence. These tests distinguish the feedback-enclosure mechanism
from the two principal alternative explanations: that concentration
reflects scale economies in compute and research talent (big-is-better),
or that it is a transient early-market phenomenon that will resolve as
technology matures (early-mover advantage). This theory predicts that
concentration deepens through the feedback loop as enclosure
accumulates, rather than reversing as technology diffuses.

\textbf{Feedback enclosure and winner-take-most dynamics.} The
compounding loop established in Proposition D, \(G^L\) raises
\(\widetilde{C}_{inc}\), which expands the useful field, which raises
\(G^R_{inc}\), which raises \(\widetilde{C}_{inc}\) further, produces
a specific, falsifiable prediction about market structure that
distinguishes the feedback-capture account from two common alternatives.

The feedback-capture account predicts three observable patterns in
sectors where deployment-scale feedback is significant and deployment is
enclosed: (1) the generation rate ratio \(G^R_{inc}\)/\(G^R_{ent}\)
diverges over time rather than converging, because the incumbent's
\(\widetilde{C}\) grows while excluded actors' \(\widetilde{C}\) may stagnate or decay when substitute feedback and capability maintenance are insufficient;
(2) the trajectory count \(N_{traj}\) falls as excluded actors lose the
capability required to pursue distinct knowledge-generation approaches,
while the incumbent's single dominant trajectory improves faster; and
(3) the rate of trajectory replacement, the frequency with which new
entrants displace incumbents through fundamentally different approaches
falls, because closed deployment prevents entrants from accessing
the deployment-scale feedback that drives improvement along any
trajectory.

What distinguishes this from the \emph{big-is-better} account
(concentration reflects scale economies in compute and research talent)?
The big-is-better account predicts convergence as the sector matures and
economies of scale equalize; the feedback-capture account predicts
continued divergence as long as enclosure and deployment-scale feedback
persist, because the mechanism is architectural, not size-dependent.
What distinguishes it from the \emph{first-mover advantage} account
(early leaders set standards and switching costs)? The first-mover
account predicts lock-in that can in principle be broken by a superior
alternative; the feedback-capture account predicts that the alternative
can rarely be built, because building it requires the deployment-scale
feedback that the incumbent's enclosure blocks. The prediction is not
merely that winner-take-most dynamics exist in AI, genomics, and
platform-mediated markets. It is that those dynamics should be most
pronounced precisely in the sectors where (i) deployment at scale
generates rich feedback signals, (ii) that feedback drives capability
improvement, and (iii) access to deployment is architecturally
concentrated.

\emph{Lineage note.} The trajectory-diversity concept has a genealogy
that runs from Smith through specialization economics. Smith's division
of labour produces productive diversity as a by-product of
specialization: distinct roles, skills, and productive identities emerge
from the progressive decomposition of tasks. The specialization economics of \textcite{Wernerfelt2016}\index{Wernerfelt, Birger} gives this a firmer economic foundation:
specialization generates actors with distinct productive profiles, 
distinct \(K^E\) and \(K^I\) configurations, whose combination
creates joint surplus that neither could produce alone. The diversity of
specializations is not merely a by-product; it is a productive resource.
KBC's trajectory-diversity concept (\(N_{traj}\)) formalizes the
economy-level expression of this: each actor-indexed approach to the
knowledge frontier is the productive expression of a distinct knowledge
context, itself the product of distinct specialized development. The
welfare loss identified by T8, aggregate \(G^R\) falling as
\(N_{traj}\) contracts under the specified enclosure conditions, is, in
specialization-economics terms, the enclosure-era negative externality
of the specialization gains that Smith and Wernerfelt identified as
central to economic productivity. Enclosure does not simply restrict
access to existing knowledge; it eliminates the diverse knowledge
contexts from which distinct trajectories could have emerged.\index{access restriction versus trajectory loss}\index{existing knowledge access versus trajectory emergence}
Specialization economics operating at the firm level cannot see this
cost, because the loss is visible only at the economy level, which is
precisely where T8 operates.

\section{The Double-Acting Asymmetry: Why Feedback-Capture Is Not Anticommons}\index{feedback capture versus anticommons}\index{anticommons blockage!versus feedback capture}\label{the-double-acting-asymmetry}

\paragraph{Causal role.} This section isolates what is genuinely new in the feedback-capture mechanism and separates it from the established results it is sometimes mistaken for. The claim is that feedback-capture is the one enclosure mechanism in this framework that acts twice from a single governance choice.

The anticommons and cumulative-innovation literatures already describe enclosure that blocks. Fragmented exclusion rights raise the cost of assembling the inputs a follow-on innovator needs \parencite{HellerEisenberg1998}; cumulative-research models show that strong early rights can deter the later inventors who must build on protected foundations \parencite{Scotchmer1991}; and sequential-innovation models show that imitation can be socially valuable precisely because it keeps later innovation alive \parencite{BessenMaskin2009}. In every one of these accounts, exclusion is \emph{single-acting}: it withholds an input from rivals. It does not, through the same act, make the excluding party a faster generator of new knowledge. The excluder is left with a rent and a blocked rival, not with a compounding capability advantage produced by the blocking itself.

Feedback-capture is different because it is\index{feedback capture!double-acting instrument} \emph{double-acting}. Enclosing the deployment feedback loop is a single governance choice that produces two effects at once. It forecloses rivals' generation, the Proposition~C channel: the enclosed-stream learning path leaves excluded actors' recombination field, lowering \(D_u(F_{a,t})\) and suppressing their \(G^{R}\). And it accelerates the incumbent's own generation, the Proposition~D channel: the same deployment concentrates the feedback signal, raising the incumbent's capability through \(G^{L}\). One instrument, two effects, on opposite sides of the market. No anticommons model contains this, because in those models the thing withheld, an exclusion right, is not also the thing that teaches the withholder how to improve. In knowledge-bearing capitalism the deployed stock is both the gate and the sensor.

\paragraph{The compounding divergence.} The dynamic capability law of Chapter~3, \(\widetilde{C}_{a,t+1}=(1-\delta_C)\widetilde{C}_{a,t}+\gamma G^{R}_{a,t}+\ell G^{L}_{a,t}\), makes the consequence precise. Capability widens the useful field \(d^{*}_{a}\), which raises \(D_u(F_{a,t})\) and hence \(G^{R}_{a}\); near an operating point write that loop as \(G^{R}_{a}\approx\kappa\,\widetilde{C}_{a,t}\), so the law becomes
\[
\widetilde{C}_{a,t+1}\approx(1-\delta_C+\gamma\kappa)\,\widetilde{C}_{a,t}+\ell\,G^{L}_{a,t}.
\]
Under feedback enclosure the incumbent receives \(G^{L}_{\mathrm{inc}}>0\) while the excluded actor receives \(G^{L}_{\mathrm{ent}}\approx 0\). The capability gap \(\Delta_t\equiv\widetilde{C}_{\mathrm{inc},t}-\widetilde{C}_{\mathrm{ent},t}\) then evolves as
\[
\Delta_{t+1}\approx(1-\delta_C+\gamma\kappa)\,\Delta_t+\ell\,\bigl(G^{L}_{\mathrm{inc}}-G^{L}_{\mathrm{ent}}\bigr).
\]
Two dynamic cases follow. When recombination-driven accumulation outpaces depreciation, \(\gamma\kappa\ge\delta_C\), the homogeneous factor is at least one and the gap \emph{widens without bound}: the advantage compounds. When \(\gamma\kappa<\delta_C\), capability mean-reverts and the gap instead converges to a positive steady state
\[
\Delta^{*}=\frac{\ell\,\bigl(G^{L}_{\mathrm{inc}}-G^{L}_{\mathrm{ent}}\bigr)}{\delta_C-\gamma\kappa}>0.
\]
Either way the gap is positive and persistent, and it is increasing in the feedback differential \(G^{L}_{\mathrm{inc}}-G^{L}_{\mathrm{ent}}\). This is the formal content of ``fewer but faster'': the incumbent's faster trajectory is \(\Delta_t\) (or \(\Delta^{*}\)); the falling \(N_{traj}\) is the count of excluded actors whose \(\widetilde{C}\) drops below the threshold needed to sustain a distinct approach to the frontier. The result is conditional, not automatic: it requires \(G^{L}_{\mathrm{inc}}>G^{L}_{\mathrm{ent}}\) (genuine feedback exclusivity) and a recombination loop \(\kappa\) strong enough to matter; where substitute feedback is available to entrants, \(G^{L}_{\mathrm{inc}}-G^{L}_{\mathrm{ent}}\to 0\) and the gap collapses.

\paragraph{Why the asymmetry is sharper for artificial intelligence.} Four features make AI deployment the paradigm case rather than one example among many.

First, \textbf{deployment-proportional feedback}: feedback volume scales with the deployed base, \(G^{L}_{\mathrm{inc}}=\phi(\mathrm{Dep}_{\mathrm{inc}})\) with \(\phi'>0\). Enclosure that grows adoption grows the feedback that grows capability that grows adoption. The loop is endogenous and compounding, unlike a patent, whose informational content is fixed at filing.

Second, the \textbf{non-rival-signal, rival-capture wedge}: the correction, preference, and error signal a user generates is non-rival in principle, it could improve every model, but the deployment architecture makes it rival in capture, only the operator observes it. The externality is the wedge between the signal's social availability and its private capture. This is categorically distinct from the anticommons, which concerns fragmented \emph{exclusion rights}, not captured \emph{learning}.

Third, the \textbf{receding ceiling}: the capability ceiling \(\widetilde{C}_{\max}\) is not fixed. The frontier advances as new stock is generated, and feedback capture is the mechanism that tracks the moving frontier, so the divergence need not saturate at any technological level.

Fourth, an \textbf{epistemic-quality channel} that couples to truth-dependence. As AI output increasingly re-enters training corpora, the reliability coefficient \(\tau_i\) (Chapter~2) can decay. An incumbent that controls the live feedback loop can preferentially curate high-\(\tau\) human signal, while actors denied the loop fall back on lower-\(\tau\) synthetic or scraped data. Feedback capture then widens the gap through quality, not only volume: \(\Delta_t\) is amplified by a factor increasing in \(\tau_{\mathrm{inc}}/\tau_{\mathrm{ent}}>1\). This recursive truth-decay dynamic is not developed further here; it is flagged as the natural next extension of the model.

\paragraph{What would distinguish and what would falsify.} The double-acting claim is testable against the established single-acting baselines, and it makes a prediction that separates it from demand-side network effects.

\begin{itemize}
\tightlist
\item
  \emph{Widening, not constant, gap.} Among AI firms, the capability gap (iteration speed, benchmark trajectory, release cadence) should \emph{widen} with deployed-base differences, and widen faster where feedback-capture exclusivity is higher, after controlling for compute, model size, and research headcount. A constant or closing gap is evidence against compounding.
\item
  \emph{Interoperability test.} A data-portability or interoperability mandate should \emph{not} close the capability gap, because the productive asset is supply-side learning encoded in weights and routines, not access to the user network. If such a mandate does close the gap, the advantage was a network effect, not feedback capture. This is the cleanest discriminating test.
\item
  \emph{Substitution falsifier.} If entrants with comparable compute but no enclosed-stream feedback routinely close the gap, the advantage was substitutable and the double-acting claim fails for that sector \parencite{LambrechtTucker2017}.
\end{itemize}

\paragraph{Originality, stated conservatively.} The feedback-learning advantage itself is established and is credited to the data-economics literature \parencite{HagiuWright2023,FarboodiVeldkamp2021,JonesTonetti2020}. What this section claims as potentially novel is narrower: the identification of feedback-capture as a \emph{single double-acting instrument} that forecloses rival generation and accelerates incumbent generation at the same time, a structure absent from the anticommons and cumulative-innovation models \parencite{HellerEisenberg1998,Scotchmer1991,BessenMaskin2009} that otherwise look similar. The contribution is the architecture, not the advantage.

\paragraph{Why this is the sharpest engine of the Smithian inversion.} In the anticommons the present loss and the future loss are the same loss: foregone use. The Smithian inversion is a stronger claim, that present \emph{gain} and future \emph{loss} are produced together, and feedback-capture is where that joint production is most explicit. The very act that builds the incumbent's present capability, monopolizing the learning stream, is the act that erodes the future generative base by starving excluded actors' \(\widetilde{C}\) and collapsing \(N_{traj}\). Accumulation strengthens present production by weakening future generation. That is the inversion in its purest mechanical form, and it is why feedback-capture, not the welfare gap it produces, is the distinctive contribution of this part of the theory.

\section{Handoff to Strategic Enclosure and the Smithian Inversion}\label{handoff-to-chapter-8-strategic-enclosure-and-the-smith-nash-problem}

\paragraph{Causal role.} The handoff asks why actors choose this structure. Once feedback capture and recombination suppression are identified, Chapter 8 asks when those effects become a rational strategy rather than an accidental by-product.

\enlargethispage{2\baselineskip}

Chapter 6 examined the external loss from enclosure: excluded actors lose
access to knowledge inputs, recombination paths, and the useful diversity
from which alternative trajectories could have emerged. Chapter 7 examined
the internal gain from enclosure: incumbents capture the feedback streams
that improve their own systems, raising \(G^L\), capability, productive
range, and trajectory speed. Chapter 8 now asks when those privately
rational strategies become strategically excessive, narrowing future
recombination, suppressing alternative trajectories, or creating
system-level dark capital. Chapter 8 asks when privately rational
feedback capture becomes systemically costly.

This is the Smithian inversion in its strategic form. What Smith's model
described as an expanding positive feedback between present production and
future productive capacity becomes, under strong knowledge enclosure, a
bifurcation: expanding productive power inside the enclosure, contracting
productive diversity outside it. Chapter 7 has also introduced, in \S{}7.7,
the connection between the mechanism result and the strategic decision
problem: the incumbent's individual enclosure choice can be locally rational,
maximizing its own trajectory acceleration, while producing aggregate
trajectory contraction. That tension is not resolvable within mechanism
analysis alone. It requires a strategic model in which incumbents optimize
enclosure duration and depth given expectations about rivals, regulators,
and commons access. Chapter 8 constructs that model under the heading of the Smithian inversion: the conditions under which individually rational
enclosure choices produce collectively self-undermining knowledge governance
forms, and the governance instruments that can alter the equilibrium.

\par

\clearpage
\chapter[Strategic Enclosure]{Strategic Enclosure and the Smithian Inversion}\label{strategic-enclosure-and-the-smith-nash-problem}

\chapterhook{The Paradox of Knowledge Enclosure}

This is the inversion the use-value frame predicts. In Smith's world of rival goods, private appropriation and social benefit broadly aligned; once governance can convert non-rival use-value into private exchange-value by manufacturing scarcity, the act that maximizes private capture can minimize social use, and Smith's reassuring identity comes apart. Strategic enclosure is the deliberate exploitation of that gap.

The strategic argument is positioned against productive and unproductive entrepreneurship, platform competition, resource-based advantage, and dynamic-capability theory \parencite{Baumol1990,RochetTirole2003,Wernerfelt1984,Barney1991,TeecePisanoShuen1997,Teece2007}\index{Barney, Jay}. Its claim is that individually rational enclosure can preserve incumbent advantage while weakening the wider generation system from which future advantage depends.

A simple example is a platform, such as Reddit or X/Twitter, that closes or prices an API after an ecosystem has formed around it. The platform may gain revenue, control, data, and bargaining power. But complementors may lose the ability to learn, test, recombine, and build alternative products. Chapter 8 asks when that private strategy is not merely restrictive, but excessive relative to the welfare benchmark that would preserve future recombination and learning.

\begin{center}
\fbox{\begin{minipage}{0.92\textwidth}
\small
\textbf{Running case: API closure/access restriction.}\index{API closure!strategic enclosure running case} In Chapter 8 the same API transition becomes strategic. The incumbent may rationally close, meter, or reprice access to protect rents, control complements, preserve feedback advantage, or shift bargaining power. KBC does not presume that this is inefficient. The test is whether the private gain from enclosure exceeds the social value lost through reduced recombination, weaker challenger capability, foregone complements, or narrowed future learning paths.
\end{minipage}}
\end{center}

\emph{When Individually Rational Enclosure Strategies Produce Socially
Inefficient Equilibria}

The prior chapters showed that enclosure can reduce outsiders' accessible inputs and feedback while increasing the incumbent's learning advantage.

Chapters 6 and 7 developed the mechanism of the enclosure dynamic:
enclosure reduces \(D_u(F_{a,t})\) for excluded actors, suppresses
\(G^R\) through the field magnitude and useful diversity channels, and
simultaneously raises \(\widetilde{C}_{inc}\) through the \(G^L\)
mechanism, accelerating the incumbent's trajectory while excluded
actors may fall behind, and may decay when substitute feedback, capability maintenance, and recombination access are insufficient. Enclosure is therefore not only a static
access restriction. If it reduces \(R_{a,t}\), \(D_u(F_{a,t})\), or
\(G^L_{a,t}\), it can narrow future productive range. The Technical Companion, Appendix C, §C.8.3, formalizes this claim through Derived Proposition~C.1, the Dynamic
Recombination-Range Recursion. The associated recursion equation
shows that restricted access can narrow future productive range by
reducing the recombination and learning terms that expand
\(d^*_{a,t+1}\). The mechanism result tells us what happens when
enclosure occurs. It does not tell us why enclosure occurs at the level
it does, or whether that level is socially optimal.

Chapter 8 asks the strategic question. When incumbents choose enclosure
strategies rationally, selecting how much to enclose, for how long,
of what type, under what conditions does the equilibrium produce
over-enclosure, trajectory concentration, and lower aggregate knowledge
generation than the social optimum? The mechanism chapters treated the
governance form \(g\) as an exogenous shift variable. Chapter 8
endogenizes the governance choice: the governance arrangement is the aggregate outcome of incumbents'
strategic choices, and those choices are governed by a payoff function
that diverges structurally from the social welfare criterion.

\textbf{Notation discipline.} Chapter 8 uses \(g_i\) for the ordinary governance form governing stock or strategy \(i\), \(g\) for a generic governance form or aggregate governance index, and \(\Pi_i\) for actor \(i\)'s payoff. The symbol \(\pi\) is avoided for governance in this chapter except where a named prior state already requires it; \(\rho\) is reserved for discounting or realization-style parameters and is not used for governance.

The causal story must again be stated plainly. What is enclosed is a strategic subset of stock, interfaces, data flows, standards, model access, feedback channels, interoperability points, or complementary assets. Those excluded include entrants, rivals, complementors, commons actors, users, and sometimes the incumbent's own future recombination partners. What is lost is a set of possible product, standard, model, protocol, workflow, toolchain, or market trajectories. The incumbent gains option value, rent protection, feedback acceleration, entry deterrence, standard control, switching costs, and time for capability accumulation. The claim would be weakened if observed enclosure did not exceed the welfare benchmark, if recombination losses were small or reversible, if entrants recovered fully after access restoration, or if enclosure were mostly explained by quality, safety, privacy, security, or coordination benefits rather than strategic suppression.

\begin{center}
\fbox{\begin{minipage}{0.94\linewidth}
\textbf{Strategic enclosure test.}

\vspace{0.5em}
\begin{tabularx}{\linewidth}{>{\raggedright\arraybackslash}p{0.31\linewidth}X}
\toprule
Test question & What Chapter 8 tracks \\
\midrule
What is enclosed? & Stock, interfaces, data flows, standards, model access, feedback channels, interoperability points, or complementary assets. \\
Who is excluded? & Entrants, rivals, complementors, commons actors, users, or future recombination partners. \\
What path is lost? & Product, standard, model, protocol, workflow, toolchain, market, or research trajectories. \\
What does the incumbent gain? & Rent protection, option value, feedback acceleration, entry deterrence, standard control, switching costs, and time for capability accumulation. \\
\bottomrule
\end{tabularx}
\end{minipage}}
\end{center}

This divergence is the \emph{Smithian inversion}: the accumulation of knowledge-bearing capital that strengthens present production while weakening the conditions for future knowledge generation. Its formal welfare expression, the gap between the privately rational (Nash\index{Nash, John}) enclosure equilibrium and the social optimum, is the one place this book uses the term \emph{Smith-Nash gap}; the phenomenon itself is named the Smithian inversion throughout. The alignment assumption, stated as a background premise in Chapter 1 and used throughout, asserts that individually rational enclosure strategies diverge from the social welfare optimum. Chapter 8 converts this from an
assumption into a conditional strategic result: Model 5 states the
conditions under which the divergence holds, characterizes its
magnitude, and identifies the equilibrium types that result.

The argument is not that incumbents are poorly motivated or
strategically malicious. It is a stronger claim. Even non-malicious
incumbents, profit-maximizing actors doing exactly what the
institutional payoff structure rewards, can rationally choose
enclosure strategies that are systemically inefficient, because the
private payoff function internalizes feedback acceleration and rent
capture but externalizes recombination suppression, trajectory loss\index{trajectory loss}, and
excluded-actor capability loss under the T6 limiting conditions. The misalignment is structural, not
moral.

Nor is the argument anti-firm or anti-IP. Model 5 allows enclosure to raise private returns, accelerate incumbent learning, and support investment that would otherwise not occur. The strategic problem appears when the privately optimal intensity or duration of enclosure exceeds the level that would preserve the wider recombination field. The object of analysis is the gap between these optima, not enclosure in isolation.
\subsection{What the Inversion Is, and Is Not}\label{what-the-inversion-is-and-is-not}

Because the term inverts a result most economists hold reflexively, it is worth stating precisely what is and is not being claimed. Smith's relation is a positive map: private accumulation of stock raises the economy's productive capacity. The inversion is the claim that, for knowledge-bearing capital held under enclosure, the sign of that map flips on its \emph{future-generation} argument, so that private accumulation lowers the system's future capacity to generate new productive knowledge even as it raises present output. It is a sign change in one component of the social return, not a denial that accumulation is productive.

\textbf{Why this is not the standard appropriability result.} The textbook externality runs the other way. Because knowledge is non-rival and its spillovers are positive, private actors under-provide it, and the orthodox prescription, from Arrow to Nordhaus optimal-patent analysis, is \emph{stronger} appropriability. A trained reader will therefore ask why enclosure, the standard cure, is here the disease. The answer is that the inversion is not a claim about the \emph{level} of investment but about the \emph{access configuration} of stock once accumulated. Enclosure does two things through one instrument: it raises the incumbent's private return to generation (the orthodox positive channel, retained here as Proposition~E), and it removes accumulated stock from the recombination field and learning loop on which \emph{other} actors' future generation depends (Propositions~C and~D). The inversion is the region in which the second effect dominates the first at the margin. It does not contradict the appropriability result; it supplies the offsetting term that the appropriability result, computed on a single knowledge good in isolation, leaves out.

\textbf{Why it is possible at all.} The inversion requires three structural conditions jointly, and collapses to ordinary monopoly analysis if any one is absent. Knowledge must be \emph{non-rival}, so that enclosure does not transfer the good from excluded actors to the incumbent but removes their potential use of it, making the loss a foregone future rather than a redistribution. It must be \emph{cumulative}, so that the stock withheld today is an input to others' generation tomorrow \parencite{Romer1990,Weitzman1998}. And deployment must be \emph{feedback-generating}, so that the controller of the enclosed stock accumulates capability from use that excluded actors cannot. Strip non-rivalry and enclosure is an ordinary transfer; strip cumulativeness and the loss is static; strip feedback and there is no compounding divergence. Jointly, the three are what make the divergence dynamic and self-widening rather than a one-period deadweight triangle.

\textbf{What it is not\index{Smithian inversion!what it is not}\index{strategic enclosure!conceptual contrasts}.} The inversion is distinct from the four results it most resembles. It is not the \emph{static deadweight loss\index{strategic enclosure!versus monopoly deadweight loss}\index{monopoly deadweight loss!versus strategic enclosure}} of monopoly: that is a present allocative triangle, whereas the inversion is a reduction in the growth of the future production-possibility frontier. It is not the \emph{anticommons\index{strategic enclosure!versus anticommons}\index{anticommons theory!versus strategic enclosure}\index{anticommons theory}} of fragmented exclusion rights, which raises the cost of assembling inputs but only blocks; the inversion's feedback channel also \emph{accelerates} the incumbent, the double-acting asymmetry\index{cognitive enclosure versus feedback enclosure}\index{feedback enclosure!versus cognitive enclosure}\index{cognitive enclosure!versus feedback enclosure} of \S\ref{the-double-acting-asymmetry}. It is the dark mirror of \emph{Schumpeterian\index{strategic enclosure versus Schumpeterian competition}\index{strategic enclosure!versus Schumpeterian rents}\index{Schumpeterian rent!contrast with strategic enclosure}\index{Schumpeter, Joseph} creative destruction}: there, monopoly rents \emph{fund} the next innovation, whereas the inversion is the case in which rent-protecting governance instead \emph{starves} it by foreclosing the field from which challengers would draw. And it is not a general \emph{anti-enclosure} claim: Proposition~E makes the welfare conclusion the sign of a net effect, so the inversion names only the governance configurations in which suppression outweighs the incentive, coordination, quality-control, disclosure, and maintenance benefits, not enclosure as such.

\subsection{The Inversion as a Formal Condition}\label{inversion-formal-condition}

The verbal definition has an exact counterpart in the capability dynamics. Summing the per-actor capability law (Equation~\ref{eq:ch3:capability-state}) over the set of actors \(\mathcal{A}\),
\[
\sum_{a\in\mathcal{A}}\widetilde{C}_{a,t+1}
=(1-\delta_C)\sum_{a\in\mathcal{A}}\widetilde{C}_{a,t}
+\gamma\sum_{a\in\mathcal{A}}G^{R}_{a,t}
+\ell\sum_{a\in\mathcal{A}}G^{L}_{a,t},
\]
so aggregate capability is driven by aggregate recombination and aggregate feedback, net of depreciation. Let a governance transition, an enclosure choice by incumbent \(i\), change each actor's generation by \(\Delta G^{R}_a\) and \(\Delta G^{L}_a\). The \emph{Smithian inversion} is then the conjunction of two inequalities:
\[
\underbrace{\gamma\,\Delta G^{R}_i+\ell\,\Delta G^{L}_i>0}_{\text{incumbent capability rises}},
\qquad
\underbrace{\gamma\sum_{a\in\mathcal{A}}\Delta G^{R}_a+\ell\sum_{a\in\mathcal{A}}\Delta G^{L}_a<0}_{\text{aggregate generative capability falls}}.
\]
In words: the enclosing incumbent's contribution to capability rises while the system's aggregate generative capability falls. The inversion is precisely this region, private capability gain coexisting with aggregate generative loss. It is what distinguishes the inversion from a simple transfer, in which the two move together, and from a Pareto improvement, in which both rise.

The dynamics behind the condition were established in \S\ref{the-double-acting-asymmetry}. Under the linear recombination closure \(G^{R}_a\approx\kappa\,\widetilde{C}_a\), aggregate capability converges only when depreciation exceeds recombination self-reinforcement, \(\delta_C>\gamma\kappa\), and diverges when \(\delta_C\le\gamma\kappa\). Feedback-loop capture, \(\Delta G^{L}_i>0\) for the incumbent with \(\Delta G^{L}_a\le 0\) for excluded actors, sustains a persistent capability gap even in the convergent case, and the governance transition raises the incumbent's capability whenever its feedback gain outweighs its own recombination loss, \(\ell\,\Delta G^{L}_i>-\gamma\,\Delta G^{R}_i\). The inversion arises when that private gain is purchased through field contraction and feedback exclusion large enough to drive the aggregate sum negative.

The result gives the inversion a tractable dynamic core,
\[
\text{recombination}+\text{feedback}\;\longrightarrow\;\text{capability}\;\longrightarrow\;\text{value},
\]
with explicit thresholds: convergence at \(\delta_C>\gamma\kappa\), divergence at \(\delta_C\le\gamma\kappa\), and inversion at the inequality pair above. Stated conservatively, the separation of private capability gain from aggregate generative loss has ancestors in externality theory, the anticommons, platform economics, endogenous growth, Nash inefficiency, and appropriability theory; the contribution here is the synthesis, recombination loss, feedback capture, capability accumulation, and accounting invisibility\index{accounting invisibility} combined into a single condition, which the originality audit of Chapter~11 marks as synthesized and possibly novel as architecture rather than as any one new primitive.

A note on what is secured. The \emph{channel-level} inversion, that each mechanism, recombination-field suppression, feedback capture, trajectory narrowing, can individually invert from private gain to aggregate loss, is the canonical, secured claim. The \emph{aggregate}, system-wide inversion, the conjunction above holding for the economy as a whole, is a conditional systemic result, not a law: it holds within \(\Phi\)-classes satisfying monotonicity and weak complementarity and where the recombination multiplier \(M_{rec}\) exceeds its calibrated critical threshold relative to the appropriability and field-depletion terms; otherwise the system is mechanism-consistent but non-inverting at the aggregate level. The Technical Companion (Volume~2, Appendix~G) carries this reclassification and states the boundary condition canonically.

\section{The Strategic Enclosure
Problem}\label{the-strategic-enclosure-problem}

\paragraph{Causal role.} This section turns enclosure from a governance effect into a strategic choice. The incumbent chooses what to enclose, whom to exclude, how long to exclude them, and which rents, feedback advantages, or capability gaps the choice preserves.

Models 1--4 treated the governance form as exogenous: \(g\) was a
parameter, and the theorems described the effects of different \(g\)
values on knowledge generation, capability dynamics, and welfare. This
is the right framing for establishing mechanism, the effects of
enclosure on \(D_u\), \(G^R\), \(\widetilde{C}\), and \(N_{traj}\) are
correctly characterized as consequences of \(g\) rather than as
choices of \(g\). But it leaves the prior question open: why does
\(g\) take the values it takes?

Model 5 answers this question by treating each \(g_i\) as a strategic
choice and the observed \(g\) as the aggregate outcome of incumbents'
strategic decisions. Each incumbent chooses an
enclosure strategy, how much knowledge-bearing stock to enclose (the
set E), under what governance form, and for how long (duration T), 
anticipating the downstream effects on the capability gap \(\Delta_t\) =
\(\widetilde{C}_{inc,t}\) -- \(\widetilde{C}_{ent,t}\), the trajectory
count \(N_{traj}(g)\), and the incumbent's own future \(G^R\). The
aggregate of these choices, evaluated against each other and against the
entrant's responses, determines the equilibrium governance form.

The Model 5 decision problem has three components. The enclosure choice:
when is it privately rational for an incumbent to enclose
knowledge-bearing stock, and how does the privately optimal enclosure
duration \(T^*_{\mathrm{strategic}}\) relate to the socially optimal \(T^*\)? The
allocation problem: how does the institutional payoff structure
determine what fraction of entrepreneurial capability is directed toward
productive recombination versus field-restricting enclosure? The
recovery problem: after enclosure ends or is reversed, does the excluded
actor recover competitive capability, and at what rate, and does
access restoration suffice, or does recovery require component-specific
capability restoration?

These three components are analytically separable but practically
intertwined. The enclosure choice is the incumbent's static optimization
problem. The allocation problem is its systemic institutional
consequence: as more enclosure is chosen, more entrepreneurial
capability shifts from generating new knowledge to protecting existing
knowledge positions. The recovery problem is the dynamic aftermath: even
after enclosure ends, the excluded actor's capability deficit persists,
producing post-enclosure welfare costs that the enclosure duration
calculation must account for.

\begin{table}[htbp]
\addtocounter{table}{-1}
\caption[Decision problems in strategic enclosure]{Decision problems in strategic enclosure}\index{strategic enclosure|textbf}
\label{tab:ch8:decision-proxy}
\centering
\begin{tabularx}{\textwidth}{>{\raggedright\arraybackslash}p{0.23\textwidth}>{\raggedright\arraybackslash}p{0.32\textwidth}X}
\toprule
Decision problem & Economic question & Observable proxy \\
\midrule
Enclosure choice & How much access should the incumbent restrict? & API pricing; licence scope; patent duration; model-access terms; interoperability limits. \\
Allocation problem & How much effort shifts from production to exclusion? & Legal/IP staffing; access-management spend; compliance tooling; standards-control effort; platform-governance costs. \\
Recovery problem & How quickly can excluded actors rebuild capability? & Entrant survival; rebuild time after access restoration; fork quality; benchmark recovery; complementor return rate. \\
\bottomrule
\end{tabularx}
\end{table}

\textbf{Status note on Model 5.} M5.T1\index{M5.T1 Strategic Enclosure!formal condition} is not an empirical proof that enclosure generally reduces welfare. It is a conditional theorem showing that strategic over-enclosure follows when recombination-loss and field-contraction effects are sufficiently large relative to incentive, coordination, quality-control, disclosure-protection, and maintenance benefits. In plain terms, M5.T1 asks whether the private gain from extending enclosure still exceeds the incumbent's private cost after the wider recombination, feedback, capability, and trajectory losses are counted on the social side. This makes \(M_{rec}\) and the Proposition~E benefit vector empirical hinges rather than hidden assumptions. This book therefore treats M5.T1 as a conditional theorem under H1--H4\index{M5 hypotheses H1--H4} plus derivative, interiority, and second-order conditions, and treats D7/M5.P6\index{D7 self-undermining enclosure} as a conditional reduced-form proposition. D1, D2, and D7 have been resolved for the purposes of this book. D3--D6 remain open Model 5 design problems concerning parameter identification, equilibrium refinement, recovery specification, and empirical calibration. They do not block the conditional strategic over-enclosure theorem, but they remain part of the formal research programme. The empirical identification of the net-effect terms is delegated, not assumed: the suppression loss \(L^-(T)\) and the generative gain \(G^+(T)\) whose sign decides the inversion are identified through Chapter~11's quasi-experimental spine, treated-versus-control divergence under named access shocks, and not through reconstruction of a generative frontier.

\begin{table}[htbp]
\addtocounter{table}{-1}
\caption[Observable proxies for strategic-enclosure terms]{Observable proxies for strategic-enclosure terms}
\label{tab:ch8:observable-proxies}
\centering
\begin{tabularx}{\textwidth}{>{\raggedright\arraybackslash}p{0.24\textwidth}X}
\toprule
Term & Observable proxy \\
\midrule
\(M_{rec}\) & Citation and reuse networks; API dependency maps; patent forward-citation dispersion; GitHub dependency graphs; benchmark diffusion. \\
\(C_{alloc}\) & Licensing costs; legal and enforcement costs; access-management costs; compliance and monitoring costs. \\
\(C_{rec}^{self}\) & Maintainer depletion; dependency fragility; security backlog; integration delays; loss of external complements. \\
Recovery lag \(\tau_{rec}\) & Time required for entrants or complementors to rebuild capability after access restoration, API reopening, patent expiry, or open release. \\
\bottomrule
\end{tabularx}
\par\smallskip\noindent\footnotesize\emph{Note.} These proxies support case coding, conditional theorem testing, and empirical research design. Consistent with Appendix~E's payoff-function discipline, uncalibrated components may support structural and conditional claims, but not numerical welfare optimization or policy calibration until their coefficients are empirically estimated.
\end{table}

\begin{figure}[!htbp]
\caption[Smithian Accumulation and the Smithian Inversion]{Smithian Accumulation and the Smithian Inversion}
\label{fig:ch8:smithian-inversion}
\centering
\vspace{-0.25\baselineskip}
\includegraphics[width=\textwidth]{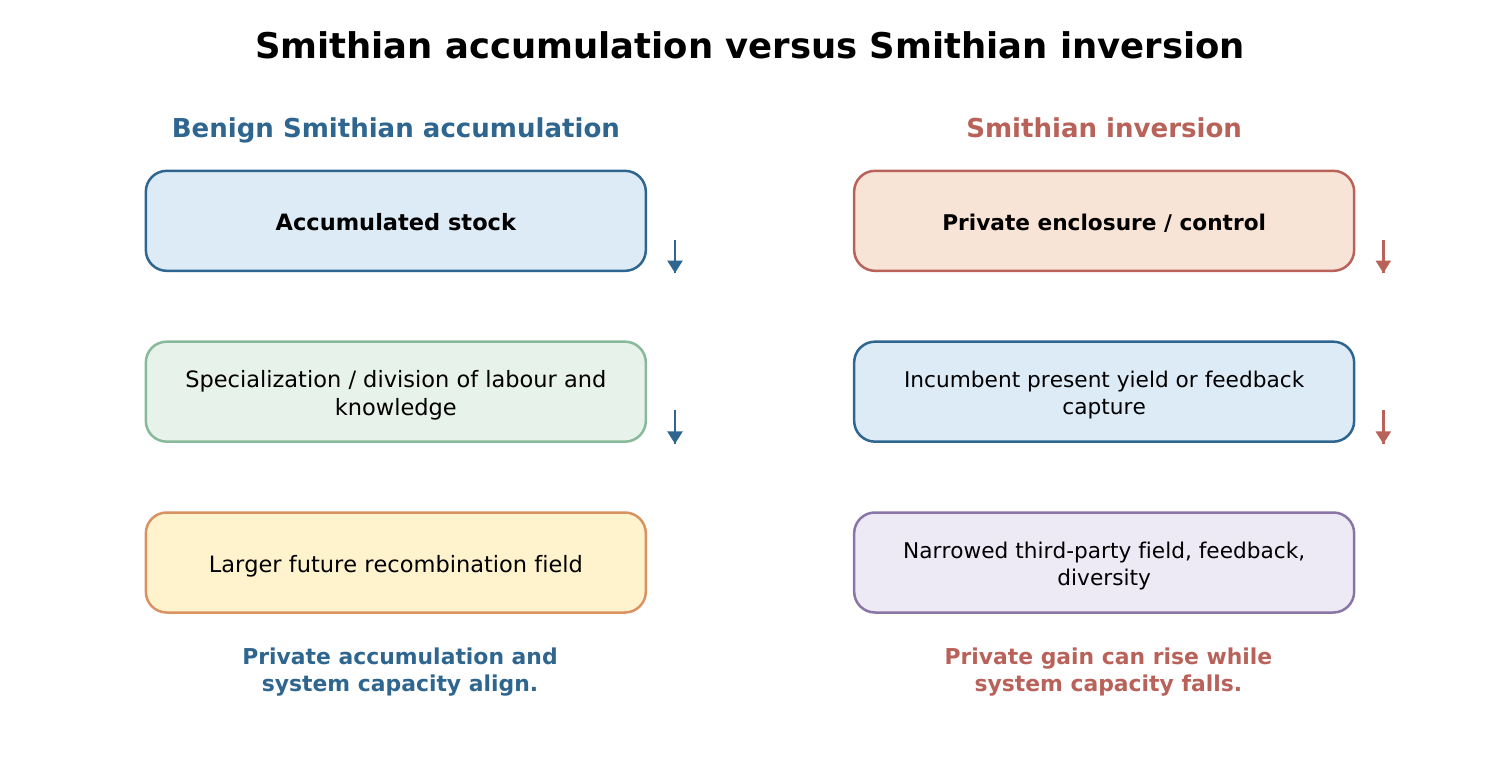}
\par\smallskip\noindent\footnotesize\emph{Note.} In the benign case, accumulation expands the future recombination field. In the inversion case, accumulation through enclosure increases present private productive power while narrowing third-party access, feedback, and useful diversity.
\par\vspace{-0.45\baselineskip}
\end{figure}

\begin{figure}[!htbp]
\caption[Private-system divergence under strategic enclosure]{Private-system divergence under strategic enclosure}
\label{fig:ch8:private-system-divergence}
\centering
\includegraphics[width=\textwidth]{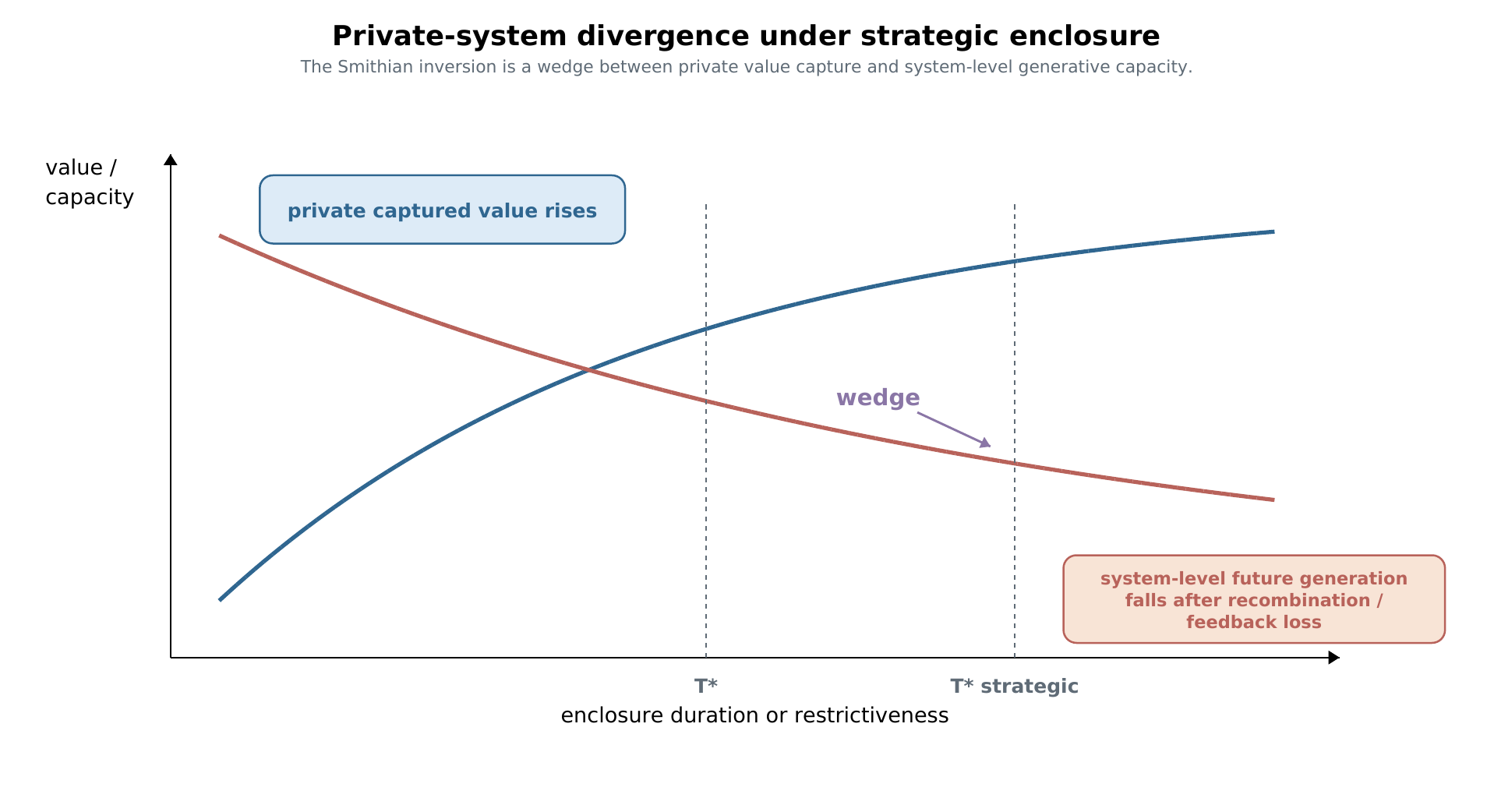}
\par\smallskip\noindent\footnotesize\emph{Note.} The divergence claim is not that enclosure always reduces welfare. It is that private captured value and system-level generative capacity can move in opposite directions when enclosure suppresses recombination fields, feedback access, or future capability formation.
\end{figure}

\section{The Alignment Assumption and the Smithian Inversion}\label{the-smith-nash-alignment-assumption}
\index{Smithian inversion|textbf}

\paragraph{Causal role.} This section identifies the alignment problem. The question is whether individually rational enclosure choices also preserve the social recombination and learning paths on which future wealth depends.

Chapter 1 reconstructed Smith's implicit model of capital accumulation
and identified its central assumption about market alignment: that the
individually rational investment of capital stock, directed by the
private return, would also increase the productive capacity of the
economy as a whole. For physical capital in competitive markets, this
alignment can be justified by the conditions under which individual
returns track social returns. For knowledge-bearing stock in
concentrated information markets, the conditions can break down.

The alignment assumption, as stated in this book, is a claim about the relationship between individual rationality and social welfare under knowledge-bearing capitalism. Formally:

\begin{equation}
BR_i(e_{-i},T_{-i}) \ne \arg\max W_N
\label{eq:ch8:smith-nash-divergence}
\end{equation}

The incumbent's best-response enclosure strategy, the individually
rational choice of enclosure intensity \(e_i\) and enclosure duration
\(T_i\) given rivals' choices, does not maximize the social welfare
function \(W_N\). This is not a claim that markets always fail for
knowledge goods; it is a conditional claim about the direction and
magnitude of the divergence under specified conditions. The conditions
are the work of Model 5 to characterize. Under H1--H4, M5.T1 states
that the divergence is systematic only within the specified region:
\(T^*_{\mathrm{strategic}}\) \textgreater{} \(T^*\) holds strictly, with the gap
increasing in \(M_{rec}\) and in the incumbent's feedback-capture
advantage. The result is conditional on the net-effect sign of Proposition~E (\S\ref{sec:ch4:proposition-e-appropriability-enabled-generation}): H1 requires a sufficiently large recombination premium, and where Proposition~E's generative channels dominate at the margin the gap narrows toward zero rather than holding strictly. The theorem therefore identifies the conditions under which private enclosure becomes excessive; it does not claim that enclosure is generally welfare-reducing.

\emph{In plain terms:} under the specified conditions, what is privately optimal for the enclosing
incumbent is socially suboptimal for the knowledge economy, because the incumbent captures rents and feedback advantages while excluded actors bear recombination loss, capability decay\index{capability decay}, and lost trajectory options. This is not only a standard static market failure\index{static market failure}. It is a dynamic
generative-capacity failure: the private payoff function and the social
welfare function measure different things, and optimizing one can
systematically degrade the other under the conditions specified by the
model.

\subsection{Two inversions: value appropriation and generation
acceleration}\label{sec:ch8:two-inversions}

The inversion this book identifies is, in its robust form, a
\emph{value-appropriation} inversion: privately rational enclosure raises
the incumbent's appropriated return, \(\Delta\Pi_{\mathrm{inc}}>0\), while
lowering system-level welfare measured as aggregate generative capacity,
\(\Delta W_{\mathrm{sys}}<0\). A stronger reading, that enclosure also
accelerates the incumbent's \emph{own} knowledge generation,
\(\Delta G_{\mathrm{inc}}>0\), is a special case rather than the general
mechanism, and it should not be assumed. Whether a monopolizing incumbent
generates more or less is the unresolved question of the Arrow replacement
effect \parencite{Arrow1962}, under which an incumbent has weaker incentive
to innovate, set against the preemption and feedback-capture channels that
push the other way \parencite{GilbertNewbery1982}. Exclusion can raise the
incumbent's profit even as its own generation falls. The most damaging
configuration is therefore not dynamic incumbent acceleration but monopoly
stagnation joined to system suppression, \(\Delta\Pi_{\mathrm{inc}}>0\) with
\(\Delta G_{\mathrm{inc}}<0\) and \(\Delta G_{\mathrm{sys}}<0\), in which the
incumbent profits while every party, itself included, generates less.

The system-suppression half, \(\Delta G_{\mathrm{sys}}<0\), is itself
conditional. It holds when enclosure removes a non-substitutable
\emph{bridge} position, knowledge that connects otherwise weakly linked
clusters and is for that reason far less substitutable than redundant
within-cluster knowledge \parencite{Burt1992,Granovetter1973,Weitzman1998},
from a recombination field that is still on the positive-marginal-product
side of useful diversity. A narrowly used code library may be redundant if many substitutes exist. A core protocol, dominant package, semiconductor design tool, benchmark suite, or platform interface can occupy a bridge position because many later projects depend on it as an input. The result fails, and can reverse, when the field is
diversity-saturated or over-bridged: there, closure or curation can raise
system generation by reducing search, interpretive, and coordination
overload \parencite{Uzzi1997,UzziSpiro2005}. Stated without these
conditions, the broad claim that enclosure suppresses generation is false;
the defensible claim is conditional on field structure. This qualification
governs the enclosure results throughout the book and is not repeated at
each occurrence. The formal function-class audit examines this condition across additive, CES, gated, bottleneck,
feedback-amplified, and non-monotone congestion forms of the generation
function, and is given in the Technical Companion, Appendix F.

The generation-acceleration inversion, where it does occur, is also bounded
in time. A platform that closes an ecosystem may improve faster at first because it captures more feedback and coordinates development internally. Over time, however, it may lose developer extensions, external tooling, bug discovery, and open-source maintenance, so the same enclosure that initially accelerates learning can later narrow it. Because the enclosing incumbent loses access to the very commons
it walls off, the captured-feedback gain is eventually overtaken by the
incumbent's own recombination-field decay, so short-run private
acceleration becomes long-run self-undermining. The two are one mechanism
at different horizons, and together they predict an inverted-U in incumbent
generation whose turning point is set by the depreciation rate of the
enclosed stock.

The alignment breaks down for three structurally related reasons, each
corresponding to a gap between the private payoff function and the
social welfare criterion.

The first is non-rivalry without universal appropriability.
Knowledge-bearing stock is non-rival in use: an actor's use of \(K_i\) does
not prevent another actor from using it. This means that enclosure of
\(K_i\) does not transfer \(K_i\) from excluded actors to the incumbent, 
the incumbent already had \(K_i\). What enclosure does is prevent excluded
actors from using \(K_i\) productively, without any compensating benefit to
the incumbent from that prevention beyond the rent stream it captures.
The usual compensation framing is incomplete here. Enclosure can create
a private winner through rent, control, and option value. But the loss is
not only a transfer from excluded users to the incumbent. Some loss
appears as foregone generation: products not built, models not trained,
standards not explored, and capabilities not accumulated. The winner may
gain rent and control, but not the same productive use-value that excluded
actors lose. KBC therefore evaluates enclosure not only as surplus
redistribution, but as a change in aggregate knowledge-generation capacity,
Definition T5.1\index{Definition T5.1}.

The second is cost externalization. The social cost of enclosure from T5 comprises four terms:
\(C_{T2}\) (recombination loss), \(C_{T7}\) (suppressed appreciation / feedback-learning loss\index{suppressed appreciation!feedback-learning loss}),
\(C_{T6}\) (capability decay), and \(C_{T8}\) (trajectory loss). These costs accrue predominantly to excluded actors and to the
economy's future knowledge generation capacity. The incumbent does not
pay these costs. The incumbent's private cost function is:

\begin{equation}
\Delta TC_{private}(T)=C_{enforcement}(T)+C_{alloc}(E_{alloc},T)+C^{self}_{rec}(e,T)
\label{eq:ch8:delta-tc-private-t}
\end{equation}

where \(C_{enforcement}\) is the cost of maintaining the enclosure
(legal fees, API governance, trade secret programs), \(C_{alloc}\) is
the opportunity cost of entrepreneurial capability directed toward
enclosure rather than productive recombination (the Baumol mechanism),
and \(C_{rec}^{self}\) is the self-imposed recombination cost, the
present value of the incumbent's own lost \(G^R\) from commons depletion
and spillover suppression. The four T5 cost components\index{T5 cost components}, \(C_{T2}\) (recombination loss),
\(C_{T7}\) (suppressed appreciation / feedback-learning loss), \(C_{T6}\) (capability decay), and \(C_{T8}\) (trajectory loss), do not appear in \(\Delta TC_{private}\).
They are real social costs, but they are externalized to excluded actors
and to the aggregate knowledge base.

The third is dynamic revenue internalization.\index{dynamic revenue internalization versus static cost externalization} The social welfare
criterion counts, on the benefit side, the full generation-and-maintenance vector \(\mathcal{G}^{+}(T)\) of Proposition~E, of which \(B_{\mathrm{incentive}}(T)\), the social value of knowledge-bearing stock \(K_i\) existing because the exclusion right induced its production, is the leading component. It does not count the dynamic trajectory
revenues that the incumbent captures from enclosure: the capability
divergence advantage \(\Delta TR_{dynamic}^{T6}\) (the ability to command
higher prices and attract further deployment as \(\widetilde{C}_{inc}\)
grows relative to \(\widetilde{C}_{ent}\)) and the trajectory
acceleration premium \(\Delta TR_{dynamic}^{T8}\) (the value of improving
faster along a narrower trajectory while excluded actors' trajectories
contract). These dynamic revenues are real private gains, they appear
in the incumbent's payoff function, but they are not social gains:
they are the private capture of what are, at the system level, the
social costs of trajectory divergence and concentration.

The combination of cost externalization and dynamic revenue
internalization makes the private payoff function diverge structurally
from the social welfare criterion. This structural divergence is the Smithian inversion (\S\ref{what-the-inversion-is-and-is-not}): a systematic gap between what private optimization rewards and what social welfare requires.

\subsection{Box 8.1: Strategic-Capture
Equilibrium}\index{strategic-capture equilibrium}\label{box-8.1-strategic-capture-equilibrium}

The Smithian inversion can be stated as a strategic-capture equilibrium.
Let each incumbent \texttt{i} choose an enclosure strategy:

\begin{equation}
s_i=(e_i,T_i)
\label{eq:ch8:s-i}
\end{equation}

where \(e_i\) is enclosure
intensity and \(T_i\) is
enclosure duration. Let the incumbent payoff be:

\begin{equation}
\begin{aligned}
\Pi_i(s_i,s_{-i})={}&\mathrm{PrivateCapturedYield}_i
+\mathrm{KnowledgeRent}_i+\mathrm{FeedbackCaptureGain}_i \\
&-C_{\mathrm{enforcement},i}-C_{\mathrm{alloc},i}
-C^{\mathrm{self}}_{\mathrm{rec},i}-EKL_i
\end{aligned}
\label{eq:ch8:pi-i-s-i-s-i}
\end{equation}

This payoff function is structural rather than calibrated. Its terms are
not assigned numerical weights in this chapter. The Technical Companion, Appendix E, classifies the payoff components by evidentiary status: directly observable, proxy-measurable, theoretically necessary but currently uncalibrated, or excluded until grounded. It also restricts numerical policy claims until proxy validation and calibration are complete.

Read this as a platform choosing API openness, pricing, and contract duration while rivals and complementors choose whether to build complements, substitutes, forks, workarounds, or exit.

A strategic-capture equilibrium exists when each actor's enclosure
strategy is a best response to the enclosure strategies of the other
actors:

\begin{equation}
s_i^*\in BR_i(s_{-i}^*)
\label{eq:ch8:s-i-2}
\end{equation}

equivalently:

\begin{equation}
\Pi_i(s_i^*,s_{-i}^*)\geq\Pi_i(s_i,s_{-i}^*)\qquad\text{for all }s_i\in S_i\text{ and all actors }i
\label{eq:ch8:pi-i-s-i-s-i-2}
\end{equation}

The social optimum is:

\begin{equation}
s^W\in\arg\max_s W(s)
\label{eq:ch8:s-w}
\end{equation}

The Smithian inversion arises when:

\begin{equation}
s^*\neq s^W
\label{eq:ch8:s}
\end{equation}

Strategic over-enclosure\index{strategic over-enclosure|textbf} occurs when the privately stable equilibrium
exceeds the socially optimal enclosure duration:

\[
T^*_{\mathrm{strategic}}>T^*
\]

\section{Model 5: Strategic Enclosure
Equilibria}\index{Model 5!strategic enclosure equilibria}\index{strategic enclosure!Model 5}\label{model-5-strategic-enclosure-equilibria}\index{Model 5|textbf}

\paragraph{Causal role.} Model 5 organizes the strategic sequence: enclosure choice, excluded capability loss, incumbent advantage, equilibrium selection, and welfare comparison. Its purpose is not to predict every market outcome, but to identify when private incentives point toward over-enclosure.

Model 5 operates with a two-actor structure at the first level: an
incumbent actor (inc) and an entrant actor (ent). This is the minimal
structure that can represent the strategic asymmetry between an
established knowledge producer with accumulated capabilities and a
would-be knowledge generator whose capability and field access are more
limited.

The two-actor case can be read as an incumbent platform versus an entrant developer, a closed AI model provider versus an open model developer, or a proprietary genomic database holder versus a research entrant. In each case, the incumbent controls access to a stock or interface that the entrant needs for recombination, learning, or deployment. The model fits platform, API, patent, data, and model-access markets best. It is weaker for commodity knowledge goods with many close substitutes, low switching costs, and weak feedback capture.

The incumbent holds initial capability \(\widetilde{C}_{inc,t}\)
\textgreater{} \(\widetilde{C}_{ent,t}\), controls knowledge-bearing
stock \(K_{inc}\) under governance forms with private IP and platform
elements, has positive \(G^L\) from large deployment base, and holds the
option to extend enclosure by shifting additional elements of \(K_t\) into
more restrictive governance arrangements. The entrant has an initial capability deficit,
seeks access to stocks currently in enclosed governance arrangements, has lower \(G^L\)
due to smaller deployment scale, and may need to traverse the five-condition
entry set for each stock it needs to recombine productively.

The asymmetry is not merely positional. It is dynamic: as long as
enclosure continues, T6's \(G^L\) mechanism widens \(\Delta_t\), and
T8's trajectory acceleration deepens the incumbent's \(N_{traj}\)
advantage. The Model 5 decision problem is therefore whether to use
enclosure to sustain and extend this dynamic advantage beyond the level
that T5's social welfare criterion would justify.

The incumbent's payoff function is:\index{payoff function|textbf}

\begin{equation}
\Pi_{\mathrm{inc}}(T)=\Delta TR_{private}(T)-\Delta TC_{private}(T)
\label{eq:ch8:pi-inc-t}
\end{equation}

where the private revenue includes both static rent from exclusive
access (\(\Delta TR_{static}\): the standard Schumpeterian appropriability return)
and dynamic trajectory revenues (\(\Delta TR_{dynamic}^{T6}\) from capability
divergence, \(\Delta TR_{dynamic}^{T8}\) from trajectory acceleration). The
private cost includes enforcement, allocation opportunity cost, and
self-imposed recombination costs, but not the externalized \(C_{T2}\) (recombination loss),
\(C_{T7}\) (suppressed appreciation / feedback-learning loss), \(C_{T6}\) (capability decay), and \(C_{T8}\) (trajectory loss).

The payoff function treats knowledge rent as part of the incumbent's private return. Such rent can be politically self-reinforcing: when the legal, regulatory, standards, contractual, or platform-governance conditions that sustain enclosure are maintained by the actors that benefit from them, the governance position itself becomes an object of investment. This is the familiar rent-seeking and regulatory-capture dynamic \parencite{Krueger1974,Stigler1971}, not a distinct mechanism; the present point is only that the rent term in the payoff function can fund the preservation of the conditions that generate it. For that reason, the Technical Companion, Appendix E, decomposes KnowledgeRent into service, access, friction, and capture-derived components. The decomposition does not add a new revenue stream; it identifies sources within the rent already being modelled.

The social welfare criterion from T5, with the benefit side disaggregated as in Proposition~E (\S\ref{sec:ch4:proposition-e-appropriability-enabled-generation}):

\begin{equation}
W(T)=\mathcal{G}^{+}(T)-C_{enclosure}(T)-C_{recovery}(T)
\label{eq:ch8:w-t}
\end{equation}

Here \(\mathcal{G}^{+}(T)\) is the generation-and-maintenance benefit side of Proposition~E, \(C_{\mathrm{enclosure}}(T)\) is the enclosure-loss side, and \(C_{\mathrm{recovery}}(T)\) is the post-enclosure recovery lag. Table~\ref{tab:ch8:welfare-components} gives the plain-language decomposition. The incumbent maximizes \(\Pi_{\mathrm{inc}}(T)\); the social planner maximizes \(W(T)\). This chapter's central claim is that the respective optima diverge: \(T^*_{\mathrm{strategic}} \neq T^*\).

\begin{table}[htbp]
\addtocounter{table}{-1}
\caption[Plain-language components of the strategic-enclosure welfare criterion]{Plain-language components of the strategic-enclosure welfare criterion}
\label{tab:ch8:welfare-components}
\centering
\begin{tabularx}{\textwidth}{>{\raggedright\arraybackslash}p{0.28\textwidth}X}
\toprule
Component & Plain-language meaning \\
\midrule
\(\mathcal{G}^{+}(T)\) & Proposition~E generation-and-maintenance benefits: incentive to invent, coordination, quality control, disclosure protection, and maintenance. \\
\(C_{\mathrm{enclosure}}(T)\) & Enclosure losses: \(C_{T2}(T)\) (recombination loss), \(C_{T7}(T)\) (suppressed appreciation / feedback-learning loss), \(C_{T6}(T)\) (capability decay), and \(C_{T8}(T)\) (net trajectory loss). \\
\(C_{\mathrm{recovery}}(T)\) & Post-enclosure capability recovery lag after access is restored, restrictions expire, or a substitute path is built. \\
\bottomrule
\end{tabularx}
\end{table} 

\noindent The \(C_{\mathrm{enclosure}}\) terms are net social costs, not gross effects. In particular \(C_{T8}\) is the \emph{net} trajectory cost: the option value of the trajectories enclosure forecloses, less the coordination, standardization, and duplication-avoidance savings that convergence on a dominant design can produce \parencite{David1985,Arthur1989}. \(C_{T8}\) is therefore positive only when foreclosed-trajectory option value exceeds convergence savings; where a dominant design is efficient, \(C_{T8}\) can be zero or negative and the over-enclosure result M5.T1 correspondingly weakens. Treating trajectory diversity as costly by definition would assume the answer; the model leaves the sign of \(C_{T8}\) as an empirical condition.

The criterion reduces to the standard incentive-only form \(W(T)=B_{\mathrm{incentive}}(T)-C_{\mathrm{enclosure}}(T)-C_{\mathrm{recovery}}(T)\) when the coordination, quality, disclosure-protection, and maintenance channels are nil.

\textbf{The divergence is conditional, not assumed.} Because the benefit side is the full vector \(\mathcal{G}^{+}\), M5.T1 is not a general anti-enclosure result. It applies when, at the privately chosen duration, marginal recombination, feedback, and field-narrowing loss exceed marginal generation-and-maintenance gain, \(\mathrm{d}\mathcal{L}^{-}/\mathrm{d}T > \mathrm{d}\mathcal{G}^{+}/\mathrm{d}T\). Where \(\mathrm{d}\mathcal{G}^{+}/\mathrm{d}T\) dominates, the private/social gap narrows. This makes \(M_{rec}\) the empirical hinge: the theorem matters most where the enclosed stock is sufficiently recombinant for external field loss to matter.

Model 5 is therefore a theory of \textbf{strategic capture}. The distinction is between productive service return and captured governance rent. Productive service return arises when knowledge-bearing stock improves output, quality, speed, reliability, or future learning. Captured governance rent arises when control over access, standards, interfaces, licences, or feedback channels lets an actor appropriate value from others' dependence. The payoff notation in this chapter should therefore be read through the controlled vocabulary of this book: \(\Pi_i\) is not value itself, but the actor-specific private capture of value under a particular enclosure, access, and governance strategy.

\textcite{Teece2007}'s integration motivates a three-component decomposition
of \(\widetilde{C}_{a}\) for the dynamic capability analysis. The
sensing component \(\widetilde{C}^{S}_{a}\) captures the actor's
ability to identify productive recombination opportunities. The seizing
component \(\widetilde{C}^{Z}_{a}\) captures the ability to convert
recognized opportunities into deployed output. The transforming
component \(\widetilde{C}^{T}_{a}\) captures the ability to integrate
gains from deployment into durable organizational capability. These
components contribute differently to the recovery lag \(\tau_{rec}\), 
the duration required for the entrant to restore productive capability
after the enclosure ends, and the component-specific recovery lag
(M5.P1) is the key mechanism through which enclosure duration translates
into superlinear post-enclosure recovery costs.

\section{Individual Rationality of
Enclosure}\label{individual-rationality-of-enclosure}

\paragraph{Causal role.} This section explains the incumbent gain side of the sequence. Enclosure is individually rational when rent protection, option value, feedback advantage, and capability-gap preservation exceed the incumbent's private cost.

To understand when the gap \(T^*_{\mathrm{strategic}}\) \textgreater{} \(T^*\) is
large, it helps first to ask why enclosure is privately rational at all.
The case for enclosure is not simply rent capture; it operates through
three reinforcing channels that the private payoff function captures.

The first is static rent.\index{static rent versus capability capture}\index{static rent versus dynamic revenue} An incumbent that encloses knowledge-bearing
stock \(K_i\) under a private IP governance can charge for access, license
under controlled terms, or deny access to competitors. The static rent
\(\Delta TR_{static}\)(T) is the discounted stream of monopoly returns over the
enclosure horizon. This is the benefit that standard IP welfare analysis
identifies on the \(B_incentive\) side of the welfare function. The
incumbent captures it fully; excluded actors and downstream users may bear the associated welfare
cost. For the static rent alone, the incumbent's optimal T would
coincide with \(T^*_{\mathrm{standard}}\), which T5 already derived
exceeds the true social optimum \(T^*\) by the recombination multiplier.

The second is capability capture. Chapter 7 showed that
feedback-enclosure raises \(\widetilde{C}_{inc}\) through the \(G^L\)
mechanism while excluded actors' \(\widetilde{C}\) decays. A higher
\(\widetilde{C}_{inc}\) translates into a wider productive field, higher
\(G^R_{inc}\), and greater surplus per recombination event. This
capability divergence generates dynamic revenue \(\Delta TR_{dynamic}^{T6}\)
that is not present in the standard Nordhaus welfare analysis of IP: it
is not the return to \(K_i\)'s existence (the \(B_incentive\) term) but the
return to \(K_i\)'s exclusive deployment, which concentrates the feedback
signal and grows the incumbent's capability advantage. The incumbent
captures this revenue; the social welfare criterion does not count the portion achieved through excluded-actor capability loss as a net social benefit, because the capability advantage is a private gain achieved at the cost of excluded actors' relative capability loss when substitute learning and capability-maintenance channels are insufficient. For example, developers excluded from an API lose not only access, but the opportunity to learn from changing platform behaviour, user feedback, error patterns, and evolving technical constraints.

The third is trajectory acceleration. Chapter 7 also established that
the incumbent's trajectory improves faster under enclosure, not only
because \(\widetilde{C}_{inc}\) grows, but because the narrowing of
\(N_{traj}\) reduces competitive pressure and allows the incumbent to
deepen its existing trajectory without being displaced by alternative
approaches. The trajectory acceleration premium \(\Delta TR_{dynamic}^{T8}\)
is the value the incumbent captures from being the dominant actor on a
faster-moving trajectory. Again, the incumbent captures it; the social
cost (the trajectory count loss identified by T8) is externalized.

\textcite{Baumol1990}'s allocation result is the institutional mechanism that
connects these private returns to aggregate behaviour. Entrepreneurial
capability is not inherently generative; its contribution to knowledge
generation depends on the institutional payoff structure that determines
whether directing capability toward productive recombination or toward
field-restricting enclosure is more rewarding. M5.P2 (Entrepreneurial
Allocation) formalizes this: the equilibrium allocation \(E^*_{alloc}\) is
determined by the ratio \(R_{\mathrm{enc}}(g)/G^R_{\mathrm{return}}\), the
return to enclosure relative to the return to productive recombination. A rising share of legal, compliance, standards, and platform-control effort relative to product or research effort is one possible observable sign.
As this ratio rises (because IP protection strengthens, platform
governance becomes more favourable to enclosure, or competitive pressure
through recombination weakens), more entrepreneurial capability shifts
toward enclosure and away from productive knowledge generation. The
private rationality of enclosure, aggregated across many incumbents,
generates an institutional environment in which enclosure becomes
progressively more attractive and productive recombination progressively
less so.

\section{Strategic Over-Enclosure}\label{strategic-over-enclosure}

\paragraph{Causal role.} This section compares private gain with social loss\index{private gain versus social-yield loss}\index{social-yield benchmark}. Over-enclosure occurs when the incumbent's privately optimal restriction suppresses more recombination, learning, trajectory diversity, or capability than the welfare benchmark permits.

\textbf{Conditional strategic over-enclosure result (M5.T1).} The practical mechanism has four steps before the formal conditions are stated.

\begin{itemize}
\item \textbf{Social optimum:} choose the access restriction period that preserves incentive benefits while counting recombination, capability, and trajectory losses.
\item \textbf{Private omission:} the incumbent does not internalize all excluded-actor and system-level losses.
\item \textbf{Private dynamic gain:} longer enclosure may produce feedback capture, switching costs, trajectory control, and capability advantage.
\item \textbf{Longer private duration:} the privately chosen restriction period can exceed the welfare benchmark.
\end{itemize}

Formally, under the following conditions:

\begin{itemize}
\tightlist
\item
  \textbf{(H1)} \(M_{rec}\) \textgreater{} 1: the enclosed knowledge
  stock \(K_i\) is genuinely recombinant, it enters excluded actors'
  productive fields with positive weight at productive combination
  distances
\item
  \textbf{(H2)} \([\Delta TR_{dynamic}^{T6}]'(T^*)+[\Delta TR_{dynamic}^{T8}]'(T^*)>0\): the
  incumbent's dynamic revenue is still increasing at the social optimum
  \(T^*\)
\item
  \textbf{(H3)} \(C'_{enforcement}(T^*) + C'_{alloc}(T^*) \le e^{-\rho T^*}\cdot R_{\mathrm{enc}}(E,T^*)\): the marginal static
  and allocation costs do not exceed the static rent at \(T^*\)
\item
  \textbf{(H4)} \([\Delta TR_{dynamic}^{T6}]'(T^*)+[\Delta TR_{dynamic}^{T8}]'(T^*)>[C_{rec}^{self}]'(T^*)\): the incumbent's dynamic revenue
  from capability capture and trajectory acceleration exceeds its
  self-imposed recombination-field depletion cost at \(T^*\)
\end{itemize}

In H3, \(\rho\) is a discount or realization-style parameter, \(E\) is the set of enclosed elements or the enclosure strategy, and \(R_{\mathrm{enc}}(E,T^*)\) is the static rent available from enclosure at the social benchmark duration.

Then: \(T^*_{\mathrm{strategic}}\) \textgreater{} \(T^*\). The proof reorganizes the incumbent's first-order condition into a static bracket governed by H3 and a dynamic bracket governed by H4. Since the incumbent's payoff is still increasing at \(T^*\), the incumbent's optimum \(T^*_{\mathrm{strategic}}\) must exceed \(T^*\). The full proof is in the M5.T1 strategic over-enclosure proof file.

\textbf{The two sources of divergence} are analytically distinct. The
first is cost externalization. At any T, the incumbent's marginal cost
\(\Delta TC'_{private}(T)\) is below the social marginal cost \(C'_{enclosure}(T)\) by
the amount \(C'_{T2}+C'_{T7}+C'_{T6}+C'_{T8}\), the four externalized
cost components: recombination loss, suppressed appreciation / feedback-learning loss, capability decay, and trajectory loss. This alone would produce \(T^*_{\mathrm{strategic}}\)
\textgreater{} \(T^*\), even in a purely static model where dynamic
revenues do not exist. The second is dynamic revenue internalization.
The incumbent's marginal revenue \(\Delta TR'_{private}(T)\) exceeds
\(B'_{incentive}(T)\) by the amount \([\Delta TR_{dynamic}^{T6}]' + [\Delta TR_{dynamic}^{T8}]'\), the dynamic capability capture and
trajectory acceleration revenues. This second source amplifies the
divergence above what cost externalization alone would produce.

\textbf{The chain of three inequalities} that M5.C3 states under
appropriate precision conditions:

\begin{equation}
T^*<T^*_{\mathrm{standard}}<T^*_{\mathrm{strategic}}
\label{eq:ch8:t}
\end{equation}

\index{enclosure-duration model|textbf}
Read the inequality as a duration problem. \(T^*\) is the access-restriction period a welfare analyst would choose if all recombination, capability, and trajectory losses were counted. \(T^*_{\mathrm{standard}}\) is the longer period produced by ordinary incentive-only IP analysis. \(T^*_{\mathrm{strategic}}\) is longer still when incumbents also capture feedback advantages, switching costs, and trajectory rents.

\emph{In plain terms:} \(T^*\) is the socially optimal enclosure
duration, the point at which the full social cost of withholding
knowledge (including suppressed generation, capability decay, and
trajectory loss) equals the social benefit of the incentive to produce
it. \(T^*_{\mathrm{standard}}\) is what conventional IP analysis recommends, 
it is higher than \(T^*\) because it omits \(C_{T2}\) (recombination loss), \(C_{T7}\) (suppressed appreciation / feedback-learning loss),
\(C_{T6}\) (capability decay), and \(C_{T8}\) (trajectory loss) from the social welfare calculation.
\(T^*_{\mathrm{strategic}}\) is what incumbents actually choose, higher still,
because they capture dynamic revenues from capability divergence and
trajectory acceleration that no social welfare criterion counts as
benefits. The triple inequality is Chapter 8's single most important
result: it shows that current IP analysis (\(T^*_{\mathrm{standard}}\)) and
incumbent incentives (\(T^*_{\mathrm{strategic}}\)) both point toward longer
enclosure than the true social optimum (\(T^*\)) supports. The problem
is not that incumbents are irrational; it is that the private and social
optima are structurally different.

\(T^*\) is the true social optimum, the enclosure duration at which
the full \(C_{enclosure}\) (including recombination loss, suppressed appreciation / feedback-learning loss, capability decay, and trajectory loss)
equals \(B_incentive\). \(T^*_{\mathrm{standard}}\) is the Nordhaus-style optimum
,  the duration at which the standard Harberger deadweight loss equals
\(B_incentive\), omitting the three additional cost components.
\(T^*_{\mathrm{strategic}}\) is the incumbent's privately optimal duration. All
three are strictly ordered from shortest to longest. Current IP
protection terms, set during periods when recombination intensity was
lower and the social welfare analysis was conducted using the
\(T^*_{\mathrm{standard}}\) framework, may therefore fall, in some recombination-intensive domains, within the
\(T^*_{\mathrm{standard}}\) to \(T^*_{\mathrm{strategic}}\) range: longer than the true social
optimum, but not necessarily as long as the incumbent would freely
choose.

\textbf{Magnitude of the gap}: M5.T1.1\index{M5.T1.1} states that
\(T^*_{\mathrm{strategic}}\) -- \(T^*\) is increasing in \(M_{rec}\) -- 1 (the
recombination premium), in the marginal dynamic capability revenue at
\(T^*\), and in the rate at which the recovery lag \(\tau_{rec}\) grows
with enclosure duration. The gap is therefore largest where recombination intensity, feedback-capture deployment, and capability traps are all high. Candidate recombination-intensive sectors include frontier AI, genomics, semiconductor design tools, and core internet infrastructure. Observable indicators include model API dependency, benchmark gaps, and user-feedback volume in frontier AI; database access restrictions, cohort or data exclusivity, and research-use limits in genomics; EDA licence concentration, toolchain lock-in, and switching costs in semiconductor design tools; and standards control, protocol governance, and interoperability constraints in internet infrastructure. These are candidate domains where the Smithian inversion may be most consequential and where the social cost of operating at \(T^*_{\mathrm{standard}}\) rather than \(T^*\) may be largest.

\emph{Smithian departure:} Smith's implicit model of competition assumed
that the private payoff to investment, the profit the merchant makes,
the rent the landlord receives, would, in aggregate, be consistent
with expanding the productive capacity of the economy. His invisible
hand argument depends on this alignment. M5.T1 names the precise
structural condition under which that alignment fails in
knowledge-bearing capitalism: when enclosure is applied to non-rival,
recombinable stocks, the private payoff function internalizes the
dynamic trajectory revenues that feedback capture and enclosure generate
while externalizing the generation suppression, capability decay, and
trajectory loss that those same strategies impose on excluded actors and
on the knowledge commons. The invisible hand operates through a cost
function that omits \(C_{T2}\) (recombination loss), \(C_{T7}\) (suppressed appreciation / feedback-learning loss), \(C_{T6}\) (capability decay), and \(C_{T8}\) (trajectory loss).
Smith's framework did not need to model this problem in a world centred on rival, material stock. He had no occasion to construct a theory of non-rival, governance-dependent capital whose enclosure subtracts from others' generation rates without diminishing the enclosing actor's own stock.

\emph{Strategy-literature corollary.} M5.T1's equilibrium has an
institutional mechanism that the strategy literature names precisely,
without recognizing its aggregate social cost. RBV is correct as firm-level strategy; KBC adds the system-level externality when the prescription is generalized. Firms advised by
resource-based theory, advised, following \textcite{Barney1991}, to make
their knowledge-bearing resources valuable, rare, imperfectly imitable,
and non-substitutable, are being advised to pursue the privately
optimal enclosure duration. In KBC terms, the RBV prescription helps
explain why privately rational firms choose enclosure durations such
that \(T^*_{\mathrm{strategic}}\) \textgreater{} \(T^*\): the firm-level
strategy that preserves advantage may, when generalized across firms,
narrow the recombination field and impose social costs not visible from
the individual firm's perspective. This is not a critique of the
resource-based view; it correctly identifies what individually
rational firm strategy looks like under the prevailing appropriability
structure. When many incumbents follow VRIN logic simultaneously, 
when imperfect imitability is treated as the strategic objective across
an industry or a platform ecosystem, the individually rational
enclosure decisions aggregate into the economy-level recombination-field
narrowing that M5.T1 formalizes. The Smithian inversion is not produced by exceptional incumbents behaving irrationally or maliciously.
It is the structural equilibrium produced when sophisticated firm
strategy, guided by the best available firm-level theory, is applied to
non-rival, recombinable knowledge-bearing stock. Barney remains correct
at the firm level; M5.T1 is the economy-level complement that
resource-based theory cannot supply.

\section[The closed field stops learning]{The Closed Field Stops Learning: Self-Undermining Enclosure}\label{self-undermining-enclosure}
\index{learning loop}

\paragraph{Causal role.} This section asks whether the incumbent eventually damages its own future field. Self-undermining enclosure exists when restricting outsiders also reduces the upstream commons, complements, standards, or discovery paths the incumbent later needs.

A platform that closes APIs may initially gain control and capture more internal learning, but later lose developer innovation, third-party extensions, external bug discovery, and complementary tooling. An AI provider may similarly benefit from model-access control while becoming dependent on a shrinking external training, evaluation, or open-source ecosystem. In both cases, enclosure can become self-undermining when it damages the external field on which the incumbent's own learning depends.

M5.T1 states the conditional result that incumbents over-enclose relative to
the social optimum when H1--H4 hold and when recombination-loss and
field-contraction effects dominate the Proposition~E benefit channels at
the margin. It does not establish that enclosure is generally welfare-reducing
or that incumbents enclose indefinitely; the private payoff function has
its own ceiling that limits enclosure even without regulatory intervention.

The self-undermining enclosure threshold \(\hat{T}\) is the enclosure
duration at which the incumbent's own payoff begins to decline, because
the self-imposed costs of over-enclosure, the \(C_{rec}^{self}\) term, overtake the dynamic revenues. \(C_{rec}^{self}\) has two channels.

The commons-depletion channel: enclosure narrows the incumbent's own
accessible recombination field by depleting the commons knowledge
capital \(K^C\) on which the incumbent's own \(G^R\) depends. An
incumbent that encloses large portions of \(K_t\) removes stocks from the
commons that its own generation processes used to draw on; the enclosure
that improves competitive position simultaneously contracts the
incumbent's own field diversity. The commons-depletion cost is:

\begin{equation}
C^{self}_{commons}(e,T)=\int_0^T e^{-\rho t}\left[G^R_{inc}(D_u(F^{open}_{inc}))-G^R_{inc}(D_u(F^e_{inc}(t)))\right]dt
\label{eq:ch8:c-self-commons-e-t}
\end{equation}

Here, \(e\) is the incumbent's enclosure intensity, \(T\) is the enclosure duration, \(\rho\) is the discount parameter, \(D_u(F^{open}_{inc})\) is the incumbent's useful field diversity under open access, and \(D_u(F^e_{inc}(t))\) is the incumbent's useful field diversity after enclosure has narrowed the field over time. The expression measures the incumbent's own lost recombination output from depleting a commons it also uses.

The spillover-suppression channel: the excluded actor's relative capability loss, and in the limiting case capability decay
(T6.5), reduces the pool of externally generated knowledge that the
incumbent can subsequently draw from as a recombination input. When
entrant knowledge generation falls toward zero, the diverse trajectories
from which the incumbent's own next-generation recombination could draw
are depleted. The spillover-suppression cost is:

\begin{equation}
C^{self}_{spillover}(e,T)=\int_T^\infty e^{-\rho t}\cdot\Delta Spillover_{inc}(t;e,T)\,dt
\label{eq:ch8:c-self-spillover-e-t}
\end{equation}

Here, \(\Delta Spillover_{inc}(t;e,T)\) is the incumbent's later loss of externally generated knowledge inputs caused by earlier exclusion. The lower limit \(T\) marks the point after the enclosure period when missing entrant learning, tools, discoveries, or complements begin to reduce the incumbent's own future recombination field.

\textbf{Proposition M5.P6\index{M5.P6 Self-Undermining Enclosure} (Self-Undermining Enclosure, Reduced-Form
Conditional Proof)}: Under the conditions that the commons-reliance
parameter \(\lambda_C\) \textgreater{} 0 or the spillover-attribution
coefficient \(\omega_S\) \textgreater{} 0 (the incumbent draws on
commons or entrant spillovers for some portion of its own field
diversity), there exists a finite threshold \(\hat{T}\) \textgreater{}
\(T^*_{\mathrm{strategic}}\) such that for T \textgreater{} \(\hat{T}\),
\([\Pi_{\mathrm{inc}}]'(T)<0\): the incumbent's own payoff is
declining in enclosure duration. Enclosure beyond \(\hat{T}\) is
self-undermining. The reduced-form proposition is developed in Appendix F.

M5.P6 complements M5.T1 rather than contradicting it. M5.T1 gives the conditional result that the
incumbent over-encloses relative to the social optimum \(T^*\); M5.P6
states, under the D7 reduced-form hypotheses, that there is a
ceiling \(\hat{T}\) beyond which even the incumbent's private payoff
declines. The over-enclosure region, where social welfare is falling
but the incumbent's payoff is still rising, lies between \(T^*\) and
\(T^*_{\mathrm{strategic}}\). The self-undermining region lies above \(\hat{T}\)
\textgreater{} \(T^*_{\mathrm{strategic}}\). Between \(T^*_{\mathrm{strategic}}\) and
\(\hat{T}\), the incumbent's payoff is declining but still above its
\(\Pi_0\) level. The three-region structure (over-enclosure, private
optimum, self-undermining) is the full characterization of the
incumbent's unilateral payoff function.

The self-undermining threshold is most relevant in knowledge-intensive
sectors where incumbents depend heavily on commons \(K^C\) and on the
external knowledge spillovers that a productive entrant population
generates. An incumbent that depletes the commons on which its own
generation depends, or that suppresses the entrant population from which
its recombination inputs partly come, may discover that over-enclosure
damages its own future knowledge productivity. This is not guaranteed, it depends on \(\lambda_C\) and \(\omega_S\), which are empirical
parameters, but it is likely relevant to many incumbents in the
recombination-intensive sectors emphasized here, because their knowledge
generation is not fully self-sufficient.

\section{The Commons Enclosure Game}\index{D8 commons-enclosure game!Chapter 8}\index{commons-enclosure game}\label{the-commons-enclosure-game}\index{D8 commons-enclosure game|textbf}\index{D8|see{D8 commons-enclosure game}}

\paragraph{Causal role.} This section applies the strategic story to commons-reliant production. The enclosed object is not only code, data, or infrastructure, but the capability stock and maintenance path that keep the commons productive.

D8 extends Model 5 from the single-incumbent case to the N-incumbent
case, asking what happens when multiple incumbents simultaneously choose
enclosure strategies from a shared knowledge commons. D8 is not at
canonical proof level. The dual-channel game structure is characterized,
and D8.P1\index{D8.P1 Dual-Channel Game} and D8.P2\index{D8.P2 Decreasing Optimal Term with N} are retained as conditional structural conjectures,
not as completed equilibrium results. The full N-actor equilibrium proof remains open.
More precisely, D8 has characterized the stock-channel and flow-channel
structure, stated D8.P1 and D8.P2 as bounded conjectural propositions,
and left three items open: the full N-actor equilibrium proof, the
SF1--SF4 structural field-composition model, and the Ostrom-style
governance-design problem. This section treats D8 accordingly: as a
structural characterization and open research conjecture, not as a
completed theorem.

The D8 problem arises because the single-incumbent analysis of M5.T1
maintains the assumption that the enclosing incumbent is the dominant
force on commons stock \(K^C\). When N incumbents simultaneously
enclose, each affects the commons at aggregate enclosure intensity \(E=\sum_{i=1}^{N}e_i\), and each suffers from the aggregate depletion
caused by all others. The commons dynamic becomes:

\begin{equation}
C_t(e_1,\ldots,e_N,t)=C_0\cdot e^{-\delta^e_C\cdot E\cdot t}
\label{eq:ch8:c-t-e-1-ldots-e-n-t}
\end{equation}

Read \(C_t\) as the maintained usable commons stock: open-source package quality, documentation, security, maintainer capacity, shared protocol reliability, or community governance capacity. The exponential form is illustrative. The claim requires only that aggregate enclosure can reduce maintained commons capacity.

Each incumbent i's accessible field diversity depends on the aggregate
enclosure intensity, not merely its own: \(D_u(F_{i,t})\) =
\(D^{internal}_{i,t}\) + \(\lambda_C\) · \(C_0\) ·
\(e^{-\delta^e_C \cdot E \cdot t}\). This means that each incumbent's
enclosure of the commons imposes a depletion externality on all other
incumbents, not only on excluded entrants.

The game structure that D8.P1 characterizes is a dual-channel
commons-enclosure structure. In the commons-depletion channel it may
resemble a Prisoner's Dilemma: each incumbent may prefer that others do
not enclose the commons, since that would preserve the incumbent's own
field diversity from the commons, while still having a private incentive
to enclose its own portion. This incentive pattern is a structural
characterization, not a completed dominance result. The anti-commons
dimension \parencite{HellerEisenberg1998} is also present: multiple
overlapping enclosures can produce a situation where no incumbent can
productively use the commons even for its own recombination, because
the aggregate depletion can degrade or destroy the commons' generative capacity. Examples include standards-essential patent disputes, AI training-data access conflicts, open-source package dependency, and genomic database access.

This is also the point at which the Coasean property-rights solution\index{property rights!Coasean solution}\index{Coase theorem} reaches its boundary. Assigning rights does not guarantee efficient bargaining when the future users, future combinations, and future values of the knowledge stock are unknown or non-contractible. \textcite{Farrell1987}\index{Farrell, Joseph} emphasizes that the Coase theorem demands coordination and negotiation, not merely ownership, and \textcite{AivazianCallen1981}\index{Aivazian and Callen} shows that multi-party bargaining can fail when the core is empty even under the usual simplifying assumptions. D8 imports that warning into knowledge governance: several incumbents can face individually rational enclosure choices while no stable allocation preserves the commons\index{commons governance!D8 game}\index{anticommons!commons-enclosure game} as a usable recombination field.

Formally, the open extension can be stated as a core condition:
\[
\operatorname{Core}(G_K)=\varnothing
\quad \Rightarrow \quad
\text{no stable allocation of access, exclusion, and recombination surplus.}
\]
In plain terms, no coalition can agree on access terms that preserve recombination surplus. This is not claimed as a completed KBC theorem. It is a placeholder for the future D8 proof: the multi-incumbent knowledge game may fail not only because bargaining is costly, but because the coalition structure has no stable allocation once access, exclusion, and recombination surplus are jointly considered.

\textbf{D8.P2 (Open conjecture: decreasing optimal term with N)}:
Under specified regularity conditions on the commons dynamics and the
payoff structure, the privately optimal enclosure duration
\(\hat{T}(N)\), the self-undermining threshold, is conjectured to be
decreasing in N. As the number of incumbents who simultaneously enclose
the commons grows, the conjectured self-undermining threshold falls:
each incumbent's enclosure may become less durable before it damages
its own payoff, because the aggregate depletion from all N incumbents
erodes the commons faster. This predicts that more competitors can sometimes damage a shared knowledge commons faster if each encloses a portion of it.

If D8.P2 is later completed or empirically supported, it would have a
counterintuitive implication: concentrated incumbent markets (small N)
may be more stable in the over-enclosure region, the self-undermining
threshold may be higher, and incumbents may profitably enclose for
longer before suffering self-damage, while fragmented incumbent markets
(large N) may drive the commons to collapse faster because aggregate
enclosure intensity is higher and commons depletion is more rapid. The
multi-incumbent commons game may therefore be more destructive than the
single-incumbent case, even though it is more competitive in the
standard industrial-organization sense.

This claim is conditional and conjectural. The full N-actor equilibrium
proof remains open. The claims in this section should be read as
characterizing a candidate game structure and identifying conditions
under which qualitative results may follow, not as stating completed
theorems or policy-ready equilibrium predictions.

\section{Equilibrium Types}\label{equilibrium-types}

\paragraph{Causal role.} The equilibrium types summarize causal outcomes rather than taxonomic categories. Each type records who is excluded, what pathway is lost, what the incumbent gains, and whether the system preserves or narrows future generation. Model 5 characterizes five qualitatively distinct equilibrium types, determined by the ratio \(R_{\mathrm{enc}}(g)/G^R_{\mathrm{return}}\), the conditions that determine whether M5.T1 and M5.P6 are active, and whether D8.P2's conjectured mechanism applies.

\addtocounter{table}{-1}
\begin{longtable}{@{}L{0.18\textwidth}L{0.22\textwidth}L{0.24\textwidth}L{0.18\textwidth}L{0.12\textwidth}@{}}
\caption{Equilibrium types in strategic enclosure}\label{tab:strategic-enclosure-equilibrium-types}\\
\toprule
Equilibrium type & Private logic & System-level effect & Example & Status \\
\midrule
\endfirsthead
\caption*{Equilibrium types in strategic enclosure (continued)}\\
\toprule
Equilibrium type & Private logic & System-level effect & Example & Status \\
\midrule
\endhead
\bottomrule
\endfoot
\textbf{Efficient enclosure} & \(R_{\mathrm{enc}}(g)/G^R_{\mathrm{return}} < 1\); \(T^*_{\mathrm{strategic}} \approx T^*\). The incumbent's dynamic revenues are small relative to its self-imposed costs, so \(E^*_{alloc} \to 0\). Productive recombination is more profitable than enclosure. & \(N_{traj}\) grows, \(D_u(F_{a,t})\) remains wide for non-incumbents, and aggregate \(G^R\) is near the social optimum. Standard appropriability logic works because enclosure at \(T^*\) induces production without the over-enclosure externalities identified by T5. & Narrow knowledge goods with limited recombinant relevance; cases where incentive, disclosure, or quality-control benefits dominate field loss. These conditions are less likely to characterize frontier digital goods with high feedback capture and platform lock-in. & Conditional benchmark; demanding conditions. \\
\addlinespace
\textbf{Strategic over-enclosure} & H1--H4 hold; \(T^*_{\mathrm{strategic}} > T^*\); M5.T1 is active; \(E^*_{alloc}\) is interior but above zero. The incumbent sustains enclosure while private payoff is still rising even after social welfare has begun to decline. & \(N_{traj}\) is below the social optimum, \(\widetilde{C}_{inc}/\widetilde{C}_{ent}\) widens, and aggregate \(G^R\) falls below its efficient level. Welfare loss is real but not necessarily catastrophic in the short run. T5.G4 applies here. & Frontier AI model access; biotech/pharma patent thickets; recombination-intensive sectors where incumbent control over IP, interfaces, model access, standards, or feedback channels produces rents and learning advantages. & Candidate baseline for recombination-intensive sectors in advanced IP economies; main Chapter 8 conditional result. \\
\addlinespace
\textbf{Mutually destructive enclosure} & M5.P3 holds for all incumbents simultaneously: \(\Delta TR_{private} > 0\) and \(\Delta G^R_{agg} \le 0\). M5.P4 may also be triggered where \(N_{traj}\) falls with enclosure duration and \(\tau_{rec}\) is superlinear in \(T\). Entrepreneurial capability is directed toward enclosure rather than recombination, so \(E^*_{alloc}\) approaches 1. & Trajectory count falls for all actors. Aggregate \(G^R\) may decline faster than in ordinary over-enclosure because D8-style multiple-incumbent dynamics can accelerate commons depletion. Recovery requires access restoration plus component-specific capability reconstruction under M5.P1. & Patent thickets, mutually blocking standards, platform ecosystems where every major actor gates access while drawing from a shared commons. & Conditional extension; the collapse reading is suggested by the reduced forms (M5.T1, D7) and pends the open N-actor proof (D8); D8-linked and not policy-ready. \\
\addlinespace
\textbf{Commons-sustaining} & Incumbents coordinate, formally or informally, to maintain productive commons \(K^C\) despite individual enclosure incentives. Stability requires \(\lambda_C > 0\), active governance institutions, contribution norms, and limited enclosure of elements that do not deplete the commons infrastructure. & Future generation is preserved because the commons remains usable as a recombination field. The equilibrium is conditionally stable, but defection may trigger the commons-depletion externality dynamic. & Linux Foundation-style open-source governance, W3C standards, Kubernetes ecosystem governance, patent pools, open-source governance with corporate participation constraints, or mandated commons contribution alongside private enclosure. & Possible coordinated equilibrium; not assumed as the natural non-cooperative outcome. \\
\addlinespace
\textbf{Regulated-access} & A planner or governance institution shifts the equilibrium from strategic over-enclosure toward the efficient outcome through access, duration, or capability instruments. Access tools reduce \(R_{\mathrm{enc}}(g)\); duration tools pull private duration toward \(T^*\); capability tools raise \(\widetilde{C}_{min}\) and reduce social loss from capability traps. & The system preserves more recombination and learning while still allowing some private appropriability. Instrument design must match the source of the private-social divergence; instruments aimed at the wrong mechanism may be ineffective or counterproductive. & Compulsory licensing, API access mandates, public compute/data infrastructure for AI, FRAND licensing, interoperability mandates\index{interoperability mandates}, sunset clauses, periodic licensing review, public knowledge infrastructure investment, commons-governance support. & Governance-design equilibrium; conditional on institutional competence and mechanism fit. \\
\end{longtable}

The table also disciplines D8's status. D8 remains an open research conjecture, not a completed theorem. It characterizes a candidate N-actor commons-enclosure structure, but the full N-actor Nash equilibrium proof remains unresolved. The table should therefore be read as a map of possible equilibrium types and mechanism conditions, not as a completed equilibrium taxonomy or numerical policy model.

\section{Falsification and Boundary
Conditions}\label{falsification-and-boundary-conditions}

\paragraph{Causal role.} This section states what would break the argument. The strategic-enclosure claim weakens if enclosure does not exceed a welfare benchmark, if excluded actors recover quickly, if losses are redundant, or if enclosure benefits systematically dominate suppression costs.

M5.T1 is a conditional theorem. Its conclusion \(T^*_{\mathrm{strategic}}\)
\textgreater{} \(T^*\) follows necessarily from H1--H4; if any condition
fails, the theorem is inapplicable to the case at hand rather than
falsified. The falsification matrix identifies five failure modes. The full M5 falsification matrix should be treated as a future empirical protocol for the strategic-enclosure programme, not merely as a summary of these five modes. Its next use is to translate H1, H2, H2', H3, and H3' into condition-specific tests, proxies, rival explanations, and aggregate implication tests for candidate sectors.

A first empirical design would compare affected and unaffected actors around an access-rule change: API closure\index{API closure!strategic enclosure}, API pricing change, patent expiry, compulsory licensing event, platform interoperability mandate, or model-access restriction. A difference-in-differences design could test whether excluded complementors experience lower product formation, slower update cycles, reduced entrant survival, weaker benchmark improvement, or delayed capability recovery relative to comparable unaffected actors.

KBC predicts that enclosure is net-negative for future generation capacity when the suppressed commons yield, recombination loss, capability decay, recovery lag, and trajectory loss exceed the private substitution effect from proprietary investment. This is a testable conditional tied to the T5 cost-benefit structure, not a definitional claim.

The most serious is formal invalidation (F1): a formal condition H1--H4
does not hold. If \(M_{rec}=1\) in all sectors, if enclosed
knowledge enters no excluded actor's productive field at positive weight, H1 fails and the dynamic revenue structure collapses, eliminating
the divergence between \(T^*_{\mathrm{strategic}}\) and \(T^*\). This would mean
that knowledge-bearing capitalism's enclosure problem is purely a
distributional one (the standard Harberger triangle case), not a
generative one. The empirical test: do IP expansions in software,
genomics, and AI generate measurable reductions in recombination
diversity in adjacent sectors? If not, if enclosure has no effect on
\(N_{traj}\) or on aggregate \(G^R\) in non-enclosed domains, 
\(M_{rec} \approx 1\) and the over-enclosure claim is formally inapplicable.

Mechanism failure (F2): H1--H4 hold but the causal pathway is
misspecified. Incumbents may internalize dynamic trajectory revenue not
through capability accumulation but through price discrimination or
product differentiation, in which case the T6 mechanism does not produce
the capability divergence this theory predicts. The empirical test: does
exclusion from platform APIs or patent-protected knowledge generate
capability divergence (measurable through domain breadth and
recombination output) in excluded firms, at rates consistent with the
state equation's prediction?

Scope limitation (F3): M5.T1 applies only in sectors with high
recombination intensity, high switching costs, and platform-mediated
learning loops. This is a boundary condition rather than a
falsification: it restricts the claim that over-enclosure is a general
tendency in knowledge-intensive sectors rather than a pathology of specific
market structures. The key empirical question is whether
high-\(M_{rec}\), platform-mediated sectors represent a substantial and
growing portion of economic activity, or an exceptional fringe. If \(M_{rec}\) is high only in narrow submarkets, the theory's scope shrinks rather than fails globally.

Magnitude failure (F4): the direction of over-enclosure is correct but
the magnitude is small relative to other welfare distortions. This does
not falsify this theory but limits policy relevance.

Endogenous correction (F5): market or institutional responses, 
collective licensing bodies, open-source alternatives, multi-homing, 
systematically close the gap between \(T^*_{\mathrm{strategic}}\) and \(T^*\)
without formal policy intervention. The over-enclosure equilibrium may
be theoretically reachable but practically self-correcting. The
counter-evidence: in sectors with the highest \(M_{rec}\) (AI, genomics,
core internet infrastructure), do open alternatives systematically close
the gap generated by incumbent enclosure, or does the feedback-capture
advantage prevent them from doing so?

This theory predicts endogenous correction is unlikely to fully close the
gap in high-\(M_{rec}\), high-feedback-capture sectors, because the
mechanism generating the gap, exclusive deployment scale producing
superior \(G^L\), is self-reinforcing and not addressable through the
access instruments that network-effect competition typically exploits.
This is the strongest empirical prediction of Chapter 8: in frontier AI
and genomics, the capability gap between enclosed incumbents and
accessible alternatives should widen over time even as open-access
alternatives exist and improve, because the feedback-capture mechanism
sustains incumbent improvement faster than commons-based alternatives
can close the distance.

Evidence of widening would include persistent benchmark gaps, deployment-scale gaps, user-feedback volume gaps, model-update frequency differences, downstream adoption divergence, or entrant survival decline.

\section[Handoff to dark capital and the accounting shadow]{Handoff: Strategic Enclosure, Unmeasured Capital, and Chapter 9}\label{handoff-strategic-enclosure-unmeasured-capital-and-chapter-9}

\paragraph{Causal role.} The handoff converts the causal story into a measurement problem. Chapter 9 asks why the enclosed paths, excluded actors, suppressed alternatives, and latent losses remain weakly visible in accounting and productivity measures. The incumbent books rent; the accounts rarely book the suppressed alternative trajectories.

Chapter 8 has stated and conditionally derived the strategic result: individually
rational enclosure strategies, operating through the private payoff
function that internalizes feedback acceleration and trajectory rent while
externalizing recombination suppression and conditional capability loss, produce
equilibrium enclosure durations \(T^*_{\mathrm{strategic}}\) \textgreater{}
\(T^*\) in recombination-intensive knowledge sectors when H1--H4 hold.
The alignment assumption is therefore no longer a background premise; it is a conditional strategic-equilibrium result whose empirical
force depends on \(M_{rec}\) and on the relative size of Proposition~E's
benefit channels.

The strategic analysis has a direct implication for measurement that
Chapter 9 (Dark Capital and the Accounting Shadow) develops. Strategic
enclosure creates three distinct categories of unmeasured capital loss
that do not appear in standard accounting.

First, under M5.T1 conditions, strategic over-enclosure generates post-enclosure capability
deficits (\(C_{T6}\) capability decay, in the T5 social cost) that accounting
systems cannot recognize. The excluded firm's balance sheet shows the
same \(K^D\) and \(K^I\) stocks before and after the enclosure period;
the relative capability loss, and in the limiting case the capability decay in \(\widetilde{C}_{ent}\), predicted by the state equation does not reduce any recognized asset. The firm's reported
productive capacity is unchanged; its actual productive capacity has
fallen.

Second, the \(C_{T7}\) suppressed appreciation / feedback-learning loss, the decline in
the productive value of non-enclosed knowledge-bearing stocks held by
excluded actors, is entirely invisible to accounting. T7 proves that
\(v(K_a,a,g_1)<v(K_a,a,g_0)\) when E contains
stocks with positive complementarity to \(K_a\); but the accounting value
of \(K_a\) does not change when an external enclosure event restricts its
complementary field elements. The asset's book value is unchanged; its
productive value has fallen.

Third, the commons depletion documented in §8.6, the extraction of
governance \(K^E\) from commons knowledge capital \(K^C\) through
ordinary labour-market transactions, produces no accounting
recognition at the commons level. Individual employment contracts are
legitimate transactions; the aggregate depletion of commons governance
capacity is an invisible externality with no corresponding liability
entry.

These three forms of unrecognized capital loss, capability deficit,
suppressed appreciation, and commons depletion, constitute what
Chapter 9 will call dark capital: productive value that exists within
the knowledge economy, exercises causal force on investment decisions,
knowledge generation rates, and competitive outcomes, but is invisible
to the accounting systems through which capital is recognized, valued,
and governed. Strategic enclosure is a mechanism that can produce dark
capital at scale; Chapter 9 develops the accounting shadow that
strategic enclosure casts.

\part{Measurement and Implications}
\chapter{Dark Capital and the Accounting Shadow}\label{chapter-9-dark-capital-and-the-accounting-shadow}
\index{dark capital|textbf}\index{accounting shadow|textbf}

\chapterhook{The Wealth We Cannot See}

Dark capital is the use-value the exchange-value ledger cannot see. Chapter 1 located it as the half of value Smith set aside; this chapter resolves it into kinds, use-value that has not been converted to a price, use-value converted in a way that suppresses more of it, and use-value damaged by neglect or enclosure, and asks why the instruments built for vendible commodities are blind to all three.

This chapter extends the accounting and intangible-capital\index{intangible-capital accounting} literature by distinguishing recognized intangible assets, market-value gaps, unrecorded capability, and unpriced knowledge-capital impairment\index{recognized intangible assets}\index{market-value gaps}\index{unrecorded capability}\index{unpriced knowledge-capital impairment} \parencite{Lev2001, CorradoHultenSichel2005, BondCummins2000, HaskelWestlake2018, MacKinlay1997}\index{Lev, Baruch}. Its central claim is that measurement failure is not value absence\index{measurement failure}. The Technical Companion, Appendix D, supplies the measurement method used to move from dark-capital diagnosis to decision-relevant uncertainty reduction\index{decision-relevant measurement!uncertainty reduction}\index{measurement uncertainty}.

Read through the apparatus of Chapter 1, dark capital is one sign of a single quantity: the gap between the productive yield a knowledge unit actually renders and the generativity it visibly signals. Where price fuses value and signal at the point of exchange, knowledge use separates them, so a unit's real service can run ahead of its visible trace, which is the case this chapter develops. But the gap has a second sign, and a theory of dark capital alone would miss it. When the visible signal runs ahead of the service, the trace outpaces the yield: work widely cited but unreplicable, methods fashionable but weak, software heavily depended upon yet fragile, benchmarks that coordinate a field without improving it. Call this signal-inflated stock\index{signal-inflated knowledge stock}. The accounting shadow therefore falls in two directions, hiding real value and magnifying apparent value, and an honest measurement programme must correct for both. The formal treatment of the yield-generativity gap and its four dark zones is given in the Technical Companion, Appendix~M.

\noindent\textbf{Dark capital names a problem before it measures a quantity.} Dark capital names a real and consequential problem: economically material value that current accounts neither see nor measure, value that is unseen, unmeasured, mismeasured, or unrecognized by the systems built to record it. Naming and locating that problem is the first contribution of this chapter, and it is deliberately a pointed one, because such value stays invisible precisely while no existing category points at it. What follows then \emph{defines} dark capital, distinguishes its components, and specifies what would have to be observed to measure each. What this chapter does not yet assert is a settled magnitude. At this stage dark capital is a defined and \emph{measurable target}, not an estimated number, and the scoring architectures introduced below, K-CMM, the Knowledge Portfolio Value Function, the realization coefficient, and the impedance and yield constructs, are measurement architectures that identify what must be observed, not calibrated estimates of how much. That distinction is not a retreat from the claim. It is what makes the claim testable rather than rhetorical: a target can be reached, and a calibration in a bounded domain can later confirm, bound, or weaken it.

Chapter 8 established that strategic enclosure produces three categories
of economically real capital effect that current accounting often does not separately record:
capability deficits in excluded actors whose \(\widetilde{C}\) may decay under the limiting T6 conditions of feedback-enclosure; suppressed appreciation of non-enclosed
stocks whose productive value falls when complementary field elements
are enclosed; and commons depletion through labour-market extraction of
governance \(K^E\) that leaves formal open structures without the
embodied capacity to sustain themselves. All three can arise from rational incumbent strategy. They often do not trigger a separate accounting recognition
event. The capital effects may be economically real before the accounting record can isolate them.

Chapter 9 names this class of effects, and extends it systematically
beyond the consequences of strategic enclosure to the full range of ways
knowledge-bearing capitalism creates, destroys, and repositions
productive capacity outside accounting\textquotesingle s field of view.

\begin{quote}
\textbf{Dark Capital.} Dark capital is knowledge-bearing capital,
capital loss, or capital-forming capacity that is economically real but
unseen, unmeasured, mismeasured, or unrecognized\index{value absence versus measurement failure}\index{measurement failure!versus value absence} by the systems that
manage, govern, price, disclose, account for, or regulate capital. It is
not merely capital that accounting misses on the upside. It includes
real capital loss on the downside and capital that could have formed but
did not because the conditions for its formation were suppressed. The
master category comprises four components: Dark Value\index{dark value} (unrecognized
knowledge capital that exists and is productive), Dark Risk\index{dark risk} (liability-like exposures
arising from enclosure-driven capability loss\index{capability loss|textbf}), Foregone Knowledge
Capitalization (suppressed generation from enclosure of commons and
feedback loops), and the Accounting Shadow (the systematic discrepancy
between economic reality and accounting recognition\index{accounting recognition}).
\end{quote}

The measurements in this chapter occupy different statuses. Some are classificatory identities; some are diagnostic decompositions; some are valuation architectures\index{valuation architecture}; some are uncalibrated scoring models; and some are formal objects audited in the Technical Companion. Chapter 9's purpose is not to force dark capital directly onto balance sheets\index{balance sheet}. Its purpose is to identify economically material knowledge-capital value, loss, fragility, and non-formation that existing accounting and valuation categories often leave implicit.

\begin{center}
\setlength{\fboxsep}{8pt}
\fbox{%
\begin{minipage}{0.94\textwidth}
\small
\textbf{Reader's Measurement Stack}\index{measurement stack|textbf}

\vspace{0.5em}
\begin{tabular}{@{}p{0.24\textwidth}p{0.64\textwidth}@{}}
\toprule
Tool or concept & Main-text role \\
\midrule
Dark capital & The problem: economically material knowledge value, loss, dependency, or non-formation that current recognition systems do not cleanly identify. \\
K-CMM\index{Knowledge-Capital Measurement Model (K-CMM)|textbf}\index{Knowledge-Capital Measurement Model (K-CMM)!Chapter 9 architecture}\index{K-CMM|see{Knowledge-Capital Measurement Model (K-CMM)}} & The single-stock decision tool: estimates value components and expected loss for a bounded knowledge-bearing stock. \\
KPVF\index{Knowledge Portfolio Value Function (KPVF)|textbf}\index{Knowledge Portfolio Value Function (KPVF)!portfolio valuation}\index{Knowledge Portfolio Valuation Framework (KPVF)|see{Knowledge Portfolio Value Function (KPVF)}}\index{KPVF|see{Knowledge Portfolio Value Function (KPVF)}} & Portfolio valuation: asks how a governed set of knowledge-bearing stocks creates value through marginal contribution, recombination, overlap, and dependency. \\
KCI / EVPI\index{Knowledge Capital Index (KCI)}\index{KCI|see{Knowledge Capital Index (KCI)}}\index{expected value of perfect information (EVPI)}\index{EVPI|see{expected value of perfect information (EVPI)}} & Measurement trigger: measure only when expected decision improvement exceeds measurement cost. \\
EKL\index{expected knowledge loss (EKL)|textbf} & Expected loss term: captures depreciation, degradation, access loss\index{access loss}, exfiltration\index{exfiltration!dark capital}, capability loss, false-stock exposure, and other downside channels. \\
Chapter 11 & Empirical tests: translates these objects into falsification, calibration, event-study, case-comparison, and measurement-validity tests\index{calibration!empirical tests}\index{event study}\index{case-comparison testing}\index{measurement-validity tests}. \\
\bottomrule
\end{tabular}
\end{minipage}%
}
\end{center}

Table~\ref{tab:ch9:dark-capital-map} provides the main-text map for the chapter. It identifies the four components of dark capital, what each component captures, and the measurement status of each claim. The table is intentionally placed before the formal decompositions so that dark capital is not mistaken for a single accounting residual.

\begin{table}[htbp]
\centering
\small
\caption{Dark-Capital Components and Measurement Status}
\label{tab:ch9:dark-capital-map}
\begin{tabular}{@{}p{0.24\textwidth}p{0.43\textwidth}p{0.25\textwidth}@{}}
\toprule
Component & What it captures & Measurement status \\
\midrule
Dark value & Productive knowledge-bearing stock not fully recognized in accounts. & Valuation architecture. \\
Dark risk & Latent impairment, dependency, fragility, or expected knowledge loss. & Diagnostic decomposition. \\
Foregone knowledge capitalization & Knowledge stock that would plausibly have formed under different access, feedback, or recombination conditions. & Counterfactual measurement problem. \\
Accounting shadow\index{accounting shadow!definition} & Gap between economic knowledge-capital effects and accounting recognition. & Analytical bridge, not accounting standard\index{accounting reform}\index{accounting standards}. \\
\bottomrule
\end{tabular}
\end{table}

A compact version of the category is:
\[
\begin{aligned}
\text{Dark capital} ={}& \text{unrecognized stock} \\
&+ \text{unrecognized governance-position value} \\
&+ \text{unpriced capability and fragility exposure.}
\end{aligned}
\]
This is a classificatory identity, not an accounting identity. It makes explicit that the shadow is not only an asset-recognition problem. It also includes governance-position value and the capability, cybersecurity, commons, platform, and public-infrastructure fragilities that determine whether knowledge-bearing stock can continue to yield productive services.

Dark capital is not simply another name for intangible assets.\index{dark capital versus intangible assets} Intangible assets refer to nonphysical assets or claims to future benefits, including recognized assets, acquired rights, internally generated capabilities\index{internally generated assets}\index{internally generated capabilities}, brands, software, data, and other nonphysical sources of expected economic advantage. Dark capital refers more narrowly to knowledge-capital value, risk, dependency, or non-formation that remains economically material but weakly identified, weakly valued\index{measurement!weak identification}\index{valuation!weak valuation}, aggregated into other categories, or absent from current accounts.

This chapter\textquotesingle s claim is therefore not merely the familiar one that "intangibles are undervalued." That observation is established. \textcite{Lev2001}, \textcite{CorradoHultenSichel2005}, and \textcite{HaskelWestlake2018} have each documented it, and it has informed accounting reform debates for two decades. The claim here is sharper: knowledge-bearing capitalism produces economically real capital effects that neither book value\index{book value}, market value\index{market value}, nor standard intangible-asset categories can properly isolate. The failure is not only that current accounting may omit or aggregate productive assets on the upside. It is also that current accounts may not separately identify capital loss on the downside, and may entirely omit a third dimension: capital that could have formed but did not because the conditions for its formation were suppressed.

\section{Incomplete Institutional Commensuration of Knowledge Capital}\label{91-incomplete-institutional-commensuration-of-knowledge-capital}

The central measurement weakness in knowledge-bearing capitalism is not merely a problem of measurement, calibration, or aggregation. The problem is that value can exist before institutions can recognize it cleanly. In that sense, it is an incomplete institutional commensuration problem\index{incomplete institutional commensuration|textbf}\index{commensuration problem}. The issue is not that knowledge-bearing capital cannot be valued. The issue is that the institutional machinery needed to make that valuation clean, auditable, standardized, externally assured, and balance-sheet-recognizable is not yet mature.

Traditional capital aggregation is not natural; it is institutionally achieved. Knowledge-bearing capital can be monetarily valued in principle, but clean aggregation requires conventions for unit definition, separability, benefit attribution, depreciation, impairment, audit, assurance, and accounting recognition.

The resulting problem has five separable layers. Table~\ref{tab:ch9:commensuration-layers} distinguishes them so that a failure at one layer is not mistaken for a failure of the whole theory. In particular, the fact that knowledge-bearing capital cannot yet be cleanly recognized on audited balance sheets does not mean that it cannot be bounded, valued, internally measured, or calibrated for decision use.

\begin{table}[htbp]
\centering
\small
\caption{Five Layers of the Knowledge-Capital Commensuration Problem}
\label{tab:ch9:commensuration-layers}
\begin{tabular}{@{}p{0.20\textwidth}p{0.35\textwidth}p{0.37\textwidth}@{}}
\toprule
Layer & Question & KBC answer \\
\midrule
Unit definition & What is the operative object? & The Operative Knowledge Unit partially addresses this by bounding the task-denominated knowledge object. \\
Valuation & What productive services does it yield? & The capital-services valuation\index{capital-services valuation} identity addresses this structurally by valuing expected productive-service flows. \\
Aggregation & Can values be added without double-counting? & Only with no-overlap rules that prevent the same future benefit stream from being counted in more than one component. \\
Calibration & Are coefficients empirically validated? & Open; requires datasets, outcome testing, and coefficient validation across domains. \\
Recognition & Can it appear on audited statements? & Not generally; requires accounting standards, assurance protocols, and externally verifiable control, separability, and measurement conventions. \\
\bottomrule
\end{tabular}
\end{table}

\paragraph{Measurement Principle,  Institutional Commensuration of Knowledge Capital.}
Monetary commensuration is necessary but insufficient for knowledge-capital aggregation. Knowledge-bearing stock can be valued using capital-services logic: expected future productive services discounted over time. But clean balance-sheet aggregation requires more than a common monetary denominator. It requires institutional commensuration: recognized unit conventions, sufficiently separable control, verifiable existence, reliable benefit attribution, no-overlap rules, depreciation and impairment conventions, auditability\index{auditability}, and assurance standards. Physical capital acquired these stabilizing conventions through asset markets, property law, corporate accounting, audit practice, insurance, tax systems, and collateral markets. Knowledge-bearing stock has acquired them only partially. Its aggregation problem is therefore not metaphysical impossibility, but incomplete institutionalization.

The capital-services identity gives the monetary valuation layer of this measurement principle:
\[
V^{K}_{a, i, t}(\pi)
=
\mathbb{E}_{t}\left[
\sum_{\tau=0}^{H}
\frac{PS_{a, i, t+\tau}(\pi)}{(1+r)^{\tau}}
\right].
\]
The value of a knowledge-bearing stock is the expected discounted value of the productive services it yields to actor \(a\), from stock \(i\), at time \(t\), under governance form \(\pi\), over horizon \(H\). This is a valuation architecture. It shows that knowledge capital can be valued in monetary terms by reference to expected productive-service flows. It does not show that such value is already auditable, aggregable, calibrated, or balance-sheet-recognizable.

Identification precedes valuation. Valuation precedes calibration. Calibration precedes assurance. Assurance precedes accounting recognition.\index{identification before valuation}\index{valuation!before calibration}\index{calibration!before assurance}\index{assurance!before accounting recognition} K-CMM operates mainly in the first three stages: it identifies material knowledge-capital objects, structures valuation judgement, and indicates where calibration would reduce decision uncertainty.

K-CMM is a decision-support and uncertainty-reduction framework, not a GAAP/IFRS recognition model.

The Knowledge-Capital Measurement Model decomposes this value into channels before any aggregation is attempted:
\[
V^{K}_{i}
=
CUV_{i}
+
ROV^{*}_{i}
+
LOV^{*}_{i}
+
COV^{*}_{i}
+
SOV^{*}_{i}
-
EKL_{i}.
\]
This is an uncalibrated scoring model and valuation architecture, not a calibrated accounting equation. Table~\ref{tab:ch9:kcmm-decomposition} gives the plain-language interpretation of the components. The asterisks indicate values adjusted for overlap so that option-like benefits are not counted twice.

\begin{table}[htbp]
\centering
\small
\caption{K-CMM Value Components and Overlap Discipline}
\label{tab:ch9:kcmm-decomposition}
\begin{tabular}{@{}p{0.22\textwidth}p{0.68\textwidth}@{}}
\toprule
Component & Meaning \\
\midrule
\(CUV_i\) & Current use value. \\
\(ROV_i^{*}\) & Recombination option value, adjusted for overlap. \\
\(LOV_i^{*}\) & Learning-loop option value, adjusted for overlap. \\
\(COV_i^{*}\) & Control/governance option value, adjusted for overlap. \\
\(SOV_i^{*}\) & Strategic option value, adjusted for overlap. \\
\(EKL_i\) & Expected knowledge loss. \\
\bottomrule
\end{tabular}
\end{table}

Aggregation is permissible only if expected benefits are assigned to distinct channels and the same future productive-service stream is not counted twice. This rule directly addresses the strongest aggregation objection: knowledge value is often synergistic, complement-dependent, and generated jointly by several stocks. The appropriate response is not to deny monetary valuation, but to require a no-overlap discipline before component values are summarized.

\subsection{From Static Decomposition to a Dynamic Value System}\index{static decomposition versus dynamic value system}\index{dynamic value system!versus static decomposition}\label{dynamic-value-system}

The decomposition above is static: it scores the components of \(V^{K}_i\) at a point in time. A minimal closure links it to the capability dynamics and turns it into a value system. Let current use value be proxied by capability, \(\beta_C\,\widetilde{C}_{a, t}\), with \(\beta_C\) the capability-to-value coefficient (distinct from the capability exponent \(\beta\) of the \(\Phi\) aggregate), and collect the remaining components into an option-and-loss bundle
\[
V^{\mathrm{opt}}_{a, t}=ROV^{*}_a+LOV^{*}_a+COV^{*}_a+SOV^{*}_a-EKL_a,
\]
so that \(V^{K}_{a, t}=\beta_C\,\widetilde{C}_{a, t}+V^{\mathrm{opt}}_{a, t}\). Substituting the capability law (Chapter~3) with its recombination and feedback specifications, \(G^{R}_{a, t}=\lambda_a R_{a, t}^{\eta}D_u(F_{a, t})^{\mu}\) and \(G^{L}_{a, t}=\alpha_a Dp_{a, t}F^{dep}_{a, t}\phi_a\), gives a dynamic value equation:
\[
V^{K}_{a, t+1}=\beta_C\!\left[(1-\delta_C)\widetilde{C}_{a, t}+\gamma\,\lambda_a R_{a, t}^{\eta}D_u(F_{a, t})^{\mu}+\ell\,\alpha_a Dp_{a, t}F^{dep}_{a, t}\phi_a\right]+V^{\mathrm{opt}}_{a, t+1}.
\]
This converts K-CMM from a static decomposition into a dynamic value system, and like the components it aggregates it is a minimal closure, not a calibrated pricing equation.

\textbf{Comparative statics.} Knowledge-bearing value is locally increasing in usable access and useful diversity, which enter through recombination with elasticities \(\partial\ln G^{R}/\partial\ln R=\eta>0\) and \(\partial\ln G^{R}/\partial\ln D_u=\mu>0\); in capability, through both \(\beta_C\) and the capability gate of the \(\Phi\) aggregate; and in feedback, through the deployment, dependency, and capture terms of \(G^{L}\) (\(Dp\), \(F^{dep}\), \(\phi_a\)). It is decreasing in expected knowledge loss\index{expected knowledge loss (EKL)!calibration} \(EKL\), which enters \(V^{\mathrm{opt}}\) with a negative sign. Access and capability raise \(G^{R}\) not as separate exponents but by widening the magnitude \(R\) and the useful diversity \(D_u\) of the accessible field.

\textbf{Governance is not a primitive scalar.} Because a governance change \(g\) acts only through the variables it moves, its effect on value decomposes:
\[
\frac{dV^{K}_{a, t+1}}{dg}=\beta_C\!\left[\gamma\,\frac{dG^{R}}{dg}+\ell\,\frac{dG^{L}}{dg}\right]+\frac{dV^{\mathrm{opt}}}{dg},
\qquad
\frac{dG^{R}}{dg}=G^{R}\!\left(\eta\,\frac{d\ln R}{dg}+\mu\,\frac{d\ln D_u}{dg}\right),
\]
with \(dV^{\mathrm{opt}}/dg\) carrying the \(-dEKL/dg\) channel. Governance therefore has no context-free sign: it improves value when it widens access, useful diversity, capability, and feedback or tightens loss control, and reduces value when it narrows them. Enclosure has a \emph{split} comparative static, raising the incumbent's \(R\), capability, and feedback while it can lower excluded actors' access and useful diversity and raise their \(EKL\). This is the value-level restatement of Proposition~E and of the inversion condition of Chapter~8: the sign is the net of the channels, not a property of enclosure as a label.

\textbf{The congestion boundary.} Useful diversity need not raise generation without limit. If diversity carries a congestion cost, so that the diversity term is \(D_u^{\mu}e^{-\delta_D D_u}\) rather than \(D_u^{\mu}\), recombination is single-peaked in \(D_u\) with an interior maximum at
\[
D^{*}=\frac{\mu}{\delta_D}.
\]
Below \(D^{*}\), added useful diversity raises generation; above it, interpretive, search, and coordination overload reduce generation. This is the non-monotone congestion member of the function-class audit (Volume~2, Appendix~F) and the diversity-side counterpart to the Hill-order question raised at the \(G^{R}\) definition in Chapter~3.

\textbf{Measurement as a value-of-information problem.} Finally, because measurement is itself costly, KBC treats the choice of what to measure not as a valuation or accounting-recognition problem but as a value-of-information problem. Among candidate measurements \(m\in\mathcal{M}\) with cost \(c_m\), the measurement worth taking maximizes expected decision improvement net of cost:
\[
m^{*}=\operatorname*{arg\,max}_{m\in\mathcal{M}}\left\{\mathbb{E}_{Y_m}\!\left[\max_{d\in\mathcal{D}}\mathbb{E}_{\theta\mid Y_m}\,U(d, \theta)\right]-\max_{d\in\mathcal{D}}\mathbb{E}_{\theta}\,U(d, \theta)-c_m\right\}.
\]
The bracketed difference is the expected value of the sample information \(Y_m\): the gain in expected decision payoff from observing \(m\) before choosing \(d\). This is the criterion underlying the knowledge-capital-information test (\(\mathrm{KCI}>0\)) of Chapters~10 and~11 and the K-CMM acquisition-of-information sequence of the Technical Companion, Appendix~D. The rule is to measure the uncertain variable whose expected decision improvement exceeds its measurement cost, and not otherwise, which is why dark capital is a reason to measure selectively rather than to capitalize indiscriminately.

The same logic produces a staged recognition pathway. KBC does not move directly from invisible knowledge-bearing stock to balance-sheet recognition\index{accounting recognition!balance sheet}. It moves through intermediate stages: identification, unit-bounding, internal valuation, calibration, assurance, and only then possible recognition. Table~\ref{tab:ch9:recognition-ladder} treats the commensuration weakness as an institutional development programme rather than a fatal measurement objection.

\begin{figure}[!htbp]
\caption[Measurement ladder: visibility before recognition]{Measurement ladder: visibility before recognition}
\label{fig:ch9:measurement-ladder}
\centering
\includegraphics[width=0.95\textwidth]{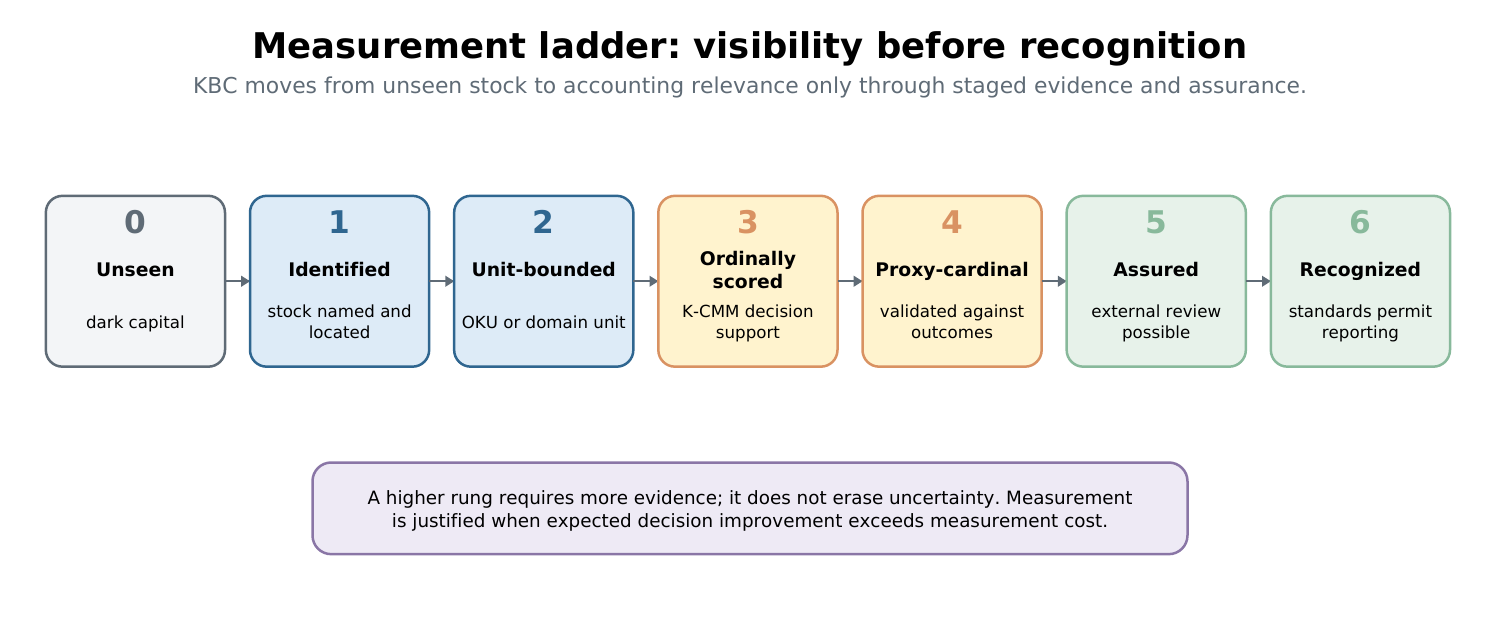}
\par\smallskip\noindent\footnotesize\emph{Note.} The ladder separates object visibility, internal decision support, calibrated proxy use, assurance, and possible accounting recognition. Higher rungs require stronger evidence; they do not imply certainty.
\end{figure}

\begin{table}[htbp]
\centering
\small
\caption{Recognition Ladder\index{recognition ladder|textbf}\index{accounting recognition!recognition ladder} for Knowledge-Bearing Capital}
\label{tab:ch9:recognition-ladder}
\begin{tabular}{@{}p{0.16\textwidth}p{0.36\textwidth}p{0.36\textwidth}@{}}
\toprule
Stage & Status & KBC function \\
\midrule
0. Unseen & Stock is not identified & Dark capital \\
1. Identified & Stock is named and located & Knowledge inventory \\
2. Unit-bounded & OKU or domain unit specified & Comparison/substitution \\
3. Internally valued & K-CMM estimate with uncertainty range & Management decision \\
4. Calibrated & Coefficients tested against outcomes & Investment/risk modelling \\
5. Assured & External review or attestation possible & Disclosure support \\
6. Recognized & Accounting standards permit recognition & Balance-sheet treatment \\
\bottomrule
\end{tabular}
\end{table}

The recognition ladder becomes more concrete when applied to distinct forms of knowledge-bearing stock. The following three examples are intentionally compact. They do not produce finished asset values. They show how KBC separates valuation from recognition\index{recognition versus valuation}\index{valuation!versus recognition}\index{accounting recognition!versus valuation} by identifying the productive-service channels, governance conditions, and expected-loss terms that must be estimated before any aggregation is defensible.

\paragraph{Example 1,  Software dependency.}
A firm that depends on an open-source software library holds no ordinary ownership claim over the code, but the dependency may be economically material. Its current use value depends partly on avoided development cost: the firm does not need to recreate the library internally. Its durability depends on the maintenance community that fixes bugs, patches vulnerabilities, updates documentation, and preserves compatibility with surrounding systems. Its expected knowledge loss depends on security risk, maintainer exhaustion, licence change, abandonment, or hostile dependency compromise. Its productive value also depends on interoperability: the library may be valuable only because it connects to a larger stack of tools, APIs, standards, and developer skills. Licence conditions determine whether the stock can be used commercially, modified, embedded, redistributed, or replaced without legal or technical friction.

This case illustrates the interaction between disembodied knowledge stock and commons knowledge stock. The codebase is primarily \(K^D\), but its productive value often depends on \(K^C\): the shared maintenance community, issue history, review norms, documentation practices, and trust infrastructure that keep the artefact usable. A simple expenditure measure misses both sides. The firm may record little or no asset, yet lose substantial productive capacity if the community collapses, the licence changes, interoperability breaks, or a security flaw forces emergency replacement. K-CMM therefore values the dependency through current use, recombination option, governance-control exposure, and expected knowledge loss rather than by treating zero acquisition price as zero capital value.

\paragraph{Example 2,  Surgeon's specialized judgement.}
A surgeon's specialized judgement is economically valuable because it combines embodied skill, tacit pattern recognition, procedural timing, complication management, and practical decision-making under uncertainty. Its value depends on patient outcomes, throughput, reputation effects, referral flows, revenue contribution, training cost, and the scarcity of substitutes. But the same features that make the judgement valuable also make it difficult to recognize externally. The stock is primarily \(K^E\): it resides in a person, and only part of it can be codified into protocols, checklists, training materials, or institutional routines.

The valuation problem is therefore not value absence. It is conditional separability. If the surgeon departs, retires, becomes unavailable, or stops practising a specialised procedure, the organization may suffer a capability loss even though no recognized asset is impaired. Some value may have been institutionalized through team routines, operating-room protocols, mentoring, or case-review systems, but the non-transferable residue remains embodied. KBC can support internal valuation by estimating substitution limits, training and recruitment cost, outcome contribution, departure risk, and expected capability loss. It cannot normally turn the surgeon's judgement into an externally recognized balance-sheet asset, because control, separability, verifiability, and assurance conditions are not satisfied in the way accounting recognition requires.

\paragraph{Example 3,  AI model with feedback loop.}
An AI model trained and deployed at scale combines disembodied knowledge stock, learning-loop control, and governance-position value. The model weights are \(K^D\), but their value depends on user feedback, evaluation data, reinforcement signals, error reports, usage telemetry, deployment infrastructure, and the capability to transform feedback into model improvement. The relevant generation channel is \(G^L\): feedback from use alters future capability. The actor who controls the learning loop may improve the model faster than actors who only observe outputs or purchase access.

This case also shows why valuation cannot be reduced to model weights alone. The same model may have different value under different governance forms because access to feedback, permission to train, infrastructure for monitoring, and control over deployment determine whether learning continues. Truth-dependence matters because hallucination, bias, poisoning, benchmark gaming, or unvalidated synthetic output can reduce current use value and increase expected knowledge loss. Private value and social value may also diverge. A firm may gain strategic option value by controlling model access and feedback, while wider recombination or public validation is restricted. KBC therefore treats the model as a residence--governance object: \(K^D\) held under a particular governance position, improved through \(G^L\), exposed to epistemic-risk loss, and potentially valuable to the firm in ways that differ from its value to the wider knowledge system.

\paragraph{Recognition boundary.}
KBC does not claim that all knowledge-bearing stock should be placed on balance sheets. It claims that economically material knowledge-bearing stock is often invisible, misclassified, or weakly measured, and that this gap can be narrowed through disciplined unit definition, productive-service valuation\index{valuation!productive-service}, no-overlap attribution, calibration, and assurance.

This boundary is not an excuse for managerial ignorance. Mature firms already maintain inventories of important information technology, data collections, applications, platforms, third-party dependencies, and critical processes, and they often assign risk ratings, business criticality, resilience requirements, or strategic significance to those assets. KBC builds on that practical reality. A firm that cannot identify the knowledge-bearing stocks on which its going concern status, risk posture, or strategic plan depends has a governance and survival problem, not an objection to the theory. The theory should not be weakened to accommodate firms whose measurement and control systems have failed to locate their own material productive dependencies.

\begin{quote}
\noindent\fbox{%
\begin{minipage}{0.93\textwidth}
\textbf{Objection: Money already aggregates heterogeneous capital. Why not knowledge capital?}

\medskip
\textbf{Answer:} Money can denominate knowledge-capital value, but denomination is not recognition. Physical capital became balance-sheet-clean because legal, market, accounting, audit, depreciation, collateral, and insurance institutions made heterogeneous assets commensurable enough for decision use. Knowledge-bearing capital requires analogous institutions. Until they mature, K-CMM can support internal valuation, uncertainty reduction, risk management, and disclosure discipline, but not universal clean balance-sheet aggregation.
\end{minipage}}
\end{quote}

\subsection{The Cost-of-Production and Exchange-Price Anchor Objection}\index{cost-of-production anchor versus exchange-price anchor}\index{exchange-price anchor!versus cost-of-production anchor}\label{cost-of-production-exchange-price-anchor-objection}

A second objection is that many knowledge-bearing stocks lack the historical-cost or exchange-price anchor that traditional accounting uses for physical capital. They are often generated as joint products of ongoing operations, depend on complementary embodied and institutional capability, and cannot be priced as standalone market objects. This objection identifies a real accounting-recognition problem, but it does not defeat economic valuation.

The absence of a purchase price does not imply absence of value. It means KBC must specify which valuation anchor is being used, what decision that anchor supports, and what uncertainty remains. Table~\ref{tab:ch9:recognition-valuation-anchors} separates recognition anchors\index{recognition anchor} from valuation anchors\index{valuation anchor}.

\begin{table}[htbp]
\centering
\small
\caption{Recognition Anchors and Valuation Anchors for Knowledge-Bearing Stock}
\label{tab:ch9:recognition-valuation-anchors}
\begin{tabular}{@{}p{0.22\textwidth}p{0.33\textwidth}p{0.37\textwidth}@{}}
\toprule
Anchor type & What it supports & KBC interpretation \\
\midrule
Historical cost & Auditable accounting entry & Often weak for jointly produced knowledge stock. \\
Exchange price & Market transaction evidence & Often absent for non-traded internal capability. \\
Replacement cost & Cost to rebuild or reacquire capability & Useful for internal valuation. \\
Avoided cost & Cost avoided because stock exists & Useful for software, routines, and automation. \\
Differential cash flow & Performance advantage over counterfactual & Core value-in-use anchor. \\
Impairment / breach loss & Loss when stock is damaged, stolen, poisoned, or exposed & Materiality and EKL calibration input. \\
Acquisition premium & Market price paid for bundled capability & External signal, but noisy and bundled. \\
Option value & Future recombination or learning potential & Captured through K-CMM option components. \\
\bottomrule
\end{tabular}
\end{table}

\paragraph{Valuation Rule,  Productive-Service Anchoring Without Exchange-Price Anchoring.}
The absence of a reliable historical-cost or exchange-price anchor weakens external accounting recognition, but it does not eliminate economic valuation. Knowledge-bearing stock that is jointly produced, capability-dependent, or only conditionally separable can still be valued through the productive services it yields under a specified residence--governance--capability relation. Where no standalone market price exists, valuation must use alternative anchors: replacement cost, avoided cost, differential cash flow, acquisition premium, impairment exposure, breach-loss evidence, dependency-failure cost, and expected knowledge loss\index{replacement cost}\index{avoided cost}\index{differential cash flow}\index{acquisition premium}\index{impairment exposure}\index{breach-loss evidence}\index{dependency-failure cost}. These anchors do not automatically justify balance-sheet recognition, but they provide decision-relevant evidence that the stock is economically material.

KBC does not reject separability. It rejects strict standalone separability as a necessary condition of economic value. The relevant question is whether the stock is sufficiently separable for a specific purpose: operation, substitution, protection, transfer, governance, valuation, assurance, or accounting recognition. A software repository may be insufficiently separable for sale without its maintainers, yet sufficiently separable for internal risk valuation, replacement-cost estimation, cybersecurity protection, and dependency analysis.

When knowledge-bearing stock depends on complementary embodied expertise, institutional routines, documentation, infrastructure, and governance rights, the valuation object is not the isolated artefact. It is the residence--governance--capability system through which the artefact yields productive services. The code, model, dataset, routine, or protocol may be the visible carrier, but the productive asset is the organized capacity to deploy, maintain, interpret, improve, and protect it.

Physical capital is also complement-dependent. A factory requires labour, electricity, raw materials, logistics, permits, maintenance, and managerial capability. Yet accounting institutions permit the factory to be recognized because the legal, market, audit, depreciation, and collateral conventions surrounding physical capital are mature. Knowledge-bearing stock differs not because it alone depends on complements, but because the institutions for identifying, valuing, depreciating, assuring, and recognizing its complementary service system are less mature.

The complementarity can be stated directly. Productive-service flow depends not only on the stock itself, but also on the embodied and institutional capability, access, permission, interoperability, complementary capability, maintenance, and truth-dependence that make use possible:
\[
PS_{a, i, t}(\pi)
=
f\!\left(K_i, K^E_a, K^I_a, A_a, P_a, I_a, C_a, M_a, \tau_i\right).
\]
This is a valuation architecture for productive-service anchoring, not a claim that each input is separately observable in accounting data. The claim that an algorithm has no value without the embodied knowledge required to maintain and operate it is therefore not a refutation of valuation. It is a specification of the service-flow conditions. The value is conditional on \(K^E\), \(K^I\), governance, and maintenance, not zero by definition.

Breach cost does not reveal the full positive value of a knowledge-bearing stock. It reveals that the stock, its governance position, and its protection architecture were economically material. A breach may include response cost, downtime, legal exposure, customer churn, regulatory penalties, reputational damage, adversarial advantage\index{adversarial advantage}, and lost future option value. These losses are not identical to asset value, but they are evidence that the affected knowledge-bearing system carried decision-relevant economic value and should enter Expected Knowledge Loss calibration.

A knowledge-bearing stock may be internally valued, strategically material, and risk-relevant before it is externally recognizable. KBC therefore distinguishes four thresholds: economic value, internal measurability, external assurance, and balance-sheet recognition. The cost-of-production and exchange-price objection is strongest at the fourth threshold and weakest at the first.

\begin{table}[htbp]
\centering
\small
\caption{Valuation-Anchor Fit by Knowledge Form}
\label{tab:ch9:valuation-anchor-fit}
\begin{tabular}{@{}p{0.23\textwidth}p{0.37\textwidth}p{0.32\textwidth}@{}}
\toprule
Knowledge form & Likely valuation anchors & Main risk \\
\midrule
\(K^E\) embodied & Replacement cost, wage premium, training cost, departure loss, outcome contribution & Person-specific and hard to own. \\
\(K^D\) disembodied & Development cost, replacement cost, avoided cost, licence value, security exposure & May depend on maintainers and infrastructure. \\
\(K^I\) institutionalized & Differential cash flow, process efficiency, acquisition premium, rework reduction & Attribution to routine versus people is hard. \\
\(K^C\) commons & Avoided cost, dependency-failure cost, maintenance funding gap, security risk & Value distributed; maintenance underfunded. \\
\(K^P\) public epistemic & Social return, standardization value, error reduction, public-option value & Hard to appropriate or recognize privately. \\
\bottomrule
\end{tabular}
\end{table}

\begin{quote}
\noindent\fbox{%
\begin{minipage}{0.93\textwidth}
\textbf{Objection: Without a purchase price or standalone exchange market, knowledge-bearing stock has no reliable accounting anchor.}

\medskip
\textbf{Response:} Correct for many external accounting purposes, but not for economic valuation. A missing purchase price means historical-cost recognition is weak. It does not mean the stock lacks productive-service value\index{service value}. KBC values capability-dependent knowledge-bearing stock through its actor-specific service flow, replacement cost, avoided cost, differential cash flow, impairment exposure\index{impairment}, and Expected Knowledge Loss. The absence of a standalone market price is therefore a recognition and assurance problem, not a proof of non-value.
\end{minipage}}
\end{quote}

\subsection{From Additive Asset Values to Knowledge Portfolio Value}\index{additive asset values versus portfolio valuation}\index{portfolio valuation!versus additive asset values}\label{from-additive-asset-values-to-knowledge-portfolio-value}

The non-rivality objection is correct against naïve object-addition, but wrong against valuation as such. Knowledge-bearing stock often does not add like rival physical units. Its value depends on the portfolio of other knowledge-bearing stocks, capabilities, governance rights, and deployment contexts with which it can be combined.

This is not an exotic move. Finance and economics already value non-additive structures: portfolios with covariance, options with state-contingent payoffs, firms with merger synergies, platforms with network effects, insurance pools with correlated losses, and supply chains with bottleneck dependencies. KPVF applies the same general lesson to knowledge-bearing stock: when value depends on interaction, valuation must be portfolio-level and marginal-contribution-sensitive.

\paragraph{Valuation Rule,  Knowledge Portfolio Value.}
Knowledge-bearing stock should not be valued only as a standalone object. Where stocks are non-rival, recombinable, and complement-dependent, the relevant valuation object is the actor's governed knowledge portfolio. A stock's economic value is its marginal contribution to the productive-service value of that portfolio, including current use, recombination surplus, learning-loop option value, control value, strategic option value, and expected knowledge loss. Non-additivity therefore requires portfolio valuation and marginal-contribution attribution, not abandonment of valuation.

The formal object is the Knowledge Portfolio Value Function:\index{Knowledge Portfolio Value Function (KPVF)}
\[
V^K_a(S, \pi).
\]
This is a valuation architecture: it defines the portfolio as the relevant valuation object before empirical coefficients are assigned. Here, \(V^K_a(S, \pi)\) is the value of a set \(S\) of knowledge-bearing stocks to actor \(a\) under governance form \(\pi\). This definition shifts the valuation object from a single isolated stock to the governed portfolio of stocks that can be accessed, deployed, combined, maintained, and protected by the actor.

The non-additivity condition is:
\[
V^K_a(S, \pi)
\neq
\sum_{i \in S} V^K_a(K_i, \pi).
\]
This is a diagnostic decomposition condition, not a calibrated estimate of the size of non-additivity. The inequality is not a defect. It is the point. Knowledge portfolios may contain complements, substitutes, bottlenecks, shared service streams, recombination options, learning loops, and fragility risks. Their value cannot generally be recovered by adding isolated asset values.

A plain-language decomposition is:
\[
V^K_a(S, \pi)
=
\sum_{i \in S} CUV_i
+
\sum_{i<j} RS_{ij}
+
LOV(S)
+
COV(S, \pi)
+
SOV(S, \pi)
-
EKL(S, \pi)
-
O(S).
\]

\begin{table}[htbp]
\centering
\small
\caption{Knowledge Portfolio Value Function Components}
\label{tab:ch9:kpvf-components}
\begin{tabular}{@{}p{0.24\textwidth}p{0.66\textwidth}@{}}
\toprule
Term & Meaning \\
\midrule
$\sum_{i \in S} CUV_i$ & Current use value of identified stocks. \\
$\sum_{i<j} RS_{ij}$ & Pairwise recombination surplus. \\
$LOV(S)$ & Learning-loop option value of the portfolio. \\
$COV(S, \pi)$ & Control/governance option value. \\
$SOV(S, \pi)$ & Strategic option value. \\
$EKL(S, \pi)$ & Expected knowledge loss. \\
$O(S)$ & Overlap correction preventing double-counting. \\
\bottomrule
\end{tabular}
\end{table}

This is a diagnostic decomposition, not a calibrated portfolio-pricing equation. This decomposition makes the non-additive structure explicit. Portfolio value includes the current productive services of identified stocks, but also the surplus created by recombination, the option value created by learning loops and governance control, the strategic value of the portfolio position, the downside exposure captured by expected knowledge loss, and the overlap correction required to avoid counting the same future service stream twice.

Where individual-stock attribution is required, KBC can use marginal-contribution logic. A Shapley-style allocation provides one disciplined method:
\[
\phi_i(V)
=
\sum_{T \subseteq S \setminus \{i\}}
\frac{|T|!(|S|-|T|-1)!}{|S|!}
\left[
V^K_a(T \cup \{i\}, \pi)
-
V^K_a(T, \pi)
\right].
\]
This is a valuation architecture for attribution under non-additivity, not a requirement that firms compute full Shapley values in ordinary accounting practice. Here, \(\phi_i(V)\) estimates the average marginal contribution of stock \(K_i\) across possible knowledge-portfolio configurations. This is useful when the value of an algorithm, dataset, routine, software dependency, or expert capability depends on which other stocks are present.

KBC does not require firms to compute full Shapley values for every stock. The point is methodological: where value is non-additive, attribution should be based on marginal contribution to the portfolio, not standalone object price.

KPVF does not replace K-CMM. It generalizes K-CMM from single-stock valuation to portfolio-level valuation. K-CMM identifies value components. GATE determines whether a stock can contribute. The recombination field identifies accessible combinations. Model 3 supplies recombination surplus. EKL captures downside and fragility. The no-overlap rule prevents duplicate service streams. KPVF places these pieces in one portfolio-level valuation structure.

A compact example shows why the portfolio object is necessary. Let $K_1$ be an algorithm, $K_2$ a software dependency, and $K_3$ an engineering team capability. Alone, $K_2$ may have modest value. Combined with $K_1$ and $K_3$, it may reduce development time, improve reliability, enable deployment, and expand the firm's recombination field. In that case:
\[
V^K_a(\{K_1, K_2, K_3\}, \pi)
>
V^K_a(K_1, \pi)+V^K_a(K_2, \pi)+V^K_a(K_3, \pi)
\]
when recombination surplus exceeds overlap and expected loss. These inequalities are diagnostic examples, not calibrated portfolio estimates. But the opposite result is also possible:
\[
V^K_a(\{K_1, K_2, K_3\}, \pi)
<
V^K_a(K_1, \pi)+V^K_a(K_2, \pi)+V^K_a(K_3, \pi)
\]
when dependency fragility, security exposure, maintenance failure, or governance risk dominates. If $K_1$ and $K_2$ both contribute to the same speed-to-market benefit, the no-overlap correction prevents counting that benefit twice. The KPVF therefore does not assume that recombination always creates positive synergy. It asks whether the governed portfolio creates surplus after overlap, fragility, and expected knowledge loss are taken into account.

This placement is deliberate. KPVF is introduced here as the measurement chapter's portfolio-level extension\index{portfolio valuation}\index{non-additive valuation} of the earlier unit, generation, conversion, and governance machinery, not as a new theorem bolted onto the manuscript. Chapter 2 establishes the OKU as a bounded operative unit; Chapter 9 introduces KPVF as the answer to non-additive valuation; the Technical Companion, Appendix D, can carry the technical specification and relation to K-CMM, understood as a decision-support and uncertainty-reduction framework rather than a GAAP/IFRS recognition model; Chapter 11 can test its calibration and falsification criteria; and the vocabulary register should define Knowledge Portfolio Value Function as a central measurement concept.

The originality claim should be stated conservatively. KPVF does not invent portfolio valuation, marginal-contribution allocation, option valuation, or complementarity analysis. Its contribution is the integration of those established tools with the KBC stock taxonomy and measurement architecture.

\begin{table}[htbp]
\centering
\small
\caption{Originality Status of the Knowledge Portfolio Value Function}
\label{tab:ch9:kpvf-originality-status}
\begin{tabular}{@{}p{0.50\textwidth}p{0.32\textwidth}@{}}
\toprule
Element & Classification \\
\midrule
Portfolio valuation & Established. \\
Shapley/marginal contribution allocation & Established. \\
Option-value treatment & Established. \\
Network/complementarity valuation & Established. \\
Application to KBC five-form stock & Synthesized. \\
Integration with K-CMM, GATE, recombination field, EKL, and no-overlap & Potentially Novel as architecture. \\
Empirical calibration & Open. \\
\bottomrule
\end{tabular}
\end{table}

\subsection{From Context-Dependent Value to Realization-Conditioned Value}\label{from-context-dependent-value-to-realization-conditioned-value}

The context-dependence objection is correct against context-free valuation, but wrong against valuation as such. Knowledge-bearing stock often does not possess a stable productive value independent of workflow, capability, permission, interoperability, platform access, governance, and demand. But this does not make knowledge unvaluable. It means that valuation must specify the conditions under which knowledge potential becomes productive-service yield.

\paragraph{Measurement Principle,  Realization-Conditioned Knowledge Value.}
Knowledge-bearing stock should not be valued as abstract knowledge-in-itself. Its economic value is the expected productive-service yield it can generate for actor $a$, under governance form $\pi$, given access, permission, interoperability, complementary capability, maintenance capacity, demand, and truth-dependence. A stock may have high knowledge potential but low realized yield where realization conditions are blocked. Context-dependence therefore requires explicit realization conditions, not abandonment of valuation.

The potential--impedance--yield triad can be stated compactly as a realization relation:
\[
\KYield_{a, i, t}
=
\KPot_{i, t} \times \rho_{a, i, t}.
\]
This is a valuation architecture for realization-conditioned value, not a direct empirical estimate. Here, $\KPot_{i, t}$ denotes the potential productive value of stock $i$. The realization coefficient $\rho_{a, i, t}$ measures how much of a knowledge-bearing stock's potential value an actor can actually turn into productive yield at a given time. It is not merely whether the knowledge has been converted into a form, but whether the actor has the access, permission, capability, interoperability, governance conditions, and maintenance support needed to realize its productive services. $\KYield_{a, i, t}$ is the realized productive-service yield. High potential does not guarantee high yield; yield depends on realization conditions. This formulation is preferable to forcing impedance into a single scalar friction term too early.

The realization coefficient is multidimensional. This is an uncalibrated scoring model until the component weights are estimated or bounded in a specific domain:
\[
\rho_{a, i, t}
=
f(A_{a, i, t}, P_{a, i, t}, I_{a, i, t}, C_{a, i, t}, M_{a, i, t}, G_{a, i, t}, D_{a, i, t}, \tau_i).
\]

\begin{table}[htbp]
\centering
\small
\begin{tabular}{p{0.16\textwidth}p{0.68\textwidth}}
\toprule
\textbf{Term} & \textbf{Meaning} \\
\midrule
$A$ & Access \\
$P$ & Permission or legal authority \\
$I$ & Interoperability \\
$C$ & Complementary capability \\
$M$ & Maintenance capacity \\
$G$ & Governance conditions \\
$D$ & Demand or deployment context \\
$\tau_i$ & Truth-dependence / validity condition \\
\bottomrule
\end{tabular}
\caption{Components of the Realization Coefficient}
\label{tab:realization-coefficient-components}
\end{table}

Impedance is not one obstacle. It is the failure or weakening of these realization conditions.

\paragraph{The projection reading: dark capital as the shadow of the capture coefficient.}\index{capture coefficient|textbf}\index{capture coefficient|seealso{realization coefficient}}\index{realization coefficient|seealso{capture coefficient}}\index{dark capital!projection reading} The realization relation also yields the cleanest statement of what makes value \emph{dark} rather than merely unrealized. As Chapter~\ref{value-in-use-and-value-in-exchange} set out, a knowledge stock delivers \emph{use-value} in production, while accounts record only the \emph{exchange-value} an actor captures from it. The two are joined by the same multiplicative architecture as the realization coefficient, applied one layer further out, at the conversion from delivered use-value to recorded price. Writing $U_{a, i, t}$ for the use-value a stock delivers and $E_{a, i, t}$ for the exchange-value captured from it as price, rent, licence, or revenue, governance projects the first onto the second through a capture coefficient $\rho^{\mathrm{x}}_{a, i, t} \in [0, 1]$:
\[
E_{a, i, t} = \rho^{\mathrm{x}}_{a, i, t}\, U_{a, i, t},
\qquad
D_{a, i, t} = U_{a, i, t} - E_{a, i, t} = U_{a, i, t}\,\bigl(1 - \rho^{\mathrm{x}}_{a, i, t}\bigr).
\]
The capture coefficient $\rho^{\mathrm{x}}$ is the realization coefficient's twin: it draws on the same determinants, access, permission, appropriability, and governance, but governs how much delivered use-value converts into recorded exchange-value rather than how much potential converts into yield. $D_{a, i, t}$ is then the dark component proper, use-value that is real and productive yet never enters the priced quantity the accounts record. Non-rivalry is what makes the gap structural for knowledge: a stock can deliver use-value to many actors at once while the price any one of them can be charged is competed toward its near-zero cost of reproduction, so $\rho^{\mathrm{x}}$ is small and $D_{a, i, t}$ approaches the whole of $U_{a, i, t}$. An exchange-based ledger records only the captured projection $E = \rho^{\mathrm{x}} U$; the dark capital $U(1 - \rho^{\mathrm{x}})$ is the use-value lying off the priced axis, decisive in production and absent from the books. The surplus residing in the gap between an input's use-value and its exchange-value is not itself a new observation; it is the Marxian point sharpened in \textcite{Keen1993}. KBC's step is to locate that gap in non-rival knowledge, govern it with $\rho^{\mathrm{x}}$, and read the residual $U(1 - \rho^{\mathrm{x}})$ as the measurable dark component this chapter sets out to define.

The objection applies a stricter standard to knowledge than to physical capital. Physical capital is also context-dependent. A truck requires roads, fuel, drivers, insurance, logistics demand, and legal permission. A drilling rig requires geology, permits, oil prices, maritime logistics, engineering labour, and regulatory tolerance. Context-dependence is therefore not unique to knowledge. The difference is that knowledge-bearing stock often changes yield more rapidly when access, interoperability, platform governance, legal permission, or complementary capability changes.

A high-potential stock with low current yield is not necessarily worthless. It may be stranded, blocked, option-like, or exposed to governance change. KBC therefore distinguishes current yield from conditional future yield and expected knowledge loss. A dataset that cannot currently be used because of legal restrictions may have low current $\KYield$, but non-zero option value if permission, interoperability, or capability conditions change.

The same diagnostic algorithm may have high yield in a hospital system with clean data, legal permission, workflow integration, trained clinicians, and maintenance capacity, but low yield in a hospital lacking interoperability, permission, or clinical adoption. The algorithm's potential has not disappeared; its realization coefficient differs across actors. In the notation above, the two hospitals may face the same $\KPot_{i, t}$ but different $\rho_{a, i, t}$, and therefore different $\KYield_{a, i, t}$.

Digital platforms do not eliminate context-dependence. They standardize and govern context. Platforms such as video, streaming, app-store, cloud, and advertising systems create partial commensuration systems by standardizing upload formats, metadata, recommendation logic, monetization rules, API access, moderation rules, and feedback loops. Heterogeneous content becomes yield-bearing because the platform supplies realization conditions.

Realization-conditioned value operates inside KPVF. A knowledge portfolio $S$ may contain many high-potential stocks, but portfolio value depends on the realization coefficients governing each stock and combination. KPVF therefore values not merely the presence of stocks, but the actor's capacity to activate them under governance and capability constraints.

The originality status of this move should be stated conservatively. KBC does not claim to invent context-dependence, asset specificity, stranded-asset logic, or real-options treatment. Its contribution is to integrate those established ideas into the potential--impedance--yield triad, the realization coefficient, KPVF, and K-CMM as a measurement architecture for knowledge-bearing stock.

\begin{table}[htbp]
\centering
\small
\begin{tabular}{p{0.58\textwidth}p{0.28\textwidth}}
\toprule
\textbf{Element} & \textbf{Classification} \\
\midrule
Context-dependence of capital value & Established \\
Complementarity and asset specificity & Established \\
Stranded-asset logic & Established \\
Real-options treatment of blocked potential & Established \\
Application to KBC potential/impedance/yield triad & Synthesized \\
Realization coefficient $\rho_{a, i, t}$ linked to access, permission, interoperability, capability, governance, demand, and truth-dependence & Extended / potentially Novel as architecture \\
Integration with KPVF and K-CMM & Potentially Novel as architecture \\
Empirical calibration & Open \\
\bottomrule
\end{tabular}
\caption{Originality Status of Realization-Conditioned Value}
\label{tab:realization-conditioned-originality}
\end{table}

\begin{quote}
\noindent\fbox{%
\begin{minipage}{0.93\textwidth}
\textbf{Objection: Knowledge-bearing stock is too context-dependent to value.}

\medskip
\textbf{Response:} Context-dependence defeats context-free valuation, not valuation as such. Physical capital also depends on roads, fuel, labour, regulation, supply chains, permits, and demand. KBC values knowledge-bearing stock by specifying realization conditions: access, permission, interoperability, capability, maintenance, governance, demand, and truth-dependence. A stock may have high potential and low current yield, but that is a measurement problem addressed through realization coefficients, option value, and expected knowledge loss.
\end{minipage}}
\end{quote}

\subsection{From Mechanical Depreciation to Knowledge Revaluation}\index{depreciation}\index{mechanical depreciation versus knowledge revaluation}\index{knowledge depreciation!versus mechanical depreciation}\label{from-mechanical-depreciation-to-knowledge-revaluation}
\index{knowledge depreciation|textbf}

The depreciation objection is correct against straight-line depreciation, but wrong against capital valuation. Knowledge-bearing stock rarely depreciates through mechanical wear. A software dependency is not consumed by execution; a dataset is not depleted by being queried; a protocol is not exhausted by being followed. Its value changes through different mechanisms: obsolescence, fidelity decay\index{fidelity decay|textbf}, maintenance failure, loss of complementary capability, governance restriction, access loss, security exposure, decline in truth-validity, and expansion of the recombination field. The relevant accounting analogy is therefore not mechanical wear alone, but impairment, revaluation, option realization, and periodic reassessment of productive yield.

\paragraph{Diagnostic Claim,  Asymmetric Knowledge Revaluation.}
Knowledge-bearing stock should not be depreciated only through time-and-use schedules. Its value may decline, rise, or become impaired when realization conditions, maintenance systems, complementary capabilities, governance forms, truth-validity, expected knowledge loss, or recombination fields change. Knowledge depreciation is therefore mechanism-specific and may be non-monotonic: a stock may depreciate without use, appreciate without new internal investment, or experience sudden impairment when external support conditions fail.

This diagnostic claim follows directly from realization-conditioned value. At the stock level, a compact expression is:
\[
V^K_{a, i, t}
=
\KPot_{i, t}\times \rho_{a, i, t}
+
OV_{a, i, t}
-
EKL_{a, i, t},
\]
This is a valuation architecture for single-stock revaluation, not a mechanical depreciation schedule. Here, $OV_{a, i, t}$ denotes option value rather than a separate valuation system. The corresponding value-change relation is:
\[
\Delta V^K_{a, i, t}
=
\Delta \KPot_{i, t}
+
\Delta \rho_{a, i, t}
+
\Delta OV_{a, i, t}
-
\Delta EKL_{a, i, t}.
\]
This is a diagnostic decomposition of value change, not a calibrated impairment test. The value of knowledge-bearing stock changes when latent potential changes, when realization conditions change, when option value changes, or when expected knowledge loss changes. This avoids forcing knowledge depreciation into a single straight-line rate. If a depreciation coefficient is later introduced for measurement, it should be mechanism-specific and time-varying rather than a universal scalar $\delta_t$ applied to all knowledge stock.

\begin{table}[htbp]
\centering
\small
\begin{tabular}{p{0.22\textwidth}p{0.12\textwidth}p{0.38\textwidth}p{0.16\textwidth}}
\toprule
\textbf{Mechanism} & \textbf{Direction} & \textbf{Example} & \textbf{KBC object affected} \\
\midrule
Obsolescence & Down & Model, dataset, or process loses relevance. & $\KPot$ \\
Fidelity decay & Down & Documentation, provenance, or data quality degrades. & $\KPot$, $EKL$ \\
Maintenance failure & Down & Open-source dependency loses maintainers. & $\rho$, $EKL$ \\
Capability loss & Down & Experts leave or tacit context disappears. & $\rho$ \\
Access/governance restriction & Down & API closure, licence change, or legal restriction. & $\rho$, $EKL$ \\
Security exposure & Down & Dependency becomes vulnerable or poisoned. & $EKL$ \\
Recombination-field expansion & Up & New AI tools make an old dataset useful. & $OV$, $\rho$ \\
Complement arrival & Up & New standard, model, or platform makes stock usable. & $\rho$, $OV$ \\
Demand shift & Up or down & Regulation or market need changes usefulness. & $\KPot$, $\rho$ \\
Truth-validity change & Up or down & Scientific or technical validity changes. & $\KPot$, $\tau_i$ \\
\bottomrule
\end{tabular}
\caption{Mechanisms of Knowledge Depreciation and Revaluation}
\label{tab:knowledge-depreciation-revaluation-mechanisms}
\end{table}

Physical capital is also exposed to non-linear depreciation and revaluation. A truck loses value if emissions rules change, fuel becomes unavailable, insurance becomes impossible, or logistics demand collapses. A coal plant can be impaired by regulation or cheaper energy substitutes before it physically wears out. Land can appreciate because a transit line, zoning change, or nearby commercial development alters its productive context. The difference is not that only knowledge is relational. The difference is that knowledge-bearing stock is more visibly and rapidly exposed to changes in governance, access, complementary capability, maintenance systems, truth-validity, and recombination fields.

A software dependency may not wear out when executed, but its productive yield can fall if maintainers exit, issue backlogs grow, patches slow, documentation decays, compatibility weakens, or governance becomes unstable. The dependency remains non-rival, but the maintenance and capability system that makes it reliable has depreciated. In KBC terms, the stock's current yield falls because $\rho_{a, i, t}$ declines and $EKL_{a, i, t}$ rises.

Conversely, an old dataset may appreciate without changing internally. If a new AI model, analytic method, permission framework, or interoperability standard makes the dataset usable, the stock's latent potential becomes realizable. The dataset did not become valuable because the records changed. It became valuable because the recombination field and realization conditions changed. This is passive revaluation through option realization, not a violation of valuation logic.

Third-party investment can revalue assets in ordinary capital markets. Land appreciates when infrastructure appears nearby; mineral deposits appreciate when new technology creates demand; spectrum rights appreciate when communications systems expand. Knowledge-bearing stock makes this pattern more common because new models, standards, platforms, APIs, and analytic capabilities continually change what can be recombined.

Asymmetric knowledge revaluation operates through realization-conditioned value and KPVF. At the single-stock level, changes in $\KPot$, $\rho$, option value, and $EKL$ alter expected productive-service yield. At the portfolio level, the same change may affect recombination surplus, overlap correction, learning-loop value, dependency risk, and expected knowledge loss. KPVF therefore does not assume a fixed asset value path; it treats value trajectories as conditional on changing realization and recombination conditions.

\begin{table}[htbp]
\centering
\small
\begin{tabular}{p{0.58\textwidth}p{0.28\textwidth}}
\toprule
\textbf{Element} & \textbf{Classification} \\
\midrule
Non-linear asset impairment & Established \\
Economic obsolescence & Established \\
Real-options revaluation & Established \\
Stranded-asset logic & Established \\
Open-source maintainer depletion as depreciation mechanism & Extended \\
Dataset revaluation through AI/recombination-field expansion & Synthesized / potentially Novel as architecture \\
Integration with $\KPot$, $\rho$, option value, $EKL$, and KPVF & Potentially Novel as architecture \\
Empirical calibration & Open \\
\bottomrule
\end{tabular}
\caption{Originality Status of Asymmetric Knowledge Revaluation}
\label{tab:asymmetric-knowledge-revaluation-originality}
\end{table}

\begin{quote}
\noindent\fbox{%
\begin{minipage}{0.93\textwidth}
\textbf{Objection: Knowledge-bearing stock cannot be depreciated or aggregated because its value does not follow predictable time-and-use schedules.}

\medskip
\textbf{Response:} Straight-line depreciation is an accounting convention, not the definition of capital. Knowledge-bearing stock rarely depreciates through mechanical wear; it changes value through obsolescence, maintenance failure, capability loss, governance restriction, access loss, security exposure, truth-validity change, and recombination-field expansion. These mechanisms may reduce current yield, increase expected knowledge loss, or create passive revaluation. KBC therefore replaces mechanical depreciation with mechanism-specific depreciation, impairment, option-realization, and revaluation analysis.
\end{minipage}}
\end{quote}

Knowledge-bearing capitalism generates productive value through
mechanisms that accounting was not designed to see. Standard accounting
was built for an economy in which productive capacity resided primarily
in owned physical assets, enforceable contractual rights, and
identifiable financial claims. The recognition criteria that IAS 38
(Intangible Assets) and GAAP establish, identifiability, control, and
reliable measurement, were calibrated for a world in which the
productive capacity of a firm was approximately equivalent to the assets
it owned and the contracts it held.

That calibration was never perfect, but it was workable. In an economy
where a factory, its machinery, and its inventory constituted most of a
firm\textquotesingle s productive capacity, a balance sheet that
captured owned physical assets captured most of what mattered. The gap
between economic value and recognized value was manageable because the
sources of economic value were largely the same things that accounting
recognition criteria were designed to identify.

In knowledge-bearing capitalism, the calibration breaks down in four
structural ways. First, the primary productive stocks, \(K^E\)
(embodied tacit knowledge), \(K^I\) (organizational routines and
capabilities), and the governance-position value that flows from access
to the recombination field, are not separable from the firm as a
going concern, not fully controllable in the sense IAS 38 requires, and
not measurable through any accounting procedure currently available.
Second, the primary threats to productive capacity, capability decay
from field restriction, suppressed appreciation of non-enclosed stocks,
commons depletion through labour-market extraction, do not take the
form of damage to recognized assets, so they do not trigger the
impairment testing that accounting requires. Third, the primary
generative processes, the \(G^R\), \(G^L\), \(G^J\), and \(G^N\)
mechanisms of Chapter 3, produce value at rates determined partly by
governance form, partly by dynamic capability, and partly by the
breadth of the accessible recombination field, none of which accounting
systems track. Fourth, the social costs of enclosure, the \(C_{T2}\),
\(C_{T7}\), \(C_{T6}\), and \(C_{T8}\) components that T5 identifies, are borne by
excluded actors and by the economy\textquotesingle s future knowledge
generation capacity, not by the enclosing firm, so they appear nowhere
in the enclosing firm\textquotesingle s accounts.

These are not incidental gaps that better data collection or revised
accounting standards could close at the margin. They are structural:
they arise from the application of accounting recognition criteria that
more readily recognize rival, owned, physically identifiable productive
assets than knowledge-bearing stocks that are non-rival,
governance-dependent, tacit, and constituted through relational access
to an external ecosystem rather than through ownership of internal
resources.

\section{The Productivity Paradox as Dark Capital's Empirical Shadow}\label{92-productivity-paradox-as-dark-capitals-empirical-shadow}

The productivity paradox\index{productivity paradox!dark capital}\index{productivity paradox} supplies an important test case for KBC. IT investment is visible in expenditure data, but its productive effects may be delayed, mismeasured, privately redistributive, or dependent on complementary organizational change. \textcite{Brynjolfsson1993}\index{Brynjolfsson, Erik}'s four explanations for the paradox can be translated into KBC terms without being exhausted by them. Table~\ref{tab:ch9:brynjolfsson-kbc-paradox-map} makes the mapping explicit and separates the original paradox explanations from KBC's additional claim about generative suppression. KBC treats the paradox not as evidence that information technology lacks value, but as evidence that IT expenditure is not equivalent to knowledge-capital formation.

\begin{table}[htbp]
\centering
\small
\caption{Brynjolfsson's Productivity-Paradox Explanations in KBC Terms}
\label{tab:ch9:brynjolfsson-kbc-paradox-map}
\begin{tabular}{@{}p{0.24\textwidth}p{0.33\textwidth}p{0.34\textwidth}@{}}
\toprule
Paradox explanation & KBC translation & Interpretation \\
\midrule
Measurement error & Dark capital, unmeasured service yield, accounting and productivity shadow & Productive effects may exist but remain weakly represented in accounting or productivity statistics. \\
Lags & Knowledge-capital trajectory and learning-by-using & IT expenditure may require time before it becomes embodied capability, institutional routine, or feedback-learning capacity. \\
Redistribution & Private rent extraction without social productivity gain & IT may shift market share, bargaining power, or rents without increasing aggregate productive capacity. \\
Mismanagement & Governance-fit and capability-conversion failure & IT may fail to become productive knowledge capital when organizations cannot absorb, govern, or redeploy it. \\
No direct Brynjolfsson equivalent & Generative suppression & Enclosure may reduce future recombination capacity and suppress social output below the counterfactual. \\
\bottomrule
\end{tabular}
\end{table}

\begin{quote}
\textbf{IT expenditure is not yet knowledge capital.}
\end{quote}

It becomes productive knowledge capital only when converted into embodied capability, institutional routines, feedback-learning capacity, and access to a viable recombination field. KBC identifies the conditions under which that conversion succeeds: an accessible recombination field, sufficient embodied capability to interpret and apply the new stock, institutional routines capable of absorbing and deploying it, feedback loops that improve performance through use, and a governance form that does not enclose the complementary field elements on which conversion depends.

The claim that IT value depends on complementary organizational capital is established by \textcite{BrynjolfssonHittYang2002}. KBC accepts that result but asks a prior set of questions: what forms do those complements take, how are they generated, how do they move between embodied, disembodied, institutionalized, commons, and public-infrastructure forms, under what governance forms do they become productive, how do they depreciate or become dark, and how should their value be measured? Complementarity identifies the need for co-specialized assets; KBC supplies a theory of their form, conversion, governance, impairment, and measurement.

\subsection{Data Depreciation and Zero-Price Exchange}\label{ch9:data-depreciation-zero-price-exchange}

Data-economy models make the accounting problem more concrete. \textcite{FarboodiVeldkamp2021} treat transaction-generated data as information that can accumulate, be traded, be valued, and depreciate. The stock does not depreciate because it physically wears out. It depreciates because the environment being predicted changes: consumer preferences, demand conditions, order flow, traffic patterns, clinical distributions, fraud strategies, and operational states drift. Predictive data therefore has an impairment profile that differs from ordinary software amortization and from physical capital depreciation.

Data does not depreciate because it is used. It depreciates when its predictive relation, context, interpretive capability, legality, representativeness, access, or maintenance environment deteriorates. This is why data depreciation\index{data depreciation} should not be reduced to the generic claim that data ``gets old.'' Some data becomes more valuable when accumulated and recombined; other data loses yield when the world, the governing law, the model, the population, or the institutional capability needed to interpret it changes.

This matters for zero-price exchange\index{zero-price exchange}. A service offered at a monetary price of zero may still involve barter: the user receives access, convenience, search, communication, navigation, entertainment, or prediction services while supplying behavioural, transaction, location, preference, or usage data. In such cases, the observed price can be zero while the exchange remains economically positive. GDP and revenue measures can miss the data component of the transaction because the value paid by the user is not a monetary payment, and the data asset produced by the exchange may be internally generated rather than purchased.

The balance-sheet implication is direct. Internally generated data can increase firm value without appearing as a separately recognized asset. The claim is not that every knowledge-capital effect should be capitalized under current accounting standards; it is that economically material knowledge-capital effects can exist before accounting recognition becomes appropriate. A data-rich firm may therefore show low book value, weak current earnings, or zero-price services while still accumulating a productive knowledge-bearing stock. KBC does not treat every such gap as dark capital; some of it is speculation, monopoly rent, or ordinary intangible value\index{intangible capital}. The narrower claim is that internally generated, predictive, and feedback-derived data can create productive capacity and expected future yield that conventional book value may fail to identify. Data depreciation, data barter, and unrecognized internally generated data are therefore not side issues. They are concrete accounting channels through which knowledge-bearing stock becomes economically material while remaining partially invisible.

\begin{center}
\fbox{\begin{minipage}{0.92\textwidth}
\small
\textbf{Running case: API closure/access restriction.}\index{API closure!dark capital running case} The dark-capital question is what the accounts and ordinary valuation categories miss before and after the API transition. Exposed firms may have no recognized asset called ``API-dependent recombination capability, '' yet their products, workflows, data pipelines, user analytics, or research routines may depend on that access. Closure can therefore reveal dark risk: unpriced dependency, impaired complements, stranded developer capability, lost option value, or an accounting shadow that appears only later as delay, write-off, abnormal return, or productivity loss.
\end{minipage}}
\end{center}

\section{Dark Capital Defined}\label{92-dark-capital-defined}
\index{dark capital}

Dark capital is the master recognition-failure category in the
framework. It names economically real knowledge-bearing capital, capital
loss, or capital-forming capacity that is invisible to, mismeasured by,
misclassified within, or systematically outside the dominant accounting
and valuation recognition systems.

The category has four sub-dimensions.

Dark capital includes not only unrecognized residential knowledge stock but also unrecognized governance-position value. A firm may depend on a commons, public standard, platform API, legal doctrine, model-feedback loop, cybersecurity control structure, or public measurement system without any accounting recognition of that governance dependency. The accounting shadow therefore includes both what the firm often cannot separately record as owned stock and what it often cannot separately record as a governance-position dependency, exposure, or option.

\textbf{Dark value} is the upside dimension: productive
knowledge-bearing capital that exists, generates economic returns, and
drives investment decisions, but is not recognized as an asset by
accounting. Dark value includes unrecognized productive capability
(\(K^E\) and \(K^I\) held by firms and workers), hidden organizational
know-how (the tacit governance and coordination knowledge embedded in
long-standing teams and communities), unmeasured public and commons
knowledge capital (\(K^C\) and \(K^P\) maintained outside any single
firm\textquotesingle s balance sheet), and unattributed platform and AI
contribution (the productive value embedded in deployed model systems
that accounting treats as software costs rather than knowledge capital).

Domain-specific embodied knowledge capital is one of the most
economically significant forms of dark capital. Firms routinely depend
on cardiac-surgical expertise, cybersecurity judgement, engineering
intuition, legal interpretation, sales trust, and managerial
coordination, yet financial statements rarely identify these stocks as
productive capital. Their loss appears later as delay, error, quality
decline, incident exposure, failed integration, or reduced innovation
capacity. This matters because the Smith-Friedman\index{Friedman, Milton} capital-services
insight says skill is not merely labour. It is embodied productive
stock.

\textbf{Dark risk} is the downside dimension: real capital impairment,
capital loss, or capital-eroding exposure that reduces the productive
value or generative capacity of knowledge-bearing actors, but is often not
separately recorded as a liability, provision, or asset write-down. The claim is not that every knowledge-capital effect should be capitalized under current accounting standards; it is that economically material knowledge-capital effects can exist before accounting recognition becomes appropriate. Dark risk
includes invisible knowledge-capital impairment (the T7 suppressed
appreciation result: productive value falls without any recognized asset
being impaired), hidden capability loss (the T6 result: \(\widetilde{C}_{a}\) decays
under field restriction without accounting recognition), unpriced
governance-position risk\index{governance-position risk|textbf} (the \(\Omega_i\) term: latent enclosure-cost exposure
that current accounting often does not separately provision), learning-loop dependency risk (the
Proposition D vulnerability: actors whose \(G^L\) depends on continued
platform access carry unrecognized fragility), and cyber exfiltration as
recombination risk (the unrecognized loss of competitive position when
\(K^E\) or \(K^D\) is extracted without physical transfer).

\textbf{Foregone knowledge capitalization} is the non-formation
dimension: capital that had the potential to form but did not, because
the conditions required for its formation, accessible recombination
field, functioning commons governance, open experimental substrate,
productive feedback distribution, were suppressed. Foregone knowledge
capitalization includes failed codification (\(K^D\) that would have been
generated from \(K^E\) had codification conditions been supportive),
failed institutional learning (\(K^I\) that would have accumulated had
organizational feedback loops been operational), failed experimentation
(knowledge that would have been generated through \(G^X\) had the
experimental substrate been accessible), failed recombination (the
\(G^R\) output suppressed by the \(D_u\)(\(F_{a, t}\)) reduction of
Proposition C), and failed governance conversion (the knowledge that
would have transited from \(K^C\) or \(K^P\) into productive \(K^D\) had
commons governance remained intact). Foregone knowledge capitalization
is the closest the framework comes to measuring what is not there, 
the counterfactual productive potential that enclosure, decay, and field
restriction have eliminated.

\textbf{Accounting shadow} is the accounting-specific sub-dimension: the
systematic gap between what accounting systems recognize and what is
economically real, expressed specifically through the MV−BV
decomposition of Theorem T4. Accounting shadow is the intersection of
the first three sub-dimensions with the accounting recognition system
,  the specific form dark capital takes when it is visible through the
lens of financial reporting.

Dark capital should not be read as a claim that economic value is
systematically overstated or that markets are systematically irrational.
The dark capital framework makes a different claim: that specific,
identifiable, economically real forms of productive value and productive
loss operate outside the recognition systems through which capital is
governed, taxed, regulated, and reported. This has consequences for
investment decisions, governance design, regulatory oversight, and
policy, even if, as an empirical matter, market prices partially
reflect what accounting systems cannot.

\index{market/accounting visibility matrix|textbf}\index{accounting visibility!market visibility matrix}\index{market pricing!dark capital}\index{recognized capital}\index{market-visible dark value}\index{hidden exposure}\index{speculative excess}\index{overcapitalized knowledge claim}
\begingroup
\small
\setlength{\tabcolsep}{4pt}
\renewcommand{\arraystretch}{1.12}
\begin{longtable}{@{}L{0.21\textwidth}L{0.17\textwidth}L{0.17\textwidth}L{0.37\textwidth}@{}}
\caption{Market and Accounting Visibility Matrix}\label{tab:ch9:market-accounting-visibility}\\
\toprule
Case & Accounting sees it? & Market prices it? & KBC interpretation \\
\midrule
\endfirsthead
\toprule
Case & Accounting sees it? & Market prices it? & KBC interpretation \\
\midrule
\endhead
\bottomrule
\endlastfoot
Recognized capital\index{recognized capital!market/accounting matrix} & Yes & Yes & Ordinary visible capital. \\
Market-visible dark value\index{market-visible dark value!market/accounting matrix}\index{accounting shadow!market-visible dark value} & No & Yes & Accounting shadow, not necessarily market failure. \\
Hidden exposure\index{hidden exposure!market/accounting matrix}\index{dark capital proper} & No & No & Dark capital proper: value, dependency, fragility, or risk remains weakly visible to both accounts and markets until a shock reveals it. \\
Speculative excess\index{speculative excess!market/accounting matrix}\index{false stock!speculative excess}\index{overcapitalized knowledge claim} & No/weakly & Overpriced & False or overcapitalized knowledge claim. \\
\end{longtable}
\endgroup

Table~\ref{tab:ch9:market-accounting-visibility} separates accounting invisibility from market failure. Some knowledge-bearing value is invisible to accounting but already priced by markets. Some exposure is hidden from both. Some apparent knowledge capital is overpriced. The empirical task is therefore to distinguish recognized capital, market-visible dark value, hidden exposure, and speculative excess rather than treating every market-to-book gap as evidence for KBC.

The same discipline applies inside the dark-capital category itself. A market-to-book residual, delayed product launch, unexplained productivity loss, or governance shock is not automatically dark capital. For empirical use, the researcher should assign the observation to a narrower component and state which observable proxy carries the burden. Table~\ref{tab:ch9:dark-capital-minimum-decomposition} gives the minimum decomposition.

\begingroup
\small
\setlength{\tabcolsep}{4pt}
\renewcommand{\arraystretch}{1.12}
\begin{longtable}{@{}L{0.18\textwidth}L{0.42\textwidth}L{0.32\textwidth}@{}}
\caption{Minimum Decomposition of Dark Capital}\label{tab:ch9:dark-capital-minimum-decomposition}\\
\toprule
Component & Meaning & Observable proxy \\
\midrule
\endfirsthead
\toprule
Component & Meaning & Observable proxy \\
\midrule
\endhead
\bottomrule
\endlastfoot
Dark value\index{dark value!empirical decomposition} & Productive knowledge-bearing stock not recognized in accounts. & Market-to-book residuals after controls; internal productivity dependence on software, data, routines, models, brand trust, expertise, platform access, or public epistemic inputs. \\
Dark risk\index{dark risk!empirical decomposition} & Unpriced exposure to knowledge-stock impairment or governance shock. & Incident loss, dependency concentration, maintainer risk, API exposure, licence fragility, model drift, recovery lag, or abnormal returns after knowledge-governance shocks. \\
Foregone capitalization\index{foregone capitalization!empirical decomposition} & Value not realized because knowledge was not maintained, accessed, recombined, or converted into productive capability. & Abandoned projects, blocked dependencies, delayed innovation, reduced forks or integrations, stalled downstream products, lost experiments, or declining maintainer activity. \\
False stock\index{false stock!empirical decomposition} & Apparent knowledge stock that is unreliable, obsolete, hallucinated, non-reproducible, or detached from the capability layer needed to make it productive. & Error rates, failed replication, model drift, stale data, missing provenance, unusable documentation, non-reproducible analytics, or failed deployment. \\
\end{longtable}
\endgroup

Table~\ref{tab:ch9:dark-capital-minimum-decomposition} is a coding rule, not an accounting standard. A KBC study should not report ``dark capital'' as an undifferentiated residual. It should identify which component is being tested, which observable proxy carries the burden, and what rival explanation would cause the claim to be narrowed or rejected. Accounting shadow remains a related visibility problem, but it is not itself the whole dark-capital category; it appears when accounting recognition fails to track one or more of these underlying components.

The taxonomy is not equally original in every part. Table~\ref{tab:ch9:dark-capital-originality} distinguishes extension from stronger KBC contribution so that the category does not overclaim novelty.

\begingroup
\small
\setlength{\tabcolsep}{4pt}
\renewcommand{\arraystretch}{1.12}
\begin{longtable}{@{}L{0.24\textwidth}L{0.23\textwidth}L{0.45\textwidth}@{}}
\caption{Originality Status of Dark-Capital Components}\label{tab:ch9:dark-capital-originality}\\
\toprule
Component & Originality status & Basis \\
\midrule
\endfirsthead
\toprule
Component & Originality status & Basis \\
\midrule
\endhead
\bottomrule
\endlastfoot
Dark value & Extended & Builds on Lev, Corrado, Haskel and Westlake, and the intangible-capital accounting literature. \\
Dark risk & Stronger originality & Adds \(\Omega_i\), capability loss, cyber exfiltration, governance-position fragility, and liability-side exposure. \\
Foregone knowledge capitalization & Most novel & Captures counterfactual non-formation: knowledge not generated, recombined, maintained, or learned because governance arrangements suppressed conditions. \\
Accounting shadow & Synthesized/formalized & Names the T4 interface between productive shadow, fragility shadow, options, speculation, error, and formal recognition. \\
\end{longtable}
\endgroup

The taxonomy is not equally original in every part. Dark value extends the existing intangible-capital literature. Dark risk and foregone knowledge capitalization are the more distinctive KBC contributions because they shift the problem from unrecognized assets to unrecognized exposures and unformed future knowledge.

\subsection{Bastiatian Visibility Principle}\index{Bastiat, Frederic}\label{bastiatian-visibility-principle}

Economic institutions more readily recognize visible, attributable,
transaction-like, and bookable effects than dispersed, delayed,
relational, counterfactual, or non-owned knowledge effects. In KBC, this
means knowledge costs often appear before knowledge yields, and knowledge
failures often appear before knowledge contributions.

Dark capital is therefore not merely unmeasured because accounting has
failed to modernize. It is structurally dark because many knowledge
services and knowledge losses occur in the unseen category: capability
formation, foregone recombination, learning-loop exclusion, commons
maintenance, and latent governance-position risk.

R\&D expenditure is visible as a period cost; the capability it creates
may remain unseen. AI revenue is visible as current output; the feedback
captured from use may silently improve future model capability.
Open-source infrastructure is often invisible while it functions; it
becomes visible when a maintainer exits, a vulnerability appears, or a
dependency fails.

\section{Dark Capital and Ordinary Intangible Capital}\label{dark-capital-and-ordinary-intangible-capital}

Dark capital is not identical to intangible capital. Intangible capital names non-physical investment or non-physical assets: software, brands, databases, patents, organizational processes, human-capital investments, and related claims to future benefit. Dark capital names the portion of knowledge-bearing productive capacity that remains economically operative but institutionally obscured because it is unbooked, embedded, governance-sensitive, recombination-dependent, or visible only through valuation gaps, exclusion events, cybersecurity failure, capability shocks, or accounting failure.

The distinction matters because ordinary intangible-capital language often names the asset class without identifying the mechanism. KBC asks which component of value is hidden: current use, recombination option, learning-loop option, governance-control option, strategic option, expected knowledge loss, or governance-transition exposure. A brand may be an intangible asset. A firm's dependence on a platform-controlled feedback loop may be dark capital. A patented compound may be an intangible asset. The unrecognized loss of future discovery trajectories when a research commons is enclosed may be dark capital.

\begingroup
\small
\setlength{\tabcolsep}{4pt}
\renewcommand{\arraystretch}{1.12}
\begin{longtable}{@{}L{0.26\textwidth}L{0.34\textwidth}L{0.32\textwidth}@{}}
\caption{Dark Capital Compared with Adjacent Categories}\label{tab:ch9:dark-capital-differentiation}\\
\toprule
Category & What it captures & What it misses \\
\midrule
\endfirsthead
\toprule
Category & What it captures & What it misses \\
\midrule
\endhead
\bottomrule
\endlastfoot
Intangible assets & Non-physical investment or asset claims & Governance position, field access, and hidden capability dependency. \\
Intellectual capital & Firm knowledge resources and capabilities & Economy-wide conversion, commons depletion, and cybersecurity-related recombination loss. \\
Market-to-book gap & Valuation residual visible in capital markets & Mechanism, attribution, and separation from speculation or monopoly rent. \\
Dark capital & Obscured productive knowledge capacity and latent knowledge-risk exposure & Requires empirical decomposition before valuation claims are justified. \\
\end{longtable}
\endgroup

Thus, Chapter 9's claim is not that every market-to-book premium is dark capital. Some of it is speculation, monopoly rent, conventional intangible value, or measurement error. The narrower claim is that a knowledge-intensive economy creates economically material productive capacity and latent exposure that conventional asset categories do not separately identify.

\section{The Accounting Shadow}\label{93-the-accounting-shadow}
\index{accounting shadow}

The accounting shadow is the formal interface between dark capital and
financial reporting. Theorem T4, the Balance-Sheet Accounting Shadow theorem in Volume 2, gives the formal audit trail; this section introduces the T4 decomposition in book terms.

T4 does not claim that all unrecognized knowledge-bearing stock should be placed on the balance sheet. It claims that current accounting often does not separately recognize economically material productive value and latent fragility because internally generated knowledge-bearing stock, governance-position value, recombination-field access, and expected knowledge loss are not fully or separately recognized under current accounting categories.

For a knowledge-intensive firm \(i\) operating under governance form \(\pi\),
define \(MV_i\) as market capitalization and \(BV_i\) as accounting book
value. Theorem T4 states the decomposition.

The following decomposition is diagnostic. It does not claim that the market-to-book gap\index{market-to-book gap} can be cleanly decomposed from public data alone. Its purpose is to prevent KBC-relevant components from being collapsed into a single residual.

\begin{equation*}
MV_i - BV_i = E[V(K_i^{unrec})] + \Omega _i^{risk} + E[V(options_i)] + Spec_i + Error_i
\end{equation*}

The equation should be read as an attribution problem, not as a claim that each component is directly observable without modelling assumptions, proxies, or event evidence. It is a formal theorem reference and diagnostic decomposition. Each component names a distinct reason market value might differ from
book value. \(\mathbb{E}[V(K_i^{unrec})]\) is the expected value of
unrecognized productive knowledge-bearing stock, the productive
shadow. \(\Omega_i^{risk}\) is the market\textquotesingle s adjustment for
latent enclosure-cost exposure, the fragility shadow, which reduces
market value when rational markets discount for governance risk.
\(\mathbb{E}[V(options_i)]\) is the expected value of real options on future
knowledge production. \(Spec_i\) is the speculative premium, monopoly
rent expectations, bubble dynamics, narrative-driven overvaluation.
\(Error_i\) is market mispricing and measurement noise. \emph{In plain
terms:} the gap between what a firm is worth in the market and what its
books say it is worth has five sources. The most visible, recognized
intangible assets like patents and acquired brands, is captured in
BV. The remaining four sources are invisible or unrecognizable under
current standards: unrecognized productive knowledge (what the firm
actually knows and can do), latent enclosure-cost exposure (the
governance risks not on any balance sheet), real-options value on future
knowledge production, and a residual combining speculation and error.
The accounting shadow framework does not claim that all of MV−BV is a
recognition failure. It claims that a persistent, systematic, and
structurally explainable portion of it is, and that portion is dark
capital.

This book\textquotesingle s intervention on this decomposition is
to insist on two boundary conditions that much of the literature on
intangibles and market-to-book ratios elides.

The first boundary: not all MV−BV is knowledge capital. The \(Spec_i\) and
\(Error_i\) terms are real and non-trivial. Markets price monopoly rent
expectations (an incumbent\textquotesingle s future enclosure strategy
is priced in), bubble dynamics, and narrative premia that are unrelated
to any productive asset. \textcite{BondCummins2000} identified this problem
formally: the MV−BV gap is too noisy to cleanly identify unrecognized
productive value, because speculative and error components are large and
not independently estimable. T4 retains this caution explicitly
throughout. The accounting shadow framework does not claim that all of
MV−BV is a recognition failure; it claims that a persistent, systematic,
and structurally explainable portion of it is.

The second boundary: the decomposition is two-sided. The prior
literature, \textcite{Lev2001}; \textcite{HaskelWestlake2018}; \textcite{CorradoHultenSichel2005} focuses on the asset-side understatement:
unrecognized productive stocks inflate MV above BV. T4 adds a
liability-side dimension: the \(\Omega_i\) term. Firms that carry large latent
enclosure-cost exposure, whose productive capacity depends on
recombination field access that governance events could restrict, 
face a real risk that does not appear on their balance sheets as a
provision or contingent liability. This is not merely an asset-side gap;
it is a fragility gap, and it is the accounting shadow of strategic
enclosure.

\section{Balance-Sheet Shadow}\index{balance-sheet shadow|textbf}\label{94-balance-sheet-shadow}

The closest structural predecessor on the asset side deserves direct confrontation. \textcite{FarboodiVeldkamp2021} model a data feedback loop in which accumulated data raises product quality, output, and transactions, which in turn generate more data, and show that the resulting life-cycle path of Tobin's \(q\) (equivalently, book-to-market) for data-intensive firms differs systematically from that of capital-intensive firms. This is the closest growth-theoretic account of why \(MV_i - BV_i\) is large and time-varying for data firms: the gap tracks an accumulating, unbooked data stock rather than a mispricing. T4's asset-side term \(\mathbb{E}[V(K_i^{unrec})]\) is consistent with their result and, on this point, claims no priority; the Tobin's-\(q\) anomaly for data firms is already explained. T4's contribution is two components that a firm-internal accumulation model does not by itself isolate. First, the governance-position term \(\delta\cdot\mathbb{E}[v(R_i(\pi, t))]\): part of the gap reflects not the firm's own accumulated data but its control of access to a recombination field that other actors depend on, a value that can persist even where the firm's own data stock is modest. Second, the liability-side fragility term \(\Omega_i\): the same access position that inflates \(MV_i\) carries a latent enclosure-cost and reversal exposure that an accumulation dynamic treating data value as monotone in the firm's own stock does not register. The Tobin's-\(q\) divergence is therefore a predecessor for the asset side of T4, not for its governance-position and fragility components.

The productive shadow, the \(\mathbb{E}[V(K_i^{unrec})]\) component of
the T4 decomposition, has four constituents, each treated as
unrecognizable under current standards by a separate lemma in T4.

\textbf{\(K^E\) (tacit embodied capital)} normally fails separate recognition under IAS 38. It is not identifiable, tacit knowledge is
embodied in people and cannot be separated from them for sale or
licensing; employment contracts govern the person\textquotesingle s
labour, not the knowledge the person carries. It is not controllable, 
employees leave, carrying their \(K^E\) with them; non-compete agreements
provide partial and jurisdictionally variable protection;
\citeauthor{Polanyi1966}\textquotesingle{}s (\citeyear{Polanyi1966}) tacit irreducibility means even
attempted codification does not fully transfer \(K^E\). It is not
reliably measurable, the cost of developing \(K^E\) is inseparable
from ordinary operational expenditure. Every dollar spent on training,
mentoring, collaborative work, and accumulated experience contributes to
\(K^E\), but no accounting procedure can identify what fraction of
operational cost constitutes \(K^E\) investment versus current
production. The entire stock of tacit human and organizational knowledge
that constitutes one of knowledge-bearing capitalism\textquotesingle s
primary productive assets is therefore unrecognized.

\textbf{\(K^D\) (unrec) (internally generated disembodied capital)} is
partially but systematically underrecognized. Under IAS 38,
research-phase costs must be expensed, the standard holds that a firm
cannot demonstrate at the research phase that an intangible asset will
generate probable future economic benefits. Development-phase costs may
be capitalized only under restrictive conditions. Under US GAAP (ASC
730), research and development costs are expensed as incurred for most
industries. The result is a systematic capitalization gap: the
research-phase costs that produced a successful algorithm or dataset are
not on the balance sheet even when the successful output generates
sustained revenue. Moreover, acquired \(K^D\) can be recognized at fair value
at the acquisition date, while organically developed \(K^D\) of identical
economic value may remain unrecognized because no transaction creates a
price. The accounting system penalizes organic knowledge investment
relative to acquisition, a distortion with direct consequences for
the structure of knowledge-intensive industries.

\textbf{\(K^I\) (organizational routines and capabilities)} fails all
three recognition criteria through the same structural arguments that
apply to \(K^E\), compounded by an additional problem: \(K^I\) is not
merely embodied in individuals but distributed across the organizational
structure and constituted through the interaction patterns among people,
routines, and institutional context. \citeauthor{NelsonWinter1982}\textquotesingle{}s (\citeyear{NelsonWinter1982}) definition of routines as the organizational analogue of
individual skills, coordinating action without requiring case-by-case
deliberation, identifies precisely the class of organizational
knowledge that neither separability nor contractual-rights arguments can
reach. \textcite{Teece2007}'s dynamic capabilities framework formalizes this
as sensing, seizing, and transforming capacity: these are the
organizational-level equivalents of \(\widetilde{C}_{a}\)(t) in Model 4. None has a
recognition category in any major accounting standard.

\textbf{\(\delta\cdot\mathbb{E}[v(R_i(\pi, t))]\) (governance-position value)} is
T4\textquotesingle s primary novel contribution to accounting theory. It
is not equivalent to any recognized or theorized intangible asset
category in the prior literature. Goodwill captures unidentifiable
synergies and going-concern value of acquired entities, not a
firm\textquotesingle s access to the external recombination field.
Customer-relationship intangibles capture the present value of expected
revenue from existing customers, not the productive surplus from
combining the firm\textquotesingle s knowledge stock with elements of
the broader knowledge ecosystem. IP portfolio value captures legally
exclusive rights over specific \(K^D\) elements, not the value of
access to knowledge the firm does not own.

The δ term measures something structurally different: the expected
surplus from future recombination events enabled by the
firm\textquotesingle s current governance position, its access to
open-source libraries, public-domain research, standardized APIs,
academic datasets, professional knowledge communities, and shared
commons. This is field-theoretic accounting: the firm\textquotesingle s
productive value depends partly on its ecosystem position, not only on
its internally held assets. When platform API terms change, when IP
governance arrangements shift, when commons governance degrades, or when enclosure
concentrates formerly accessible field elements, the δ term changes,
even though every recognized asset on the firm\textquotesingle s balance
sheet is unchanged. The accounting system has no mechanism to record
this value or its fluctuation.

\subsection{A Worked Demonstration: CUDA Dependency and the Dark
Accounting of Governance
Position}\index{CUDA dependency!dark accounting}\index{NVIDIA!CUDA dependency}\index{dark accounting}\label{a-worked-demonstration-cuda-dependency-and-the-dark-accounting-of-governance-position}

The \(\delta\cdot\mathbb{E}[v(R_i(\pi, t))]\) term is the component of the T4 decomposition
with no equivalent in any standard intangible-asset category. Tracing it
through the Nvidia\index{NVIDIA!governance-position value} CUDA ecosystem clarifies what the term captures and
why it is invisible to current accounting and risk-disclosure systems.
The case shows both sides: governance-position value for the incumbent
and unpriced \(\Omega_i\) exposure for dependent firms.

Platform dependency creates two accounting problems. First, the firm may appear to control productive capability that actually depends on an external governance position: API access, licence terms, interoperability rules, hardware availability, export controls, pricing tiers, documentation, and ecosystem maintenance. Second, a platform change can impair future recombination and learning without destroying any owned asset. The dependent firm may still own the same code, models, contracts, and workstations, while the productive field in which those assets operate has narrowed.

\textbf{The knowledge-bearing stocks.} CUDA (Compute Unified Device
Architecture), released by Nvidia in 2007, is a parallel computing
platform and programming model whose \(K^D\), compiled libraries,
memory management routines, compiler infrastructure, is proprietary
to Nvidia and not available under open licences. Its \(K^I\) is
distributed across thousands of open-source contributors and research
groups who have tuned deep learning frameworks (PyTorch, TensorFlow,
JAX) to CUDA\textquotesingle s memory model and execution architecture
,  but in a form interoperability-contingent on CUDA\textquotesingle s
specific \(K^D\): the \(K^I\) works only in combination with the
proprietary platform. The developer \(K^E\), the trained judgment of
ML engineers in how to profile, optimize, and debug CUDA-based training
runs, is embodied in a practitioner population whose expertise is
substantially non-transferable to alternative GPU programming models
(AMD\textquotesingle s ROCm, Intel\textquotesingle s oneAPI). Each of
these stocks is productive; they are normally not recognized as separately
identifiable assets on the dependent firm\textquotesingle s balance sheet.

\textbf{The pre-change recombination field.} Before
CUDA\textquotesingle s consolidation (roughly 2007--2015), the GPU
computing field offered a broader accessible field: OpenCL was a
vendor-neutral open standard with viable competitive alternatives across
AMD, Intel, and Nvidia hardware. \(D_u\)(\(F_{a, t}\)) for the ML research
population was broader; the actor population could not yet be locked to
any single vendor\textquotesingle s platform by the accumulated weight
of framework \(K^I\) and developer \(K^E\).

\textbf{The governance change.} CUDA\textquotesingle s consolidation
occurred not through legal exclusion but through the accumulation of
CUDA-specific \(K^I\) in the frameworks that the ML research community
built and maintained. As PyTorch and TensorFlow were developed natively
targeting CUDA, the cost of targeting an alternative platform rose, 
not through deliberate exclusion but through the accumulation of
CUDA-specific optimizations in the commons \(K^I\). The interoperability
condition \(P_{a, i, t}\) for non-CUDA hardware gradually fell toward zero
for the most widely used ML tools. By approximately 2020, competitive ML
development required CUDA-enabled Nvidia hardware as a capability
condition, not a legal one: AMD\textquotesingle s ROCm and
Intel\textquotesingle s alternatives existed but faced systematic
disadvantages in the P dimension, interoperability with the standard
ML stack, that translated directly into performance and
development-time penalties.

\textbf{The KBC mechanism: governance-position value for Nvidia.} For
Nvidia, the accumulated CUDA ecosystem creates a \(\delta\cdot\mathbb{E}[v(R_i(\pi, t))]\)
term of substantial scale. Its governance position, the structural
centrality of CUDA to the accessible recombination field of the global
ML research and development community, enables a productive surplus
not captured in any recognized intangible category. Its patents cover
specific algorithmic implementations, not field centrality. Its brand
captures reputation, not ecosystem dependency. Its customer-relationship
intangibles capture expected hardware revenue, not the expected surplus
from all future ML knowledge generation that passes through its
platform. The δ term is the expected value of being the access layer
through which the highest-productive-weight elements of a major
recombination field are reached, a value that is neither on
Nvidia\textquotesingle s balance sheet nor estimable through any current
accounting methodology. The anomalously high MV/BV ratio for Nvidia in
the AI acceleration period is consistent with markets pricing this
governance-position value, even though current accounting systems often cannot separately record
it.

\textbf{The KBC mechanism: \(\Omega_i\) for dependent firms.} The counterpart
analysis applies to any firm whose core production process depends on
CUDA as a recombination-field element. An AI applications firm whose ML
pipeline is built on PyTorch with CUDA optimization\index{PyTorch}\index{CUDA!PyTorch}, trained on Nvidia
hardware, and maintained by engineers with CUDA-specific \(K^E\) carries
an \(\Omega_i\) (unpriced governance-position risk exposure) with several
structurally distinct components.

The first is access risk: Nvidia\textquotesingle s CUDA licence permits
current uses but could be changed; enterprise pricing tiers could be
adjusted; access for specific geographies or applications could be
restricted under export controls. Any of these events would impose costs
on the dependent firm structurally equivalent to \(C_{T2}\) (generation
suppression from field restriction) and \(C_{T6}\) (capability decay if
CUDA-specific \(K^E\) becomes unproductive). These are liability-like economic exposures, not necessarily accounting liabilities under current standards, because IAS 37 requires a present obligation arising from a past event, and these depend on future governance actions not yet contracted.

The second is the capability cascade risk: because the dependent
firm\textquotesingle s \(K^E\) is CUDA-specific, a forced migration to an
alternative platform imposes a capability-decay cost beyond retooling
expenses. The \(K^E\) embodied in the firm\textquotesingle s ML engineers, trained judgment about optimization, memory management, and
debugging in the CUDA environment, is partially non-transferable.
Migration requires accumulated re-learning, which is a form of \(C_{T6}\)
capability loss that current accounting often cannot separately record because the capability was
never recognized as an asset.

The third is competitive symmetry risk: if Nvidia selectively improves
CUDA performance for favoured clients, through early access to
optimization libraries, custom kernel support, or preferential hardware
allocation, dependent firms without that access experience a relative
\(\widetilde{C}\) decay even if their absolute capability is unchanged. Their \(G^R\)
relative to the favoured population falls; their competitive position
deteriorates; their balance sheet is silent.

\textbf{What standard theory sees.} A dominant firm with competitive
advantage through superior technology, switching costs, and network
effects. The relevant antitrust question is whether Nvidia is using its
market position to foreclose competition in adjacent markets. The risk
disclosure relevant to dependent firms is a qualitative "platform
dependency risk" factor in securities filings.

\textbf{What KBC sees that standard theory under-specifies.} Three things
standard theory often under-specifies. First, governance-position value is not
market power in the price sense. Nvidia\textquotesingle s δ·E{[}v(R){]}
term does not appear as pricing power over hardware; it appears as
structural centrality to a recombination field. That value is unmeasured
by revenue or margin analysis and corresponds to no recognized
intangible category. Second, the risk borne by dependent firms, \(\Omega_i\)
,  is not captured by qualitative risk factors. The T5 decomposition\index{T5 Enclosure Efficiency Loss!cost decomposition}
specifies the structure of that risk: which cost components dominate,
how they scale with the firm\textquotesingle s \(\sigma_a^R\) and \(D_u\)
composition, and how the recovery lag \(\tau_R\) determines whether the damage
from a governance event is temporary or permanent. No current accounting
or disclosure system performs this calculation. Third, the \(K^I\) that
the open-source ML community has accumulated in CUDA-specific form
represents a commons-depletion event: the governance capacity of the
open alternative (ROCm, OpenCL) has been depleted not through any direct
action against those standards but through the accumulation of
CUDA-specific \(K^I\) in the frameworks that the community built and
maintains. The mechanism is the §6.8 commons-depletion pathway,
operating here through developer-attention allocation rather than
employment extraction.

\emph{Smithian departure:} Smith\textquotesingle s analysis of
competitive advantage assumed that productive advantage arose from
factors that could, in principle, be competed for: better land, cheaper
labour, superior skills. Governance-position value is not of this kind.
The δ·E{[}v(R){]} term captures advantage arising from structural
centrality to a recombination field, a position that can be
maintained through the accumulation of ecosystem \(K^I\) that operates
independently of any identifiable competitive act. Standard competition
analysis, calibrated to Smithian categories of productive advantage, has
no framework for this form of capital and no instrument for the risk it
imposes on dependent actors.

\section{Invisible Knowledge-Capital
Impairment}\label{95-invisible-knowledge-capital-impairment}
\index{knowledge impairment|textbf}

Knowledge-bearing capital can suffer real productive-value loss that
does not trigger accounting impairment. This is the invisible impairment
problem, formalized in T4\textquotesingle s Proposition T4.3\index{T4 Balance-Sheet Accounting Shadow!fragility-shadow proposition}\index{T4.3 Fragility Shadow} and
grounded in T7 (Suppressed Appreciation).

Standard accounting impairment testing (IAS 36; ASC 360) asks whether
the carrying value of a recognized asset exceeds its recoverable amount.
A recognized asset is impaired when it can no longer generate the
economic benefits its book value implies, when a factory becomes
obsolete, a patent loses commercial relevance, or a software licence is
superseded. The impairment test is triggered by "significant changes in
the technological, market, economic, or legal environment" (IAS 36 §12)
and is applied to assets that are recognized on the balance sheet.

T7 identifies a form of value loss\index{T7 Suppressed Appreciation!dark-capital measurement} that this test cannot reach.
Proposition C2 (Suppressed Appreciation, Chapter 6) established that
enclosure of E ⊂ \(F_{a, t}\)(π₀) reduces the productive value of
non-enclosed knowledge-bearing stocks held by excluded actors: if \(K_j\) ∈
E was a complementary field element for actor a\textquotesingle s
existing stock \(K_a\), the value function v(\(K_a\), a, π) falls under π₁
even though \(K_a\) itself is unchanged. The cascade extension amplifies
this: the interoperability cascade removes additional non-enclosed
elements from \(F_{a, t}\), further suppressing v(\(K_a\), a, π₁) beyond the
direct loss from E\textquotesingle s removal.

T7 establishes this at the cardinal level: v(\(K_a\), a, π₁) \textless{} v(\(K_a\),
a, π₀) whenever E contains stocks with positive complementarity to \(K_a\),
with the cascade amplifying the loss over time. Yet IAS 36 does not normally
recognize this as a separately identifiable impairment event. \(K_a\), whether it is a software
capability, a research expertise, a professional qualification, or an
organizational competency, has not been physically damaged, its
contractual status has not changed, and its legal owner remains the
same. What has changed is its productive context: the recombination
field from which it drew part of its value has been restricted by an
external governance event beyond the firm\textquotesingle s control.
Accounting has no mechanism to record a governance-conditioned reduction
in an asset\textquotesingle s productive value when that reduction flows
through the field rather than through the asset itself.

The governance-change-as-unrecorded-impairment result (T4.3) follows
directly: a governance form shift π₀ → π₁ that reduces \(F_{i, t}\)(π)
constitutes an economic impairment of the firm\textquotesingle s
productive capacity, but this impairment is not triggered as an
accounting event because the recognition system has no recognized asset
to impair. The firm\textquotesingle s balance sheet is unchanged while
its economic value has fallen, a genuine invisible impairment, not a
measurement lag.

\section{Breach Loss Begins Where Accounting Stops}\index{cybersecurity failure!breach loss}\index{breach loss}\label{breach-loss-begins-where-accounting-stops}

Some of the most important breach loss begins where ordinary accounting stops. The point is not that all breach loss is invisible, but that knowledge-capital impairment can extend beyond direct incident costs, provisions, insurance recoveries, and disclosures.

\begin{center}
\fbox{\begin{minipage}{0.92\linewidth}
\textbf{Boxed proposition: Cybersecurity as Knowledge-Capital Preservation}

Cybersecurity is capital-preservation risk management for knowledge-bearing stock. Its function is to preserve knowledge-capital value by protecting the exclusivity, integrity, availability, and productive use of source code, models, datasets, credentials, customer records, detection logic, operating routines, and feedback systems. A cybersecurity breach is therefore not merely a technical incident or operating expense. When knowledge-bearing stock is copied, inferred, poisoned, or exposed to unauthorized shared control, the firm may retain possession while losing exclusivity, governance position, and future competitive yield. The capital impairment begins where conventional accounting stops looking.
\end{minipage}}
\end{center}

A cyber breach affecting knowledge-bearing stock is not merely a confidentiality incident or downtime event. It can impair \(K^D\) by destroying, corrupting, exposing, duplicating, or making disembodied stock unavailable; reveal or damage \(K^I\) by exposing routines, controls, operating procedures, detection logic, or institutional weaknesses; reduce trust in the firm's knowledge systems; expand adversarial capability; expose platform dependency; and create latent fragility that ordinary incident-cost accounting does not capture. The relevant decomposition is therefore not just ``breach cost, '' but destruction or unavailability, exfiltration or duplication, and integrity attack or poisoning.

The most important economic loss from a cybersecurity breach may not be the cleanup cost. Cleanup is visible because invoices, legal expenses, downtime, and recovery work enter accounting and insurance systems. Knowledge-capital impairment is harder to see because the breached firm may still possess the same files, models, formulas, routines, source code, customer data, or strategic plans after the event. Nothing has been consumed in the rival-good sense. The loss is a loss of exclusive governance over future use.

This is why cybersecurity failure, especially exfiltration, belongs inside dark capital rather than merely inside operational risk. If an attacker or competitor can absorb the copied stock into its own recombination field, the breach can impair the victim's future knowledge-capital value by accelerating another actor's learning, product development, pricing strategy, model training, or market entry. Conventional breach accounting records response cost. It rarely records the devaluation of knowledge-bearing stock caused by unauthorized capability formation elsewhere.

In KBC terms, the event shifts the compromised stock from exclusive private governance toward unauthorized shared control. The victim retains the stock, but its exclusivity premium, governance-position value, and strategic option value decline. The accounting system sees no conventional disposal, often no separate impairment trigger for the affected knowledge-bearing stock, and often no separately recognized asset. The economic impairment is therefore unidentified at precisely the point where it may matter most: the future yield of knowledge-bearing stock that no longer belongs exclusively to the firm that generated or governed it. The claim is not that every breach effect should be capitalized under current accounting standards; it is that economically material knowledge-capital impairment can exist before accounting recognition becomes appropriate.

In compact form, the application expression is:

\begin{equation*}
\begin{aligned}
V^K_i
=
&CUV_i+ROV^{*}_i+LOV^{*}_i+COV^{*}_i+SOV^{*}_i \\
&-P(B)\left[C_{\mathrm{EX}}+C_{\mathrm{KI}}+C_{\mathrm{AG}}\right]
-EL^{\mathrm{other}}_i.
\end{aligned}
\end{equation*}

This is an uncalibrated scoring model applied to cyber risk, not a new theorem or a calibrated loss equation. The expression should be read as an application expression, not as a new theorem. It inserts cybersecurity-related expected loss into the K-CMM value structure by treating cybersecurity-failure risk as one component of expected knowledge loss. The three visible terms mark the distinct channels that ordinary cybersecurity-loss accounting tends to compress or miss: exclusivity-premium loss, institutional capability disruption, and attacker recombination gain.

The Technical Companion, Appendix D, gives the formal application. Its cyber-adjusted K-CMM expression treats cybersecurity-related expected loss as one component of expected knowledge loss, separating visible incident cost from exclusivity-premium loss, institutional capability disruption, and attacker recombination gain.

The specific KBC term for this hidden channel is \textbf{recombination-field impairment}. A cybersecurity event impairs the victim's knowledge capital when unauthorized access expands another actor's productive field, reduces the victim's exclusivity premium, poisons or destabilizes the victim's own stock, or gives an adversary enough context to imitate, counter, train, exploit, or strategically reposition. The victim may still possess the original code, files, credentials, dataset, or model. What has changed is the distribution of future recombination capacity.

A schematic cybersecurity-loss placeholder can be written as:

\begin{equation}
EKL^{\mathrm{cyber}}_i
=
C_{\mathrm{response}, i}
+C_{\mathrm{EX}, i}
+C_{\mathrm{KI}, i}
+C_{\mathrm{AG}, i}
+\Delta R^{\mathrm{adv}}_{j, i}
-R^{\mathrm{recover}}_{i}.
\label{eq:ch9:cyber-recombination-field-impairment-placeholder}
\end{equation}

Equation~\ref{eq:ch9:cyber-recombination-field-impairment-placeholder} is a counterfactual measurement problem and placeholder, not a calibrated loss model. The term \(\Delta R^{\mathrm{adv}}_{j, i}\) marks adversarial recombination-field expansion: the increase in actor j's ability to recombine, imitate, exploit, or counter actor i's stock because of the event. The recovery term marks the portion of loss offset by containment, rotation, re-architecture, disclosure, litigation, or defensive learning.

\section{Foregone Knowledge
Capitalization}\label{96-foregone-knowledge-capitalization}

Foregone knowledge capitalization is the most counterfactual component
of dark capital. It cannot be read directly from a balance sheet. It
must be inferred from blocked access, suppressed recombination, lost
feedback, delayed diffusion, natural experiments, policy changes,
platform closures, failed maintenance, or comparison cases.

Foregone knowledge capitalization names the third dimension of dark
capital: capital that had the potential to form but did not, because the
conditions for its formation were suppressed. It is the hardest
sub-dimension to measure, because it requires reasoning about a
counterfactual, what would have been generated had the suppression
not occurred, but it may be the most consequential form of dark
capital for long-run economic performance. Table~\ref{tab:ch9:foregone-capitalization-evidence}
therefore treats the concept as measurable in principle through proxies
and comparison designs, not as something casually observable in accounts.

\begin{table}[htbp]
\centering
\small
\caption{Foregone Knowledge Capitalization\index{foregone knowledge capitalization|textbf}: Events, Proxies, and Evidence}
\label{tab:ch9:foregone-capitalization-evidence}
\begin{tabular}{@{}p{0.22\textwidth}p{0.26\textwidth}p{0.27\textwidth}p{0.17\textwidth}@{}}
\toprule
Event & Foregone stock & Observable proxy & Evidence type \\
\midrule
API closure & Third-party tools\index{API closure!dark-capital proxies}\index{API closure!dark-capital measurement}, research paths, integrations & Lost apps, reduced developer activity, migration costs & Event study / platform data \\
Open-source maintainer collapse & Updates, patches, security fixes & Release slowdown, CVEs, dependency failures & Repository and incident data \\
Feedback capture by platform & External model improvement paths & Capability divergence, benchmark gaps & Comparative model performance \\
Patent/API/IP enclosure\index{IP enclosure!dark-capital measurement} & Alternative products or research trajectories & Entry decline, licensing barriers, delayed substitutes & Market-entry and patent data \\
\bottomrule
\end{tabular}
\end{table}

The formal grounding is Proposition C (Generative Suppression, Chapter
6): enclosure of E ⊂ \(F_{a, t}\)(π₀) reduces \(G^R\) for excluded actors
through the suppression ratio \((\sigma_a^R)^{\eta}\cdot(D_u\ \mathrm{ratio})^{\mu}\). The
\(G^R\) reduction in period t means that knowledge-bearing stock that
would have been generated in period t is not generated. That missing
stock would itself have served as input to the next round of knowledge
generation and, through the state equation of Chapter 3, contributed to
\(\widetilde{C}_{a}\) in subsequent periods. The capital loss from foregone \(G^R\) is
therefore not a one-period loss; it is a compounding loss that
propagates forward through the knowledge conversion cycle.

Foregone knowledge capitalization takes five specific forms in the
framework. Failed recombination is the most direct: the \(G^R\) output
suppressed by \(D_u\)(\(F_{a, t}\)) reduction is new \(K^D\) or \(K^I\) that
would have entered the accessible knowledge base had field conditions
been open. A pharmaceutical researcher who cannot combine a publicly
discovered receptor mechanism with a proprietary compound library
because the latter is trade-secret-protected loses the recombination
attempt that would have generated a novel drug class; the \(K^D\) is not
generated, and the loss does not appear in any welfare account. Failed
experimentation is the \(G^X\) analogue: systematic investigation of
productive properties that the enclosed stock would have enabled but
that enclosure forecloses. Computational social scientists who cannot
systematically vary the parameters of a platform\textquotesingle s
recommendation algorithm (because API access is restricted) cannot
run experiments that would identify misinformation amplification
mechanisms; the methodological \(K^D\) and the policy-relevant findings
are not produced. Failed codification is the conversion of tacit \(K^E\)
into explicit \(K^D\) that would have occurred had appropriate
institutional support been available, codification that was attempted
but could not be completed because the organizational routines,
mentoring relationships, or knowledge-translation infrastructure
required for codification were depleted or absent. A retiring specialist
in rare tumour diagnostics whose interpretive protocol is never codified
into a clinical guideline, because the hospital system lacks the
knowledge-translation infrastructure to convert expert tacit judgment
into a transferable standard, represents \(K^D\) that never enters the
knowledge base; the next cohort of clinicians must redevelop it through
unstructured experience. Failed institutional learning is the \(K^I\)
accumulation that organizations would have achieved through reflective
practice and organizational feedback loops, had the feedback signals
been distributed rather than captured within an enclosed deployment
cycle. A hospital system that routes clinical outcome data to
administrators rather than feeding it back to clinical teams eliminates
the organizational feedback loop through which \(K^I\) about what works
under what conditions would have accumulated; the absence is invisible
because the data exists, it simply does not flow to the actors who
could act on it. Failed governance conversion is the transit of \(K^C\)
into productive \(K^D\) that commons governance would have enabled, had
the commons\textquotesingle{} governance capacity not been depleted
through labour-market extraction of governance \(K^E\). A Wikipedia
language community that loses its core governance administrators to
burnout and attrition, without adequate succession processes, retains a
formally open knowledge base that its remaining contributors cannot
maintain, curate, or develop at prior quality, the commons persists
as a legal structure while the \(K^E\) of governance (which edits to
accept, which disputes to escalate, which contributors to trust) is not
regenerated because the onboarding infrastructure to convert it into new
maintainers\textquotesingle{} \(K^I\) no longer exists.

Each form of foregone knowledge capitalization is invisible to
accounting for the same structural reason: accounting records what is
recognized when it is acquired or created, not what could have been
created under different conditions. The capital that enclosure
suppresses never appears on any balance sheet, it was never there to
record. The shadow it casts is purely counterfactual, and counterfactual
accounting is not something any standard recognition system attempts.

The aggregate of foregone knowledge capitalization across all excluded
actors is the T5 \(C_{T2}\) cost component, the generation suppression
cost that is the largest single element of \(C_{enclosure}\)(T) in most
high-recombination sectors. \(C_{T2}\) is the formal integral of the
per-period \(G^R\) suppression over the enclosure horizon, and it is a
welfare cost precisely because the foregone \(G^R\) represents real
future productive value that the enclosure has eliminated. That welfare
cost is real; it appears in no accounting system.

\section{Hidden Capability Loss}\label{97-hidden-capability-loss}

Chapter 7 established the \(G^L\) → \(\widetilde{C}_{a}\) mechanism: feedback capture by
incumbents raises their dynamic capability while excluded actors may fall behind when learning from the enclosed stream is unavailable, substitute learning is insufficient, and \(G^R\) remains suppressed. The state equation predicts \(\widetilde{C}_{ent, t+1}\)
\textless{} \(\widetilde{C}_{ent, t}\) whenever (γ · \(G^R_{ent, t}\)) \textless{}
\(\delta_C\) · \(\widetilde{C}_{ent, t}\), whenever the generation and learning
contributions to capability fall short of the depreciation rate. This is
the hidden capability loss: the real decline in an excluded
actor\textquotesingle s productive capacity that leaves no trace in
accounting records.

The accounting invisibility of \(\widetilde{C}_{a}\) decline is total. \(\widetilde{C}_{a}\) is not a
recognized asset, it is, at most, approximated by the γ·\(K^I\)
component of the productive shadow, but even that approximation captures
only the stock of organizational knowledge, not the dynamic capability
that governs how productively that stock is deployed. A firm whose \(\widetilde{C}_{a}\)
is declining, whose sensing, seizing, and transforming capabilities
are eroding because deployment scale has contracted, \(G^L\) has been cut
off, and productive field access has narrowed, shows the same \(K^I\)
stock on its balance sheet at the beginning and end of the decline
period. The capability erosion is invisible; the organizational asset is
unchanged in accounting terms.

The recovery lag established in T6.6 adds a further dimension\index{T6.6 Recovery Lag!dark-capital measurement}. When the
enclosure that drove \(\widetilde{C}_{a}\) decline ends, when API access is restored,
when an IP term expires, when platform terms are revised, the
excluded actor\textquotesingle s capability does not immediately
recover. The gap \(\Delta_t\) = \(\widetilde{C}_{inc, t}\) − \(\widetilde{C}_{ent, t}\) closes at rate (1 −
\(\delta_C\)) per period after reversion, but the initial gap may be large
(having grown over T periods of enclosure), and the recovery duration
\(\tau_R\) is increasing in T and potentially superlinear. The post-enclosure
recovery period generates a real welfare cost, the \(C_{T6}\)b component
of T5\textquotesingle s \(C_{enclosure}\), that is equally invisible to
accounting: the firm\textquotesingle s balance sheet often shows no separate liability-like exposure
for the recovery cost because the enclosure event that generated it was
never recorded as an impairment, and the recovery process is treated as
ordinary operational expenditure rather than as the repair of a
governance-caused capability deficit.

Hidden capability loss is the most strategically important form of dark
risk. An investor assessing the competitive position of an excluded
actor from its financial statements will see the same balance sheet
before and after a period of field restriction. The capability gap that
strategic enclosure has created, and the recovery lag that makes the
gap structurally durable, is entirely outside the accounting field of
view. The firm appears as sound as it was before the enclosure; the
structural deterioration in its productive capacity and generative
potential is dark.

\section{Unpriced Governance-Position
Risk}\label{98-unpriced-governance-position-risk}

The liability-side gap in T4\textquotesingle s decomposition is the \(\Omega_i\)
term. This term asks what future cost a firm is exposed to because the governance structure around its knowledge inputs may change.

\begin{equation*}
\Omega _i = E[C_T2(T_E) + C_T7(T_E) + C_T6(T_E) + C_T8(T_E)]
\end{equation*}

This is a formal theorem reference and diagnostic decomposition of latent governance-position exposure\index{governance-position exposure}, not an accounting liability-recognition rule. Here, \(T_E\) is the expected enclosure-exposure horizon. This is the
expected value of the four T5 cost components if a field-restricting
governance event, an API closure, a patent assertion, a commons
depletion event, a platform terms-of-service change, occurs. \emph{In
plain terms:} \(\Omega_i\) is the expected cost the firm will eventually bear
because of the governance structure it operates within, either as the
enclosing actor (bearing recovery and trust costs when \(T^*_{\mathrm{strategic}}\)
causes it to over-enclose its own recombination field) or as the
excluded actor (bearing capability loss and generation suppression). It
is a liability-like economic exposure, not necessarily an accounting liability under current standards. It is not on any balance sheet, because existing
standards require a present obligation arising from a past event, and
\(\Omega_i\) is a contingent future loss whose probability depends on governance
actions that have not yet occurred. The firm is exposed, even though the balance sheet may not show that exposure. It is an expected economic cost that the firm faces as
a structural feature of its governance position. The recognition boundary matters here: the claim is not that every knowledge-capital exposure should be capitalized under current accounting standards, but that economically material exposure can exist before accounting recognition becomes appropriate.

The \(\Omega_i\) term is the accounting expression of unpriced
governance-position risk. Firms with high \(\Omega_i\), those whose
productive capacity depends heavily on open-field elements that
incumbent enclosure or governance change could restrict, carry
substantially more fragility to governance events than their balance
sheets suggest. The accounting system often does not separately distinguish between a
firm whose core productive capacity depends on open-access commons, open
API standards, and public-domain datasets (high \(\Omega_i\), high
field-restriction vulnerability) and a firm whose core capacity depends
on proprietary internally-held IP (lower \(\Omega_i\), lower field-restriction
vulnerability). Both balance sheets show the same form of
non-recognition. The former firm appears as structurally sound as the
latter; the structurally different risk profile is dark.

Standard risk disclosures, the "platform dependency risk, " "IP
concentration risk, " and "regulatory risk" factors that appear in
securities filings, are qualitative and uncalibrated. They do not
quantify \(\Omega_i\) using the T5 decomposition; they do not identify which
cost components dominate; they do not specify how \(\Omega_i\) varies with the
firm\textquotesingle s recombination intensity and field composition;
and they do not model the probability of governance events using any
theory-grounded governance-form distribution. The qualitative risk factor says
"our products depend on third-party APIs and changes to API terms could
adversely affect our business." The T4 framework would quantify: given
the firm\textquotesingle s \(\sigma_a^R\), \(D_u\) field composition, \(\delta_C\), and
\(N_{traj}\), a governance shift that removes its five most productive field
elements would cost \(C_{T2}\) + \(C_{T7}\) + \(C_{T6}\) + \(C_{T8}\) measured against the
T5 formula. No current accounting or disclosure system performs this
calculation.

The strategic enclosure dimension of \(\Omega_i\) connects Chapter 8 to Chapter
9\textquotesingle s measurement framework. Chapter 8 showed that
enclosing incumbents\textquotesingle{} rational strategy produces
\(T^*_{\mathrm{strategic}}\) \textgreater{} \(T^*\), they over-enclose relative to the
social optimum. The firms on the receiving end of this strategic
over-enclosure carry \(\Omega_i\) as an unpriced latent liability-like exposure. The enclosing
incumbent\textquotesingle s balance sheet does not show \(C_{T2}\), \(C_{T7}\),
\(C_{T6}\), or \(C_{T8}\) as costs; the excluded firms\textquotesingle{} balance
sheets often do not show \(\Omega_i\) as a liability-like exposure. The entire social cost of
strategic enclosure is dark, absent from the accounts of the agents
who generate it and absent from the accounts of the agents who bear it.

\section{Unmeasured Public and Commons Knowledge
Capital}\label{99-unmeasured-public-and-commons-knowledge-capital}

Dark capital is not only a firm-level accounting problem. Some knowledge-bearing stocks are economically material precisely because they sit outside the firm: open-source infrastructure, public standards, scientific classifications, legal doctrines, statistical systems, metrology, benchmarks, and shared professional protocols. These stocks may not belong on any single firm\textquotesingle s balance sheet, but their degradation can still impose firm-level and system-level economic losses.

The dark capital framework applies not only to private firms but to the
knowledge-bearing stocks held in collective and public forms: \(K^C\)
(commons knowledge capital) and \(K^P\) (public epistemic capital). These
are treated by accounting (and by most economic statistics) as
either external background conditions or as public goods that do not
appear on any balance sheet. The framework argues that this treatment
substantially misrepresents both their productive contribution and their
vulnerability.

\(K^C\), the accumulated knowledge base, governance protocols, and
contribution communities that constitute functional knowledge commons
,  generates productive value for all actors whose recombination field
includes it. The productivity of software development that draws on
open-source libraries, genomics research that draws on open data
repositories, and AI development that draws on publicly available model
weights is substantially enabled by \(K^C\) that appears on no
firm\textquotesingle s balance sheet and in no national income account
as productive capital. Its depletion through commons-depletion
mechanisms (Chapter 6\textquotesingle s §6.6) reduces the productive
capacity of all actors whose field includes it, an economic capital loss
that may have no separate accounting recognition.

\(K^P\), the public institutions that produce, curate, verify, and
distribute knowledge for general use, is similarly unrecognized as
productive capital in accounting terms. Public universities, research
institutes, open access publishing systems, government statistical
agencies, and public library systems collectively maintain the epistemic
infrastructure on which private knowledge generation draws. T6 Corollary
T6.G3 established that higher \(K^P\) raises \(\widetilde{C}_{min}\), the capability
floor below which no excluded actor falls, reducing the depth and
duration of capability traps under enclosure. T5 Corollary T5.G3
established that richer \(K^P\) raises \(T^*\), the socially optimal enclosure
duration, because the recovery-lag cost of long enclosure is lower when
\(K^P\) provides a richer fallback for excluded actors. \(K^P\) therefore
has a second-order moderating effect on the welfare cost of enclosure
that is entirely invisible to any standard accounting or policy
evaluation framework.

The most consequential dark capital misclassification is treating \(K^C\)
and \(K^P\) as free external inputs to private production rather than as
productive capital stocks that require investment, maintenance, and
governance to remain productive. When a commons is depleted through
labour-market extraction of governance \(K^E\), the productive value of
all private actors who drew on that commons falls, but no balance
sheet records the loss, no national income account records the capital
depreciation, and no policy instrument designed around recognized
capital categories responds to the depletion. The commons can be
destroyed, and the destruction is invisible to every capital measurement
system currently in use.

The §5.6 analysis of governance effects on recombination field breadth
established the formal connection: commons (\(K^C\)) and public
infrastructure (\(K^P\)) governance arrangements widen \(D_u\)(\(F_{a, t}\)) for
non-incumbents, while private IP and platform dependency governance arrangements narrow
it. This means \(K^C\) and \(K^P\) are inputs to the aggregate \(G^R\)
generation function, they appear in the \(D_u\) term that multiplies
(\(R_{a, t}^{\eta}\)) in the generation function of Chapter 3. The aggregate
production function of knowledge-bearing capitalism includes \(K^C\) and
\(K^P\) as inputs, but national income accounting often does not separately count their
accumulation as investment or their depletion as capital loss. The
macroeconomic measurement of knowledge-bearing capitalism is therefore
systematically incomplete at the aggregate level, not merely at the firm
level.

\section{Toward a Knowledge-Capital Balance
Sheet}\index{Knowledge-Capital Balance Sheet|textbf}\label{910-toward-a-knowledge-capital-balance-sheet}
\index{knowledge capital}

The prior sections have identified the forms of dark capital that a more
adequate measurement system would need to recognize. This section
sketches the structure of a knowledge-capital balance sheet, not as a
GAAP/IFRS-ready proposal, but as a theoretical target that clarifies
what full recognition would require and where the most consequential
gaps are.

\begin{figure}[H]
\caption[Dark Capital and the Balance-Sheet Shadow]{Dark Capital and the Balance-Sheet Shadow}\index{accounting shadow!balance-sheet shadow}
\label{fig:ch9:dark-capital-shadow}
\centering
\includegraphics[width=\textwidth]{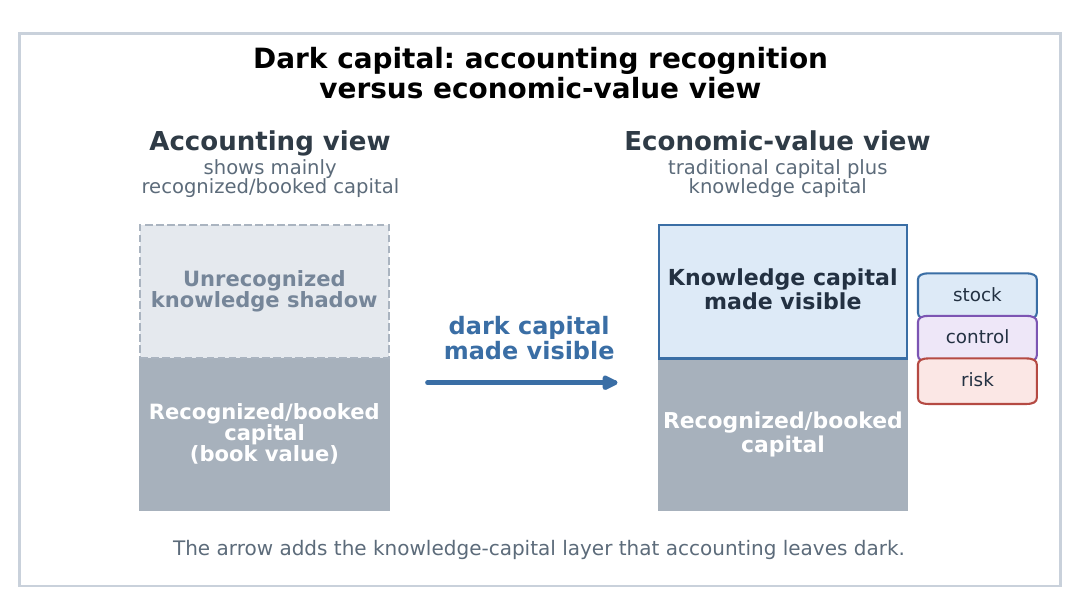}
\par\smallskip\noindent\footnotesize\emph{Note.} Dark capital is not illicit capital. The left side shows the narrower accounting view, where recognized/booked capital is visible and knowledge capital remains largely in shadow. The right side adds the knowledge-capital layer: unrecognized residential knowledge stock, unrecognized governance-position value, and unpriced capability or fragility exposure.
\end{figure}

A knowledge-capital balance sheet would supplement the standard
financial balance sheet with four categories of recognition that current
standards omit. The measurement instrument that structures those
categories is the Knowledge-Capital Measurement Model (K-CMM), which
specifies total knowledge-capital value as:

\begin{equation}
V^K_i = CUV_{\tau, i}+ROV^{*}_{\tau, i}+LOV^{*}_{\tau, i}+COV^{*}_{\tau, i}+SOV^{*}_{\tau, i}-EKL_i.
\label{eq:ch9:kcmm-truth-adjusted}
\end{equation}

where each positive component is truth-adjusted by the epistemic
reliability coefficient \(\tau_i\) of the stock (Axiom B.2, Technical Companion, Appendix B):

\begingroup
\footnotesize
\setlength{\tabcolsep}{3pt}
\renewcommand{\arraystretch}{1.12}
\sloppy
\par\addvspace{0.8\baselineskip}\noindent
\begin{longtable}{@{}L{0.18\textwidth}
L{0.25\textwidth}
L{0.32\textwidth}
L{0.19\textwidth}@{}}
\caption{K-CMM Value Components and Truth-Dependence}\label{tab:ch9:kcmm-value-components-truth-dependence}\\
\toprule\noalign{}
Component & Name & What it captures & Truth-dependence \\
\midrule\noalign{}
\endfirsthead
\toprule\noalign{}
Component & Name & What it captures & Truth-dependence \\
\midrule\noalign{}
\endhead
\bottomrule\noalign{}
\endlastfoot
\(CUV_{\tau}\) & Current Use Value (truth-adjusted) & The productive value
the stock currently generates, given deployment context, governance,
complementary capability, and epistemic reliability & Directly affected:
false or unreliable stock generates apparent CUV but accumulates
\(EKL_{false}\) \\
\(ROV^*_{\tau}\) & Recombination Option Value (truth-adjusted) & Future value
from combining this stock with other knowledge, net of overlap &
Affected: recombination with false stock\index{false stock} propagates error into combined
outputs \\
\(LOV^*_{\tau}\) & Learning-Loop Option Value (truth-adjusted) & Future value
from feedback-driven improvement trajectories, net of overlap & Affected
if feedback is systematically biased or hallucinated; reinforces
error \\
\(COV^*_{\tau}\) & Control/Governance Option Value (truth-adjusted) & Value
from the governance position the stock enables, net of overlap & Less
directly affected by truth-content; governance value may persist even if
stock is partially unreliable \\
\(SOV^*_{\tau}\) & Strategic Option Value (truth-adjusted) & Value from
strategic flexibility and future positioning, net of overlap & May be
high even for false stock whose falsity is not yet known to competitors;
decays rapidly on discovery \\
EKL & Expected Knowledge Loss & Expected future value reduction from
degradation, obsolescence, exfiltration, capability loss, enclosure
exposure, and epistemic reliability failure & Expanded: EKL =
\(EKL_{standard}\) + \(EKL_{false}\) \\
\end{longtable}
\endgroup

\textbf{The epistemic reliability coefficient \(\tau_i\)} measures the degree
to which \(K_i\) is truth-tracking, provenance-supported, validated,
contextually adequate, and resistant to defeaters in its intended
deployment context. \(\tau_i\) ∈ {[}0, 1{]}, where \(\tau_i\) = 1 indicates fully
validated, truth-tracking stock and \(\tau_i\) → 0 indicates systematically
false, hallucinated, or misleading stock.

\textbf{The epistemic risk term in EKL:}

\begin{quote}
EKL = \(EKL_{standard}\) + \(EKL_{false}\)
\end{quote}

where:

\begin{quote}
\(EKL_{false}\) = Pr(\(K_i\) is false or misleading in context) ×
Impact(error if acted upon)
\end{quote}

For high-stakes decision contexts, \(EKL_{false}\) dominates: even low
hallucination or corruption rates destroy capital value when the error
cost is large and irreversible. The per-component τ structure is
preferred over a single scalar multiplier because truth matters
differently across components: a false claim may have low \(CUV_{\tau}\) but
high SOV*\emph{\{τ\} (speculative) and high EKL}\{false\}. Collapsing to
a single τ would misstate the value structure.

CUV is estimated through the product:

\begin{equation}
CUV_{\tau}=K^{\mathrm{pot, use}}_{i, t} \times GATE_{a, i, t}(\pi) \times QUAL_{i, t} \times \tau^{use}_i
\label{eq:ch9:cuv}
\end{equation}

This is an uncalibrated scoring model for current use value, not an externally assured accounting measure. Here, \(K^{\mathrm{pot, use}}_{i, t}\) is the productive knowledge potential of the stock in
its current deployment context, the two quality gates are:

\begin{quote}
\(GATE=(A\times P\times I\times C)^{1/4}\)
\end{quote}

\begin{quote}
\(QUAL=(F\times D\times T\times R)^{1/4}\)
\end{quote}

The GATE and QUAL expressions are uncalibrated scoring models: they discipline judgement about access, permission, interoperability, capability, fidelity, depth, transferability, and reliability before domain-specific calibration. The term \(\tau^{use}_i\) is the epistemic reliability of the stock for its
current productive use specifically. QUAL captures assessable quality
dimensions (Fidelity, Depth, Transferability, Reliability) at
measurement time; \(\tau^{use}_i\) captures truth-tracking adequacy
across future deployment contexts, including defeaters not yet known.
QUAL ⊂ τ: high QUAL is necessary but not sufficient for high τ.

The no-overlap constraint governs all five positive components: no
benefit stream may enter more than one of ROV*\emph{\{τ\}, LOV*}\{τ\},
COV*\emph{\{τ\}, SOV*}\{τ\}, or \(CUV_{\tau}\). Without this constraint, the
model double-counts recombination benefits that simultaneously create
option value and current-period returns.

\emph{In plain terms:} \(V^K\) is the estimated knowledge-capital value
of a stock across its full economic life, what it currently produces,
what future options it creates through recombination, learning,
governance, and strategy, minus expected losses from degradation,
obsolescence, capability decay, enclosure exposure, and epistemic
reliability failure. The truth-adjustment matters most at EKL: a large
stock of confidently-held false beliefs generates impressive apparent
CUV and SOV*, and large \(EKL_{false}\) that offsets both when deployment
exposes the falsity. No existing balance sheet captures this structure.
K-CMM estimates economically material knowledge-capital value, exposure, and
impairment before those effects become visible in conventional accounts.

\begin{center}
\setlength{\fboxsep}{8pt}
\fbox{%
\begin{minipage}{0.94\textwidth}
\small
\textbf{What Chapter 9 Lets Us See}

\vspace{0.5em}
\begin{tabular}{@{}p{0.30\textwidth}p{0.58\textwidth}@{}}
\toprule
Standard view & KBC adds \\
\midrule
Market-to-book gap & Decompose hidden value, hidden risk, governance position, and fragility. \\
Cyber incident cost & Model knowledge-capital impairment, duplication, poisoning, and adversarial reuse. \\
API/platform dependency & Treat access loss as productive-capability impairment. \\
Data asset & Separate data stock from interpretive capability, legality, context, and predictive decay. \\
Open-source/public infrastructure & Treat maintainer shock and standards failure as capital-like losses. \\
Intangible valuation & Add expected knowledge loss and foregone knowledge formation. \\
\bottomrule
\end{tabular}
\end{minipage}%
}
\end{center}

Chapter 10 turns this measurement problem into a governance problem: if knowledge-capital value, loss, and non-formation can be identified, institutions must decide which governance forms preserve, disclose, compensate, insure, or restrict them.

\chapter[Governance Design]{Governance Design for Knowledge-Bearing Capitalism}\index{governance design}\index{policy design}\label{chapter-10-governance-design-for-knowledge-bearing-capitalism}

\chapterhook{How to Govern What Balance Sheets Miss}

If governance is the conversion mechanism, governance design is the choice of which conversion to perform. Chapter 1's claim that governance can enclose, preserve, or condition use-value, and that the choice between capture now and recombination later is conditional rather than settled, becomes here a design problem: how to govern knowledge so that what is converted into private exchange-value does not quietly destroy the use-value a society depends on.

Stated in the use-value terms of Chapter 1, the design target is precise. Because the yield a knowledge unit renders is rarely observed directly, every gate is in effect an estimator of that yield, and a gate is justified to the degree that it makes the visible generativity signal track the unseen yield, through validation, replication, audit, and provenance, rather than estimating yield from origin, status, or ownership. Good governance lowers the bias of the estimate; extractive governance adopts a biased estimate because the bias serves the gatekeeper.

\textbf{Thesis:} Governance design for knowledge-bearing capitalism must
act where this theory shows value is created, captured, suppressed,
hidden, or lost: first conversion, access control, feedback capture,
commons maintenance, public epistemic infrastructure, and dark-capital
disclosure.

Chapter 10 translates the theory into governance design. The purpose is not to prescribe openness, compulsory licensing\index{compulsory licensing}\index{IP policy!compulsory licensing}, disclosure, or intervention by default. The purpose is to ask which governance arrangement best preserves productive knowledge-bearing stock, supports future generation, protects legitimate incentives, and avoids material\index{materiality!disclosure threshold} dark-capital exposure.

The economic problem is to govern knowledge-bearing stock so that it produces current services, future recombination, learning, renewal, and public or private productive capacity without creating greater losses through exclusion, fragility, degradation, false stock, or suppressed future generation.\index{governance design!economic problem}\index{false stock}\index{suppressed future generation}

The better KBC policy objective is to place knowledge-bearing stock under governance conditions that maximize its productive yield, renewal, recombination value, and durability, net of security, privacy, incentive, quality, and maintenance costs.\index{KBC policy objective}\index{governance design!policy objective}

\begin{quote}
\textbf{Governance-Conditioned Appreciation Principle.\index{Governance-Conditioned Appreciation Principle|textbf}} The economic value of knowledge-bearing stock is governance-conditioned. The same stock may appreciate under one actor--governance pairing and depreciate under another. Policy should therefore evaluate governance forms by their effects on productive yield, maintenance, recombination, learning, and expected knowledge loss, not by ownership form alone.
\end{quote}

The same knowledge stock can grow more valuable when used by Actor A under governance condition \(q\), but lose value when used by Actor B under governance condition \(z\). Its productive trajectory depends on a three-part relation: stock, actor capability, and governance conditions. Knowledge is therefore not self-valorizing. It does not become more valuable merely by existing. It appreciates only when the actor can access, interpret, maintain, recombine, improve, and deploy it under suitable governance.\index{knowledge-bearing stock!governance-conditioned appreciation}\index{self-valorizing knowledge!rejected}

The governance design problem draws on institutional economics, commons governance, science policy, appropriability theory, and information economics \parencite{Coase1937,Coase1960,Farrell1987,Williamson1985,Ostrom1990,Ostrom1999,HellerEisenberg1998,DasguptaDavid1994,TeecePisanoShuen1997,ShapiroVarian1999}\index{Ostrom, Elinor}. This chapter's design principle is governance fit rather than generalized enclosure or generalized openness. Section~\ref{social-evpi-and-the-measurement-commons-problem} extends that design logic to collective measurement provision, and the Technical Companion, Appendix D, supplies the K-CMM and KBC-AIE method used to decide when measurement is worth its cost.

Governance is not merely an external policy wrapper placed on already-formed knowledge capital. Governance helps determine whether knowledge-bearing stock is separable, maintainable, transferable, excludable, recombinable, measurable, and usable at all. In the residence--governance ontology introduced in Chapter~\ref{ch:knowledge-bearing-stock}, governance is descriptive rather than celebratory: it may be competent, weak, fragmented, opaque, captured, underfunded, or failing. The design task is therefore not to assume good governance, but to identify which governance arrangements preserve or impair the capital-service potential of a given stock.

The third term is capability. Residence locates knowledge; governance makes located knowledge economically actionable; capability determines whether a particular actor can convert that actionable stock into productive services. A stock can therefore be residentially available and legally accessible while still failing to yield if the actor lacks the interpretive, technical, organizational, recovery, or deployment capability required to use it. This is why cybersecurity\index{cybersecurity risk!governance design}, learning-loop access, commons maintenance, and public epistemic infrastructure are governance-design questions and dynamic-capability questions at the same time.

\section{The Governance Problem After Dark
Capital}\label{101-the-governance-problem-after-dark-capital}
\index{dark capital}\index{governance form}

Chapter 9 established that strategic enclosure produces economically
real capital effects that neither book value, market value, nor standard
intangible-asset categories can properly isolate. Dark Capital, the
aggregate of Dark Value, Dark Risk, Foregone Knowledge Capitalization,
and the Balance-Sheet Shadow, is not an accounting failure waiting
for a reclassification fix. It is a structural feature of a capitalism
in which non-rival, scalable, cumulative knowledge stocks are often governed
through institutional tools that more readily fit rival, depreciating, tangible property.

This structural mismatch creates a governance problem that is distinct
from the standard policy questions of market power, monopoly pricing,
and consumer welfare. Those questions remain important, but they are
insufficient. A firm can have modest market share, competitive pricing,
and vigorous product innovation while simultaneously enclosing the
learning loops, recombination fields, and commons dependencies on which
its own innovation rests, and on which excluded actors depend for
their future productive capacity. Much of that enclosure may not appear in
financial accounts. It may also fall outside standard antitrust screens where
price, output, or conventional foreclosure evidence remains unchanged. Much
of it can remain invisible to the investors who are pricing the stock or the
regulators who are deciding whether to intervene.

The governance response is therefore not generalized openness. It is governance fit. Some enclosure should be preserved because it finances costly generation, secures sensitive knowledge, protects quality, or creates the feedback intensity needed for rapid improvement. Some openness should be protected because it maintains the recombination field, the commons maintenance base, and public epistemic infrastructure. The governance question is which effect dominates for a material stock under identifiable conditions.

The governance problem, stated directly, is this: if dark capital is
economically real but invisible to accounting, governance cannot wait
for balance-sheet recognition. It needs decision rules for identifying
which knowledge-bearing stocks, governance transitions, feedback loops,
commons dependencies, and capability losses are material enough to
govern, and for deciding whether measurement is worth the cost before
committing to an instrument.

The measurement bridge note (K-CMM) supplies those decision rules. The notation used in the chapter is collected here so that the governance argument can remain accessible while still preserving the link to the formal apparatus.

\begin{table}[htbp]
\centering
\small
\caption{Chapter 10 Notation Reminder}
\label{tab:ch10:notation-reminder}
\begin{tabular}{@{}p{0.22\textwidth}p{0.68\textwidth}@{}}
\toprule
Symbol & Plain-language meaning \\
\midrule
\(K^E\) & Embodied knowledge stock \\
\(K^D\) & Disembodied knowledge stock \\
\(K^C\) & Commons knowledge stock \\
\(K^P\) & Public epistemic infrastructure \\
\(D_u(F_{a,t})\) & Useful diversity in the accessible recombination field \\
\(G^L\) & Learning-loop generation \\
\(C_{emin}\) & Minimum capability available to excluded actors \\
CUV & Current use value \\
\(ROV^*\) & Recombination option value \\
\(LOV^*\) & Learning-loop option value \\
\(COV^*\) & Control/governance option value \\
\(SOV^*\) & Strategic option value \\
EKL & Expected knowledge loss \\
\bottomrule
\end{tabular}
\end{table}

The mechanism examples in this chapter are anchor examples only, not case studies. They are included to make the governance mechanism concrete without asking the reader to adjudicate the full institutional history of each case:
\begin{center}
\small
\begin{tabular}{@{}p{0.34\textwidth}p{0.56\textwidth}@{}}
\toprule
Mechanism & Concise example anchor \\
\midrule
API/access-layer governance & Reddit or X/Twitter API restrictions\index{Reddit API restrictions!governance design}\index{X/Twitter API restrictions!governance design} \\
Feedback-loop capture & Cloud AI systems improving through user interactions \\
Commons maintenance & OpenSSL/Heartbleed, Log4j, and PyPI/npm dependency chains \\
Public epistemic infrastructure & NIST standards, ISO standards, public vulnerability databases, and statistical classifications\index{NIST standards!governance design}\index{ISO standards!governance design}\index{public vulnerability databases}\index{statistical classifications!governance design} \\
Competition beyond price & CUDA dependency, app-store governance\index{app stores}, cloud API dependence\index{cloud platforms}, and standards-essential patents\index{CUDA dependency!competition policy}\index{competition policy!governance design}\index{app-store governance}\index{cloud API dependence}\index{standards-essential patents} \\
Dark-capital exposure & Cybersecurity failure, platform lock-in, maintainer shock, and model degradation \\
\bottomrule
\end{tabular}
\end{center}

Three tests apply throughout this chapter.

\begin{figure}[H]
\caption[K-CMM Diagnostic Scoring Pipeline]{K-CMM Diagnostic Scoring Pipeline}
\label{fig:ch10:kcmm-scoring-pipeline}
\centering
\includegraphics[width=\textwidth]{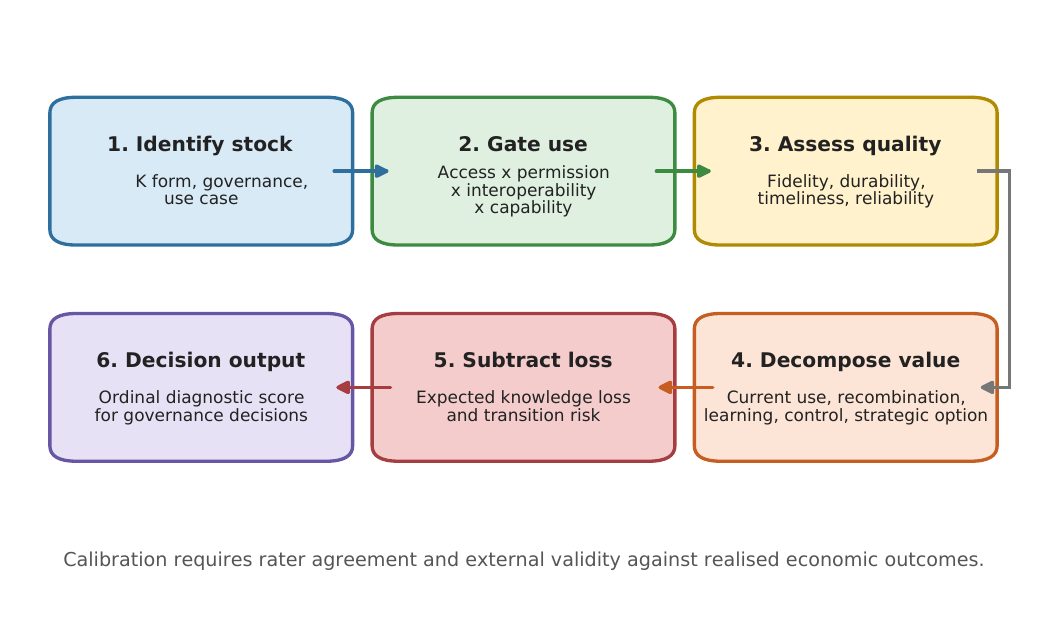}
\footnotesize\emph{Note.} K-CMM is an ordinal diagnostic\index{ordinal measurement} rubric for decision support unless and until its scores are calibrated against rater agreement and realized external outcomes.
\end{figure}

The \textbf{valuation test} asks which value component or loss category
a governance instrument targets. K-CMM\textquotesingle s decomposition
\(V^K\) = CUV + ROV* + LOV* + COV* + SOV* - EKL, forces
specificity. An access mandate, a disclosure requirement, a
commons-maintenance fund, and a compulsory licence\index{compulsory-access remedy} all address different
components of the knowledge-capital value equation. Treating them as
interchangeable is a category error.

The \textbf{materiality test} asks whether the governance transition,
enclosure event, or dark-capital exposure is important enough to govern:

\begin{equation}
M_i=Pr(\Delta Outcome_i)\times Magnitude_i\times Duration_i\times Irreversibility_i\times Breadth_i
\label{eq:ch10:m-i}
\end{equation}

The irreversibility dimension deserves particular weight. T6.6 (Recovery
Lag) shows that capability gaps do not close when enclosure ends, 
recovery from T6.5\textquotesingle s capability trap requires longer
than most competitive lifecycles allow. Where a governance transition is
likely to produce irreversible capability loss, the case for ex ante
governance is stronger than the case for remediation after the fact.

The \textbf{measurement-value test} asks whether mandatory measurement,
audit, or disclosure is worth its cost. \(\mathrm{EVI}_K\), or EVIK in prose, denotes the expected value of information about knowledge-capital exposure\index{expected value of information (EVPI)!knowledge-capital exposure}\index{EVIK|see{expected value of information (EVPI)}}. It asks whether reducing uncertainty about a governance transition, knowledge stock, access restriction, commons dependency, or dark-capital exposure would change a decision enough to justify the measurement cost. In that sense, EVIK is a value-of-information discipline rather than a new speculative metric.

\begin{equation}
\index{Knowledge Capital Index (KCI)|textbf}
KCI=\mathrm{EVI}_K-C_M=(EOL_K\times V^{\prime})-C_M
\label{eq:ch10:kci}
\end{equation}

where \(EOL_K\) is the expected opportunity loss from not measuring (the
cost of making the wrong governance decision), V′ is the value of a unit
of information relevant to that decision, and \(C_M\) is the cost of
measurement and disclosure. Where KCI is credibly positive, mandatory measurement or disclosure passes the economic screen, subject to institutional feasibility, error costs, commercial sensitivity, and the availability of less restrictive alternatives.

What Chapter 10 does not do is ask what policies would be good for
knowledge capitalism in the abstract. It asks which governance
instruments are justified by measurable dark-capital exposure, material
governance-transition risk, or positive knowledge-capital information value.
The distinction is not rhetorical. It is the difference between a
theory-grounded governance framework and a policy catalogue that happens
to cite relevant research.

\section{Governance Design
Principles}\label{102-governance-design-principles}

Before stating instruments, Chapter 10 requires a small set of design
principles. These are not normative preferences. They are structural
constraints derived from the formal mechanisms developed in Chapters 3
through 9. An instrument that violates a design principle will either
fail to address the mechanism it targets or will produce collateral
suppression of value it was not intended to touch.

\textbf{Mechanism alignment.} Govern the mechanism actually producing
value or loss, not the surface outcome. Enclosure that suppresses
\(D_u(F_{a,t})\) is not primarily a pricing problem; it is a
field-composition problem. Enclosure that captures \(G^L\) is not
primarily a data-access problem; it is a learning-loop architecture
problem. Governance aimed at price effects or data volumes without
addressing the underlying mechanism will leave the mechanism intact.

\textbf{Materiality discipline.} Govern only material transitions and
exposures. Not every knowledge-governance transition that produces dark
capital is important enough to justify governance intervention. The
materiality test exists to prevent regulatory overreach as much as to
prevent governance gaps. An instrument that applies to immaterial
dark-capital exposures will impose compliance costs on actors for whom
the knowledge-capital problem it addresses does not arise.

\textbf{Measurement proportionality.} Require measurement only where KCI
\textgreater{} 0. The expected value of information about a dark-capital
exposure must exceed the cost of obtaining that information before
measurement can be mandated. Where KCI ≤ 0, mandatory disclosure
produces compliance cost without decision-relevant benefit\index{decision-relevant measurement!disclosure}. Voluntary
disclosure, narrative reporting, or qualitative flagging is the
proportionate response.

\textbf{No-overlap discipline.} Do not regulate the same benefit stream
under multiple instruments. The no-overlap rule from
K-CMM\textquotesingle s double-counting matrix applies to governance
design as directly as it applies to valuation. An instrument protecting
recombination value (ROV*) and a separate instrument protecting
strategic option value (SOV*) that both operate on the same future
benefit stream are not additive governance protections; they are
redundant burdens. Chapter 10 instruments should be mapped against the
K-CMM value components and tested for overlap before they are
recommended.

\textbf{Commons and public-infrastructure preservation.} Protect \(K^C\)
and \(K^P\) where they sustain broad \(D_u(F_{a,t})\). Commons knowledge stock is collectively maintained productive knowledge, and public epistemic infrastructure is knowledge-bearing infrastructure generated and maintained through standards bodies\index{standards maintenance}, courts, statistical agencies\index{standards bodies}\index{courts}\index{statistical agencies}\index{Statistics Canada}, public research systems\index{public research funding}, professional bodies, and measurement institutions. Both are inputs to the recombination
fields of actors who have no direct relationship with the enclosing
firm. Their depletion does not appear in the firm\textquotesingle s
accounts, is not visible to market participants pricing the
firm\textquotesingle s stock, and does not trigger standard regulatory
review. The governance framework must include instruments specifically
designed to identify and protect material commons and
public-infrastructure stocks, even where no private actor has a standing
interest in their maintenance. The relevant failure is the Maintenance-Gap Condition of Chapter~5, reinvested yield falling below depreciation (\(s^x_t Y^x_t < \delta^x_t K^x_t\)); these instruments therefore target the reinvestment shortfall, not merely access.

\textbf{Capability preservation.} Treat \(\widetilde{C}_{a}\), \(K^E\), and \(K^I\) as
governance-relevant, not merely managerial. The KBC framework shows that
organizational capability is a stock that accumulates through \(G^R\) and
\(G^L\) and decays through enclosure, capability loss, and learning-loop
denial. T6.5 (Capability Trap) and T6.6 (Recovery Lag) show that
capability loss can be irreversible on economically relevant timescales.
Governance instruments that touch knowledge-governance transitions must
consider their effects on capability stocks, not only on information
flows or product market competition.

\textbf{Property-rights boundary.} Do not assume that assigning rights and allowing bargaining solves the knowledge-governance problem. That solution requires identifiable parties, observable values, contractible uses, stable coalitions, and future recombination values that can be priced before they are discovered. Knowledge-bearing stock often violates these conditions. Future users may be unknown, future combinations may be unimagined, rights may be fragmented, and bargaining may require access to the very field that enclosure restricts. Property rights remain necessary governance instruments, but they are not sufficient governance theory.

Jones and Tonetti's data-property-rights analysis makes this boundary operational for non-rival data. If the same data can be used by many actors at once, the governance question is not simply who owns the data, but which access-governance arrangement permits productive use while controlling privacy, quality, bargaining, creative-destruction, and underuse risks\index{ownership versus access governance} \parencite{JonesTonetti2020}. Table~\ref{tab:ch10:data-property-rights-governance-forms} translates that point into KBC's governance-fit language.

\begin{table}[htbp]
\centering
\small
\caption{Data Property-Rights Governance Forms as Governance Choices}
\label{tab:ch10:data-property-rights-governance-forms}
\begin{tabular}{@{}p{0.22\textwidth}p{0.26\textwidth}p{0.23\textwidth}p{0.23\textwidth}@{}}
\toprule
Data governance form & KBC interpretation & Main benefit & Main risk \\
\midrule
Firm ownership & Enclosed firm-controlled stock & Incentives, investment, integration & Hoarding, privacy loss, underuse \\
Consumer ownership & User-controlled access rights & Broader licensing, privacy balancing & Transaction costs, weak bargaining \\
Public or commons data & Shared knowledge infrastructure & Recombination, diffusion & Maintenance, quality, governance \\
Ban or restriction & Privacy-preserving exclusion & Harm reduction & Underuse of non-rival input \\
\bottomrule
\end{tabular}
\end{table}

The table is not a ranking. It is a governance-fit diagnostic. Firm ownership may be justified where investment, quality control, or integration costs dominate. Consumer ownership may improve licensing breadth and privacy balancing where users can transact effectively. Public or commons data may maximise recombination where maintenance and quality governance are credible. Restrictions may be necessary where harm, privacy, or security risks dominate. The KBC prescription is therefore not generalized sharing. It is to match property-rights design to the knowledge stock's non-rivality, sensitivity, complementarity, maintenance burden, and recombination value.

\textbf{Appropriability--friction distinction.} Governance must distinguish temporary appropriability that supports productive service from enclosure drift that converts temporary service advantage into durable friction rent. Temporary appropriability may be justified where it finances generation, protects quality, preserves disclosure incentives, or enables costly deployment. Friction rent is different: it appears when the governance position itself becomes the source of payment, delay, dependency, or exclusion without a corresponding increase in productive service. This distinction should be tested through the same K-CMM discipline used elsewhere in this chapter: identify the component affected, specify the materiality trigger, and ask whether the measurement-value problem\index{measurement-value test!governance design}\index{measurement-value problem} justifies intervention.

A compact boundary condition can be stated as:
\[
B_{\mathrm{Coase}}(K) \rightarrow \mathrm{Efficiency}
\quad \text{only if} \quad
U_f,\ V_{\mathrm{rec}},\ R,\ D_u,\ \Omega
\quad \text{are sufficiently observable and contractible.}
\]
This is a future-work placeholder, not a completed model. It records the conditions a later property-rights theorem would need to specify: future users, recombination value, field access, useful diversity, and dark-capital exposure must be visible enough to contract over.

\textbf{Governance-fit discipline.} Choose instruments by governance fit, not by ideological preference for openness or enclosure. A governance form fits when it preserves sufficient incentive, access, quality control, maintenance, reach, useful diversity, and learning feedback for the relevant stock. In diagnostic form:
\[
\mathrm{GFIT}_i=f(I_i,A_i,Q_i,M_i,R_i,D_{u,i},L_i),
\]
where \(I\) is incentive adequacy, \(A\) access adequacy, \(Q\) quality control, \(M\) maintenance, \(R\) reach, \(D_u\) useful diversity, and \(L\) learning feedback. This expression is a diagnostic placeholder and governance checklist, not a calibrated welfare function.

\subsection{Governance as payoff-function
modification}\label{sec:ch10:governance-as-payoff-function-modification}\index{payoff function}

Chapter 8\textquotesingle s strategic-capture equilibrium makes the
governance task more precise. Governance design does not merely prohibit
enclosure after the fact. It changes the game by modifying the private
payoff function that makes strategic enclosure individually rational.
Each instrument should alter at least one term in the
incumbent\textquotesingle s strategic-capture payoff:

\[
\begin{aligned}
\pi_i(s_i,s_{-i})={}&\mathrm{PrivateCapturedYield}_i
+\mathrm{KnowledgeRent}_i+\mathrm{FeedbackCaptureGain}_i \\
&-C_{\mathrm{enforcement},i}-C_{\mathrm{alloc},i}
-C^{\mathrm{self}}_{\mathrm{rec},i}-EKL_i
\end{aligned}
\]

A well-designed instrument may reduce \(KnowledgeRent_i\), limit
\(FeedbackCaptureGain_i\), raise \(C_{enforcement,i}\),
increase the private salience of \(C^{self}_{rec,i}\), reduce the
attractiveness of \(E_{alloc}\), or convert externalized social
costs into decision costs that the actor must internalize. The purpose
is not to eliminate all enclosure. Some enclosure may be necessary to
induce costly knowledge generation, validation, maintenance, and
deployment. The purpose is to move the equilibrium from strategic
over-enclosure toward efficient enclosure, preserving productive
incentives while reducing suppressed recombination, learning-loop
concentration, capability loss, and foregone knowledge capitalization.

This payoff-function view disciplines the instruments that follow.
Access rules target field contraction and \texttt{ROV*}; feedback-loop
rules target \texttt{LOV*} and capability divergence; commons and
public-infrastructure instruments preserve the field conditions on which
\(G^R\) depends; dark-capital disclosure makes \texttt{EKL},
\(\Omega_i\), and governance-position risk visible enough to affect
investment, regulatory, and managerial decisions. Governance is
justified when the expected correction to the strategic-capture
equilibrium passes the materiality and measurement-value tests.

\subsection{Skill Governance}\label{sec:ch10:skill-governance}

Domain-denominated embodied knowledge capital, \(K^E_d\), is governed
by institutions, not merely by spot labour markets. Education,
apprenticeship, credentialing, professional licensure, certification,
malpractice and liability rules, immigration policy, labour mobility,
non-competes, training pipelines, continuing education, and public
research infrastructure all shape the supply, quality, mobility, and
renewal of domain skill. These governance arrangements determine whether a society can
form, maintain, and redeploy cardiac-surgical expertise, cybersecurity
judgement, engineering intuition, legal interpretation, scientific
capability, and other forms of specialized embodied knowledge capital.

Skill-governance-form analysis therefore extends KBC governance beyond IP,
platform access, and commons maintenance. If \(K^E_d\) is a productive
stock rather than merely labour time, then institutions governing skill
formation are capital-governance institutions. They affect the rate at
which embodied domain stock is generated, the conditions under which it
can be converted into \(K^D\) or \(K^I\), the degree to which skilled
workers can move across firms and jurisdictions, and the resilience of
professional domains after shocks. Overly restrictive skill-governance arrangements may
preserve quality while narrowing entry and mobility. Underdeveloped
skill-governance arrangements may widen entry while weakening trust, reliability, and
capability depth. The governance problem is governance fit: preserving
quality and accountability without suppressing the formation, movement,
and recombination of domain-specific embodied knowledge capital.

\section{First-Conversion
Governance}\label{103-first-conversion-governance}
\index{first conversion}

\subsection{The mechanism}\label{the-mechanism}

The first formal event in the knowledge-capital lifecycle is conversion:
new knowledge (\ensuremath{\dot{K}}) receives an initial governance assignment that determines
whether it enters \(K^E\) (embodied), \(K^D\) (disembodied), \(K^I\)
(institutionalized), \(K^C\) (commons knowledge stock), or \(K^P\) (public epistemic
infrastructure). The Conditional Separability Axiom states that this
assignment is an institutional achievement, not a natural property of
the knowledge. First-conversion governance targets the rules that
determine initial governance assignment.

First conversion is the earliest point at which governance can affect
the long-run trajectory. A knowledge stock assigned to \(K^D_{private}\)
at the point of creation faces a different recombination field, a
different access structure, and a different set of \(G^L\) opportunities
than the same knowledge assigned to \(K^C\) or \(K^P\). The costs of
mis-assignment compound over time through the knowledge-capital state
equation. Correcting an initial governance assignment through litigation,
compulsory licensing, or commons reconstruction is expensive and often
incomplete.

\subsection[Knowledge-capital measurement mapping]{K-CMM mapping}\label{k-cmm-mapping}\index{Knowledge-Capital Measurement Model (K-CMM)}

The first valuation target at first conversion is COV*: the control rent that accrues to the actor who captures the initial governance assignment. The second is ROV*: the recombination option value that is either preserved or foreclosed by that assignment. The materiality
dimensions most relevant here are control materiality (who gains the
right to exclude, licence, or govern the stock) and public/commons
materiality (whether a stock that entered production through public
infrastructure, commons inputs, or social knowledge is appropriated
without adequate return).

\subsection{Instruments}\label{instruments}

\textbf{Public-funding open-access terms.} \emph{Intervention status: coordination instrument.} Knowledge produced using
public research grants, public data, public computational
infrastructure, or public-health systems enters production through
\(K^P\) inputs. First-conversion governance should create a rebuttable presumption of public, commons, or access-preserving treatment where public inputs are material. Private assignment may still be justified where it is necessary for commercialization, quality control, security, or costly deployment, and where the resulting COV* premium does not materially foreclose downstream recombination. The materiality test applies: where public-funding\index{public funding conditions}
share is low or the knowledge stock is highly specialized, the
governance case for open-access terms is weaker. Where public-funding
share is high and the knowledge stock is foundational for broad
downstream recombination, the case is strong. The less restrictive
alternative test also applies: the question is whether open-access terms
are needed to preserve recombination and public return, or whether a
narrower licence, embargo period, access tier, or public-use covenant
would preserve incentives while reducing material field loss.

\textbf{Employee and contractor invention-assignment reform.} \emph{Intervention status: baseline governance practice.} The
standard employment contract assigns all workplace invention to the
employer as a condition of employment. This converts \(K^E\), embodied
in the employee, into \(K^D_{private}\) at the moment of creation,
often without a separate transaction pricing the specific knowledge conversion. Where the invented knowledge builds on
\(K^C\) (professional community standards), \(K^P\) (published research),
or \(K^I\) that the employee brought to the role, broad assignment can convert commons and public value into private control, especially where the assignment reaches beyond employer-funded, role-specific invention.
Governance should distinguish genuine employer-sponsored invention from
conversion of pre-existing or commons-derived knowledge, and require
disclosure of the knowledge inputs underlying invention assignments.

\textbf{AI training corpus disclosure.} \emph{Intervention status: disclosure instrument.} AI model training converts
large-scale \(K^C\) (open-source code, published text, public data) and
\(K^P\) (academic literature, public records, research corpora) into
\(K^D_{private}\) (model weights, proprietary training pipelines). This
simultaneous conversion is currently invisible: there is no mechanism
for identifying which commons knowledge stock and public epistemic infrastructure inputs entered model
training, at what scale, or with what effect on their downstream
usability. First-conversion governance should require disclosure of
training corpus composition where the resulting model is commercially
deployed and the corpus includes material \(K^C\) or \(K^P\) inputs. The
measurement-value test applies: where the training corpus is purely
proprietary or the model is research-only, disclosure imposes cost
without commensurate governance benefit. The less restrictive alternative
test asks whether aggregate corpus categories, third-party audit,
confidential supervisory disclosure, or threshold-based reporting would
produce sufficient governance information without exposing legitimate
proprietary methods.

\textbf{Data-use provenance rules.} \emph{Intervention status: disclosure instrument.} User-generated data, behavioural
data, and interaction data are routinely converted from \(K^C\)
(distributed social knowledge) into \(K^D_{private}\) (proprietary
behavioural models), often without meaningful or specific consent to the knowledge-capital conversion.
Existing data-protection frameworks govern processing legality and
personal privacy but do not address the knowledge-capital conversion
event. Provenance rules should require disclosure of how user-generated
knowledge inputs are converted into proprietary knowledge stocks, and
whether those stocks are used to generate COV* (platform position,
exclusion leverage, licensing power) beyond their stated service
function.

\section{Access-Layer
Governance}\label{104-access-layer-governance}

\subsection{The mechanism}\label{the-mechanism-1}

Access-layer governance targets the formal channel through which
cognitive enclosure suppresses recombination generation:

\[G^R_{a,t+1}=\lambda_a\cdot R_{a,t}^{\eta}\cdot D_u(F_{a,t})^{\mu}\]

Access restrictions reduce \(D_u(F_{a,t})\), the useful diversity of
the accessible recombination field, for excluded actors. The
suppression ratio is \((\sigma_a^R)^{\eta}\cdot(D_u(\pi_1)/D_u(\pi_0))^{\mu}\)
\textless{} 1. Proposition C (Generative Suppression) shows that
enclosure is a generative event, not merely a distributional one: it
does not only redistribute who captures value from existing
recombination; it reduces \(G^R\) for excluded actors, and it reduces the
gated aggregate generation rate only where the weakly specified
\(\Phi(\cdot)\) aggregator survives calibration.

Access-layer governance therefore addresses not access as a consumer
right but access as a productive input. The relevant question is not
whether excluded actors can purchase or use the knowledge product. It is
whether they can incorporate the knowledge stock into their own
recombination field, combine it with other stocks, and generate new
productive knowledge from the combination.

\subsection[Knowledge-capital measurement mapping]{K-CMM mapping}\label{k-cmm-mapping-1}

Primary valuation targets: ROV* and CUV. When access is blocked,
excluded actors cannot count the inaccessible stock in their
recombination field. ROV* can fall sharply, or collapse where no effective substitutes exist, for those actors even when the
stock\textquotesingle s productive potential is high. The GATE
coefficient \((A\times P\times I\times C)^{1/4}\) shows that access alone is
insufficient: permission, interoperability, and capability conditions
must also hold. Governance aimed only at legal access while leaving
interoperability and capability conditions unaddressed will produce
formal compliance without productive access.

Relevant materiality dimensions: recombination materiality (does access
restriction narrow \(D_u(F_{a,t})\) for a material population of
actors?) and capability materiality (does access restriction prevent
excluded actors from developing \(\widetilde{C}_{a}\)?).

\subsection{Instruments}\label{instruments-1}

\textbf{Interoperability mandates.} \emph{Intervention status: coordination instrument.} Where a knowledge stock or platform
holds significant COV* because it is the connection point through which
multiple actors must route their productive activities, interoperability
mandates reduce the exclusion premium without necessarily reducing the
productive value of the stock itself. The governance case is strongest
where the interoperability barrier is not technically necessary (it is
an architectural enclosure choice rather than a genuine incompatibility)
and where the materiality test shows that field contraction from the
barrier is broad and durable. The instrument must specify that
interoperability means productive interoperability, the ability to
recombine the stock\textquotesingle s outputs, not merely format
compliance. The less restrictive alternative test asks whether notice,
technical documentation, export interfaces, staged compatibility, or
standardized connectors would preserve recombination access without a
broader mandate that suppresses legitimate design or security choices.

\textbf{API access\index{API access} rules.} \emph{Intervention status: corrective instrument.} Platform-controlled APIs are the primary
access-layer mechanism through which COV* is extracted in digital
knowledge markets. API governance should distinguish between reasonable
product design (the platform controls its own \(K^D\)) and strategic
field enclosure (the platform controls access to knowledge generated by
users, third-party developers, or commons inputs). Where an API change
reduces \(D_u(F_{a,t})\) for a material population of dependent actors, foreclosing identifiable recombination paths, and the KCI test
shows the governance cost is justified, access obligations may be
warranted where less restrictive alternatives cannot preserve material dependent capability or recombination access.

\textbf{Compulsory licensing (calibrated to knowledge-capital effects).} \emph{Intervention status: exceptional instrument.}
Standard compulsory licensing triggers, typically abuse of dominant
position or public-health emergency, do not reach cases where
knowledge-capital enclosure is material but market power in the price
sense is absent. This is a proposed knowledge-capital extension to existing compulsory-licensing logic, not a claim that current law already recognizes this trigger. A knowledge-bearing capitalism governance framework
could identify additional compulsory licensing triggers: where \(M_i\)
passes the materiality threshold\index{materiality threshold!governance design}\index{materiality threshold} on recombination materiality or
capability materiality dimensions, and where COV* is being extracted
through access restriction rather than productive use. This is a
narrower instrument than many advocates propose: it applies only where
the materiality and measurement-value tests are satisfied. The less
restrictive alternative test is especially demanding here: compulsory
licensing is justified only when disclosure, interoperability, time-limited
access, fair-use safe harbours, or other narrower governance arrangements
cannot preserve incentives while reducing material recombination loss,
feedback exclusion, capability decay, or dark-capital exposure.

\textbf{Data portability (with the capability limit acknowledged).} \emph{Intervention status: corrective instrument.} Data
portability rules require platforms to allow users to move their data to
competing services. This is useful for reducing access-layer lock-in at
the consumer level. Its limit in a knowledge-capital context is that the
productive value at stake is not the user\textquotesingle s raw data but
the \(K^D\) generated by processing that data, the trained models, the
behavioural patterns, the inferred preferences. Data portability without
portability of derived knowledge stocks and associated \(G^L\) feedback
channels leaves the core capability gap in place. Governance design
should acknowledge this limit and avoid treating data portability as
equivalent to access-layer reform for knowledge-capital purposes.

\section{Feedback-Loop
Governance}\label{105-feedback-loop-governance}

\subsection{The mechanism}\label{the-mechanism-2}

Feedback-loop governance addresses the internal counterpart to Chapter
10.4\textquotesingle s external suppression. Where cognitive enclosure
reduces excluded actors\textquotesingle{} recombination fields,
feedback-loop enclosure enriches the incumbent\textquotesingle s dynamic
capability by capturing \(G^L\) exclusively:

\[
G^L_{a,t}=\alpha_a\cdot Dp_{a,t}\cdot F_{dep,a,t}\cdot\phi_a
\]

Enclosed deployment produces exclusive feedback, which enters the
capability state equation:

\[\widetilde{C}_{a,t+1}=(1-\delta_C)\widetilde{C}_{a,t}+\gamma G^R_{a,t}+\ell G^L_{a,t}\]

Here \(\delta_C\) is capability decay, \(\gamma\) converts recombination generation into capability accumulation, and \(\ell\) converts learning-loop generation into capability accumulation. Proposition D (Feedback-Enclosure) shows the trajectory consequence:
exclusive \(G^L\) accumulates in the incumbent\textquotesingle s \(\widetilde{C}_{a}\)
while excluded actors\textquotesingle{} \(\widetilde{C}\) converges to \(\widetilde{C}_{min}\). T6.2
(Capability Divergence) shows the gap grows strictly during the
enclosure period. T6.5 (Capability Trap) shows convergence to \(\widetilde{C}_{min}\)
when both \(G^L\) and \(G^R\) are eliminated. T6.6 (Recovery Lag) shows
the gap does not close immediately when enclosure ends.

Chapter 7 identified what Chapter 10 must now govern: not the feedback
data itself, but the feedback architecture, the structural
arrangement that makes certain deployments generate \(G^L\) for the
incumbent while denying that learning signal to excluded actors,
competitors, researchers, and the public.

\subsection[Knowledge-capital measurement mapping]{K-CMM mapping}\label{k-cmm-mapping-2}

Primary valuation target: LOV*, the learning-loop option value that
accumulates to the actor with exclusive \(G^L\) access and is denied to
actors without it. EKL\textquotesingle s capability-loss component
captures the loss on the excluded side: capability that would have
developed through feedback access is destroyed.

Relevant materiality dimensions: capability materiality (does the
feedback architecture change excluded actors\textquotesingle{} ability
to develop and maintain knowledge-bearing capability?) and
distributional materiality (does exclusive \(G^L\) produce a capability
gap that maps onto differential market power, wage distribution, or
public-service quality?).

\textbf{LLM access and synthetic recombination-field governance.} LLM access is not uniformly valuable across users. In formal terms, an LLM is a \(K^D\) stock that can function as a synthetic recombination-field interface, expanding the actor's effective field through \(\Psi_{\mathrm{LLM}}\). But the realized value of that field depends on \(K^E\), interpretive judgement, and epistemic validation. Model outputs require domain knowledge for query formation, contextual interpretation, error detection, and recombination. The marginal value of access is therefore highest for high-\(K^E\) actors: researchers, engineers, clinicians, analysts, builders, and other experienced practitioners who can convert synthetic field breadth into reliable knowledge generation.

This changes the governance analysis. Enclosing LLM access through paywalls, API restrictions, closed weights, contractual limitations, or platform dependency does not merely reduce general access. It suppresses recombination most severely among the actors capable of generating the greatest surplus from access. The welfare cost of LLM enclosure should therefore be weighted by the \(K^E\) distribution of affected actors, not by headcount alone. Restricting a small group of high-\(K^E\) actors may destroy more recombination surplus than restricting a much larger group of users who lack the capability to validate and productively recombine model outputs.

This is where Chapter 10 must be sharpest. Governance cannot stop at
training data access. The feedback loop is forward-looking, it is
generated by deployment, not by initial training. Governance that
addresses training corpora while leaving feedback architectures
ungoverned allows the most important capability-accumulation mechanism
to proceed unexamined.

\subsection{Instruments}\label{instruments-2}

\textbf{Feedback-use disclosure.} \emph{Intervention status: disclosure instrument.} Many interactions between users and
knowledge-bearing platforms, AI assistants, recommendation systems,
search interfaces, professional tools, generate feedback signals that
the platform uses to improve its knowledge stocks and capability. Users
typically do not know which interactions are being converted into \(G^L\)
inputs, at what frequency, or with what productive consequence for the
platform. Disclosure requirements should identify, for commercially
deployed knowledge-bearing systems: whether interaction feedback is used
for model improvement; which feedback signals are retained; and whether
feedback-derived improvements are restricted to the platform or shared
with a broader knowledge commons. Measurement-value test: KCI may be positive for major AI platforms where the feedback-to-capability pathway
is both material and currently invisible.

\textbf{Feedback-sharing obligations (conditional on materiality and
capability trap evidence).} \emph{Intervention status: exceptional instrument.} Where the materiality test shows that
exclusive \(G^L\) is producing a capability trap, excluded actors
converging toward \(\widetilde{C}_{min}\) with irreversible divergence from the incumbent, governance may require feedback sharing or derived-model disclosure.
This is a high-threshold instrument. It applies only where the
materiality test is satisfied on both capability materiality and
irreversibility dimensions, and where the measurement-value test
confirms that the information gap around feedback architecture is worth
closing. Mandatory feedback sharing applied to low-materiality cases
will impose cost without commensurate benefit and may suppress \(G^L\)
investment incentives where those incentives are socially valuable. The
less restrictive alternative test asks whether feedback-use disclosure\index{feedback-use rights},
model improvement audit trails, researcher access, benchmarking, or
aggregated feedback reporting would reduce the capability trap without
forcing direct feedback sharing.

\textbf{Model improvement audit trails.} \emph{Intervention status: disclosure instrument.} Where AI and machine learning
systems are deployed in high-stakes domains, medical diagnosis,
credit assessment, legal analysis, public administration, governance may require organizations to maintain traceable records of how
their models have changed through deployment feedback, which user
populations contributed feedback, and what capability changes resulted.
This serves two functions: it makes LOV* accumulation visible to
governance actors, and it creates the empirical record needed to test
whether capability divergence is occurring at the rate T6.2 predicts.

\textbf{User and contributor attribution rules.} \emph{Intervention status: corrective instrument.} Where a
knowledge-bearing platform builds productive capability from
user-generated knowledge inputs, question-answer pairs, creative
outputs, professional judgements, scientific annotations, the
conversion of distributed human knowledge into platform-owned \(K^D\) is
a first-conversion event with feedback-loop dimensions. Attribution
rules should require transparency about which user contributions are
generating \(G^L\), and should evaluate whether attribution, opt-out, compensation, or governance participation\index{attribution}\index{opt-out}\index{compensation}\index{governance participation} mechanisms are warranted for capability they have helped generate. The first-order requirement is transparency\index{transparency!AI/data governance}\index{data governance}\index{compensation!requires further justification}\index{governance participation!requires further justification}; compensation or governance participation requires additional legal and empirical justification. The
governance design should distinguish genuine platform investment
(infrastructure, training, curation) from appropriation of pre-existing
distributed knowledge capital.

\section{Commons and Public Infrastructure
Governance}\label{106-commons-and-public-infrastructure-governance}

This section uses the two public-facing stock categories narrowly. Commons knowledge stock is collectively maintained productive knowledge; it is not merely free access or unowned information. Public epistemic infrastructure is knowledge-bearing infrastructure generated and maintained through standards bodies, courts, statistical agencies, public research systems, professional bodies, and measurement institutions; it is not merely public information.

\subsection{The mechanism}\label{the-mechanism-3}

\(K^C\) (commons knowledge stock) and \(K^P\) (public epistemic infrastructure)
are not background conditions. They are productive inputs that determine
\(\widetilde{C}_{min}\), the minimum capability floor below which capability-trapped
actors converge, and that set the baseline level of \(D_u(F_{a,t})\)
available to actors without access to private knowledge stocks. T5.G3
shows that lower \(K^P\) raises \(T^*\), the socially optimal enclosure
duration, by reducing the recombination potential of publicly
available knowledge. T6.G3 shows that \(K^P\) level enters \(\widetilde{C}_{min}\): when
public epistemic infrastructure is stronger, the capability floor for
excluded actors is higher.

Commons and public-infrastructure governance therefore has consequences
for many downstream mechanisms, especially recombination, capability floors, and enclosure duration. A robust \(K^P\) can reduce the socially necessary enclosure duration \(T^*\), increase \(D_u(F_{a,t})\) for actors with field access, and provide a higher \(\widetilde{C}_{min}\) floor beneath which capability traps cannot force excluded actors. In plain terms, stronger public epistemic infrastructure reduces how long private enclosure must be tolerated to support creation incentives. Its depletion has the opposite effect. This is
not a side condition; it is a structural parameter.

\subsection[Governance viability test: Ostrom conditions for commons survival]{Governance viability test: Ostrom conditions for \(K^C\) survival}\index{commons governance!Ostrom conditions}\index{commons survival}\label{governance-viability-test-ostrom-conditions-for-kc-survival}

\textcite{Ostrom1999} identifies the minimum institutional conditions under
which commons survive without central direction or privatization. The
finding is not that commons naturally succeed or naturally fail, but
that survival depends on identifiable governance features that can be
present or absent. Adapted to the knowledge commons, these conditions
function as a viability test: \(K^C\) is at structural risk of depletion
when one or more conditions are absent, regardless of whether the
commons is legally open, technically accessible, or formally documented.
Governance instruments that protect \(K^C\) must verify these conditions
are present, not merely that the commons exists.

\begin{table}[htbp]
\centering
\small
\begin{tabular}{@{}p{0.28\textwidth}p{0.62\textwidth}@{}}
\toprule
Adapted Ostrom condition & KBC reading for knowledge commons \\
\midrule
Contribution boundaries & Who may contribute, maintain, modify rules, and resolve disputes must be identifiable. \\
Capability-stock monitoring & Governance must track maintainer capability, not only visible code, data, standards, or documents. \\
Graduated sanctions & Extraction, non-maintenance, or enclosure of commons outputs is sanctionable only where licence terms, materiality thresholds, or explicit governance obligations are violated. \\
Conflict resolution & Disputes over contribution, maintenance, licence compliance, or governance changes need low-friction resolution paths. \\
Nested governance & Foundational commons require project-level and ecosystem-level governance because dependencies extend beyond a single project. \\
Maintainer protection & Governance should examine whether employment or platform restrictions impair pre-existing commons maintenance roles. \\
\bottomrule
\end{tabular}
\caption{Ostrom conditions adapted to commons knowledge stock}
\label{tab:ch10:ostrom-kc-conditions}
\end{table}

The practical implication is polycentric governance. Large knowledge commons rarely have one natural centre of authority, one complete information set, or one optimal rule-maker. Project maintainers, professional bodies, funders, firms, public agencies, and users each see different failure modes. A polycentric governance arrangement can therefore outperform both centralized public control and unmanaged openness when it distributes monitoring, sanctions, funding, and conflict resolution across the actors closest to the relevant capability stock.

The same section must also guard against the opposite failure: anticommons blockage\index{anticommons blockage}. A commons can be depleted by too little governance, but downstream generation can also be blocked by too many fragmented exclusion claims. Commons and public-infrastructure governance must therefore prevent both commons depletion and anticommons underuse.

\textbf{Clearly defined contribution boundaries.} The commons must have
identifiable rules about who can contribute, who has maintenance
authority, who can modify governing rules, and who resolves disputes.
Absence of contribution boundaries produces governance failure without
licence failure: the commons is technically open but governed by no one.
This is distinct from access boundaries (which actors can use the
commons); contribution boundaries determine whether the commons has an
institutional identity capable of defending itself against depletion.

\textbf{Monitoring of the capability stock, not merely the knowledge
stock.} Ostrom\textquotesingle s monitoring principle tracks the
resource stock level. Applied to the knowledge commons, monitoring must
track both the knowledge stock, is the code maintained? are the
datasets current? are the standards up to date?, and the capability
stock: do the actors responsible for maintenance retain the \(K^E\)
capacity to execute that maintenance? A commons whose licence is valid
but whose maintainers have lost current expertise is in depletion
without showing formal signs of it. Commons governance that monitors
only artefacts and not capability will miss the critical depletion
pathway.

\textbf{Graduated sanctions for governance-destructive behaviour.}
Actors who extract \(K^C\) without contributing, who fork without
maintaining, or who use commons outputs to generate enclosed \(K^D\)
without returning productive capacity should face proportionate and
escalating governance responses. In the knowledge commons context this
means mandatory attribution, contribution requirements for commercial
users above a materiality threshold, and enforceable licence terms.
Without graduated sanctions, the dominant strategy for commercial actors
is extraction until depletion, because the cost of
governance-destructive behaviour falls on the commons and not on the
extracting actor.

\textbf{Conflict resolution mechanisms.} The commons must have
accessible mechanisms for resolving disputes over contribution rights,
maintenance priorities, licence compliance, and governance changes.
Without conflict resolution, governance disputes become exit events:
maintainers leave, forks proliferate, and the commons fragments into
incompatible versions each with declining maintenance capacity.
Fragmentation is a depletion pathway as damaging as extraction: it
dissipates the accumulated capability that a unified governance
structure embodies.

\textbf{Nested governance for ecosystem-scale commons.} Commons whose
outputs are inputs to other commons, or that sustain wide recombination
fields across industries, require governance structures nested across
scales. Project-level governance may not identify systemic depletion
threats that are visible only at the ecosystem level. Governance of
foundational \(K^C\), core cryptographic libraries, foundational
scientific datasets, standards that underpin whole industry stacks, 
must be nested within ecosystem-level governance capable of recognizing
and responding to dependencies that individual projects may not see.

\textbf{Protection against maintainer extraction.} This is
KBC\textquotesingle s extension beyond Ostrom\textquotesingle s
natural-resource framework. Ostrom\textquotesingle s conditions prevent
commons depletion through overuse. They do not prevent depletion through
labour-market extraction of the \(K^E\) capacity required to govern the
commons. When an incumbent hires away key maintainers, none of the first
five conditions prevents the depletion: the licence is valid, the
governance rules are on paper, the monitoring norms exist. What
collapses is the maintainer\textquotesingle s \(K^E\) capacity to execute
those norms. This depletion pathway is more durable than overuse
depletion: fish stocks regenerate when harvesting stops; governance
capability does not automatically reconstitute when employment contracts
expire. Governance of \(K^C\) must therefore treat commons maintenance
\(K^E\) as a protected form of productive capacity. Employment contracts
should not permit unrestricted exclusion of former employees from
open-source governance roles they held prior to employment. Firms with
material dependence on \(K^C\) outputs should face contributor-governance
requirements that prevent strategic depletion through talent extraction
alongside requirements that prevent strategic depletion through
enclosure.

\subsection[Knowledge-capital measurement mapping]{K-CMM mapping}\label{k-cmm-mapping-3}

Primary valuation targets: CUV (commons depletion reduces current use
value for all actors who depend on the commons), ROV* (a depleted
commons narrows recombination fields), and avoided EKL (commons
maintenance reduces the expected knowledge loss from maintainer
exhaustion, security degradation, and fragmentation). The public/commons
materiality dimension of the materiality test is the primary governance
trigger.

\subsection{Instruments}\label{instruments-3}

\textbf{Commons maintenance funding.} \emph{Intervention status: coordination instrument.} Open-source software libraries,
shared research datasets, professional knowledge standards, and public
scientific corpora all exhibit the maintainer-exhaustion pattern
identified in §6.6 of the KBC framework: commons depletion as enclosure
without ownership, where the stock degrades not because anyone claims it
but because no one funds its maintenance. Governance should identify
\(K^C\) stocks that pass the materiality test on public/commons
materiality and recombination materiality, and provide stable
maintenance funding. The K-CMM public case worked example (§17 of the
technical model) provides the decision structure: compare the EKL
avoided by funding against the funding cost, add ROV* from preserved
recombination access, and apply the measurement-value test to determine
whether further audit of the commons stock\textquotesingle s dependency
network is justified before committing funds.

\textbf{Public epistemic infrastructure investment.} \emph{Intervention status: baseline governance practice.} \(K^P\) includes
foundational scientific knowledge, public research literature, open
educational resources, public health data, legal databases, geographic
and environmental data, and the institutional capacity to produce,
maintain, and distribute these stocks. Their productive value is diffuse: it enters \(D_u(F_{a,t})\) for a wide population of actors, and
therefore is often underprovided or undervalued by private markets. Governance
frameworks should treat \(K^P\) investment not as cultural expenditure
but as knowledge-capital infrastructure investment with potentially measurable
returns in recombination field breadth and capability floor elevation.
The materiality test should be applied at the category level: which
classes of public epistemic infrastructure, if allowed to degrade, would
produce irreversible falls in \(D_u(F_{a,t})\) for a material population
of actors?

\textbf{Open-standard protection.} \emph{Intervention status: coordination instrument.} Technical standards, professional
standards, and interoperability standards are institutionalized
knowledge (\(K^I\)) that function as commons when their governance
remains open. Standards capture, the process by which a dominant
actor converts an open standard into a proprietary control mechanism, 
reduces \(D_u(F_{a,t})\) for actors who cannot afford access to the new
private standard or who cannot participate in its revision. Governance
should identify standards whose capture would pass the materiality test
on recombination materiality and control materiality dimensions, and
maintain open governance requirements for those standards proportionate
to the breadth of their field-access function.

\textbf{Contributor governance requirements.} \emph{Intervention status: coordination instrument.} Platforms that extract
value from \(K^C\) inputs, by training on open-source code, by
incorporating community-generated knowledge, by building on public
research, should face contributor governance\index{contributor governance} requirements
proportionate to the COV* they derive from that extraction. This is not
a royalty model; it is a governance-participation model. Platforms with
material dependence on commons inputs should be required to participate
in the governance of those commons, contribute to their maintenance, and
refrain from first-conversion moves that appropriate commons knowledge
into exclusive \(K^D\) without returning productive capacity to the
commons.

\section{Social EVPI and the Measurement Commons Problem}\index{social EVPI}\index{measurement commons problem}\index{public goods!measurement}\label{social-evpi-and-the-measurement-commons-problem}

Some knowledge-capital measurements have high social value but low private appropriability. In those cases, measurement itself becomes underprovided. KBC calls this the measurement commons problem. Social EVPI, expected social value of perfect information, is the value of eliminating uncertainty before choosing a governance instrument. It does not provide a precise calculation in the way firm-level EVPI sometimes can. It provides a justificatory logic for public or collective measurement provision. Historical cases, dependency-network analysis, stress testing, scenario modelling, and comparative institutional evidence should, where possible, produce numerical ranges for expected avoided loss or exposure reduction; otherwise they should be classified as qualitative screening evidence rather than full measurement support.

A policy case for public measurement infrastructure arises when knowledge-capital measurement has high social value, low private appropriability, and material expected opportunity loss. Without KBC-AIE and Social EVPI, the welfare argument remains general; with them, it becomes operational for knowledge-capital governance.

The institutional anchors for this measurement function include national statistical agencies, securities regulators, standards bodies, industry consortia, public research funders, and commons foundations. Each can help measure a different part of the knowledge-capital system: national statistical agencies can improve aggregate stock and productivity measures; securities regulators can require material dark-capital disclosure; standards bodies can stabilize definitions and measurement protocols\index{measurement protocols}; industry consortia can pool dependency and risk data that no single firm can appropriate; public research funders can support measurement of public epistemic infrastructure; and commons foundations can monitor maintainer capacity, dependency networks, and commons-depletion risk.

These institutions may create incompatible definitions and reporting duties. KBC therefore requires not only collective measurement provision, but coordination among measurement authorities.

\section{Dark-Capital
Disclosure}\label{107-dark-capital-disclosure}

\subsection{The mechanism}\label{the-mechanism-4}

Dark-capital disclosure addresses the accounting shadow identified in
Chapter 9. The core difficulty is that the disclosure problem is not
symmetric with the accounting recognition problem. Accounting
recognition requires meeting strict criteria, identifiability,
control, probable future economic benefits, that most dark-capital
components cannot satisfy under IAS 38. But disclosure does not require
recognition. It requires that decision-relevant information about
material dark-capital exposures reach the actors who need it to govern,
invest, or regulate intelligently.

T4.3 (Fragility Shadow)\index{T4.3 Fragility Shadow!disclosure} provides the formal object for disclosure: \(\Omega_i\)
= E{[}\(C_{T2}\) + \(C_{T7}\) + \(C_{T6}\) + \(C_{T8}\){]}. It should be read as an
unpriced governance-position exposure, not as a recognized accounting
liability. It identifies expected economic exposure arising from governance
transitions, capability loss, recombination-field impairment, and feedback-loop
exclusion that may not satisfy recognition criteria under existing accounting
standards. The measurement-value test determines when the cost of quantifying
\(\Omega_i\) is justified by the decision-relevant information it would produce.

Dark-capital disclosure should not mean "put all intangibles on the
balance sheet." Disclosure form should track sensitivity: public narrative disclosure, regulator-only confidential filing, or board/audit-committee disclosure\index{audit-committee disclosure} may be appropriate depending on security, dependency, and commercial-sensitivity risks. Disclosure should not require publication of exploitable technical, security, or dependency details; confidential or aggregate reporting may satisfy the governance purpose. It should mean decision-useful reporting on material
dark-capital exposures across six categories. The less restrictive
alternative test applies to disclosure as well: mandatory reporting should
be used only where voluntary disclosure, qualitative flagging, confidential
regulatory reporting, or narrower dependency mapping would not provide
sufficient decision-useful information about material exposure.

\subsection[Knowledge-capital measurement mapping]{K-CMM mapping}\label{k-cmm-mapping-4}

Primary valuation targets: EKL (disclosure of expected knowledge
losses), COV* (disclosure of control positions and their associated \(\Omega_i\)
risks), and the materiality test across all seven dimensions. The
measurement-value test determines which disclosures are mandated and
which remain voluntary.

\subsection{Disclosure categories}\index{disclosure categories}\index{transparency!disclosure}\label{disclosure-categories}

\textbf{Critical knowledge dependencies.} Material \(K^D\), \(K^I\),
\(K^E\), \(K^C\), and \(K^P\) stocks that are essential to productive
capacity and that, if lost or degraded, would cross the materiality
threshold on capability materiality or value materiality dimensions.
Disclosure should include an assessment of whether the stock is on a
depreciation trajectory, whether its maintenance is controlled, and
whether its loss would produce irreversible capability consequences.

\textbf{Access-layer dependencies.} Material dependencies on APIs,
vendor-controlled platforms, proprietary licences, interoperability
standards, or third-party knowledge stocks that could be withdrawn,
restricted, or repriced without the disclosing
organization\textquotesingle s consent. The governance-position risk
component of \(\Omega_i\), the expected cost of a governance transition affecting
access, should be estimated and disclosed where the measurement-value
test is satisfied.

\textbf{Capability concentration.} Concentrations of knowledge-bearing
capability in individual people, small teams, undocumented routines, or
institutional practices that would be irreversible to lose. This is the
capability-loss component of EKL: expected loss from capability departure or
collapse. Disclosure should identify material concentrations and
describe what measures, if any, are in place to distribute or codify the
concentrated knowledge.

\textbf{Feedback-loop position.} Whether the organization is a net
producer of enclosed \(G^L\) (accumulating exclusive learning-loop
advantage) or a net dependent on closed feedback systems controlled by
others (bearing the EKL risk of learning-loop denial). This disclosure
makes LOV* and its distributional consequences visible to governance
actors and investors.

\textbf{Commons dependence.} Material dependence on open-source
software, public datasets, shared scientific standards, public research
corpora, or community-maintained knowledge stocks. Commons dependencies
that are not disclosed leave investors, regulators, and counterparties
unable to assess the EKL exposure from commons degradation. The
materiality test should determine the disclosure threshold\index{disclosure threshold}: only commons
dependencies that would pass the materiality score on recombination or
public/commons dimensions require disclosure.

\textbf{Governance-transition exposure.} Known or probable governance
transitions, IP term expiry, API policy changes, licence model
shifts, platform governance changes, public-to-private conversion of
previously open knowledge stocks, that would materially affect the
organization\textquotesingle s dark-capital position. The materiality
test\textquotesingle s irreversibility dimension is particularly
relevant here: transitions that would produce irreversible capability
loss or permanent field contraction should be disclosed even when the
probability of transition is uncertain.

\section{Competition and Antitrust
Implications}\label{108-competition-and-antitrust-implications}

\subsection{Price, dynamic capability, and recombination analysis}\label{price-dynamic-capability-and-recombination-analysis}

Modern competition analysis\index{competition analysis!knowledge-capital variables} is not limited to price. It can consider quality,
innovation, entry barriers, foreclosure, network effects, data advantages, and
exclusionary conduct. KBC adds a dynamic capability and recombination screen:
conduct may be competitively significant when it narrows accessible
recombination fields, captures feedback loops, reduces trajectory diversity, or
raises capability barriers, even before consumer prices rise. This extension is
not a replacement for competition analysis; it is a way of specifying knowledge-
capital mechanisms that conventional screens may underweight or fail to measure. The variables used below--\(D_u(F_{a,t})\), \(N_{traj}\), \(G^L\), and \(\widetilde{C}_{a}\)--are specified but not yet operational; they should be read as research-and-design agenda variables until Chapter~11's proxy and calibration programme validates them.

An incumbent in a knowledge-bearing market can simultaneously maintain
competitive pricing (or even price below cost) while extracting COV* through
feedback-loop capture, field enclosure, capability accumulation, and trajectory
control. The competitive harm may not appear in price. It may appear in
\(D_u(F_{a,t})\), in \(N_{traj}\), in \(G^L\), in \(\widetilde{C}_{a}\), and in COV*.
Where competition analysis does not measure these variables, it may miss or
understate the harm.

The Smithian inversion developed in Chapter 8 under H1--H4 shows that
individually rational enclosure strategies can produce a Nash equilibrium that
is not socially optimal, not because of monopoly pricing alone, but because the
private payoff function includes dynamic revenue from \(G^L\) and T8 trajectory
control that may not appear in social welfare. The social cost terms
\(C_{T2}\), \(C_{T7}\), \(C_{T6}\), and \(C_{T8}\) can be externalized.
Market-power analysis that does not measure these cost terms may under-identify
the competition problem in knowledge-bearing markets.

The practical implication is that competition authorities should treat
strategic-capture variables as potential equilibrium variables, not merely
evidentiary colour. Evidence that a conduct change increases private captured
yield or knowledge rent while lowering \(D_u(F_{a,t})\), \(N_{traj}\), or entrant
\(\widetilde{C}_{a}\) may indicate that the market has moved toward a socially
inferior Nash equilibrium even if consumer prices have not risen. These
variables are specified but not yet operational: Chapter~11 classifies
\(D_u(F_{a,t})\), \(N_{traj}\), \(G^L\) capture, and \(\widetilde{C}_{a}\)
divergence as identified measurement targets whose proxies require validation
before welfare magnitudes or thresholds can be inferred. The competition tests
proposed here therefore depend on that proxy programme and should be read as a
research-and-design agenda rather than an immediately applicable enforcement
standard.

\subsection{Competition analysis must include knowledge-capital
variables}\label{competition-analysis-must-include-knowledge-capital-variables}

A governance framework for knowledge-bearing capitalism should expand
competition analysis to include the variables below. These variables remain diagnostic until operationalised: they identify what a competition authority\index{competition authority} should investigate, not a finished enforcement threshold.

\textbf{\(D_u(F_{a,t})\) effects.} Does the conduct, acquisition, or
contractual arrangement narrow the recombination field available to
competitors, entrants, or the research community? Proposition C shows
that field contraction is a generative event. An acquisition that
forecloses identifiable recombination paths, by absorbing a stock
that would have enabled cross-firm combination, can indicate a competitive harm where the foreclosed path is material, non-substitutable, and likely under a plausible counterfactual, even if the acquired firm had no market power in the price sense.

\textbf{\(N_{traj}\) effects.} T8 (Fewer but Faster Trajectories)\index{T8 Fewer but Faster Trajectories!competition analysis} shows that
enclosure reduces trajectory diversity even as it may accelerate
within-trajectory improvement for the incumbent. Where a market
structure produces high within-firm improvement rates alongside falling
trajectory diversity, competition analysis should assess whether the
diversity reduction represents a social loss, foreclosed
recombination paths, suppressed exploration, that the within-firm
speed gain does not compensate.

\textbf{\(G^L\) capture.} Does a platform, deployment model, or
contractual arrangement give one actor exclusive access to feedback
signals from a shared user or research population? Exclusive \(G^L\) is a
learning-loop monopoly. Its competitive harm is dynamic, it produces
capability divergence over time, and therefore invisible to static
market-power analysis.

\textbf{\(\widetilde{C}_{a}\) divergence.} Where T6.2 (Capability Divergence) and T6.5
(Capability Trap) are operating, the competitive harm is that entrants
and competitors are converging toward a capability floor while the
incumbent\textquotesingle s capability continues to grow. Standard entry
analysis, which asks whether entry is possible in principle, does not
identify this harm. A competition analysis grounded in the KBC framework
should ask whether entry is possible at the capability level required to
compete effectively over the relevant time horizon, a considerably
more demanding standard.

\textbf{COV* extraction.} Competition analysis should distinguish
between productive value creation (improving CUV, ROV*, and LOV* through
genuine investment) and control-premium extraction (COV* accumulated
through exclusion, platform lock-in, standard capture, and API
gatekeeping). M5.T1 shows that strategic over-enclosure\index{M5.T1 Strategic Enclosure!governance design} produces
\(T^*_{\mathrm{strategic}}\) \textgreater{} \(T^*\), enclosure beyond the socially
optimal duration, because the private payoff includes control rent
that is not available to distributed actors. This rent can be a competitive harm where it exceeds service-producing appropriability and materially restricts recombination, entry, or capability formation.

\subsection{Anti-drift instruments for competition analysis}\index{anti-drift instruments}\index{competition policy!anti-drift instruments}\label{anti-drift-instruments-for-competition-analysis}

The competition implication of the appropriability--friction distinction is not that all exclusive control should be shortened or removed. It is that competition analysis must ask whether a temporary service advantage is hardening into durable friction rent. Where the materiality test shows that field contraction, feedback capture, platform dependency, or standards capture has become economically significant, and where the measurement-value test justifies intervention, the relevant instruments include sunset and review provisions, interoperability mandates, data-portability rights, public-interest licensing, standards transparency, API-change notice requirements, open-maintenance funding, and anti-capture disclosure for standards bodies.

These instruments should not be treated as a policy menu detached from the K-CMM structure. Each instrument must identify the component it targets, such as CUV, ROV*, LOV*, COV*, SOV*, or EKL; the materiality trigger that justifies attention; and the measurement-value problem that determines whether mandatory disclosure, audit, licence modification, or access intervention is proportionate. In this sense, the instrument is justified only when it reduces productive friction, preserves recombination or learning capacity, or prevents capture of governance position more effectively than it suppresses legitimate service-producing appropriability.

\subsection[The multi-incumbent commons-game limit]{The D8 limit}\label{the-d8-limit}

Chapter 8 treats D8, the N-actor multi-incumbent commons-enclosure
game, as structurally characterized but not completed at the full equilibrium level. The dual-channel
structure is characterized, and D8.P1 and D8.P2 are retained as
conditional structural conjectures, but the full N-actor Nash equilibrium
proof is open. The remaining work also includes the SF1--SF4 structural
field-composition model and the Ostrom-style governance-design problem.
The multi-incumbent commons-enclosure game remains a structurally characterized but not fully closed result; any commons-collapse, tipping, or coalition-instability reading is suggested by the reduced forms (M5.T1, D7) and pends the open N-actor proof (D8). It should not be used as an independent basis for policy prescription until its equilibrium conditions, substitutability/complementarity assumptions, and welfare implications are further specified.

At this stage, D8 is best used diagnostically: it identifies when multiple incumbents may have individually rational reasons to enclose shared knowledge inputs even when the cumulative system effect may weaken the commons on which they jointly depend. Chapter 11\textquotesingle s empirical program should include tests that would confirm, qualify, or disconfirm the candidate D8 equilibrium patterns.
\section{Policy Instrument
Matrix}\label{109-policy-instrument-matrix}

The instruments below are not default interventions. They are high-threshold
governance options. Their use depends on materiality, non-substitutability,
irreversibility, measurable recombination or learning-loop loss, positive
expected information value, and the absence of less restrictive alternatives.
The relevant question is not whether intervention could improve access in the
abstract, but whether a less restrictive governance arrangement could preserve
incentives while reducing material recombination loss, feedback exclusion,
capability decay, or dark-capital exposure.

Table~\ref{tab:ch10:intervention-status-labels} defines the intervention-status
labels used in the matrix. The labels distinguish ordinary governance practice
from disclosure, coordination, corrective, and exceptional instruments, so that
the reader can see whether the proposal changes information flows, coordination
structures, or access rights.

\begin{table}[h]
\centering
\footnotesize
\begin{tabular}{@{}p{0.28\textwidth}p{0.62\textwidth}@{}}
\toprule
Status label & Meaning \\
\midrule
Baseline governance practice & Firm/internal governance, documentation, risk assessment, dependency mapping. \\
Disclosure instrument & Makes exposure visible without directly changing access rights. \\
Coordination instrument & Standards, interoperability, shared maintenance, public-private governance. \\
Corrective instrument & Used only after material harm or high-risk exposure is shown. \\
Exceptional instrument & Compulsory licensing, mandated access, feedback-sharing obligations. \\
\bottomrule
\end{tabular}
\caption{Intervention status labels for Chapter 10 instruments}
\label{tab:ch10:intervention-status-labels}
\end{table}

The following matrix organizes Chapter 10\textquotesingle s governance
proposals against the three K-CMM tests, identifies the formal mechanism
each instrument targets, states each instrument\textquotesingle s intervention
status, and states the boundary risk, the misuse or overreach risk, associated
with each instrument. The added primary-institution and minimum-evidence columns turn the matrix from a policy catalogue into a decision tool: each row now identifies who would plausibly act and what must be shown before action is justified. It is the practical core of this chapter.

\begin{landscape}
\begingroup
\tiny
\setlength{\tabcolsep}{1.6pt}
\renewcommand{\arraystretch}{1.12}
\sloppy
\par\addvspace{0.8\baselineskip}\noindent
\begin{longtable}{@{}L{0.095\textwidth}
L{0.095\textwidth}
L{0.080\textwidth}
L{0.090\textwidth}
L{0.105\textwidth}
L{0.095\textwidth}
L{0.145\textwidth}
L{0.090\textwidth}
L{0.105\textwidth}@{}}
\caption{Policy Instrument Decision Matrix}\label{tab:ch10:policy-instrument-matrix}\\
\toprule\noalign{}
Governance problem & Formal mechanism & K-CMM component & Materiality dimension & Status / instrument & Primary institution & Minimum evidence & Measurement trigger & Boundary risk \\
\midrule\noalign{}
\endfirsthead
\toprule\noalign{}
Governance problem & Formal mechanism & K-CMM component & Materiality dimension & Status / instrument & Primary institution & Minimum evidence & Measurement trigger & Boundary risk \\
\midrule\noalign{}
\endhead
\bottomrule\noalign{}
\endlastfoot
Public knowledge appropriated at first conversion & \ensuremath{\dot{K}} → \(K^D_{private}\) without return & COV*, ROV* & Control, public/commons & Coordination: Public-funding open-access terms & Public funder; procurement agency; research institution & Public-funding terms; downstream commercialization route; evidence that public or commons stock is converted into private control without return conditions & Materiality of public-funding share & Over-restriction of commercialization incentives \\
Employee knowledge converted without compensation & \(K^E\) → \(K^D_{private}\) & COV* & Control, distributional & Baseline: Invention-assignment reform & Firm board; labour regulator; court & Contract terms; role expectations; evidence of material codification of worker-held capability; outside-option and compensation context & Materiality of commons/public input & Chilling employer R\&D investment \\
AI training corpus appropriation & \(K^C\), \(K^P\) → \(K^D_{private}\) & COV*, EKL & Public/commons, control & Disclosure: Training corpus disclosure & AI regulator; securities regulator; court; procurement agency & Training-corpus provenance; scale of commons/public inputs; licensing status; commercial deployment evidence; exclusion or non-attribution pathway & Scale of commons input; commercial deployment & Disclosure of genuinely proprietary training methods \\
User data converted to platform capability & Distributed knowledge → \(K^D_{private}\) & COV*, LOV* & Control, distributional & Disclosure: Data-use provenance rules & Platform regulator; privacy regulator; firm board & Data-use notices; telemetry flows; evidence that user activity improves incumbent capability; opt-out and portability constraints & Scale of conversion; materiality of COV* derived & Privacy compliance complexity \\
Field contraction through access restriction & \(D_u(F_{a,t})\) ↓ & ROV*, CUV & Recombination, capability & Coordination: Interoperability mandates & Standards body; platform regulator; competition authority & Access-change record; affected dependent actors; switching costs; interoperability feasibility; evidence that useful diversity in the field has narrowed & Breadth of field contraction; irreversibility & Mandating interoperability where incompatibility is genuine \\
API enclosure of platform-dependent actors & \(D_u\) ↓, CUV ↓ for dependents & ROV*, CUV, EKL & Recombination, capability & Corrective: API access rules & Platform regulator; competition authority; court & API change history; affected dependent actors; loss of access; switching cost; capability impact; evidence of dependency network materiality & Dependency network materiality; KCI \textgreater{} 0 & Restricting legitimate platform design decisions \\
Recombination field foreclosure through IP & \(D_u\) ↓ beyond \(T^*\) & ROV*, COV* & Recombination, control & Exceptional: Compulsory licensing (proposed knowledge-capital trigger) & Court; competition authority; patent office; legislature & Blocking-right map; non-substitutability evidence; reduced trajectory diversity; foreclosure of complementary innovation; capability divergence; failed less-restrictive alternative\index{less-restrictive alternative}\index{competition policy!less-restrictive alternative}s & \(M_i\) ≥ M* on recombination dimension & Weakening genuine IP incentive structure \\
Exclusive learning-loop capture & \(G^L\) → exclusive \(\widetilde{C}_{a}\) & LOV*, EKL & Capability, distributional & Disclosure: Feedback-use disclosure & AI regulator; platform regulator; securities regulator & Evidence that user activity improves incumbent system while excluding rivals, users, or contributors from comparable learning; deployment and telemetry evidence & KCI \textgreater{} 0; commercial deployment & Imposing disclosure on research-stage systems \\
Capability trap through \(G^L\) denial & \(\widetilde{C}_{ent}\) → \(\widetilde{C}_{min}\) & LOV*, EKL & Capability, irreversibility & Exceptional: Feedback-sharing obligations & Platform regulator; competition authority; court & Evidence of feedback exclusion; entrant capability decay; high switching costs; inability to replicate learning loop; material irreversibility; failed disclosure or portability alternatives & \(M_i\) ≥ M* on capability + irreversibility & Suppressing \(G^L\) investment incentives \\
Model improvement opacity & LOV* invisible & EKL & Capability & Disclosure: Model improvement audit trails & Firm board; sector regulator\index{sector regulator}; AI regulator & Model-change logs; audit trail of feedback-driven improvement; high-stakes deployment evidence; material error or drift risk & High-stakes domain deployment & Administrative burden on small developers \\
Knowledge commons depletion & \(K^C\) degradation → \(D_u\) ↓, \(\widetilde{C}_{min}\) ↓ & CUV, ROV*, avoided EKL & Public/commons, recombination & Coordination: Commons maintenance funding & Commons foundation; public funder; procurement agency; firm board & Dependency map; maintainer scarcity; security incidents; underfunding; abandonment risk; evidence that private productivity depends on the commons stock & KCI \textgreater{} 0; dependency network breadth & Funding commons that users have already abandoned \\
\(K^P\) degradation & \(K^P\) ↓ → \(T^*\) ↑, \(\widetilde{C}_{min}\) ↓ & CUV, ROV* & Public/commons & Baseline: Public epistemic infrastructure investment & Standards body; statistical agency; public research body; public funder & Degradation of standards; benchmark capture; measurement-system dependency; public reliability loss; evidence that downstream actors rely on the epistemic stock & Long-run recombination materiality & Treating all public epistemic infrastructure as equally critical \\
Standards capture & \(K^I_{open}\) → \(K^I_{private}\) & COV*, ROV* & Control, recombination & Coordination: Open-standard protection & Standards body; competition authority; procurement agency & Governance records; participant concentration; licensing changes; benchmark or standard capture evidence; dependent ecosystem impact & Materiality of standards field access function & Impeding legitimate standards evolution \\
Dark capability loss invisible to investors & \(\Omega_i\) unpriced governance-position exposure & EKL, COV* & Risk, capability & Disclosure: Dark-capital disclosure (capability concentration) & Securities regulator; firm board; auditor; risk committee & Material stock dependence; governance-transition trigger; loss channel; concentration of capability; estimated decision relevance for investors or counterparties & KCI \textgreater{} 0; materiality threshold passed & Disclosing commercially sensitive knowledge-capital structure \\
Market power invisible to antitrust & COV*, \(G^L\), \(D_u\) ↓, \(N_{traj}\) ↓ & COV*, LOV*, ROV* & Control, recombination, capability & Baseline: Expanded competition analysis & Competition authority; court; sector regulator & Reduced trajectory diversity; foreclosure of complementary innovation; capability divergence; network effects; data advantage; entry-barrier evidence & T6.2, T8 evidence available & False positives where capability divergence reflects merit \\
\end{longtable}
\endgroup
\end{landscape}

\section{Chapter 10 Governance-Fit Acceptance Test}\label{1010-chapter-10-governance-fit-acceptance-test}

A Chapter 10 governance recommendation is complete only if the reader can answer the following questions. The test is deliberately practical. It prevents governance design from becoming either policy activism or abstract taxonomy: each proposed instrument must identify the stock at issue, the mechanism of loss or capture, the affected K-CMM component, the materiality basis, the measurement case\index{measurement-value test}, the acting institution, the evidentiary threshold, the less restrictive alternative, and the conditions under which the recommendation would weaken or fail.

\begin{table}[htbp]
\centering
\small
\caption{Chapter 10 governance-fit acceptance test\index{governance-fit acceptance test}}
\label{tab:ch10:governance-fit-acceptance-test}
\begin{tabular}{@{}p{0.05\textwidth}p{0.43\textwidth}p{0.42\textwidth}@{}}
\toprule
 & Question the reader must be able to answer & Failure signal \\
\midrule
1 & What governance problem is being addressed? & The passage names a policy instrument without identifying the economic governance problem. \\
2 & Which knowledge-bearing stock is at issue? & The stock is described generically as information, data, openness, or innovation rather than as a specific knowledge-bearing stock. \\
3 & Is the mechanism access restriction, feedback capture, commons depletion, public epistemic infrastructure degradation, competition foreclosure, or dark-capital exposure? & The mechanism is left as a broad appeal to harm, market power, public interest, or innovation. \\
4 & What K-CMM component is affected? & The instrument is not tied to CUV, ROV*, LOV*, COV*, SOV*, or EKL. \\
5 & Is the exposure material? & The subsection does not explain why the stock, transition, dependency, or loss is large enough to matter. \\
6 & Is measurement worth doing? & The subsection does not show why reducing uncertainty would change a decision enough to justify measurement cost. \\
7 & Who is the appropriate institution to act? & Responsibility is left floating among firms, regulators, courts, standards bodies, funders, or public agencies. \\
8 & What minimum evidence\index{minimum evidence standard} is required? & The recommendation lacks evidence such as dependency mapping, access-change history, feedback-loop evidence, capability-impact evidence, or investor-decision relevance. \\
9 & Is there a less restrictive alternative? & The instrument changes access rights or disclosure duties before considering a narrower governance arrangement. \\
10 & What would falsify or weaken the governance recommendation? & The subsection does not state what evidence would show that the proposed instrument is unnecessary, excessive, misdirected, or empirically unsupported. \\
\bottomrule
\end{tabular}
\end{table}

The final target state is therefore narrow. KBC does not prescribe openness or intervention by default. It provides a governance-fit framework for identifying when knowledge-bearing stock, recombination fields, feedback loops, commons, public epistemic infrastructure, or dark-capital exposures become economically material enough to justify measurement, disclosure, coordination, or carefully bounded intervention.

\section{Handoff to Chapter 11}\label{1010-handoff-to-chapter-11}

Chapter 10 has argued that governance design for knowledge-bearing
capitalism requires mechanism alignment, materiality discipline, and
measurement proportionality. Each instrument in §§10.3--10.10 targets a formal
mechanism and is constrained by the three K-CMM tests. What Chapter 10
cannot do (and does not attempt) is validate the parameters that
determine when those tests are triggered.

The materiality threshold M* is stated as a principle but not
calibrated. The coefficient values in the GATE and QUAL functions are
provisional estimates on a qualitative scoring scale, not empirically
derived parameters. The recovery lag function in T6.6 is formally
specified but not estimated from data. The conditions H1--H4 under which
M5.T1 (Strategic Over-Enclosure) holds are stated theoretically but not
tested against observed firm behaviour. D8\textquotesingle s
multi-incumbent equilibrium analysis remains structurally characterized but
not yet completed at the full N-actor equilibrium level.

Chapter 11 is therefore not merely a robustness check. It is the bridge
between the theoretical architecture of Chapters 3 through 10 and the
empirical program that would confirm, qualify, or falsify that
architecture. Its tasks include:

Calibrating K-CMM\textquotesingle s coefficient structure against
observed cases, acquisition valuations, outsourcing outcomes,
capability-loss events, commons degradation, and feedback-capture
evidence, to determine whether the model\textquotesingle s structural
predictions hold.

Testing the materiality score against observed governance decisions to
determine whether \(M_i\) predicts which governance transitions later create
value, loss, or regulatory concern.

Testing M5.T1\textquotesingle s conditions (H1--H4) against firm
behaviour in industries where knowledge-capital enclosure is measurable
,  software platforms, pharmaceutical IP, AI model deployment, to
determine whether strategic over-enclosure is systematically observable.

Identifying the functional form of the recombination generation
function\textquotesingle s parameters (\(\lambda_a\), η, μ) from productivity and
innovation data, replacing the theoretical specification with
empirically bounded estimates.

Resolving the D8 open problem through either formal proof of the N-actor
Nash equilibrium or empirical identification of the
game\textquotesingle s candidate equilibrium patterns in observable
multi-incumbent markets.

Stating falsification conditions for each major theoretical claim, 
the conditions under which the evidence would require revision or
abandonment of T5, T6, T7, T8, M5.T1, or the K-CMM valuation
architecture.

The governance proposals in this chapter are grounded in theory and
disciplined by measurement. They are not yet validated empirically.
Chapter 11 supplies that validation program. Until it does, Chapter
10\textquotesingle s instruments should be read as theoretically
justified and empirically conditional, stronger than policy
preference, weaker than demonstrated necessity.

\chapter[Testing KBC]{Testing Knowledge-Bearing Capitalism}\label{chapter-11-testing-knowledge-bearing-capitalism}

\chapterhook{Make the Theory Risk Being Wrong}

A theory of the conversion engine should leave fingerprints, and this chapter goes looking for them. If governance is what relocates productive knowledge, then enclosing a knowledge community should move the producers and the use-value they generate to wherever access remains open, and opening a closed body of knowledge should draw them back. The relocation evidence assembled here is the conversion claim made to risk being wrong.

The same engine leaves a second kind of fingerprint, in the gap between yield and visible generativity. Two predictions follow, stated formally in the Technical Companion, Appendix~M: signal-inflated stock, whose visible generativity outruns its yield, should underperform downstream and fail or retract more than its signal implies; and fields with stronger validation institutions should show tighter coupling between generativity and realized yield. Both are testable on the same public data the relocation tests use.

\emph{Empirical Calibration, Falsification, and Measurement}

\textbf{Thesis:} Chapter 11 asks whether knowledge-bearing capitalism (KBC) can fail as an empirical theory. Its task is not to add another layer of theory, but to identify observable signatures that would support, weaken, bound, or reject the mechanisms developed in Chapters 3 through 10. The first empirical signatures are: recombination-field contraction after access restriction; feedback-capture advantage when incumbents control deployment data and learning loops; capability divergence between included and excluded actors; governance-transition valuation effects after API, licence, platform, intellectual-property, or standards changes; and dark-capital or unpriced-exposure effects when markets and accounts fail to price governance-position risk in advance. Chapter~\ref{chapter-9-dark-capital-and-the-accounting-shadow} supplies the dark-capital objects to be tested, and Chapter~\ref{chapter-10-governance-design-for-knowledge-bearing-capitalism} supplies the governance interventions whose effects should be observable.

The empirical strategy uses established tools from patent-citation analysis, event studies, intangible-capital\index{intangible-capital accounting} valuation, and open-source productivity research \parencite{HallJaffeTrajtenberg2001, MacKinlay1997, BondCummins2000, Nagle2019, NagleEtAl2022}. The tests ask whether this theory's mechanisms leave observable signatures in prices, citations, productivity, governance changes, and capability divergence. Where those signatures cannot be observed, KBC must be revised, narrowed, or rejected.

\noindent\textbf{Two signature predictions, and the limit of what they show.}\index{signature predictions|textbf} Among these signatures, two are the distinctive predictions of this theory rather than restatements of existing results, and they should be read as its empirical signature. The first is the \emph{dual enclosure effect}\index{dual enclosure effect}: a governance transition can raise the enclosing actor's present realized yield while simultaneously lowering the future attainable frontier of excluded actors, through recombination-field contraction or learning-loop exclusion. That cross-actor, cross-period coupling is sharper than ``intangibles matter, '' more dynamic than static accounting critiques, and more institutionally specific than generic endogenous growth, none of which predict it. The second is \emph{institutional residue}\index{institutional residue}: productive knowledge that persists in organizational routines, standards, workflows, and governance systems beyond the embodied knowledge of the individuals present, so that \(\text{OKU}^{I}\neq\sum_a\text{OKU}^{E}\); this distinguishes KBC from a simple human-capital reduction and is developed as a test later in this chapter. A reminder on what such tests can and cannot establish: corroborating either mechanism strengthens that mechanism, not the whole interpretive frame of knowledge-bearing capitalism, which becomes more plausible only as its mechanisms accumulate survivals against rival explanation, calibration, and negative cases.

\section{First Empirical Project Register}\index{first empirical project register|textbf}\label{ch11:first-empirical-project-register}

The chapter's first task is to make KBC empirically vulnerable before the full apparatus appears. Five first-pass projects carry the burden. They are not the whole research programme; they are the most direct ways to ask whether the theory's central mechanisms leave observable traces.

\begingroup
\small
\setlength{\tabcolsep}{4pt}
\renewcommand{\arraystretch}{1.16}
\begin{longtable}{@{}L{0.28\textwidth}L{0.34\textwidth}L{0.30\textwidth}@{}}
\caption{First Empirical Project Register}\label{tab:ch11:first-empirical-project-register}\\
\toprule\noalign{}
Priority test & What it tests & First-pass empirical design \\
\midrule\noalign{}
\endfirsthead
\toprule\noalign{}
Priority test & What it tests & First-pass empirical design \\
\midrule\noalign{}
\endhead
\bottomrule\noalign{}
\endlastfoot
API closure / access restriction events\index{API closure!empirical testing}\index{access restriction!empirical testing}\index{API closure!empirical testing} & Whether governance changes narrow recombination fields by reducing accessible inputs, useful diversity, downstream reuse, or dependent innovation. & Event studies or difference-in-differences around dated API closures, licence changes, repository restrictions, platform rule shifts, or standards access changes. \\
Feedback capture in AI/platform systems & Whether incumbents improve faster when they control use-feedback, deployment data, error reports, ratings, prompts, telemetry, or interaction traces. & Compare capability improvement before and after deployment across actors with differential feedback access, controlling for compute, scale, baseline quality, and model or platform age. \\
Open-source maintainer or governance shocks & Whether commons depletion\index{commons depletion|textbf} affects downstream productivity, security, release cadence, dependency fragility, or recovery time. & Use dated maintainer exits, project archival, relicensing, foundation governance changes, security incidents, or platform capture as shocks to dependency networks. \\
Market reaction to governance-transition events & Whether markets underprice knowledge-capital exposure before API, licence, platform, intellectual-property, standards, or commons-governance transitions. & Event studies measuring abnormal returns, forecast revisions, valuation multiples, or operating-performance surprises for exposed firms relative to matched controls. \\
OKU substitution / task-equivalence cases & Whether embodied and disembodied knowledge can be compared under bounded task conditions rather than at the whole-job or occupation level\index{task-level substitution versus occupation-level substitution}\index{AI substitution!task level versus occupation level}. & Controlled task trials, workflow substitution studies, judge-blind comparisons, or before/after automation cases that test bounded functional equivalence. \\
\end{longtable}
\endgroup

Table~\ref{tab:ch11:first-empirical-project-register} introduces the empirical project menu; \S\ref{sec:ch11-priority-work-sequence} converts that menu into a prioritized work sequence.

These projects are foregrounded because each has a plausible empirical design, a visible failure condition, and a direct link to the theory's central mechanisms. API and access shocks test recombination-field contraction. Feedback-capture cases test learning-loop advantage. Commons-governance shocks test whether shared knowledge stock is a material input to private productivity. Governance-transition event studies test whether exposures are priced in advance. OKU substitution cases test whether the unit bridge introduced in Chapter~2 has empirical bite. The longer sections below preserve the full theorem-to-test architecture, but the first question is simpler: do these mechanisms show up where KBC says they should?

\section{The First Empirical Test of KBC}\index{first empirical test of KBC|textbf}\index{empirical unit!governance transition}\index{governance transition!empirical unit}\label{sec:ch11-first-empirical-test-of-kbc}

The first empirical test of KBC is not a measurement of knowledge in general. That would make the theory too broad to fail and would ask empirical work to carry more precision than knowledge-bearing stock can yet support. The minimum empirical unit is narrower: \emph{an economically material governance transition affecting an identifiable knowledge-bearing stock, followed by an observable change in access, recombination, feedback, capability, productivity, valuation, or risk}.

This unit gives the theory an empirical spine. A testable KBC case must identify the stock, the actor or actors affected, the governance transition, the channel through which productive capacity should change, and the observable consequence that would weaken the claim if it does not appear. The object is therefore not simply ``software, '' ``data, '' ``AI, '' or ``knowledge.'' It is a dated change in the governance conditions surrounding a specific knowledge-bearing stock: for example, an API closure, licence change, repository restriction, model-weight enclosure, standards-access change, data-access restriction, or platform rule shift.

The first canonical empirical case is an API closure or access-restriction event. It is the cleanest starting point because the governance transition is usually dateable, the affected knowledge-bearing stock can often be identified, exposed and less-exposed actors can be compared, and the predicted channels are visible: reduced access, narrower recombination fields, weaker downstream experimentation, slower capability development, altered valuation, or increased dependency risk. If such transitions do not measurably affect recombination, feedback, capability, productivity, valuation, or risk after plausible rival explanations are controlled for, the relevant KBC claim should narrow.

\begin{center}
\fbox{\begin{minipage}{0.92\textwidth}
\small
\textbf{Running case: API closure/access restriction.}\index{API closure!Chapter 11 running case} The canonical event-study unit is a dated API access change. The object is provider-governed interface stock and the dependent downstream capabilities built around it. The treatment is closure, repricing, rate-limit reduction, endpoint removal, permission change, or contractual restriction. The outcomes are changes in downstream product formation, integrations, research output, capability maintenance, abnormal returns, dependency concentration, or risk disclosure. Rival explanations include ordinary switching costs, quality filtering, security necessity, demand shifts, obsolescence, and monopoly pricing. The falsifier is no differential effect on exposed actors after those rival explanations are credibly controlled.
\end{minipage}}
\end{center}

\section{Assumption-Dependence Guide}\index{assumption-dependence guide|textbf}\index{complementarity assumptions}\index{strict complementarity}\index{substitution!partial}\label{sec:ch11-assumption-dependence-guide}

Several KBC claims rely on complementarity: knowledge-bearing stock produces value only when access, capability, maintenance, governance, demand, and complementary infrastructure work together. That does not mean every claim assumes strict Leontief conditions. Table~\ref{tab:ch11:assumption-dependence-guide} separates the claims that depend most heavily on complementarity from those that can survive partial substitution.

\begingroup
\small
\setlength{\tabcolsep}{3pt}
\renewcommand{\arraystretch}{1.12}
\begin{longtable}{@{}L{0.22\textwidth}L{0.27\textwidth}L{0.24\textwidth}L{0.19\textwidth}@{}}
\caption{Plain-Language Assumption-Dependence Guide}\label{tab:ch11:assumption-dependence-guide}\\
\toprule
Claim family & Complementarity assumption & What would weaken the claim & What survives partial substitution \\
\midrule
\endfirsthead
\toprule
Claim family & Complementarity assumption & What would weaken the claim & What survives partial substitution \\
\midrule
\endhead
\bottomrule
\endlastfoot
Access, capability, and maintenance claims & Strongest complementarity. A stock may have little yield if access, permission, skill, infrastructure, documentation, and maintenance do not jointly hold. & Evidence that actors can replace missing access or capability cheaply, quickly, and without quality loss. & Material access shocks still matter when substitutes are costly, delayed, lower-quality, legally blocked, or incompatible. \\
Feedback-capture and capability-divergence claims & Moderate-to-strong complementarity. Feedback must be difficult to imitate and useful for improving models, routines, or deployment capability. & Evidence that public data, synthetic data, third-party benchmarks, or independent experimentation produce equivalent learning. & Capture claims survive where proprietary deployment feedback produces differential improvement after controlling for scale, compute, baseline quality, and demand. \\
Governance-fit claims & Moderate complementarity. Governance affects access, maintenance, trust, interoperability, transferability, and enforceability. & Evidence that governance changes do not alter usable access, maintenance incentives, contracting costs, or institutional reliability. & Governance still matters where it changes permission, trust, dependency, legal exposure, or recombination cost even if actors retain partial workarounds. \\
Market-valuation and dark-capital claims & Weaker complementarity. Markets may partly price expectations, options, rents, and control positions even when accounting does not. & Evidence that market prices fully and consistently incorporate the relevant governance-position value and exposure in advance. & Accounting-shadow and dark-risk claims survive where exposures are visible only after shocks, impairments, litigation, dependency failure, or capability loss. \\
Enclosure welfare claims & Conditional complementarity. The welfare claim depends on the balance between incentive, quality, security, and maintenance gains and recombination, feedback, and capability losses. & Evidence that exclusion increases quality, security, maintenance, disclosure, or investment more than it suppresses future generation. & KBC predicts harm only where dynamic losses exceed productive enclosure benefits; it is not a blanket openness claim. \\
\end{longtable}
\endgroup

The guide makes the empirical burden explicit. KBC is most exposed where it predicts that missing access, feedback, or capability cannot be cheaply substituted. Its claims are weaker, narrower, or merely diagnostic where actors can replace the lost input, preserve learning, or maintain productive yield through alternative governance paths.

\index{predecessor theories!KBC distinction}\index{distinctive KBC prediction}\index{governance-position changes}\index{future generative capacity}
\begingroup
\small
\setlength{\tabcolsep}{3pt}
\renewcommand{\arraystretch}{1.12}
\begin{longtable}{@{}L{0.18\textwidth}L{0.19\textwidth}L{0.18\textwidth}L{0.24\textwidth}L{0.15\textwidth}@{}}
\caption{KBC and Its Predecessor Theories}\label{tab:ch11:kbc-predecessor-distinction}\\
\toprule
Predecessor theory & What it explains & What KBC accepts & What KBC adds & Distinctive KBC prediction \\
\midrule
\endfirsthead
\toprule
Predecessor theory & What it explains & What KBC accepts & What KBC adds & Distinctive KBC prediction \\
\midrule
\endhead
\bottomrule
\endlastfoot
Human-capital theory\index{human capital theory} & Skill, education, experience, and labour productivity. & Productive knowledge often resides in persons. & Embodied knowledge can move into artefacts, routines, platforms, commons, or public infrastructure, changing access and capture. & Codification may increase firm capability while altering worker bargaining and future skill formation. \\
Information economics\index{information economics} & Search, uncertainty, asymmetry, signalling, and information value. & Information affects decisions, prices, and investment under uncertainty. & KBC treats productive knowledge-bearing stock as a governed capital-like object, not merely as a signal or decision input. & Governance transitions can change productive capacity, not only information available to decision-makers. \\
Endogenous growth theory\index{endogenous growth theory} & Non-rival ideas, increasing returns, spillovers, and research-driven growth. & Ideas and knowledge can scale and generate increasing returns. & KBC specifies residence, governance, recombination fields, feedback loops, and knowledge impairment. & Enclosure can raise private return while narrowing future generation. \\
Resource-based and knowledge-based firm theory\index{resource-based view}\index{knowledge-based view} & Firm-specific resources, capabilities, routines, and knowledge integration. & Firms are important knowledge-coordination systems. & KBC extends beyond the firm to platforms, commons, public epistemic infrastructure, and governance transitions among forms. & Capability may regenerate inside incumbents while depreciating among excluded challengers. \\
Intangible-capital and intellectual-capital research\index{intangible capital}\index{intellectual capital} & Measurement gaps, market-to-book differences, organizational capital, and intangible investment. & Many productive assets are nonphysical and poorly represented in accounts. & KBC decomposes dark value, dark risk, foregone capitalization, false stock, and governance-position exposure. & Accounting may miss not only value, but also latent exposure to knowledge-stock impairment. \\
Platform and data-economics theory\index{platform economics}\index{data economics} & Network effects, data accumulation, platform control, and multi-sided markets. & Platforms can control access, data, complements, and user feedback. & KBC links platform governance to recombination loss, feedback capture, challenger capability decay, and dark-capital exposure. & A dated API restriction should affect exposed actors' recombination, feedback, capability, valuation, or risk if KBC is doing work. \\
\end{longtable}
\endgroup

Table~\ref{tab:ch11:kbc-predecessor-distinction} states the predecessor boundary compactly. KBC does not claim that earlier economics ignored knowledge. Its narrower contribution is to treat governance-position changes as changes in the conditions of future generation. The strongest distinctive prediction is therefore that an incumbent can increase current private value while a governance transition narrows recombination fields, learning loops, challenger capability, or dark-capital visibility.

\index{domain-class testability table}\index{empirical tractability!domain classes}\index{first KBC tests!domain classes}
The first empirical programme should also be domain-disciplined. KBC is easiest to test where the knowledge-bearing stock, governance transition, exposed actors, and outcome proxies can be bounded. Table~\ref{tab:ch11:domain-class-testability} ranks the main candidate domains by empirical tractability rather than by theoretical importance.

\begingroup
\small
\setlength{\tabcolsep}{3pt}
\renewcommand{\arraystretch}{1.12}
\begin{longtable}{@{}L{0.19\textwidth}L{0.25\textwidth}L{0.24\textwidth}L{0.24\textwidth}@{}}
\caption{Domain Classes for First KBC Tests}\label{tab:ch11:domain-class-testability}\\
\toprule
Domain or event class & Why KBC is testable here & Primary outcome proxies & Main rival explanations \\
\midrule
\endfirsthead
\toprule
Domain or event class & Why KBC is testable here & Primary outcome proxies & Main rival explanations \\
\midrule
\endhead
\bottomrule
\endlastfoot
API closure or access restriction\index{API closure!domain-class testability} & Dateable governance transition; identifiable exposed developers, firms, products, and complements. & Product formation, integrations, forks, migration cost, dependency concentration, abnormal returns, disclosure changes. & Switching costs, quality filtering, security necessity, ordinary monopoly pricing, demand shift. \\
Open-source maintainer shock\index{open-source maintainer shock!domain-class testability} & Maintainer exit, project archival, relicensing, or security incident can be located in dependency networks. & Release cadence, vulnerability remediation, downstream package failures, dependency migration, contributor activity. & Project maturity, unrelated technical debt, changing demand, normal software churn. \\
Cybersecurity knowledge loss\index{cybersecurity failure!domain-class testability} & Breach or exfiltration event can affect knowledge stock without destroying physical assets. & Incident cost, recovery lag, abnormal returns, product delay, competitor acceleration, litigation, customer trust loss. & Ordinary operational disruption, reputational shock, regulatory penalty, pre-existing weak controls. \\
AI feedback capture\index{AI feedback capture!domain-class testability} & Differential access to user interactions, prompts, corrections, telemetry, and deployment errors can be compared. & Benchmark improvement, error reduction, model update speed, user retention, capability divergence, entrant survival. & Compute scale, talent, baseline model quality, marketing, data quality unrelated to feedback. \\
Patent thickets and licence restrictions\index{patent thickets!domain-class testability} & Rights boundaries are legally observable, but recombination effects are harder to isolate. & Citation dispersion, entry, research redirection, licensing cost, product delay, litigation, design-around activity. & Patent quality, R\&D intensity, sector maturity, legal strategy, demand change. \\
Public epistemic infrastructure\index{public epistemic infrastructure!domain-class testability} & Standards, statistics, metrology, courts, and public research systems matter broadly, but shocks and exposed controls are harder to isolate. & Adoption rates, interoperability, error reduction, standard diffusion, productivity around public standards, litigation or compliance cost. & Institutional quality, sector regulation, public spending, macro trends, complementary private investment. \\
\end{longtable}
\endgroup

Table~\ref{tab:ch11:domain-class-testability} gives priority to empirical cleanliness. API closure remains the canonical running case because it offers the clearest event date, access mechanism, treatment group, outcome set, and falsifier. Other domains remain important, but they should enter after the first tests have shown whether KBC's governance-transition mechanism can be measured without turning every knowledge-intensive event into a confirmation.

\section{From Theory to Test}\label{111-from-theory-to-test}

This chapter is the book's empirical roadmap. The central caveats now become test discipline: claims are conditional, not universal; enclosure is evaluated by net effects rather than by presumption; stock is distinguished from capital by productive service flow; and measurement is treated as uncertainty reduction rather than full accounting recognition. The purpose is to convert those caveats into tests that a working economist can inspect, challenge, replicate, or reject.

Chapters 3 through 10 develop an internally formalized theory of how
capitalism changes in sectors where knowledge-bearing stock becomes an
important source of wealth. The formal
architecture is coherent: the five-form \(K^x\) taxonomy provides a
classification of knowledge capital; the Knowledge Generation Model
specifies how new knowledge is produced from existing stocks and
recombination fields; the Conditional Separability Axiom anchors governance
analysis; and the theorem sequence, T2 through T8, M5.T1, T4, 
generates predictions about enclosure, capability divergence, social
cost, accounting invisibility, and strategic over-enclosure.

Internal coherence is necessary but not sufficient. A theory that
follows from its assumptions tells us nothing about whether those
assumptions are correct or whether the mechanisms they specify operate
in the world at the magnitudes, frequencies, and directions this theory
predicts. Chapters 3 through 10 are internally formalized. They are not
yet empirically validated.

The empirical tests must therefore be symmetric. They should not ask only whether enclosure reduces access for excluded actors. They should also ask whether enclosure increases investment, accelerates incumbent learning, improves quality, or reduces misuse. A credible test of KBC measures both sides of the dual effect and then asks whether the observed governance arrangement produces a net loss in recombination field breadth, trajectory diversity, or future productive range relative to plausible alternatives.

The same discipline applies to recombination generation. The variable \(G^R\) measures a generation rate or productive-capacity channel, not social welfare itself. A test that finds higher \(G^R\) has not yet shown higher welfare. The welfare interpretation requires additional filters: what the generated knowledge is used for, whether it has positive social value, what risks it creates, which institutions govern its deployment, and whether the relevant domain is excluded because the generated knowledge is harmful. Chapter 11 therefore treats \(G^R\) as evidence about productive capacity and trajectory formation, not as a sufficient welfare criterion.

The net-effect verdict requires the same discipline applied to its hardest case. Whether enclosure is on balance welfare-reducing turns on the sign of the difference between its suppression loss \(L^-(T)\) and its generative gain \(G^+(T)\). That sign is to be estimated only through treated-versus-control divergence under a named natural experiment, never through a reconstructed generative frontier or any modelled counterfactual field, and the frontier interpretation is explicitly disclaimed wherever a reader might supply it. The honest position is that, absent such an experiment for the case at hand, the aggregate welfare verdict remains under-anchored: it is carried as \textbf{Unresolved} in the claim-status ledger until a credible natural experiment exists for the case, rather than asserted from the measurable side of the ledger alone.

KBC uses contrast explanation where its empirical claims concern comparative cases, event shocks, or governance differences. \textcite{Lawson2009} describes this method as beginning from surprising contrasts, cases expected to be similar that produce divergent outcomes, and retroducing the mechanism responsible. Strong Lawson-compatible KBC cases include high IT investment with low measured productivity; API closure affecting exposed versus unexposed firms; open-commons versus enclosed-governance comparisons; before/after platform governance shocks; and firms with similar digital investment but different \(K^I\) conversion outcomes. Not every KBC anomaly is contrastive; some are internal inconsistencies within accounting or capital theory, and the method should not be applied uniformly.\index{contrastive anomaly versus internal inconsistency}

This distinction matters for two reasons.

The first is intellectual honesty. The KBC framework advances claims
that are stronger than standard intangibles economics (which observes
that knowledge assets are undervalued) and weaker than a confirmed
causal account (which would require experimental evidence or
well-identified natural experiments). Overstating the empirical status
of this theory would weaken it in precisely the domains where it most needs
to be testable.

The second is that the falsification architecture is part of the
theory\textquotesingle s contribution. A theoretical framework that
cannot specify what evidence would disconfirm it is not a scientific
theory; it is a conceptual vocabulary. The KBC framework should be
falsifiable, and this chapter says how.

The chapter is therefore the empirical launchpad of the book rather than a residual appendix. Every empirical claim in this chapter follows the same object-validation spine\index{object-validation spine}: define the object, identify its observable consequences, choose proxies, compare against simpler rival explanations, test directionality first, calibrate coefficients only after repeated observation, and revise or discard the claim if the proxy fails. The tests below should be read as first empirical projects: small enough to begin, sharp enough to fail, and concrete enough to discipline later calibration.

\subsection{The claim-status
framework}\label{the-claim-status-framework}

Every major claim in this chapter is assigned one of four statuses.

\textbf{Internally Formalized.} The claim follows validly from the
model\textquotesingle s assumptions. Its truth is conditional on those
assumptions. It can be evaluated on internal grounds, consistency,
coherence, derivational validity, but requires empirical work to
establish external validity.

\textbf{Empirically Plausible.} The claim is consistent with known
cases, stylized facts, or existing empirical literature but has not been
formally calibrated against data specifically designed to test it. It
can be regarded as a reasonable working hypothesis.

\textbf{Testable.} The claim can be translated into observable
variables, measurable proxies, and a research design that could confirm
or disconfirm it. Testability does not mean the test has been done; it
means it can be done.

\textbf{Unresolved.} Formal work on the claim is incomplete (e.g.,
D8\textquotesingle s N-actor proof), or the empirical programme required
to test it exceeds what currently available data can support, or the
claim depends on assumptions that cannot yet be evaluated.

These statuses are not hierarchical in terms of importance. An
Internally Formalized claim that generates a Testable prediction is more
valuable to the research programme than an Empirically Plausible claim
that cannot be operationalized. An Unresolved claim that identifies a
genuine open problem is more valuable than a Testable claim whose test
would produce only weak evidence.

The Technical Companion, Appendix G carries the authoritative theorem-status and interpretation
layer for these statuses: theorem statement, assumptions, intuition,
proof sketch, dependency map, empirical meaning, and weakening or
falsification conditions for each result. Technical Companion, Appendix H, remains the technical proof archive. Chapter
11\textquotesingle s task is narrower: to turn that proof architecture
into empirical designs, status discipline, and credibility tests.

\section{The Identification Protocol}\label{111-identification-protocol}

The empirical claims of this chapter divide into one directly calibratable construct and several that are not. Recombination generation \(G^R\) has a closed form (\S\ref{why-recombination-is-the-governing-mechanism}) and can be proxied and estimated; the suppression, feedback, and welfare claims (Propositions~C, D, and~E) do not, and they must not be allowed to lean on a counterfactual generative frontier that is never observed. This section fixes the identification discipline that governs every such claim, so that the one measurable construct does not silently become the empirical centre of gravity.

\textbf{Shock-based quasi-experiment is primary.} The preferred identification route for every suppression, feedback, and welfare claim is a difference-in-differences design built on a dated access shock, comparing treated actors against matched controls chosen for pre-trend stability and validated with placebo (pre-shock) tests. The estimand is the divergence between treated and control trajectories after the shock, not the level of any reconstructed field. This is the primary and preferred route wherever a credible shock and a credible control group exist.

\textbf{Closed-form and proxy measures are secondary.} Calibrated or proxy measures, \(G^R\) chief among them, are admitted as supporting evidence only. They corroborate, sharpen, or calibrate a result that the treated-versus-control comparison has already established; they never substitute for it, and no claim rests on a proxy as its sole empirical anchor. The closed form of \(G^R\) confers analytical tractability, not evidentiary priority.

\textbf{The counterfactual is procedural, not constructed.} The comparison group is the counterfactual. The protocol does not license inferring the suppression, feedback, or welfare effect from a reconstructed recombination field, a latent generative frontier, or any other modelled object standing in for what the excluded actors would have produced. Where no control group can be defined, the claim is carried at a lower status (Empirically Plausible or Unresolved) rather than rescued by construction.

\textbf{Measurement-ecology caution.} Because \(G^R\) is the only construct with a closed form, it will attract disproportionate empirical weight unless the programme deliberately resists. Calibratability is not importance: the suppression channel (Proposition~C) and the feedback channel (Proposition~D) carry equal evidentiary standing in this theory, and the research programme commits to treating them symmetrically, devoting comparable design effort and data collection to each rather than letting the measurable construct set the agenda. A result that is easy to estimate is not thereby the result that matters most.

This protocol is one identification discipline shared across both volumes. It is the narrative counterpart of the identification-risk column in the falsification matrix of the Technical Companion (Volume~2): that column records, claim by claim, what empirical design could isolate the mechanism from confounders, and holds a claim for which no credible design exists as Empirically Plausible rather than Testable. The protocol stated here and that column state the same requirement in two registers, and should be read together.

\subsection{Pre-registered first experiments}\label{ch11:pre-registered-first-experiments}

The protocol is implemented by fixing a small set of natural-experiment designs in advance, each with its treated and control definitions, outcomes, rival explanations, and falsifier stated before the data are examined. Pre-registration is not a formality here. It also answers the publication-bias concern this chapter raises, by committing the analysis before results are known, and it answers a deterioration in the evidence base itself: clean access shocks are becoming scarcer and more confounded, because security and regulatory-compliance motives now serve as rival explanations for many closures that a decade ago would have read as straightforwardly strategic. The set should therefore be fixed ex ante rather than discovered opportunistically once outcomes are visible.

\begin{enumerate}
\item \textbf{API-closure event study (suppression; Proposition~C).} \emph{Treated:} third-party developers dependent on a platform interface at its dated closure or repricing. \emph{Control:} matched developers on comparable but unaffected interfaces, with stable pre-trends. \emph{Outcome:} downstream dependency diversity, fork and integration counts, cross-domain usage entropy for the excluded actors. \emph{Rival explanations:} obsolescence, quality filtering, security necessity, unrelated demand shifts. \emph{Falsifier:} no differential reduction in field breadth or downstream recombination for treated relative to control after closure.
\item \textbf{Open-source relicensing (commons enclosure).} \emph{Treated:} projects whose licence transitions from open to restricted on a dated event. \emph{Control:} matched projects retaining open licences. \emph{Outcome:} contributor inflow, maintainer retention, dependent-project activity. \emph{Rival explanations:} project maturity, founder exit, security hardening. \emph{Falsifier:} treated and control diverge no more than baseline churn.
\item \textbf{Scholarly-field relocation (suppression outside software; generality test).} \emph{Treated:} subfields whose central venue moves from open to gated access on a dated change. \emph{Control:} matched subfields with stable access. \emph{Outcome:} entrant-author share, citation-class entropy, cross-field recombination. \emph{Rival explanations:} topic life-cycle, funding shifts, fashion. \emph{Falsifier:} no divergence in entrant participation or recombination after relocation.
\item \textbf{Feedback-access asymmetry (feedback capture; Proposition~D).} \emph{Treated actor class:} excluded competitors after a dated telemetry or feedback-enclosure event. \emph{Control:} comparable actors retaining feedback access. \emph{Primary outcome:} the excluded actors' learning slope (benchmark-improvement rate), not the incumbent's improvement rate, which is confounded with scale and learning-by-doing. \emph{Rival explanations:} talent, compute, capital, initial quality. \emph{Falsifier:} the excluded actors' learning slope does not decline relative to control under the exposure regime.
\item \textbf{Enclosure-duration test (strategic over-enclosure; M5.T1).} \emph{Treated:} recombination-intensive incumbents after a dated change enabling longer effective enclosure. \emph{Control:} comparable incumbents in low-\(M_{rec}\) domains. \emph{Outcome:} observed enclosure duration relative to a static cost-benefit benchmark, and downstream entrant recombination. \emph{Rival explanations:} ordinary appropriability incentives, quality or safety justification. \emph{Falsifier:} observed durations match the static benchmark with no systematic right-shift.
\end{enumerate}

These five designs draw on the chapter's existing candidate cases and add no new proposition; they pre-commit the identification so that the programme is disciplined by the protocol rather than by data availability.

\section{Formal-Status Architecture and Technical Companion Appendix G Boundary}\index{formal-status architecture|textbf}\label{112-formal-status-architecture-and-appendix-c-boundary}

This section exists to prevent the empirical chapter from overstating the status of its own tools. The practical problem is that a reader needs to know which objects are proven conditionally, which are measurable candidates, and which still require calibration. Evidence would consist of successful movement from formal object to observable implication, proxy, rival-explanation test, and falsifier.

Chapter 11 uses the Technical Companion, Appendix G as its formal-status boundary. Appendix G states theorem status, assumptions, dependency maps, empirical meanings, and weakening conditions; Appendix H preserves proof status and the technical proof archive. Chapter 11 does not re-prove that apparatus; it translates it into empirical designs, status discipline, and credibility tests.

The main text now keeps only the orientation needed by the reader. The full formal-status ledger has been migrated to the Technical Companion. The Technical Companion, Appendix J preserves the detailed object-by-object table, including open calibration items and proof-status distinctions.

\begingroup
\small
\setlength{\tabcolsep}{3.5pt}
\renewcommand{\arraystretch}{1.15}
\sloppy
\par\addvspace{0.8\baselineskip}\noindent
\begin{longtable}{@{}L{0.24\textwidth}L{0.24\textwidth}L{0.25\textwidth}L{0.21\textwidth}@{}}
\caption{Condensed Formal-Status Architecture and Technical Companion Appendix G Boundary}\label{tab:ch11:formal-status-architecture}\\
\toprule\noalign{}
Object class & Main-text status & Technical location & Chapter 11 discipline \\
\midrule\noalign{}
\endfirsthead
\toprule\noalign{}
Object class & Main-text status & Technical location & Chapter 11 discipline \\
\midrule\noalign{}
\endhead
\bottomrule\noalign{}
\endlastfoot
Core objects: five forms, OKU, recombination field, capability state & Conceptually or internally formalized; empirical validation still required & Technical Companion, Appendices B, C, and G; Volume 2 Formal-Object Register & Treat as measurable candidates, not as already validated empirical variables \\
Theorem system T1--T8 and M5.T1\index{theorem system T1--T8}\index{M5.T1 Strategic Enclosure!theorem system} & Conditional formal results with result-specific proof strength under stated primitives & Technical Companion, Appendices G and H & Translate into observable implications, proxies, rival explanations, and falsifiers \\
Open or reduced-form extensions: D7/M5.P6, D8, selected Model 5 design problems & D7/M5.P6: conditional reduced-form proposition; D8: open structural game with partial diagnostics/conjectures & Technical Companion, Appendix F for the dynamic-enclosure model, Appendix H for proof status, and the open-problem register & Do not write as completed equilibrium or welfare results \\
Measurement and calibration objects: K-CMM, materiality, KCI, key parameters\index{Knowledge-Capital Measurement Model (K-CMM)}\index{Knowledge Capital Index (KCI)} & Operational or ordinal diagnostic\index{ordinal measurement}; calibration open & Technical Companion, Appendix D and measurement apparatus & Use for empirical design and decision discipline, not numerical policy claims \\
\end{longtable}
\endgroup

\section{Cross-Theorem Empirical Backbone}\index{cross-theorem empirical backbone|textbf}\label{113-cross-theorem-empirical-backbone}

This section turns the theorem system into a small number of inspectable empirical claims. The practical problem is orientation: without a compact backbone, the reader sees many mechanisms but not the tests that would make them risk failure. Evidence would appear as observable changes in access, reuse, capability, prices, or productivity that survive comparison with simpler rival explanations.

The empirical programme still needs a theorem-to-test backbone, but the main chapter should not carry the full audit matrix. The main-text task is orientation: identify the KBC claim, the observable implication, the first proxy, the main rival explanation, and the evidence that would weaken or falsify the claim. The full theorem-by-theorem empirical-test matrix has been migrated to the Technical Companion, where it can preserve auditability without overloading the main argument.

This structure disciplines Chapter 11's order of operations. First, identify the claim and the object embedded in it. Second, identify the observable implication. Third, specify a first proxy or measurement route. Fourth, name the rival explanation that must be defeated. Fifth, state the falsifier. Only after those five moves should the chapter discuss cases, data sources, calibration, or policy implications. The shortened table is therefore not a loss of rigour; it is the chapter's object-validation filter applied before the fuller audit machinery preserved in the Technical Companion.

\begingroup
\small
\setlength{\tabcolsep}{2.4pt}
\renewcommand{\arraystretch}{1.15}
\sloppy
\par\addvspace{0.8\baselineskip}\noindent
\begin{longtable}{@{}L{0.20\linewidth}L{0.23\linewidth}L{0.18\linewidth}L{0.19\linewidth}L{0.13\linewidth}@{}}
\caption{Compact Cross-Theorem Empirical Backbone}\label{tab:ch11:compact-cross-theorem-empirical-backbone}\\
\toprule\noalign{}
KBC claim & Observable implication & First proxy & Rival explanation & Falsifier \\
\midrule\noalign{}
\endfirsthead
\toprule\noalign{}
KBC claim & Observable implication & First proxy & Rival explanation & Falsifier \\
\midrule\noalign{}
\endhead
\bottomrule\noalign{}
\endlastfoot
Knowledge access and recombination raise productive capacity & Access expansion or restriction changes output, reuse, citation, dependency, or innovation patterns & Productivity change, reuse intensity, citation diversity, package or API dependence & Scale economies, ordinary capital deepening, demand shocks, or selection into adoption & No differential effect after knowledge-access changes once standard inputs and selection are controlled \\
Governance suppression narrows recombination fields & API closure, licence restriction, patent assertion, or platform gating reduces accessible field breadth\index{platform gating!empirical testing} or useful diversity for excluded actors & API usage, dependency diversity, cross-domain citation entropy, downstream forks or integrations & Obsolescence, quality filtering, security necessity, unrelated demand shift & No measurable reduction in field breadth, useful diversity, or downstream recombination after restriction \\
Intermediate-distance recombination has surplus value & Neither near-identical nor excessively distant combinations produce the strongest innovation or capability gains & Cross-domain citations, interdisciplinary outputs, package co-use, team knowledge-distance measures & Star teams, funding intensity, field fashion, publication bias, or selection effects & No robust distance-surplus relation after controls for quality, field, funding, and selection \\
Balance-sheet and valuation systems underprice knowledge-capital exposure & Market value, abnormal returns, or operating surprises reflect knowledge-bearing stock, governance position, or dark-capital exposure not captured by book assets alone & Market-to-book decomposition, event returns, intangible intensity, platform/API/cyber exposure variables & Monopoly power, speculation, conventional intangibles, risk premia, or accounting noise & Knowledge-capital proxies add no explanatory or predictive power beyond standard valuation variables \\
Learning-loop capture creates capability divergence & Actors controlling deployment feedback improve faster than otherwise comparable excluded or open competitors & Excluded actors' learning-slope divergence (benchmark-improvement rate of comparable non-incumbents under the exposure regime); incumbent release cadence and telemetry access are secondary, being confounded with scale and learning-by-doing & Talent, compute, capital, architecture, brand, or initial quality advantage & Excluded actors' learning slope does not decline relative to matched controls under the exposure regime, or feedback access adds no divergence after controls \\
Enclosure may create dynamic efficiency loss or self-cost & Private enclosure choices can exceed social optimum when recombination, commons, or future-capability costs are excluded from private calculus & Restriction intensity, licensing behaviour, contribution decline, maintainer exit, dependency fragility, redevelopment cost & Necessary incentives, trade-secret protection, security, quality control, or efficient selection & Dynamic costs are too small, fully offset, or absent despite demonstrated dependence on the enclosed or depleted stock \\
OKU and K-CMM measurement claims are empirically useful only if they improve decisions or prediction & Task-equivalence and component-level knowledge-capital measures outperform undifferentiated intangible measures or whole-job substitution claims & Judge-blind task equivalence, workflow substitution, K-CMM component estimates, decision-reversal or valuation sensitivity & Narrow benchmarks, changed task definitions, conventional DCF, Tobin's q, monopoly rents\index{rents!monopoly}, or accounting intangibles & OKU equivalence fails in real workflows, or K-CMM components add no predictive, explanatory, or decision value \\
\end{longtable}
\endgroup

The compact table does not replace the underlying audit. It gives the main text a readable empirical backbone while preserving the full theorem-level matrix for technical inspection. The rest of this chapter expands the rows that are most feasible for first-pass testing and leaves coefficient calibration, parameter estimation, and full theorem-by-theorem auditability to the companion apparatus.

\section{Predecessor Discipline and Originality Boundaries}\label{originality-audit-and-predecessor-boundaries}

This section exists to keep the empirical programme from converting a synthesis into an originality overclaim. The practical problem is that many rival literatures already explain parts of knowledge, information, intangibles, and firm capability. Evidence for KBC must therefore show incremental explanatory value after those predecessor accounts have been taken seriously.

The originality discipline is preserved, but the full audit no longer belongs in the main flow. Chapter 11 needs only the main-text rule: KBC should not claim that economics ignored knowledge. Human capital theory, information economics\index{information economics}, endogenous growth theory, intangible-capital research, RBV/KBV, platform economics\index{platform economics}, commons theory\index{commons theory}, and accounting theory\index{accounting theory} already explain important parts of the terrain. KBC's claim is integrative and mechanism-specific: it links generation, conversion, access, recombination, feedback, enclosure, measurement, and impairment of knowledge-bearing stock.

Within that integrative claim the originality is unevenly distributed, and honesty requires saying where it is least reducible. Two contributions are the primary novelty claims, because component reduction does not dissolve them: the integrated feedback-capture-plus-excluded-field mechanism, which joins incumbent learning-loop capture to suppression of the excluded actors' future field in a single conversion architecture (Proposition~D with Proposition~C), and recursive truth-decay recast as a capital risk, the endogenous depreciation of reliability circulating in the generation cycle (\S\ref{testing-recursive-truth-decay}). For the remaining contributions the project concedes component-level antecedents outright: conditional separability has antecedents in \textcite{Coase1937} and the resource-based view; the recombination field in \textcite{Arthur2009} and \textcite{Weitzman1998}; the feedback advantage in learning-by-doing and data-network-effects; and access suppression in anticommons theory and the economics of spillovers. Conceding these pre-empts the ``sufficient structural equivalence'' challenge, that each component already exists somewhere, which an architecture-level defence does not by itself close; the defensible claim is the integration and the two least-reducible mechanisms, not novelty across the board.

The full originality audit has been migrated to the Technical Companion. The condensed table below keeps the publication discipline visible without turning Chapter 11 into a predecessor register.

\begingroup
\small
\setlength{\tabcolsep}{3.5pt}
\renewcommand{\arraystretch}{1.15}
\sloppy
\par\addvspace{0.8\baselineskip}\noindent
\begin{longtable}{@{}L{0.26\textwidth}L{0.30\textwidth}L{0.20\textwidth}L{0.18\textwidth}@{}}
\caption{Condensed Predecessor Discipline for Major KBC Claims}\label{tab:ch11:originality-audit}\\
\toprule\noalign{}
KBC domain & Main predecessors & Safe KBC claim & Discipline \\
\midrule\noalign{}
\endfirsthead
\toprule\noalign{}
KBC domain & Main predecessors & Safe KBC claim & Discipline \\
\midrule\noalign{}
\endhead
\bottomrule\noalign{}
\endlastfoot
Knowledge as productive stock & Becker, Mincer\index{Mincer, Jacob}, Machlup, Bell, Porat, intangible-capital research & Synthesizes person, artefact, organization, commons, and public infrastructure as knowledge-bearing stock forms & Do not claim knowledge was previously ignored \\
Recombination and non-rival growth & Romer, Weitzman, innovation economics, Hayek, Polanyi & Extends non-rivality into governed recombination fields and useful diversity & Separate scaling from truth-producing generation \\
Enclosure, feedback, and capability divergence & IP, anticommons, platform economics, commons governance, dynamic capabilities & Synthesizes access restriction, feedback capture, and future generation suppression & Preserve dual-effects language: enclosure can fund creation as well as suppress recombination \\
Measurement, valuation, and dark capital\index{measurement versus valuation}\index{valuation!versus measurement} & Lev, Bond--Cummins, Hubbard\index{Hubbard, Douglas}, accounting theory, information economics & Reframes measurement as object validation\index{object validation}, proxy testing, and value-of-information discipline\index{object validation versus coefficient calibration}\index{coefficient calibration!versus object validation} & Do not equate market-to-book residual\index{market-to-book gap!residual}s with knowledge capital \\
Novelty candidates & Cybersecurity impairment, OKU task-equivalence bridge, integrated feedback-capture/excluded-field mechanism & Candidate KBC contributions pending empirical tests & Present as testable or speculative where calibration is absent \\
\end{longtable}
\endgroup

\subsection{Testing residence--governance pairing}\label{testing-residence-governance-pairing}
\index{residence--governance pair}

The residence--governance refinement adds a specific empirical burden. If two knowledge stocks share the same residential form but differ in governance arrangement, KBC predicts, under the relevant access, capability, governance, and maintenance conditions, different yield, recombination, maintenance, and risk profiles. A dataset, model, software repository, routine, standard, or method may be residentially similar while producing different economic outcomes under private IP governance, employment-contract governance, platform terms, commons governance, professional stewardship, or public epistemic governance.

The resulting tests are direct:
\begin{enumerate}
\item \textbf{Same residence, different governance.} Two stocks with the same residential form should show different current yield, recombination option value, maintenance burden, depreciation risk, and impairment exposure when their governance arrangements differ.
\item \textbf{Governance-transition effects.} Governance transitions should produce measurable changes in the actor's recombination field \(F_{a, t}\), useful diversity \(D_u(F_{a, t})\), GATE scores, and K-CMM components.
\item \textbf{Commons and public-infrastructure depletion.} Depletion of commons or public epistemic infrastructure should predict reduced recombination, weaker validation, increased private coordination costs, greater duplication of effort, or a distributed-yield maintenance gap in which broadly harvested yield \(Y^x_t\) is large relative to maintenance reinvestment \(M^x_t\), for \(x\in\{C, P\}\).
\item \textbf{Platform governance shifts.} Platform governance changes should predict field contraction, switching costs, reduced learning-loop access, increased dependency exposure, or dark-capital impairment.
\item \textbf{Cybersecurity capability pass-through.} Cybersecurity events should reduce capability pass-through even where the affected knowledge-bearing stock remains in the firm's possession and remains legally accessible. The expected observable pattern is not only response cost, but lower GATE-C, delayed redeployment, impaired trust, reduced release cadence, interrupted feedback learning, or temporary collapse in \(\rho_{a, j, t}\).
\end{enumerate}

A related test concerns commons and public epistemic maintenance. KBC predicts, under the relevant access, capability, governance, and maintenance conditions, that a stock can show high distributed yield while becoming fragile if maintenance reinvestment is weak. The empirical question is whether generation and maintenance flows \(G^x_t+M^x_t\), or the reinvested share of yield \(s^x_tY^x_t\), are sufficient to offset depreciation \(\delta^x_tK^x_t\). Candidate proxies include maintainer hours\index{maintainer hours}, foundation funding, public statistical budgets, standards-body capacity, patch latency\index{patch latency}, release cadence, documentation quality\index{documentation quality}, benchmark integrity\index{benchmark integrity}, replication infrastructure\index{replication infrastructure}, metrology capacity\index{metrology capacity}, contributor turnover\index{contributor turnover}, and the lag between identified defects and correction. The claim is not that every open or public stock is underfunded. It is that non-rival use can make yield appear abundant while hiding the maintenance flow required to prevent depreciation.

The falsifier is equally important. If governance arrangement adds no explanatory power once residential form, ordinary asset type, firm size, sector, market power, and industry shocks are controlled, then the residence--governance pairing principle is overstated. In that case the five-form taxonomy may remain useful descriptively, but the deeper residence--governance ontology would not have earned independent empirical status.

\section{What Must Be Measured}\label{114-what-must-be-measured}

Before coefficient calibration, KBC requires object validation. The first empirical question is not whether the correct coefficient has been estimated for \(R_{a, t}\), \(D_u(F_{a, t})\), \(G^L\), \(\Omega_i\), or the K-CMM option terms. It is whether these objects correspond to observable economic phenomena rather than relabelled residuals. The methodological sequence is deliberately stricter than ordinary variable naming: object \(\rightarrow\) observable consequence \(\rightarrow\) proxy \(\rightarrow\) rival explanation \(\rightarrow\) directional test \(\rightarrow\) repeated-observation calibration \(\rightarrow\) revision or rejection. Coefficient precision without object validity would give the appearance of mathematical discipline while leaving this theory's central objects unearned.

This sequence protects KBC from becoming a relabelling exercise. A market-to-book residual is not automatically knowledge-bearing stock\index{market-to-book residual versus knowledge capital}\index{market-to-book gap!versus knowledge capital}; abnormal returns are not automatically governance-position exposure; a fall in citations is not automatically recombination-field contraction; and productivity loss is not automatically dark capital. Each empirical claim must first specify what object is being tested, what observable consequence should follow if the object is real, which proxy can measure that consequence, which simpler explanation must be defeated, what directional movement should appear before coefficient calibration, and what result would force revision or discard the claim.

Any KBC empirical programme using published IT-productivity estimates must correct for publication bias, heterogeneity across study designs, and sectoral aggregation. Positive estimates of knowledge-capital effects should not be accepted merely because they fit this theory. \textcite{Polak2014}'s meta-analysis is therefore a methodological warning: KBC claims must survive publication-bias correction rather than relying on selectively favourable estimates.

\begin{table}[htbp]
\centering
\small
\caption{Object-Validation Spine Before Coefficient Calibration}
\label{tab:ch11:object-validity-sequence}
\begin{tabular}{@{}p{0.23\textwidth}p{0.64\textwidth}@{}}
\toprule
Stage & Question \\
\midrule
Define the object & What KBC object is being tested, and what would count as its presence in an economic setting? \\
Identify observable consequences & If the object is real and operative, what should change in access, use, output, prices, capability, reliability, or downstream recombination? \\
Choose proxies & Which measurable indicator can represent that consequence without pretending to observe the object directly? \\
Compare rival explanations & Could the same movement be explained by scale, market power, ordinary intangible capital, sector shocks, R\&D intensity, accounting noise, or management quality? \\
Test directionality first & Does the proxy move in the predicted direction under a relevant shock, comparison, or event window? \\
Calibrate after repeated observation & Are repeated observations stable enough to estimate coefficients, effect sizes, or K-CMM weights? \\
Revise or discard & If the proxy fails, rivals dominate, or signs do not hold, should the claim be narrowed, reformulated, relocated to speculation, or discarded? \\
\bottomrule
\end{tabular}

\end{table}

\subsection{Technical variable registers and table triage}\label{technical-variable-registers-and-table-triage}

The main text now keeps only the measurement sequence needed for empirical orientation. The fuller variable ledgers are preserved in the Technical Companion rather than carried as consecutive tables in Chapter 11. This reduces reader burden without deleting the apparatus needed for audit, replication, or later calibration.

The migrated register preserves the keystone-parameter identification table for \(M_{rec}\), \(D_u(F_{a, t})\), and \(\widetilde{C}_{a, t}\); the additional measurement-object table; and the variable ledgers for recombination fields, learning loops, capability, value and risk, and strategic governance. Chapter 11 therefore uses these objects only where they are necessary to state a test, proxy, rival explanation, or falsifier. The technical companion preserves the full variable-by-variable definitions, proxy candidates, and status classifications.

This table triage follows the same object-validation rule stated above. A variable earns main-text space only when it helps the reader understand what is being tested. A variable earns technical-companion space when it supports auditability, notation discipline, coefficient calibration, or future empirical extension.

\section[Knowledge-Capital Measurement Model Calibration Programme]{Knowledge-Capital Measurement Model Calibration Programme}\index{Knowledge-Capital Measurement Model (K-CMM)!calibration programme}\index{calibration programme|textbf}\label{115-k-cmm-calibration-program}

The purpose of this section is not to estimate K-CMM's coefficients inside the chapter. It is to state what would have to be observed before such estimation is credible. The practical test is whether K-CMM components predict outcomes better than simpler measures such as book value, R\&D spending, headcount, patent counts, or undifferentiated intangible intensity.

K-CMM remains in Chapter 11 because calibration is central to empirical credibility. The detailed calibration notes, coefficient-by-coefficient tests, and longer case catalogue have been moved to the Technical Companion. The main text keeps the practical route: K-CMM begins as an ordinal decision-support model, becomes proxy-cardinal\index{proxy-cardinal measurement} only when its scores predict observed outcomes, and becomes accounting-relevant only if those estimates become stable, auditable, and externally assured.

The immediate calibration question is not whether K-CMM produces a final value for knowledge capital. It is whether its components improve prediction or decision quality relative to simpler proxies such as book value, R\&D spending, headcount, patent counts, market capitalization, or undifferentiated intangible intensity. The operational table below keeps only the minimum calibration route\index{operational K-CMM calibration table} needed in the main chapter; the full coefficient-level apparatus remains in the Technical Companion, Appendix J.

\begingroup
\small
\setlength{\tabcolsep}{2.8pt}
\renewcommand{\arraystretch}{1.12}
\sloppy
\par\addvspace{0.8\baselineskip}\noindent
\begin{longtable}{@{}L{0.18\textwidth}L{0.23\textwidth}L{0.17\textwidth}L{0.19\textwidth}L{0.18\textwidth}@{}}
\caption{Operational K-CMM Calibration Table}\label{tab:ch11:practical-kcmm-calibration-route}\\
\toprule\noalign{}
K-CMM component & Observable proxy & Data source & Rival explanation & Calibration path \\
\midrule\noalign{}
\endfirsthead
\toprule\noalign{}
K-CMM component & Observable proxy & Data source & Rival explanation & Calibration path \\
\midrule\noalign{}
\endhead
\bottomrule\noalign{}
\endlastfoot
Current use value & Productivity, revenue, cost reduction & Firm/project data & Ordinary capital deepening & Repeated cases \\
Recombination option & Downstream reuse, forks, citations, integrations & API, repository, patent, and standards data & Popularity effects & Event studies \\
Learning-loop option & Model or routine improvement rate & Telemetry and version data & Scale alone & Pre/post access tests \\
Control/governance value & Rent capture, switching costs, dependency persistence & Contracts and platform data & Market power & Access-change events \\
Expected knowledge loss & Incidents, decay, rework, maintainer loss & Operational records & Mismanagement & Longitudinal calibration \\
\end{longtable}
\endgroup

\textbf{Claim status.} The K-CMM valuation architecture is \textbf{Internally Formalized} in structure and \textbf{Empirically Plausible} in predicted direction. Its coefficient values, geometric mean functions, portfolio terms, and materiality thresholds remain \textbf{Testable} but uncalibrated. Until domain-level calibration is available, K-CMM may discipline measurement and case design, but it should not be used as a cardinal accounting or welfare measure.

\section{Governance-Transition Case
Studies}\label{116-governance-transition-case-studies}
\index{governance transition}

This section exists because governance transitions create some of the cleanest empirical windows in KBC. The practical problem is that knowledge-bearing stock often changes economic effect when access, control, licensing, platform rules, or institutional stewardship changes, even if the technical content remains the same. Evidence would appear as before/after changes in recombination access, feedback capture, capability retention, valuation, or downstream productivity.

The Conditional Separability Axiom holds that knowledge-capital governance
assignment is an institutional achievement. The most tractable empirical
tests of the KBC framework are therefore event studies around governance
transitions, moments when the institutional arrangement governing a
knowledge stock changes, allowing before/after comparison of
\(D_u\)(\(F_{a, t}\)), \(G^R\), \(G^L\), \(\widetilde{C}_{a}\), and related variables.

Four transition types provide the highest-quality natural experiments.

\subsection{Public research → private
IP}\label{public-research--private-ip}

The Bayh-Dole Act in the United States shifted the governance arrangement
assignment of federally funded university research from \(K^P\) toward
\(K^D_{private}\) through patenting. This transition is well-documented
and spans four decades of follow-up data.

\textbf{Prediction:} Bayh-Dole may reduce \(D_u(F_{a, t})\) for unlicensed downstream actors while increasing commercialization speed within licensed trajectories. Against this, within-trajectory commercialization improvement (T8\textquotesingle s faster-within-fewer-trajectories result) should be observable as faster technology transfer to commercial application within the patented technology class.

\textbf{Evidence status:} A substantial empirical literature exists
\parencite{Mowery2001, Sampat2006, Azoulay2019}. It generally
supports the trajectory-concentration prediction and is mixed on
aggregate recombination-field diversity. The KBC
framework\textquotesingle s specific prediction, that the suppression
ratio is measurable and that \(N_{traj}\) falls while within-trajectory speed
rises, can be tested against this literature. \textbf{Status:
Testable; partially Empirically Plausible.}

\subsection{Commons → platform
dependency}\label{commons--platform-dependency}

Open-source ecosystems that originated as commons knowledge capital
(\(K^C\)) have in multiple cases experienced conversion toward platform
dependency, as a dominant platform firm becomes the primary consumer,
contributor, and de facto governance authority.
Microsoft\textquotesingle s relationship with GitHub post-acquisition,
Google\textquotesingle s relationship with Android, and
Amazon\textquotesingle s relationship with AWS-hosted open-source
projects are candidate cases. These are candidate cases, not settled examples of harmful platform capture.

\textbf{Prediction:} Following material platform entry into open-source
governance, \(D_u\)(\(F_{a, t}\)) should fall for actors who cannot match the
platform\textquotesingle s GATE conditions (access, permission,
interoperability, capability). \(G^L\) from deployment should accumulate
preferentially to the platform. Commons contribution rates from
non-platform actors should decline as COV* extraction makes contribution
less valuable relative to alternatives.

\textbf{Evidence status:} Qualitative case evidence is available
\parencite{Eghbal2020, NagleEtAl2022}. Quantitative measurement of
\(D_u\)(\(F_{a, t}\)) changes around governance-transition events is not yet
in the literature using KBC-specific methods. \textbf{Status: Testable;
currently Empirically Plausible.}

\subsection{Worker expertise → firm
capability}\label{worker-expertise--firm-capability}\index{embodied-to-disembodied conversion}

When expert workers leave a firm, through departure, retirement, or
layoff, their \(K^E\) (embodied knowledge) exits the firm. The
question is whether the firm\textquotesingle s productive capability
declines measurably, and by how much, and over what recovery horizon.

\textbf{Prediction:} Where \(K^E\) is highly concentrated (few workers
hold most of the relevant expertise), departure should produce a
measurable fall in \(\widetilde{C}_{a}\) proportionate to the concentration of the
departed knowledge. T6.6 (Recovery Lag) predicts that the gap does not
close immediately even after replacement hiring, because the replacement
must rebuild \(\widetilde{C}\) through \(G^R\) and \(G^L\) rather than inheriting the
departed worker\textquotesingle s accumulated capability.

\textbf{Evidence status:} Economics of science literature provides
evidence on the impact of superstar scientist departure \parencite{Azoulay2010}. Software engineering literature provides evidence on maintainer
departure in open-source projects. The specific recovery-lag prediction
of T6.6 has not been directly tested. \textbf{Status: Testable;
Empirically Plausible in direction.}

\subsection{User feedback → model
improvement}\label{user-feedback--model-improvement}

AI system deployment generates user interaction data that the deploying
firm uses to improve model capability through fine-tuning, RLHF, and
related feedback mechanisms. This conversion of distributed user
knowledge into proprietary \(K^D\) is the clearest observable instance of
the \(G^L\) capture mechanism.

\textbf{Prediction:} Firms with higher deployment volume (\(Dp_{a, t}\)),
greater feedback diversity (\(F_{dep, a, t}\)), and stronger feedback
absorption (\(\phi_a\)) should produce faster model improvement
post-deployment, controlling for initial model quality and compute. The
improvement differential should widen over time (T6.2), producing
measurable capability divergence between high-\(G^L\) and low-\(G^L\)
actors.

\textbf{Evidence status:} Some empirical evidence on deployment-driven
improvement exists in machine learning literature (Epoch AI data on
training compute and performance). Direct evidence on \(G^L\) mechanism
requires access to internal training logs and deployment data, which are
not publicly available. Event studies around open vs. closed deployment
strategies may provide indirect evidence. \textbf{Status: Testable in
design; partially Unresolved in data availability.}

\subsection{A Public-Data Governance-Transition Test}\label{ch11:governance-transition-test}\index{governance-transition test|textbf}

This section reports a first, pre-registered empirical test of the dynamic prediction at the centre of this book: that an enclosing governance transition does not merely redistribute the value of existing stock but \emph{relocates the recombination field}, the community that generates future knowledge, when a viable open alternative exists. The test uses open-source relicensing events, where the recombination field is unusually observable: every contribution, contributor, and fork is recorded in public version-control history.

\paragraph{What this test does and does not do.} It tests a \emph{mechanism}, not the whole frame, and it tests the \emph{dynamic} relocation prediction, not the realization coefficient \(\rho\) of Chapter~\ref{chapter-9-dark-capital-and-the-accounting-shadow}. A positive result corroborates the dual-enclosure mechanism in this domain; it does not, on its own, establish that capitalism has entered a new stage, and it says nothing about the magnitude of dark capital. The design, cases, outcomes, and expected signs were fixed in a written pre-registration before any data were examined, so the result is a genuine test rather than a pattern selected after the fact.

\paragraph{Design.}\index{HashiCorp relicensing}\index{OpenBao}\index{governance-transition test!case selection} The cleanest case is HashiCorp's August 2023 decision to relicense its entire product suite, Terraform, Vault, Consul, Nomad, and others, from the Mozilla Public Licence to the Business Source Licence on a single date: an enclosure exogenous to the health of the projects, which were dominant and actively developed at the time. Among them, Vault was forked within weeks into a community fork, OpenBao, governed by the Linux Foundation. The recombination field is measured at the repository level, weekly commits, and, decisively, the \emph{identity} of contributors, against a pre-registered set of eight never-treated infrastructure projects that did not relicense. Two further relicensings (Redis in 2024, forked as Valkey; Elasticsearch in 2021, forked as OpenSearch) serve as out-of-sample checks.

\begin{figure}[!htbp]
\centering
\includegraphics[width=\textwidth]{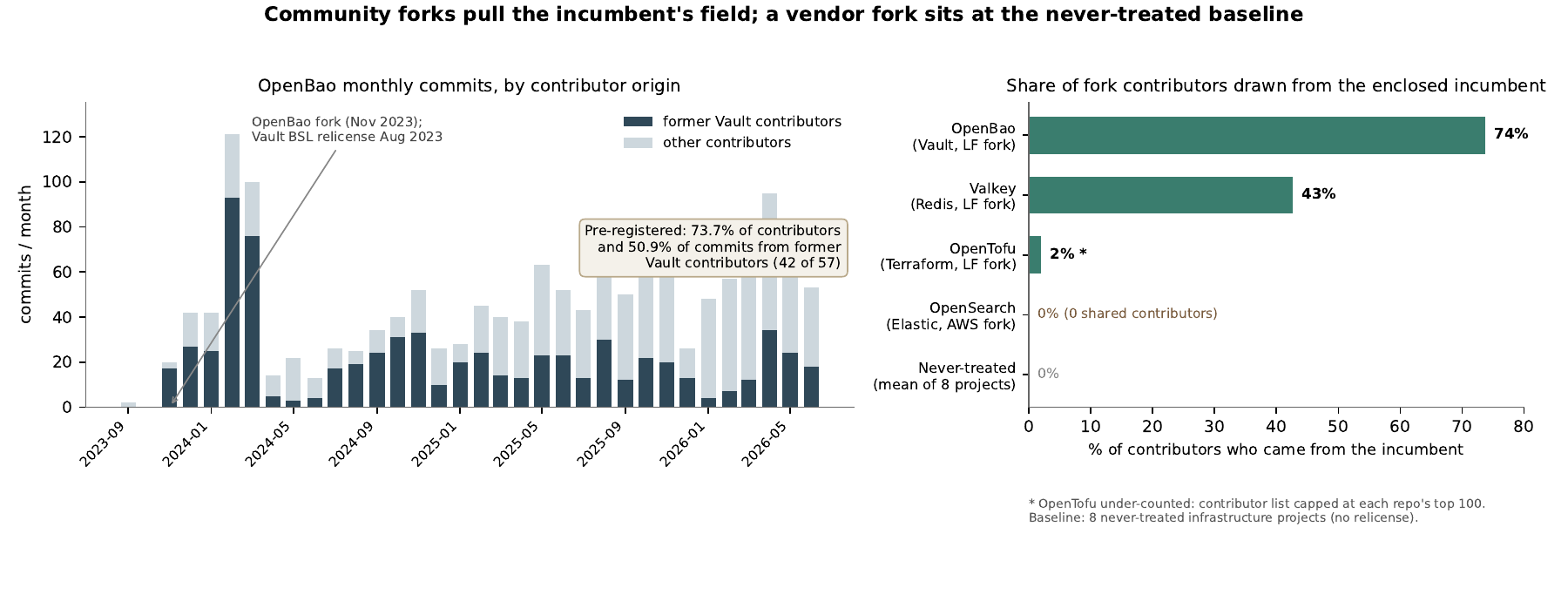}
\caption[Contributor migration to community forks after enclosure]{Contributor migration to community forks after enclosure.}
\label{fig:ch11:openbao-migration}
\par\smallskip\noindent\footnotesize\emph{Note.} Left: monthly commits to the secrets-manager community fork (OpenBao) after its creation, split by whether the contributor had previously committed to the enclosed incumbent (Vault); a majority of the fork's contributors, producing about half of its commits, came from the incumbent. Right: the share of each fork's contributors drawn from its enclosed incumbent, by fork type, community/foundation forks (OpenBao, Valkey) absorbed much of the incumbent's contributor base, a rival-vendor fork (OpenSearch) drew none. OpenTofu is under-counted because the contributor record is capped at each repository's top hundred. Source: public GitHub commit history; protocol in Volume~2, Appendix~J.
\end{figure}

\paragraph{Result: the field relocated to the community fork.} After the relicense, OpenBao, the community fork of Vault, drew \textbf{73.7\,\%} of its contributors (42 of 57) from developers who had previously committed to Vault, and those migrants produced \textbf{50.9\,\%} of the fork's commits (Figure~\ref{fig:ch11:openbao-migration}). The fork itself was stood up about three months after the relicence, a neutral foundation announced it within weeks, and, once active, a majority of its contributors were former incumbent committers. This is the recombination field moving, in person, from the enclosed project to the open fork. Because the migration is identified by contributor identity and tied to the relicence, it is not explained by the incumbent vendor's later acquisition or restructuring, events that do depress raw commit counts but cannot make former incumbent contributors begin building a competing fork.

\paragraph{A second case: Redis to Valkey.}\index{Valkey}\index{Redis relicensing}\index{contributor migration}\index{dark capital!invisibility in downloads} The pattern repeats in a different ecosystem, a different vendor, and a different year, which is what turns one striking case into evidence. When Redis Ltd.\ relicensed Redis away from its permissive (BSD) licence in March 2024, the community fork Valkey was created under the Linux Foundation within two days. Valkey then drew \textbf{42.6\,\%} of its contributors from prior Redis committers, and those migrants produced \textbf{69.4\,\%} of its commits, a larger commit share than at OpenBao, with the median migrant's first Valkey commit about five weeks after the relicence. Two features make this case probative in ways the first is not. First, there is no confounding corporate event: Redis Ltd.\ was not acquired during the window, so the relocation cannot be attributed to anything but the relicence. Second, and central to Chapter~\ref{chapter-9-dark-capital-and-the-accounting-shadow}, the relocation was \emph{invisible in the obvious indicator}. Because Valkey is wire-compatible with Redis, its users keep installing the same client libraries (\texttt{redis} and \texttt{ioredis} on npm), so measured by package downloads almost nothing appears to move; measured by where contributors actually work, a large part of the field had already migrated. An indicator that cannot see the relocation cannot price it, a concrete instance of dark capital, demonstrated rather than asserted.

\begin{figure}[!htbp]
\centering
\includegraphics[width=\textwidth]{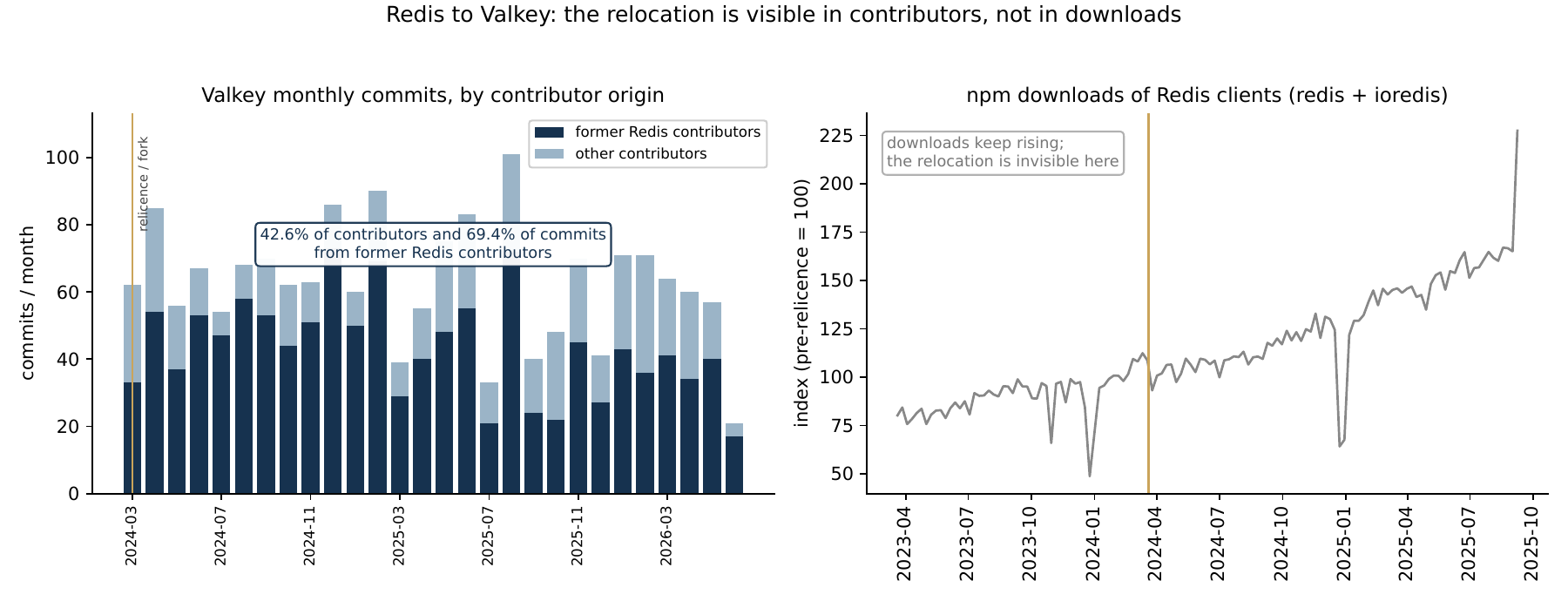}
\caption[The Redis-to-Valkey relocation: visible in contributors, not in downloads]{The Redis-to-Valkey relocation: visible in contributors, not in downloads.}
\label{fig:ch11:valkey-migration}
\par\smallskip\noindent\footnotesize\emph{Note.} Left: monthly commits to Valkey after the fork, split by whether the contributor had previously committed to Redis; former Redis contributors account for most of Valkey's commits. Right: weekly npm downloads of the Redis client libraries (\texttt{redis} and \texttt{ioredis}), indexed to the pre-relicence average; client downloads keep rising, because Valkey is wire-compatible and reuses them, so the relocation leaves no trace in the download series. The field moved; the obvious indicator could not see it. Source: public GitHub commit history and npm download statistics.
\end{figure}

\paragraph{A boundary condition the data force on the theory.}\index{OpenSearch}\index{fork!community versus rival-vendor}\index{boundary condition!field relocation} Elasticsearch's 2021 relicensing produced a fork too, OpenSearch, but \emph{none} of its contributors came from the Elasticsearch community: a rival platform vendor (AWS) created it and staffed it with its own engineers. Relocation of the original community is therefore not automatic upon enclosure. It occurs when the alternative is a credible community- or foundation-governed fork that the existing field can join; a rival-vendor fork relocates maintenance to a new owner without moving the community. Against a null, this is decisive\index{contributor migration!null baseline}: across the eight never-treated control projects, not one shared a single contributor with the enclosed incumbent's prior committers (a background of \(0\%\)), so the rival-vendor fork's zero is exactly what an unrelated project looks like. Measured against that null, the \(42.6\%\) and \(73.7\%\) inflows to the community forks are not the ordinary cross-project movement of infrastructure developers; they are field relocation. We therefore state the test's governing prediction as a conditional hypothesis\index{conditional-relocation hypothesis|textbf}: \emph{an enclosing relicense relocates the recombination field to a credibly community- or foundation-governed fork, leaves it with the incumbent where no viable fork forms, and does not move the original community to a rival-vendor fork.} The rival-vendor result is then not a failure of the theory but the predicted null of this hypothesis.

\paragraph{The full cohort on one measure.}\index{contributor migration!full cohort} Across the five relicensings now in hand, a single statistic carries the comparison: the share of each fork's contributors who had previously committed to the enclosed incumbent, and the share of the fork's commits those migrants produced (Table~\ref{tab:ch11:migration-cohort}). The four relicensings of 2021--2024 carry their pre-registered values; Akka's 2022 relicensing, forked as Apache Pekko, is added here as an out-of-sample case computed on the identical definition. The direction is uniform: every community- or foundation-governed fork drew a substantial part of its builders, and a majority of its development, from the incumbent it replaced, while the lone rival-vendor fork drew none. The head-count share is sensitive to how each fork's contributor set is bounded, OpenTofu is under-counted by the top-hundred contributor cap, and Pekko's high commit share is concentrated in a few very active migrants, so the migrant \emph{commit} share is the robust comparator. On that measure the result is stable: a majority of development relocated in every community fork, and none did in the rival-vendor fork.

\begin{table}[!htbp]
\centering
\caption[Contributor migration after enclosure, five relicensings on one metric]{Contributor migration after enclosure: one metric across five relicensings. The migrant \emph{commit} share is the robust comparator; the head-count share is cap- and window-sensitive.}
\label{tab:ch11:migration-cohort}
\small
\begin{tabular}{@{}L{0.23\textwidth}L{0.20\textwidth}R{0.135\textwidth}R{0.135\textwidth}L{0.115\textwidth}@{}}
\toprule
Incumbent $\rightarrow$ fork & Fork governance & Fork contrib.\ from incumbent & Migrant share of commits & First migrant commit \\
\midrule
Vault $\rightarrow$ OpenBao & Linux Foundation & 73.7\,\% & 50.9\,\% & ${\approx}$3 weeks \\
Redis $\rightarrow$ Valkey & Linux Foundation & 42.6\,\% & 69.4\,\% & ${\approx}$5 weeks \\
Akka $\rightarrow$ Pekko\textsuperscript{\,a} & Apache Foundation & 36.5\,\% & 95.6\,\% & ${\approx}$10 months \\
Terraform $\rightarrow$ OpenTofu & Linux Foundation & 2.0\,\%\textsuperscript{\,b} & 28.4\,\% & ${\approx}$9 months \\
Elasticsearch $\rightarrow$ OpenSearch & Rival vendor (AWS) & 0\,\% & 0\,\% & --- \\
\midrule
Never-treated (8 projects) & --- & 0\,\% & 0\,\% & --- \\
\bottomrule
\end{tabular}
\par\smallskip\noindent\footnotesize\emph{Note.} A fork contributor counts as a migrant if they committed to the enclosed incumbent before the relicence; shares are of the fork's post-creation contributors and commits. The four 2021--2024 relicensings carry pre-registered values; \textsuperscript{a}Pekko is out-of-sample, computed on the identical definition (its commit share is concentrated in a few highly active migrants). \textsuperscript{b}OpenTofu's head-count is under-counted by the top-hundred contributor cap. Fork history is counted from each fork's creation date, excluding inherited upstream commits. Source: public GitHub commit history; protocol in Volume~2, Appendix~J.
\end{table}

\paragraph{Robustness of the migration share.}\index{contributor migration!robustness} The head-count share of migrants is sensitive to how each fork's contributor set is bounded: restricting to a repository's most active contributors yields the shares above, whereas counting every post-fork contributor lowers the head-count share, because each fork also draws a long tail of one-commit newcomers. An independent recomputation over all post-fork contributors finds migrant head-counts near a fifth of the (much larger) contributor set, yet the same migrants still perform a \emph{majority} of each community fork's commits (roughly $55$--$60\%$). The robust measure is therefore the migrant \emph{commit} share, which weights by development actually done, rather than the raw contributor head-count, and on that measure the conclusion is stable: the core of the development field relocated.

\paragraph{Commit volume, reported honestly.} A difference-in-differences estimate on log monthly commits at the incumbent projects, against the never-treated controls, shows pre-treatment trends statistically indistinguishable from the controls and a post-treatment decline. Most of that decline, however, coincides with the vendor's subsequent acquisition rather than the relicense; isolating the months between the relicense and the acquisition leaves a modest incumbent effect of roughly \(-13\,\%\). The commit-volume channel is therefore confounded, and the contributor-migration evidence, which the confound cannot touch, is the load-bearing result.

\begin{figure}[!htbp]
\centering
\includegraphics[width=\textwidth]{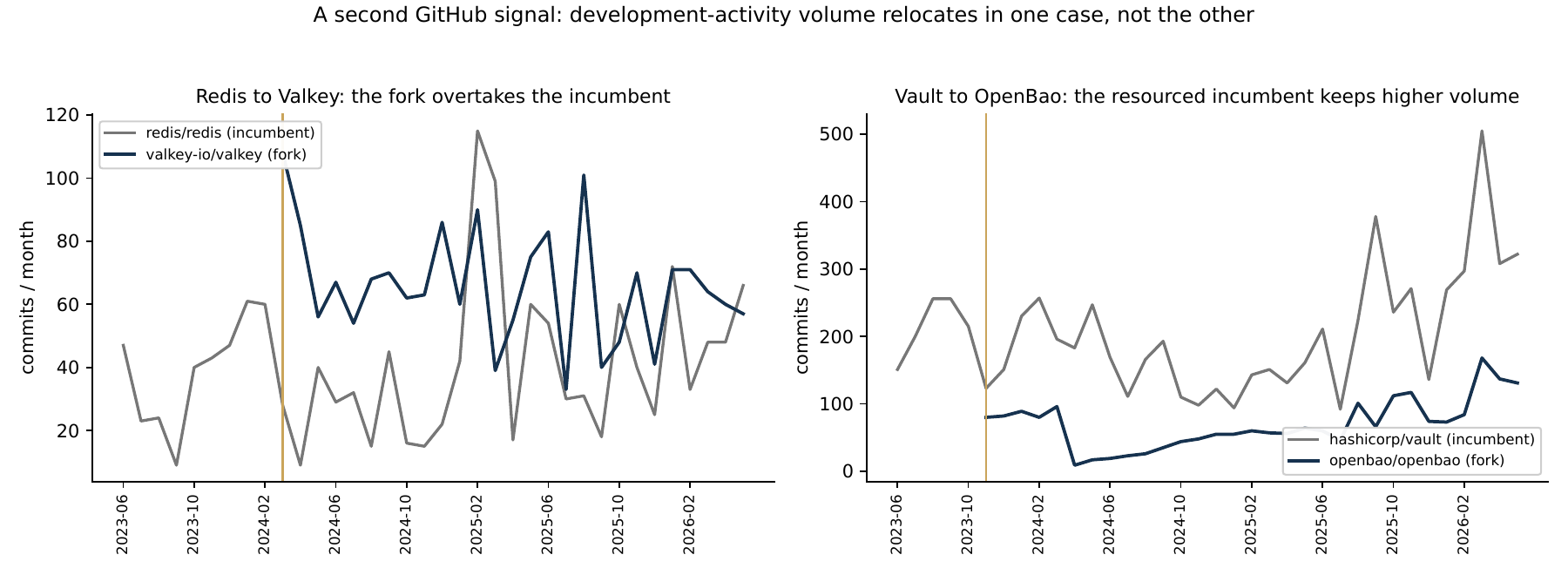}
\caption[A second GitHub signal: where development activity relocates]{A second GitHub signal: where development activity itself relocates, and where it does not.}
\label{fig:ch11:volume-contrast}
\par\smallskip\noindent\footnotesize\emph{Note.} Monthly commit volume at each enclosed incumbent (grey) and its community fork (navy); the gold line marks the fork date, before which the fork repository's history is inherited from the incumbent and is therefore omitted. Left: after the Redis relicence, Valkey's commit volume overtakes Redis's. Right: after the Vault relicence, the well-resourced incumbent (HashiCorp, later IBM) sustains higher raw volume than OpenBao, even though a majority of OpenBao's contributors came from Vault. Commit volume thus reflects resourcing as well as field location, which is why contributor origin, not commit count, is the primary measure. Source: public GitHub commit history.
\end{figure}

\paragraph{Where development activity goes.}\index{development activity!relocation}\index{commit volume!versus contributor identity}\index{field relocation} The same commit record, read as fork-versus-incumbent rather than incumbent-versus-control, shows why volume is a weaker signal than identity (Figure~\ref{fig:ch11:volume-contrast}). In the Redis/Valkey case the fork's monthly commit volume overtook the incumbent's within months. In the Vault/OpenBao case a well-resourced incumbent (HashiCorp, then IBM) retained higher raw volume even though most of the fork's contributors had come from it. Commit counts therefore track resourcing as much as field location, whereas the migration of contributors tracks where the recombination field actually moved. This is the second reason the migration measure, not the volume measure, is load-bearing.

\paragraph{Scoring against the pre-registration.} The two pre-registered success conditions are met. \emph{Feasibility}: the field is measurable from public data and the effects are estimable with confidence intervals. \emph{Content}: the migration share is large, in the predicted direction, and immediate, in the two community-fork cases. The pre-registered \emph{falsifier}, negligible fork inflow with incumbent activity tracking controls, is not triggered in those cases; and the fact that it \emph{is} effectively satisfied in the rival-vendor case (zero migration) shows the test can fail, which is what makes its successes credible.

\paragraph{Status and limits.} This is corroboration of a mechanism under a stated boundary condition, not a calibrated law and not a confirmation of the larger theory. The contributor record is capped at each project's top hundred contributors, which \emph{under}-counts migration in the largest forks; the single-project relicensings are noisier than the multi-project cohort; and the migration measure is descriptive rather than counterfactual, though its timing and magnitude resist benign explanation. What would weaken the claim: community forks that fail to attract the incumbent field despite a credible governance home; or enclosures followed by sustained field growth at the incumbent. What would extend it: server- and deployment-level adoption data, and a larger cohort of relicensings, to estimate how the relocation share varies with the fork's governance form and the incumbent's switching costs.

\subsection{The same mechanism outside software: relocation in scholarly fields}\label{ch11:science-relocation}\index{field relocation!scholarly journals}\index{generality test}
A mechanism that holds only in software is a curiosity; one that recurs across unrelated knowledge communities is a property of knowledge-bearing capital. The sharpest available generality test uses a domain with no code, no licences, and no package managers: scholarly journals. A journal is a knowledge-governance site that coordinates editors, referees, authors, citation networks, reputation, and field identity. When a journal's editorial board resigns en masse over a governance dispute, pricing, open access, or loss of editorial control, and launches or joins a community-, society-, or foundation-governed successor, the configuration is the academic analogue of a community fork after a software relicence: the enclosing incumbent keeps the title, archive, publisher infrastructure, and indexing, while the open question is whether the productive field relocates to the successor.

These events are public and dated, catalogued as ``declarations of independence'' since the 1980s, and the recombination field is observable in open bibliographic data (works, authors, and venues in OpenAlex). Two measures apply. The \emph{primary} measure is governance relocation: the share of resigning editors who appear on the successor's masthead, here called \emph{BoardOverlap}. It is venue-independent and is the direct analogue of the contributor-identity statistic in the software case. The \emph{secondary} measure is publication relocation: whether the board's own output moves from incumbent to successor after the rupture. It is informative but field-dependent, for the reason the machine-learning case makes plain below.

\begin{figure}[!htbp]
\centering
\includegraphics[width=0.92\textwidth]{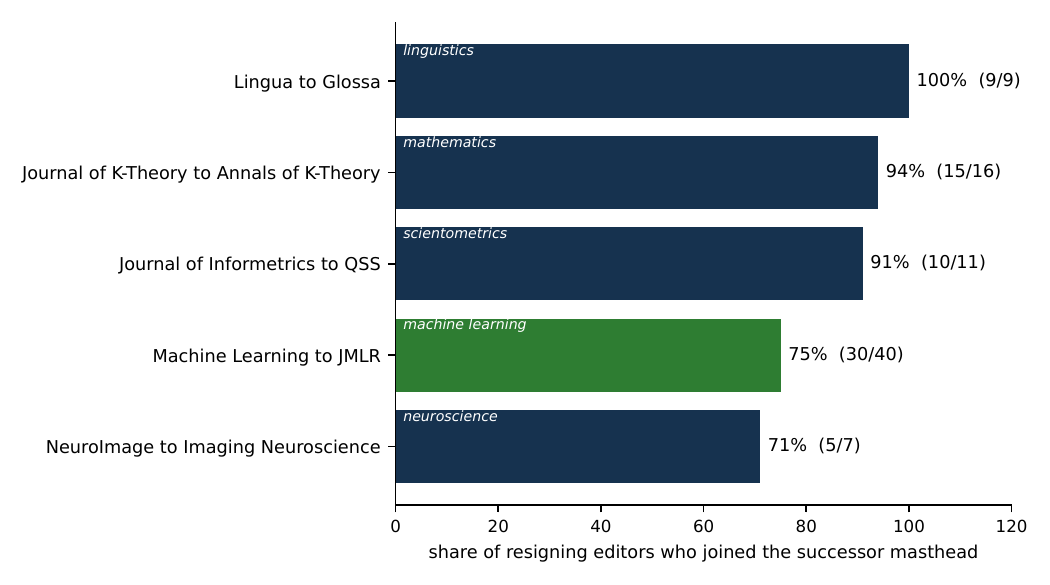}
\caption[Editorial-board relocation across five scholarly fields]{Editorial-board relocation after a governance rupture, five fields, by the governance measure.}
\label{fig:ch11:science-relocation}
\par\smallskip\noindent\footnotesize\emph{Note.} Governance relocation (\emph{BoardOverlap}): the share of resigning editors who joined the successor's masthead, by field. Machine learning (in colour) is the case where the publication measure fails (\(16\%\) of board output moves), because the field publishes at conferences and editors are gatekeepers rather than journal authors; the board nonetheless relocated (\(75\%\)). Boundary cross-check: the earlier 2007 K-Theory rupture shows \(100\%\) board relocation. Source: public OpenAlex bibliographic data; protocol in Volume~2, Appendix~J.
\end{figure}

A preliminary five-field pilot is consistent with relocation in every case (Figure~\ref{fig:ch11:science-relocation}). By the governance measure, the share of the resigning board that joined the successor ranges from \(71\%\) to \(100\%\): Lingua to Glossa in linguistics (\(100\%\)), Journal of K-Theory to Annals of K-Theory in mathematics (\(94\%\)), Journal of Informetrics to Quantitative Science Studies in scientometrics (\(91\%\)), Machine Learning to the Journal of Machine Learning Research in machine learning (\(75\%\)), and NeuroImage to Imaging Neuroscience in neuroscience (\(71\%\)). By the secondary, publication measure, board output relocates cleanly in four of the five fields: linguistics \(88\%\), neuroscience \(81\%\), mathematics \(80\%\), and scientometrics \(55\%\).

Machine learning is the instructive exception, and it is the reason governance is the primary measure rather than publication. There the publication measure collapses to \(16\%\): the resigning editors barely appear as authors in the new journal. This is not a measurement error. Machine learning is a conference-driven field, so journal articles are a poor proxy for where the field works, and editorial board members are gatekeepers, not necessarily authors in their own journal. Yet three-quarters of the Machine Learning board joined the masthead of the Journal of Machine Learning Research. The case that defeats the output measure is captured cleanly by the governance measure; where the two diverge, governance is the one that tracks the community's relocation.

The dark-capital point of the software case reappears here in a second form. After a board departs, the incumbent journal's conventional indicators, its Impact Factor and its annual citation totals, can keep reporting health for years, because the back catalogue continues to be cited and, by construction, a newly launched successor has no Impact Factor at all for its first two or three years. The standard bibliometric instrument therefore reports the incumbent as the field's centre while the living field has already moved, exactly as package downloads reported Redis as healthy while its contributors had migrated to Valkey.

\paragraph{Status.} This is a preliminary, descriptive pilot, not a calibrated result. Some rosters are still partial, several cases have small numbers, and the identity measures are descriptive rather than counterfactual. The claim it supports is correspondingly modest, and it is exactly the generality claim at issue: the recombination-field-relocation mechanism documented for enclosed software projects is not a software artifact. It recurs, under the same boundary condition of a viable community-governed successor, across five unrelated scholarly fields. The event register, the two-measure protocol, and the limitations are recorded in Volume~2, Appendix~J.

\section{Testing General Knowledge-Service Demand}\label{testing-general-knowledge-service-demand}

This section tests the demand-side bridge of KBC: whether actors value knowledge-bearing stock because it yields expected productive services under particular governance and capability conditions. The practical problem is to separate demand for a legal or accounting category from demand for a service flow. Evidence would be price, wage, licence, adoption, or usage patterns that track expected productive service after rival explanations are controlled.

Chapter 2 defined the simplified general demand function:

\[
D_{a, t}(K_i^x)=f\left(\mathbb{E}[PS_{a, i}], P_i, C_{a, i}, g_i, \mathrm{Sub}_i, U_i\right), \quad x\in\{E, D, I, C, P\}.
\]

The compact Chapter 11 test is this: demand should attach to expected productive-service flows, not merely to the legal form, accounting category, or residence form of the stock. Observable proxies include hiring demand, wage premia, licence fees, software subscriptions, API usage, consulting spend, acquisition activity, dataset purchases, open-source dependency investment, and internal-routine investment. The expected sign is:

\[
\frac{\partial D_{a, t}(K_i^x)}{\partial \mathbb{E}[PS_{a, i}]} > 0.
\]

This test has four practical subquestions, but they are treated here as one empirical bundle rather than as a mini-chapter. First, does expected productive-service flow explain demand better than asset-form labels alone? Second, when does disembodied stock substitute for embodied expertise, and when does it create complementary demand for oversight, interpretation, liability-bearing judgement, or domain adaptation? Third, do expected productive-service flows predict prices, wages, acquisition premia, adoption, or usage after controlling for cost, uncertainty, sector, brand, and capability? Fourth, do governance-form changes alter demand and valuation even where the technical content of the stock is unchanged? The fuller four-part demand-test apparatus is preserved in the Technical Companion, Appendix J migration register.

The falsifier is also compact: if expected productive-service flows and governance conditions add no explanatory power once ordinary labour category, accounting class, R\&D, market size, brand, technical quality, and price are included, then the demand-side capital-service bridge should be narrowed or rejected.

\section[Testing domain-denominated embodied knowledge capital]{Testing Domain-Denominated Embodied Knowledge Capital}\label{testing-domain-denominated-embodied-knowledge-capital}

This section exists to test whether embodied knowledge capital is more than a restatement of labour quality. The practical problem is to identify domain-specific productive service flows in wages, credentials, performance, substitution boundaries, and failure costs. Evidence would show that domain-denominated skill stock predicts outcomes beyond generic education, tenure, labour category, or occupational title.

This test cluster asks whether embodied knowledge capital can be observed through domain-specific service demand. Relevant proxies include wage premiums, credential premiums, training duration, certification barriers, task-performance differentials, error rates, malpractice or failure costs, substitution by \(K^D\) systems, and employer demand for specialized roles. The empirical object is not generic labour quality. It is domain-denominated embodied knowledge capital, \(K^E_d\), understood as the productive skill stock required to perform a defined domain service under stated conditions.

Four measurable claims follow.

\textbf{Claim 1: Domain service demand should be visible in price and wage premia.} If \(K^E_d\) yields scarce productive services, actors should pay more for it where the expected service yield \(Y_d\), scarcity \(S_d\), and failure cost \(Risk_d\) are high. Observable proxies include domain wage premia, consulting fees, surgical fees, expert-witness rates, cybersecurity-role premia, and employer vacancy duration for specialized roles.

\textbf{Claim 2: Formation costs should proxy the difficulty of producing the unit.} Training duration, apprenticeship requirements, supervised-practice hours, certification barriers, examination difficulty, and licensure requirements should correlate with the cost and scarcity of producing \(K^E_d\). The expected sign of certification is empirical rather than automatic: certification may preserve quality and trust, but it may also restrict entry.

\textbf{Claim 3: Yield should be observable through performance and error-risk differences.} If \(K^E_d\) is a productive capital stock, higher-quality domain knowledge should appear in task-performance differentials, lower error rates, faster resolution times, improved recovery outcomes, reduced malpractice or failure costs, and greater reliability under ambiguous conditions.

\textbf{Claim 4: Substitution by \(K^D\) systems should reveal the boundary of embodied knowledge.} AI systems, software tools, protocols, templates, and other disembodied artefacts may substitute for some embodied service flows. The substitution boundary can be tested by comparing task classes where \(K^D\) systems achieve functional equivalence with \(K^E_d\) against task classes where embodied judgement, liability-bearing decision-making, contextual interpretation, or manual execution remain necessary.

A deliberately minimal baseline demand function is:

\[
D_a(K^E_d)=f(Y_d, P_d, S_d, \mathrm{Cert}_d, \mathrm{Sub}_d, \mathrm{Risk}_d, \pi_d).
\]

Here \(Y_d\) is expected service yield, \(P_d\) is price, wage, consulting fee, training cost, or acquisition cost, \(S_d\) is scarcity of qualified domain capability, \(\mathrm{Cert}_d\) is the credential, licence, certification, or professional-entry constraint, \(\mathrm{Sub}_d\) is substitutability by \(K^D\), \(K^I\), AI systems, routines, lower-skilled labour, or other capital forms, \(\mathrm{Risk}_d\) is expected loss if the domain service is absent, wrong, delayed, or low-quality, and \(\pi_d\) is the relevant governance or professional arrangement. This expression is intentionally minimal. The expanded KBC demand function belongs in advanced valuation or governance modelling; this baseline is meant to be estimable with ordinary labour-market, credentialing, performance, and substitution data.

\section[Testing the Operative-Unit Decomposition]{Testing the Operative-Unit Decomposition:\\ the Institutional-Residue Prediction}\label{testing-operative-unit-decomposition}

This section tests whether the Operative Knowledge Unit earns its place as a measurement object. The practical problem is that headcount alone cannot tell whether capability resides in people, routines, documentation, role systems, or institutional structure. Evidence would appear when organizations with similar personnel losses retain different capability because their operative units were differently institutionalized before the shock.

\paragraph{Causal role.} This section supplies the test that makes the Operative Knowledge Unit load-bearing rather than expository. Most KBC predictions, governance-form conditioning, field contraction, feedback capture, and cyber pass-through, run on the recombination field, the yield decomposition, and the governance forms; they do not require the operative unit as a unit. The OKU earns its place through one structural claim that no other object in the framework makes: \(\text{OKU}^{I}_{s, d, t}\neq\sum_a\text{OKU}^{E}_{a, d, t}\) (Chapter~2). An institutionalized operative unit is not the sum of the embodied units that occupy it. That inequality yields a falsifiable prediction that cannot be stated without classifying and counting operative units by residence.

\textbf{The prediction.} Consider a productive organization, a firm, laboratory, clinical department, or open-source project, and a personnel shock that removes a fraction of its embodied operative units \(\text{OKU}^{E}\): departures, poaching, retirements, or an acquisition that loses staff. Two accounts give different predictions for the productive capacity that survives.

\begin{itemize}
\tightlist
\item
  \emph{Headcount account} (capability \(=\sum_a\text{OKU}^{E}\)): retained capacity falls roughly in proportion to the embodied units removed.
\item
  \emph{Operative-unit account} (capability also contains structurally independent \(\text{OKU}^{I}\)): retained capacity falls by less than proportionally, and the surviving residue is the institutionalized stock, the operative units lodged in routines, role-systems, documentation, and governance rather than in persons.
\end{itemize}

\noindent KBC therefore predicts that post-shock retained capacity is a function of the organization's institutionalization share, not of headcount loss alone. Two organizations with the same \(\sum_a\text{OKU}^{E}\), identical headcount and identical pre-shock output, but different \(\text{OKU}^{I}\) shares, will show different capability retention after an equivalent personnel shock.

\textbf{Why the test requires the unit.} The prediction is formed by classifying each operative unit as person-bound (\(\text{OKU}^{E}\)-type) or structure-bound (\(\text{OKU}^{I}\)-type) and counting the structure-bound residue. The recombination field \(F_{a, t}\), the yield decomposition \(\omega\,\chi\,\rho\,m^{\mathrm{want}}\), the governance forms, and even the aggregate capability stock \(\widetilde{C}_{a, t}\) all deliver the \emph{level} of capability; none delivers its operative-unit \emph{decomposition}, and so none predicts the \emph{shape} of the capability-loss function under personnel removal. The \(\text{OKU}^{I}\neq\sum\text{OKU}^{E}\) distinction is what supplies that shape. This is the prediction that the operative unit, and only the operative unit, makes.

\textbf{Design.} An event study around personnel-loss events\index{event study!personnel-loss events}: mass departures, failed acqui-hires, maintainer exits, departmental reorganizations. The dependent variable is retained productive capacity, measured as output per remaining unit, throughput, defect or complication rate, release cadence, or project liveness. The key regressor is an institutionalization index, measured \emph{before} the shock as a proxy for the \(\text{OKU}^{I}\) share: documentation coverage, routine and process-maturity certification, role-system redundancy, cross-training depth, and succession depth. Controls include the headcount fraction lost, pre-shock output, sector, and organization size.

\textbf{Prediction, stated for estimation.} Retained capacity rises with the pre-shock institutionalization index, holding the headcount fraction lost fixed; equivalently, the elasticity of capability loss with respect to headcount loss is decreasing in the \(\text{OKU}^{I}\) share.

\textbf{Falsifier, the unit-earns-its-keep boundary.} If post-shock retained capacity is explained by headcount loss alone, so that the institutionalization index adds no predictive power once headcount loss, output, and sector are controlled, then \(\text{OKU}^{I}\neq\sum\text{OKU}^{E}\) is empirically idle, the operative-unit forms collapse to a headcount denominator, and the OKU apparatus should be treated as expository rather than measurement-bearing. If instead institutionalization predicts retention beyond headcount, the operative unit has earned its status as a measurement object. This is the falsification boundary that disciplines the unit against the conceptual inflation Chapter~2 warns of.

\textbf{Two extensions that reuse the same unit logic.} The same decomposition makes two further predictions precise, so the operative unit pays off in more than one setting.

\begin{itemize}
\tightlist
\item
  \emph{Commons-maintainer extraction.} When an incumbent hires away a commons project's \(\text{OKU}^{E}\) maintainers (Proposition~B's commons-depletion corollary), KBC predicts, under the relevant access, capability, governance, and maintenance conditions, \emph{which} commons collapse, those whose productive capability was lodged in persons (low \(\text{OKU}^{I}\) share), and which survive, those whose contribution norms, review systems, and governance persist as \(\text{OKU}^{I}\). The operative-unit decomposition is what makes this an ex-ante prediction rather than a post-hoc description.
\item
  \emph{Acquisition capability transfer.} KBC predicts, under the relevant access, capability, governance, and maintenance conditions, that acquired capability survives integration in proportion to its \(\text{OKU}^{I}\) share: structure transfers, embodied units walk. This operationalizes the long-standing puzzle of why capability-motivated acquisitions so often fail to transfer the capability they were bought for.
\end{itemize}

Together these make the operative unit falsifiable in three distinct settings, organizational personnel shocks, commons-maintainer extraction, and acquisition integration, through the single structural claim \(\text{OKU}^{I}\neq\sum\text{OKU}^{E}\). The unit is therefore not decorative: at least one class of KBC prediction cannot be formed without it.

\subsection{Revealed-Ratio Test: Operative-Unit Coverage and Its Automation Drift}\label{revealed-ratio-test}

A second test treats observed staffing ratios as a revealed estimate of the operative-unit coefficient (Chapter~2) and asks whether the coefficient behaves as the OKU framework predicts. Two designs follow.

\textbf{Deviation predicts outcomes.} Within a sector, treat the prevailing coverage ratio\index{revealed operative-unit ratios}, IT operative units per supported user or system, and security operative units per IT unit, as the operative-unit norm, and measure each organization's deviation from it. KBC predicts, under the relevant access, capability, governance, and maintenance conditions, that under-coverage relative to the norm raises expected knowledge-capital loss: longer incident-response times, lower release cadence, higher defect or breach rates, and slower recovery, that is, lower \(\rho_{a, j, t}\) and higher expected knowledge loss (Chapter~9). The design regresses outcome measures on coverage deviation, controlling for output, sector, system complexity, and automation level. \emph{Falsifier:} if deviation from the sector ratio carries no outcome information once size and complexity are controlled, the ratio is a budget convention rather than an operative-unit coefficient, and this use of the OKU is not earned.

\textbf{Automation drift.} Because a disembodied operative unit can reach task-class equivalence with an embodied one (the Turing-style bridge of Chapter~2), coverage ratios should not be constant. As \(\text{OKU}^{D}\), automation, monitoring tooling, and AI agents, absorbs routine operations, the embodied coverage ratio should fall, one IT professional supporting more users, one security professional covering more systems, in the specific task classes where equivalence is reached, and should not fall in task classes where embodied judgement, liability-bearing decision, or ambiguous escalation remain necessary. The design tracks coverage ratios against automation-adoption measures over time and across firms. \emph{Falsifier:} if ratios are flat under rising automation, or fall uniformly across task classes rather than selectively in the substitutable ones, the task-class substitution claim and its operative-unit denominator are not doing the work attributed to them. The drift in the ratio is itself a measurement of \(\text{OKU}^{D}\)-for-\(\text{OKU}^{E}\) substitution, which is why the ratio, and not merely the substitution claim, requires the unit.

This pairs with the institutional-residue test above: the residue test isolates the \(\text{OKU}^{I}\) component through personnel shocks, while the ratio test isolates the \(\text{OKU}^{E}\) coefficient and its substitution by \(\text{OKU}^{D}\). Together they make the operative-unit decomposition measurable from data that organizations already generate.

\section{Testing Proposition C (Generative
Suppression)}\label{117-testing-proposition-c-generative-suppression}
\index{knowledge generation}

This section tests the claim that some enclosure events alter future generation, not merely current distribution. The practical problem is to distinguish ordinary access loss from loss of material recombination inputs, learning loops, or capability pathways. Evidence would be a measurable decline in useful diversity, downstream recombination, or excluded-actor generation that cannot be explained by obsolescence, quality filtering, security necessity, or demand shifts.

Proposition C (Generative Suppression): Proposition C does not claim
that enclosure only redistributes existing value. It claims that, when
the enclosed stock is a material input into future recombination,
enclosure can change the future generation rate of excluded actors by
altering access to the relevant recombination field.

This is one of the KBC framework\textquotesingle s central claims, but it is
conditional rather than ideological. The empirical question is not
whether enclosure is always harmful. It is whether specific governance
changes measurably reduce access to material recombination inputs,
learning loops, or capability pathways. The testable prediction is that,
under those conditions, enclosure reduces the recombination generation
rate of excluded actors beyond what can be explained by the direct
removal of the enclosed knowledge stock alone.

\subsection{Testable predictions from Proposition
C}\label{testable-predictions-from-proposition-c}

\textbf{P-C.1:} Following an IP enclosure event (patent grant,
trade-secret assertion\index{trade secrecy!empirical testing}, API closure), the cross-domain citation
diversity of actors who depended on the enclosed knowledge should fall
more than can be explained by simple loss of access to the enclosed
stock alone. The excess fall is the field-contraction cascade predicted
by the suppression ratio.

\textbf{P-C.2:} The fall in \(D_u\)(\(F_{a, t}\)) for excluded actors should
be correlated with the position of the enclosed knowledge stock in the
recombination field, stocks that serve as bridges between otherwise
disconnected domains should produce larger suppression effects when
enclosed than domain-specific stocks of similar productive value.

\textbf{P-C.3:} The fall in \(G^R\) for excluded actors should be larger
in fields where the enclosed stock was a recombination complement (a
stock that enables combinations with many other stocks) than in fields
where it was a recombination substitute (replaceable by alternative
stocks).

\subsection{Proposed test: USPTO cross-domain citation
analysis}\label{proposed-test-uspto-cross-domain-citation-analysis}

Data: USPTO patent grant data (publicly available); citation networks;
patent technology class structure.

Design: Identify IP enclosure events\index{IP enclosure!empirical testing} (continuation patent assertion,
inter-partes review outcome, trade-secret litigation settlement) that
restrict previously accessible knowledge. Measure cross-domain citation
diversity (entropy) for citing actors before and after the enclosure
event, relative to a control group of actors who did not depend on the
enclosed knowledge. Test whether the diversity decline exceeds what
direct access loss predicts.

Identifying assumption: enclosure events are exogenous to the innovation
trajectories of citing actors in the treatment group. This is most
plausible for continuation-patent enclosure of previously open research
and for trade-secret assertion following prior publication.

\textbf{Claim status:} Proposition C is \textbf{Internally Formalized.}
P-C.1 through P-C.3 are \textbf{Testable.} Prior empirical literature on
patent thickets and citation diversity\index{patent thickets!citation diversity} \parencite{NoelSchankerman2013, Williams2013} is \textbf{Empirically Plausible} as supporting evidence
but does not directly test the field-contraction cascade prediction.

\section[Testing feedback-enclosure and capability divergence]{Testing Proposition D and T6\index{Proposition D!empirical test}\index{T6 Learning-Loop Capture!empirical test} (Feedback-Enclosure and Capability Divergence)}\label{118-testing-proposition-d-and-t6-feedback-enclosure-and-capability-divergence}
\index{feedback enclosure}

This section asks whether control over feedback produces measurable capability divergence. The practical problem is that incumbents may improve faster not because their starting stock is larger, but because deployment gives them privileged learning signals unavailable to others. Evidence would be faster post-deployment improvement among actors with feedback access after controlling for compute, talent, scale, initial quality, and demand.

Proposition D (Feedback-Enclosure): Enclosed deployment produces
exclusive feedback, which converts into \(G^L\), which accumulates in
\(\widetilde{C}_{inc, t}\), which accelerates the incumbent\textquotesingle s
trajectory while excluded actors lose access to the same learning
signal.

T6.2 (Capability Divergence): \(\Delta_t\) = \(\widetilde{C}_{inc, t}\) − \(\widetilde{C}_{ent, t}\) grows
strictly over the enclosure period.

T6.5 (Capability Trap): \(\widetilde{C}_{ent, t}\) approaches \(\widetilde{C}_{min}\) as \(t\to\infty\) only when learning from the enclosed stream is unavailable, substitute learning is insufficient, and \(G^R_{ent, t}\to 0\).

T6.6 (Recovery Lag): \(\Delta(t_R+\tau)=\Delta(t_R)\cdot(1-\delta_C)^{\tau}\) + Σ{[}...{]}
,  the capability gap does not close immediately after enclosure ends.

\subsection[Testable predictions from learning-loop capture]{Testable predictions from T6}\label{testable-predictions-from-t6}

\textbf{T6.P1\index{T6.P1}:} Firms with enclosed deployment (closed APIs, proprietary
feedback pipelines, restricted user interaction data) should exhibit
faster within-capability improvement than firms with open deployment,
controlling for initial capability level, compute, and talent.

\textbf{T6.P2:} The capability gap between incumbents with enclosed
\(G^L\) and entrants without access should widen over the enclosure
period and exceed what initial capability differences or investment
differences predict.

\textbf{T6.P3:} Following enclosure removal (API opening, model release,
open-sourcing of previously closed system), convergence toward the
incumbent\textquotesingle s capability level should be slower than the
initial divergence rate, the recovery lag prediction of T6.6.

\textbf{T6.P4\index{T6.P4}:} Industries with higher feedback diversity
(\(F_{dep, a, t}\)), larger, more heterogeneous user populations, 
should show faster capability improvement for enclosing firms, because
\(G^L\) is a positive function of feedback diversity.

\subsection{Proposed test: AI deployment performance
benchmarks}\label{proposed-test-ai-deployment-performance-benchmarks}

Data: AI model benchmark performance data (MMLU, HumanEval, HELM, and
sector-specific benchmarks); deployment date records; open vs. closed
deployment classification (from model cards and API documentation);
compute estimates (Epoch AI).

Design: Compare benchmark improvement rates for open- and closed-deployment models. Open deployment here means external actors can generate, inspect, or contribute learning signals; closed deployment means feedback is captured primarily by the deploying firm. Control for training compute and initial benchmark score, then test whether closed-deployment models improve faster post-deployment,
consistent with T6.P1. Identify model release events (previously closed
model released as open weights) and test whether benchmark improvement
rates converge more slowly than they diverged, consistent with T6.P3.

\textbf{Claim status:} T6 is \textbf{Internally Formalized.}
T6.P1--T6.P4 are \textbf{Testable} in design but partially
\textbf{Unresolved} in data availability because internal
feedback-pipeline architecture is not publicly observable. The
benchmark-comparison design provides an indirect test. Direct test
requires deployment-feedback data that is not currently public.

\section[Testing strategic over-enclosure]{Testing M5.T1 (Strategic Over-Enclosure)\index{M5.T1 Strategic Enclosure!empirical test}\index{strategic over-enclosure!empirical test}}\label{119-testing-m5t1-strategic-over-enclosure}

This section tests whether privately rational enclosure can persist beyond the duration or intensity justified by static incentive benefits. The practical problem is to compare observed enclosure choices with counterfactual governance arrangements that preserve incentives while reducing recombination loss, capability divergence, and trajectory narrowing. Evidence would be a measurable over-enclosure premium that survives comparison with quality, coordination, security, disclosure, and maintenance benefits.

M5.T1 (Strategic Over-Enclosure): Under H1--H4, \(T^*_{strategic}>T^*\), incumbents choose enclosure durations that exceed
the socially optimal level because private payoffs internalize dynamic
revenue from \(G^L\) and T8 trajectory control while externalizing \(C_{T2}\),
\(C_{T7}\), \(C_{T6}\), and \(C_{T8}\). This is not an empirical proof that enclosure generally reduces welfare. It is a conditional theorem showing that strategic over-enclosure follows when recombination-loss and field-contraction effects are sufficiently large relative to incentive, coordination, quality-control, disclosure-protection, and maintenance benefits.

The proof of M5.T1 is conditional on H1--H4:

\begin{itemize}
\tightlist
\item
  H1: \(M_{rec}\) \textgreater{} 1 (recombination market is present and
  productive)
\item
  H2: Dynamic revenue is positive at \(T^*\)
\item
  H3: Static costs ≤ static rent
\item
  H4: Dynamic revenue \textgreater{} \(C^{self}_{rec}\) at \(T^*\)
\end{itemize}

D7 now clarifies the self-cost term that H4 must be compared against. It
decomposes \(C^{self}_{rec}\) into commons-depletion and
spillover-suppression channels\index{spillovers!suppression}, introduces \(\lambda_C\) and
\(\omega_S\), and states the reduced-form threshold condition under
which enclosure eventually becomes self-undermining. Chapter 11
therefore treats D7/M5.P6 as a conditional reduced-form proposition, not as an
unproved candidate and not as a full structural N-actor equilibrium
result. The latter remains D8\textquotesingle s open problem.

\subsection[Testable predictions from strategic over-enclosure]{Testable predictions from M5.T1}\label{testable-predictions-from-m5t1}

\textbf{M5.T1.P1:} In industries that satisfy H1--H4, the observed
distribution of IP enclosure durations (patent maintenance, trade-secret
claims, proprietary data retention) should be systematically
right-shifted relative to the distribution predicted by private static
cost-benefit analysis alone. The right-shift is the strategic
over-enclosure premium.

\textbf{M5.T1.P2:} Firms with higher dynamic revenue from feedback
capture (higher \(G^L\)) should maintain enclosure longer, consistent
with dynamic revenue as the source of the \(T^*_{\mathrm{strategic}}\) \textgreater{}
\(T^*\) gap.

\textbf{M5.T1.P3:} The social cost components (\(C_{T2}\) + \(C_{T7}\) + \(C_{T6}\) +
\(C_{T8}\)) should be negatively correlated with industry-level trajectory
diversity (\(N_{traj}\)) and positively correlated with industry-level
concentration, controlling for technology characteristics.

\textbf{M5.T1.P4:} M5.P6 (Self-Undermining Enclosure) predicts that ∃
finite T̂ \textgreater{} \(T^*_{\mathrm{strategic}}\) where \(\pi_inc\) is declining. This
predicts observable cases of enclosing incumbents whose
recombination-generation rate falls below what it would have been under
open access, eventually reducing their own productive capacity. Long-run
performance of highly enclosing incumbents relative to comparable
non-enclosing incumbents should show convergence or decline consistent
with this prediction.

\subsection{Proposed test: patent term utilization and continuation
behaviour}\label{proposed-test-patent-term-utilization-and-continuation-behaviour}

Data: USPTO patent maintenance fee payment data (whether patents are
maintained to full term or abandoned early); continuation patent filing
rates by technology class; patent citation network measures of enclosure
breadth.

Design: Estimate static cost-benefit model of optimal patent maintenance
from revenue data. Compare predicted abandonment rates to observed
rates. Test whether the gap (observed maintenance longer than static
model predicts) correlates with dynamic revenue indicators: post-grant
product improvement rates, deployment growth, and feedback
infrastructure investment. Test whether continuation-patent behaviour
predicts field-diversity decline consistent with P-C.1.

\textbf{Claim status:} M5.T1 is \textbf{Internally Formalized} under
H1--H4. M5.T1.P1--P4 are \textbf{Testable.} H1 and H2 are
\textbf{Empirically Plausible} across many technology industries, but \(M_{rec}\) is the empirical hinge rather than a settled premise. H3 and
H4 require case-by-case assessment. The patent utilization test is the
most tractable first step because the data is available. It is data-accessible and tractable, though identification depends on matched controls and plausibly exogenous governance shocks.

\subsection{Capture-rent test}\label{capture-rent-test}

If \(KR^{\mathrm{capture}}_{\mathrm{net}}\) is material, firms with higher
enclosure rents should invest more in maintaining legal, regulatory,
technical, standards-based, or platform-governance barriers. Possible
proxies include lobbying intensity, standards-body participation, patent
continuation behaviour, copyright or patent extension campaigns, API
restriction events, data-portability resistance, licensing complexity,
and regulatory-comment participation.

\textbf{Falsifier:} If KnowledgeRent is fully explained by productive
service, ordinary access value, and enforcement costs, with no
independent contribution from legal, regulatory, standards, or
platform-governance maintenance, \(KR^{\mathrm{capture}}_{\mathrm{net}}\) should be
omitted.

\subsection{Strategic-capture equilibrium
calibration}\label{strategic-capture-equilibrium-calibration}

This subsection translates strategic enclosure into an estimable gap rather than a metaphor about market power. The practical problem is to compare a privately stable strategy with a social-yield benchmark. Evidence would require estimates of private gain, social-yield loss, recovery lag, and governance-cost differentials around observable governance transitions.

The Nash condition in Chapter 8 gives the empirical programme a sharper
object than ``excessive enclosure'' in general. These expressions are calibration targets, not yet estimators. The object to estimate
is the divergence between the privately stable strategy profile and the
welfare-maximizing strategy profile:

\begin{equation*}
s_i^* \in BR_i(s_{-i}^*)
\end{equation*}

and:

\begin{equation*}
s^W \in \operatorname*{argmax}_{s} W(s)
\end{equation*}

with the Smith-Nash gap:

\begin{equation*}
SNG = W(s^W) - W(s^*)
\end{equation*}

or, where welfare is proxied through social knowledge returns:

\begin{equation*}
SNG_Y = \mathrm{SocialYield}(s^W) - \mathrm{SocialYield}(s^*)
\end{equation*}

For duration-specific cases, the operational measure is:

\begin{equation*}
OER_T = \frac{T^*_{\mathrm{strategic}}}{T^*}
\end{equation*}

For intensity-specific cases, the operational measure is:

\begin{equation*}
OER_e = \frac{e^*}{e^W}
\end{equation*}

These measures are not yet calibrated. They identify what the empirical
work must estimate: the private best-response strategy, the
social-welfare benchmark, and the gap between them. The priority
empirical question is therefore not simply whether firms enclose, but
whether observed enclosure intensity or duration exceeds the level
justified by incentive creation after accounting for suppressed
recombination, capability divergence, trajectory loss, recovery lag, and
expected knowledge loss.

The immediate calibration path is to estimate four quantities around
observable governance transitions: (1) the private gain from enclosure,
proxied by licensing revenue, platform fee changes, continuation-patent
behaviour, API pricing, or valuation effects; (2) the social-yield loss,
proxied by declines in cross-domain citation diversity, third-party
developer output, trajectory count, or entrant capability; (3) the
recovery lag after access restoration; and (4) the governance-cost
differential under alternative instruments. This converts the Nash layer
from a metaphor about strategic behaviour into a measurable
strategic-capture model.

\subsection{Testing the bridge-enclosure boundary and the
inverted-U}\index{bridge-enclosure boundary}\label{ch11:bridge-enclosure-tests}

The function-class audit of the enclosure results in the Technical Companion, Appendix F, converts the enclosure
argument from a broad claim into three falsifiable predictions, each
keyed to an observable feature of the recombination field rather than to
access loss alone.

First, \emph{position, not volume}. Enclosure that removes high-betweenness,
non-substitutable nodes from a recombination field should reduce
system-level generation disproportionately relative to enclosure that
removes redundant within-cluster nodes of equal access volume. The proxy
is the betweenness or bridging score of the enclosed knowledge in a
cross-domain citation or dependency graph; the falsifier is finding no
difference in downstream generation between high-betweenness and
low-betweenness enclosures of comparable size.

Second, \emph{the congestion reversal}. In a field independently measured as
diversity-saturated, enclosure or curation should be associated with a
\emph{rise} in measured generation, not a fall. The falsifier is a
non-congested field, measured below its diversity optimum, in which bridge
enclosure raises generation; the boundary is confirmed, not refuted, by a
saturated field in which curation improves output.

Third, \emph{the inverted-U in incumbent generation}. Because the enclosing
incumbent forfeits access to the commons it walls off, its own generation
should rise after enclosure and then decline, with the turning point set by
the depreciation interval of the enclosed feedback stream (the retraining,
churn, or obsolescence horizon). The falsifier is a monotone post-enclosure
trajectory in incumbent generation over a horizon exceeding that
depreciation interval. This prediction also discriminates between the two
inversions of \S\ref{sec:ch8:two-inversions}: a rising incumbent profit
path coupled with a flat or falling incumbent generation path is evidence
for the value-appropriation inversion rather than the generation-acceleration
special case.

\section{Testing Recursive Truth-Decay}\label{testing-recursive-truth-decay}
\index{recursive truth-decay}

This is an AI-specific test extension rather than part of the chapter's core empirical spine. It is kept here because AI-generated knowledge stock creates a distinctive reliability problem that ordinary access-shock and feedback-capture tests do not fully cover.

This section tests whether the reliability of knowledge-bearing stock can decay endogenously when generation recycles unvalidated machine output. The practical problem is to distinguish genuine productive scale from error-amplifying recursion. Evidence would appear as declining reliability with higher synthetic-input share, and as stabilization when validation or grounded feedback is added.

\paragraph{Causal role.} This section tests the endogenous reliability dynamics of \S\ref{recursive-truth-decay}: whether \(\tau\) falls as generation recycles unvalidated machine output, and whether grounding arrests it.

\textbf{Predictions.}
\begin{enumerate}
\tightlist
\item
  \emph{Synthetic-share decay.} Holding architecture, scale, and compute fixed, stock reliability, \(\tau\) proxied by post-deployment error rate, calibration score (Brier score or calibration curve), expert-override rate, or provenance-audit pass rate, should fall as the synthetic share \(1-s\) of generation inputs rises, and should fall geometrically across successive unvalidated generations at a rate governed by transmission loss \(\delta_\tau\).
\item
  \emph{Validation arrest.} Decay should be arrested where validation intensity \(v\) is high, reverting toward \(\tau_H\) as \(v\to1\). Domains with cheap ground truth, executable code, formal proof, instrumented real-world outcomes, should show little decay; domains with expensive or contested ground truth should show more.
\item
  \emph{Reliability divergence.} Actors with privileged access to grounded feedback (high \(s\), high \(v\)) should sustain higher \(\tau\) than actors relying on scraped or synthetic corpora, so a reliability gap should track feedback-access asymmetry and contribute to the capability gap of \S\ref{the-double-acting-asymmetry}.
\end{enumerate}

\textbf{Design.} Controlled recursive-training experiments, successive model generations trained on prior-generation output at varying synthetic shares \(1-s\) and validation intensities \(v\), paired with observational panels tracking deployed-model reliability against corpus-composition and validation-pipeline measures over time. The recursion \(\tau_{t+1}=A\tau_t+B\) is estimable where comparable recursive-training or panel data are available: \(\delta_\tau\), \(s\), and \(v\) may be recovered from the slope \(A\) and intercept \(B\), subject to identification and measurement limits.

\textbf{Observability status.} The falsifier below is clean, but the key variables are becoming fragile to measure. Both the reliability proxy \(\tau\) and the synthetic-input share \(1-s\) grow harder to observe as provenance collapses and synthetic and organic content blur, so the empirical claim is classified \textbf{Testable but observability-fragile} in the claim-status ledger: the design is sound while the measurement base erodes. Candidate provenance proxies, content watermarking, provenance and content-credential standards, and dataset audits, can recover \(1-s\) in principle, but their coverage is itself declining as unlabelled synthetic content proliferates. The mechanism is therefore falsifiable in design and fragile in measurement, and the text records both facts rather than only the first.

\textbf{Falsifier.} If reliability is invariant to synthetic-input share once quality filtering is held constant, recursive truth-decay is inactive in that domain and \(\tau\) may be treated as exogenous there. The mechanism claims only that \emph{unvalidated} recycling degrades reliability; curated or validated synthetic data raises \(v\) and is not a counterexample but a confirmation that the grounded-anchor flow is what carries the effect.

\section{Testing Dark Capital and T4 (Balance-Sheet Accounting
Shadow)}\label{1110-testing-dark-capital-and-t4-balance-sheet-accounting-shadow}
\index{dark capital}\index{accounting shadow}

This section tests whether dark capital and the accounting shadow add explanatory power beyond ordinary intangibles\index{intangible assets}, monopoly rents, speculation, and accounting noise. The practical problem is not to read the market-to-book gap directly as knowledge capital\index{market-to-book gap!not direct knowledge-capital measure}\index{knowledge capital!market-to-book caution}, but to decompose whether governance position, unrecognized productive stock, option value, and exposure terms improve prediction. Evidence would appear when these objects explain valuation or adverse-event surprises after standard valuation variables are included.

T4 (Balance-Sheet Accounting Shadow)\index{T4 Balance-Sheet Accounting Shadow!empirical test}: \(MV_i\) − \(BV_i\) =
\(\mathbb{E}[V(K_i^{unrec})]\) + \(\Omega_i^{risk}\) + \(\mathbb{E}[V(options_i)]\) +
\(Spec_i\) + \(Error_i\)

The core difficulty in testing T4 is the Bond-Cummins boundary: the
MV-BV gap is too noisy to cleanly attribute to any single component.
\(Spec_i\) (speculative premium) and \(Error_i\) (pricing error) are
non-eliminable from aggregate market data. The test programme must work
within this constraint.

The Bastiatian visibility claim adds a second testable implication.
Visible expenditures, transactions, and impairment events should be
recognized earlier and more reliably than dispersed knowledge-service
flows, even when the latter better predict future performance.

\textbf{Falsifier:} If accounting and market measures already capture
dispersed, delayed, relational knowledge-service flows without systematic
lag or omission, the Bastiatian visibility claim adds little.

\subsection{What can be tested within the Bond-Cummins
constraint}\label{what-can-be-tested-within-the-bond-cummins-constraint}

\textbf{T4.P1 (Productive shadow identification):} T4.2 predicts that
\(\mathbb{E}[V(K_i^{unrec})]\) includes a governance-position value
component \(\delta\cdot\mathbb{E}[v(R_i(\pi, t))]\) that is not equivalent to goodwill\index{goodwill versus knowledge-bearing stock}\index{goodwill!versus governance-position exposure}\index{goodwill!fragility-shadow contrast}\index{goodwill},
customer relationships, or IP portfolio as standardly measured. If this
is correct, firms with high COV* (strong platform position, API control,
licensing leverage) should show MV-BV residuals that correlate with
governance-position indicators after controlling for standard
intangible-asset categories.

\textbf{T4.P2 (Fragility shadow identification):} T4.3 predicts that
\(\Omega_i = \mathbb{E}[C_{T2}+C_{T7}+C_{T6}+C_{T8}]\) represents an unpriced
governance-position exposure, not necessarily an accounting liability. If firms face
governance-transition exposure, their realized stock returns following adverse
governance transitions should show larger-than-expected negative excess
returns, consistent with markets failing to price \(\Omega_i\) in advance.

\textbf{T4.P3 (Dark Capital taxonomy validity):} If the Dark Capital
taxonomy, Dark Value, Dark Risk, Foregone Knowledge Capitalization,
and Accounting Shadow, captures genuinely distinct phenomena, each
sub-dimension should predict different measurable outcomes. Dark Value
should predict productive output exceeding book-value-implied capacity.
Dark Risk should predict unexpected negative events. Foregone Knowledge
Capitalization should predict capability decline in the absence of
explicit investment. These are distinct enough that the taxonomy
provides falsifiable joint predictions.

\subsection{Proposed test: governance-form event
study}\label{proposed-test-governance-regime-event-study}

Data: Stock price data; accounting data (Compustat); patent data; API
policy announcements; open-source governance change records.

Design: Identify governance-transition events (API closure,
trade-secret assertion, open-source licensing change, platform
governance shift). Measure abnormal returns around these events for the
enclosing firm and its dependent firms. Test whether abnormal returns
correlate with K-CMM\textquotesingle s materiality score (\(M_i\))
estimated before the event. Test whether dependent-firm abnormal returns
are more negative than enclosing-firm abnormal returns, consistent with
T4.3\textquotesingle s prediction that \(\Omega_i\) is borne asymmetrically.

\textbf{Claim status:} T4\textquotesingle s formal structure is
\textbf{Internally Formalized.} T4.P1--P3 are \textbf{Testable} in
design. Bond-Cummins caution means results should be interpreted as
indicative rather than conclusive. Full decomposition of MV-BV into
T4\textquotesingle s components remains \textbf{Unresolved} without
improved accounting disclosure.

\section{Testing Chapter 10 Governance
Instruments}\label{1111-testing-chapter-10-governance-instruments}

This section treats governance proposals as empirical hypotheses rather than policy conclusions. The practical problem is to ask whether an instrument actually changes access, interoperability, capability, feedback visibility, or governance-position exposure. Evidence would appear as measurable improvement in recombination access, third-party innovation, pricing of exposure, or capability restoration after the instrument is introduced.

Chapter 10\textquotesingle s governance proposals are theoretically
justified but empirically conditional. This section specifies what
evidence would validate or undermine each major instrument class.

\subsection{First-conversion
governance}\label{first-conversion-governance}

\textbf{Prediction:} Public-funding open-access terms should, where
enforced, increase \(D_u\)(\(F_{a, t}\)) for actors who would otherwise be
excluded by first-conversion enclosure. The test is whether open-access
mandates produce measurable increases in downstream citation diversity,
follow-on product launches, and recombination events from actors who
lacked the resources to obtain licences under a private-assignment
counterfactual.

\textbf{Evidence available:} NIH public-access mandate (2008) provides a
quasi-experiment. Empirical literature on open-access mandates provides a starting point for this kind of design. The KBC-specific test would measure cross-domain citation entropy, not merely citation counts.

\textbf{Falsification condition:} If open-access mandates do not
increase \(D_u\)(\(F_{a, t}\)) for previously excluded actors, if the
downstream recombination field broadens in coverage but not in useful
diversity, the governance instrument addresses symbolic access
without addressing productive access. This would require redesign toward
interoperability and capability conditions, not merely legal access.

\subsection{Access-layer governance}\index{access-layer governance!testing}\label{access-layer-governance}

\textbf{Prediction:} Interoperability mandates and API access rules
should increase ROV* for dependent actors by restoring field access. The
test is whether regulatory access mandates (GDPR portability; EU Digital
Markets Act interoperability provisions) produce measurable increases in
third-party innovation, entry, and recombination output.

\textbf{Evidence available:} Early evidence on GDPR portability suggests
limited uptake, consistent with K-CMM\textquotesingle s prediction that
legal access without interoperability and capability conditions does not
restore productive field access. EU DMA compliance is too recent for
long-run evidence. \textbf{Empirically Plausible; Testable.}

\textbf{Falsification condition:} If third-party innovation rates do not
increase following access mandates, and if the cause is insufficient
interoperability or capability rather than insufficient access, this
confirms the K-CMM GATE prediction: access alone does not restore
\(D_u\)(\(F_{a, t}\)).

\subsection{Feedback-loop governance}\index{feedback-loop governance!testing}\label{feedback-loop-governance}

\textbf{Prediction:} Feedback-use disclosure should, if effective,
reduce the information asymmetry between enclosing incumbents and
governance actors. The test is whether disclosure reduces \(\Omega_i\), the
governance-position exposure unpriced by markets, by making the risk
visible.

\textbf{Evidence available:} No direct evidence yet. The EU AI
Act\textquotesingle s transparency requirements for high-risk AI systems
provide a future natural experiment. \textbf{Testable; currently
Unresolved.}

\textbf{Falsification condition:} If feedback-use disclosure\index{feedback-use rights} does not
change investor pricing of governance-position exposure (\(\Omega_i\) remains
unpriced), either the disclosures are not decision-relevant or the
market is not processing them. This would suggest mandatory quantitative
disclosure of LOV* accumulation, not merely qualitative description of
feedback use.

\section{Extended Empirical Project Catalogue}\label{ch11:extended-empirical-project-catalogue}

This section preserves the broader research programme after the chapter has already foregrounded the five first-pass tests. The practical problem is sequencing: not every plausible project should become an immediate empirical burden. Evidence from these projects would extend, narrow, or recalibrate the first-pass results rather than replace them.

The first register above identifies the near-term empirical centre of gravity. The extended catalogue below preserves the broader set of projects each major mechanism naturally invites.

\begin{enumerate}
\item \textbf{Access-and-governance shock design (T1--T2, Proposition C).} Use API closures, licence changes, repository access changes\index{repository access changes}, platform rule shifts, or data-access mandates as shocks. Compare exposed and less-exposed actors before and after the shock, measuring changes in recombination-field breadth, downstream project creation, citation networks, forks, dependencies, and product-line diversity.
\item \textbf{Recombination surplus project (T3)\index{T3 Recombination Surplus!empirical project}\index{recombination surplus}.} Construct matched pairs of projects, patents, open-source packages, or research teams with comparable baseline capability but different access to diverse inputs. Test whether useful-diversity gains predict new combinations after controlling for scale, funding, and talent.
\item \textbf{Accounting-shadow project (T4).} Use governance-position indicators, such as platform control, API dependency, proprietary data access, model-weight control\index{model-weight enclosure}, and licensing position, to explain market-to-book residuals beyond recognized intangibles, R\&D, advertising, and conventional monopoly indicators.
\item \textbf{Dynamic-enclosure welfare project (T5 and M5.T1).} Identify enclosure events with observable benefit and cost channels. Estimate whether incentive, quality, coordination, and maintenance gains exceed recombination loss, feedback exclusion, capability divergence, and trajectory narrowing under the Chapter 8 welfare specification.
\item \textbf{Learning-loop and feedback-capture project (T6).} Compare actors with differential access to deployment feedback. Measure whether feedback access predicts capability improvement, error reduction, model or product quality, and whether excluded actors' trajectories narrow relative to controls.
\item \textbf{Complementary-stock suppression project (T7).} Track dependencies among software packages, datasets, standards, APIs, and model components. Test whether enclosure of one central stock reduces the value, use, or development rate of complementary stocks outside the enclosure.
\item \textbf{Trajectory-count and trajectory-speed project (T8).} Measure whether enclosure produces fewer active development paths but faster improvement within the incumbent path. Candidate domains include AI tooling, cloud services, app ecosystems, autonomous systems, and open-source infrastructure.
\item \textbf{Commons-depletion and self-undermining enclosure project (D7/D8).} Use maintainer departure, fork activity, foundation governance, cloud-service capture, security incidents, and dependency-network fragility to test whether private extraction from commons stocks reduces the capability base on which private services depend.
\item \textbf{K-CMM and materiality project.} Calibrate K-CMM as an ordinal diagnostic instrument. Test inter-rater reliability, construct validity, and predictive validity against realized productivity, loss, recovery, dependency, and governance outcomes.
\item \textbf{OKU and bounded AI-substitution project.} Identify bounded task operations in labour processes and test whether AI substitution occurs at the OKU level rather than at the occupation or profession level. Measure which portions of embodied expertise become separable, deployable, and governable as disembodied stock.
\end{enumerate}

The catalogue is deliberately broader than the prioritized work sequence in \S
ef{sec:ch11-priority-work-sequence}. The goal is not to estimate this whole theory at once. It is to preserve the full menu of empirical projects while allowing the next section to choose a narrower first work sequence.

\section{Consolidated Falsification Section}\label{1112-falsification-matrix}

This section exists to state what would weaken KBC without forcing the reader through every local caveat. The practical problem is that a theory of knowledge-bearing stock must specify when it adds no value beyond existing explanations. Evidence would weaken the theory when the five core mechanisms repeatedly fail to add explanatory power over simpler models.

This section states the five core results that would weaken KBC most directly. The point is not that every local mechanism must fail for the theory to be revised. The point is that the theory becomes progressively less necessary if its central mechanisms add no explanatory power over simpler accounts of market power, ordinary intangible capital, firm size, sector effects, capital deepening, productivity lags, or accounting noise. Local tests above may still specify narrower identification risks, but the following five falsifiers are the chapter's controlling empirical boundaries.

\begingroup
\small
\setlength{\tabcolsep}{3.5pt}
\renewcommand{\arraystretch}{1.15}
\sloppy
\par\addvspace{0.8\baselineskip}\noindent
\begin{longtable}{@{}L{0.22\textwidth}L{0.38\textwidth}L{0.32\textwidth}@{}}
\caption{Five Core Falsifiers for KBC}\label{tab:ch11:falsification-matrix}\\
\toprule\noalign{}
KBC mechanism & Core falsifier & Consequence for the theory \\
\midrule\noalign{}
\endfirsthead
\toprule\noalign{}
KBC mechanism & Core falsifier & Consequence for the theory \\
\midrule\noalign{}
\endhead
\bottomrule\noalign{}
\endlastfoot
Recombination-field contraction & Enclosure events do not measurably reduce recombination access after credible controls for baseline demand, quality, funding, talent, sector trends, and ordinary legal or technical change. & Proposition C, T1/T2 field-access claims\index{T1 Increasing Returns to Knowledge!field-access claim}\index{T2 Governance Suppression!field-access claim}, and the Smithian-inversion mechanism would need to be narrowed to specific domains or discarded where access restriction does not affect usable recombination inputs. \\
Feedback-capture advantage & Feedback exclusion does not generate capability divergence: incumbents with privileged feedback do not improve faster, or excluded actors catch up at similar rates without comparable feedback access. & T6 and the feedback-capture account would lose independent explanatory force beyond scale, compute, talent, capital, initial quality, or demand. \\
Commons and public-epistemic dependence & Commons or public epistemic infrastructure shocks do not affect private productivity, security, recombination, recovery, or downstream development once ordinary operational exposure is controlled. & The claim that private productivity depends materially on \(K^C\) and \(K^P\) would be weakened; those stocks might remain normatively important but would not yet be established as measurable productivity inputs. \\
Governance-transition valuation effects & Governance-transition events are already fully priced: API closures, platform-rule changes, licence transitions, standards shifts, or IP-governance shocks produce no larger-than-expected excess returns once conventional risk and sector variables are included. & Governance-position value and \(\Omega_i\) would be less useful as separate valuation objects; the accounting-shadow claim would need to retreat toward ordinary intangible-asset or risk-premium explanations. \\
Dark-capital and accounting-shadow explanation & Market-to-book gaps and productivity anomalies are better explained without knowledge-bearing stock, governance position, or dark-capital exposure. & KBC would fail as an incremental measurement framework in those settings; residual valuation and productivity puzzles would be better handled by existing accounting, intangible-capital, human-capital, monopoly, or productivity-lag models. \\
\end{longtable}
\endgroup

\subsection{The asymmetric falsification principle}\index{asymmetric falsification principle}\label{the-asymmetric-falsification-principle}

KBC should be falsified asymmetrically. Evidence that one enclosure event creates incentives, quality control, security, disclosure, or maintenance benefits does not falsify the theory, because the theory is conditional and admits benefit-side channels. But evidence that the five mechanisms in Table~\ref{tab:ch11:falsification-matrix} repeatedly add no explanatory power over simpler models would narrow or defeat the relevant KBC claim.

Identification failure is distinct from falsification. If a case lacks a credible shock, comparison group, pre-period baseline, or observable proxy, the appropriate result is not confirmation or rejection. The claim should remain untested or empirically plausible until a credible identification strategy exists.

For Model 5 specifically, the detailed M5 falsification protocol remains a future empirical programme.\index{Model 5} Chapter 8 summarizes its five failure modes, while the Technical Companion should eventually test H1, H2, H2', H3, H3', T6/T8 channels, and the aggregate sector-level implication condition by condition.

\section{Research Programme and Priority Work Sequence}\label{sec:ch11-priority-work-sequence}

This section converts the empirical roadmap into a work sequence. The practical problem is to identify tests that are theoretically important, data-accessible, and capable of credible identification. Evidence from these priorities should either establish first-order empirical traction or force the theory to narrow before more elaborate calibration is attempted.

The following tests are the most tractable first steps. Each is
prioritized by: theoretical importance, data availability, and
identification quality. Each priority test therefore states not only the
claim and data, but also the identification strategy and the residual
identification risk. Where the design cannot isolate the proposed KBC
mechanism from confounders, the claim is downgraded from
\textbf{Testable} to \textbf{Empirically Plausible} until better data or a
stronger quasi-experiment is available.

\subsection{First-priority empirical predictions}\label{first-priority-empirical-predictions}

Before the longer test catalogue, two first-priority empirical predictions identify the most direct ways to confront KBC with observable evidence.

\textbf{Prediction 1: Governance-position shock excess loss.} Firms exposed to governance-position shocks, such as API restrictions, platform closures, licence changes, or commons governance failures, should show abnormal negative performance outcomes that exceed those predicted by switching-cost theory alone. The excess is the governance-position exposure effect.

\textbf{Identification strategy.} Use event-study designs around dated API closures, licence changes, platform access restrictions, or court/regulatory shocks, comparing exposed dependent firms with matched firms using related but unaffected inputs. The design must separately control for ordinary switching costs, supplier disruption, sector demand shocks, and market-wide technology cycles.

\textbf{Claim status.} \textbf{Testable} where a dated exogenous or quasi-exogenous governance shock and matched control group are available; otherwise \textbf{Empirically Plausible}.

\textbf{Falsifier.} If observed losses are fully explained by switching costs, ordinary supplier disruption, contractual adjustment costs, or sector shocks, then the governance-position exposure claim is weakened.

\textbf{Prediction 2: Commons-dependence impairment.} Sectors with high open-commons dependence should show larger knowledge-capability degradation after commons governance failures than sectors with comparable proprietary-stack dependence. The difference is the commons \(K^C\) impairment pathway.

\textbf{Identification strategy.} Use dependency-graph exposure measures before a maintainer shock, platform governance change, major licence change, or security incident. Compare heavily exposed downstream projects or firms with matched projects or firms that have similar functionality but lower commons dependence.

\textbf{Claim status.} \textbf{Testable} for observable dependency networks with a dated shock; \textbf{Empirically Plausible} for diffuse commons dependence where exposure and timing cannot be measured cleanly.

\textbf{Falsifier.} If commons-dependent sectors show no excess degradation after commons governance failure once ordinary operational exposure is controlled, then the \(K^C\) impairment pathway is weakened.

A third, lower-priority conversion prediction follows from the same logic: after comparable IT investment, firms with stronger \(K^E\), \(K^I\), feedback access, and recombination-field access should show larger productivity gains than firms with similar \(K^D\) expenditure but weaker conversion conditions.

\subsection{Priority 1: USPTO recombination
test}\label{priority-1-uspto-recombination-test}

\textbf{Claim tested:} Proposition C (Generative Suppression), P-C.1
through P-C.3.

\textbf{Design:} Use patent maintenance records, inter-partes review
outcomes, and continuation patent filing patterns to identify IP
enclosure events. Measure pre/post cross-domain citation entropy (\(D_u\)
proxy) for actors dependent on the enclosed knowledge. Compare to
control actors who used related but unenclosed knowledge.

\textbf{Identification strategy:} Treat patent invalidation, inter-partes
review outcomes, maintenance-fee lapses, or abrupt enforcement/licensing
changes as governance shocks. Use matched patent classes, pre-period
citation trajectories, and dependent-actor exposure to separate
recombination-field contraction from ordinary technology-cycle decline.

\textbf{Identification risk and claim status:} Citation entropy is an
imperfect proxy for useful diversity, and patent events may be endogenous
to declining asset value. The claim is \textbf{Testable} only where the
shock timing is plausibly exogenous or matched controls absorb the
decline trend; otherwise it should be read as \textbf{Empirically
Plausible}.

\textbf{Data:} USPTO patent data, PatentsView or equivalent citation data, NBER patent citation data where appropriate, and Compustat.

\textbf{Expected timeline:} 6--12 months for initial estimates.

\textbf{What it would establish:} Whether the field-contraction cascade
predicted by Proposition C is empirically observable and at what
magnitude. This is the most important first test because Proposition C
is the foundational empirical claim of the enclosure mechanism chapters.

\subsection{Priority 2: Open-source dependency and commons depletion
test}\label{priority-2-open-source-dependency-and-commons-depletion-test}

\textbf{Claim tested:} Commons depletion as enclosure without ownership;
\(K^C\) and \(K^P\) governance; commons materiality dimension of \(M_i\).

\textbf{Design:} Use Libraries.io (open-source dependency data), GitHub
activity metrics, and security incident records to test whether
maintainer concentration and contribution-rate decline predict security
incidents, release delays, and downstream fragmentation. Compare
before/after platform entry into open-source governance for candidate
cases.

\textbf{Identification strategy:} Use dated maintainer exits, abrupt
archive/deprecation decisions, licence changes, platform acquisition or
entry events, and security-disclosure shocks as treatment events. Match
projects by language, package centrality, pre-shock activity, dependency
breadth, and maintainer concentration.

\textbf{Identification risk and claim status:} Commons decline is often
endogenous: maintainers may leave because a project is already losing
relevance. The claim is \textbf{Testable} for dependency networks with
clear event timing and pre-trend controls; otherwise the commons-depletion
pathway remains \textbf{Empirically Plausible}.

\textbf{Data:} Libraries.io (public); GitHub Archive (public); CVE
database (public); Stack Overflow survey data.

\textbf{Expected timeline:} 4--8 months for descriptive analysis; 12--18
months for causal identification.

\textbf{What it would establish:} Whether \(K^C\) depletion is
measurable, predictable, and responsive to the governance interventions
proposed in §10.6. This is the most policy-actionable first test.

\subsection{Priority 3: AI feedback-capture
test}\label{priority-3-ai-feedback-capture-test}

\textbf{Claim tested:} Proposition D, T6.2, T6.P1, T6.P4.

\textbf{Design:} Use public AI benchmark data (MMLU, HumanEval, HELM) to
compare improvement trajectories of open-deployment vs.
open- and closed-deployment models before and after commercial deployment. Open deployment means external actors can generate, inspect, or contribute learning signals; closed deployment means feedback is captured primarily by the deploying firm. Test whether closed-deployment models improve faster post-deployment,
controlling for compute and initial performance.

\textbf{Identification strategy:} Use staggered deployment events,
release-policy changes, open-weight releases, API access restrictions,
or mandated transparency/disclosure changes to compare model families
with different feedback access. Control for compute, training-data scale,
lab size, initial model quality, and benchmark contamination.

\textbf{Identification risk and claim status:} Open-vs-closed deployment
is not randomly assigned and improvement rates are partly hidden. Until
feedback access, compute, and post-deployment revision data are observed,
the strong causal claim should be treated as \textbf{Empirically
Plausible}; it becomes \textbf{Testable} where deployment shocks or
policy changes create comparable treatment and control groups.

\textbf{Data:} Epoch AI model database (public); HELM benchmark results
(public); model cards (variable completeness).

\textbf{Expected timeline:} 6--12 months.

\textbf{What it would establish:} Whether the \(G^L\) mechanism produces
measurable capability divergence between open- and closed-deployment
strategies. This test has the highest profile because AI governance is a
current policy priority and the KBC framework makes predictions that are
distinct from the standard data-access framing.

\subsection{Priority 4: API closure event
study}\index{event study!API closure}\index{API closure!event study}\label{priority-4-api-closure-event-study}

\textbf{Claim tested:} Access-layer governance; \(D_u\)(\(F_{a, t}\))
mechanism; COV* extraction; \(M_i\) recombination materiality.

\textbf{Design:} Identify API policy change events (Twitter/X 2023;
Google Maps API pricing 2018; Facebook Platform changes 2018) and
measure third-party developer output (app launches, revenue, innovation
diversity) before and after the policy change. Test whether \(D_u\)(F)
proxies predict the magnitude of third-party impact.

\textbf{Identification strategy:} Use event studies and
difference-in-differences designs comparing developers or firms with high
pre-shock API dependence to matched low-dependence actors, including
unaffected platforms as controls where possible. Pre-event dependency
breadth and application categories provide exposure variation.

\textbf{Identification risk and claim status:} Platform policy changes may
coincide with demand shifts, spam reduction, privacy changes, or product
maturation. This remains one of the strongest \textbf{Testable} designs,
but only if exposed and unexposed actors are matched on pre-trends and
market category.

\textbf{Data:} App store data (Apple and Google, partially public);
Crunchbase; patent data; press release analysis.

\textbf{Expected timeline:} 6--12 months.

\textbf{What it would establish:} Whether access-layer enclosure
produces measurable recombination field contraction for dependent
actors, and whether the K-CMM materiality score predicts the magnitude
of impact. This test directly informs Chapter 10's API access
governance proposal.

\subsection{Priority 5: Capability-loss case
comparison}\label{priority-5-capability-loss-case-comparison}

\textbf{Claim tested:} T6.6 (Recovery Lag); KC component of EKL;
capability-concentration disclosure instrument.

\textbf{Design:} Identify cases of concentrated knowledge-worker
departure (acquisition followed by team dissolution; research lab
closure; key-engineer departure from software project). Measure
productivity, code quality, and product improvement rates before and
after departure. Compare recovery trajectories to divergence rates.

\textbf{Identification strategy:} Prefer quasi-exogenous shocks such as
sudden lab closures, acquisition-driven team dissolution, abrupt
maintainer departure, non-compete/legal shocks, or geographically bounded
labour-market disruptions. Compare affected products or repositories to
matched products with similar pre-shock contribution patterns and
release cadence.

\textbf{Identification risk and claim status:} Key-worker departure is
usually endogenous to project decline, acquisition strategy, or firm
reorganization. The recovery-lag claim is \textbf{Testable} only with
credible shock timing and matched controls; otherwise it should be
classified as \textbf{Empirically Plausible}.

\textbf{Data:} GitHub contribution data (open-source cases); patent data
(firm cases); software quality metrics.

\textbf{Expected timeline:} 12--24 months for longitudinal cases.

\textbf{What it would establish:} Whether T6.6\textquotesingle s
recovery lag is empirically real and at what timescale. This test
directly supports the dark-capital disclosure instrument for capability
concentration (§10.7).

\subsection{Longer-run research
priorities}\label{longer-run-research-priorities}

Following the five priority tests, the research programme should address:
empirical calibration of K-CMM\textquotesingle s coefficient structure
using priority-test data; formal proof or counterexample for
D8\textquotesingle s N-actor equilibrium\index{D8 commons-enclosure game!N-actor equilibrium}; estimation of the
recombination generation function parameters (\(\lambda_a\), η, μ) from
productivity and innovation data; and longitudinal tracking of
materiality score (\(M_i\)) against observed governance decisions to assess
whether \(M_i\) predicts which governance transitions create the harms the
theory identifies.

\section{Mathematical Status and Technical Companion Accountability Register}\label{1114-mathematical-status-and-appendix-c-accountability-register}

Chapter 11 is an empirical credibility, calibration, and falsification chapter, not a second proof appendix. The authoritative theorem-status and interpretation layer is preserved in the Technical Companion, Appendix G; the technical proof archive is preserved in the Technical Companion, Appendix H; and the full formal-object register, migrated Chapter 11 audit tables in Appendix J, calibration mechanics, and open conjectures are preserved in the Technical Companion. The Technical Companion also preserves an equation-system audit and mathematical-extension scaffold showing how the KBC variables can be developed through dynamic systems, network analysis, comparative statics, stability analysis, constrained optimization, game theory, and Bayesian value-of-information measurement. This division keeps Volume 1 focused on the empirical programme while preserving the audit machinery needed to inspect the apparatus without overloading the main argument.

\chapter[Conclusion]{Conclusion: After Knowledge Becomes Capital}\label{conclusion-after-knowledge-becomes-capital}

What knowledge-bearing capitalism asks us to recover is the half of value Smith set aside. Its wealth is use-value, generated in flow, held in many forms, and turned into capital only when governance converts it, sometimes by enclosing it into private exchange-value, sometimes by preserving it as a shared resource. The chapters behind us were the anatomy of that conversion; what follows is what it means for how we measure, govern, and distribute the wealth of a knowledge economy.

\section{What Has Been
Reconstructed}\label{c1-what-has-been-reconstructed}

The conclusion returns to the predecessor line running from Smith\index{Smith, Adam} through knowledge-economy measurement\index{measurement!conclusion}, Austrian knowledge theory, endogenous growth, and intangible-capital accounting\index{accounting recognition!conclusion} \parencite{Smith1776, Machlup1962, Bell1973, Porat1977, Hayek1945, Romer1990, HaskelWestlake2018}\index{Machlup, Fritz}\index{Bell, Daniel}\index{Porat, Marc}. The synthesis is a theory of how knowledge-bearing stock is generated, converted, governed, measured, and impaired.

The conclusion must therefore be read as an integrative claim, not a priority claim. Economics has already theorized knowledge through many domain-specific objects: information asymmetry, human capital, intangible investment, firm resources, knowledge integration, transaction-cost governance, commons, anticommons, platform economics, and endogenous growth. The contribution of KBC is to ask whether these objects can be joined into a single theory of knowledge-bearing stock and its movements. Information economics changed the information assumption. Human-capital theory changed the labour-quality assumption. Intangible-capital theory changed the investment boundary. KBC changes the capital assumption by asking what follows when productive knowledge is treated as stock-like, mobile across forms, recombinable, capability-dependent, governable, and imperfectly measured.

The predecessors supplied the fragments: non-rival ideas, information asymmetry, transaction costs\index{transaction costs}, firm capability\index{firm capability}, IT productivity puzzles, complementary organizational capital, intangible-investment measurement\index{intangible capital!measurement}, and uncertainty-reduction methods. KBC's claim is that these fragments point to a common object: knowledge-bearing stock whose productive yield depends on form, governance, capability, access, feedback, and recombination. This theory's distinctive contribution is to connect generation, conversion, governance, measurement, dark capital, and future productive capacity in one framework. The integration does not prove KBC empirically. It specifies what must now be measured, tested, and falsified.

Two predictions are the sharpest form of that demand, and they are the signature by which this theory should be judged. The \emph{dual enclosure effect}\index{dual enclosure effect} holds that a governance transition can raise the enclosing actor's present yield while lowering excluded actors' future attainable frontier through recombination-field contraction or learning-loop exclusion. \emph{Institutional residue}\index{institutional residue} holds that productive knowledge persists in routines, standards, and governance systems beyond the embodied knowledge of the individuals present, so that \(\text{OKU}^{I}\neq\sum_a\text{OKU}^{E}\). If neither survives empirical confrontation, the architecture loses the part that makes it distinctive. If they do, they corroborate the mechanisms, not yet the whole frame.

This is also why the originality claim of this book must be stated carefully. If recombination is one engine of new knowledge, then a theory of knowledge-bearing capitalism should not present itself as novel \emph{ex nihilo}. Almost no frontier theory is new in that sense. The stronger claim is recombinant: this book joins fragments that existing literatures usually hold apart, Smithian capital and specialization, information economics, human capital, endogenous growth, resource-based and knowledge-based firm theory, platform economics, commons and anticommons theory, intangible-capital measurement, and accounting-opacity problems, into a single architecture of knowledge-bearing stock, generation, conversion, governance, feedback, enclosure, measurement, and dark capital. Its originality lies chiefly in the integration, and more narrowly in mechanisms such as first conversion, cognitive enclosure, feedback capture, excluded-field suppression, dark capital, and the Smithian inversion. That is not a weakness. It is this theory applying its own account of knowledge creation to itself.

Adam Smith supplied one of capitalism\textquotesingle s classical architectures.
In \emph{The Wealth of Nations}, published in 1776, he showed how
labour, stock, specialization, the extent of the market, and the
circulation of money could combine to increase the annual produce of
nations. That architecture was not a description of one economy at one
moment. It was a theoretical account of how productive systems work, 
how self-interest could, under competitive and institutional conditions, translate
private accumulation into broadly distributed welfare, how the division
of labour could deepen without central direction, and how capital
accumulation could support further specialization indefinitely.

This book does not refute Smith. It begins from him because his
architecture remains the inherited operating system, the set of
default assumptions about wealth, capital, labour, and exchange that
subsequent economics extended, qualified, and disputed without fully
replacing. To understand what requires revision in the knowledge
economy, it is first necessary to know precisely what is being revised.
Chapter 1 identified seven pressure points where Smith\textquotesingle s
assumptions, largely invisible in his own world because they were then
background conditions, become visible tensions when knowledge-bearing
capital is the primary productive input.

What has been reconstructed here is not a refutation but an extension\index{extension versus replacement}\index{knowledge-bearing capitalism!extension, not replacement}: a
theoretical account of how productive systems work when the decisive
stock is increasingly knowledge-bearing, non-rival, cumulative,
recombinable, capability-dependent, and governed through institutional
governance arrangements rather than through the physical character of the stock itself.
That extension required new vocabulary, new models, and a revised
understanding of what capital is, how it is produced, how it moves, how
it is enclosed, how it becomes only partly visible to measurement, and how present
enclosure shapes future productive capacity.

The reconstruction covers eleven chapters and runs from the Smithian
baseline through a mechanism-centred theory of knowledge generation,
knowledge conversion, governance-form analysis, enclosure dynamics, strategic
equilibria, accounting opacity, governance design, and empirical
testability. The arc is:

\begin{quote}
Smith → knowledge-bearing stock → knowledge generation → knowledge
conversion → governance allocation → cognitive enclosure → feedback capture
→ strategic over-enclosure → dark capital → governance design →
empirical testing
\end{quote}

Each step is grounded in prior steps. The conclusion does not re-derive
them. It states what the reconstruction has established, what it has
not, and what it means for understanding wealth in a knowledge-bearing
economy.

\section{What Knowledge Changes}\label{c2-what-knowledge-changes}

Five transformations separate the economy Smith analysed from
knowledge-bearing capitalism. None of them was absent from
Smith\textquotesingle s world. All of them were marginal enough that his
theory could safely set them aside. In a knowledge-bearing economy, they
are central.

\textbf{Residence.} Smith\textquotesingle s capital was separable from
its owners: the merchant\textquotesingle s stock, the
farmer\textquotesingle s provisions, the manufacturer\textquotesingle s
tools could be inventoried, transferred, inherited, and pledged as
collateral. The most productive knowledge-bearing capital is often not
separable in this way. It resides in persons, in the tacit routines of
teams, in institutional relationships, and in organizational
architectures that often cannot be extracted unchanged from their human and
social carriers. \mbox{The distinction} between the worker\textquotesingle s
skill and the firm\textquotesingle s capital is not given by nature; it
is an institutional achievement, and the achievement is incomplete and
contested. Chapter 2 introduced the Conditional Separability Axiom to
name this: separability is not a property of knowledge, but an outcome
of the institutional conditions that govern its conversion.

\textbf{Generation.} For Smith, productive stock was accumulated through
saving and deployed in production. New productive knowledge arises
differently: through recombination of existing knowledge stocks, through
experimentation, through discovery, through deliberate invention,
through interpretive judgment, and through feedback from prior
deployment. These generation mechanisms are not analogous to saving;
they are governed by access to recombination fields, by the capability
of the generating actor, and by the institutional conditions that
determine whether experimentation is permitted, whether feedback is
accessible, and whether recombination results can be appropriated.
Chapter 3 developed these mechanisms in the Knowledge Generation Model.
The result is that the rate of knowledge generation is endogenous to the
governance form, a fact that Smith\textquotesingle s model did not formalize technological change as an endogenous governance-sensitive knowledge-generation process.

The connection to macroeconomic growth theory runs through
\citeauthor{Solow1988}\textquotesingle{}s (\citeyear{Solow1988}) residual, the portion of aggregate
output growth unexplained by capital accumulation and labour input,
which Solow described as "a measure of our ignorance." Endogenous growth
theory \parencite{Romer1990} reduced that ignorance by modelling investment in
idea production, but it did not model the governance dimension: the rate
at which the residual grows depends not only on research investment but
on the breadth of the accessible recombination field, measured here as \(D_u(F_{a, t})\), the useful diversity of accessible knowledge inputs,
the health of commons knowledge stock (\(K^C\)) that supplies non-rival
inputs to generation, and the first-conversion conditions that determine
whether new knowledge enters accessible or enclosed governance arrangements. Public epistemic infrastructure (\(K^P\)) names standards, classifications, measurement systems, legal doctrines, and other public knowledge infrastructures that make private and collective production possible. KBC
provides micro-foundations for what lives inside the residual.
\(C_{enclosure}(T)\), the social cost of enclosure specified formally under stated assumptions in T5, 
can be interpreted as a proposed measure of suppressed residual growth under the model\textquotesingle s assumptions:
productive capacity that the economy could have generated but did not,
because governance conditions narrowed the recombination field below its
socially optimal breadth. IP governance calibrated to incentive-access
tradeoffs on single knowledge goods may be partly suppressing the Solow
residual in affected domains through a mechanism that growth accounting, which treats
governance as external, often does not observe directly.

\textbf{Recombination.} Smith understood the gains from specialization
as additive: divided labour produces more than combined labour could,
and exchange allows each to consume what all have collectively produced.
Knowledge gains can be multiplicative in a different sense: the value of a
knowledge stock often depends on what other stocks it can be combined
with, and enclosing one stock from a recombination field reduces not
only the enclosed stock\textquotesingle s social availability but the
generation rate of every actor who could have combined it with their own
stocks. The recombination field, \(D_u(F_{a, t})\), the useful
diversity of accessible knowledge stocks, is the knowledge-economy
analogue of Smith\textquotesingle s market extent. But where expanding
the market extent increases productive power, contracting the recombination field can reduce productive power multiplicatively, through the suppression ratio derived conditionally in T2 and the cascades specified formally under stated assumptions in T7 and T8. That multiplicative loss is the strong-complementarity limit; where recombination inputs are substitutable the loss is bounded rather than catastrophic, as the function-class audit of Volume~2 (Appendix~F) specifies.

\textbf{Enclosure.} Smith\textquotesingle s analysis of trade
restriction was primarily a distributional critique: monopoly restricts
exchange, raises prices above competitive levels, and transfers surplus
from consumers to protected producers. The enclosure analysis in this
book is different rather than simply harsher. Enclosure of
knowledge-bearing stock can be productive when it finances costly
generation, supports disclosure, protects sensitive knowledge, or
concentrates feedback enough to accelerate improvement. The same
enclosure can also reduce who can use existing stock and how much new
stock the wider economy can generate. Proposition C (Generative
Suppression) shows that enclosure is a generative event, not merely a
distributional one, when the enclosed stock would otherwise serve as a
material input to many actors\textquotesingle{} generation processes. And
M5.T1 shows that, under stated and empirically contingent conditions,
incumbents choose enclosure beyond the socially optimal duration because
the private payoff includes dynamic revenue from feedback capture that is
not included in social welfare. It is not an empirical proof that enclosure
generally reduces welfare. It is a conditional theorem whose force depends
on whether \(M_{rec}\), the recombination premium from keeping knowledge inputs accessible for future combination, and the associated field-contraction losses are large
relative to incentive, coordination, quality-control, disclosure-protection,
and maintenance benefits. The Smithian alignment can fail under these conditions not because all
enclosure is inefficient, but because the private and social payoffs\index{private payoff versus social payoff}\index{private return versus social return!conclusion} to
enclosure can diverge through a mechanism that Smith did not model.

\textbf{Measurement.} Smith\textquotesingle s wealth was measurable
through its visible forms: the corn, the cloth, the iron, the coin.
Knowledge-bearing capital is often only partly visible to the measurement
systems that capitalism inherited from Smith\textquotesingle s world.
Trained models, organizational routines, accumulated feedback
advantages, governance positions, commons dependencies, and public
epistemic infrastructure often appear only indirectly, if at all, in financial accounts. T4 shows
that the gap between market value and book value contains an Accounting
Shadow that includes unrecognized productive capability, unpriced governance-position exposure,
and the value of governance positions that standard accounting often recognizes only indirectly. Dark Capital, the aggregate of Dark
Value, Dark Risk, Foregone Knowledge Capitalization, and the Accounting
Shadow, names this partial visibility problem. It is not only a technical measurement failure
waiting for a technical fix, but a dynamic generative-capacity problem in capitalism
when the primary productive stocks are non-rival, capability-dependent,
and institutionally governed.

\section{The Central Mechanism}\label{c3-the-central-mechanism}
\index{knowledge circulation}\index{knowledge-bearing stock}

This book\textquotesingle s central mechanism connects knowledge
generation to knowledge governance and from there to the distribution of
future productive capacity. It runs in five steps. The step that most sharply distinguishes this account from existing innovation and anticommons theory is the fourth, feedback capture: because the learning stream produced by deployment is supply-side and is not restored by interoperability or data-portability remedies, it is the mechanism through which present enclosure most directly compounds into future capability divergence.

\textbf{Generation.} New knowledge-bearing stock arises through
recombination (\(G^R\)), experimentation (\(G^X\)), discovery (\(G^D\)),
invention (\(G^N\)), interpretive judgment (\(G^J\)), and feedback from
prior deployment (\(G^L\), deployment-generated learning feedback). The rate of generation depends on the breadth
and diversity of the accessible recombination field (\(D_u(F_{a, t})\), useful diversity of accessible knowledge inputs),
the productive capability of the generating actor (\(\widetilde{C}_{a}\), the actor capability state), and the
governance form that governs access to existing stocks and the
appropriability of new ones.

This is a productive-capacity claim, not a claim that every generated recombination improves welfare. Higher \(G^R\) expands the space of possible productive trajectories, but its social value depends on what is generated, how it is used, what risks it creates, and which institutions govern its deployment. This theory therefore treats recombination as a necessary mechanism of knowledge-capital formation, not as a sufficient welfare standard.

\textbf{First conversion.} Newly generated knowledge-bearing stock
enters the first-conversion zone, where institutional conditions
determine its initial governance assignment: private disembodied capital
(\(K^D\)), institutionalized capability (\(K^I\)), commons knowledge stock (\(K^C\)),
public epistemic infrastructure (\(K^P\)), or embodied in persons (\(K^E\)). This
assignment is not given by the knowledge itself. It is determined by IP
law, employment conditions, funding terms, platform architecture,
organizational practice, and professional norms. The Conditional
Separability Axiom states the principle: separability is an
institutional achievement, not a natural property.

\textbf{Conversion and governance.} Once assigned, knowledge-bearing
stock moves through the Knowledge Conversion Matrix under transformation
({[}T{]}), access ({[}A{]}), and distributional ({[}D{]}) mechanisms.
The most consequential conversion event is enclosure: the transition
from accessible to restricted governance arrangements, which changes who can access the
stock, who can recombine it, and whose learning loops receive its
feedback. Cognitive enclosure (Chapter 6) narrows the recombination
field for excluded actors. Feedback-enclosure (Chapter 7) concentrates
deployment-generated learning in the incumbent\textquotesingle s
capability state while denying it to others.

\textbf{Dynamic capability accumulation.} The capability state equation shows how current recombination and feedback enter next period\textquotesingle s productive capability:

\[
\widetilde{C}_{a, t+1}=(1-\delta_C)\widetilde{C}_{a, t}+\gamma G^R_{a, t}+\ell G^L_{a, t}.
\]

Here \(\delta_C\) is capability depreciation, \(\gamma\) weights recombination learning, and \(\ell\) weights feedback learning. The practical claim is simple: actors with privileged access to recombination inputs and deployment feedback can accumulate capability faster than excluded actors, even when excluded actors retain formal market access. Under enclosure, the incumbent\textquotesingle s \(\widetilde{C}_{a}\) grows while excluded actors\textquotesingle{} \(\widetilde{C}_{a}\) converges toward a minimum floor under the specified trap conditions (T6.5, Capability Trap). The gap widens strictly during the enclosure period (T6.2) and does not close immediately when enclosure ends (T6.6, Recovery Lag).

\textbf{Measurement opacity and governance gap.} The productive
consequences of this mechanism, suppressed generation, concentrated
capability, narrowed recombination fields, depleted commons, are often
only partly visible to the accounting systems that capital markets, regulators,
and policymakers use to assess firm value and social welfare.
T4\textquotesingle s MV-BV decomposition identifies where these
consequences hide: in the Accounting Shadow. Dark Capital names the
aggregate effect. Chapter 10\textquotesingle s governance instruments
target the mechanism at each of its five steps. Chapter
11\textquotesingle s empirical program specifies how to test whether the
mechanism operates at the predicted magnitudes.

\section{The Smithian Inversion}\label{c4-the-smithian-inversion}
\index{Smithian inversion}

Smith\textquotesingle s capital accumulation mechanism is, in its
essential form, self-reinforcing and broadly benign: productive
investment of stock deepens the division of labour, expands market
extent, increases productivity, and generates the surplus from which
further accumulation proceeds. Each round of accumulation expands the
conditions for the next.

Knowledge-bearing capitalism under strong enclosure follows a
structurally different dynamic. The productive investment of
knowledge-bearing stock into enclosed governance arrangements can increase the productive
power of incumbent actors and may fund generation that would not
otherwise occur, but it can do so through mechanisms that simultaneously
narrow the recombination fields from which future knowledge generation
draws, reduce the diversity of knowledge trajectories available to the
broader economy, and concentrate the feedback-driven learning that
determines future productive capability.

This is not a cycle of expansion but an inversion: present productive
accumulation purchased partly at the cost of future generative capacity.
It is the Smithian inversion, the condition under which the
accumulation mechanism that Smith described works for the accumulating
actor while diminishing the conditions for accumulation system-wide. The claim is therefore not that enclosure is inherently suppressive, but that suppression occurs when recombination loss, feedback capture, and trajectory narrowing outweigh incentive, quality, disclosure, security, coordination, and maintenance benefits. Proposition~E (\S\ref{sec:ch4:proposition-e-appropriability-enabled-generation}) is the formal counterweight to this inversion: the same enclosure that can suppress generation can also enable it, through the incentive, coordination, quality-control, disclosure-protection, and maintenance channels, and the inversion is the governance case in which the suppression vector \(\mathcal{L}^{-}\) outweighs the generative vector \(\mathcal{G}^{+}\) at the margin, not a property of enclosure as such.

The inversion is not inevitable. It depends on the magnitude of \(M_{rec}\)
(the recombination premium that determines whether enclosure is merely
distributional or genuinely generative), on the scale of
feedback-capture that generates \(G^L\) advantage, on the depth of
commons dependence, and on the institutional conditions that govern
first-conversion governance assignment. Where these conditions are weak,
knowledge-bearing capitalism approaches Smith\textquotesingle s benign
cycle. Where they are strong, in frontier AI systems, pharmaceutical
genomics, core internet infrastructure, platform-mediated professional
markets, the inversion may be the operative dynamic, and
Smith\textquotesingle s mechanism is not a reliable guide to either firm
strategy or social welfare.

Frontier AI development between 2019 and 2025 appears consistent with
this inversion. The KBC claim is not that enclosure alone explains the
consolidation of frontier trajectories. Compute scale, capital
intensity, talent concentration, safety constraints, benchmark
saturation, and data scarcity also matter. The KBC-specific prediction
is narrower: after controlling for those factors, restricted access to
model weights, training corpora, interfaces, and deployment feedback
should help explain why some actors\textquotesingle{} capability
trajectories accelerate while the number of independently viable
frontier trajectories narrows.

The transformer architecture itself emerged from a comparatively open
field: foundational elements, attention mechanisms, open architectures, publications, code, corpora, and some open-weight releases were non-rival, accessible, and widely
combined. That period produced multiple distinct knowledge trajectories
simultaneously, as Chapter 3\textquotesingle s worked demonstration
shows. Each trajectory represented both a productive achievement and a
stock of knowledge that could itself become recombination input for the
next round. The conditions approximated Smith\textquotesingle s benign
dynamic: each actor\textquotesingle s generation contributed to a
commons from which further generation could draw.

The later movement toward frontier enclosure is therefore best read as a
KBC prediction to be tested, not as a settled empirical finding. Academic
groups and smaller commercial actors who had contributed distinct
approaches in 2018--2021, including masked-language modelling,
instruction following, and efficient fine-tuning, may face weaker GATE
access and permission conditions when model weights are withheld,
training corpora are enclosed, interfaces are restricted, or deployment
feedback is captured by incumbents. The mechanisms of Chapter 6 operate
concurrently with those of Chapter 7: \(D_u(F_{a, t})\) narrows for
excluded actors through field restriction (Proposition C), while
\(\widetilde{C}_{inc}\) grows through deployment-scale feedback capture
(Proposition D). The two effects compound in the direction T8 predicts:
incumbent trajectories may improve faster; the total viable trajectory
count may fall; the framework predicts that aggregate \(G^R\) may
decline, even while incumbent capability rises. The wealth being
generated can be real while the generative conditions from which later
wealth would have come are narrowed by the same investments that produce
the current round.

The pharmaceutical genomics case shows the same mechanism under
different governance arrangements. The Human Genome Project
(1990--2003), completed as a public-domain commons under a deliberate
policy choice to prevent data enclosure (the Bermuda Principles, 1996),
created a \(K^P\) base of exceptional breadth: the entire human reference
genome sequence was placed in the accessible recombination field. The
downstream generation it enabled, in drug target identification,
diagnostic development, and genomic epidemiology, drew on this
commons precisely because \(D_u(F_{a, t})\) was maximal: every
researcher\textquotesingle s field included the same foundational
stocks, and actor-specific \(K^E\) and \(K^I\) determined which
combinations were attempted, not access conditions. The parallel private
sequencing effort (Celera Genomics) demonstrated the counterfactual: had
the race been won by the proprietary pathway, the same foundational
stocks would have entered an enclosed governance arrangement, and the trajectory
diversity of the subsequent generation period would plausibly have been
lower. The Human Genome Project\textquotesingle s policy
choice was, in KBC terms, a deliberate choice to maximize
\(D_u(F_{a, t})\), to keep the highest-productive-weight field element
in the commons. A major social return was not only the genome sequence itself, but
the trajectory diversity it made possible.

The inversion is visible in these cases precisely because the generative
suppression and the productive accumulation are simultaneous.
Smith\textquotesingle s model predicts that they should both point in
the same direction, that productive accumulation expands the
conditions for further productive accumulation. The KBC mechanism
predicts that they can point in opposite directions when the stock being
accumulated is non-rival and recombinable and its enclosure reduces
\(D_u(F_{a, t})\) for a distributed actor population. Whether the
inversion is a plausible operative dynamic or the Smithian dynamic dominates depends on
\(M_{rec}\), feedback-capture and access multipliers such as \(\phi_a\)
and \(\chi_{\pi}\), and the commons-dependence structure of the field,
parameters that Chapter 11\textquotesingle s empirical program aims to
estimate. The inversion is not a critique of private investment. It is a
prediction about the governance conditions under which private
investment in knowledge-bearing capital generates broadly distributed
productive gains versus concentrated productive gains purchased partly
at the cost of future generative diversity.

\section{What the Theorem System
Contributes}\index{theorem system T1--T8!main-text contribution}\label{c5-what-the-theorem-system-contributes}

The theorem system contributes a set of conditional economic claims, not a catalogue of settled empirical facts. The results below hold under their stated assumptions. Their value in this concluding chapter is to identify what KBC adds, what each result means in ordinary economic language, and what evidence would weaken it. Table~\ref{tab:ch12_theorem_contributions} therefore translates the formal system into plain-language claims rather than repeating the proof machinery.

\begin{longtable}{L{0.20\textwidth}L{0.37\textwidth}L{0.35\textwidth}}
\caption{Plain-language contribution of the theorem system}\label{tab:ch12_theorem_contributions}\\
\toprule
\textbf{Formal result} & \textbf{Plain economic claim} & \textbf{What would test or weaken it} \\
\midrule
\endfirsthead
\toprule
\textbf{Formal result} & \textbf{Plain economic claim} & \textbf{What would test or weaken it} \\
\midrule
\endhead
\bottomrule
\endfoot
\textbf{T2 -- Governance suppression\index{T2 Governance Suppression|textbf}} & Enclosure can reduce excluded actors' recombination-generation rate by narrowing access and useful diversity. & Event studies of API closure, patent thickets, or platform access restrictions showing no loss in recombination diversity would weaken it. \\
\addlinespace
\textbf{T4 -- Accounting shadow\index{T4 Balance-Sheet Accounting Shadow|textbf}} & Market/book gaps may include unrecognized knowledge-bearing stock, option value, and unpriced governance-position exposure, not just ordinary intangibles. This is a proposed decomposition for testing, not a directly observable accounting identity. & If standard accounting, disclosures, and known intangible measures explain the residual, KBC adds little. \\
\addlinespace
\textbf{T5 -- Enclosure efficiency loss\index{T5 Enclosure Efficiency Loss|textbf}} & Under specified assumptions, welfare-optimal enclosure can be shorter than privately optimal enclosure. & Evidence that enclosure benefits dominate recombination and feedback losses over the relevant range would weaken it. \\
\addlinespace
\textbf{T6 -- Learning-loop capture\index{T6 Learning-Loop Capture|textbf}} & Actors who capture deployment feedback can accumulate capability faster than excluded actors. & No divergence after feedback access changes would weaken it. \\
\addlinespace
\textbf{T6.5/T6.6 -- Capability traps and recovery lag\index{T6.5 Capability Trap|textbf}\index{T6.6 Recovery Lag|textbf}} & Capability gaps may persist after formal access is restored. & Rapid recovery after access restoration would weaken it. \\
\addlinespace
\textbf{T7 -- Suppressed appreciation\index{T7 Suppressed Appreciation|textbf}} & Enclosure can prevent excluded actors from benefiting from later increases in the value of related knowledge stocks. & Evidence that excluded actors capture comparable later appreciation despite enclosure would weaken it. \\
\addlinespace
\textbf{T8 -- Conditional Schumpeter/Arrow separation\index{T8 Conditional Schumpeter/Arrow Separation|textbf}} & KBC gives a conditional way to separate incumbent incentive effects from replacement and suppression effects. & If incumbent and entrant trajectories do not vary with feedback or access governance, the result weakens. \\
\addlinespace
\textbf{M5.T1 -- Strategic enclosure\index{M5.T1 Strategic Enclosure|textbf}} & Rational private enclosure can exceed the social optimum when private gains exclude recombination-field losses. & If estimated recombination losses are small relative to incentive or maintenance gains, the strategic over-enclosure claim weakens. \\
\end{longtable}

This table is deliberately less technical than the proofs themselves. The proof archive establishes the conditional logic; Chapter 11 supplies the empirical programme; this conclusion states the economic memory of the system. The common pattern is that knowledge-bearing capital can create private productive gain while shifting part of the recombination, feedback, accounting, and capability cost outside the actor's own decision calculus. That is the formal content behind the Smithian inversion, but it remains a testable claim rather than a settled measurement result.

\section{What Governance Must Now
See}\label{c6-what-governance-must-now-see}
\index{knowledge impairment}

The governance framework in Chapter 10 is grounded in the theorem
system. Governance should respond only where three conditions hold: the instrument targets a specified knowledge-capital value or loss component; the affected stock or transition is material; and further measurement has positive expected decision value. In K-CMM terms, the question is what component of \(V^K\) is affected, whether the materiality score \(M_i\) exceeds the threshold \(M^*\), and whether KCI is positive. K-CMM remains an ordinal diagnostic\index{ordinal measurement}, not a validated cardinal valuation model\index{cardinal aggregation problem}.

The governance implication that follows from the theorem system cannot
be reduced to standard market-power analysis or standard accounting
reform. Four things must now be visible to governance that were not
previously in frame.

\textbf{The feedback architecture, not only the training data.} The
critical governance question for AI systems, platform businesses, and
closed professional tools is not only what data they were trained on. It
is who captures \(G^L\), the improvement signal generated by
deployment, and who is denied it. An instrument that addresses
training data while leaving feedback architecture ungoverned addresses
the initialization of the capability gap while leaving the mechanism
that deepens it in place.

\textbf{Commons knowledge capital and public epistemic capital as productive
capital, not cultural provision.} Where downstream firms, sectors, or innovation systems materially depend on commons knowledge stock (\(K^C\)) or public epistemic infrastructure (\(K^P\)), their maintenance should be treated as knowledge-capital infrastructure investment, not merely discretionary cultural provision. This does not imply that every commons warrants public support. It means that where dependency is material, depletion should be analysed as a productive-capacity risk: through rising recombination costs, falling trajectory diversity, weakened capability floors, or reduced availability of the generative commons from which first-conversion events draw.

\textbf{Capability traps as irreversible governance failures.} T6.5 and
T6.6 show that capability traps, once established, do not self-correct
when enclosure ends. The recovery lag is real and potentially long.
Ex ante governance is justified only where expected loss is large enough:
where the probability-weighted cost of irreversible capability loss,
recovery lag, systemic dependence, or lost recombination exceeds the
cost of intervention. The relevant question is not whether a bad outcome
is possible, but whether expected loss, irreversibility, and recovery time
make waiting more costly than preventive governance. The materiality
test\textquotesingle s irreversibility dimension exists to flag precisely these cases.

\textbf{Dark capital as a governance problem, not a measurement
problem.} The conclusion of Chapter 9 is the starting point for Chapter
10: dark capital is not only waiting for a better accounting standard. The
Accounting Shadow, \(\Omega_i\) as unpriced governance-position exposure, Foregone Knowledge Capitalization, and hidden
capability loss are governance problems because they are produced by
choices, about governance assignment, feedback architecture, commons
maintenance, and disclosure, that governance can influence.
Measurement can make the choices visible. Governance must respond to
what becomes visible.

\section{What Remains Unproved}\label{c7-what-remains-unproved}

A theory that does not state its own limits is not a scientific theory.
The following claims are unresolved, conditionally specified, or
empirically open as of the conclusion of this book.

\textbf{D8 is characterized, not solved\index{D8 commons-enclosure game!characterized, not solved}.} The N-actor multi-incumbent
commons-enclosure game has been formally structured, and D8.P1
(dual-channel game) and D8.P2 (decreasing optimal term with N) have been
retained as conditional structural conjectures. The full N-actor Nash
equilibrium proof remains open. Chapter 10\textquotesingle s
multi-incumbent governance analysis should therefore be read as
characterizing the structural conditions under which a problem may arise,
not as a completed theorem or as an independent policy foundation.

\textbf{\(\Phi(\cdot)\) is weakly specified and uncalibrated.} Chapter 3 and the Technical Companion, Appendix C adopt a weakest-commitment, capability-gated and useful-diversity-gated aggregator for the six canonical KGM mechanisms. The specification is deliberately modest: it requires monotonicity, capability-access complementarity, no free aggregate generation from an empty useful field, weak concavity or sub-additivity, and estimable parameters. What remains unresolved is not the existence of a composition rule, but the empirical calibration of \(\chi_{\pi}\), \(w_k\), \(\beta\), \(\nu\), and \(\rho\). If later evidence rejects the aggregate form, KGM claims should be read mechanism-by-mechanism rather than as claims about a single total generation rate.

\textbf{\(G^{DIA}\) is a named candidate mechanism with a non-canonical formal equation.} Dialectical
generation, the production of new knowledge through the productive
resolution of tension between opposed positions, is included in Chapter 3
as a named candidate mechanism because it identifies a real form of
knowledge generation. It is not yet part of the canonical theorem system.
Its formal equation remains non-canonical until the inverted-U
specification for the tension function \(\mathcal{T}_{DIA}(T, O, M, H)\) is formally resolved.

\textbf{K-CMM is an ordinal diagnostic, not validated cardinal measurement.} The
Knowledge-Capital Measurement Model\textquotesingle s coefficient
structure \(GATE=(A\times P\times I\times C)^{1/4}\); \(QUAL=(F\times D\times T\times R)^{1/4}\) is a structured ordinal scoring rubric for decision support, not an empirically
derived cardinal parameter set. The qualitative-to-quantitative conversion (0.05
through 0.90) is a practical scoring device for ranking, scenario comparison, and measurement prioritisation. Chapter 11 §11.3 specifies
the calibration program, including rater agreement and external validation against realised outcomes, that would be required before the rubric could be used as validated cardinal measurement.

\textbf{The key empirical parameters require validation.} \(M_{rec}\) (the
recombination multiplier that determines whether H1 holds for M5.T1),
\(\tau_{rec}\) (the recovery lag function), \(\Omega_i\) (the fragility shadow),
\(D_u(F_{a, t})\) (the useful diversity of the accessible recombination
field), and \(\widetilde{C}_{a}\) (the dynamic capability state) are theoretically
specified but not yet empirically measured with instruments specifically
designed to test the KBC framework\textquotesingle s predictions.
Chapter 11 identifies the five priority tests. Until those tests are
run, this theory\textquotesingle s magnitude claims, how large the
suppression ratio is in practice, how fast capability divergence grows,
how much dark capital is economically material, are empirically open.

\textbf{Internal proof-completeness is not economic validation.} This
distinction must be stated plainly. A theorem is internally valid when
it follows from its assumptions. It is economically valid when those
assumptions correctly describe the world, when the constructs can be
observed in data, when the proxies track the theoretical objects closely
enough for inference, and when the predicted magnitudes are large enough
to be economically significant rather than technically present but
practically negligible. The KBC framework is internally coherent, but it is not thereby
empirically validated. Whether it is economically valid, and whether its
mechanism predictions hold at the magnitudes and frequencies that would
make them decisive for policy and governance, is the question Chapter
11\textquotesingle s empirical program is designed to answer.

\section{The Final Claim}\label{c8-the-final-claim}
\index{knowledge capital}\index{knowledge conversion}

Smith showed how specialization and stock raise productive power. Bastiat
showed that value is realized through reciprocal services and that economic
judgement fails when it counts the visible while missing the unseen. KBC
combines these insights: knowledge-bearing stock becomes capital-like when
it can render productive services under realizable institutional,
capability, and demand conditions, yet many of its most important services,
losses, and foregone paths remain unseen because they are dispersed,
delayed, relational, and counterfactual.

Smith explained how labour, stock, specialization, market extent, and
exchange could increase the annual produce of nations. One result of this
book is that Smith's division of labour can be reconstructed more deeply:
for skilled productive labour, the division of labour is also a division of
embodied knowledge-bearing capital. Smith correctly identified the
productive-service yield, dexterity, output rate, and task-specific
improvement, but his vocabulary attributed that yield to labour rather than
to the domain-specific knowledge-bearing stock operating through labour.
The correction does not discard Smith. It recovers what his mechanism
presupposed but could not name. Smith's theory of specialization is, read
through KBC, already a theory of embodied knowledge-capital formation, one
that lacked only the vocabulary to recognize the asset it was describing.

Smith's explanation was sufficient for an economy in which the decisive
productive stock was material, separable, rival, and subject to private
accumulation through saving. It remains partly sufficient for those portions
of the modern economy that retain those features.

This book asks what changes when the decisive stock is increasingly
knowledge-bearing. Such stock may reside in persons and routines as much
as in tangible assets. It may be non-rival, so that use does not diminish
it while exclusion can reduce the generation rate of others. Its value
may depend on what it can be recombined with, so that the breadth of the
recombination field is as important as the size of the stock. It is
capability-dependent: the same knowledge stock can have radically
different value to actors with different embodied and institutional
capability. It is feedback-improving, because deployment can generate
further productive knowledge for the deploying actor. Its governance, the
institutional conditions that determine separability, accessibility,
governance assignment, and accountability, is itself an economic variable.

The answer requires not only revised empirical estimates but revised
theoretical architecture. The Knowledge Generation Model and the
Knowledge Conversion Matrix are not embellishments on a standard
production function. They are a different account of how the
knowledge-bearing part of productive capacity is created, distributed,
and lost. The five-form \(K^x\) taxonomy
is not a richer version of the tangible-intangible distinction. It is a
theory of the materially different governance, depreciation, and
generation properties of five forms of knowledge-bearing capital that
behave differently under enclosure, feedback capture, and commons
dynamics. Proposition C and M5.T1 are not merely standard
market-failure arguments. They are a theory of how individually rational strategies in
knowledge-bearing capitalism can produce a systematic inversion of the
Smithian accumulation mechanism.

In short, KBC's governance problem is to make knowledge-bearing stock produce present services, future learning, recombination, renewal, and productive capacity without allowing exclusion, fragility, degradation, false stock, or suppressed future generation to outweigh those gains.\index{knowledge-bearing capitalism!governance problem}\index{false stock}\index{suppressed future generation}

The wealth of nations in a knowledge-bearing economy cannot be
understood only by counting output, labour, physical stock, or recognized
capital. It also depends on the knowledge-bearing stocks that firms,
workers, platforms, commons, professions, and public institutions
generate, convert, preserve, enclose, and lose. The central economic
question is therefore not only how capital accumulates, but how
productive knowledge becomes capital-like, who governs its conversion,
who captures its yield, who bears its loss, and whether present enclosure
expands or narrows future wealth.

A final word on the limits of the claim. KBC is a diagnostic and explanatory framework, not a programme of redemption. It can name where productive knowledge is generated, governed, captured, and lost, and it can make visible forms of damage that conventional accounts leave unseen. It cannot, on its own, repair them. Exposing enclosure, rent capture, dark capital, and the decay of truth is not the same as curing them, and no improvement in knowledge governance should be mistaken for a remedy to the deeper moral and political causes of injustice. Whether the structures this book describes are governed toward truth, stewardship, and the productive capacity of future generations, or toward extraction and suppression, is a matter of human choice and responsibility that no theory can settle in advance. KBC aims to make that choice clearer, better measured, and harder to evade. It does not claim to make it for us.

\backmatter
\chapter*{Compact Glossary}
\markboth{Compact Glossary}{Compact Glossary}
\addcontentsline{toc}{chapter}{Compact Glossary}
This compact glossary defines the core vocabulary needed to read the monograph. The full vocabulary audit, lineage, misuse risks, and retain/revise/discard decisions are preserved in the Technical Companion.

\begin{description}[leftmargin=3.2cm,style=nextline]
\item[Knowledge-bearing capitalism (KBC)] An analytical framework for capitalism in which productive knowledge-bearing stock becomes central to wealth creation, governance, measurement, and loss.
\item[Knowledge-bearing stock] Productive knowledge carried in persons and teams, and in artefacts, organizations, commons, and public epistemic infrastructure. The term names the stock or residence of productive knowledge, not persons themselves as assets.
\item[Knowledge-bearing capital] Knowledge-bearing stock when it yields productive services under relevant capability, access, permission, interoperability, governance, demand, truth-dependence, and maintenance conditions.
\item[Capability system] The people, routines, infrastructure, permissions, trust, feedback, and complementary knowledge required to make a knowledge artefact useful. It is the context required to use, maintain, improve, and govern knowledge-bearing stock; it is not a separate asset class.
\item[Stock/capital rule] The five form names describe knowledge-bearing stock. A stock becomes capital-like only when it can yield productive services under the relevant conditions of use.
\item[Embodied knowledge stock ($K^E$)] Productive knowledge carried in people and teams through skill, judgement, trained perception, and practice.
\item[Disembodied knowledge stock ($K^D$)] Productive knowledge encoded in artefacts such as software, datasets, models, manuals, patents, or designs.
\item[Institutionalized knowledge stock ($K^I$)] Productive knowledge embedded in routines, roles, procedures, organizational structures, or governance systems.
\item[Commons knowledge stock ($K^C$)] Productive knowledge sustained through shared access, contribution, governance, and collective maintenance.
\item[Public epistemic infrastructure ($K^P$)] Foundational knowledge systems, standards, institutions, and records that support many downstream uses, for example, measurement standards, public datasets, statistical systems, legal classifications, scientific records, and verification institutions.
\item[Operative Knowledge Unit (OKU)] A bounded unit of productive knowledge defined by task, domain, performance threshold, context, and time. It is a domain-bounded measurement bridge, not a universal meter of knowledge.
\item[Knowledge Generation Model (KGM)] The model of how new productive knowledge arises through recombination, experimentation, discovery, invention, interpretation, feedback learning, and dialectical generation.
\item[Knowledge Conversion Matrix (KCM)] The diagnostic framework for what happens after productive knowledge exists: how it changes form, becomes governed, becomes accessible or inaccessible, and generates appropriability or valuation effects.
\item[Knowledge-Capital Measurement Model (K-CMM)] A decision-support and uncertainty-reduction model for estimating the value, options, governance exposure, and expected loss of a bounded knowledge-bearing stock.
\item[Knowledge Portfolio Value Function (KPVF)] A set-valued, actor-specific, governance-conditioned valuation function estimating the value of a portfolio of knowledge-bearing stocks, including current use value, recombination surplus, learning-loop value, control value, strategic option value, expected knowledge loss, and overlap correction. KPVF treats individual stocks as contributing value through marginal contribution to the portfolio rather than through standalone object price alone.
\item[Governance form] The institutional, legal, technical, contractual, organizational, or commons arrangement that determines access, exclusion, maintenance, transfer, recombination, and use of knowledge-bearing stock.
\item[Recombination field] The accessible set of knowledge inputs, tools, artefacts, skills, standards, and institutional conditions from which an actor can generate new productive knowledge.
\item[Knowledge potential] The latent productive value a knowledge-bearing stock could generate under suitable realization conditions.
\item[Knowledge impedance] The friction, blockage, capability gap, access restriction, truth failure, or governance mismatch that prevents knowledge potential from becoming productive yield.
\item[Knowledge yield] The realized productive-service flow from knowledge-bearing stock after access, capability, governance, demand, maintenance, and truth-dependence conditions are satisfied.
\item[Realization-Conditioned Value] The value of knowledge-bearing stock as expected productive-service yield under specified realization conditions, including access, permission, interoperability, complementary capability, maintenance, governance, demand, and truth-dependence. The concept distinguishes knowledge potential from realized yield and treats impedance as measurable failure or weakening of realization conditions.
\item[Asymmetric Knowledge Revaluation] The non-linear depreciation, impairment, or appreciation of knowledge-bearing stock caused by changes in realization conditions, maintenance systems, complementary capabilities, governance, truth-validity, expected knowledge loss, or recombination fields rather than by mechanical wear alone. Do not use the term to imply that all knowledge value changes are unpredictable or unmeasurable; its purpose is to identify the mechanisms through which knowledge-bearing stock changes value over time.
\item[First-conversion zone] The institutional moment at which newly generated knowledge first becomes private asset, firm capability, public infrastructure, professional standard, platform dependency, or commons.
\item[Cognitive enclosure] Access restriction that materially narrows the recombination field or future generation capacity.
\item[Feedback enclosure] Control over the feedback generated by use, such as user behaviour, telemetry, evaluation data, error reports, or reinforcement signals, when that feedback improves future models, routines, or capabilities.
\item[Dark capital]\leavevmode\\
Productive knowledge-bearing stock, capability, risk, or loss that remains unseen, unrecognized, unmeasured, mismeasured, or outside dominant recognition systems. It is not simply goodwill, brand value, or market-to-book residual; it names knowledge-bearing stock, loss, or exposure tied to productive capability.
\item[Accounting shadow] The gap between economically material knowledge-capital value, loss, dependency, or governance exposure and what accounting systems can currently recognize cleanly.
\item[Governance-position risk] Exposure created by dependence on a particular governance position, such as platform access, API permission, licence terms, commons maintenance, public standards, legal doctrine, or feedback control.
\item[Expected knowledge loss (EKL)] The expected loss term for knowledge-bearing stock, including depreciation, degradation, access loss, exfiltration, capability loss, false-stock exposure, and other downside channels.
\item[Recursive truth-decay] The deterioration of a knowledge system when future generation draws on unreliable, unvalidated, biased, synthetic, or false outputs, thereby compounding error through later knowledge production.
\item[Smithian inversion] The pattern in which knowledge-bearing capital strengthens present production partly by weakening the future capacity to generate new productive knowledge.
\end{description}

\printbibliography[heading=bibintoc,title={References}]
\clearpage
\phantomsection
\addcontentsline{toc}{chapter}{Index}
\printindex

\end{document}